\documentclass[
12pt, 
english, 
onehalfspacing, 
nolistspacing, 
liststotoc, 
toctotoc, 
parskip, 
headsepline, 
]{MastersDoctoralThesis} 

\usepackage{mathpazo} 
\usepackage{graphicx}
\usepackage{hyperref,doi}
\usepackage{subcaption}
\usepackage{float}
\usepackage{mathtools}
\usepackage{algorithm}
\usepackage[bottom]{footmisc}
\usepackage{amsmath,amssymb,amsfonts,mathrsfs}
\usepackage{bm}
\usepackage{enumitem}
\usepackage{braket}
\usepackage{dcolumn}
\usepackage{algpseudocode}
 \usepackage{isodate}
\usepackage[utf8]{inputenc} 
\usepackage[T1]{fontenc} 
\usepackage[c]{esvect}
\usepackage[normalem]{ulem}
\usepackage{manfnt,nicefrac}
\usepackage{scalerel, stackengine}
\usepackage{bbm}    
\usepackage{accents} 
\usepackage[makeroom]{cancel}
\usepackage[framemethod=tikz]{mdframed}
\usepackage[most]{tcolorbox}
\usepackage{cleveref}
\usepackage{color}
\usetikzlibrary{arrows}
\usepackage{hyphenat}
\usepackage{anyfontsize}

\usepackage[numbers, sort, compress]{natbib}

\usepackage{lipsum}

\renewcommand*{\thechapter}{\Roman{chapter}}

\newcommand{\ignore}[1]{}

\newcommand{\gu}{g^{\mu\nu}}
\newcommand{\gd}{g_{\mu\nu}}

\newcommand{\Ll}{{\mathscr L }}
\newcommand{\scrip}{{\mathscr I}^+}
\newcommand{\scri}{{\mathscr I}}
\newcommand{\scrim}{{\mathscr I}^-}
\newcommand{\well}{\widetilde{\ell}}
\newcommand{\wm}{\widetilde{m}}
\newcommand{\wmbar}{\widetilde{\Bar{m}}}

\newcommand{\lie}{\mathcal{L}}
\newcommand{\D}{\mathcal{D}}
\newcommand{\R}{\mathcal{R}}
\newcommand{\Hh}{\mathcal{H}}
\newcommand{\F}{\mathcal{F}}
\newcommand{\bF}{\mathbf{F}}
\newcommand{\bt}{\boldsymbol{\theta}}
\newcommand{\bT}{\boldsymbol{\Theta}}
\newcommand{\bw}{\boldsymbol{\omega}}
\newcommand{\Ss}{\mathcal{S}}

\newcommand{\wn}{\widetilde{n}}
\newcommand{\RNum}[1]{\uppercase\expandafter{\romannumeral #1\relax}}
\newcommand{\Msol}{\ensuremath{M_{\odot}}}
\newcommand{\dd}{\mathrm{d}}
\newcommand{\ud}[2]{^{#1}{}_{#2}}
\newcommand{\du}[2]{_{#1}{}^{#2}}

\newcommand{\udd}[3]{^{#1}{}_{#2}{}_{#3}}

\renewcommand{\Im}[1]{\operatorname{\text{Im}}\left[#1\right]}
\renewcommand{\Re}[1]{\operatorname{\text{Re}}\left[#1\right]}
\newcommand{\T}[1]{\text{#1}}

\newcommand{\hfull}{h^\text{eff}}
\newcommand{\hecho}{h^\T{echo}}
\newcommand{\hnorm}{h^\infty}

\newcommand{\Hp}{\mathscr H^+}
\newcommand{\TQBH}{\mathcal{T}^{\text{QBH}}}
\newcommand{\Teco}{\mathcal{T}^{\text{ECO}}}
\newcommand{\Reco}{\mathcal{R}^{\text{ECO}}}
\newcommand{\RBH}{\mathcal{R}^{\text{BH}}}
\newcommand{\RQBH}{\mathcal{R}^\T{QBH}}
\newcommand{\Hm}{\mathscr H^-}
\newcommand{\Zecho}{Z^{\text{echo}}_{\ell m \omega}}

\newcommand{\Zin}{Z^{\text{in}}_{\ell m \omega}}
\newcommand{\Zouteco}{Z^{\text{out ECO}}_{\ell m \omega}}

\newcommand{\Zoutqbh}{Z^{\text{out QBH}}_{\ell m \omega}}

\newcommand{\ZBH}{\mathcal{Y}^{\text{horizon}}_{\ell m \omega}}
\newcommand{\Zout}{Z^{\text{out}}_{\ell m \omega}}
\newcommand{\Din}{D^{\text{in}}_{\ell m \omega}}
\newcommand{\Dout}{D^{\text{out}}_{\ell m \omega}}
\newcommand{\Cin}{C^{\text{in}}_{\ell m \omega}}
\newcommand{\Cout}{C^{\text{out}}_{\ell m \omega}}
\newcommand{\Zinf}{Z^{\infty}_{\ell m \omega}}
\newcommand{\Yin}{\mathcal{Y}^{\text{in}}_{\ell m \omega}}
\newcommand{\Yineco}{\mathcal{Y}^{\text{in ECO}}_{\ell m \omega}}
\newcommand{\Yinqbh}{\mathcal{Y}^{\text{in QBH}}_{\ell m \omega}}
\newcommand{\Yout}{\mathcal{Y}^{\text{out}}_{\ell m \omega}}
\newcommand{\Yinf}{\mathcal{Y}^{\infty}_{\ell m \omega}}
\newcommand{\hmem}{h_\text{mem}}

\newcommand{\modelfamily}[1]{\textit{#1}} 
\newcommand{\NR}{\modelfamily{NR}}
\newcommand{\SXS}{\modelfamily{SXS}}
\newcommand{\Surr}{\modelfamily{Surrogate}}
\newcommand{\Phen}{\modelfamily{Phenom}}
\newcommand{\EOB}{\modelfamily{EOB}}
\newcommand{\TEOB}{\modelfamily{TEOB}}
\newcommand{\LALSuite}{\textit{LALSuite}}

\makeatletter
\DeclareRobustCommand\Dcirc{%
  \mathrel{
    \vphantom{\square}%
    \mathpalette\aw@Dcirc\relax
  }%
}
\newcommand{\aw@Dcirc}[2]{%
  \sbox\z@{$#1\mathcal{D}$}%
  \sbox\tw@{$#1\tiny{\text{ $\circ$}}$}%
  \dimen@=.5\dimexpr\ht\z@-\ht\tw@\relax
  \ooalign{%
    $\m@th#1\mathcal{D}$\cr
    \hidewidth\raise\dimen@\box\tw@\hidewidth\cr
  }%
}

\makeatletter
\newcommand*{\pgfunderleftarrow}{%
  \@ifstar
    {\let\ifpgf@depth\iftrue\mathpalette\@pgfunderleftarrow}
    {\let\ifpgf@depth\iffalse\mathpalette\@pgfunderleftarrow}%
}
\newcommand*{\@pgfunderleftarrow}[2]{%
  #2%
  \edef\pgf@math@fam{\the\fam}%
  \pgfpicture
    \pgfsetbaseline{0pt}
    \pgf@relevantforpicturesizefalse      
    \pgfsetroundcap                       
    \pgfsetarrowsend{left to}
    \pgfutil@tempdima=0.28pt%
    \advance\pgfutil@tempdima by.8\pgflinewidth%
    \pgfutil@tempdima-4\pgfutil@tempdima
    \sbox\pgfutil@tempboxa{$\m@th\fam\pgf@math@fam#1#2$}%
    \advance\pgfutil@tempdima-\dp\pgfutil@tempboxa
    \pgfutil@tempdimb\wd\pgfutil@tempboxa
    \pgfpathmoveto{\pgfqpoint{0pt}{\pgfutil@tempdima}}%
    \pgfpathlineto{\pgfqpoint{-\pgfutil@tempdimb}{\pgfutil@tempdima}}%
    \pgfusepath{stroke}
    \ifpgf@depth
      \pgf@relevantforpicturesizetrue
      \pgfpathmoveto{\pgfqpoint{0pt}{-\pgfutil@tempdimb}}%
      \pgfusepath{use as bounding box}%
    \fi
  \endpgfpicture
}
\makeatother

\makeatletter
\newcommand*{\pgfunderrightarrow}{%
  \@ifstar
    {\let\ifpgf@depth\iftrue\mathpalette\@pgfunderrightarrow}
    {\let\ifpgf@depth\iffalse\mathpalette\@pgfunderrightarrow}%
}
\newcommand*{\@pgfunderrightarrow}[2]{%
  #2%
  \edef\pgf@math@fam{\the\fam}%
  \pgfpicture
    \pgfsetbaseline{0pt}
    \pgf@relevantforpicturesizefalse      
    \pgfsetroundcap                       
    \pgfsetarrowsend{right to}
    \pgfutil@tempdima=0.28pt%
    \advance\pgfutil@tempdima by.8\pgflinewidth%
    \pgfutil@tempdima-4\pgfutil@tempdima
    \sbox\pgfutil@tempboxa{$\m@th\fam\pgf@math@fam#1#2$}%
    \advance\pgfutil@tempdima-\dp\pgfutil@tempboxa
    \pgfutil@tempdimb\wd\pgfutil@tempboxa
    \pgfpathmoveto{\pgfqpoint{-\pgfutil@tempdimb}{\pgfutil@tempdima}}%
    \pgfpathlineto{\pgfqpoint{0pt}{\pgfutil@tempdima}}%
    \pgfusepath{stroke}
    \ifpgf@depth
      \pgf@relevantforpicturesizetrue
      \pgfpathmoveto{\pgfqpoint{0pt}{-\pgfutil@tempdimb}}%
      \pgfusepath{use as bounding box}%
    \fi
  \endpgfpicture
}
\makeatother

\newcommand{\newpb}[1]{\pgfunderleftarrow{#1}}

\newcommand{\newpf}[1]{\pgfunderrightarrow{#1}}

\usepackage[acronym]{glossaries}
\usepackage{amsthm}
\newtheorem*{definition}{Definition}
\newtheorem*{proposition}{Proposition}
\newtheorem*{lemma}{Lemma}
\newacronym{inrep}{INREP}{Initial Noise REduction Pipeline}
\newacronym{tdi}{TDI}{Time Delay Interferometry}
\newacronym{ttl}{TTL}{Tilt-To-Length couplings}
\newacronym{dfacs}{DFACS}{Drag-Free and Attitude Control System}
\newacronym{ldc}{LDC}{LISA Data Challenge}
\newacronym{lisa}{LISA}{Laser Interferometer Space Antenna}
\newacronym{emri}{EMRI}{Extreme Mass Ratio Inspiral}
\newacronym{ifo}{IFO}{Interferometry System}
\newacronym{gws}{GWs}{Gravitational Waves}
\newacronym{gw}{GW}{Gravitational Wave}
\newacronym{bbh}{BBH}{Binary Black Hole}
\newacronym{bms}{BMS}{Bondi-Metzner-Sachs}
\newacronym{pta}{PTA}{Pulsar Timing Array}
\newacronym{gr}{GR}{General Relativity}
\newacronym{nr}{NR}{Numerical Relativity}
\newacronym{snr}{SNR}{Signal-to-Noise Ratio}
\newacronym{bh}{BH}{Black Hole}
 \newacronym{ssb}{SSB}{Solar System Barycentric}
\newacronym{flrw}{FLRW}{Friedmann-Lema\^itre-Robertson-Walker}
\newacronym{ns}{NS}{Neutron Star}
\newacronym{grs}{GRS}{Gravitational Reference Sensor}
\newacronym{tmdws}{TM-DWS}{Test-Mass Differential Wavefront Sensing}
\newacronym{ldws}{LDWS}{Long-arm Differential Wavefront Sensing}
\newacronym{cmb}{CMB}{Cosmic Microwave Background}
\newacronym{sgwb}{SGWB}{Stochastic Gravitational Waves Background}
\newacronym{ngc}{NGC}{null geodesic congruence}
\newacronym{qnm}{QNM}{Quasinormal Mode}
\newacronym{qbh}{QBH}{Quantized Black Hole}
\newacronym{cs}{CS}{Cosmic Strings}
\newacronym{eco}{ECO}{Exotic Compact Object}
\newacronym{e2e}{E2E}{End-To-End}
\newacronym{pt}{PT}{Phase Transition}
\newacronym{tcb}{TCB}{Barycentric Coordinate Time}
\newacronym{bcrs}{BCRS}{Barycentric Celestial Reference System}
\newacronym{psd}{PSD}{Power Spectral Density}
\newacronym{scird}{SciRD}{Science Requirement Document}
\newacronym{oms}{OMS}{Optical Metrology System}

\algnewcommand\algorithmicforeach{\textbf{for each}}
\algdef{S}[FOR]{ForEach}[1]{\algorithmicforeach\ #1\ \algorithmicdo}

\graphicspath{{images/}}

\usepackage[autostyle=true]{csquotes} 
\usepackage{ragged2e}
\usepackage{csvsimple,longtable,pdflscape,booktabs}

\thesistitle{Testing Gravity with Gravitational Waves} 
\supervisor{Professor Lavinia Heisenberg} 
\secondsupervisor{Professor Matteo Matturi} 
\thirdsupervisor{Third \textsc{supervisor}} 

\examiner{} 
\degree{Doctor of Natural Sciences} 
\author{David Maibach} 
\addresses{} 

\subject{Theoretical Physics} 
\university{\textcolor{mdtRed}{University of Heidelberg}} 
\department{\textcolor{mdtRed}{\large of Heidelberg University, Germany}} 
\group{\textcolor{mdtRed}{\large Combined Faculty of Mathematics, Engineering and Natural Sciences}} 
\faculty{\textcolor{mdtRed}{Faculty Name}} 

\AtBeginDocument{
\hypersetup{pdftitle=\ttitle} 
\hypersetup{pdfauthor=\authorname} 
\hypersetup{pdfkeywords=\keywordnames} 
}

\begin{document}

\setlength{\cftchapnumwidth}{2.5em}


\pagestyle{plain} 


\begin{titlepage}
\begin{center}

\textsc{\huge Dissertation}\\[0.5cm] 

\large \textit{submitted to the}\\[0.4cm]
\groupname \\ \deptname\\[0.4cm] 

\large \textit{ for the degree of \\ \degreename}\\[0.2cm]\vspace{13.7cm} 



 


\large \textcolor{mdtRed}{Put forward by}\\[0.1cm]
\large \textit{David Maibach}\\[0.1cm]
\large \textit{born in: Usingen, Germany}\\[0.1cm]
\large \textit{Oral examination: 23. July 2025}\\[0.1cm]


\vfill
\end{center}
\end{titlepage}


\begin{titlepage}
\begin{center}
\vspace{3.4cm}
\begin{spacing}{1.8}
\textsc{\huge \bfseries Across the Horizon: On Gravitational Wave Flux Laws and Tests of Gravity}


\end{spacing}
\vspace{4.4cm} 

\includegraphics[width=9cm]{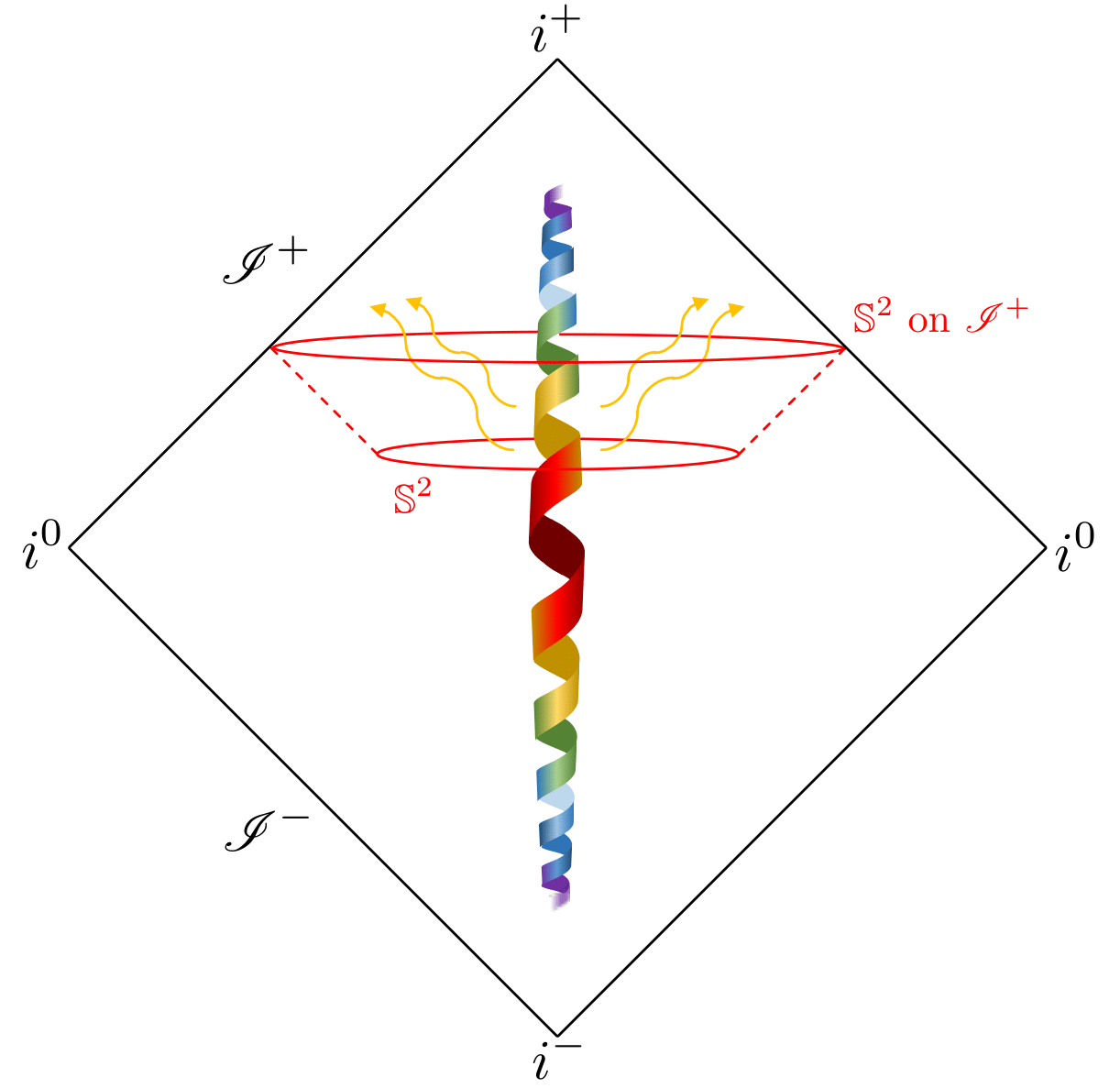}\vspace{4.7cm} 

 
\begin{center} \large
\emph{Referees:} \\[0.4cm]
\textcolor{mdtRed}{\supname} 

\textcolor{mdtRed}{\secondsupname} 

\end{center} 


\vfill
\end{center}
\end{titlepage}

    




\frontmatter
\begin{topics}
    \begin{flushleft}
        \LARGE \textsc{Abstract}
    \end{flushleft}\vspace{0.5cm}
    \begin{justify}
    Motivated by the recent first detection of a gravitational wave signal, this dissertation reviews and develops analytical, numerical, and data analysis techniques to address the remaining blind spots in the current understanding of gravity.

    Beginning with the definition of asymptotically flat spacetimes and the mathematical framework of null geodesic congruences, the derivation of the shear tensor—as the carrier of the radiative content of the gravitational field—is revisited. This is followed by an application of the covariant phase space formulation of General Relativity to derive a non-conservation law associated with the symmetry group at null infinity in asymptotically flat spacetimes. This approach is shown to generalize previous formulations of the radiative phase space in General Relativity, recovering consistent results. The physical interpretation of these flux laws is then employed to derive constraint equations, which serve as the foundation for evaluating and comparing state-of-the-art numerical waveform models. This analysis yields novel insights into commonly used models and establishes a robust algorithm for assessing future improvements in waveform modeling.

    The flux laws are subsequently applied to compute quantum corrections to the gravitational waveform strain, arising from the gravitational wave echo effect. A detailed discussion of the echo effect is presented, with focus on two leading phenomenological scenarios involving echoes from binary black hole merger events. Both the original echo signal in the strain and quantum corrections to its nonlinear structure are analyzed for their potential observability by the future space-based LISA detector. The results indicate that the echo effect lies within LISA’s sensitivity range, and that the mission could potentially probe black hole area quantization through these measurements.

    Finally, in light of recent evidence towards a detection of a stochastic gravitational wave background from Pulsar Timing Array data, the theoretical motivation for such a background is reviewed. Several scenarios contributing significant astrophysical and cosmological components of this background are examined. This comprehensive study culminates in a forecast for the detection prospects of the gravitational wave background with the LISA instrument, using a modern, ready-to-use data analysis pipeline. The findings suggest that LISA will be capable of constraining the extra-galactic stochastic gravitational wave background to levels below a dimensionless spectral energy density of at least $\Omega_\T{GW} \lesssim 10^{-8}$.

    \end{justify}
    \vspace{2cm}
    \newpage
    \begin{flushleft}
        \LARGE \textsc{Zusammenfassung}
    \end{flushleft}\vspace{0.5cm}
    \begin{justify}
    Motiviert durch die erste Messung eines Gravitationswellensignals, untersucht und entwickelt diese Dissertation analytische, numerische sowie datenanalytische Methoden mit dem Ziel, bestehende Mysterien im derzeitigen Verständnis der Gravitation zu adressieren.

    Ausgehend von der Definition asymptotisch flacher Raumzeiten und dem mathematischen Rahmenwerk gegeben durch nullartiger Geodätenkongruenzen wird die Herleitung des Shear-Tensors – als Träger der strahlenden Freiheitsgrade des Gravitationsfeldes – aufgefrischt. Darauf folgt die Anwendung der kovarianten Phasenraum-Methode auf die Allgemeinen Relativitätstheorie, die zur Herleitung eines Fluss-Erhaltungsgesetzes eingesetzt wird, das mit der Symmetriegruppe in nullartiger Unendlichkeit in asymptotisch flachen Raumzeiten verknüpft ist. Dieser Methode generalisiert frühere Formulierungen des Phasenraums der strahlenden Freiheitsgerade in der Allgemeinen Relativitätstheorie und liefert konsistente Resultate. Die physikalische Interpretation der erhaltenen Flussgesetze wird anschließend genutzt, um einschränkende Bedingungen herzuleiten, die als Grundlage der Evalu-ierung und des Vergleiches modernster numerischer Wellenformmodelle dienen. Diese Analyse liefert neue Einsichten in verbreitet verwendete Modelle und etabliert einen robusten Algorithmus zur Bewertung zukünftiger Verbesserungen in der Wellenmodellierung.
    
    Die besagten Flussgesetze werden in der Folge angewendet, um Quantenkorrekturen der Gravitationswelle zu berechnen, die aus dem sogenannten Echoeffekt resultieren. Eine detaillierte Diskussion dieses Effekts wird präsentiert, mit Fokus auf zwei phänomenologische Szenarien, in denen Echosignale aus Verschmelzungen binärer Schwarzer Löcher entstehen. Sowohl das ursprüngliche Echosignal als auch Quantenkorrekturen des nichtlinearen Anteils der Gravitationalwelle werden im Hinblick auf ihre potenzielle Messbarkeit mit dem zukünftigen LISA-Detektor analysiert. Die Ergebnisse zeigen, dass der Echoeffekt innerhalb der Sensitivität des Instrumentes liegt und die Mission möglicherweise die Quantisierung der Oberfläche Schwarzer Löcher durch diese Messungen untersuchen könnte.
    
    Abschließend wird im Licht jüngster Hinweise auf eine Detektion eines stochastischen Gravitationswellenhintergrunds in Daten des Pulsar-Timing-Arrays die theoretische Motivation für ein solches Hintergrundsignal beleuchtet. Verschiedene Szenarien, die signifikante astrophysikalische und kosmologische Beiträge zu diesem Hintergrund leisten könnten, werden untersucht. Diese umfassende Studie mündet in einer Prognose der Messbarkeit des Gravitationswellenhintergrunds mit dem LISA-Instrument unter Verwendung einer modernen Datenanalyse-Pipeline. Die Ergebnisse deuten darauf hin, dass LISA in der Lage sein wird, den extragalaktischen stochastischen Gravitationswellenhintergrund auf ein Niveau unterhalb einer dimensionslosen spektralen Energiedichte von mindestens 
    $\Omega_\T{GW} \lesssim 10^{-8}$
    zu be-schränken.
    \end{justify}
\end{topics}




\begin{acknowledgements}
\addchaptertocentry{\acknowledgementname}\vspace{0.5cm} 
First of all, I wish to express my inmost gratitude to my advisor, Prof. Lavinia Heisenberg, for her pertinent support, academically and personally. Her kindness, exceptional mentoring qualities and motivated attitude have made my Ph.D. a successful and extremely enjoyable journey. I am forever grateful to have had the opportunity to get to know and work together with her.

Moreover, I am especially thankful to my second supervisor, Dr. Fabio D'Ambrosio, from whom I learned so much and was fortunate to work together for a short but memorable time. I would also like to extend my gratitude to Prof. Matthias Bartelmann, who has been a well of wisdom and support my Ph.D., as well as Prof. Matteo Maturi, who has volunteered to the burden of evaluating this thesis. I further wish to thank my previous supervisor Prof. Robert Brandenberger for supporting my career, countless productive discussions and connecting me to other motivated researchers. My sincere gratitude also goes out to Prof. Nicol\`o Defenu from whom I learned to thrive in an academic environment and who I consider a good friend as well as an esteemed collaborator.

I am deeply thankful to have collaborated with an international team of researchers at University of Heidelberg, many of whom I also consider as friends by now. Prime examples of such are Dr. Henri Inchausp\'e, Dr. Do\u{g}a Veske, and Dr. Francesco Gozzini, with whom I had a great time personally and academically, Shaun Fell and Jann Zosso, with whom I enjoyed the best conferences, and of course my collegues Carlos Pastor Marcos, Nadine Nussbaumer, Ricardo Waibel, Marvin Sipp, Hannes Heisler. All of you have made my Ph.D. a wonderful time of my life and I am unimaginably happy to have had the honors to meet you, let alone calling you friends.

I would also like to thank all colleagues at the ITP and at the STRUCTURES Office for fostering my career. I am especially grateful for the support of the YRC who has enabled me to advertise my research at international conference. 

Naturally, I would be remiss if I did not mention my friends and family who supported me throughout the years of my doctorate. My gratitude belongs in particular to my family whose unending support has made my studies in higher
education possible. Without their love and support, I would not have been here today, let alone being able to achieve all that I have in the past 3 years. Additionally, I am proud to say that I have been equally supported by my friends at home, who have been there for me since the beginning of time, expressed indestructible support and have enriched my life by priceless moments of joy. I am truly thankful to have such wonderful human beings as friends, my second family. 

Though this thesis has only one author, it is the joint efforts of many that has made it possible and I am thankful for each individual person who has participated through either a thought-provoking discussion, a supportive gesture, or simply a good talk about life. Nonetheless, I wish to dedicate my final acknowledgment to myself for not giving up in face adversity and for being ever-curious. And so, I may finish the acknowledgments with the words of a great man facing heavy weight: ``Light weight, baby!''.


\end{acknowledgements}

{
\hypersetup{linkcolor=black}

    \newpage
    \tableofcontents 
    \newpage
    
    \phantomsection
    \addcontentsline{toc}{chapter}{List of Figures}
    \listoffigures 
    
    \phantomsection
    \addcontentsline{toc}{chapter}{List of Tables}
    \listoftables 
    }



\begin{abbreviations}{ll} 

\textbf{BBH} & \textbf{B}inary \textbf{B}lack \textbf{H}ole\\
\textbf{BCRS} & \textbf{B}arycentric \textbf{C}eleestial \textbf{R}eference \textbf{S}ystem \\
\textbf{BH} & \textbf{B}lack \textbf{H}ole\\
\textbf{BHP} & \textbf{B}lack \textbf{H}ole \textbf{P}erturbation\\
\textbf{BMS} & \textbf{B}ondi-\textbf{M}etzner-\textbf{S}achs\\
\textbf{CCE} & \textbf{C}auchy-\textbf{C}haracteristic \textbf{E}xpansion\\
\textbf{CMB} & \textbf{C}osmic-\textbf{M}icrowave \textbf{B}ackground\\
\textbf{CS} & \textbf{C}osmic \textbf{S}tring\\
\textbf{CS} & \textbf{C}osmic \textbf{S}trings\\
\textbf{E2E} & \textbf{E}nd-\textbf{2}-\textbf{E}nd\\
\textbf{ECO} & \textbf{E}xotic \textbf{C}ompact \textbf{O}bject \\
\textbf{EMRI} & \textbf{E}xtreme \textbf{M}ass \textbf{R}atio \textbf{I}nspiral \\
\textbf{ET} & \textbf{E}stein \textbf{T}elescope\\
\textbf{FLRW} & \textbf{F}riedmann-\textbf{L}ema\^itre-\textbf{R}obertson-\textbf{W}alker\\
\textbf{GW} & \textbf{G}ravitaional \textbf{W}ave \\
\textbf{LIGO} & \textbf{L}aser \textbf{I}nterferometer \textbf{G}ravitational-Wave \textbf{O}bservatory \\
\textbf{LISA} & \textbf{L}aser \textbf{I}nterferometer \textbf{S}pace \textbf{A}ntenna \\
\textbf{NGC} & \textbf{N}ull \textbf{G}eodesic \textbf{C}ongruence\\
\textbf{NR} & \textbf{N}umerical \textbf{R}elativity\\
\textbf{NS} & \textbf{N}eutron \textbf{S}tar\\
\textbf{NPS} & \textbf{N}ewman \textbf{P}enrose \textbf{S}calars \\
\textbf{PBH} & \textbf{P}rimordial \textbf{B}lack \textbf{H}ole \\
\textbf{PN} & \textbf{P}ost \textbf{N}ewtonian \\
\textbf{PT} & \textbf{P}hase \textbf{T}ransition \\
\textbf{PTA} & \textbf{P}ulsar \textbf{T}iming \textbf{A}rray \\
\textbf{QBH} & \textbf{Q}uantized \textbf{B}lack \textbf{H}ole \\
\textbf{QCD} & \textbf{Q}uantum \textbf{C}hromo \textbf{D}ynamics\\
\textbf{QNM} & \textbf{Q}uasinormal-\textbf{M}ode\\
\textbf{SciRD} & \textbf{S}cience \textbf{R}equirement \textbf{D}ocument \\
\textbf{SGWB} & \textbf{S}tochastic \textbf{G}ravitaional \textbf{W}ave \textbf{B}ackground \\
\textbf{SNR} & \textbf{S}ignal to \textbf{N}oise \textbf{R}atio\\
\textbf{SSB} & \textbf{S}olar \textbf{S}ystem \textbf{B}arycentric \\
\textbf{SVT} & \textbf{S}calar-\textbf{V}ector-\textbf{T}ensor\\
\textbf{TCB} & \textbf{B}arycentric \textbf{C}oordinate \textbf{T}ime \\
\textbf{TDI} & \textbf{T}ime \textbf{D}elay \textbf{I}nterfereometry\\
\textbf{TT} & \textbf{T}ransverse-\textbf{T}raceless

\end{abbreviations}






\begin{symbols}{lll} 


$\scrip$ & Future null infinity \\
$\scrim$ & Past null infinity \\
$\scri$ & All of null infinity \\
$i^+$ & (Future) Timelike infinity \\
$i^-$ & (Past) Timelike infinity \\
$i^\circ$ & Spacelike infinity\\
$\Hp$ & Future black hole horizon \\
$\Hm$ & Past black hole horizon \\
$\newpb{w}$ & Pullback of a form $w$\\
$\newpf{v}$ & Pushforward of a vector $v$\\
$\mathcal L _{v}$ & Lie derivative along a vector field $v$\\
$\boldsymbol{\omega}$ & Differential $n$-form \\\\

$g_{\mu\nu}$ & Spacetime metric tensor \\
$\eta_{\mu\nu}$ & Minkowski metric (flat spacetime) \\
$\nabla_\mu$ & Covariant derivative operator\\
$\D_\mu$ & Covariant derivative operatpr at $\scrip$ \\
$\sigma_{\mu\nu}$ & Shear tensor \\
$S_{\mu\nu}$ & Schouten tensor \\
$R^\mu{}_{\nu\rho\sigma}$ & Riemann curvature tensor \\
$R_{\mu\nu}$ & Ricci tensor \\
$R$ & Ricci scalar \\
$G_{\mu\nu}$ & Einstein tensor \\
$\Gamma^\mu_{\nu\rho}$ & Christoffel symbols \\
$\Box$ & D'Alembert operator: $\Box = g^{\mu\nu} \nabla_\mu \nabla_\nu$ \\\\

$h_{\mu\nu}$ & Metric perturbation \\
$\bar{h}_{\mu\nu}$ & Trace-reversed metric perturbation \\
$h_{ij}^{\mathrm{TT}}$ & Transverse-traceless metric perturbation \\
$\Psi_i$ & Weyl scalar (Newman–Penrose formalism) \\\\

$T_{\mu\nu}$ & Energy–momentum tensor \\
$\phi$ & Scalar field \\
$\mathcal{L}$ & Lagrangian density \\
$S$ & Action functional \\\\

$a(t)$ & Scale factor \\
$H = \dot{a}/a$ & Hubble parameter \\
$\Omega_{\mathrm{GW}}(f)$ & Gravitational wave energy spectrum \\
$\rho_c$ & Critical energy density of the universe \\\\

$G$ & Newton's gravitational constant \\
$c$ & Speed of light \\
$\hbar$ & Reduced Planck constant \\
$M_\mathrm{Pl}$ & Planck mass \\
$\ell_\mathrm{Pl}$ & Planck length \\
$\Msol$ & Solar mass \\
$D_L$ & Luminosity distance \\

\end{symbols}

\addcontentsline{toc}{chapter}{Notation}
\chapter*{Notation}

\textbf{Spacetime manifold:} The spacetime manifold $(\mathcal M, g)$ is $4$-dimensional, oriented and differentiable. It is a topological manifold with a tangent and cotangent space on which vectors and forms are defined, respectively. 

\textbf{Metric signature:} For the Minkowski background, the metric signature $\eta_{\mu\nu}\equiv\text{diag}(-1,+1,$ $+1,+1)$ is adopted. 

\textbf{Spacetime vectors:} Spacetime vectors are denoted by Greek indices $\alpha, \beta, ... = 0,1,2,3$, while spatial indices are denoted by Latin letters $i,j,.. = 1,2,3$. A vector $v=v^\mu\partial_\mu$ is denoted by its components $v^\mu$ and the base vectors $\partial_\mu$ of the tangent space on $\mathcal M$, $\mathcal T_x \mathcal{M}$, where $x\in \mathcal{M}$. A differential form on $\mathcal{M}$ is denoted by $\omega = \omega_\mu \dd x^\mu$ where $\omega_\mu$ are the components and $\dd x^\mu$ denote the basis of the cotangent space $\mathcal T^*_x \mathcal M$. Spatial vectors $x^i$ are denoted by $\mathbf{x}$. At future null infinity, that is a $3$-dimensional surface, one maintains the notation of intrinsic quantities with Greek indices (in contrast to other literature). 

\textbf{Units:} Unless stated otherwise, we set $c=G=\hbar = 1$. 

\textbf{Definition of Gravitational Waves:} Unless stated otherwise, this work defines \textit{Gravitational Waves} in the sense of linearized gravity. That is, in the absence of matter, the spacetime metric $\gd$ can be decomposed into small perturbations over a fixed Minkowski background,
\begin{align}\label{equ:fundamental}
    \gd(x^\mu) = \eta_{\mu\nu} + h_{\mu\nu}(x^\mu)\,, && |h_{\mu\nu}|\ll1\,.  
\end{align}
Restricting the coordinate choice to instances where the latter is true, a general diffeomorphism on $x^\mu$ transforms the metric perturbation $h_{\mu\nu}$ as $h'_{\mu\nu}(x'^\mu) = h_{\mu\nu}(x^\mu) - \partial_\mu\xi_\nu-\partial_\nu\xi_\mu$, where $\xi_\mu$ is an arbitrary infinitesimal vector field on the spacetime manifold $\mathcal M$. The validity of equation \eqref{equ:fundamental} requires $|\partial_\mu\xi_\nu|\lesssim|h_{\mu\nu}|$. In particular, it follows that for Lorentz transformations $\Lambda\ud{\mu}{\nu}$ it must hold that $|\Lambda\du{\mu}{\alpha}\Lambda\du{\nu}{\beta}h_{\alpha\beta}(x^\mu)|\ll 1$, restricting the selection of boosts to such compliant with the setup \eqref{equ:fundamental}. In the linearized regime, the affine connection is defined as 
\begin{align}
    \Gamma\ud{\alpha}{\mu\nu}= \frac{1}{2}\left(\partial_\nu h\ud{\alpha}{\mu} + \partial_\mu h\ud{\alpha}{\nu} - \partial^\alpha h_{\mu\nu}\right)\,,
\end{align}
while the Riemann tensor, Ricci tensor, and Ricci scalar in this definition read 
\begin{align}
    R\ud{\alpha}{\mu\nu\beta}&=\partial_\nu \Gamma\ud{\alpha}{\mu\beta}-\partial_\beta \Gamma\ud{\alpha}{\mu\nu}= \frac{1}{2}\left(\partial_\mu\partial_\nu h\ud{\alpha}{\beta} + \partial_\beta\partial^\alpha h_{\mu\nu} - \partial_\nu\partial^\alpha h_{\mu\beta} - \partial_\beta \partial_\mu h\ud{\alpha}{\nu}\right)\\
    R_{\mu\nu}&\equiv -R\ud{\alpha}{\mu\nu\alpha}= \frac{1}{2}\left(\partial_\nu\partial^\alpha h_{\alpha\mu} + \partial_\beta\partial^\alpha h_{\mu\nu} - \partial_\nu\partial^\alpha h_{\mu\beta}-\partial_\beta\partial_\mu h\ud{\alpha}{\nu}\right)\\
    R &= R\ud{\mu}{\mu} =(\partial^\alpha\partial^\beta h_{\alpha\beta}-\Box h) \,.
\end{align}
Here, $h\equiv h\ud{\mu}{\mu}$, $h\ud{\alpha}{\nu}=\eta^{\mu\alpha}h_{\mu\nu}$, $\partial^\alpha \equiv \eta^{\alpha\mu}\partial_\mu$ and $\Box \equiv\partial_\mu\partial^\mu$. The latter linearized tensors and scalars enable the computation of the linearized Einstein equation in vacuum, 
\begin{align}
    G_{\mu\nu}&\equiv R_{\mu\nu} - \frac{1}{2}\eta_{\mu\nu}R = \frac{1}{2}\left(\partial_\nu\partial_\alpha h\ud{\alpha}{\mu} + \partial_\mu\partial^\alpha h_{\nu\alpha} - \partial_\mu\partial_\nu h -\Box h_{\mu\nu} -\eta_{\mu\nu}\partial^\alpha\partial_\beta h\du{\alpha}{\beta} + \eta_{\mu\nu} \Box h \right)\notag\\
    &=0\,.
\end{align}
To remove the gauge freedom in the definition of $h_{\mu\nu}$, a suitable gauge condition can be chosen. Adapting the most frequently used gauge choice in literature, i.e., the Lorentz gauge $\partial^\mu\bar h_{\mu\nu}=0$, with $\bar h_{\mu\nu}= h_{\mu\nu}-\frac{1}{2}\eta_{\mu\nu}h$ being the trace-reversed metric perturbation, the linearized Einstein equation simplifies to 
\begin{align}\label{equ:fundamental_2}
    G_{\mu\nu} = \Box \bar h_{\mu\nu}=0\,.
\end{align}
Equation \eqref{equ:fundamental_2} establishes a wave equation for metric perturbations. Exhausting the residual gauge freedom to describe only real physical degrees of freedom in terms the wave equation \eqref{equ:fundamental_2}, the Lorentz gauge is further refined into the transverse-traceless gauge
\begin{align}\label{equ:TT_Gauge}
    h_{\mu0}=0\, &&h=h\ud{i}{i}=0\,,&& \partial_i h_{ij}=0\,,
\end{align}
which is inherently trace-reversed and allows only two propagating degrees of freedom while completely saturating the gauge freedom. Throughout this work, the term ``gravitational waves'' refers to the two propagating polarizations obtained from the solution of equation \eqref{equ:fundamental_2} in convolution with the gauge \eqref{equ:TT_Gauge}.

\textbf{Pullback:} The pullback is defined through a differentiable map $\varphi$ from manifold $\mathcal N$ to $\mathcal M$. It maps between co-tangent spaces, $\varphi^*:T^*_{\varphi(p)}\mathcal M\rightarrow T^*_p \mathcal N$, where for any point $p\in\mathcal{N}$ it acts on objects in the co-tangent space of $\mathcal M$ at $\varphi(p)$ and returning an object in the co-tangent space of $\mathcal N$ at $p$. Thereby, $\mathcal N$ can be a submanifold of $\mathcal M$. The pullback is denoted by $\newpb{\,\cdot\,}$. The pullback of a covector $\omega_\mu$ on $\mathcal M$, $\newpb{\omega_\mu}$ defines a covector on $\mathcal N$ with all free indices restricted to $\mathcal{N}$. That is, the index of $\newpb{\omega_\mu}$ has to be contracted with an index of a tangent vector intrinsic to $\mathcal N$. Fully contracted quantities are not subject to the pullback.

\textbf{Pushforward:} The pushforward is, analogously to the pullback, defined via the differentiable map $\varphi$ between two manifolds manifold $\mathcal N$ to $\mathcal M$. It maps between tangent spaces, $\varphi_*: T_p\mathcal M \rightarrow T_{\varphi(p)}\mathcal N$. The pushforward is denoted by $\newpf{\, \cdot\, }$.

\textbf{Equality at $\scrip$:} An equality at $\scrip$ is indicated by both parts of the equation being evaluated at $\scrip$, denoted as $\cdot |_{\scrip}$. Unless a tensor or operator is explicitly defined on $\scrip$, equality at $\scrip$ requires its evaluation at this boundary. 

\textbf{Limit to $\scrip$:} The limit at $\scrip$ is defined as the limit from the bulk in null-like direction towards future null infinity. In coordinates $(u,r,\theta,\phi)$ the limit is denoted as $\lim_{r\rightarrow \infty} \cdot$.

\textbf{Differential forms:} An arbitrary $k$-form on an $n$-dimensional Manifold is denoted by a bold letter, e.g., $\mathbf{L}$, and defined as $\mathbf{L} = \sum_{1\leq \mu_1 ...\mu_k\leq n} L_{\mu_1...\mu_k} \dd x^{\mu_1}\wedge ... \wedge \dd x^{\mu_k}$. For the Lagrangian density $L$ for instance, the corresponding $n$-form Lagrangian density is given by $\mathbf{L}= \epsilon L$, where $\epsilon$ is the Levi-Civita symbol. A $k$-form $\boldsymbol \omega$ is \textit{closed} if $\dd \boldsymbol{\omega} =0$ where $\dd$ is the exterior derivative. A $k$-form is \textit{exact} if there exists a $(k-1)$-form $\boldsymbol{\nu}$ such that $\boldsymbol{\omega}= \dd \boldsymbol{\nu}$.




\mainmatter 

\pagestyle{thesis} 


\parindent=1cm

\chapter{Introduction} 
\label{chap:Intro}


\section{A brief History of Gravitational Waves}
\label{sec:intro_hist}
\begin{figure}[t!]
    \centering
    \includegraphics[width=0.6\textwidth]{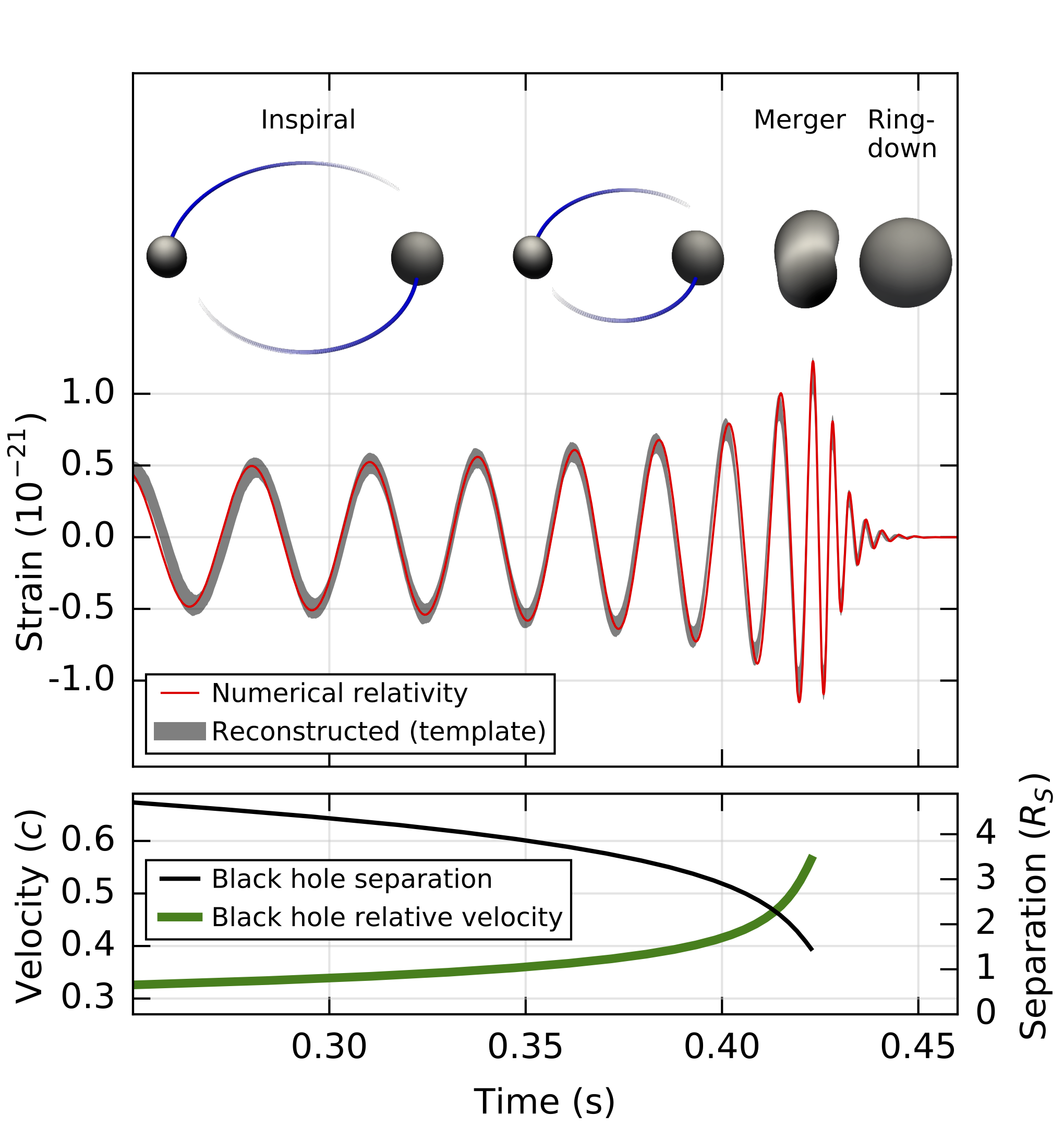}
    \caption{Illustration of the first GW detection, denoted by event number GW150914, \cite{First_LIGO_detection}. The upper sketch shows the three phases of the BBH merger. The middle plot illustrates the gravitational waveform. The lower panel highlights the high speeds and small distances of the depicted process.} 
    \label{fig:GW_waveform}
\end{figure}

Without a doubt, the recent detection of the first \gls{gw} by the LIGO collaboration \cite{First_LIGO_detection, ALIGO_2015} (see Fig. \ref{fig:GW_waveform}) marks one of the most profound breakthroughs in modern physics, and the beginning of a new era for astronomy. Shortly before 10 am UTC, on September 14th, 2015, mankind's first observation of ripples in spacetime traversing the known Universe at the speed of light took place. The received signal, so evidence hints, originated from a \gls{bbh} system that merged roughly 1.4 billion years ago with the heavier of the two orbiting black holes (BHs) weighing in at $62$ $\Msol$\footnote{The estimated weight itself was another sensation as, up to this point, the stellar-mass \gls{bh} known to researchers reached only around $30$ $\Msol$ \cite{narayan2014observationalevidenceblackholes}.}. The initial measurement was followed by an manifold of detections by the LIGO and VIRGO \cite{AVIRGO_2014} collaboration, such that, until today, a total of 90 merging events have been recorded and analyzed over multiple observing runs \cite{LIGO_GWTC_1, LIGO_GWTC_2,LIGO_GWTC_3}. Astonishingly, the latter events include not only \gls{bbh} systems but also binaries consisting of one \gls{ns} and one BH \cite{LIGO_NS_BH} as well as \gls{ns}-\gls{ns} combinations \cite{LIGO_Binary_NS} orbiting each other until their gravitational attraction causes a collision, forming a remnant \gls{bh} in the process. It goes without saying that the knowledge extracted from these detections has been invaluable to \gls{bh} as well as \gls{ns} physics. For \gls{bh}s, insights gained from \gls{gw} data concern their mass distribution (e.g., \cite{BH_mass_1,BH_mass_2, BH_mass_3, BH_mass_4}), formation channels (e.g., \cite{BH_formation_1, BH_formation_2}), as well as other properties of inspiraling \gls{bh}s (e.g., \cite{BH_astrophysics_1,BH_astrophysics_2, BH_astrophysics_3}). Observing merging \gls{ns} binaries as well as \gls{bh}-\gls{ns} pairs, on the other hand, has led to a vast improvement in understanding the nature of these objects, in particular, regarding their general composition, tidal deformability, and all-determining equation of state (e.g., \cite{NS_1, NS_2, NS_3, NS_4}). At the current stage, \gls{gw} observations mainly benefit astrophysical models as well as gravity theories. However, as the collected data and its precision constantly increases, \gls{gw}s have the potential to uncover further mysteries of extremely compact objects in the Universe, reaching as far as the \gls{bh} horizon's quantum properties or the particle physics theory behind the hot and dense cores of \gls{ns}s. Thus, \gls{gw}s can have a far wider-reaching impact in physics, beyond the astronomy and cosmology community, potentially providing new attack angles on complex problems of quantum gravity, solid state theory, or particle physics. 
 
Yet, there is even more information extractable from gravitational radiation besides the individual, resolved \gls{gw} resulting from the merger of compact objects. Just as LIGO found the first evidence of a \gls{gw} in their data in 2015, 8 years later the Nanograv collaboration \cite{PTA_nanograv} as part of the \gls{pta} consortium \cite{PTA_CPTA,PTA_ipta,PTA_europe,PTA_parkes,PTA_meerkat,PTA_nanograv}, after 15 years of data collection and analysis, reported first evidence for a \gls{sgwb} \cite{Nanograv_SGWB_I, Nanograv_SGWB_II} based on the observed correlation of pulsars across the sky \cite{Principles_PTA}. Opposed to the clearly resolved mergers measured by LIGO, the \gls{sgwb} contains all unresolved sources of gravitational radiation, including both astrophysical and cosmological origins, that are either too faint or simply do not match inspiral waveforms in shape. Collectively, they create a stochastic signal \cite{SGWB_1,SGWB_2} rich in phenomenology. Therefore, the \gls{sgwb} can serve as compelling evidence for new physics on cosmological scales, offering a distinct avenue of exploration beyond high-energy physics at the TeV scale and probing the early Universe prior to the \gls{cmb}.

Given the, by now, overwhelming amount of data serving as evidence for the existence of GWs, it is hard to believe that they remained a wildly debated theoretical concept for decades, eluding direct observation due to their incredibly subtle effects on the fabric of spacetime. Even before Einstein published his theory of \gls{gr} \cite{Einstein} in 1915, the phrase \textit{gravitational wave} was first mentioned in Henri Poincar\'e's attempt to formulate a theory of gravity transmitted through waves (\textit{onde gravifique}) \cite{Poincare} in 1905. Ten years after Poincar\'e's work, Einstein established a completely new viewpoint of how gravity might be connected to the construct of spacetime itself. His equations match the curvature of local spacetime on the one side, with the local energy and momentum within spacetime on the other side. Only one year after publishing his theory, Einstein conjectured, just as Poincar\'e, the existence of \gls{gw}s similar to electromagnetic waves. Despite his own works on linearizing the equations of \gls{gr} in the presence of an external, time-dependent source \cite{Einstein_GW} pointing in the right direction, the conceptual differences between electromagnetism and his theory of gravity made him uncertain about his view. To Karl Schwarzschild, he wrote in 1916:

\textit{``Since then [November 14] I have handled Newton's case differently, of course, according to the final
theory [the theory of General Relativity]. Thus, there are no gravitational waves analogous to light waves. This
probably is also related to the one-sidedness of the sign of the scalar T, incidentally [this implies the
nonexistence of a ``gravitational dipole''].''}\cite{Einstein_Schwarzschild}

Einstein intensified his efforts to manipulate his equations such that their form resembles those of Maxwell's equations for electromagnetic waves in the months after publishing his field equations. Despite his efforts, he had to make several approximations to move forward in his efforts that were open to criticism from other researchers as well as Einstein himself. With a particular choice of coordinate system, Einstein was able to find three types of \gls{gw}s, later named longitudinal-longitudinal, transverse-longitudinal, and transverse-transverse waves by Hermann Weyl. Despite these preliminary results, the puzzle around the existence of \gls{gw}s was often avoided by Einstein and other scientists in the field due to the widespread skepticism regarding the choice of coordinates. In 1922, Arthur Eddington brought the topic back to the table by demonstrating that two out of three types of waves Einstein found were in fact coordinate artifacts. According to Eddington's calculations, these artificial waves could travel at speeds which are coordinate-dependent and arise solely due to the choice of particular ``wavy'' coordinates. His findings spread distrust regarding the authenticity of the remaining transverse-transverse waves, whose speed is constant at the speed of light for all coordinate choices. The skepticism Eddingtons paper triggered led Einstein to investigate the issue of coordinate choice together with his student Nathan Rosen in 1933. Their collaboration culminated in a paper titled ``Are there any gravitational waves?'' for which no original version exists today. Based on a letter Einstein wrote to Max Born, however, it is very likely that Einstein and Rosen argued against their existence. To Born, Einstein wrote in 1936:

\textit{``Together with a young
collaborator [Rosen], I arrive at the interesting result that gravitational waves do not exist, though they
have been assumed a certainty to the first approximation. This shows that the non-linear general relativistic field equations can tell us more or, rather, limit us more than we have believed up to now.''} \cite{Einstein_Born}

As it is well established by now, fortunately, Einstein and Rosen had made a mistake in their calculation, which was first pointed out by the reviewer of the manuscript Einstein and Rosen drafted, Howard Robertson. Due to Robertson's review, the Editor of the Journal \textit{Physical Review}, John T. Tate, requested Einstein to respond to the comments. This infuriated Einstein, who sent a letter to Tate criticizing him for showing the unpublished manuscript to another expert in the field. As Rosen departed for the Soviet Union in 1936, Einstein hired a new assistant named Leopold Infeld, who befriended Robertson at Princeton, where Einstein had worked since 1933. Robertson and Infeld discussed Robertson's criticism of the Einstein-Rosen paper. Infeld subsequently informed Einstein about the discussion shortly before Einstein's scheduled talk at Princeton titled ``Nonexistence of gravitational waves''. The discussion with Infeld led Einstein to conclude his presentation with, \textit{``If you ask me whether there are gravitational waves or
not, I must answer that I don’t know. But it is a highly interesting problem''} \cite{Einsteins_talk}. The Einstein's paper with Rosen was submitted to another journal, which accepted the version including the mistakes pointed out by Robertson. Einstein later corrected the paper, including fundamental changes to their proofs and renaming it ``On gravitational waves''. While Einstein's initial skepticism regarding the existence of \gls{gw}s was cured by discussions with Robertson and his collaboration with Infeld, Rosen continued to believe that they resulted from mathematical artifacts without a physical foundation. The subject was heavily debated within the community for decades, and the confusion about the physicality of gravitational radiation remained until the 1960s. A consensus was reached when Hermann Bondi, Rainer Sachs, and others published a series of paper \cite{Bondi_Origin_1, Bondi_Origin_2, Sachs_Origin_1, Sachs_Origin_2, Sachs_Origin_3}, in which they demonstrated that \gls{gw}s are in fact physical and a direct consequence of Einstein's theory of \gls{gr}. Their main advantage was the usage of frame fields and expansions in the null direction. Thereby, they rigorously showed that \gls{gw}s radiate energy in null direction, towards future null infinity, $\scrip$, which implies that they are indeed physical. Although from a theoretical point of view, Bondi and Sachs applied flawless mathematics, hesitation in the community partially remained. Simultaneously with the analytical efforts, physicists tried to develop experiments that could, once and for all, settle the discussion. It took another $20$ years until the instruments were advanced enough such that Joseph Taylor and his collaborators were able to provide the first indirect confirmation of the existence of \gls{gw}s. Together with Russell Hulse, Taylor detected the pulsar \textsc{PSR J1915+1606} forming a binary system with another \gls{ns}. They measured the orbital decay rate of the period of the binary system and found it matched precisely the prediction derived from \gls{gr} when emission of gravitational radiation is assumed to be one of the major carriers of energy away from the pulsar \cite{First_indirect_detection}. The ultimate confirmation of the previous analytical and experimental results was, naturally, provided by LIGO's direct detection in 2015, settling the $100$ year-long discussion once and for all.  

Although the dispute around GWs predicted by Einstein's theory of \gls{gr} is resolved by the abundance of analytical and experimental evidence, \gls{gr} is still somewhat peculiar compared to other established theories in physics, such as Maxwell's theory of electrodynamics, for instance. To blame for this outsider role is the non-linear nature of Einstein's field equation, which makes them incredibly hard to solve while, at the same time, gives rise to most interesting phenomena, such as the \textit{gravitational wave memory effect}. For the few exact solutions that do exist, most prominently the Schwarzschild and Kerr solutions \cite{Schwarzschild_metric, Kerr_metric}\footnote{Note that almost 50 year passed between the two exact solutions presented by Schwarzschild and Kerr, demonstrating the immense complexity of finding analytical solutions to Einstein's equation.}, spacetime contains singularities inside the \gls{bh}s. For these singular points, energy density and pressure become infinite, indicating a failure of Einstein's theory. In principle, such singularities should be addressed and resolved by quantum physics. Yet, every attempt to incorporate the relevant ``quantumness'' into \gls{gr} leads to deviations from Einstein's theory. This motivated investigations of other (modified) theories of gravity. In fact, shortly after Einstein's initial paper \cite{Einstein}, Eddington tried to modify his theory, searching for alternative theories already in 1923. Additional incentives to consider Einstein's \gls{gr} to be somewhat incomplete come from the cosmological observation that spacetime is expanding in an accelerated manner. The latter was first established through the observation of accelerated expansion-caused dimming of Type Ia supernovae luminosities \cite{Dark_Energy} in 1998. The driver of this expansion was named \textit{Dark Energy}, indicating that its nature is completely unknown. Dark Energy is commonly incorporated into Einstein's equations as a cosmological constant. This particular way of fitting cosmological observations into \gls{gr} came with multiple conceptual problems and tension \cite{Tensions}, seeking resolution through building new models and questioning fundamental assumptions such as the underlying theory of gravity. For instance, while so far there have been no definite proofs, it is possible that the cosmological constant is in fact the wrong approach to marry the observed acceleration with \gls{gr} and, instead, the accelerated expansion constitutes an observable deviation of gravity from Einstein's theory of \gls{gr}. The resolution of this and many more mysteries of gravity and its prediction is the prime objective of ongoing and future \gls{gw} and multi-messenger research initiatives. 

The effort of decoding \gls{gr}'s intricacies and solving its equations for more and more complex scenarios over the past 70 years has fostered the development of remarkable mathematical tools enabling the computability of complex spacetime configurations. Instances of such include the \gls{bms} framework developed around asymptotically Minkowskian space times \cite{Bondi_Origin_2,Sachs_Origin_2}, the Newman Penrose formalism \cite{Newman_Penrose}, Weinberg's Soft Theorems \cite{Weinberg_ST}, and Twistor theory \cite{Twistor_Theory}. With the steadily increasing availability of reliable \gls{gw} measurement data, these tools enable stringent tests of the underlying theory whilst facilitating the precision of \gls{gw} models relevant for the data analysis.
Demonstrating the utility of selected analytical tools in the context of future \gls{gw} precision measurements, primarily focused on the asymptotically flat \gls{bms} formalism valid in full, non-linear \gls{gr}, is one of the major focal points of this dissertation. Thereby, an emphasis is put on the so-called \textit{balance flux laws} derived in the BMS framework acting as constraint equations for gravitational waveforms of simulated BBH mergers (applied in Chapter \ref{chap:asymptotics} and \ref{chap:quantum}). An in-depth exploration of the \gls{bms} formalism and the flux laws, including a derivation, working assumptions, and key properties, is subject of chapter \ref{chap:asymptotics}. To fully understand the motivation leading to the set of analytical tools used in this thesis, as well as the domain of applicability, a closer look at \gls{gr}s fundamental properties, first and foremost, its nonlinearity, is crucial. The latter, as it will be demonstrated below, gives rise to the \gls{gw} memory effect, which is tightly related to the energy flux of gravitational radiation.



\section{Why Gravity is resentful}
\label{sec:intro_memory}
Einstein's theory of gravity relies on promoting the metric tensor $g_{\mu\nu}$, describing the geometry of spacetime, to a dynamical variable characterizing the gravitational field. It is governed by the field equations, which are the equations of motion derived from the action principle and the Einstein-Hilbert action 
\begin{align}
    S = \frac{1}{16 \pi} \int \dd^4x \sqrt{-g}(R[g]-2\Lambda)\,,
\end{align}
where the determinant of the spacetime metric tensor $g_{\mu\nu}$ is denoted as $g$ and the Ricci scalar $R$ is computed w.r.t. this spacetime metric. Throughout this thesis, the convention $\hbar = c = G = 1$ is applied and one sets $\Lambda=0$ \footnote{Interestingly, the presence of $\Lambda$ leads to enormous complications regarding the methodology of the BMS framework and $\Lambda=0$ embodies a necessary assumption. Where it is relevant, issues are highlighted and appropriate literature is referenced.}, unless otherwise stated. With the spacetime metric $g_{\mu\nu}$ acting as the dynamical filed, the variational principle yields
\begin{align}
\label{equ:einstein_field_eq}
    G_{\mu\nu}= R_{\mu\nu}-\frac{1}{2}R g_{\mu\nu} = 8\pi T_{\mu\nu}\,.
\end{align}
The left-hand side of this equation contains at highest second derivatives of the metric, i.e., terms such as $g \partial \partial g$ and $ \partial g\partial g$. There also exist higher derivative theories including terms with more than two derivatives and leading to similar field equations. These theories are tightly constrained by physical viability and often subject to instabilities in the form of negative energy states. Most prominently, they can be plagued by the Ostrogradsky instability \cite{Ostrogradsky_I, Ostrogradsky_II}. More importantly, unlike many other theories, all terms on the left-hand side of Eq. \eqref{equ:einstein_field_eq} are at least second order in $g_{\mu\nu}$. Therefore, working with Einstein's equations, one is facing a non-linear system of differential equations that is incredibly hard to solve unless further symmetry assumptions are to enter the arena. In fact, due to their complexity, a new branch of physics, called \gls{nr}, aiming for numerical solutions of this non-linear equation, emerged. \\
Another way to untie the complex mesh of equations is through a perturbative expansion. Motivated by the fact that earth is solely subject to the weak field regime of gravity due to the large separation with respect to sources of strong gravitational fields, the spacetime metric can be expanded around the flat Minkowski metric $\eta_{\mu\nu}$, that is $g_{\mu\nu} \approx \eta_{\mu\nu} + h_{\mu\nu}$. Here, the metric perturbations $h_{\mu\nu}$ are of small amplitude, i.e., $|h_{\mu\nu}|= \mathcal{O}(\alpha)$ and $\alpha\ll 1$. Einstein's field equation can be expanded correspondingly, such that the left-hand side of the linearized field equations reads
\begin{align}
\label{equ:lin_Einstein}
    \frac{1}{2}(\partial_\sigma\partial_\mu h^\sigma_\nu + \partial_\sigma\partial_\nu h^\sigma_\mu - \partial_\mu\partial_\nu h - \square h_{\mu\nu} - \eta_{\mu\nu}\partial_\rho\partial_\lambda h^{\rho\lambda} + \eta_{\mu\nu}\square h)\,.
\end{align}
The (flat space) d'Alembert operator is denoted as $\square = \eta^{\mu\nu} \partial_\mu \partial_\nu$ and the diffeomorphism invariance of \gls{gr} is translated to a gauge transformation that leaves Eq. \eqref{equ:lin_Einstein} invariant, namely $h_{\mu\nu}  \rightarrow h_{\mu\nu} + \partial_\mu\xi_\nu + \partial_\nu\xi_\mu$. The gauge can be used to eliminate unphysical degrees of freedom from the metric perturbations $h_{\mu\nu}$, leaving only 6 physical ones. One can easily show that 2 are scalar modes, 2 are incorporated in a transverse vector, and the final two are encoded as a transverse-traceless tensor modes. Of particular interest here is the transverse-traceless (TT) part of the field equation, which determines the time evolution of the tensor modes,
\begin{align}
\label{equ:wave_eq}
    \square h^{TT}_{\mu\nu} = -16\pi T^{TT}_{\mu\nu}\,.
\end{align}
It is the only part of the Einstein equations involving tensor modes and, coincidentally, also the only part involving time derivatives. Thus, the transverse-traceless part of the metric perturbation $h_{\mu\nu}$ includes the only truly dynamical and gauge-invariant degrees of freedom. The latter corresponds to the two polarizations of the \gls{gw}. The notion of ``wave'' already implies that Eq. \eqref{equ:wave_eq} resembles the wave equation in vacuum. The same equation can be derived straightforwardly for the trace-reversed part $\overline{h}_{\mu\nu}= h_{\mu\nu} - \frac{1}{2}\eta_{\mu\nu} h$ in combination with the Lorentz gauge $\partial^\mu\overline{h}_{\mu\nu}=0$. Interestingly, while the tensor modes satisfy the wave equation in the absence of any matter or stress, i.e., $T_{\mu\nu}=0$, the other components of the metric perturbations can be set to zero in vacuum by an appropriate gauge choice. 

The linearized theory was long believed to be satisfactory for observers far away from the source. Indeed, the linearized Einstein equations well describe \gls{gw}s in vacuum and the presence of matter. It took until 1990 to change this perception\footnote{A first indication of relevant contribution at second order was made in \cite{First_hint_Mem}.}: It was shown that there is a non-negligible contribution at second order in $h_{\mu\nu}$ which leads to a permanent displacement of test masses in a laser interferometer after it has been traversed by gravitational radiation \cite{Christodoulou_Mem, Memory_paper_I}. The latter overruled the assumption that a strictly linearized treatment of \gls{gr} suffices for gravitational radiation at a detectable level. This displacement becoming apparent at second order in the metric perturbation is, thus, a direct manifestation of the nonlinearity of Einstein's equations and now known as the \textit{non-linear gravitational wave memory effect}, or \textit{null memory}. There exists also another type of memory effect which is generally subdominant compared to the non-linear memory \cite{Christodoulou_Mem, Ordinary_Mem}. It is called the \textit{linear} or \textit{ordinary memory} and associated with unbound compact objects of a system and the merger's remnant kick velocity, for instance a binary on a hyperbolic orbit \cite{Lin_Mem_0}. In its linear form, the memory was already discovered 20 years prior, in the 1970s \cite{First_hint_Mem,Lin_Mem_I,Lin_Mem_II}. It is a product of the linearized theory and appears as a result of a changing second (time) derivative of the source's quadrupole moment. Perhaps more intuitively, this overall change in the second time derivative is equivalent to an overall change in the linear momenta of the compact objects involved in the merger. Until the discovery of the non-linear memory, it was commonly believed that due to the negligible linear momentum radiated away by typical \gls{gw} sources, the (linear) memory would not be observable. This point of view, however, changed with the introduction of its non-linear counterpart, proven to be largely dominant throughout merging events \cite{Christodoulou_Mem}. Both types of memory have been subject to broad investigations approaching the subject from various angles, among others \cite{Memory_paper_I,Memory_paper_II,Memory_paper_III,Memory_paper_IV,Memory_paper_V,Memory_paper_VI,Memory_paper_VII}. One of the most frequently used approaches to a concrete computation of the memory based on a \gls{gw} strain relies on the BMS framework. It associates the memory with symmetry transformations, resulting in flux conservation laws that relate non-linear memory to energy flux sent away from the spacetime bulk to null infinity \cite{Bondi_Origin_2,Sachs_Origin_2,Mem_BMS_I,Mem_BMS_II}. The approach has been used in the context of modified gravity theories as well, where BMS (flux) laws have been derived for \textit{Brans-Dicke} \cite{Brans_Dicke_I,Brans_Dicke_II,Brans_Dicke_III} and \textit{Chern-Simions theory} \cite{Chern_Simons_I,Chern_Simons_II}. 

\begin{figure}[t!]
    \centering
    \includegraphics[width=0.85\textwidth]{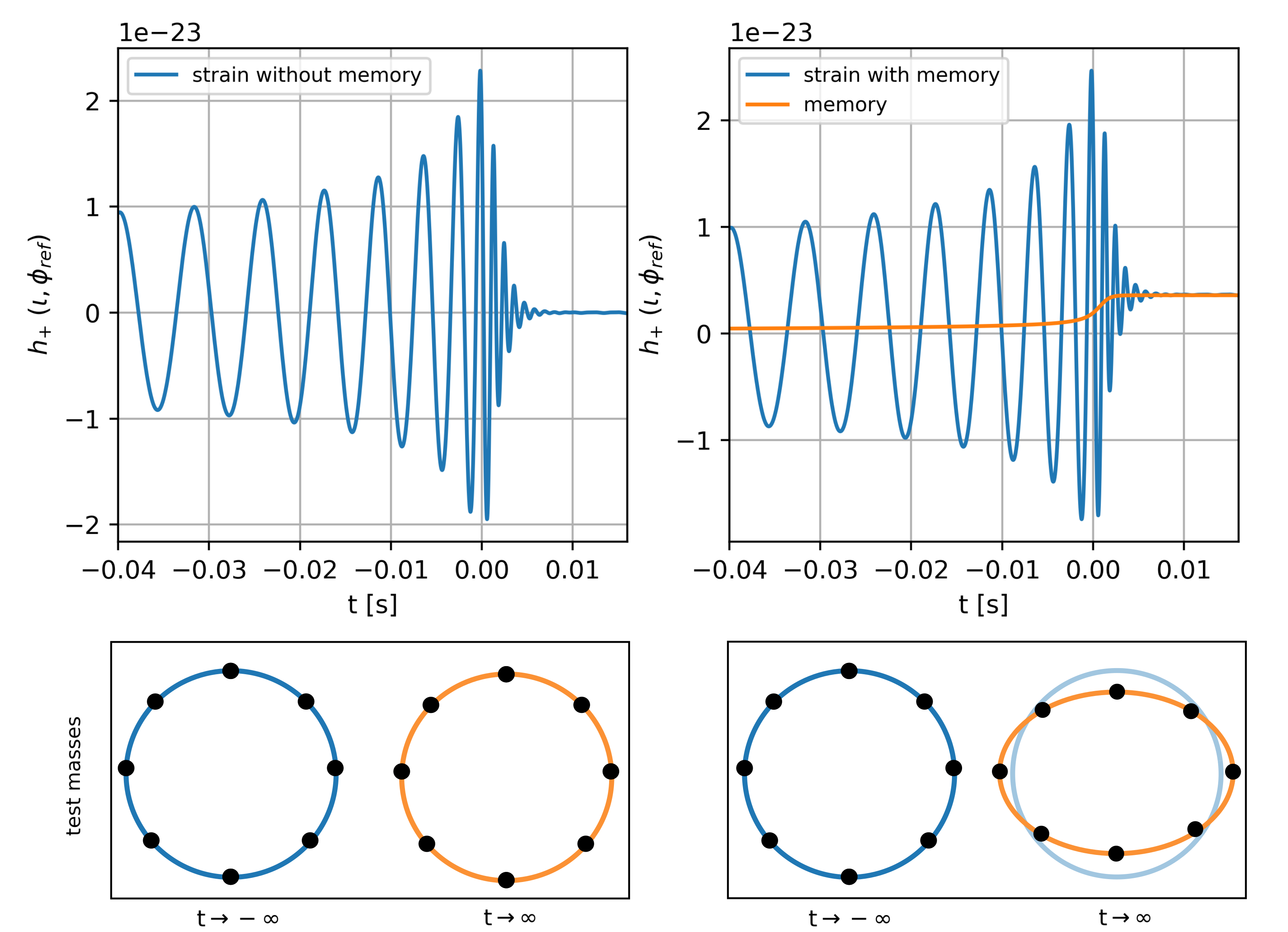}
    \caption{Simulated gravitational waveform with (right) and without (left) memory. The top plot depicts the strain time series. The bottom plot sketches the deformation of test masses before and after the GW has passed an ideal detector.} 
    \label{fig:mem_sketch}
\end{figure}

Generally, each source of \gls{gw}s will carry some form of \gls{gw} memory, which is easiest expressed as at least one of the polarizations remaining permanently altered due to the passage of gravitational radiation, or
\begin{align}
    \label{equ:memory_definition_easy}
    \Delta h^\text{mem}_{+,\times} = \lim_{t\rightarrow +\infty} h_{+,\times}(t) - \lim_{t\rightarrow -\infty} h_{+,\times}(t)\,.
\end{align}
The two polarizations $+,\times$ thereby account for the two dynamical degrees of freedom of the transverse-traceless metric perturbation $h^{TT}_{\mu\nu}$. For the $h_+$, the memory is sketched in Figure \ref{fig:mem_sketch}: When a \gls{gw} hits the detector consisting of freely falling test masses, it induces oscillatory deformations from which a \gls{gw} signal can be reconstructed. After the wave has passed through, the test masses settle into the initial state. This process can be readily computed in a fully linearized theory as it was outlined above, finding only a negligible permanent displacement due to the linear memory. Adding the second order in $h_{\mu\nu}$, one finds that after the oscillations decay off, a significant permanent deformation remains. 
The \gls{gw} that has passed through the idealized detector has left a (permanent) ``memory'' of the gravitational radiation. Therefore, it is fair to say that gravity is indeed resentful.\\
So far, this phenomena has not been confirmed experimentally as the current ground-based detectors respond to \gls{gw}s with timescales much shorter than that of typical memory signals. They also lack the capability of accumulating or storing memory as their test masses are not truly free. Instead, their internal hardware is designed to force the test masses back into their equilibrium shape. Due to its truly freely-floating test masses, the \gls{lisa} space-based instrument \cite{LISA} could maintain a permanent displacement, resulting in promising prospects of detecting the memory of supermassive binaries \cite{Henris_Mem}. 

For the exact analytical expressions of the memory contributions, there exist many versions. In Sections \ref{sec:Gravitational_radiation_at_Scri} to \ref{sec:Covariant_PS} of this thesis, the memory and its derivation in the BMS framework are presented in more detail. In this paragraph, to illustrate the quadratic strain dependence of the non-linear memory, the memory definitions from \cite{Memory_paper_V} are adapted featuring direct solutions of wave Eq. \eqref{equ:wave_eq} in terms of the trace-reversed strain $\overline{h}_{\mu\nu}$ in Lorentz gauge as well as the TT part. \\
Before starting with the non-linear memory, consider first the asymptotic nature of the memory: It was mentioned before that this contribution is sourced by the energy flux of the \gls{gw}s passing a \gls{gw} detector. Naturally, the detector is located far away from the original source, so the energy flux and thus the memory will decay to some degree on their way to the detector. To account for this decay, the following notation is adapted (throughout the rest of this work): At $\scrip$, the asymptotic strain polarizations will be defined as
\begin{align}
\label{equ:asymptotic_strain}
    h_+^\circ = \lim_{r\rightarrow \infty} rh_+, && h_\times^\circ = \lim_{r\rightarrow \infty} rh_\times\,.
\end{align}
The proportionality in $r$ thereby follows form the asymptotic properties of \gls{gr} and is rigrously derived in Chapter \ref{chap:asymptotics}. The rule \eqref{equ:asymptotic_strain} can be applied when asymptotic strains need to be converted to finite distances. In fact, many derivations of the memory and, in particular, the mentioned balance flux laws in the BMS framework heavily rely on asymptotic quantities (as will be thoroughly explained in Section \ref{sec:Gravitational_radiation_at_Scri}). However, the actual detectors are not infinitely far away from the source but only a large finite distance. Hence, the rescaling is predominantly applied when dealing with actual detection data. For instance, the \gls{gw} energy per unit time and angle is given by
\begin{align}
\label{equ:GW_energy_def}
    \frac{\dd E^\text{GW}}{\dd t \dd \Omega} = \frac{1}{16 \pi} \braket{(\dot{h}^\circ_+)^2 +(\dot{h}^\circ_\times)^2} \equiv \frac{r^2}{16 \pi}\sum_{\ell,m}\sum_{\ell',m'}\braket{\dot{h}_{\ell m}\dot{\overline{h}}_{\ell' m'} } {}_{-2}Y_{\ell m}{}_{-2}Y_{\ell' m'}\,,
\end{align}
where the right-hand side is scaled to a finite distance $r$, which, for a real measurement, would be replaced with the estimated luminosity distance $D_L$. The average $\braket{.}$ can be understood as being computed over a spacetime region much larger than the wavelength of the \gls{gw}s involved. The harmonic strain modes in the latter were expanded in terms of spin-weighted spherical harmonics. The strain is of spin weight $-2$, which uniquely fixes its behavior under $U(1)$ gauge transformations. For details on this categorization, see \cite{Our_Review} or Section \ref{subsec:shear_and_GW}. \\
With definitions \eqref{equ:asymptotic_strain} and \eqref{equ:GW_energy_def} at hand, the non-linear memory can be directly obtained from the Einstein equation with some slight modifications. In contrast to Eq. \eqref{equ:lin_Einstein}, the Einstein tenor $G_{\mu\nu}$ is now expanded to second order in $h_{\mu\nu}$. The non-linear terms are subsequently shifted to the energy stress tensor. Then, applying the Lorentz gauge for the trace-reversed part of $h_{\mu\nu}$, one obtains
\begin{align}
\label{equ:Einstein_equ_Lorentz_gauge}
    \Box \overline{h}_{\mu\nu} = -16\pi \tau_{\mu\nu}\,,
\end{align}
where $\tau_{\mu\nu}$ depends on the stress-energy tensor $T_{\mu\nu}$ and other terms quadratic in $h_{\mu\nu}$, including a term proportional to the energy of \gls{gw}s. In the next Section, it is demonstrated that in spacetimes with a dynamical gravitational field, i.e., some form of gravitational radiation present, the inclusion of the \gls{gw} energy term causes conventional energy conservation approaches to fail for Einstein's gravity. At this point, however, the latter can be ignored and the Green's function $\Box^{-1}$ can be applied to all terms in $\tau_{\mu\nu}$ on the right-hand side of Eq. \eqref{equ:Einstein_equ_Lorentz_gauge}. For each term in $\tau_{\mu\nu}$, this yields a correction to the \gls{gw} strain and, in particular, for the term proportional to the \gls{gw} energy, one finds \cite{Memory_paper_V}
\begin{align}
\label{equ:non_lin_mem_Christ}
    \delta h_{jk}^{TT} =\frac{4}{r} \int_{-\infty}^{u}\dd t \left[\int \dd \Omega \frac{\dd E^\text{GW}}{\dd t \dd \Omega} \frac{n_jn_k}{1-n_jN_j}\right]^{TT}\,.
\end{align}
The contribution was reduced to its transverse-traceless part again. In Eq. \eqref{equ:non_lin_mem_Christ}, the vector $N_i$ points from the source to the observer, and $n_i$ is a unit radial vector. The factor involving these quantities  accounts for the radial dependence of radiation emitted by a binary. The time component $u=t-r$ is simply the retarded time. Eq. \eqref{equ:non_lin_mem_Christ} represents the non-linear memory contribution as derived directly from Einstein's equations by including non-linear terms. In a similar fashion, the term proportional to the stress-energy tensor $T_{\mu\nu}$ in $\tau_{\mu\nu}$ can be solved. For that particular calculation, all non-linear terms can be neglected, reducing the field equation to the known
\begin{align}\label{equ:britt_kaffee}
    \Box \overline{h}_{\mu\nu} = -16\pi T_{\mu\nu}\,.
\end{align}
This equation can be meaningfully solved by choosing a particular instance for $T_{\mu\nu}$. For illustrative purposes, consider the stress-energy tensor to describe $N$ gravitationally unbound particles with masses $M_i$ and constant velocities $\mathbf{v_i}$. Solving the field equation and projecting out the transverse-traceless part, as above, the linear memory for this particular system is described by the solution of Eq. \eqref{equ:britt_kaffee} given by \cite{Lin_Mem_II}
\begin{align}\label{equ:130_outbound}
    \Delta h_{jk}^{TT} = \left(\lim_{t\rightarrow + \infty} - \lim_{t\rightarrow - \infty}\right) \sum_{i=1}^{N} \frac{4M_i}{r\sqrt{1-v_i^2}}\left[\frac{v_i^jv_i^k}{1-{v}_j^i N_j}\right]^{TT}\,.
\end{align}
Here, again, $N_i$ points in the direction of the source. The equation for the linear memory demonstrates the correlation to velocities and masses involved in the studied system. Generally, however, there are many interpretations of the latter \cite{Memory_paper_V}. To prevent spoiling the reader's impartiality, a more concrete interpretation is not given at this stage. Other formulations of Eq. \eqref{equ:130_outbound} in the context of \gls{bbh} mergers are presented to Sections \ref{sec:Covariant_PS}.\\
In conclusion, both linear and non-linear memory can be directly extracted from solutions of Einstein's field equations. By now, it is established that for the most relevant cases, i.e., binary \gls{ns} and \gls{bh} mergers, the non-linear contribution far exceeds the linear memory. Coincidentally, as aforementioned, it is also exactly this component that plays a significant role in the conservation laws that can be formulated in \gls{gr}. In the subsequent Subsection, a different perspective on the non-linear terms encapsulating the \gls{gw} energy is taken, demonstrating the fundamental issue faced when trying to formulate conservation laws for dynamical spacetimes and answering the question of why conventional approaches cannot resolve it. 

Before continuing on that line, it is instructive to highlight another formulation of gravitational memory that is often overlooked in the community but finds its basic motivation in the definition of energy and has been invaluable to properly distinguish between oscillatory and non-oscillatory features of the memory. This so-called \textit{Isaacson approach} \cite{Isaacson_origin_I, Isaacson_origin_II} relies on the separation of scales in metric perturbation and is thoroughly addressed in multiple standard textbooks on \gls{gw}s and gravity \cite{Isaacson_textbook_I,Isaacson_textbook_II,Isaacson_textbook_III}. In principle, Isaacson's approach is simple, yet powerful: One requires the separation between slowly varying background perturbations with frequency $f_L$ and high-frequency perturbations at frequency $f_H$ (\gls{gw}s), where  $f_L\ll f_H$. For \gls{gr} on a four-dimensional manifold $\mathcal{M}$ with a Lorentzian metric $\gd$, one thus writes 
\begin{align}
    \gd = \overline{g}_{\mu\nu}^L + h_{\mu\nu}^H\,,
\end{align}
where $\overline{g}_{\mu\nu}^L$ corresponds to the slowly varying background and $h_{\mu\nu}^H$ encodes the high frequency \gls{gw} content. This decomposition manifests the separation of physical scales and does not depend on a specific coordinate choice. Together with the assumption that metric perturbations are of small amplitude $|h^H_{\mu\nu}|=\mathcal{O}(\alpha)$, where $\alpha\ll 1$, one finds two perturbation parameters in which the equations of motion can be expanded in, namely $\alpha$ and $f_L/f_H$. Thereby, $\overline{g}_{\mu\nu}^L  = \mathcal{O}(1)$, $\partial\overline{g}_{\mu\nu}^L\leq \mathcal{O}(f_L)$ and $\partial h_{\mu\nu}^H = \mathcal{O}(\alpha f_H)$. Without providing any details (see \cite{Isaacson_textbook_I,Isaacson_textbook_II,Isaacson_textbook_III} if interested), one can jump right ahead and write down the leading order set of equations (assuming the absence of matter, i.e., the full equation of motion reads $R_{\mu\nu} =0$), arriving at
\begin{align}
    {}_{(0)}R_{\mu\nu}[\overline{g}^L]&= - \braket{{}_{(2)}R_{\mu\nu}[\overline{g}^L,h^H]}\label{equ:isaacson_0} \,,\\
    {}_{(1)}R_{\mu\nu}[\overline{g}^L,h^H]&=0\,,\label{equ:isaacson_1}
\end{align}
where the notation ${}_{(N)}O[A,B]$ is read as the expanded operator $O$ at $N$th order in perturbation field $B$ and computed for background $A$. With this notation, it is clear that Eq. \eqref{equ:isaacson_1} simply denotes the leading-order \gls{gw} propagation equation. Eq. \eqref{equ:isaacson_0}, on the other hand, establishes a leading-order, low-frequency equation relating the background curvature to the backreaction of the coarse-grained operator $\braket{{}_{(2)}R_{\mu\nu}}$. In fact, the right-hand side of Eq. \eqref{equ:isaacson_0} is proportional to the energy-momentum tensor of \gls{gw}s, while the left-hand side is determined by the leading-order \gls{gw} operator that contributes to the low-frequency equations. As it is shown, for instance, in \cite{Janns_paper_I}, Eq. \eqref{equ:isaacson_0} directly relates to the memory definition given in Eq. \eqref{equ:non_lin_mem_Christ}. This result is somehow remarkable as it is obtained purely from the separation argument in conjunction with an expansion of the equations of motion in terms of the relevant small scales. Yet, a plentiful of information is encoded in the governing equations. For this dissertation, Isaacson's approach is outlined for completeness and does not find any application here. It should be mentioned, however, that a comparison of its results with the other approaches to the memory (e.g., outlined in Section \ref{sec:Covariant_PS}) is subject to ongoing projects involving the author of this thesis.

Finally, teasing what will be more closely derived in Section \ref{sec:BMS_metric_and_Conserved_Quanties}, it is worth pointing out that the memory obtains a clear interpretation within the BMS framework. The latter is built based on the BMS metric, parametrizing the group of asymptotically flat metrics and giving rise to the BMS symmetry group at null infinity $\scrip$. Applying a version of Noether's theorem, non-conservation laws can be derived stating how much the passing gravitational radiation changes each charge corresponding to a symmetry generator in the BMS group, respectively. Of particular interest is thereby the supertranslation subgroup of the BMS group associated with angle-dependent spacetime translations. In words, these flux laws may be expressed as 
\begin{align}
    \T{GW strain} = \T{change in angle-dependent mass} + \T{flux of anlge-dependent energy}\,.\notag
\end{align}
While the change in the angle-dependent mass is nothing else but a change in the mass multipole moment (i.e., the very definition of gravitational radiation, see Chapter 36 and 37 in \cite{Bible}), this equation indicates that the \gls{gw} strain is also sourced by an angle-dependent energy flux. Therefore, the memory, defined as the net change in \gls{gw} strain (as in Eq. \eqref{equ:memory_definition_easy}), is sourced by two contributions resulting from the angle-dependent mass and energy, respectively. These contributions are directly related to the linear and non-linear memory as shown by \cite{Memory_paper_VI}. The former is sourced by systems with unbound masses (massive bodies approaching timelike infinity), such as supernovae events or hyperbolic BHs, while the latter is produced when null radiation escapes towards $\scrip$ \footnote{For a more detailed elaboration including a brief mathematical fundament, the reader is referred to \cite{Mitman:2024uss}.}. \\
The interpretation of the memory in terms of non-conservation laws, i.e., flux laws for the change of charges corresponding to the symmetries present at $\scrip$, also allows for the definition of BMS vacuum transitions \cite{IR_triangle_III}. Each vacuum state thereby describes the instance before or after the passage of a \gls{gw} has sourced a non-trivial flux, e.g., an asymptotically flat Schwarzschild solution defined solely by its total mass. Passing gravitational radiation now changes this vacuum state by altering total mass and sourcing an energy flux across $\scrip$. The transformation characterizing the change of vacuum states is related to the supertranslation subgroup and thus acquires an angular dependence. The difference in the charge quantifying the corresponding initial and final vacuum state is proportional to the \gls{gw} memory and can be explicitly related to gravitational strain, see Section \ref{sec:BMS_metric_and_Conserved_Quanties} and \cite{IR_triangle_III}.



\section{To be conserved or not conserved}
\label{sec:intro_conserved}
Unlike with Maxwell's theory of electromagnetism, one typically does not encounter full \gls{gr} in a perceivable way on a daily basis. Even for astrophysical and cosmological purposes, it is mostly satisfactory to consider static spacetimes such as that of an isolated \gls{bh} or other compact objects. In those cases, energy (mass and angular momentum) conservation is well-defined and conserved quantities for \gls{gr} are straightforwardly derived. However, mixing some gravitational radiation into the relevant spacetime, this picture changes drastically. The reason for that resides, again, in \gls{gr}'s self-interactions as well as its fundamental relation to the (background) geometry of spacetime, making it resistant against the well-established techniques of (Quantum) Field Theory (QFT).\\
For a given Lagrangian field theory in an arbitrary background, the first quantity to compute when searching for energy conservation is the stress-energy tensor. In the context of the action principle, the latter is computed via
\begin{align}
    \label{equ:action_stress_energy}
    T^{\mu\nu} = -\frac{2}{\sqrt{-g}}\frac{\delta S}{\delta g_{\mu\nu}}= -\frac{2}{\sqrt{-g}}\frac{\delta (\sqrt{-g}\Ll)}{\delta g_{\mu\nu}}\,,
\end{align}
and it follows that
\begin{align}
    \nabla_\mu T\ud{\alpha\beta}{} = 0\,.
\end{align}
The latter indicates a conservation of momentum and energy, and, indeed, the stress-energy tensor corresponds to the conserved Noether current associated with spacetime translations. What has purposely not been indicated in Eq. \eqref{equ:action_stress_energy} is that the Lagrangian $\Ll$ describes the field theory only on top of the background and not the background itself. \gls{gr}, however, describes exactly that background, and thus, Eq. \eqref{equ:action_stress_energy} cannot include \textit{gravitational energy}. This raises the question whether a ``full'' stress-energy tensor for \gls{gr}, including gravitational energy, can generally be defined in the first place. \\
Eq. \eqref{equ:einstein_field_eq} might mislead to the conclusion that this particular stress-energy tensor on the right-hand side includes gravitational energy. This is not the case and one rather interprets Einstein's equations in the context of ``extra'' matter fields being present, i.e., as all energy having mass, and all mass acting gravitationally. However, a closer look at the left-hand side of \eqref{equ:einstein_field_eq} reveals that, gravitational energy is included in Einstein's field equation; however, as written down in \eqref{equ:einstein_field_eq}, it is not yet apparent. The secret lies, as above, in the nonlinearity of the field equation, which translates to the fact that \gls{gw}s interact with each other. In other words, the gravitational field self-interacts. It is the total stress-energy that sources spacetime curvature, so if the gravitational field carries stress-energy, it will source itself. This is in stark contrast to many other well-known field theories, such as, for instance, Maxwell's (linear) theory of electrodynamics. \\
The nonlinearity of full \gls{gr} is hidden in $G_{\mu\nu}$. For the full theory, it is generally impossible to dismantle the Einstein tensor to extract a gravitational energy tensor $T^\text{GW}_{\mu\nu}$. However, sufficiently far from the source, i.e., when the metric is nearly Minkowskian, perturbation theory can be applied. In this region, let the metric again be approximated by $g_{\mu\nu} = \eta_{\mu\nu} + h_{\mu\nu}$ and $|h_{\mu\nu}|\ll 1$. In this approximation, the full Einstein tensor is non-linear in $h_{\mu\nu}$ and, by virtue of Eq. \eqref{equ:einstein_field_eq}, proportional to the matter stress-energy tensor
$T^\text{matter}_{\mu\nu}$. To obtain Einstein's equation for the background, i.e., an equation relating the change of the background due to the presence of gravitational energy and matter, $G_{\mu\nu}$ is separated into linearized and non-linear (in $h_{\mu\nu}$) parts. The difference between the latter two,
\begin{align}
    t_{\mu\nu} = -\frac{1}{8\pi}(G_{\mu\nu}-G^\text{bkg}_{\mu\nu})\,,
\end{align}
is symmetric, but not generally covariant. This is due to the linearized Einstein tensor, $G^\text{bkg}_{\mu\nu}$, being a coordinate-dependent quantity. As a result, $t_{\mu\nu}$ defines a pseudo-tensor that is non-linear in $h_{\mu\nu}$ but contains, as $G_{\mu\nu}$, at most two derivatives. Given this definition, the background equation reads
\begin{align}
\label{equ:bkg_einstein_equation}
    G_{\mu\nu}^\text{bkg} = 8\pi (T^\text{matter}_{\mu\nu} + t_{\mu\nu})\equiv 8\pi T^\text{tot}_{\mu\nu}\,.
\end{align}
As aimed, the total stress-energy tensor, $T^\text{tot}_{\mu\nu}$, now contains two contributions, matter and gravitational energy. \\
In this particular case of an observer being located far away from the observer, the gravitational energy is essentially given by the \gls{gw} energy, i.e., $t_{\mu\nu}\rightarrow T^\text{GW}_{\mu\nu}$. The latter is obtained when all terms higher than second order in $h_{\mu\nu}$ are neglected. By selecting out the (gauge invariant) transverse-traceless part of $h_{\mu\nu}$, the effective stress-energy tensor of \gls{gw} can be computed as \cite{Bible}
\begin{align}
\label{equ:GW_stress_energy}
    T^\text{GW}_{\mu\nu} = \frac{1}{32\pi}\braket{h^{TT}_{\alpha \beta, \mu}h^{TT\,\alpha\beta}{}_{,\nu}}\,,
\end{align}
where $\braket{.\,}$ indicate an average over a region of space much larger than a wavelength of the \gls{gw}. In the vacuum, $T^\text{GW}_{\mu\nu}$ is divergence-free and so is $T^\text{tot}_{\mu\nu}$. By explicitly writing out $T^\text{GW}_{\mu\nu}$ in Eq. \eqref{equ:bkg_einstein_equation} for the effective \gls{gw} stress-energy tensor $t_{\mu\nu}$, one obtains a definition of a total stress-energy tensor in the (almost) linearized theory, which receives a partial contribution from the energy of gravitational radiation. Naively, one could use Eq. \eqref{equ:bkg_einstein_equation} in combination with \eqref{equ:GW_stress_energy} to argue that a valid stress-energy tensor can be computed for \gls{gr}, which implies the definition of a conserved quantity, in particular since $\nabla_\mu T\ud{\alpha\beta}{} = 0$. In fact, for any theory with a locally conserved stress-energy tensor $T_{\mu\nu}$ in a background spacetime with at least one Killing field $k^{\mu}$, a locally conserved current can be defined as $J^\mu = T\ud{\mu}{\nu}k^{\nu}$. Integrating the latter over an admissible Cauchy surface $\Sigma$ then yields the conserved quantity associated to $k^\mu$\,\footnote{Depending on the orientation of the Cauchy surface, one can interpret the integral $\int_\Sigma J^\mu \dd \Sigma_\mu$ as a flux through $\Sigma$. This interpretation holds true in particular for timelike or null Cauchy surfaces.}.\\
One of the main problems in \gls{gr}, however, as it was emphasized above by the averaging in Eq. \eqref{equ:GW_stress_energy}, is that the stress-energy tensor of \gls{gw}s cannot be localized and is only meaningful after averaging over several wavelengths and periods. More generally, all diffeomorphism covariant theories, including \gls{gr}, fail to define a notion of local stress-energy. Thus, even when Killing fields are present, conserved quantities and fluxes cannot be defined in the ``traditional'' sense. A well-known exception to this is the ADM mass, momentum, and angular momentum \cite{ADM_book, ADM_Regge_Teitelboim}, which are, however, derived at spatial infinity where gravitational radiation is never to be seen. A version of these can be obtained for any diffeomorphism covariant theory that admits a spacelike slice $\Sigma$ that can be extended to spatial infinity in a suitable manner with a vector field $\xi^\mu$ acting as the time evolution and representing a suitable infinitesimal asymptotic symmetry \cite{Wald_Zoupas}. \\
The latter suggests that when restricting to a particular domain or limit, the issue following the notion of a local stress-energy tensor may be somehow circumvented. Indeed, there is another, perhaps more practical limit for which a conservation law can be formulated in \gls{gr}: For a lot of observationally interesting applications, especially in an astrophysical context, gravitational radiation is received long after the triggering event. This means not only that the detectors picking up the signal are located very far away from the bulk spacetime at whose origin the triggering event is placed, but also that the receiver of the signal stays mostly ignorant of what has happened in the bulk. An exception is constituted by the information that is extractable purely from the received signal. Wrapping these assumptions into a mathematical language, it is reasonable to assume that space- and ground-based gravitational wave interferometers act as observers located at future null infinity $\scrip$ in \textit{asymptotically flat spacetimes}\footnote{A formal definition of these type of spacetimes and the delimitation of those from asymptotically Minkowskian spacetimes is discussed in Section \ref{sec:Gravitational_radiation_at_Scri}.}. Restricting to these types of space times is already enough to advance in the question of conserved quantities. In fact, in the 1960s, a definition of mass-energy and radiated energy at $\scrip$ was first derived by Trautman and Bondi \cite{Bondi_Origin_2,Trautman}. Both arrived at their respective definitions by studying the behavior of the metric in the asymptotic limit. Over the following years, multiple attempts to reproduce or re-derive similar conserved quantities based on asymptotic symmetries at $\scrip$ lead to inequivalent results\footnote{Most prominently the so-called \textit{linkage flux} \cite{Geroch:1981ut}.}, leaving an unsatisfactory gap in literature. In the 1980s, the pioneering work by Abhay Ashtekar and others \cite{Ashtekar_Streubel, Dray_Streubel, Ashtekar_Literature} filled this gap by obtaining a general, fully satisfactory definition of conserved quantities derived from the proper derivation of radiative modes in combination with relevant symmetry transformations at future null infinity. While Ashtekar and collaborators made heavy use of the notion of \textit{radiative phase space}, the same result was obtained almost 20 years later, relying fully on a Hamiltonian perspective generally applicable to arbitrary diffeomorphism covariant theories by Robert Wald and collaborators \cite{Wald_Zoupas}. Regarding conservation laws in spacetimes including gravitational radiation, the latter two approaches, to date, form the literature benchmark.\\

Evidently, gravitational radiation plays a significant role in the dilemma presented by the formulation of conserved quantities in \gls{gr}. In particular, in \cite{Ashtekar_Streubel} an emphasis is put on the connection between the asymptotic symmetries of physical spacetimes (and their corresponding conserved quantities) and the radiative phase space at $\scrip$. This radiation dependence arises rather intuitively as \gls{gw}s are the only form in which gravitational energy can be transported across spacetime without moving the sourcing matter itself. Analogously, Maxwell's theory profits from electromagnetic waves in this regard. Thus, to define conserved quantities in \gls{gr} in the presence of gravitational radiation, it is pivotal to discern the dynamical components of the gravitational field that are due to moving celestial bodies or due to the ``actual'' radiation. As it turned out, the latter posed a non-trivial task and required a large body of the prerequisite methodology crucial to the success of Ashtekar and collaborators' work of the 1980s. 



\section{Radiation being problematic}
\label{sec:intro_radiation}
\begin{figure}[t!]
    \centering
    \includegraphics[width=0.65\textwidth]{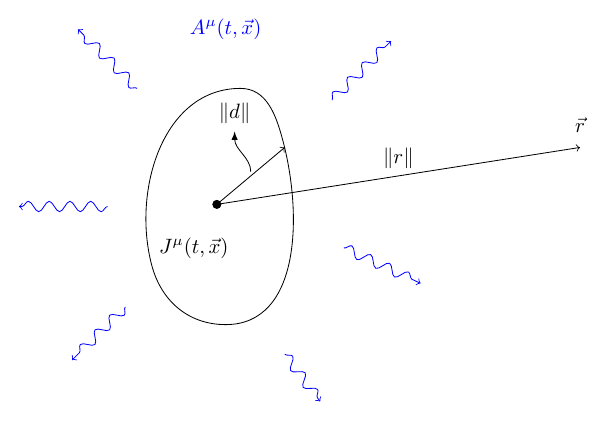}
    \caption{An electromagnetic source $J^\mu$ with finite, characteristic spatial extent $d$ producing a field $A^\mu$ \cite{Our_Review}. The vector $\vec{r}$ indicates the position of the observer.} 
    \label{fig:EM_source}
\end{figure}

To briefly illustrate the issue of properly defining gravitational radiation and to set the mood for chapter \ref{chap:asymptotics}, the less complex but structurally similar problem can be considered for Maxwell's theory (see also \cite{Our_Review}). To start, once consider the following, simple exercise: Given an confined electromagnetic source $J^\mu:= (\rho, \vec{j})^T$ which generates a vector potential $A^\mu$, an observer is tasked to locally determine whether this source generates a radiation field, or whether they are simply located within the reach of a time-dependent electromagnetic source as sketched in Figure \ref{fig:EM_source}. Thanks to Maxwell's well-understood theory of electromagnetism, for an observer located at $\vec r$, one can immediately write down the vector potential $A^\mu$ as a function of time and space as
\begin{align}\label{eq:formal_solution}
	A^\mu(t, \vec{x}) = \frac{1}{4\pi} \int_\Omega\dd^3 x'\int_{\mathbb R}\dd t'\,\frac{J^\mu(t', \vec{x}')}{\|\vec{x}-\vec{x}'\|}\,\delta\left(t'+\|\vec{x}-\vec{x}'\|-t\right)\,.
\end{align}
Despite exaggerating with the temporal integration domain, based on equation \ref{eq:formal_solution} alone, one cannot determine whether or not the vector potential has radiative contributions . The latter is fundamentally encapsulated in $J^{\mu}(t,\vec x)$, which may describe electromagnetic currents as well as radiating contributions. Even if the radiative contribution to $J^\mu(t,\vec x)$ is monochromatic, i.e., the simplest form of radiative contribution, the entanglement between the static and radiative part remains intact. To solve this issue, one needs to develop an approach that suppresses one of the parts in $J^\mu(t, \vec{x}) = J^\mu_\text{rad}(t, \vec{x}) + J^\mu_\text{stat}(\vec{x})$ such that the respective other can be faithfully measured. An immediate idea that has been successful in many related problems is to consider the system at different distances from the source. After all, many problems in electromagnetism can only be analytically solved within a certain domain of validity that is determined by a scale appearing in the problem. In this case, the scale corresponds to the wavelength of the monochromatic radiation, $J_\text{rad}^\mu(t, \vec x) \sim J_0(\vec x) e^{i\omega t}$, i.e., $\lambda = 2\pi /\omega$. Assuming a small spatial extent, $d\ll r$, of the source and $d\ll \lambda$ at all points in space and time, one can differentiate between two domains: i) very close to the source, $r\ll \lambda$, and ii) far away from it, $\lambda \ll r$. For both limits, Eq. \eqref{eq:formal_solution} can be expanded to obtain different contributions attached to powers of $1/r$.\\
For Maxwell's theory, it is well-established that radiative contributions decay as $1/r$; therefore, searching for radiation in $A^\mu$, one simply isolates the relevant terms in the expansion. As it turns out, this is only possible far away from the source, as otherwise, one ends up expanding a quasi-static ($\sim e^{-i\omega t}$) harmonic series in $1/r^{\ell+1}$, where $\ell$ denotes the spherical harmonic expansion coefficient.
Even with the knowledge of the correct power in $1/r$ supposedly associated with radiation, the expansion of Eq. \eqref{eq:formal_solution} far away from the source (neglecting contributions of order $\mathcal O(1/r^2)$ and higher),
\begin{equation}\label{eq:FarZone}
	\lim_{r\to r'\gg \lambda} A^\mu(\vec{x}, t) = \frac{1}{4\pi} \frac{e^{i \omega (r-t)}}{r}\int_\Omega\dd^3 x'\, J^\mu_0(\vec{x}')\,e^{-i \omega\vec{n}\cdot \vec{x}'} + A^\mu_\text{stat}(\vec{x})\,,
\end{equation}
still contains a static contribution, which cannot be disentangled by a local measurement. The efforts of expanding, however, are not in vain as Eq. \eqref{eq:FarZone} now clearly exhibits a crucial difference between the contributions: the radiative part decays as $1/r$ while the coulombic part (i.e., $J^\mu_\text{stat}$) decays faster (namely as $1/r^2$). Given this distinction, there are multiple approaches to suppressing the coulombic part far away from the source. It is clear, however, that this limit has to be taken inevitably, otherwise a separation of contributions cannot be enforced. It remains to clarify how, in the large-$r$ limit, one can avoid erasing the radiative information. Here, integration seems to be a straightforward solution. 

With this mindset, one is reminded of another key feature of radiation in electromagnetism, that is, in contrast to coulombic sources, it carries a flux of energy (and momentum). As delineated above, this flux has far-reaching implications, including, among others, memory effects, and it provides valuable information about the dynamics of the source of radiation. The flux marks a fundamental property of the electromagnetic field and is commonly described by the Poynting vector $\vec S =\vec E \times \vec B$, the cross product of electric and magnetic field, and its associated surface integral. It is crucial, however, to remember that the Poynting vector is not a Lorentz invariant and highly depends on the choice of reference frame. For instance, from the point of view of a boosted observer in the vicinity of a static point charge, a current is present. The latter, in turn, produces both an electric and magnetic field, yet the only source present is a static charge. Even when computing the actual flux by integrating $\vec S$ over a given surface, the issue of Lorentz invariance remains, and while a static observer will measure zero flux, a boosted one finds a non-vanishing integral despite being close to a point charge. Thus, for a local measurement alone, it is not suitable to determine the presence of radiation via the Poynting vector and its associated flux.
\\
Although the Poynting flux is Lorentz dependent, one can demonstrate that, in the right limit $r\rightarrow\infty$, it is only the \textit{true} radiative contribution that survives and contributes meaningfully to its flux. Leaving the conceptual issues of how to define the right limit and how Lorentz transformations or area elements behave in this limit aside for now, one can argue as follows: For a static source the multipole expansion's lowest contributions are the monopole for the electric field, and the dipole for the magnetic field (as, on physical grounds, magnetic monopoles cannot exist). Their corresponding scaling is $1/r$ and $1/r^2$, respectively. The lowest expansion order is independent of any frame-related transformation. That is, one cannot create magnetic monopoles by transforming our coordinate system. Thus, when combining the electric and magnetic static field in the pointing vector, the resulting expression falls off at least with $1/r^n$ with $n>2$. In the Poynting flux integral (over a 2-sphere), the static source therefore yields an expression scaling at least as $r^\epsilon$ with $\epsilon<0$, resulting in a vanishing integral when $r\rightarrow\infty$. For the radiative part of the source, it was already established that electromagnetic radiation decays as $1/r$. Combining electric and magnetic radiaton fields therefore yields a scaling of $1/r^2$ such that the flux integral is constant in $r$ and non-vanishing in the infinite-distance limit. In conclusion, the Poynting flux in the limit $r\rightarrow\infty$ reads
\begin{align}\label{eq:RadiationFlux}
\lim_{r\to\infty}\oint_{\mathbb S^2}\left(\vec{E}_\text{rad}+\vec{E}_\text{stat}\right)\times\left(\vec{B}_\text{rad}+\vec{B}_\text{stat}\right)\,\dd^2\Omega = \lim_{r\to \infty}\oint_{\mathbb S^2}\vec{E}_\text{rad}\times\vec{B}_\text{rad}\,\dd^2\Omega\,,
\end{align}
and, thus, is able to isolate radiative contributions to the vector potential $A^\mu$. Therefore, the initial question can now be answered affirmatively. Located infinitely far away from the source, an observer can, in fact, single out the radiative field for a given source.

As aforementioned, in this conceptual illustration, many technical details are neglected. For instance, it is not per se clear why the limit of $r\rightarrow\infty$ is a good choice for isolating radiation or whether it is unique in this sense. Depending on the model of spacetime under consideration, there is a manifold of ``infinities'' one could arrive at by taking the large-$r$ limit. Moreover, this limit usually involves the appearance of divergences that have to be cautiously handled. For an in-depth treatment of these subtleties, the reader is referred to \cite{Our_Review} and the soon-to-be-finished updated version thereof that goes even further into detail. The main takeaway message of this illustration is that, conceptually, it is not trivial to differentiate between what part of a measured field value of a given theory carrying radiative degrees of freedom can be associated with physical radiation, even for the well-understood Maxwell's theory. Naturally, the issues with \gls{gr}, being arguably the more complex theory, are even more severe. To be overcome, Penrose, Newman, and others had to develop a completely new framework (e.g., \cite{Newman_Penrose}) to study the independent components of the Weyl-tensor in \gls{gr}. Newman and Penrose cast these components into complex scalars and, with the interpretation of Szekeres \cite{Szekeres_interpretation_NPscalars}, linked them to different types of gravitational radiation. Just as for the example above, one key ingredient of their identification was the scaling of the different scalars in terms of powers of $1/r$. This resulted in the so-called \textit{Peeling Theorem} \cite{Our_Review,Wald_book} describing the asymptotic behavior of the Weyl tensor as one approaches null infinity, see Section \ref{subsec:NP_scalars_and_peeling}. In the illustration above, a similar moral has been applied, and, in fact, a Peeling Theorem for electromagnetism can be derived analogously. In the Chapter \ref{chap:asymptotics}, the necessary technical tools will be introduced to rigorously derive the radiative degrees of freedom in \gls{gr} far away from the source and connect them to components of the (asymptotic) Weyl tensor. In this context, peeling in \gls{gr} and Maxwell's theory of electromagnetismis discussed in more detail.



\section{Drowning in Gravitational Waves}
\label{sec:Intro_SGWB}

In the historical discussion above, one important aspect of \gls{gw}s has been neglected so far: Just as earth's ocean does not only host single waves propagating from coast to coast but instead accommodates a variety of individual waves of different sizes interfering which each other to form a chaotic conglomerate, spacetime is not traversed by sporadic \gls{gw} events either. Instead, there are, in fact, many sources causing the emission of \gls{gw}s throughout cosmic history and reaching humanity's detectors just now. However, in contrast to the well-known ``loud'' waves from BBH mergers, these signals cannot be resolved by current instruments, thus establishing an unresolved background of gravitational radiation, the so-called \textit{Stochastic Gravitational Wave Background} mentioned at the very beginning of this introduction. More technically, the SGWB refers to a random, persistent GW signal generated by the superposition of a large number of independent and unresolved sources—cosmological or astrophysical in origin—analogous in spirit to the cosmic microwave background, but probing the spacetime metric itself rather than the photon field. The concept of a GW background was first introduced in the context of cosmological models. The earliest discussions can be traced back to Starobinsky in 1979, who demonstrated that quantum fluctuations of the metric during an inflationary epoch could give rise to a relic gravitational radiation spectrum \cite{1979ZhPmR..30..719S}. These primordial tensor perturbations, stretched to cosmological scales by inflation, form a stochastic background whose statistical properties reflect the dynamics of the early Universe. Shortly thereafter, Rubakov and collaborators \cite{1982PhLB..115..189R} (see also \cite{FABBRI1983445}) offered complementary derivations, confirming that the amplification of vacuum fluctuations during inflation generically produces GWs with an approximately scale-invariant power spectrum. In addition to its primordial origin, the SGWB may also arise from a variety of astrophysical mechanisms, such as the cumulative signal from binary black hole mergers throughout cosmic history or core-collapse supernovae for instance. More details are provided in Section \ref{sec:SGWB_astro}. 

The \gls{sgwb} is typically characterized by its dimensionless spectral energy density
\begin{align}
    \Omega_\T{GW} (f) = \frac{1}{\rho_c} \frac{\dd \rho_\T{GW}}{\dd \ln f}\,,
\end{align}
where $\rho_c$ is the critical energy density and $f$ denotes the observed frequency. The spectral density is determined by the sourcing process of the SGWB and can, thus, become arbitrarily complex. Some instances of $\Omega_\T{GW}$ most relevant to this thesis are outlined in Sections \ref{sec:SGWB_astro} and \ref{sec:SGWB_cosmo}. $\Omega_\T{GW}$ further constitutes the quantity that one wishes to constrain or even measure, ideally over a range of frequencies.\\
So far, however, this endeavor has been without result for the ground-based interferometers. PTAs recently published data indicating the first sign of a gravitational background. Yet, their data is neither conclusive nor predicts $\Omega_\T{GW}$ frequency dependence \cite{Nanograv_SGWB_I} due to their particular analysis technique, see Section \ref{sec:SGWB_detection_astro}. Nonetheless, it marks the onset of a new era in \gls{gw} astronomy: A detection of the SGWB would constitute a profound probe of high-energy astrophysics and early Universe cosmology, given its accumulation of early- and late-time signatures. In particular, it offers insights inaccessible through electromagnetic observations limited by the time of recombination in the early Universe. In this dissertation, Chapter \ref{chap:analysis} extensively reviews the \gls{sgwb} as well as selected astrophysical and cosmological contributions. For readers predominantly interested in this field of \gls{gw} research, it is recommended to directly skip to this part of the thesis.



\section{Outline}
The introductory discussion above highlights the complications associated with the identification, definition and characterizations of radiation in \gls{gr}. Fortunately, by the time the first \gls{gw} was detected in 2015, the majority of conceptual issues were overcome, ultimately resulting in this measurement. Yet, there is an ensemble of residual challenges subject to cutting-edge research in gravitational physics. Most prominently, this includes efforts of extending the balance flux laws and peeling theorems to more general setups that either include another version of gravity, e.g., \cite{Brans_Dicke_I,Brans_Dicke_II, Brans_Dicke_III, Chern_Simons_I, Chern_Simons_II,Brans_Dicke_IV}, or modify the constraints on spacetime, e.g., \cite{Different_Spacetimes_I,Different_Spacetimes_II,Different_Spacetimes_III,Different_Spacetimes_IV, Different_Spacetimes_V}. In fact, only very recently were the flux laws generalized for binary events in an asymptotically de Sitter-like spacetime \cite{Different_Spacetimes_II}. \\
On the other hand, the recent discovery that the gravitational memory is related to the Weinberg's Soft Theorems and the \gls{bms} symmetries \cite{IR_triangle_I, IR_triangle_II, IR_triangle_III, IR_triangle_IV} started a completely new research initiative focusing on flat space holography \cite{Celestial_I,Celestial_II,Celestial_III,Celestial_IV,Celestial_V} (see, among others, \cite{pre_celestial_I,pre_celestial_II,pre_celestial_III,pre_celestial_IV,pre_celestial_V} for major preceding works). Another large class of publications deals with measurement-related questions, such as, for instance, the testing and constraining of theories based on present-day and future \gls{gw} detector data. In particular memory detection prospects are considered in, e.g., \cite{Henris_Mem, Mem_outlook_I, Mem_outlook_II, Mem_outlook_III, Mem_outlook_IV, Mem_outlook_V}, while predictions and implications regarding a \gls{gw}s \gls{qnm} content are analyzed in, e.g., \cite{Ringdwon_I,Ringdown_II,Ringdown_III,Ringdown_IV,Ringdown_V,Ringdown_VI,Ringdown_VII,Ringdown_VIII,Ringdown_IX, Ringdown_X}. With prospect of continuously increasing data quality from existing and futures space- and ground-based instruments, effects like higher harmonics \cite{Higher_modes} and other high-precision-effect-related features, e.g., \cite{Other_features_VIII,Other_features_I,Other_features_II,Other_features_III,Other_features_IV,Other_features_V,Other_features_VI,Other_features_VII}, gain great attention in recent literature. Furthermore, many resources are invested in improving the quality (and quantity) of simulated waveforms suitable for matched-filtering real detector data, e.g., \cite{waveform_test_BL_I,waveform_test_BL_II,waveform_test_BL_III,waveform_test_BL_IV,waveform_test_BL_V} among many other works (see also \cite{waveform_test_II, waveform_test_III, waveform_test_IV,waveform_test_V,waveform_test_VI,waveform_test_VII, waveform_test_VIII}).\\ Beyond the resolved waveform detection efforts, a large group of researchers is currently involved in the analysis of (future) data from the unresolved \gls{gw} sources, i.e., the \gls{sgwb}. Instances include \cite{SGWB_analysis_I,SGWB_analysis_II,SGWB_analysis_III,SGWB_analysis_IV,SGWB_analysis_V,SGWB_analysis_VI,SGWB_analysis_VII}, and in particular the most recent results of the \gls{pta} consortium \cite{Nanograv_SGWB_I, Nanograv_SGWB_II}. Combining both resolved and unresolved \gls{gw} sources, researchers involved in the LISA-Global-Fit initiative currently work on a versatile data processing pipeline to correctly distinguish and detect different sources of \gls{gw} that are mixed together into final instrument's data output, see for instance \cite{Lisa_global_fit_I,Lisa_global_fit_II,Lisa_global_fit_III,Lisa_global_fit_IV}. 

The goal of this thesis is to provide an comprehensive study of tools and techniques applied to present-day and future interferometer data extracting valuable insights across the full spectrum of \gls{gw} signals, i.e., resolved (Chapter \ref{chap:asymptotics} and \ref{chap:quantum}) and unresolved (Chapter \ref{chap:analysis}) \gls{gw} measurements. This includes in particular a revision and unification of the diffuse literature on balance flux laws and the whole \textit{asymptotic spacetime formalism}\footnote{Throughout this thesis, ``asymptotic spacetime formalism'' identifies all theoretical developments derived based o the notion of an asymptotically flat spacetime. This includes in particular (but not exclusively) the notion of the BMS group, Peeling theorem, radiative modes at $\scrip$, the flux balance laws.} (including the above-mentioned BMS framework) in Section \ref{sec:Gravitational_radiation_at_Scri} to \ref{sec:Covariant_PS}. To present an inclusive discussion, the main approaches towards the balance laws based on radiative degrees of freedom at future null infinity $\scrip$ \cite{Ashtekar_Streubel} and the covariant phase space of the underlying theory \cite{Wald_Zoupas} are united in Section \ref{sec:Covariant_PS}. With the theory laid out, Sections \ref{sec:Paper_BL} as well as Sections \ref{sec:Paper_LISA} to \ref{sec:Paper_Mem} demonstrate two pathways of utilizing the balance laws: In the first, strategies for the evaluation of \gls{nr} waveforms are constructed. In Chapter \ref{chap:quantum}, high-precision (quantum) corrections to the gravitational wave memory are derived and their measurement prospects analyzed. 
In Chapter \ref{chap:analysis}, the focus is switched to unresolved GW sources, providing a broad survey of potential sources, their derivation and characteristics. Section \ref{sec:SGWB_detection} outlines some challenges and opportunities for future space-based \gls{gw} interferometers regarding a potential detection of the SGWB, culminating in an in-depth investigation of LISA's capabilities to observe cosmological GW backround contributions in Section \ref{sec:SGWB_detection_Paper_HI}. In the final chapter \ref{chap:discussion}, the dissertation is concluded with a brief summary and outlook on future research on the experimental and theoretical side of \gls{gw} physics.





\chapter{Asymptotics of Gravitational Radiation}
\label{chap:asymptotics}
\graphicspath{{Figures}}
\label{lit_rev} 


\parindent=1cm
\newcommand{\keyword}[1]{\textbf{#1}}
\newcommand{\tabhead}[1]{\textbf{#1}}
\newcommand{\code}[1]{\texttt{#1}}
\newcommand{\file}[1]{\texttt{\bfseries#1}}
\newcommand{\option}[1]{\texttt{\itshape#1}}
\newcommand{\emphasis}[1]{\textit{#1}}
\textit{The Chapter introduces large parts of what is denoted as ``asymptotic spacetime formalism'' in the introduction. It thereby makes heavy use of the author's publications [G] and [D] listed under ``Publications'' below. In particular, Sections \ref{sec:Gravitational_radiation_at_Scri}, \ref{subsec:Radiative_modes}, and \ref{sec:BMS_metric_and_Conserved_Quanties} draw a lot of content from the joint work [G] and its (yet) unpublished extension. Section \ref{sec:Paper_BL} presents the results of [D]. More precise references are given in the main text. The joint work [D] will be referenced as \cite{waveform_test_BL_I} in the following, [G] as \cite{Our_Review}.} 

\noindent As teased in the introduction, defining gravitational radiation, let alone defining flux equations for such, is a non-trivial task. Luckily, the analysis of gravitational radiation in the asymptotic regime—far from isolated sources—offers a unique geometric perspective on the radiative content of spacetime. In the context of \gls{gr}, outgoing radiation is most naturally captured by considering the asymptotic structure of the metric near null infinity, $\scrip$, where radiative information can be unambiguously defined in terms of frame fields or, equally, specific metric components. As mentioned in the introduction, the latter has been intensively studied already in the 1960s, resulting in valuable insights that are applied to gravitational measurement data up until today. Yet, despite being well-aged, the results by Ashtekar, Bondi, Sachs, Metzner, Wald, Geroch, Winicour, and others have never been comprehensively united, with one exception given by \cite{Our_Review}. This Chapter aims to complement the latter by providing an alternative, more mathematical viewpoint starting from the bulk of spacetime, ultimately leading to the same outcomes. This alternative framework is developed in Section \ref{sec:Gravitational_radiation_at_Scri}. Subsequently, the definition of radiative modes at null infinity is discussed in analogy to \cite{Our_Review} in Section \ref{subsec:Radiative_modes}. Complementing this result, the construction of the Bondi-Metzner-Sachs metric is outlined in Section \ref{sec:BMS_metric_and_Conserved_Quanties} before connecting the overall results to literature on flux equations in asymptotically flat spacetimes in Section \ref{sec:Covariant_PS}. An application of the collective knowledge of this Chapter is found in Section \ref{sec:Paper_BL}.

\section{Gravitational Radiation in Asymptotically Flat Spacetime}
\label{sec:Gravitational_radiation_at_Scri}
In this Section, all tools and notations relevant to the identification of radiative degrees of freedom at $\scrip$ and the derivation of the balance flux laws are introduced. Most of the content has been studied within the scope of \cite{Our_Review} and other works in literature. Generally, there are many distinct pathways to introduce asymptotic metrics and associated symmetries. Most works start either with the introduction of a specific structure at null infinity imposed by (predominantly) Penrose's definition of asymptotic flatness or directly present a class of metrics satisfying relevant conditions (e.g., the BMS metric) \cite{Sachs_Origin_2, Sachs_Origin_3}. In the following, a derivation from scratch is presented, guided by the principle of \textit{null geodesic congruence}, which is of significant importance for various notions of (non-expanding) horizons and trapped surfaces. This approach naturally selects a set of frame fields throughout all of physical spacetime and null infinity. The presented contents are largely inspired by \cite{NGC_asym_flat, yannuk} as well as \cite{M_dler_2016, ashtekar_geometryphysicsnullinfinity} and \cite{Geroch_1977}. For more details on specific individual steps and related exercises, see also \cite{Our_Review} and the follow-up work.

\subsection{Null Geodesic Congruence}
\label{subsec:NGC}
Before starting with the formal definition of \gls{ngc}, it is instructive to motivate the approach in the first place. 
As aforementioned, the experienced reader might note that some works on asymptotically flat spacetime construct the physical manifold off $\scrip$. The \gls{ngc} paves the way to obtain the same structure by starting within the bulk. It thereby relies on a specific foliation of spacetime that naturally leads to null infinity and, at $\scrip$, can be adapted to yield the universal background structure. In a sense, one can think of this congruence to follow light rays towards $\scrip$ where it generates a desirable foliation. This picture helps in particular w.r.t. the understanding of the tetrad adapted to the \gls{ngc}. This is demonstrated explicitly in the following.

One starts with some general definitions: Let $(M, \gd)$ be a $4$-dimensional Lorentzian with the signature $(-,+,+,+)$. On this manifold $M$, let there be a nowhere vanishing vector field $\ell^\mu$.
\begin{definition}
    The set of integral curves of $\ell^\mu$ is called the congruence of $\ell^\mu$. The vector field is said to generate a null congruence if
    \begin{align}
        \ell^\mu\ell^\nu \gd = 0\,,
    \end{align}
    and a geodesic congruence, if
    \begin{align}
        \ell^\mu\nabla_\mu\ell^\nu = 0\,.
    \end{align}
\end{definition}
\noindent As integral curves of a vector field are the family of non-intersecting parametrized curves filling up spacetime, one might think that the congruence is, to some extent, parametrized as well. However, the congruence exists between the curves themselves without any particular parametrization. Therefore, the same congruence arises from $\ell^\mu$ and any $f(x^\mu)\ell^\mu$, given that the scalar function $f$ is nowhere vanishing. Assume, from now on, that $\ell^\mu$ generates a \gls{ngc}. Physically, these can be interpreted as freely propagating light rays and thus are of particular relevance for this work, aiming at an improved understanding of gravitational radiation. In fact, historically, the first ideas to characterize the presence of gravitational radiation by Sachs and others were based on observing the behavior of the gravitational field as one moves along the \gls{ngc} to infinity. Even without this connection, \gls{ngc}s are useful to define null surfaces and related structure for certain (asymptotic) spacetimes, as is demonstrated now.

To pursue this, it is instructive to introduce a coordinate system, at least locally, to visualize the congruence. With the \gls{ngc} generated by $\ell^\mu$ at hand, a preferred choice emerges. 
\begin{proposition}
    Given an \gls{ngc} generated $\ell^\mu$, one can locally always find a coordinate system $(u,r,y^A)$, with $A=1,2$, such that 
    \begin{align}
        \ell^\mu \partial_\mu = \partial_r\,,
    \end{align}
    and
    \begin{align}
    \label{equ:line_element}
        \dd s ^2 = V \dd u^2 + 2 \dd u \dd r + H_{AB}(\dd y^A - U^A \dd u)(\dd y^B - U^B \dd u) + D_A \dd y^A \dd r\,,
    \end{align}
    where $V, H_{AB}, U$ are functions of $(u,r,y^A)$ and $D_A = D_A(u,y^A)$.
\end{proposition}
\noindent \textit{Proof.} Let $\ell^\mu$ be a vector field and $r$ define the affine time along the congruence generated by $\ell^\mu$, i.e., $\ell = \partial_r$. Then, at each point, one can locally find a hypersurface $\Sigma$ transverse to $\ell^\mu$. On this hypersurface, one can always choose a local coordinate frame $(u,y^A)\in \mathbb R^3$. Provided this slicing, a metric can be constructed in full generality as
\begin{align}
    \dd s^2 &= V\dd u ^2 + 2B\dd u \dd r + A\dd r^2 + H_{AB}(\dd y^A - U^A \dd u)(\dd y^B -U^B \dd u)+ D_A \dd y^A \dd r\notag\\
    &=:\gd^\text{NGC}\dd x^\mu\dd x^\nu,
\end{align}
where $A,B,D_A,U^A,V$ are functions of $(u,r,y^A)$. Using the Levi-Civita connection for this metric, computed in Appendix \ref{app_sec:Levi_Civita}\footnote{Note that the connection displayed in Appendix \ref{app_sec:Levi_Civita} holds for the whole class of metrics that are explored in this Section. In particular, it can be used to compute the connection associated to the BMS metric defined below.}, one finds that $\ell$ being null, i.e., $\ell^\mu\ell^\nu\gd^\text{NGC}=0$, implies that $A=0$. Further, $\ell$ being geodesic, $\ell^\mu\nabla_\mu\ell^\nu = 0$ implies that $\partial_r D_A = 0$ as well as $\partial_r B = 0$. Finally, one can reabsorb $B$ via a suitable coordinate choice $\partial_{u'}/\partial_u = B(u,y^A)$.

\begin{figure}[t!]
    \centering
    \includegraphics[width=0.5\textwidth]{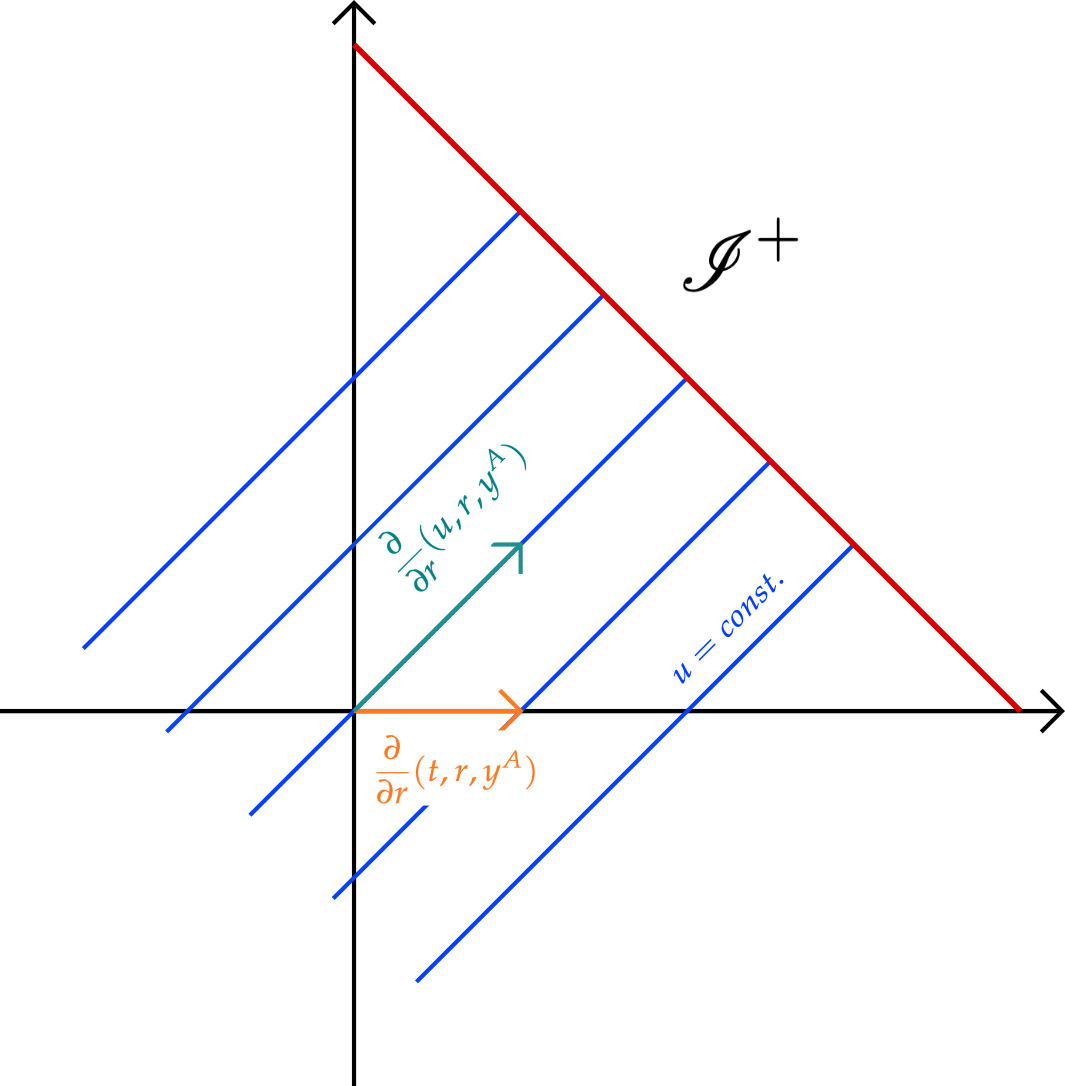}
    \caption{Sketch demonstrating the direction of $\ell^\mu$ in $(u,r,y^A)$ coordinates. In the ``classic'' $(t,r)$ grid, $\partial_r$ points in the radial direction when $t$ and angular coordinates are fixed. In the $(u,r)$ chart, however, when $u$ and angular coordinates are fixed, $\partial_r$ points along $u=const.$ hypersurfaces. Since $\ell^\mu \sim \partial_r$, $\ell^\mu$ is null in this chart and generates an NGC tracing the path of light rays. Concretely, in $(u,r,y^A)$, one has $\ell^\mu = (0,1,0,0)$. } 
    \label{fig:sketch_dr}
\end{figure}

\noindent A couple of remarks on the line element and the choice of coordinates: First, note that the angular coordinates, $y^A$, are often expressed in terms of angles $\theta,\phi$ or complex conjugates $z,\Bar{z}$. The functions $V, H_{AB}, U$ capture the essence of the relevant spacetime configuration but will not be specified in the following. From the fact that $\ell^\mu$ is geodesic, it directly follows that $\partial_r D_A = 0$. Generally, in this coordinate system, $r$ was chosen as an affine parameter along the geodesic congruence, and it holds that $\ell^\mu\gd \dd x^\mu = \dd u + D_A\dd y^A$. Therefore, a fixed $r$ establishes a (space-like) hypersurface, $\Sigma$, transverse to $\ell^\mu$ with local coordinates $(u,y^A)$ on $\Sigma$. It is to be emphasized that with this coordinate choice, the vector $\partial_r$ does not have the same meaning as in the $(t,r,y^A)$ coordinates anymore. To obtain some intuition about its meaning in $(u,r,y^A)$ coordinates and to convince oneself that it indeed follows light rays to future null infinity $\scrip$ as claimed at the beginning of this Subsection, see Fig. \ref{fig:sketch_dr}.

\begin{figure}[t!]
    \centering
    \includegraphics[width=0.9\textwidth]{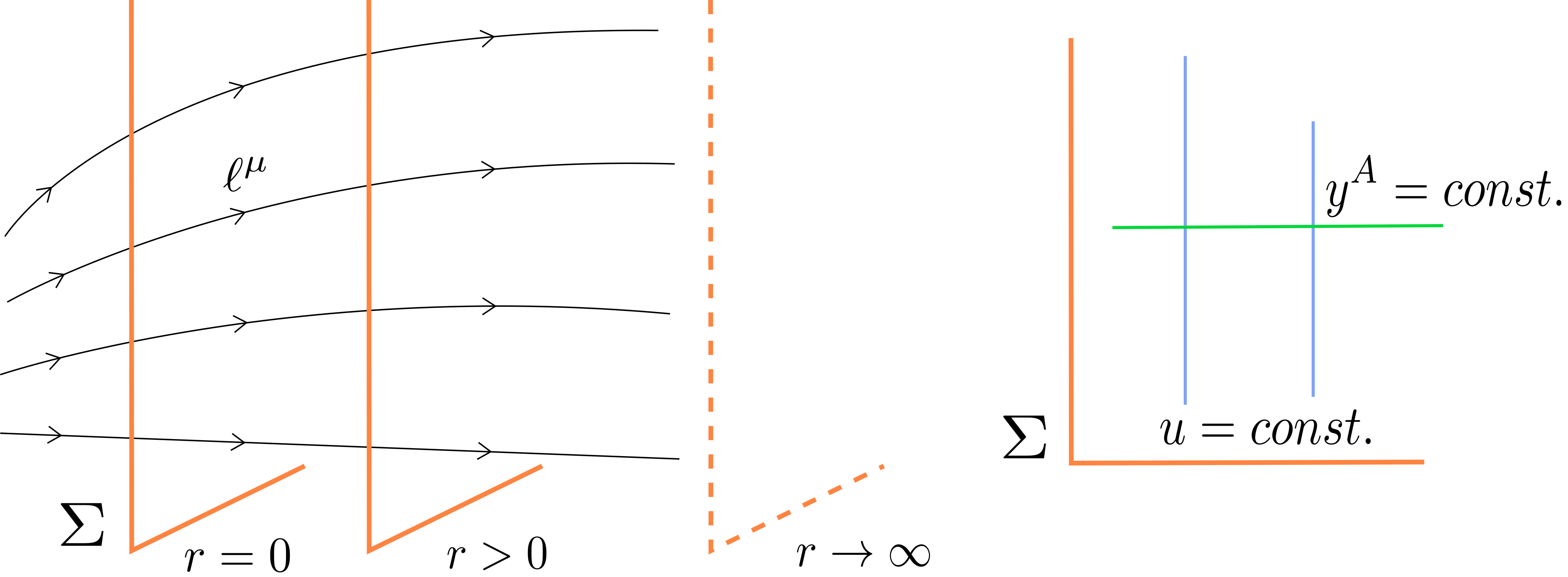}
    \caption{Sketch of the NGC generated by $\ell^\mu$ and its foliation of spacetime in hypersurfaces $\Sigma$.} 
    \label{fig:super_sketch}
\end{figure}

If one would choose another parametrization along the geodesics, i.e., not the affine parametrization, then $\ell^\mu= e^{-\beta} \partial_r$ and the term proportional to $\dd u \dd r$ gains a factor $e^{\beta}$. Upon restricting the choice of $\ell^\mu$ to make its congruence hypersurface orthogonal, the line element \eqref{equ:line_element} simplifies further by dropping the term proportional to $D_A$. To that end, it is instructive to refine the construct of \gls{ngc}s.
\begin{definition}
    A \gls{ngc} is hypersurface orthorgonal if there exists a function $u:M\rightarrow \mathbb{R}$, such that
    \begin{align}
        \ell^\mu\gd = \dd u &&\Leftrightarrow&& \ell^\mu = \gu \partial_\nu u\,.
    \end{align}
\end{definition}
\noindent Let $\ell^\mu$ be a hypersurface orthogonal \gls{ngc}, then one finds that, in the chosen coordinate system, $\ell^\mu$ is orthogonal to surfaces of constant $u$ as for any vector field $X^\mu$ along the $u=const.$ hypersurface, $X^\mu\ell^\nu \gd = 0$. However, given $u$ corresponds to the retarded time $u=t-r$, the hypersurface $u=const.$ are null hypersurfaces. As $\ell^\mu$ is null, it is both tangential and orthogonal to these hypersurfaces. A depiction of the hypersurface orthogonal \gls{ngc} is provided in Fig. \ref{fig:super_sketch}. It should be clear at this point that not every \gls{ngc} is hypersurface orthogonal. The restriction requires that the generating field's contraction with the metric yields a total derivative of some function. This can be converted into a rigorous condition for $\ell^\mu$ using Frobenius' theorem. 
\begin{proposition}
    A \gls{ngc} is hypersurface orthogonal if and only if
    \begin{align}
    \label{equ:hso}
        \ell_{[\mu}\nabla_\nu\ell_{\sigma]}=0\,.
    \end{align}
\end{proposition}
\noindent \textit{Proof.} Define a one-form $\theta = \ell_\mu \dd x^\mu$. According to Frobenius theorem, for integrable one-forms, $\theta$ is a total derivative, i.e., $\exists u$ such that $\theta = \dd u$, if $\theta \wedge \dd \theta = 0$. Inserting its definition in the latter equation, one obtains the above expression \eqref{equ:hso}.

\noindent Given a \gls{ngc}, one can construct frame fields $\{\ell^\mu, m^\mu,$ $ \Bar{m}^\mu, n^\mu\}$, also called tetrad, adapted to this particular congruence such that 
\begin{align}
\label{equ:tetrad_norm}
    \ell^\mu n^\mu \gd=-1\,, && m^\mu\Bar{m}^\mu \gd = 1\,, && \ell^2= n^2 = m^2=\Bar{m}^2=0\,.
\end{align}
Both $\ell^\mu, n^\mu$ are real vector fields, $m^\mu$ is complex. At every point $x^\mu$ of $M$ the tetrad $\{\ell^\mu, m^\mu, \Bar{m}^\mu, n^\mu\}$ forms a basis of the tangent space $\mathcal T_xM$.
\begin{proposition}
    Given a \gls{ngc} generated by $\ell^\mu$, a null tetrad $\{\ell^\mu, m^\mu, \Bar{m}^\mu, n^\mu\}$ is adapted to the \gls{ngc} if it contains the generating vector field. Then, any other adapted null tetrad $\{\ell^\mu, \hat m^\mu, \hat{\Bar{m}}^\mu, \hat n^\mu\}$ must be of the form 
    \begin{align}\label{equ:tetrad_rescale}
        \hat m = e^{i\varphi(x^\mu)}(m^\mu + f(x^\mu)\ell^\mu)\,,&& \hat n^\mu = n^\mu + f(x^\mu)\Bar{m}^\mu + \Bar{f}(x^\mu)m^\mu + \frac{1}{2}|f(x^\mu)|^2\ell^\mu\,,    
    \end{align}
    where the functions $\varphi(x^\mu), f(x^\mu) \in \mathbb{C}$.
\end{proposition}
\noindent \textit{Proof.} Given that it forms a basis of the tangent space $\mathcal T_x M$, in any new tetrad $m^\mu$ must be of the form $\hat m^\mu = e^{i\varphi(x^\mu)}(m^\mu + a(x^\mu) \bar m + b(x^\mu)\ell^\mu+c(x^\mu)n^\mu)$ with $a,b,c,\varphi\in \mathbb C$. Constraint by the (cross-)normalizations of the null tetrad, $n^\mu, \bar m$ drop out of the latter since otherwise $\ell^\mu \hat m_\mu\neq0$ and $\hat m^2\neq 0$. The normalization $\hat m ^\mu\hat{\bar{m}}_\mu =1 $ prevents a rescaling of $\hat m^\mu$. For $n^\mu$ being real nullifies the complex phase. The remaining cross-normalization equations result in a system of coupled equations that is solved by \eqref{equ:tetrad_rescale}.

\noindent In other words, the vector field $ m^\mu$ is defined up to an addition of $\ell^\mu$ while $n^\mu$ is uniquely fixed by $m^\mu, \Bar{m}^\mu$. Generally, a tetrad can be characterized through invariants of the \gls{ngc}. To obtain a simplified version of these invariants, it is intrinsic to separate the complex frame fields $m^\mu, \Bar{m}^\mu$ as 
\begin{align}\label{equ:reisen_kann}
    m^\mu = \frac{1}{\sqrt{2}} (m_1^\mu + i m_2^\mu)\,, && \Bar{m}^\mu = \frac{1}{\sqrt{2}}(m_1^\mu -i m_2^\mu)\,.
\end{align}
For the vector fields $m_A^\mu$, it holds from the cross-normalizations that $m_A^\mu m_B^\nu \gd = \delta_{AB}$. Then, the so-called deviation tensor $B_{AB}$ can be defined, $A,B\in 1,2$, as
\begin{align}
    B_{AB} := m_A^\mu m_B^\nu \nabla_\mu \ell_\nu\,,
\end{align}
and can be separated into \textit{expansion}, \textit{shear}, and \textit{twist}
\begin{align}
    \theta = \delta^{AB}m_A^\mu m_B^\mu \nabla_\mu\ell_\nu\,,&&\sigma_{AB}=m_{(A}^\mu m_{B)}^\nu\nabla_\mu \ell_\nu\,,&& w_{AB}=m_{[A}^\mu m_{B]}^\nu \nabla_\mu \ell_\nu\,.
\end{align}
The deviation tensor is an invariant of the \gls{ngc}, i.e., it does mot depend on the choice of $m^\mu$. Its components can be used to characterize the function in the line element \eqref{equ:line_element} with the right choice of tetrad.
\begin{proposition}
    For a hypersurface orthogonal \gls{ngc} and a coordinate system $(u,r,y^A)$, one can write 
    \begin{align}
    \label{equ:NGC_metric}
        \dd s^2 = V\dd u^2 + 2\dd u \dd r + H_{AB}(\dd y^A - U^A\dd u)(\dd y^B - U^B\dd u)\,,
    \end{align}
    with $\ell^\mu = \partial_r$ and $m^\mu_C = (E^A_C \partial/\partial y^A)^\mu$, where $E_C^A E^B_D H_{AB}= \delta_{CD}$.
\end{proposition}
\noindent \textit{Proof.} One mildly modifies the proof of the metric \eqref{equ:line_element} by choosing a new coordinate $\hat u$ such that  $\ell^\mu\gd = \dd u$. Effectively, this yields an additional constraint on the metric obtained by computing Eq. \eqref{equ:hso} using the Levi-Civita connection in Appendix \ref{app_sec:Levi_Civita}. The exact form of $m^\mu$ is chosen for convenience and satisfies all (cross-)normalization conditions. Note that $E_{AB}$ is used translate between the distinct sets of base vectors of the $2$-sphere displayed in Eq. \eqref{equ:reisen_kann}.

\noindent In this case, one finds that $w_{AB} = 0$, $\theta = \frac{1}{2}H_{AB}\partial_r H^{AB}$ and $\sigma_{CD}= \frac{1}{2}E_{C}^AE_{D}^B(\partial_r H_{AB})$\footnote{In fact, one can show that the condition for hypersurface orthogonality \eqref{equ:hso} is equivalent to a vanishing twist.}. This parametrization is known as \textit{Newman-Unti coordinates}. Residual gauge freedoms in this parametrization are fixed via particular fall-off conditions \cite{Newman_Unti_Gauge, Barnich_2012}. In literature, these conditions are known as the Newman-Unti gauge. \\
For some computations, it is instructive to relate the metric with the tetrad as they encapsulate metric-related information. In fact, as it is demonstrated in Section \ref{subsec:Radiative_modes}, a part of the tetrad is fundamentally involved in the definition of radiation at $\scrip$. 
\begin{proposition}
    For any spacetime manifold ${M}$ equipped with a metric $\gd$ and a tetrad composed out of frame fields satisfying Eq. \eqref{equ:tetrad_norm}, the metric can be written as
    \begin{align}
    \label{equ:tetrad_decomp}
        g_{\mu\nu} = -2\ell_{(\mu}n_{\nu)} +2m_{(\mu}\bar{m}_{\nu)}\,.
    \end{align}
\end{proposition}
\noindent \textit{Proof.} Start by writing the metric in terms of the most general combination of tetrad components, i.e.,
\begin{align}
	g_{\mu\nu} &= 2\lambda_1\,\ell_{(\mu}\ell_{\nu)} + 2\lambda_2\,\ell_{(\mu}n_{\nu)} + 2\lambda_3\, \ell_{(\mu}m_{\nu)} + 2\lambda_4\,\ell_{(\mu}\bar{m}_{\nu)} +2\lambda_5\, n_{(\mu}n_{\nu)}\notag\\
	&\qquad + 2\lambda_6\, n_{(\mu}m_{\nu)} + 2\lambda_7\, n_{(\mu}\bar{m}_{\nu)} + 2\lambda_8\, m_{(\mu}m_{\nu)} + 2\lambda_9\, m_{(\mu}\bar{m}_{\nu)} + 2\lambda_{10}\, \bar{m}_{(\mu}\bar{m}_{\nu)} \,.
\end{align}
The symmetrization in each term ensures that the metric is symmetric. One can immediately exclude the terms $\ell_{(\mu}\ell_{\nu)}$, $n_{(\mu}n_{\nu)}$, $m_{(\mu}m_{\nu)}$, and $\bar{m}_{(\mu}\bar{m}_{\nu)}$ since these terms all need to have a zero coefficient in order to satisfy~\eqref{equ:tetrad_norm}. Hence, one has $\lambda_1 = \lambda_5 = \lambda_8 = \lambda_{10} = 0$. Moreover, one can exclude the terms $\ell_{(\mu}m_{\nu)}$, $\ell_{(\mu}\bar{m}_{\nu)}$, $n_{(\mu}m_{\nu)}$, and $n_{(\mu}\bar{m}_{\nu)}$ based on the same argument. This reduces the previous expression to just two terms:
\begin{equation}
	g_{\mu\nu} = 2\lambda_2\,\ell_{(\mu}n_{\nu)} + 2\lambda_9\, m_{(\mu}\bar{m}_{\nu)}.
\end{equation}
The coefficients $\lambda_2$ and $\lambda_9$ are determined from the only two contractions which are not zero:
\begin{align}
	g_{\mu\nu}\ell^{\mu}n^{\nu} &= \lambda_2 \overset{!}{=} -1\notag\\
	g_{\mu\nu}m^{\mu}\bar{m}^{\nu} &= \lambda_9 \overset{!}{=} 1.
\end{align}
It follows that the metric can indeed be written as
\begin{equation}
	g_{\mu\nu} = -2\ell_{(\mu}n_{\nu)} + 2m_{(\mu} \bar{m}_{\nu)}.
\end{equation}

\noindent The above proof only used the abstract properties of the tetrad and, thus, the result is generally valid for an arbitrary realization given a particular choice of coordinates, provided the (cross-)normalization properties remain intact. It is noted without proof that the second term in \eqref{equ:tetrad_decomp} corresponds to a conformal rescaling of the unit $\mathcal{S}^2$ metric and fundamentally determines the metric at $\scrip$. An elaboration of the latter is found in the following Section. Similar but less relevant descriptions of, for instance, the Levi-Civita symbol and the volume $2$-form can be derived as well. For details, see Chapters 1 and 2 in \cite{Our_Review}. Finally, a note to avoid confusion here: There is another null vector $n^\mu$, by definition of the tetrad, which is actually pointing in null direction, i.e., along the light cone. On $\scrip$ itself, this null normal spans a \gls{ngc} as well and therefore provides a well-posed foliation of future null infinity into $u=const.$ hypersurfaces. Note, however, that in the bulk spacetime, not only do the integral curves of $n^\mu$ not lead to future null infinity, but they also generally fail to be geodesic except on $\scrip$ where the affine parameter is given by $u$\footnote{An intuition for integral lines generating the \gls{ngc} associated with $n^\mu$ is provided by the green lines in Fig. \ref{fig:sketch_conformal}}.\\
In practice, one can think about the relation between $\ell^\mu$ and $n^\mu$ in the following way. Let the asymptotic region of spacetime $M$ be foliated by outgoing null hypersurfaces of constant $u$. The geodesic null normal $\ell^\mu$ of this foliation has an affine parameter $r$ such that each of these null surfaces $u=const.$ is foliated by a (space-like) $2$-sphere of constant $r$. These $2$-spheres possess another null normal $n^\mu$ that on $\scrip$ coincides with the null normal $\partial_u$.

\subsection{Conformal Compactification}
\textit{From this Section on, the notation of the metric and its associated covariant derivative changes w.r.t. previous Sections. The change in notation is motivated by the introduction of an unphysical spacetime by compactification. }

\noindent
With the tetrad adapted to the \gls{ngc}, one can span out spacetime in a very particular way that is well-suited to describe radiation that travels along the null direction. What has been left untouched so far are the functions in the line-element \eqref{equ:line_element}. To determine those, one has to specify the content of spacetime locally. As mentioned before, for \gls{gw} physics, there is no particular interest in describing the bulk of spacetime. It is of much higher importance to describe the asymptotic behavior as, morally, the \gls{gw} detectors of mankind are infinitely far away from the source of radiation, e.g., a \gls{bbh} or \gls{ns} binary. Before the chosen spacetime tetrad is extended far away from a potential source of gravitational radiation, one has to define the structure of the asymptotic region. To that end, one starts with plain Minkowski space,
\begin{align}
\label{equ:Minkowski}
    \widetilde{\dd s}^2 = - \dd t^2 + \dd r^2 +\dd \Sigma ^2\,,
\end{align}
where $\dd\Sigma^2$ is the metric on the sphere $\mathcal S^2$.
\begin{proposition}
    Upon redefinition of coordinates, explicitly $u=t-r$, $v=t+r$, $u=\tan U$, $v= \tan V$, the Minkowski metric can be rewritten as 
    \begin{align}
    \label{equ:minkowski_in_weird_coordinates}
        \widetilde{\dd s}^2= \widetilde{g}_{\mu\nu}\dd x^\mu \dd x^\nu = \underbrace{\frac{1}{4 \cos^2U\cos^2V}}_{=:1/\Omega^2}\left(-4\dd U\dd V + \sin^2(V-U)\dd \Sigma^2\right)\,,
    \end{align}
    where $U,V\in (-\pi/2, \pi/2)$ and $V-U\geq 0$.
\end{proposition}
\noindent \textit{Proof.} Eq. \eqref{equ:minkowski_in_weird_coordinates} follows by direct replacement of $\dd t = 1/2 (\dd u + \dd v)$, $\dd r = 1/2 (\dd v - \dd u$ and $\dd u = 1/\cos^2{U} \dd U$, $\dd v =  1/\cos^2{V} \dd V$.

\begin{figure}[t!]
    \centering
    \includegraphics[width=0.35\textwidth]{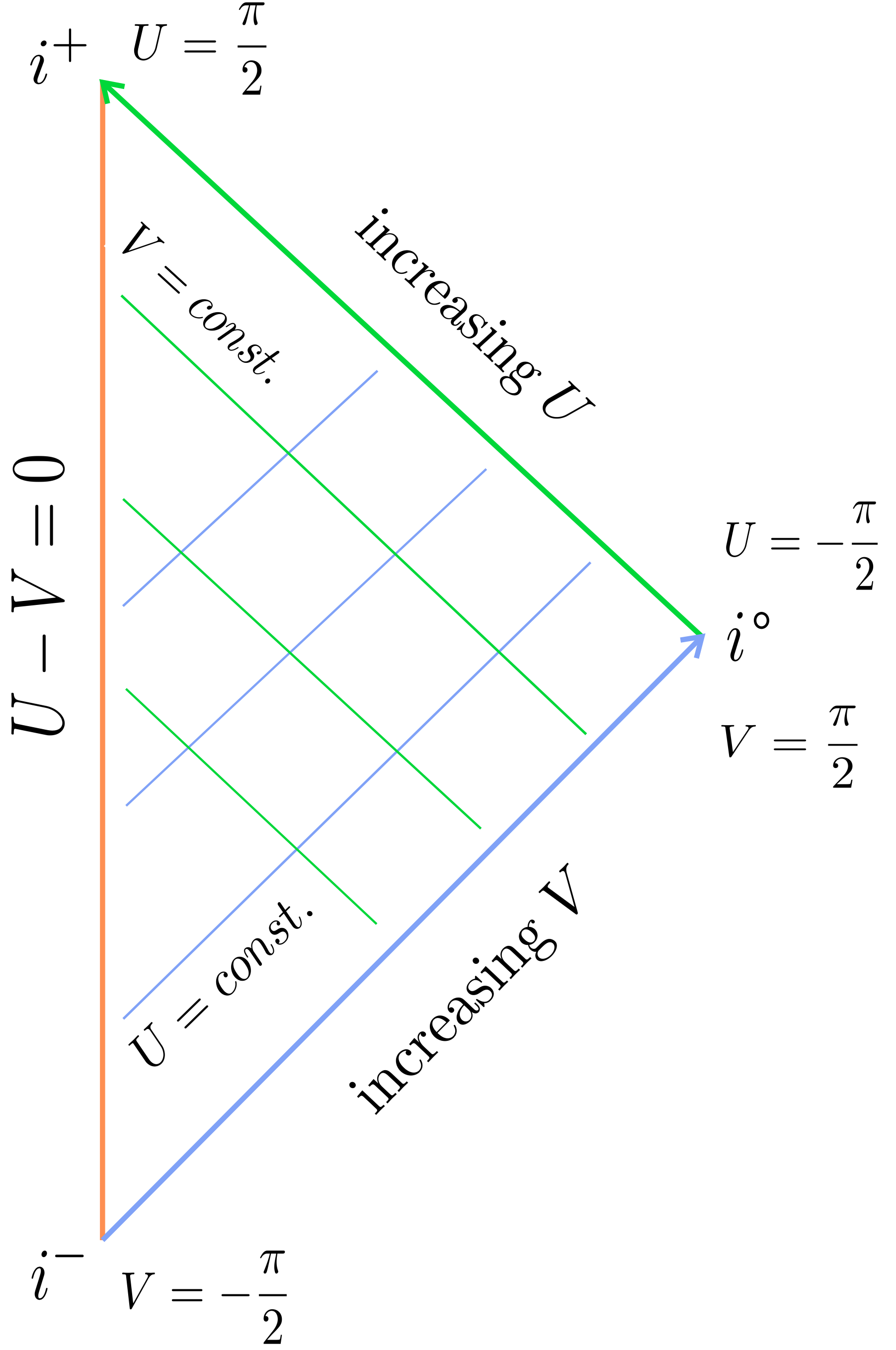}
    \caption{Sketch of conformal Minkowki space in coordinates $U,V$ defined in the main text.} 
    \label{fig:sketch_conformal}
\end{figure}

\noindent In Eq. \eqref{equ:minkowski_in_weird_coordinates}, the prefactor constitutes a conformal factor and the expression in brackets can be seen as an unphysical metric that extends to the boundary of a newly defined spacetime manifold $\mathcal{M}$ such that $\dd s^2 = \gd \dd x^\mu \dd x^\nu = (\Omega^2 \widetilde{g}_{\mu\nu} \dd x^\mu \dd x^\nu)$. In this picture, the physical metric $\widetilde{g}_{\mu\nu}$ (and its corresponding covariant derivative $\widetilde{\nabla}_\mu$) is only defined in the interior of $\mathcal{M}$, which is equivalent to Minkowski space, $\mathbb{M}$. The unphysical metric $\gd$ is defined on full $\mathcal{M}=\mathbb{M}\bigcup\scrip \bigcup \scrim $, where $\scrip$ is future and $\scrim$ past null infinity. It constitutes a \textit{conformal compactification} of Minkowski space. Note here that the added boundary does not correspond to the full boundary of $\mathbb{M}$. The latter is not a manifold but rather made up of five different pieces each of which individually is a manifold. The first two pieces are future and past timelike infinity, $\iota^+$ and $\iota^-$, defined by $U=\pi/2= V$ and $U=-\pi/2=V$ respectively. Together with spacelike infinity, $\iota^0$, these boundaries are actually points, i.e., $\dd s^2|_{\iota^{0,\pm}}=0$. The remaining null infinities $\scrip,\scrim$, defined by $V=\pi/2, U\in(-\pi/2,\pi/2)$ and $U=-\pi/2, V\in(-\pi/2,\pi/2)$ respectively, are $3$-dimensional manifolds of topology $\mathcal S^2\times \mathbb{R}$. A comment on the role of the topology is postponed to later stages. A depiction of the coordinate choice $U,V$ is provided in Fig. \ref{fig:sketch_conformal}.

\begin{figure}[t!]
    \centering
    \includegraphics[width=0.65\textwidth]{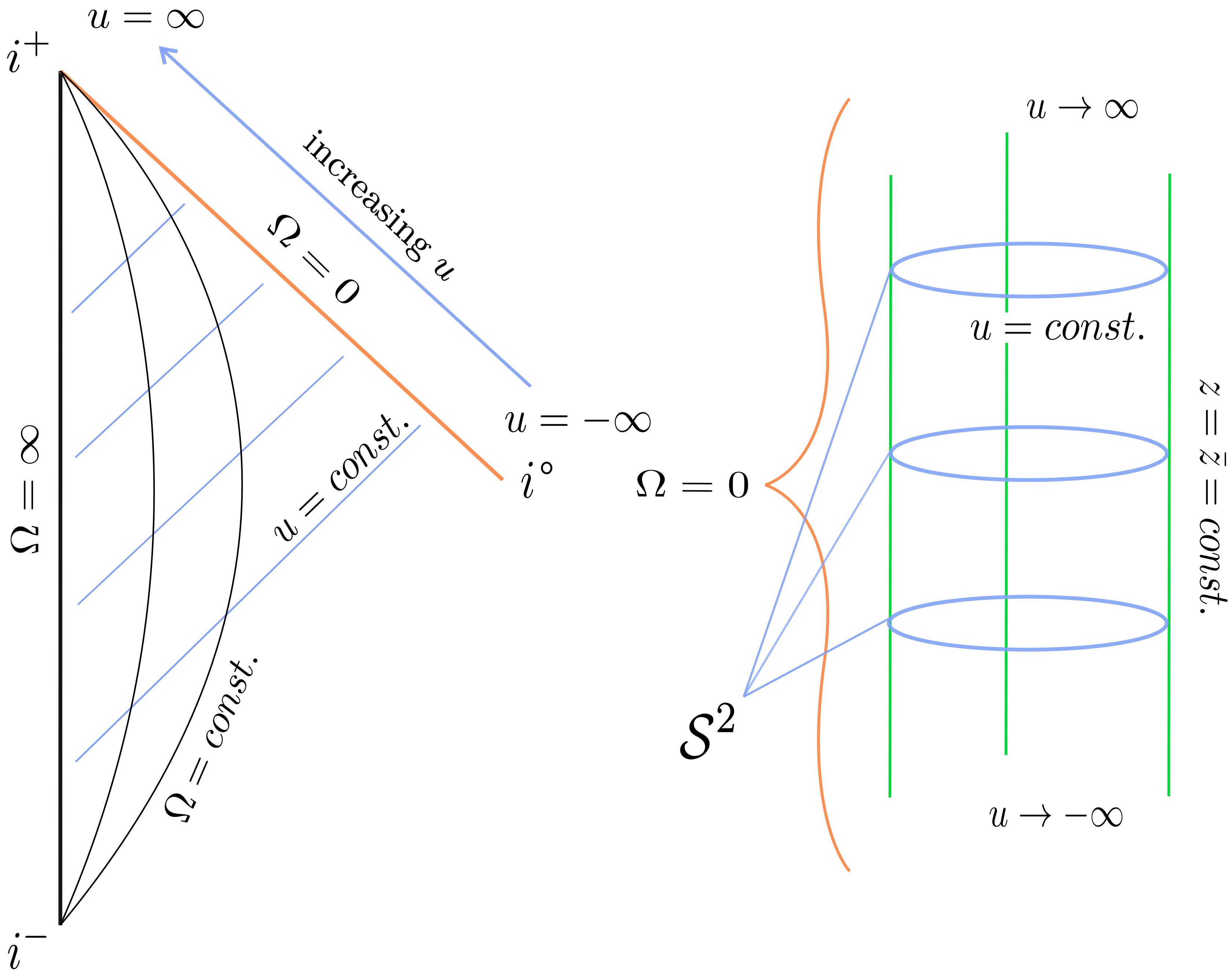}
    \caption{Sketch of the slicing of the asymptotic region of spacetime into $u=const.$ slices. The geometry of $\scrip$ defined by $\Omega = 0$ is given by $\mathcal S^2 \times \mathbb{R}$ as illustrated. $\scrim$ is not covered by this coordinate system.} 
    \label{fig:sketch_stuff}
\end{figure}

The above procedure can be summarized into the following remarks: Starting from Minkowski space $\mathbb{M}$, one can construct a manifold $\mathcal{M}=\mathbb{M}\bigcup\scrip \bigcup \scrim $ with boundary $\partial\mathcal{M}= \scrip \bigcup\scrim$. On $\mathcal{M}$, there exists a scalar field, also called \textit{conformal scale}, and a metric, $(\Omega, \gd)$ such that $\widetilde{g}_{\mu\nu}= \Omega^{-2}\gd$ is the metric on the interior manifold $\mathbb{M}$. On the boundary, $\Omega=0$ and $\nabla_\mu\Omega \neq 0$. The latter naturally defines a normal to the hypersurface that is the boundary, $n_\mu \sim \nabla_\mu \Omega$. Other normals are defined by $n'^\mu = f(x^\mu) n^\mu$ where $f$ is a smooth, nowhere vanishing function. Note that the choice of $(\Omega, \gd)$ is not unique. In fact, the same physical metric can be recovered for any pair $(f(x^\mu)\Omega, f(x^\mu)^2\gd)$ with $f\in C^\infty(\mathcal{M})$. Therefore, on this constructed manifold $\mathcal{M}$, one can only define a conformal class of metrics $[\gd]$ and a conformal density $[\Omega]$.\\
The ambiguity in the conformal factor plays a central role when defining asymptotically flat/Minkowskian spacetimes, as it is demonstrated in detail in the following Sections. On the boundary itself, the ambiguity in the conformal metric transfers over to the induced metric on $\scrip$, $q_{\mu\nu}$. The latter can be defined as the pullback\footnote{For the definition of the pullback, see the ``Notations'' at the beginning of this dissertation.} of $\gd$ to $\scrip$, $\newpb{\gd}$, and is degenerate with signature $(0,+,+)$. In particular, it can only be defined in equivalence classes as well. One particular favorable choice thereby is $q_{\mu\nu}\dd x^\mu \dd x^\nu := \gd \dd x^\mu \dd x^\nu|_{\scrip}=0\cdot \dd u^2+ \dd\Sigma^2$ such that the metric describes the unit two-sphere. Similarly, the normal to $\scrip$, $n^\mu$, given by
\begin{align}
    n^\mu \partial_\mu = \gu \partial_\mu (\Omega) \partial_\nu = \cos U \frac{\partial}{\partial U}
\end{align}
is defined only up to a coordinate dependent factor, i.e., $(n^\mu, q_{\mu\nu})\sim(f(u,z,\bar z)^{-1}n^\mu,$ $ f(u,z,\bar z)^2q_{\mu\nu})$\footnote{The choice of naming the normal to $\scrip$ as one of the previously discussed tetrad is no coincidence. A detailed explanation follows below. Note further that the rescaling function $f$ at $\scrip$ does not depend on the radial coordinate.}. The latter is referred to as \textit{universal (carrollian) structure} of null infinity, as it is common to all spacetimes that are asymptotically flat, independent of the bulk's inner complexity, as it will be demonstrated below. Note that the angular coordinates $y^A$ were swapped for a complex combination $z,\bar z$ encoding the same coordinate information.
The $3$-manifold described by $(n^\mu, q_{\mu\nu})$ can be further sliced of cross sections $\mathcal C$ determined by constant $u$ that have topology $\mathcal S^2$. Each cross section can be equipped with a metric $s_{\mu\nu}$ which is obtained by pulling back $q_{\mu\nu}$ to the cross section $\mathcal C$ on $\scrip$. For a depiction, see Fig. \ref{fig:sketch_stuff}.

To connect the constructed unphysical metric with the previous discussion of \gls{ngc}s, one simply chooses $\Omega = 1/r$ in the coordinate system $(u,r,y^A)$ for compactified Minkowski space. Comparing the resulting line element,
\begin{align}\label{equ:line_element_important}
    \widetilde{\dd s}^2=\Omega^{-2}\underbrace{(-\Omega^2\dd u^2 - 2\dd u\dd \Omega + \dd \Sigma^2)}_{=\dd s^2}\,, &&u,r,\Omega \in \mathbb{R}\,,&& \Omega \geq 0\,, &&r>0\,,
\end{align}
with the previous results, one finds that the line element \eqref{equ:line_element_important} defines a hypersurface orthogonal \gls{ngc} spanned by $\ell^\mu$ with $\ell^\mu\partial_\mu = \partial_r$ \footnote{Strictly speaking, as $\ell^\mu$ is both tangential and orthogonal to the hypersurface, one has found a hypersurface orthonormal \gls{ngc}.}. Note that here again, $\dd \Sigma^2$ denotes the metric on $\mathcal S^2$. The line element $\dd s^2$ describes the unphysical conformally rescaled metric. The congruence has vanishing shear and twist, and for the expansion, one finds $\theta = r$. Note that the restriction of $r>0$ excludes $\scrim$ from the coordinate system. Fig. \ref{fig:sketch_stuff} depicts the physical manifold and the boundary, which is cut into $u=const.$ cross sections of 
$\scrip$. With this coordinate choice, the so-called \textit{BMS coordinates}, the normal to $\scrip$ belongs in fact to the tetrad adapted to the \gls{ngc} and is defined by $n^\mu\partial_\mu = \gu (\nabla_\mu \Omega) \partial_\nu |_{\scrip}= \partial_u$ with $q_{\mu\nu}\dd x^\mu \dd x^\nu := \left. \gd \dd x^\mu \dd x^\nu\right|_{\scrip} = 0\cdot \dd u^2 + \dd \Sigma^2$ on $\scrip$. The vector fields $\ell^\mu, n^\mu$ can be associated with the tetrad 
adapted to the hypersurface orthonormal \gls{ngc} of the physical metric. This is a particular desirable results as the choice of tetrad is generally not unique, and any set of vector fields satisfying the above (cross)-normalization conditions is technically a valid tetrad. However, given the connection to \gls{ngc}s, a particular choice has been favored regarding $\ell^\mu, n^\mu$. For this choice the normal $n^\mu$ to $\scrip$, which is also null by definition (as $\scrip$ is null), is both tangential and normal to $\scrip$, as 
$n^\mu \nabla_\mu \Omega =0$. 

\begin{figure}[!t]
    \centering
    \includegraphics[width=0.45\textwidth]{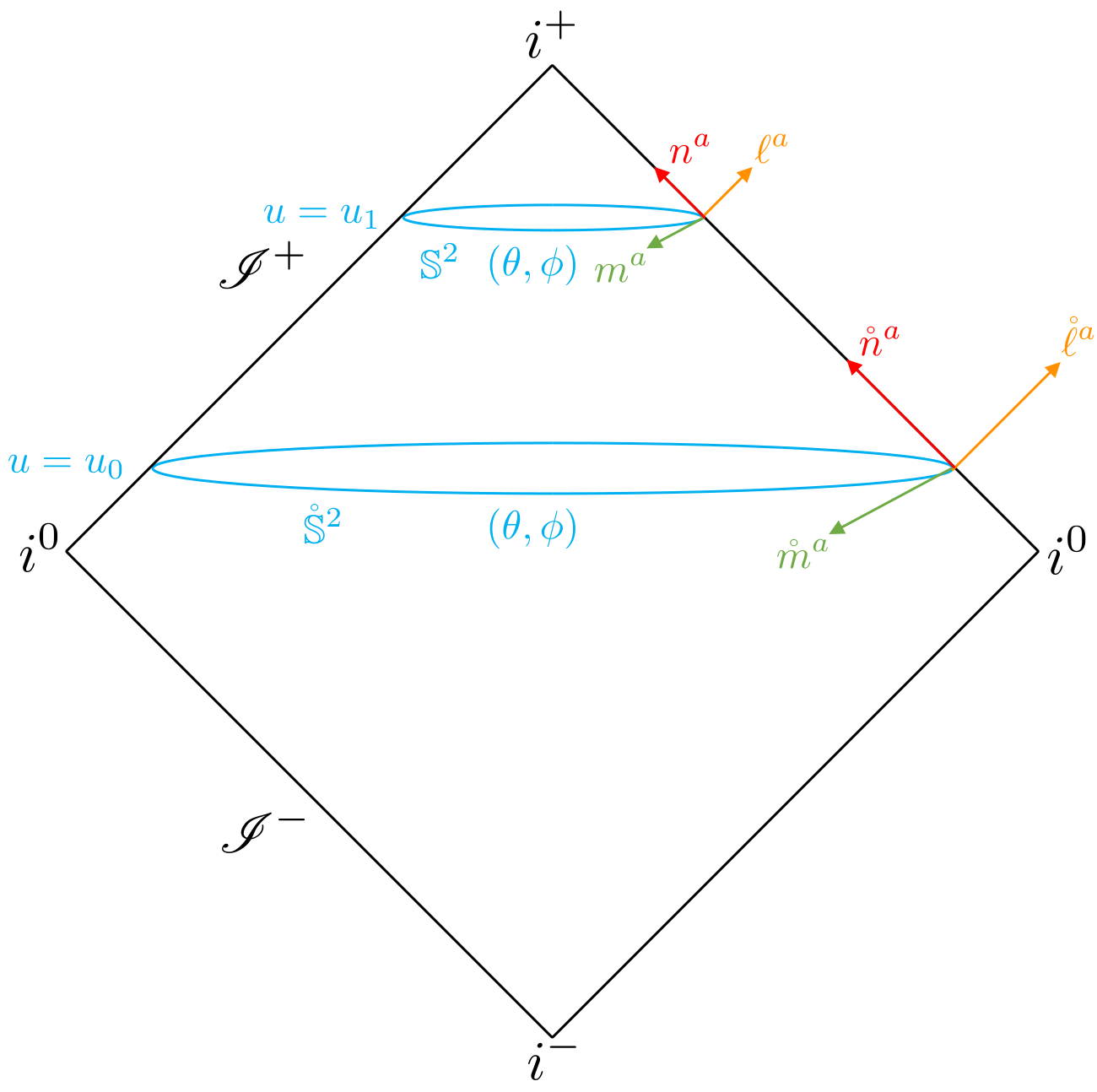}
    \caption{Sketch illustrating the definition of the frame fields as defined by the null tetrad adapted to the NGC as chosen in the main text \cite{Our_Review}. The tetrad is displayed on $\scrip$ for two instances of time.} 
    \label{fig:Pyzel_Gremlin}
\end{figure}

At this point, it is crucial to highlight that there is a clear distinction between the tetrad spanning physical spacetime and the one that is well-defined also at the boundary. For clarification, for the remainder of this work, the (tetrad) vector fields associated with the physical metric will be denoted as $\well^\mu,\wn^\mu, \wm^\mu, \wmbar^\mu$, where, according to the definitions above, $\wn^\mu=\widetilde{g}^{\mu\nu}\nabla_\nu \Omega$ and $\well^\mu \partial_\mu = \partial_r$. For the vector fields of the tetrad extending over the full conformally rescaled metric, one drops the tilde in their notation. \\
The tetrad associated with the unphysical spacetime metric can then be systematically constructed from the bulk tetrad: Starting with the null normal, by definition, $\wn^\mu$ has a smooth limit to $\scrip$ and is finite. Thus, without restriction of generality, $\wn^\mu \equiv n^\mu$. Being a null normal and tangential qualifies $n^\mu$ to be geodesic. As the chosen coordinate $u$ affinely parametrizes the integral curves of $n^\mu$ on $\scrip$, being geodesic manifests as $n^\mu\nabla_\mu n^\nu |_{\scrip}= 0$. Regarding the remaining tetrad vector fields, it is found that since $\gd = \Omega^2 \widetilde{g}_{\mu\nu}$ and $\well^\mu n^\nu \widetilde{g}_{\mu\nu}=-1$, there exists an $\ell^\mu$ such that $\ell^\mu|_{\scrip}$ is finite and $\well^\mu = \Omega^2 \ell^\mu$. Similarly, since $\wm^\mu \wmbar_\mu = 1$, there exists an $m^\mu$ such that $m^\mu|_{\scrip}$ is finite and $\wm^\mu = \Omega m^\mu$. For the induced metric at $\scrip$ one finds $n^\mu q_{\mu\nu}=0$ and $\lie_n q_{\mu\nu}= \alpha q_{\mu\nu}$ 
\footnote{By choosing a divergence-free coordinate frame, one can show that $\nabla_\mu n_\nu = 0$ at $\scrip$ such that the intrinsic metric is Lie dragged by the null normal, $\lie_n q_{\mu\nu}|_{\scrip}=0$. Note that the latter implies that on each cross section $\mathcal C$, the induced metrics $s_{\mu\nu}$ look the same and all metric information is stored in this unique $s_{\mu\nu}$.}. Note that the metric at $\scrip$ is degenerate, i.e., has signature $(0,+,+)$, and describes a $3$-dimensional manifold. Thus, some works in literature denote quantities defined at $\scrip$ by Roman letters, while Greek letters indicate a definition on the $4$-dimensional physical or unphysical spacetime. Here, the Greek letter indexing remains standard while working at $\scrip$, and it is always clearly indicated if an index is defined on or off $\scrip$ the following. The reader is, however, reminded that at $\scrip$, one deals with a lower-dimensional manifold. Fig. \ref{fig:Pyzel_Gremlin} illustrates the definition of the tetrad at two different instances of time, $u_0,u_1$. For the definition on and off $\scrip$, consider Fig. 1 in \cite{waveform_test_BL_III}. Note the different notations. 

\subsection{Asymptotical Flatness and its Consequences}
\label{subsec:asymptotical_flatness}
The previous definition of BMS coordinates generally allows for the derivation of BMS-related quantities, including the desired BMS symmetry group on which large parts of the BMS framework rely. However, so far, the discussions were restricted to (completely) flat spacetimes without any interesting physics in the bulk. Now, the aim is to generalize the previous considerations to arbitrary spacetimes that, at the boundary, match the descriptions above. In this sense, in the above a very particular case has been treated considering an everywhere-flat spacetime, which, naturally, is also asymptotically flat. In the following, the notation of the previous paragraphs is adapted. 

Generally, the notion of \textit{asymptotical flatness} is characterized by the following definition pioneered in \cite{Penrose_1962}\footnote{Note that, in literature, the expression ``asymptotically flat'' and ``asymptotically Minkowskian'' may seem to be used interchangeably. Indeed, they are equivalent in large part. The subtle but important difference is that for asymptotically Minkowskian spacetimes, it is assumed that the vector $n^\mu$ (as defined above) is complete when choosing a divergence-free frame, i.e., $\nabla_\mu n^\mu|_{\scrip}=0$.}:

\begin{definition}
    A spacetime $({M}, \widetilde{g})$ is asymptotically flat if
    
    i) there exists a manifold $\mathcal M = M \bigcup \scri $ with boundary $\partial \mathcal M = \scri := \scrip \bigcup\scrim$,
    
    ii) there exists metric on $\mathcal M$, $(\Omega, \gd)$, with a scalar field $\Omega$ such that $\widetilde{g}_{\mu\nu} = \Omega^{-2} \gd$ on the interior of $\mathcal M$,
    
    iii) on the boundary $\scri$, $\Omega|_\scri = 0$ and $\nabla_\mu \Omega |_\scri \neq 0$,
    
    iv) $\widetilde{g}_{\mu\nu}$ satisfies the Einstein's equations $\widetilde{R}_{\mu\nu} - \frac{1}{2}\widetilde{R}\widetilde{g}_{\mu\nu} = \widetilde{T}_{\mu\nu}$ with $\widetilde{T}_{\mu\nu} = \mathcal{O} (\Omega^2)$.
\end{definition}
\noindent As before, this definition comes with the caveat that the unphysical metric and the scalar $\Omega$ are defined only up to a Weyl rescaling $(\Omega, \gd)\sim (f(x^\mu)\Omega, f(x^\mu)^2\gd)$. Observe that it also covers a large class of metrics and includes an (almost completely) unspecified matter content. The only constraint on the energy-stress tensor in the bulk results from its asymptotic behavior given by iv). Based on i) - iv) alone, a structure equal to the one above can be derived. As it encompasses a large class of metrics, it deserves the attribute ``universal''. To prove this claim, one starts by defining the normal to $\scri$ via $n^\mu = g^{\mu\nu}\nabla_\nu \Omega|_{\scrip}$ and the ($4$-dimensional) Schouten tensor as\footnote{The definition can vary up to a factor of $1/2$ w.r.t. other notations in literature.}
\begin{align}
    S_{\mu\nu} = R_{\mu\nu} - \frac{1}{6} R \gd\,,
\end{align}
which, upon extension of physical to unphysical spacetime, transforms as 
\begin{align}
    \widetilde{S}_{\mu\nu} = S_{\mu\nu} + \Omega^{-1} \nabla_{\mu}\nabla_{\nu}\Omega -\frac{1}{2} \Omega^{-2}(\nabla_\sigma\Omega)(\nabla^\sigma\Omega)\gd\,.
\end{align}
Given condition iv), Einstein's equations now imply that the physical Schouten tensor vanishes at the boundary, $\widetilde{S}_{\mu\nu}|_{\scrip}=0$ which implies that $(\nabla_\sigma\Omega)(\nabla^\sigma\Omega)|_{\scrip} = n^\mu n^\nu \gd |_{\scrip} = 0$ and $\nabla_\mu n_\nu|_{\scrip} = \frac{1}{4}(\nabla_\sigma n^\sigma)\gd|_{\scrip}$. The latter is equivalent to $\nabla_{(\mu}n_{\nu)}|_{\scrip}$ $ = \lie_n \gd |_{\scrip}$ where $\lie_n \gd |_{\scrip} $$= \lie _n q_{\mu\nu}$. Thus, $\lie _n q_{\mu\nu} \sim q_{\mu\nu}$. In turn, $n^\mu n^\nu \gd = 0$ implies that $n^\mu\nabla_\mu \Omega|_{\scrip}=0$, i.e., the normal $n^\mu$ is also tangential to $\scrip$. The remaining relations i) - iii) then indicate that the induced metric $q_{\mu\nu}\dd x ^\mu \dd x ^\nu = \gd \dd x ^\nu\dd x^\mu|_{\scrip}$ is degenerate and satisfies $n^\mu q_{\mu\nu}=0$. As before, the Wely rescaling changes the normal and metric on $\scrip$.  The rescaled tuple then becomes the center of the universal structure definition according to \cite{Geroch_1977} based solely on the above assumptions:
\begin{definition}
    For spacetimes following the definition of asymptotic flatness above, the universal structure of null infinity is given by 
    \begin{align}\label{equ:definition_of_the_universal_structure}
        (n^\mu, q_{\mu\nu})\sim(f(u,z,\bar z)^{-1}n^\mu, f(u,z,\bar z)^2q_{\mu\nu}) && \text{with }\,\,n^\mu q_{\mu\nu}=0 &&\text{and }\,\,\lie _n q_{\mu\nu} \sim q_{\mu\nu} \,,
    \end{align}
    where $n^\mu, q_{\mu\nu}$ are defined above. 
\end{definition}
\noindent It is worth pointing out that in a large part of literature, this rescaling is defined with an additional condition that $\lie_n f =0$. This choice is generally relevant when it comes to computing the vector fields preserving this structure. For now, however, one does not necessarily require this constraint. \\
The structure \eqref{equ:definition_of_the_universal_structure} describes the null normal and the metric at $\scrip$ universally, i.e., it holds for all metrics matching the definition of asymptotic flatness. As such, it is denoted as the \textit{universal structure}. Recently, the structure was also linked to \textit{conformal Carrollian geometries} \cite{Carrolian_geometry} which plays a central role in the context of flat space holography. For another way of thinking about $(n^\mu, q_{\mu\nu})$ as a universal structure of $\scrip$, consider the following proposition, for which the $\mathcal S^2$ metric is defined in terms of complex parameters, i.e., $\dd \Sigma^2 = 4(1+|z|^2)^{-2} \dd z \dd \bar z$:
\begin{proposition}
    On $\scrip$, there always exists a unique choice of coordinate system $(u,z,\bar z )$ and Weyl rescaling $(n^\mu, q_{\mu\nu})\sim(f(u, z, \bar z)^{-1}n^\mu, f(u, z, \bar z)^2q_{\mu\nu})$, such that
    \begin{align}
        n^\mu \partial_\mu= \partial_u && \text{and} && q_{\mu\nu}\dd x^\mu \dd x^\nu = 0\cdot \dd u^2 + \frac{4\dd z\dd \bar z}{(1+|z|^2)^2}\,.
    \end{align}
\end{proposition}
\noindent \textit{Proof.} One starts out with an arbitrary metric $q_{\mu\nu}$ and tangent $n^\mu$ on $\scrip$. Then, one can always choose a coordinate system, $(\hat u, \hat y^A)$\footnote{The angular coordinates $y^A$ and $z,\bar z$ are used interchangeably.} such that $n\sim \partial_ {\hat u}$. From $n^\mu q_{\mu\nu}=0$ it then follows that $q_{\mu\nu}= 0 \cdot \dd \hat u^2 +  h_{AB}(u,y^A) \dd \hat y^A \dd \hat y^B$. Then, $\lie_n q_{ab}\sim q_{\mu\nu}$ implies that one can rewrite $h_{AB}(\hat u,\hat y^A) = f(\hat u,\hat y^A)q_{AB}(\hat y^A)$. By applying a suitable Weyl rescaling, one eliminates the prefactor $f$ that is a function of the coordinates $(\hat u,\hat y^A)$. As any $2$-dimensional metric is conformal to the $\mathcal S^2$ metric, one finds that $h_{AB}(\hat u,\hat z,\hat{\bar z}) = k(\hat z,\hat{\bar z}) \frac{4\dd \hat z \dd \hat{\bar z}}{(1+|\hat z|^2)^2}$. A suitable coordinate transformation $(\hat u, \hat y^A)\rightarrow (u, y^A)$ finally eliminates $k(\hat z, \hat{\bar z})$ and yields $n_\mu=\partial_u$.

The universality of this coordinate frame allows for the deduction of universal features of asymptotically flat spacetimes\footnote{For an illustration of the universal structure and its corresponding space of generators $\mathbb G$, the reader is referred to Fig. 9 in \cite{Our_Review}. Each integral line of $n^\mu$ corresponds to a point in $\mathbb G$, and one can define a non-degenerate metric $\underline{q}{}_{\mu\nu}$ on $\mathbb G$.}. As outlined in the introduction, this is of particular interest for \gls{gw} research as, in first approximation, the observing instruments can be considered as being asymptotically far away from a source of gravitational radiation, in a region of spacetime that can be considered (roughly) flat. This statement holds independently of the source's structure, mass, and dynamics. Thus, the universality of the asymptotic spacetime structure holds significant power in the context of the analysis of real experimental data. To be more concrete, it is the symmetry group preserving the universal structure that yields the most significant advantage, i.e., (non)-conservation laws. Before the symmetries preserving the universal structure are discussed in more detail, it is important to highlight the utility of the asymptotic structure, or, more concretely, asymptotic flatness with a concrete example.

\subsection{Newman-Penrose Scalars and the Peeling Theorem}
\label{subsec:NP_scalars_and_peeling}
The example refers to the definition of the so-called \textit{Newman-Penrose Scalars} (NPS). These scalars not only formulate invariants of the \gls{ngc} but also encode the independent components of the Weyl tensor and can therefore be linked (asymptotically) to different parts of radiative information of the underlying gravity theory. Their asymptotic behavior is formulated through the \textit{Peeling} theorem, which can be approached via two distinct pathways: The first one (also historically) was developed by Sachs \cite{Sachs_Origin_1,Sachs_Origin_2}. Relaying upon earlier definitions of Newman and Penrose, Sachs utilized the description of the 10 independent components of the Riemann tensor (in the vacuum case $R_{\mu\nu}=0$) in terms of 5 complex scalars $\Psi_0,...\Psi_4$. Using the definition of the (physical) tetrad above, these scalars can be realized as invariants of the \gls{ngc} generated by $\widetilde \ell^\mu$ in a spacetime without cosmological constant. Here, being an invariant to the congruence generated by $\widetilde \ell^\mu$ translates to an independence of the choice of $\widetilde m^\mu,\widetilde{\bar m}^\mu$. In the original notation, the scalars read
\begin{align}
	\widetilde{\Psi}_4 &:= \widetilde R_{\mu\nu\rho\sigma}\,\widetilde n^{\mu} \widetilde{\bar{m}}^{\nu} \widetilde n^{\rho} \widetilde{\bar{m}}^{\sigma}\,,\\
	\widetilde{\Psi}_3 &:=\widetilde R_{\mu\nu\rho\sigma}\,\widetilde \ell^{\mu} \widetilde n^{\nu} \widetilde{\bar{m}}^\rho \widetilde n^\sigma\,,\\
	\widetilde{\Psi}_2 &:=\widetilde R_{\mu\nu\rho\sigma}\,\widetilde\ell^{\mu} \widetilde m^{\nu} \widetilde{\bar{m}}^\rho \widetilde n^\sigma\,,\\
	\widetilde{\Psi}_1 &:= \widetilde R_{\mu\nu\rho\sigma}\,\widetilde\ell^{\mu} \widetilde m^{\nu} \widetilde \ell^{\rho} \widetilde n^{\sigma}\,,\\
	\widetilde{\Psi}_0 &:=\widetilde  R_{\mu\nu\rho\sigma}\, \widetilde\ell^{\mu} \widetilde m^{\nu}\widetilde \ell^{\rho} \widetilde m^{\sigma} \,,
\end{align}
where the Riemann tensor has to be replaced with the Weyl tensor $C_{\mu\nu\rho\sigma}$ if one considers off-shell solutions and deviations from vacuum, i.e., $\widetilde R_{\mu\nu}=0$. Note that the tilde here as well denotes the definition of the corresponding quantity on physical spacetime. $\widetilde\Psi_0$ is an invariant of the congruence except if $\widetilde\Psi_0=0$, then $\widetilde\Psi_1$ is one. If now $\widetilde\Psi_0,\widetilde\Psi_1$ are both vanishing, the invariant is given by $\widetilde\Psi_2$ and so on. If all scalars vanish, one can immediately conclude that the metric is flat. It is, however, not the scalars themselves but rather their asymptotic information that counts. It is found that, for large radii, these scalars ``peel'' as
\begin{align}
    \widetilde\Psi_4&\sim r^{-1}\Psi^\circ_4+ \mathcal{O}(r^{-2})\,,\label{equ:PSI4}\\
    \widetilde\Psi_3&\sim r^{-2}\Psi^\circ_3+ \mathcal{O}(r^{-3})\,,\\
    \widetilde\Psi_2&\sim r^{-3}\Psi^\circ_2+ \mathcal{O}(r^{-4})\,,\\
    \widetilde\Psi_1&\sim r^{-4}\Psi^\circ_1+ \mathcal{O}(r^{-5})\,,\\
    \widetilde\Psi_0&\sim r^{-5}\Psi^\circ_0+ \mathcal{O}(r^{-6})\,.\label{equ:PSI0}\\
\end{align}
The latter can be easily derived from the explicit insertion of an asymptotically flat metric and corresponding tetrads. A possible parametrization encompassing a large class of such metrics is described in detail in Section \ref{sec:BMS_metric_and_Conserved_Quanties}. While this approach is commonly noted as \textit{Sachs Peeling}, for an easy derivation without the explicit use of any metric or tetrad component, one can resort to the so-called \textit{Penrose Peeling} \cite{Penrose_1962}. Ultimately leading to the same result, Penrose Peeling relies on the observation that the Weyl tensor, in contrast to the Riemann tensor, is conformally invariant and vanishes at $\scrip$, i.e., $C_{\mu\nu\rho\sigma}|_{\scrip} = 0$, for asymptotically flat spacetimes. For a derivation of this statement, the reader is referred to \cite{Our_Review}. Generally, one can show that 
\begin{align}\label{equ:Weyl_zero}
    C_{\mu\nu\rho\sigma}|_{\scrip}= \begin{cases}
		0 & \text{for }\Lambda\neq 0\,,\\
		0 & \text{for }\Lambda = 0\text{ if the topology of $\scrip$ is $\mathcal S^2\times\mathbb R$}\,,
	\end{cases}
\end{align}
where $\Lambda$ denotes the cosmological constant. Penrose then proves the for an unphysical metric that is at least $\mathcal C^4$ on the unphysical manifold $\mathcal{M}$, it follows that $\widetilde C\ud{\mu}{\nu\rho\sigma}|_{\scrip}=C\ud{\mu}{\nu\rho\sigma}|_{\scrip}=0$. This statement can be straightforwardly converted to the explicit Sachs Peeling by constructing NPS directly out of the Weyl tensor. Assuming that the Weyl tensor is a smooth function within the full manifold, including a neighborhood of $\scrip$, one can apply the following Lemma.
\begin{lemma}
    Let $f$ be a function which is smooth in a neighborhood of $\scrip$ and which satisfies $\left.f\right|_{\scrip} = 0$. Its Taylor expansion looks like
\begin{equation}
	f = \left.\frac{\partial f}{\partial \Omega}\right|_{\Omega=0}\Omega + \mathcal O(\Omega^2).
\end{equation}
This implies that $\Omega^{-1} f$ is also \textit{smooth} and its limit is given by
\begin{equation}
    \lim_{\Omega\to 0}\Omega^{-1}f = \left.\frac{\partial f}{\partial\Omega}\right|_{\Omega = 0}.
\end{equation}
\end{lemma}
\noindent Given the observation \eqref{equ:Weyl_zero} and using the above Lemma, one can define the asymptotic Weyl tensor 
\begin{equation}
	K_{\mu\nu\rho\sigma} := \Omega^{-1} C_{\mu\nu\rho\sigma}\,,
\end{equation}
which is has a smooth non-trivial limit at $\scrip$. Analogously to Sachs, one then constructs (unphysical) NPS using the asymptotic Weyl tensor as
\begin{align}
	{\Psi}_4 &:= K_{\mu\nu\rho\sigma}\,n^{\mu} \bar{m}^{\nu} n^{\rho} \bar{m}^{\sigma}\,,\\
	{\Psi}_3 &:= K_{\mu\nu\rho\sigma}\,\ell^{\mu} n^{\nu} \bar{m}^\rho n^\sigma\,,\\
	{\Psi}_2 &:= K_{\mu\nu\rho\sigma}\,\ell^{\mu} m^{\nu} \bar{m}^\rho n^\sigma\,,\\
	{\Psi}_1 &:= K_{\mu\nu\rho\sigma}\,\ell^{\mu} n^{\nu} \ell^{\rho} m^{\sigma}\,,\\
	{\Psi}_0 &:= K_{\mu\nu\rho\sigma}\, \ell^{\mu} m^{\nu}\ell^{\rho} m^{\sigma} \,.
\end{align}
The latter complex scalars are defined on the conformally completed spacetime and w.r.t. the asymptotic Weyl tensor. It is crucial to note that these are generally not equivalent to the NPS defined w.r.t. the ``normal'' Weyl tensor. One can, however, extract the physical scalars defined on physical spacetime by converting back the asymptotic Weyl tensor and all vectors of the tetrad. Thereby, it is important to acknowledge that for the Weyl tensor to be conformally invariant, the first or last index has to be raised\footnote{For a demonstration of the latter, the reader is again referred to \cite{Our_Review}.}, i.e., $C\du{\mu\nu\rho}{\sigma}= \widetilde C\du{\mu\nu\rho}{\sigma}$. For $\Psi_4$, for instance, one writes 
\begin{equation}
	\Psi_4 = \Omega^{-1} C\du{\mu\nu\rho}{d} n^{\mu} \bar{m}^{\nu} n^{\rho} \bar{m}_\sigma\,,
\end{equation}
which is equivalent to 
\begin{align}
	\Psi_4 = \Omega^{-1} \widetilde{C}\du{\mu\nu\rho}{d} n^{\mu} \bar{m}^{\nu} n^{\rho} \bar{m}_\sigma\,.
\end{align}
Given the relation between physical and unphysical tetrad above, i.e.,
\begin{align}
	n^{\mu} &= \widetilde{n}^{\mu}, & \ell^{\mu} &= \Omega^{-2}\, \widetilde{\ell}^{\mu},  & m^{\mu} = \Omega^{-1}\, \widetilde{m}^{\mu}.
\end{align}
one finds that the scalar $\Psi_4$ can be written in terms of the \textit{physical} fields as
\begin{align}
	\Psi_4 &= \Omega^{-1} \widetilde{C}_{\mu\nu\rho}{}^\sigma \widetilde{n}^{\mu} \left(\Omega^{-1}\,\widetilde{\bar{m}}^\nu\right)\widetilde{n}^{\rho}\left(\Omega\,\widetilde{\bar{m}}_d\right)\notag\\
	&= \Omega^{-1} \widetilde{C}_{\mu\nu\rho}{}^\sigma \widetilde{n}^{\mu} \widetilde{\bar{m}}^\nu\widetilde{n}^{\rho}\widetilde{\bar{m}}_\sigma,
\end{align}
wher $\bar{m}_\mu = \bar{m}^{\nu} g_{\mu\nu} = \left(\Omega^{-1} \widetilde{\bar{m}}^\nu\right)\Omega^2 \widetilde{g}_{\mu\nu} = \Omega\, \widetilde{\bar{m}}_\mu$ is used. One therefore obtains a relation between the physical scalar $\widetilde{\Psi}_4$ and the Weyl scalar w.r.t. the asymptotic Weyl tensor, $\Psi_4$,
\begin{equation}\label{eq:NPscalarpeeling}
	\Psi_4 = r\, \widetilde{\Psi}_4 \,,
\end{equation}
where $r := \frac{1}{\Omega}$. This implies that the physical scalar $\widetilde{\Psi}_4$ decays like
\begin{equation}
    \widetilde{\Psi}_4(r, u,z,\bar z) = \frac{\Psi^\circ_4(u,z,\bar z)}{r} + \mathcal{O}(r^{-2})\,,
\end{equation}
where the coordinate dependence in the chart $(r,u,z,\bar z)$ is made explicit. The latter equation recovers the Sachs Peeling for $\Psi_4$, and an analog computation can be conducted for all other scalars, resulting in \eqref{equ:PSI4}-\eqref{equ:PSI0}. The scalar $\Psi_i^\circ$ always denotes the value of $\Psi_i$ at $\scrip$, i.e., it always stands for the zeroth order of the Taylor expansion of $\Psi_i$ (the scalar w.r.t. the asymptotic Weyl tensor) around $\Omega=0$. Generally, Penrose Peeling implies Sachs peeling along any hypersurface orthogonal \gls{ngc} reaching $\scrip$. Here, $\Psi_4^\circ:= rC_{\mu\nu\rho\sigma}n^\mu m^\nu n^\rho m^\sigma |_{\scrip}$ is the component invariant under $m^\mu$.

As for the Weyl tensor, measuring the curvature of spacetime, the complex scalars \eqref{equ:PSI4}-\eqref{equ:PSI0} can be assigned to a unique physical interpretation \cite{Szekeres_interpretation_NPscalars}. While $\Psi_2$ encompasses mass multipole moments, $\Psi_0, \Psi_4$ correspond to in- and outgoing traverse radiation. $\Psi_1,\Psi_3$ can be interpreted as in- and outgoing longitudinal radiation terms. For asymptotically flat spacetimes containing radiation, however, these two scalars can be trivialized by an appropriate choice of null tetrads. Hence, in this case, $\Psi_1,\Psi_3$ can be viewed as gauge quantities. It is instructive to highlight that the NPS generally depend on the definition of the tetrads, and whether they can be trivialized or not is fundamentally determined by the underlying spacetime. In the above context, the spacetime is chosen to be asymptotically flat, and the tetrad is partially fixed by adapting it to the \gls{ngc}. More generally, however, the NPS can be used to rigorously classify spacetimes according to Petrov \cite{Petrov:2000bs}. While the transformation freedom in the null tetrad boils down to the Lorentz transformation for all spacetimes, the set of Newman-Penrose scalars that can be trivialized through these transformations may vary. The classification into \textit{Petrov types} requires the definition \textit{principal null directions}:
For a general spacetime admitting the notion of a Weyl tensor and a null tetrad $\{\widetilde{\ell},\widetilde{n},\widetilde{m},\widetilde{\bar m}\}$, its Petrov type is determined by the distinct roots $B$ determined by 
\begin{align}
    \widetilde\Psi_0+4B\widetilde\Psi_1 + 6B^2\widetilde\Psi_2 + 4B^3\widetilde\Psi_3 + B^4\widetilde\Psi_4 = 0,
\end{align}
which is computed in a Lorentz frame with $\widetilde\Psi_4\neq0$. In general, the latter equation allows for four complex roots, which might not be distinct and hence can have higher multiplicities. For each root, one can define a principal null direction using the null tetrad as
\begin{align}\label{equ:petrovI}
    \widetilde{k}^\mu = \widetilde{\ell}^\mu + \bar B \widetilde{m}^\mu + B \widetilde{\bar m}^\mu + B\bar B \widetilde{n}^\mu,
\end{align}
 and which satisfies
\begin{align}\label{equ:petrovII}
    \widetilde{k}^\mu \widetilde{k}^\nu \widetilde{k}_{[\rho}\widetilde{C}_{\sigma]\mu\nu[\alpha}\widetilde{k}_{\beta]}=0.
\end{align}
There are six different Petrov types uniquely characterizing a spacetime by the number and multiplicity of distinct principal null directions \cite{Petrov:2000bs}. The number of principal null directions determines which Newman-Penrose scalars can be trivialized via an adequate choice of frame.
\begin{align}
    &\textbf{Type I: } \text{four simple principal null directions} \Leftrightarrow \widetilde \Psi_0=0 \notag \\
    &\textbf{Type II: } \text{one double and two simple principal null directions} \Leftrightarrow \widetilde \Psi_0=\widetilde \Psi_1=0 \notag \\
    &\textbf{Type D: } \text{two double principal null directions} \Leftrightarrow \widetilde \Psi_0=\widetilde \Psi_1=\widetilde \Psi_3=\widetilde \Psi_4=0 \notag \\
    &\textbf{Type III: } \text{one triple and one simple principal null direction} \Leftrightarrow \widetilde \Psi_0=\widetilde \Psi_1=\widetilde \Psi_2=0 \notag \\
    &\textbf{Type N: } \text{one quadruple principal null direction} \Leftrightarrow \widetilde \Psi_0=\widetilde \Psi_1=\widetilde \Psi_2=\widetilde \Psi_3=0 \notag \\
    &\textbf{Type O: } \text{the Weyl tensor vanishes completely} \Leftrightarrow \widetilde \Psi_0=\widetilde \Psi_1=\widetilde \Psi_2=\widetilde \Psi_3=\widetilde \Psi_4=0 \notag 
\end{align}
For \gls{bh} in GR (for instance, the Kerr vacuum), for instance, one usually deals with scenarios in which two distinct principal null vectors can be constructed with multiplicity $2$. These geometries fall into the Petrov type D. The \gls{flrw} models are type O. \\
Generally, there exists another approach to determining the Petrov type of a given spacetime, which yields an equivalent description as above. It amounts to finding the frame for which $\widetilde\Psi_0$ vanishes. This frame can be found by an adequate rotation of $\widetilde \ell^a$ to a vector of the form \eqref{equ:petrovI} satisfying \eqref{equ:petrovII}. Thus, by definition of the principal null directions, given a tetrad such that $\widetilde \Psi_0=0$, one can immediately identify $\widetilde \ell^a$ as a principal null direction. Petrov types and principal null directions are not frame dependent and thus provide an invariant way to classify spacetimes. The Petrov type indicates the maximal set of Newman-Penrose scalars that can be trivialized. However, for a certain spacetime geometry together with an arbitrary null tetrad satisfying specific normalization and orthogonality criteria,  the Petrov type is not automatically exposed. Only once the tetrad is transformed into a suitable frame does the Petrov type become apparent through the vanishing of the maximal set of Newman-Penrose scalars. The simplest instances of asymptotically flat spacetimes containing gravitational radiation fall into Petrov type N. However, more complex solutions including non-trivial dynamics may fall into type II.

In the context of NPS, it is to be emphasized that in general, the definition of the NPS is not unique in the sense that distinct contractions of the Weyl tensor with vectors of the tetrad can result in the same information encapsulated in a given scalar. In Appendix \ref{app_sec:amb_NPS}, it is demonstrated that 
\begin{align}
        \Psi_0 &:= C_{\mu\nu \rho \sigma}n^\mu\Bar{m}^\nu  n^\rho  \Bar{m}^\sigma \\
        \Psi_1 &:= C_{\mu\nu \rho \sigma}\ell^\mu m^\nu \Bar{m}^\rho m^\sigma = C_{\mu\nu \rho \sigma}\ell^\mu n^\nu \ell^\rho m^\sigma \,,\\
        \Psi_2 &:= \frac{1}{2}C_{\mu\nu \rho \sigma}\ell^\mu n^\nu (\ell^\rho n^\sigma-m^\rho \Bar{m}^\sigma)= \frac{1}{2}C_{\mu \nu \rho \sigma}m^\mu \Bar{m}^\nu (m^\rho \Bar{m}^\sigma- \ell^\rho n^\sigma) \notag\\&= C_{\mu \nu \rho \sigma}m^\nu  \ell^\mu n^\rho \Bar{m}^\sigma,\label{equ:pups}\\
        \Psi_3 &:= C_{\mu \nu \rho \sigma}n^\mu \Bar{m}^\nu n^\rho \ell^\sigma = C_{\mu \nu \rho \sigma}n^\mu \Bar{m}^\nu m^\rho  \Bar{m}^\sigma,\\
        \Psi_4 &:= C_{\mu \nu \rho \sigma}n^\mu \Bar{m}^\nu  n^\rho  \Bar{m}^\sigma \,.\label{equ:double_pups}
    \end{align}
The NPS, in particular $\Psi_4$, hold a key role in \gls{gw} physics as they can be related to direct observational quantities in \gls{gw} interferometer. In Section \ref{subsec:Radiative_modes}, the interplay between $\Psi_4$ and the \gls{gw} shear is elaborated. The scalar $\Psi_2$ encompasses information relevant for the computation of the \gls{gw} memory, as shown in Section \ref{sec:Covariant_PS}. 

Before continuing with the analysis of the universal structure, it is worth mentioning that Peeling Theorem as a concept does not exclusively pertain to \gls{gr}. To stay within the realm of massless gauge theories, another instance of peeling is provided in Maxwell's theory of electromagnetism. There, the Peeling Theorem is a direct consequence of the smoothness of the Maxwell 2-form $F_{\mu\nu}$ regarding asymptotic limits. The latter is a result of the conformal invariance fundamental to Maxwell's theory\footnote{Note that in \gls{gr}, as demonstrated above, smoothness of the Weyl tensor does not immediately follow based on its fundamental properties but instead can only be derived using additional assumptions.}. Given this simplified setup, one can define the NPS of Maxwell's theory as
\begin{align}\label{equ:NPS_EM_Maxwell}
	\Phi_2 &:=  F_{\mu\nu}  n^{\mu} \bar{m}^\nu\notag\\
	\Phi_1 &:= \frac12  F_{\mu\nu}\left( n^{\mu} \ell^{\nu} +  m^{\mu}\bar{m}^{\nu}\right)\notag\\
	\Phi_0 &:=  F_{\mu\nu}  m^{\mu}  \ell^{\nu}\,.
\end{align}
These three complex scalars represent the information contained in the six components of $F_{\mu\nu}$. Defining the functions at $\scrip$, one simply finds
\begin{align}
	\Phi^\circ_i(u,z, \bar z) &:= \left.\Phi_i(u, \Omega,z,\bar z)\right|_{\scrip}\,, & \text{for } i\in\{0,1,2\}\,.
\end{align}
The latter functions capture the leading order behavior of~$\Phi_i$ at~$\scrip$ which can be seen by performing a Taylor expansion of $\Phi_i$ around $\Omega=0$, which, by the Lemma above, gives us
\begin{equation}
	\Phi_i = \Phi^\circ_i + \left.\frac{\dd\Phi_i}{\dd \Omega}\right|_{\Omega=0}\cdot\Omega + \mathcal{O}(\Omega^2)\,.
\end{equation}
Relating the Newman-Penrose scalars $\Phi_i$ of the conformally completed spacetime $(\mathcal{M}, \eta_{\mu\nu})$ to the analogously defined physical scalars $\widetilde{\Phi}_i$ of the physical spacetime $(M, \widetilde{\eta}_{\mu\nu})$ is straightforward due to Maxwell's conformal invariance. One solely needs to rescale the null tetrad appearing in the definition of the scalars. It is immediately found that 
\begin{align}\label{eq:PeelingMaxwell}
	\widetilde \Phi_2 &= \widetilde F_{\mu\nu} \widetilde{n}^{\mu} \widetilde{\bar{m}}^{\nu} = \frac{1}{r} F_{\mu\nu}  n^{\mu} {\bar m}^{\nu} = \frac{\Phi_2}{r} = \frac{\Phi^\circ_2(u, z, \bar z)}{r} + \O(r^{-2})\,,\notag\\
	\widetilde \Phi_1 &= \frac{{\Phi}^\circ_1(u,z, \bar z)}{r^2} + \O(r^{-3})\,,\notag\\
	\widetilde \Phi_0 &=  \frac{{\Phi}^\circ_0(u,z, \bar z)}{r^3} + \O(r^{-4})\,.
\end{align}
The functions $\widetilde\Phi_i$ are the physical Newman-Penrose scalars which carry the same information as the physical Maxwell $2$-form $\widetilde{F}_{\mu\nu}$ and fall-off in a characteristic manner. One could say that as one approaches $\scrip$, the components of $\widetilde F_{\mu\nu}$ are ``peeled off'' at different rates. This is the Peeling Theorem for electromagnetism. The unique scaling in $1/r$ for each scalar offers a first clue that $\Phi_2(u, r, z, \bar z)$ encodes the radiative modes while $\Phi_1(u,r,z, \bar z)$ carries information about Coulombic modes. This is because the radiation field decays like $\frac{1}{r}$ in the radiation zone while the Coulomb field behaves like $\frac{1}{r^2}$, as discussed already in Chapter \ref{chap:Intro}. Further, evidence for this claim results from rewriting Maxwell's equations at $\scrip$ in terms of these scalars, i.e., 
\begin{align}\label{equ:maxwell_NPS}
	&\partial_u{\Phi}^\circ_1(u,z, \bar z) = \eth {\Phi}^\circ_2(u,z, \bar z) \notag\\
	&\partial_u{\Phi}^\circ_0(u,z, \bar z) = \eth {\Phi}^\circ_1(u,z, \bar z)\,,
\end{align}
where $\eth$ is the spin-weighted angular derivative operator defined by \cite{Newman_Penrose},
\begin{align}
    \eth\,_s Y_{\ell m} = \sqrt{(l-s)(l+s+1)}\,_{s+1}Y_{\ell m}\,,\\
    \bar \eth\,_s Y_{\ell m} = -\sqrt{(l+s)(l-s+1)}\,_{s-1}Y_{\ell m}\,,
\end{align}
with $\,_{s}Y_{\ell m}$ being the spin-weighted\footnote{A more thorough explanation of the spin-weight is given after Eq. \eqref{equ:ferrari_porsch}. See also \cite{Our_Review}.} spherical harmonics.
In this form, Eqs. \eqref{equ:maxwell_NPS} are first-order equations for $\Phi^\circ_0$ and $\Phi^\circ_1$ which, in order to be solved, require initial data. Further, $\Phi^\circ_2$ is not determined by the Maxwell equation and needs to be specified by hand everywhere on $\scrip$ for a valid solution. Given  $\Phi^\circ_2$ is undetermined by the equations, it has to represent the radiative mode content of the theory. A more elaborate analysis of Eqs. \eqref{equ:maxwell_NPS} and its meaning in the context of electromagnetic radiation is found in \cite{Our_Review}.

\subsection{Symmetries at Null Infinity}
\label{subsec:BMS_Group}
Continuing with the discussion around the universal structure after this interlude, a natural next step is the identification of the transformation under which it is invariant. To find the symmetries of the universal structure one, thus, has to simply identify the class of vector fields that, upon Lie dragging, preserve the metric and null vector on $\scrip$ up to a rescaling as defined in Eq. \eqref{equ:definition_of_the_universal_structure}. Concretely, first postulated by Geroch \cite{Geroch_1977}, these vector fields $X^\mu$ have to satisfy
\begin{align}
\label{equ:lie_algebra_BMS}
    \lie _X q_{\mu \nu} = 2 k q_{\mu\nu}\,, && \lie _X n^\mu = -k n^\mu\,,
\end{align}
for some function $k(x^\mu)$ on $\scrip$ \footnote{A motivation for the form of Eq. \eqref{equ:lie_algebra_BMS} is provided in \cite{Our_Review} via introducing a one-parameter family of diffeomorphisms generated by the vector field $X$ in question. This way of rewriting the vector field $X$ will become important at a later stage again, see Section \ref{subsec:Ashtekar_Streubel}.}. Upon restricting to the case where $\lie_n f=0$ (were $f$ appears in the definition of the universal structure \eqref{equ:definition_of_the_universal_structure}), one finds that $k |_{\scrip}= k(z, \bar z)$ \cite{Our_Review}. With the above choice of coordinates, the vector fields satisfying Eqs. \eqref{equ:lie_algebra_BMS} can be elegantly rewritten \cite{Geroch_1977}:
\begin{proposition}
    The Lie algebra of infinitesimal symmetries of the universal structure satisfying Eq. \eqref{equ:lie_algebra_BMS} is generated by 
    \begin{align}
    \label{equ:lie_algebra_vector}
        X = \left(\alpha(z,\bar z) + \frac{u}{2}(\partial_z \chi + \partial_{\bar z} \bar \chi)\right)\partial_u + \chi(z)\partial_z+ \bar \chi (\bar z) \partial_{\bar z}\,,
    \end{align}
    where $\alpha(z,\bar z)$ is a function on $\mathcal{S}^2$ and $\chi(z)\partial_z = (a+bz+cz^2)\partial_z$ is a generator of a Möbius transformation $PSL(2,\mathbb C)\simeq SO(3,1)$. The Lie algebra generated by Eq. \eqref{equ:lie_algebra_vector} is isomorphic to 
    \begin{align}
    \label{equ:algebra_BMS}
        \text{Lie}\left(\text{so}(3,1)\right) \ltimes \mathcal{C}^\infty (\mathcal S^2)\,.
    \end{align}
\end{proposition}
\noindent \textit{Proof.} The functional expression for the generator $X$ is obtained by direct computation in said coordinate system. Setting $k =0$ in Eq. \eqref{equ:lie_algebra_BMS}, an ansatz for $X$ is $X^\mu = \alpha n^\mu$. While $\lie_X q_{\mu\nu}=0$ is automatically satisfied for this ansatz, $\lie_X n^\mu = 0$ yield $n^\mu\D_\mu \alpha = 0$ which results in $\alpha = \alpha(z,\bar z)$. For the remaining contributions, one considers $\alpha= \alpha(z,\bar z)$. As $z\in \mathbb C$, $\alpha$ can be expanded as a polynomial up to second order. A similar argument holds for $\chi(z),\bar \chi(\bar z)$, which are general complex functions of $z,\bar z$ respectively. One can expand the remaining part of $X^{\mu}(k\neq 0)$ in terms of $n^\mu,m^\mu,\bar m^\mu$ with corresponding prefactors $\alpha_1,\alpha_2,\alpha_3$ being functions of $(u,z,\bar z)$. Computing first the prefactors of $m^\mu$ and $\bar m^\mu$ one finds $\alpha_2 = \chi(z)$ and $\alpha_3$ respectively. For the part proportional to $n^\mu$, $\lie_X q_{\mu\nu}$ then yields $\alpha_1 \sim [\partial_z \chi(z)+ \partial_{\bar z} \chi(\bar z)]$ and $\lie_X n^\mu$ results in the linear dependence on $u$. Given \eqref{equ:lie_algebra_vector}, one separates into $X_\alpha = \alpha(z,\bar z)\partial_u$ and $X_\chi = \frac{u}{2}(\partial_z \chi + \partial_{\bar z} \bar \chi)\partial_u + \chi(z)\partial_z+ \bar \chi (\bar z) \partial_{\bar z}$. Then, one finds that $[X_{\xi_1},X_{\xi_2}]=0$ and, thus, the generators $X_\xi$ for an infinite dimensional abelian Lie algebra isomorphic to $\mathcal{C}^\infty (\mathcal S^2)$. Further, $[X_{\chi_1},X_{\chi_2}] = X_{(\chi_1 \partial_z \chi_2 - \chi_2 \partial_z \chi_1)}$. Therefore, the generators $X_\chi$ form an $\text{so}(3,1)$ Lie algebra. Finally, $[X_\chi, X_\alpha] = \left((\chi \partial_z + \bar \chi \partial_{\bar z})\xi - \frac{1}{2}(\partial_z \chi + \partial_{\bar z}\bar \chi)\alpha) \right) \partial_u$, which implies that the algebra generated by $X_\alpha$ forms an ideal of the full algebra. Thus, one finds the semidirect product above. 

\noindent In Section \ref{sec:Covariant_PS}, the vector field $X^\mu$ generating the symmetries of the universal structure at $\scrip$ is denoted as $\xi^\mu$. The algebra \eqref{equ:algebra_BMS} is commonly referred to as the BMS (Lie) algebra, which acts as the generator for the BMS group. As a consequence, the BMS group can be written as a semidirect product of the Lorentz group $\mathcal L$ and the group of \textit{supertranslations} $\mathcal S$, generated by $X_\alpha$, i.e.,
\begin{equation}
    \mathcal B = \mathcal S \rtimes \mathcal L \,.
\end{equation}
Note at this point that on $\scrip$, one has $X_\alpha^\mu = \alpha n^\mu$ where $n^\mu$ is part of the chosen tetrad. The name ``supertranslations'' thereby is no coincidence as this subgroup is directly related to the symmetry group of flat spacetime, i.e., the Poincar\'e group. Analogous to the BMS group, the latter represents the semidirect product of the Lorentz group with the group of translations. For the Poincar\'e case, the algebra associated with translations, in fact, is also abelian and ideal, just as the supertranslation subgroup of the BMS group. The Lorentz transformations in both cases can then be obtained by taking the quotient group of the full symmetry group, removing the associated (super-)translation ideal. \\
Curiously, when the BMS group was first obtained by Bondi, Metzner, and Sachs, it was quite a surprise to find an infinite-dimensional enlargement of the Poincar\'e group. Naively, it was expected that the bulk symmetry group of flat spacetime, i.e., the Poincar\'e group, would simply be recovered at the boundary of asymptotically flat spacetimes. Evidently, this is not the case, but instead an infinite tower of new symmetries presents itself at $\scrip$ when the metric is asymptotically flat. This fact intuitively presents itself when decomposing $X_\alpha$ into spherical harmonics. Since $\alpha$ is a function on $\mathcal S^2$, one can simply write
\begin{align}\label{equ:shiat}
    X = \sum_{\ell,m} \alpha_{\ell,m} Y_{\ell,m} \partial_u
\end{align}
Evidently, the first for components in the above sum are given by $\{-Y_{0,0}, Y_{1,1}, Y_{1,-1},$ $ Y_{1,0}\}$ which, in real angular coordinates correspond to $\{-1, \sin\theta\cos\phi, \sin\theta\sin\phi, \cos\theta\}$ where $\{-1, \sin\theta\cos\phi, \sin\theta\sin\phi, \cos\theta\}\partial_u$ are the translation Killing vector fields of Minkowski space in the coordinate system $(u,\theta,\phi)$.
Therefore, the algebra of translations known from the Poincar\'e algebra is a subalgebra of the newly discovered supertranslation algebra.

At this point, it is worth pointing out that in literature one can find multiple attempts of extending the symmetry algebra of the BMS group to obtain more generalized and even richer symmetry structures. One of the most prominent generalizations is the extension of the Lorentz subgroup to the superrotations subgroup \cite{Superrrotations}. Analogously to the promotion of the translations to supertranslations, one finds an infinite extension of the rotation subgroup to the superrotations. This extension is a direct consequence of loosening the restriction to globally well-defined generators, as it was intrinsically done above\footnote{The choice of Killing vector fields is closely related to the ``choice'' of topology. In the above, it was assumed that $\scrip$ takes the topology $\mathcal S^2 \times \mathbb{R}$, which is a direct result of Minkowski spacetime. However, for asymptotically flat spacetimes, this can in fact be generalized (consider, for instance, a flat manifold $\mathcal M = \mathbb R \times \mathbb R^2 \times \mathbb S^1$). Note however that this particular choice transforms the confinement of the Killing vector fields from arbitrary holomorphic function (which would lead to Virasoro algebra) to global conformal transformations on the sphere (leading to $SL(2,\mathbb C)/\mathbb Z_2$, which is isomorphic to the desired Lorentz group) \cite{Ashtekar_topo_killing}.}. If one drops this assumption, the symmetry algebra of the BMS group becomes a semi-direct sum of the infinitesimal \textit{local} conformal transformations of the unit $2$-sphere with the abelian ideal of supertranslations, both being infinite-dimensional. In their original works, Bondi, Metzner and Sachse evaded the consideration of local conformal transformations by requiring a gauge \cite{Bondi_Origin_2, Sachs_Origin_2} in which the determinant of $\dd \Sigma^2$ corresponds to the determinant of the unit $\mathcal{S}^2$ sphere. However, one could keep the non-radial part of the determinant arbitrary as it was justified geometrically by Penrose \cite{Penrose_1962}. Then one finds similar Killing vector fields as above, but distinguished into two cases: 

i) Restricted to globally well-defined transformations on the unit sphere, one singles out the group of global conformal transformations isomorphic to $SL(2, \mathbb{C})/\mathbb{Z}_2$, which in turn is isomporphic to the proper, ortochronous Lorentz transformation $\text{SO}(3,1)$. In this case, the generators of the BMS group contain the Lorentz and supertranslation generators.  

ii) One can also allow for not necessarily invertible holomorphic mappings. Focusing on the local properties, it is found that an infinite-dimensional extension of the Lorentz subgroup, the \textit{superrotations}, enters the BMS group. In fact, with this choice, the BMS algebra modulo the abelian ideal of supertranslations is not the Lorentz algebra, as in i), but the infinite-dimensional Virasoro algebra.

In Section \ref{sec:Covariant_PS}, the connection between the symmetries i) and their physical manifestation as the \gls{gw} memory is discussed. The derivation there analogously applies to ii), resulting in a new type of (spin) memory. In this context, recent works \cite{Mem_outlook_V} proposed an experimental investigations with future \gls{gw} instruments to distinguish between i) and ii), i.e., to determine which symmetry group best describes the real astrophysical scenarios probed by state-of-the-art detectors based on observation or non-observation of different types of memory.


%
%
%


\section{Radiative Modes at Null Infinity}
\label{subsec:Radiative_modes}
The above treatment concludes the discussion of the universal structure obtained by conformally compactifying the physical spacetime to include a horizon $\scri$. The tetrad is constructed for both the physical and unphysical metric. The group of killing vector fields for the unphysical metric at $\scrip$ is determined, and physical information of the corresponding spacetime is retrieved by the analysis of the peeling properties of the independent components of the Weyl tensor. Yet, the challenge regarding the identification of the radiative degrees of freedom that is thoroughly described in Section \ref{sec:intro_radiation} is not addressed in a satisfactory manner. One could, in principle, argue that, as for the NPS in Maxwell's theory, the fall-off condition of the physical scalars and the structure of the equations of motion expressed in terms of NPS suggest that $\Psi_4$ captures gravitational radiation. However, deriving a formal mathematical argument requires further elaboration. In this Section, the issue regarding a formal definition of gravitational radiation is thoroughly addressed.

Intuitively, it is clear that radiative information cannot be encapsulated in any way in the universal structure outlined in previous Sections, as this structure also admits spacetimes which are devoid of gravitational radiation. What is left then? Given a standard \gls{gr} spacetime, the main geometric structure is indeed given by $(\mathcal M, \gd)$, which includes information about the universal structure under the premise that $\gd$ satisfies the conditions outlined in Section \ref{subsec:asymptotical_flatness}. There is, however, another piece of information crucial to deriving any dynamics on the said spacetime manifold: the derivative operator $\nabla$. Mathematically, the choice of derivative operator includes a lot of freedom, which has immediate physical consequences. See \cite{Heisenberg:2018vsk} for a thorough review. In \gls{gr}, this freedom is partially eliminated by choosing $\nabla$ to be the covariant derivative associated with the Levi-Civita connection. In fact, in \gls{gr} a given metric completely determines the Levi-Civita connection and thus also $\nabla$ and the Riemann curvature tensor. Hence, one might argue that the choice of $\gd$, at least in \gls{gr}, completely erases the freedom in the choice of the covariant derivative operator. This is true for physical spacetime $(M,\widetilde{g}_{\mu\nu})$. For the conformally compactified one, however, there exists a regime where this is not the case. The boundary $\scrip$ is endowed with a universal structure $(n^\mu,q_{\mu\nu})$, which, as before, would fully determine also the covariant derivative operator at $\scrip$. However, unlike $\gd$, $q_{\mu\nu}$ is degenerate. How this degeneracy impacts the definition of a derivative operator at $\scrip$, denoted as $\D$, is the subjective of the following Subsection.\\
Before diving into this matter, it is instructive to realize how the information is split between $q_{\mu\nu}$ and $\D$ as one moves towards $\scrip$: In the bulk spacetime, all information about spacetime geometry is encoded in $\gd$. Pulling $\gd$ back to $\scrip$, results in a degenerate metric, and, thus, information is lost. This loss of information is, however, essential for formulating the universal structure, as it allows for classifying spacetimes in a broader sense. Given that the universal structure does not contain information about the presence of radiation, it seems that this information is extracted from the metric when pulling back to $\scrip$. Another pathway to transport metric information to $\scrip$ is via the pullback of the covariant derivative $\newpb{\nabla} =: \D$. Below, it is demonstrated that this particular information being ``transferred'' to $\scrip$ via the derivative operator complements $q_{\mu\nu}$ by the information about the radiative degrees of freedom and thus, the covariant derivative operator at $\scrip$ is able to distinguish between spacetimes with and without gravitational radiation present.

\subsection{Covariant Derivative at $\scrip$}
Starting from conformally compactified spacetime $(\mathcal M, \gd)$ endowed with derivative operator $\nabla$, the metric intrinsic to $\scrip$ is computed via the pullback $\newpb{\gd}=:q_{\mu\nu}$. Due to its degeneracy, it is immediately clear that the $q_{\mu\nu}$ cannot define a covariant derivative based on its corresponding Levi-Civita connection as $q_{\mu\nu}$ does not possess a unique inverse. Instead, one starts the construction by $\newpb{\nabla} =: \D$ and checks for well-posedness: From $\nabla$, the derivative $\D$ inherits its action as a directional derivative, i.e., acting on a scalar $f$ it holds that $v^\mu\D_\mu f = \lie_v f =v^\mu\partial_\mu f$ (where $v^\mu$ is a vector intrinsic to $\scrip$), its linearity and the Leibniz property. It can be further shown that $\D$ has a well-defined action on 1-forms at $\scrip$, i.e., $\D_\mu\omega_\nu$ is a tensor field intrinsically defined at $\scrip$ \cite{Our_Review}. For $\D$ to be a valid candidate for the covariant derivative operator at $\scrip$, it must further be generalized to map $(p,q)$ to $(p+1,q)$ tensor fields on $\scrip$. The 
latter is a non-trivial generalization in the presence of a degenerated metric.
\begin{proposition}
    The covariant derivative operator at $\scrip$ defined by $\newpb{\nabla_\mu} =: \D_\mu$ defines a map from $(p,q)$ to $(p+1,q)$ tensor fields on $\scrip$ that is solely determined by the operator's action on a 1-from on $\scrip$.
\end{proposition}
\noindent \textit{Proof.}~As an arbitrary tensor on $\scrip$ contains components from tangent and cotangent space, it is left to show that the action of $\D$ on tangent vectors is fully determined by its action on elements of the cotangent space. To that end, consider a $1$-form $\omega_a$ and assume its derivative on $\scrip$, $\mathcal{D}_\mu\omega_\nu$, is well-defined and known. By definition, $\omega_\mu$ belongs to the cotangent space and hence is always tangent to $\scrip$ in the sense that $n^\mu\omega_\mu=0$. Then, one finds that
\begin{align}
    q_{\nu \rho}n^\rho\mathcal{D}_\mu v^\nu=0,
\end{align}
which immediately follows from $n^\rho$ being the null direction of $q_{\mu\nu}$, i.e. $n^\mu q_{\mu\nu}=0$. However, this also implies
\begin{align}
    n^\nu\mathcal{D}_\mu v_\nu=0,
\end{align}
since $\mathcal{D}_\mu q_{\nu\rho}=0$. Thus, the derivative of a vector tangent to $\scri^+$ is itself intrinsic to $\scri^+$. This dramatically simplifies the proof as now one only has to show that the derivative of $v^\mu$ contracted with a member of the cotangent space of $\scri^+$ is fully determined by the derivative of such a $1$-form. That is, 
\begin{align}
    \omega_\nu\mathcal{D}_\mu v^\nu
\end{align}
is fully determined by the derivative $\mathcal{D}_a$ acting on a 1-form.
One can rewrite
\begin{align}
    \omega_\nu\mathcal{D}_\mu v^\nu =\omega_\nu\mathcal{D}_\mu\left(q^{\nu \rho }v_\rho\right)=
    \omega_\nu q^{\nu\rho}\mathcal{D}_\mu v_\rho =\omega^\rho \mathcal{D}_\mu v_\rho
\end{align}
where $v_\rho$ is a well defined $1$-form as $v_\rho n^\rho=v^\mu q_{\mu\rho}n^\rho=0$ from $q_{\mu\rho}n^\rho=0$ and similarly $\omega^\mu$ is a tangent vector to $\scri^+$. Thus, the derivative of a tangent vector can be traced back to the derivative of an $1$-form. The latter equation holds for every pseudo-inverse $q^{\mu\nu}$ as
\begin{align}
\mathcal{D}_\mu q'^{\nu\rho}=\mathcal{D}_\mu \left(q^{\nu\rho}+t^{(\nu}n^{\rho)}\right)=\mathcal{D}_\mu q^{\nu\rho}+\mathcal{D}_\mu{t^{(\nu}n^{\rho)}}=n^{(\rho}\,\mathcal{D}_\mu t^{\nu)}=0.
\end{align}
Here, $t^\mu$ is again a vector tangent to $\scri^+$. The last step follows from the action of the derivative $\mathcal{D}$ on a tangent vector trivializing, which was shown above. 

\noindent Therefore $\D$ defines a well-defined derivative operator at $\scrip$. Note, however, that it is not the covariant derivative w.r.t. the Levi-Civita connection. Despite being metric-compatible 
\begin{align}
    0 = \newpb{\nabla_\mu g_{\nu\rho}}  = \D_\mu q_{\nu\rho}\,,
\end{align}
and torsion-freeness,
\begin{equation}
    0 = \newpb{\nabla_{[\mu}\nabla_{\nu]}f } = \D_{[\mu}\D_{\nu]} f\,,
\end{equation}
there is still no non-degenerate inverse of $q_{\mu\nu}$. In fact, any tensor $q^{ab}$ which satisfies
\begin{align}\label{eq:PseudoInv}
    q^{\mu\nu} q_{\mu\rho} q_{\nu\sigma} = q_{\rho\sigma}
\end{align}
is a \textit{pseudo-inverse} of $q_{\mu\nu}$. Suppose one finds one tensor $q^{\mu\nu}$ which satisfies condition~\eqref{eq:PseudoInv}. Then it follows immediately that $q'^{\mu\nu} = q^{\mu\nu} + t^{(\mu} n^{\nu)}$ is also a solution to~\eqref{eq:PseudoInv}, where $t^{\mu}$ is an arbitrary vector tangential to $\scrip$, i.e., $t^{\mu}n_\mu = 0$. The reason for this ambiguity is, of course, the degeneracy of $q_{\mu\nu}$ and the fact that $n^{\nu}$ is the null direction of $q_{\mu\nu}$ in the sense that $q_{\mu\nu}n^{\nu} = 0$.

So far, the uniqueness of the operator $\D$ has yet to be established. Summarizing the progress so far, one finds that any covariant derivative operator on $\scrip$ necessarily satisfies 
\begin{align}\label{equ:criteria_D}
        \D_{[\mu}\D_{\nu]}f &= 0, & && \D_\mu q_{\nu\rho} &= 0, & && \D_\mu n^{\nu} &= 0\,.
\end{align}
The latter restricts the form of $\D$, but does not fix it completely. To determine the ambiguity left in the definition of $\D$, one first considers its action on 1-forms $\omega_\mu$ which are transverse to $\scrip$, i.e., $\omega_\mu n^{\mu}|_{\scrip} =0$.
\begin{proposition}
    For any two derivative operators $\D,\D'$ defined at $\scrip$ via $\newpb{\nabla_\mu} =: D_\mu$ and satisfying \eqref{equ:criteria_D}, any 1-form $\omega_\mu$ transverse to $\scrip$, i.e., satisfying $\omega_\mu n^\mu=0$, yields
    \begin{align}
        \left(\D_\mu- \D'_\mu\right) \omega_\nu = 0\,.
    \end{align}
\end{proposition}
\noindent \textit{Proof.}~Consider an arbitrary $1$-form $\omega_\mu$ living on the co-tangent space of $\scrip$ satisfying only $n^\mu\omega_\mu=0$. Since $\{\ell_\mu,m_\mu,\Bar{m}_\mu\}$ span a basis on this space, it follows from the cross normalization properties of the tetrad that $\omega_\mu=fm_\mu+\Bar{f}\Bar{m}_\mu$ where $f$ is an arbitrary complex functions. As before, one can write
\begin{align}\label{eq:ex5.4Lie}
\D_\mu\omega_\nu=\D_{(\mu}\omega_{\nu)}+\D_{[\mu}\omega_{\nu]}=\lie_{\omega'}q_{\mu\nu}+\D_{[\mu}\omega_{\nu]},
\end{align}
where $\omega'^\mu=q^{\mu\nu}\omega_\nu$. First, notice that with $q'^{\mu\nu}=q^{\mu\nu}+t^{(\mu}n^{\nu)}$ one finds that 
\begin{align}
    \Tilde{\omega}'^\mu= q'^{\mu\nu}\omega_\nu= \omega'^\mu+t^{(\mu}n^{\nu)}\omega_\nu = \omega'^\mu + \underbrace{t^\nu\omega_\nu}_{=h}n^\mu\,,
\end{align}
where $h$ is some arbitrary function. Since $\lie_n q_{\mu\nu}=0$, so is
\begin{align}
    \lie_{f n}q_{\mu\nu} = f\underbrace{\lie_n q_{\mu\nu}}_{=0} + \underbrace{q_{\mu\rho}n^{\rho}}_{=0}\left(\D_\nu f\right) + \underbrace{q_{\rho\nu}n^{\rho}}_{=0}\left(\D_\mu f\right) = 0\,.
\end{align}
Hence, there is no ambiguity in $\omega'_\mu$. Having a closer look at $\lie_{\omega'}q_{\mu\nu}$ one finds
\begin{align}
   \lie_{\omega'}q_{\mu\nu} &=\omega'^\rho \underbrace{\D_\rho q_{\mu\nu}}_{=0}+\D_\mu\omega'_\nu+\D_\nu\omega_\mu\notag\\
   &=\partial_\mu\omega'_\nu+\partial_\nu\omega'_\mu-2\Gamma^c_{\mu\nu}\omega'_\rho\notag\\
   &=\partial_\mu\omega'_\nu+\partial_\nu\omega'_\mu -q^{\rho \sigma}(\partial_\mu q_{\nu\sigma}+\partial_\nu q_{\mu\sigma}-\partial_\sigma q_{\mu\nu})\omega'_\rho  \notag\\
   &=\partial_\mu\omega'_\nu+\partial_\nu\omega'_\mu +\omega'^\rho \partial_\rho q_{\mu\nu}-\omega'^\rho(\partial_\mu q_{\nu\rho}+\partial_\nu q_{\mu\rho})  \notag\\
   &=\partial_\mu \omega'_\nu+\partial_\nu\omega'_\mu+\omega'^\rho\partial_\rho q_{\mu\nu}-\partial_\mu(\omega'^\rho q_{\nu\rho})-\partial_\nu (\omega'^\rho q_{\mu\rho }) +q_{\mu\rho}\partial_\nu\omega'^\rho+q_{\nu\rho}\partial_\mu\omega'^\rho  \notag\\
   &=\omega'^\rho\partial_\rho q_{\mu\nu}+q_{\mu\rho}\partial_\nu\omega'^\rho+q_{\nu\rho}\partial_\mu \omega'^\rho.
\end{align}
Here, the connection drops out completely, and one essentially obtains the Lie derivative written with partial instead of covariant derivatives. This can generally be done for any Lie derivative, but here it nicely demonstrates the connection dropping out. Thus, the first part of \eqref{eq:ex5.4Lie} is indeed connection independent.\\
The same holds trivially for the second part as 
\begin{align}
    \D_{[\mu}\omega_{\nu]}=\partial_{[\mu}\omega_{\nu]}-\Gamma^\rho_{[\nu\mu]}\omega_\rho=\partial_{[\mu}\omega_{\nu]},
\end{align}
since the connection is metric-compatible and torsion-free. 

\noindent Thus, the tensor $\D_\mu\omega_\nu$ is independent of the choice of the covariant derivative operator at $\scrip$, i.e.,
\begin{align}\label{equ:ambig_D}
    \left(\D_\mu- \D'_\mu\right) \omega_\nu = 0\,.
\end{align}
Therefore, $\D$'s action on vectors transverse to $\scrip$ does not fix the remaining ambiguity in the definition of $\D$ as any of such derivative operators applied on $\omega$ gives a trivial result. Thus, there must be another condition complementing \eqref{equ:criteria_D}. Given the basis $\{\ell_\mu,n_\mu,m_\mu,\bar m_\mu\}$, it is clear that discarding the transversality condition of $\omega_\mu$ inevitably results in the computation of $\D_\mu\ell_\nu$, for which there is no immediate solution as above. Note, however, that in principle the choice of the metric fixes $\ell_\mu$ and the pullback of the covariant derivative operator $\nabla_\mu$ gives the operator $\D$. In this sense, $\D_\mu\ell_\nu$ is fully determined. At this point, the reader is reminded that since $\ell_\mu$ is not part of the universal structure, one can intuitively think of $\D_\mu\ell_\nu$ as breaking $\D$'s universality with regard to its action on 1-forms (which is determined by $q_{\mu\nu}$). 
Therefore, the suspicion that $\ell_\mu$ carries the desired (radiative) information to $\scrip$ arises. If true, $\D_\mu\ell_\nu$ would fix the derivative operator $\D$ at $\scrip$ completely (in combination with \eqref{equ:criteria_D}) and allows for the distinction of different asymptotically flat spacetimes. The ambiguity in \eqref{equ:ambig_D} can be seen as a result of selecting different conformal compactifications for a given spacetime. The latter leaves the universal structure invariant as it admits the fundamental rescaling freedom $(n^{\mu},q_{\mu\nu})\mapsto ( n'^{\mu\nu},q'_{\mu\nu}) = ( f(u,z,\bar z)^{-1}\,n^{\mu}, f(u,z,\bar z)^2 q_{\mu\nu})$, but changes the derivative operator's action. By nature of the conformal compactification, this change of the derivative operator does not affect transverse vectors.

\subsection{Identifying Radiative Modes at $\scrip$}
\label{subsec:radi_degrees}
Recapitulating the results so far, evidence hints that the derivative operator on null infinity incorporates information about the radiative modes of the gravitational field in the bulk. Finding a well-defined derivative operator on $\scrip$ leads to the identification of many such operators. In fact, any torsion-free and metric-compatible operator that satisfies $\D_\mu n^{\nu} = 0$ is admissible. 
In an attempt to quantify the ambiguity, the operator's action on $1$-forms $\omega_\mu$ on $\scrip$ transverse to $n^{\mu}$, i.e., which satisfy $\omega_\mu n^{\mu}=0$, is analyzed. In doing so, it is found that all derivative operators on $\scrip$ have the same action on such $1$-forms. However, the actions of $\D$ and $\D'$ on $\ell_\mu$ (pointing off $\scrip$, see Fig. \ref{fig:Pyzel_Gremlin}), which is a $1$-form which is not transverse to $n^{\mu}$, are generally different. Thus, $\D_\mu\ell_\nu$ being ``non-universal'' allows for the distinction between spacetimes with and without radiation. It is where radiative modes are revealed. 

Proving the latter statement, one proceeds as follows: Let $\D$ and $\D'$ be two distinct, torsion-free, and metric-compatible covariant derivative operators on $\scrip$ which satisfy $\D_\mu n^{\nu} = 0 = \D'_\mu n^{\nu}$. Furthermore, let $\alpha_\mu$ be any $1$-form on the cotangent space of $\scrip$. As the difference between any two covariant derivative operators is a tensor, one can write
\begin{align}\label{eq:DifferenceInDerivativesI}
    (\D'_\mu - \D_\mu)\alpha_\nu = C\du{\mu\nu}{\rho} \alpha_\rho\,,
\end{align}
where $C\du{\mu\nu}{\rho}=C\du{(\mu\nu)}{\rho}$ is a tensor and symmetric because the connections used to construct $\D$ and $\D'$ are torsion-free. The action of $\D$ and $\D'$ on $1$-forms $\omega_\mu$ which are transverse to $\scrip$ is universal in the sense that $\left(\D'_\mu-\D_\mu\right)\omega_\nu = 0$ can further be used to conclude
\begin{align}\label{7.3}
    \underbrace{(\D'_\mu - \D_\mu)\omega_\nu}_{=0} &= C\du{\mu\nu}{\rho}\omega_\rho &\Longleftrightarrow&&  C\du{\mu\nu}{\rho} = \Sigma_{\mu\nu} n^\rho\,,
\end{align}
i.e., the tensor $C\du{\mu\nu}{\rho}$ is the product of a symmetric tensor $\Sigma_{\mu\nu}$ and the null normal $n^{\rho}$. By its very construction, this ensures that $C\du{\mu\nu}{\rho}\omega_\rho = 0$ for all transversal $1$-forms. In addition, using $\D_\mu n^{\nu} = 0 = \D'_\mu n^{\nu}$ one further finds
\begin{align}\label{7.333333}
    \underbrace{(\D'_\mu - \D_\mu) n^\nu}_{=0} = C\du{\mu\rho}{\nu} n^\rho &= \left(\Sigma_{\mu\rho} n^\rho\right)n^\nu & \Longleftrightarrow&& \Sigma_{\mu\rho} n^\rho = 0\,,
\end{align}
hence, $\Sigma_{\mu\nu}$ itself is also transverse to $n^{\mu}$. Ignoring for this moment the existence of $\D_\mu\ell_\nu$, the previous relations indicate that the freedom in choosing a derivative operator on $\scrip$ is reduced to choosing a symmetric tensor which is transverse to $\scrip$. Thus, there are as many distinct covariant derivative operators on $\scrip$ as there are tensors of this type. Given that $\Sigma_{\mu\nu}$ is a rank-$2$ tensor which is defined on a three-dimensional manifold, i.e., $\scrip$, it has $3\times 3$ components. Its symmetry reduces the number of independent components to $\frac{3(3+1)}{2} = 6$. Taking into account the transversality, which imposes the three constraint equations $\Sigma_{\mu\nu}n^{\nu} = 0$ (remember that $\scrip$ is a 3-dimensional manifold), finally leaves three independent components which completely specify $\Sigma_{\mu\nu}$. In a lengthy but straightforward computation, it can be shown that one of the degrees of freedom associated with $\Sigma_{\mu\nu}$ is actually a gauge artifact associated with the rescaling freedom of the universal structure  $(n^{\mu},q_{\mu\nu})\mapsto ( n'^{\mu\nu},q'_{\mu\nu}) = ( f(u,z,\bar z)^{-1}\,n^{\mu}, f(u,z,\bar z)^2 q_{\mu\nu})$ \cite{Our_Review}. Thereby, one finds that the operators $\D$ admit equivalence classes whose equivalence relation is the conformal rescaling. Concretely, for a conformal rescaling selected within the universal structure with $f|_{\scrip}= 1$ and $f\neq 1$ off $\scrip$, one finds that for an arbitrary vector field on $\mathcal M$
\begin{align}\label{equ:porsche_auf_malle}
    \newpb{(\nabla_\mu-\nabla'_\mu)\alpha_\nu} &=(\D'_\mu-\D_\mu)\newpb{\alpha_\nu}\notag \\
    &= -\omega^{-1}\left(2 [\delta^{\rho}{}\newpb{\prescript{}{(\mu}{\nabla}_{\nu)}\omega]}\alpha_\rho - (\nabla^\rho\omega)\alpha_\rho \,\newpb{g_{\mu\nu}}\right)\notag \,,
\end{align}
which on $\scrip$ gives
\begin{align}
    &\left.-\omega^{-1}\left(2 [\delta^{\rho}{}\newpb{\prescript{}{(\mu}{\nabla}_{\nu)}\omega]}\alpha_\rho - (\nabla^\rho\omega)\alpha_\rho \,\newpb{g_{\mu\nu}}\right)\right|_{\scrip}= -1\left(0 - (\nabla^\rho\omega)\alpha_\rho\, q_{\mu\nu}\right)\notag\\
    &=\left(\nabla^c\omega\right)\alpha_c\, q_{\mu\nu} = \lambda n^\rho\alpha_\rho q_{\mu\nu}\,,
\end{align}
where $\nabla^\mu f|_{\scrip}=\lambda n^\mu$. The latter result is obtained by considering the rescaling behavior of $ C\du{\mu\rho}{\nu}$ which follows rescaling the derivative operators on the right hand sight of \eqref{equ:porsche_auf_malle} using the rules outlined in Appendix \ref{App:ConformalTransformations} as well as $\delta^{\rho}\newpb{{}_{(\mu}\nabla_{\nu)}}\omega|_{\scrip}=0$ \cite{Our_Review}. Given Eq. \eqref{equ:porsche_auf_malle}, a comparison with previous results yields
\begin{equation}
    \Sigma_{\mu\nu} = \lambda \, q_{\mu\nu}\,.
\end{equation}
This result indicates that the change in the derivative operator is proportional to the scalar $\lambda$, which results from the rescaling. Thus, $\lambda$ can be seen as a gauge artifact associated with the rescaling freedom and is completely independent of $\alpha_\mu$. The difference of covariant derivative operators at $\scrip$, therefore, is rather unphysical and merely indicates an ambiguity in the choice of the conformal frame. This observation leads to the introduction of equivalence classes of covariant derivatives. One denotes these equivalence classes by $[\D]$ and defines the equivalence relation as
\begin{align}
    &\D \sim \D' &\Longleftrightarrow && \left(\D'_\mu-\D_\mu\right)\newpb{\alpha_\nu} = \lambda\, q_{\mu\nu} n^\rho \alpha_\rho.
\end{align}
Ultimately, when computing physical quantities using $\D$, they have to be gauge-independent, meaning that they cannot depend on $\lambda$. Generally speaking, physical quantities do not depend on the arbitrary choice of a conformal frame. Thus, $\Sigma_{\mu\nu}$ cannot be (completely) physical. The unphysical degree of freedom associated with $\lambda$ can, however, be extracted by computing the trace of 
$\Sigma_{\mu\nu}$. This implies that the trace-free part of $\Sigma_{\mu\nu}$ indeed carries only two degrees of freedom, which are fully independent of the choice of conformal completion. It is, therefore, instructive to define a new tensor
\begin{equation}\label{ShearTensor}
    \sigma_{\mu\nu} := \Sigma_{\mu\nu} - \frac12 q_{\mu\nu} q^{\rho\sigma}\Sigma_{\rho\sigma}\,
\end{equation}
which is subsequently called the \textit{shear tensor}. Note here that the trace-free part \eqref{ShearTensor} is in fact not dependent on the pseudo-inverse $q^{\mu\nu}$ as $\Sigma_{\mu\nu}$ is transverse to $n^\mu$. \\
The shear tensor enables the distinction between equivalence classes. Namely, if two derivative operators belong to the same one, $\sigma_{\mu\nu}=0$. On the other hand, if $\D$ and $\D'$ do not ``differ'' by $\lambda\, q_{\mu\nu}$ they belong to two different equivalence classes and their distinction is quantified by $\sigma_{\mu\nu}$. Thus, it follows that two equivalence classes are disjoint if and only if $\sigma_{\mu\nu}\neq 0$,
\begin{align}
    [\D]\cap [\D]' &= \emptyset &\Longleftrightarrow && \sigma_{\mu\nu} \neq 0\,.
\end{align}
The tensor $\sigma_{\mu\nu}$ can thereby be computed with any representative of $[\D]$ and $[\D]'$. Further, it is gauge-invariant by construction and inherits the following properties from $\Sigma_{\mu\nu}$ and $q_{\mu\nu}$:
\begin{align}\label{eq:propertiesshear}
       \sigma_{\mu\nu} &= \sigma_{(\mu\nu)}, & && &&  \sigma_{\mu\nu} n^\nu &=0, & && && q^{\mu\nu} \sigma_{\mu\nu} &=0
\end{align}
It properties imply that $\sigma_{\mu\nu}$ carries two independent degrees of freedom. Given that it is gauge-independent suggests these two degrees of freedom correspond to the ones of the physical metric $\widetilde g_{ab}$.

After this somewhat lengthy derivation, it is left to show that the degrees of freedom carried by $\sigma_{\mu\nu}$ indeed capture the radiative modes and are thus $\sigma_{\mu\nu}$ able to distinguish between radiative and non-radiative configurations. To do so, consider (finally) $\D_\mu\ell_\nu$: Given a universal structure, one can define derivative operators $\D,\D'$ within the equivalence class $[\D]$ pertaining to this class of spacetimes. It is, however, also possible to define a fiducial derivative operator $\Dcirc$ which has trivial curvature and for which
\begin{align}\label{fiducialDerOperator}
    \Dcirc_\mu \ell_\nu \overset{!}{=} 0\,.
\end{align}
This operator does not necessarily belong to the same equivalence class $[\D]$. In fact, by determining whether or not $\D$, induced by the physical spacetime via $\newpb{\nabla} =: \D$, resides in the same equivalence class as $\Dcirc$, one obtains information about whether said spacetime encompasses radiation or not. To do so, one computes
\begin{align}\label{eq:ShearAndDerivativeEll}
    (\Dcirc_\mu-\D_\mu) \ell_\nu = \Sigma_{\mu\nu} n^\rho \ell_\rho = -\Sigma_{\mu\nu} = \D_\mu\ell_\nu\,.
\end{align}
Inserting \eqref{eq:ShearAndDerivativeEll} into \eqref{ShearTensor}, one finds
\begin{align}\label{eq:RunningOutOfNames}
    \sigma_{\mu\nu}=\D_\mu \ell_\nu - \frac12 q_{\mu\nu} q^{\rho\sigma} (\D_\rho \ell_\sigma)\,.
\end{align}
The latter equation suggests that, in fact, the gauge-invariant degrees of freedom are captured in $\D_\mu\ell_\nu$, which agrees with the intuitive picture from above. The tensor $\D_\mu\ell_\nu$ isolates the information in $\D$ which is not part of the universal structure but describes physical information of the underlying spacetime that can be associated with this structure. At this point, it is also worth highlighting that the shear tensor is intrinsically defined at $\scrip$ and has no correspondent off $\scrip$. This is an important factor when trying to generalize the frameworks built around \gls{gr} in asymptotically flat spacetime to other spacetime geometries. Evidently, the asymptotic structure plays a crucial role, being home to the radiative degrees of freedom of the underlying theory of gravity. Note further that the name ``shear'' results from the fact that $\sigma_{\mu\nu}$ computes the shear of $\ell^\mu$ as defined earlier in this Chapter.  

The final connection between the shear tensor and radiation can be constructed by considering the relation between the Riemann tensor and derivative operators. Namely, for a torsion-free and metric-compatible derivative operator $\nabla$ acting on the conformally compactified spacetime $\mathcal{ M}$, it holds that 
\begin{equation}\label{equ:DD_R}
    2\nabla_{[\mu}\nabla_{\nu]}\alpha_\rho = R\du{\mu\nu\rho}{\sigma} \alpha_\sigma=C\du{\mu\nu\rho}{\sigma}\alpha_\sigma+\left(g_{\rho[\mu}S\du{\nu]}{\sigma}+S_{\rho[\mu}\delta\du{\nu]}{\sigma}\right)\alpha_\sigma \,.
\end{equation}
In the second equation, the Riemann tensor is replaced with the Weyl and Schouten tensors. The latter is defined as $S_{\mu\nu}=R_{\mu\nu}-\frac{1}{6}g_{\mu\nu}R$. Pulling the expression \eqref{equ:DD_R} back to $\scrip$ where the unphysical Weyl tensor vanishes, see Section \ref{subsec:NP_scalars_and_peeling}, one finds that, at $\scrip$,
\begin{align}\label{RiemannOnScri}
    \R\du{\mu\nu\rho}{\sigma} = \left(q_{\rho[\mu}\Ss \du{\nu]}{\sigma}+\Ss_{\rho[\mu}\delta\du{\nu}{\sigma}\right) \,,
\end{align}
where $\R\du{\mu\nu\rho}{\sigma}:=\newpb{R\du{\mu\nu\rho}{\sigma}}$, $\Ss\du{\mu}{\nu}:=\newpb{S\du{\mu}{\nu}}$, and $\Ss_{\mu\nu}:=\Ss\du{\mu}{\rho}q_{\rho\nu}$\footnote{Note that because $q_{\mu\nu}$ is degenerate $\Ss_{\mu\nu}=\Ss\du{\mu}{\rho}q_{\rho\nu}$ looses some information in comparison to $\Ss\du{\mu}{\rho}$.}. Note that, using $\R_{\mu\nu} = q^{\mu\nu}\R_{\rho\mu\sigma\nu}\quad $ and $\quad \R=q^{\mu\nu}\R_{\mu\nu}$, on $\scrip$ it holds that $\Ss_{\mu\nu}q^{\mu\nu} = \R $. In this context, it is important to highlight that the pullback, as defined in ``Notations'', results in the Riemann tensor $\R_{\rho\mu\sigma\nu}$ and the Schouten tensor $\Ss_{\mu\nu}$ on $\scrip$ being transverse to this very hypersurface. Consequently, they are not affected by the ambiguity introduced by operating with a pseudo-inverse to contract indices. Moreover, the tensors can be understood as being defined on the 2-dimensional cross sections orthogonal to $n^\mu$. In this case, the Bianchi identities indicated that the Riemann tensor contains only one independent degree of freedom and can be rewritten as $\R_{\mu\nu\rho\sigma}=\R \,q_{\mu[\rho}\,q_{\sigma]\nu}\,$. This perhaps surprising result illustrates a very fundamental issue that occurs when working at $\scrip$: Extracting physical information from conformally completed spacetimes requires careful considerations of the residual rescaling freedom. Physical quantities should generally be invariant w.r.t. this rescaling, $(n^{\mu},q_{\mu\nu})\mapsto ( n'^{\mu\nu},q'_{\mu\nu}) = ( f(u,z,\bar z)^{-1}\,n^{\mu}, f(u,z,\bar z)^2 q_{\mu\nu})$, which the Riemann tensor does not. Thus, it comes as no surprise that the Riemann tensor on the boundary $\scrip$ does not encode the physical modes that are gravitational radiation. In fact, neither is the Schouten tensor, as it fails to be conformally invariant as well. For a review of relevant conformal transformation properties, the reader is referred to Appendix \ref{App:ConformalTransformations}.\\
To define a conformally invariant quantity, one can exploit the topology of the integral curves of $n^\mu$. The latter enters the definition of asymptotic flatness explicitly and guarantees the existence of a symmetric tensor $\rho_{\mu\nu}$ on $\scrip$ such that \cite{Geroch_1977}
\begin{equation}\label{RhoDefiningProperties}
    \rho_{\mu\nu}n^\nu=0\,,\quad \rho_{\mu\nu}q^{\mu\nu}=\R\,,\quad \D_{[\mu}\rho_{\nu]\rho}=0\,.
\end{equation}
With the transformation rules outlined in Appendix \ref{App:ConformalTransformations}, one can show (see Appendix \ref{app:rho_conformal}) that $\rho_{\mu\nu}$ transforms as 
\begin{equation}
    \rho'_{\mu\nu} =\rho_{\mu\nu}-\frac{2}{\omega}\D_\mu\D_\nu\omega+\frac{4}{\omega^{2}}(\D_\mu\omega)(\D_\nu\omega)-\frac{q_{\mu\nu}}{\omega^{2}}q^{\rho\sigma}(\D_\rho\omega)(\D_\sigma\omega)\,,
\end{equation}
which, coincidentally, in large parts, corresponds to the conformal transformation behavior of the Schouten tensor. It thus motivates the definition of the \textit{Bondi news tensor}, 
\begin{equation} \label{equ:Bondiiiiiii}
    N_{\mu\nu}=\underbrace{\Ss_{\mu\nu}-\rho_{\mu\nu}}_{\text{conformally invariant}}\,,
\end{equation}
which captures the conformally invariant bits of information within the Riemann tensor at $\scrip$. Therefore, this tensor, in literature often denoted as \textit{Bondi tensor} or \textit{news tensor}, indicates the presence of radiation ``passing'' through an observer located at $\scrip$. Correspondingly, it vanishes in the absence of radiation. It inherits its symmetry and other properties from $\Ss_{\mu\nu}$ and $\rho_{\mu\nu}$, i.e., 
\begin{equation}\label{PropertiesBondiNews}
    N_{\mu\nu}=N_{(\mu\nu)}\;,\quad N_{\mu\nu}n^\nu=0\;,\quad N_{\mu\nu}q^{\mu\nu}=0\,.
\end{equation} 
In summary, the Bondi news tensor is a symmetric, trace-free tensor living on the two dimensions of $\scrip$ orthogonal to $n^\mu$. It thus encompasses 2 physical degrees of freedom. The situation now is therefore similar to the previous discussion about the shear tensor. However, for the Bondi news tensor, it is explicitly demonstrated that its degrees of freedom are of radiative origin. As anticipated above, one can show explicitly that Bondi news and shear tensor are related and thus their degrees of freedom agree in their physical interpretation. To do so, it is convenient to work in the Bondi frame, that is, the frame on $\scrip$ in which $q_{\mu\nu}$ is the metric of the unit $2$-sphere, since then $\R$ is constant such that 
\begin{equation}\label{equ:rhooooo}
    \rho_{\mu\nu}=\frac{1}{2}q_{\mu\nu}\R\,.
\end{equation}
and thus
\begin{align}\label{eq:DefinitionBondi}
    N_{\mu\nu} = S_{\mu\nu}-\frac12 q^{\rho\sigma} S_{\rho\sigma} q_{\mu\nu}.
\end{align}
\begin{proposition}
    In a Bondi frame with $\D_\mu n^\nu=0$, $\ell^\mu n_\mu=-1$, $\lie_n \ell^\mu=0$ and using the decomposition of the asymptotic Riemann tensor $\R\du{\mu\nu\rho}{\sigma} = \left(q_{\rho[\mu}\Ss\du{\nu]}{\sigma}+\Ss_{\rho[\mu}\delta\du{\nu]}{\sigma}\right)$ and the shear $\sigma_{\mu\nu} = \D_\mu \ell_\nu - \frac12 q_{\mu\nu} q^{\rho\sigma} (\D_\rho \ell_\sigma)$, it holds that 
    \begin{align*}
    N_{\mu\nu} = 2\lie_n \sigma_{\mu\nu}\,,
    \end{align*}
    where $N_{\mu\nu}$ is the Bondi news tensor as defined in \eqref{eq:DefinitionBondi}.
\end{proposition}
\noindent \textit{Proof.}~First, note that the shear tensor relates to the covariant derivative of $\ell_\mu$ at $\scrip$. The Lie derivative of such a tensor w.r.t. an arbitrary vector field $\xi^\mu$ can be computed as
\begin{align}\label{equ:some_result_for_some_equation}
    (\lie_\xi\D_\nu-\D_\nu\lie_\xi)\ell_\mu=&\,\xi^\rho\D_\rho\D_\nu\ell_\mu+\D_\rho\ell_\mu\D_\nu\xi^\rho+\D_\nu\ell_\rho\D_\mu\xi^\rho-\D_\nu\xi^\rho\D_\rho\ell_\mu - \xi^\rho \D_\nu\D_\rho\ell_\mu \phantom{(E.123)}\notag\\
    &-\D_\nu\ell_\rho\D_\mu\xi^\rho-\ell_\rho\D_\nu\D_\mu\xi^\rho\notag\\
    =&\,2\xi^\rho\D_{[\rho}\D_{\nu]}\ell_\mu-\ell_\rho\D_\nu\D_\mu\xi^\rho\notag\\
    =&\,\xi^\rho\left(q_{\mu[\rho}\Ss\du{\nu]}{\sigma}+\Ss_{\mu[\rho}\delta\du{\nu]}{\sigma}\right)\ell_\sigma-\ell_\rho\D_\nu\D_\mu\xi^\rho\,,
\end{align}
where in the last step the definition of the Riemann tensor on $\scrip$ is used, i.e.,
\begin{equation}\label{equ:kack_equation}
    2\D_{[\rho}\D_{\nu]}\ell_\mu= \left(q_{\mu[\rho}\Ss\du{\nu]}{\sigma}+\Ss_{\mu[\rho}\delta\du{\nu]}{\sigma}\right)\ell_\sigma= \R\du{\rho\nu\mu}{\sigma}\ell_{\sigma}\,.
\end{equation}
Now on replaces $\xi^\mu= n^\mu$ and consider $\scrip$ in a divergence-free conformal frame such that $\D_\mu n^\nu=0$. Lie dragging $\ell^\mu$ along $n^\mu$ yields $\lie_n\ell_\mu|_{\scrip}= 0$. Further recalling that the Schouten tensor on $\scrip$ is transverse $\Ss_{\mu\nu}n^\nu=0$, one computes
\begin{align}
    \lie_n \D_\nu\ell_a&=n^\rho\left(\frac{1}{2}\left[q_{\mu\rho}\Ss\du{\nu}{\sigma}-q_{\mu\nu}\Ss\du{\rho}{\sigma}+\Ss_{\mu\rho}\delta\du{\nu}{\sigma}-\Ss_{\mu\nu}\delta\du{\rho}{\sigma}\right]\ell_\sigma\right)\notag\\
    &=-\frac{1}{2}q_{\mu\nu}n^\rho\Ss\du{\rho}{\sigma}\ell_\sigma-\frac{1}{2}(n_\rho \ell^\rho)\Ss_{\mu\nu}=\frac{1}{2}\Ss_{\mu\nu}-\frac{1}{2}q_{\mu\nu}n^\rho\Ss\du{\rho}{\sigma}\ell_\sigma\,.
\end{align}
Thus, $\lie_n\sigma_{\mu\nu}$ can be computed using \eqref{eq:RunningOutOfNames} obtaining
\begin{align}
    \lie_n\left(\D_\mu\ell_\nu-\frac{1}{2}q_{\mu\nu}q^{\rho\sigma}(\D_\rho\ell_\sigma)\right)&=\frac{1}{2}\left(\Ss_{\mu\nu}-q_{\mu\nu}n^\rho\Ss\du{\rho}{\sigma}\ell_\sigma\right)-\frac{1}{2}\left(\frac{1}{2}q_{\mu\nu}q^{\rho\sigma}\Ss_{\rho\sigma}-q_{\mu\nu}n^\tau\Ss\du{\tau}{\alpha}\ell_\alpha\right)\notag\\
    &=\frac{1}{2}\left(\Ss_{\mu\nu}-\frac{1}{2}q_{\mu\nu}q^{\rho\sigma}\Ss_{\rho\sigma}\right)\,,
\end{align}
which is exactly $\frac{1}{2}N_{\mu\nu}$ as defined in the Bondi frame.
    
\noindent The fact that Bondi news and shear tensor are related is rather intuitive if one acknowledges their interpretation in the sense the information content they are representing. While the Bondi news tensor captures the gauge invariant information content of the Riemann tensor at $\scrip$, the shear tensor encompasses the gauge invariant information contained in the equivalence class $[\D]$ of covariant derivative operators at $\scrip$.\\
Based on the tensors' intrinsic properties \eqref{eq:propertiesshear} and \eqref{PropertiesBondiNews}, they can be written in a convenient form on $\scrip$.
\begin{proposition}
    Let $u_0$ parametrize a cross section of $\scrip$ in the coordinate chart $( u, z, \bar z)$. Then, in the basis $\{\ell_\mu,n_\mu,m_\mu,\bar m_\mu\}$, the shear tensor $\sigma_{\mu\nu}$ satisfying \eqref{eq:propertiesshear} can be written as
    \begin{align}\label{eq:AsymptoticShear}
    \sigma_{\mu\nu} = -(\bar{\sigma}^\circ m_\mu m_\nu + \sigma^\circ \bar{m}_\mu \bar{m}_\nu)\,.
    \end{align}
\end{proposition}
\noindent \textit{Proof.}~Expanding an arbitrary rank-2 tensor $T_{ab}$ satisfying \eqref{eq:propertiesshear} into the tetrad basis, it is straightforward to check that the expansion must read
\begin{align}T_{\mu\nu}=\lambda_1\ell_\mu\ell_\nu+\lambda_{2}m_{(\mu}\Bar{m}_{\nu)}+\bar\lambda_3m_\mu m_\nu+ \lambda_3\Bar{m}_\mu\Bar{m}_\nu+\bar\lambda_4\ell_{(\mu}m_{\nu)}
    &+\lambda_4\ell_{(\mu}\Bar{m}_{\nu)}\,,
\end{align}
where the symmetry of $T_{\mu\nu}$ is incorporated. The expansion coefficients $\lambda_{1,2}\in \mathbb R$ and $\lambda_{3,4}\in \mathbb C$ correspond exactly to the $6$ degrees of freedom that one expects to find for a real symmetric tensor in $3$-d space prior to applying any further constraints. Now requiring that $T_{\mu\nu}$ is transverse to $n^\mu$ imposes that $\lambda_1$ and $\lambda_4$ have to be set to zero, which follows by simply multiplying $T_{\mu\nu}$ with $n^\mu$ while knowing that $n^\mu\ell_\mu=-1$ whereas $n^\mu m_\mu=0$. Furthermore, the tracelessness condition of $T_{\mu\nu}$ immediately implies that that $\lambda_{2}=0$ as well. Hence, one is left with 
\begin{align}
   T_{\mu\nu}=\bar \lambda_3 m_\mu m_\nu+ \lambda_3\Bar{m}_\mu \Bar{m}_\nu\,,
\end{align}
which, after relabeling $\lambda_3=T$, corresponds to the desired result. The complex coefficient $\sigma^\circ$ is called asymptotic shear, defined by 
\begin{equation}\label{eqr:shearDefNoStrain}
    \sigma^\circ(u,\theta,\phi) = - \lim_{r\to\infty} \left(m^{\mu}m^{\nu}\nabla_\mu \ell_\nu\right)\,.
\end{equation}
and carrying spin-weight $-2$.

\noindent Naturally, given their intimate relation, a similar expansion can be found for the Boni news tensor, i.e.,
\begin{align}\label{eq:bondiNewsexpansion}
    N_{\mu\nu} = 2\left(N^\circ m_\mu m_\nu + \bar{N}^\circ \bar{m}_\mu \bar{m}_\nu\right)\,,
\end{align}
where the factor of two is pure convention and where $N^\circ$ is a complex function of spin-weight $-2$ which is simply called the \textit{Bondi news}. Recalling that, in a Bondi frame, $n^\mu$ is the generator of pure time translations with $u$ being the affine parameter such that $n^\mu\D_\mu u |_{\scrip}=1$, one can rewrite the Bondi news tensor as
\begin{align}\label{eq:bondiI}
    N_{\mu\nu} = 2\lie_n \sigma_{\mu\nu} = 2\partial_u\sigma_{\mu\nu}\,
\end{align}
such that, in conjunction, the relation between shear and news yields
\begin{equation}\label{eq:bondiII}
    N^\circ=-\partial_u\bar \sigma^\circ=: - \dot{\bar{\sigma}}^\circ\,.
\end{equation}
Eq. \eqref{eq:bondiII} establishes one of the most profound results of the early works on asymptotically flat spacetimes. It connects gauge invariant geometric information of the Bondi news tensor with the shear of the vector field $\ell^\mu$ on $\scrip$, i.e., the derivative of the generating vector field of the \gls{ngc}. As such, a relation between information pulled back from the bulk and information intrinsically defined at $\scrip$ is manifested. Therefore, one can finally conclude that the two propagating degrees of freedom encapsulated in the shear tensor indeed describe \gls{gw}s, i.e., gravitational radiation, in \gls{gr}. It is thereby important to highlight that throughout this Section, no linearization or perturbation theory is applied. Therefore, the treatment above, including all interim results, is valid for the full theory of \gls{gr}.\\
To conclude the discussion about the identification of radiative degrees of freedom, the shear tensor is connected to the previously outlined NPS and the linearized version of the \gls{gw} strain. In this way, the next Section will connect the previous theoretical discussions with the investigations of the subsequent Chapters. 

\subsection{The Shear Tensor and Linearized Gravity}
\label{subsec:shear_and_GW}
In the context of \gls{gw}s, the classical approach to describe radiation is a perturbative one. That is, for a physical spacetime $(M, \widetilde g_{\mu\nu})$, one expands the metric linearly in the perturbative parameter
\begin{align}
    \widetilde g_{\mu\nu}= \eta_{\mu\nu}+\lambda\widetilde h_{\mu\nu}\,.
\end{align}
This turns Einstein's field equations into a 1-parameter family of equations
\begin{align}\label{eq:LinearizedEinstein}
    0 = \left.\frac{\dd}{\dd\lambda}G_{\mu\nu}(\lambda)\right|_{\lambda = 0} = -\frac12 \Box\, \bar{h}_{\mu\nu} + \nabla^\rho \nabla_{(\nu}\bar{h}_{\mu)\rho} - \frac12\eta_{\mu\nu} \nabla^\rho \nabla^\sigma \bar{h}_{\rho\sigma}.
\end{align}
where the trace-free metric perturbation is described as
\begin{align}
\bar{h}_{\mu\nu} := \widetilde h_{\mu\nu} - \frac12 (\eta^{\rho\sigma} \widetilde  h_{\rho\sigma}) \eta_{\mu\nu}\,,
\end{align}
and the derivative operators are defined w.r.t. the Minkowski metric $\eta_{\mu\nu}$ (to restore linear order). Note that the Einstein tensor on the left-hand side of Eq. \eqref{eq:LinearizedEinstein} does as well only depend on quantities defined in the bulk spacetime. The linear (physical metric) perturbation $\widetilde h_{\mu\nu}$ inherits a linearized version of the gauge freedom of the physical metric, 
\begin{align}
    \widetilde h_{\mu\nu}\,\mapsto\, \widetilde h'_{\mu\nu} =  \widetilde h_{\mu\nu} + \lie_\zeta\eta_{\mu\nu}= \widetilde h_{\mu\nu} + 2\nabla_{(\mu}\zeta_{\nu)},
\end{align}
where $\zeta_\mu$ is an arbitrary but infinitesimal vector. This freedom can be used to fix the transverse-traceless gauge (TT gauge) to reduce the freely propagating degrees of freedom in $\widetilde h_{\mu\nu}$ to two, by choosing $\zeta_\mu$ such that
\begin{align}
   \widetilde h_{0\mu}=0 \,, \qquad  \eta^{\mu\nu}\widetilde h_{\mu\nu}=0 \,, \qquad  \nabla^\nu  \widetilde h_{\mu\nu} =0.
\end{align}
It follows immediately that $\bar h_{\mu\nu}= \widetilde h_{\mu\nu}$ and the metric can be expanded into the $+,\times$ orthogonal polarization basis 
\begin{align}
   \widetilde h_{\mu\nu}^{TT}=h_\times \epsilon^\times_{\mu\nu} + h_+\epsilon^+_{\mu\nu}.
\end{align}
To connect this well-known description of \gls{gw}s, i.e., $\widetilde h_{\mu\nu}^{TT}$, with the definition of the shear, it is instructive to exploit the fact that the latter can be constructed out of quantities living on the (physical) bulk spacetime. The shear, as defined above, can be recovered by pulling back the tensor field
\begin{align}\label{equ:Toledo}
    \widetilde\sigma_{\mu\nu}=\widetilde\nabla_\mu \widetilde\ell_\nu - \frac12 \widetilde g_{\mu\nu}\widetilde g^{\rho\sigma} (\widetilde \nabla_\rho \widetilde \ell_\sigma)\,
\end{align}
to $\scrip$ where $\widetilde\nabla$ is the covariant derivative w.r.t. $\widetilde g_{\mu\nu}$. Pulling back to $\scrip$, one finds that the shear (as defined by the shear corresponding to the frame field $\ell^\mu$) is not defined on a $2$-sphere of constant $u$ and $r$. However, as the shear of $\ell^\mu$ on a Minkowski background vanishes, one can define the linearized version similar to the expansion above, as demonstrated for instance in \cite{Ashtekar_2017}, 
as
\begin{align}
   \sigma^\text{lin}_{\mu\nu} := \frac{\dd}{\dd\lambda}\left.\left(\widetilde\nabla_\mu \widetilde\ell_\nu - \frac12 \widetilde s_{\mu\nu} \widetilde s^{\rho\sigma} (\widetilde\nabla_\rho \widetilde\ell_\sigma)\right)\right|_{\lambda=0}\,,
\end{align}
where $\widetilde s_{\mu\nu}$ denotes the metric of said $2$-sphere of constant $u,r$. Note that $\widetilde\nabla_\mu \widetilde\ell_\nu $ as well as $\widetilde s_{\mu\nu}$ depend on the $\lambda$ parametrizing a 1-parameter family of metrics $\widetilde g_{\mu\nu}$ in the bulk spacetime where $\widetilde h_{\mu\nu} = \frac{\partial}{\partial \lambda} \widetilde g_{\mu\nu}|_{\lambda=0}$. Using the identity $2(\widetilde\nabla_{(\mu} \widetilde\ell_{\nu)} )(\lambda) = \mathcal L _\ell \widetilde g_{\mu\nu}(\lambda)$ as well as $\widetilde s_{\mu\nu} \widetilde m^\mu \widetilde m^\nu = 0$ and $\widetilde\nabla_\mu \widetilde\ell_\nu \widetilde s ^{\mu\nu} = \sqrt{2}/r$ one finds 
\begin{align}
    \label{equ:surfskate}
    \sigma_\T{lin} = \frac{1}{2}\widetilde m^\mu \widetilde m^\nu \mathcal L_\ell \widetilde h_{\mu\nu} - \frac{1}{\sqrt{2}r}(\widetilde m^\mu \widetilde m^\nu \widetilde h_{\mu\nu})\,.
\end{align}
Expanding the metric perturbation in powers of $1/r$,
\begin{align}
   \widetilde h_{\mu\nu}^{TT} = \frac{\widetilde h^{\circ\, TT}_{\mu\nu}}{r} + \frac{\widetilde h^{(1)\, TT}_{\mu\nu}}{r^2} + ...\,,
\end{align}
and inserting the result into Eq. \eqref{equ:surfskate}, one finds that 
\begin{align}\label{eq:FinalResult}
    \bar\sigma_\text{lin} &=\frac{1}{2}\widetilde{\bar m}^\mu\widetilde{\bar m}^\nu\frac{h_{\mu\nu}^{\circ\,TT}}{r^2}+\O\left(r^{-3}\right)\,, &\text{so that}&& \bar\sigma_\text{lin}^\circ &= \frac12 \widetilde{\bar{m}}^\mu\widetilde{\bar{m}}^\nu h^{\circ\, TT}_{\mu\nu}.
\end{align}
Thus, using $\widetilde{\bar{m}}_\mu\widetilde{\bar{m}}_\nu=\frac{1}{2}(e^+_{\mu\nu}-i\, e^\times_{\mu\nu})$ it follows
\begin{align}\label{eq:linearizedshearandmetricTT}
    h^\circ := 2\bar\sigma_\text{lin}^\circ  = \widetilde h^{\circ\, TT}_{\mu\nu} \widetilde{\bar{m}}^\mu \widetilde{\bar{m}}^\nu=h_+^\circ-ih_\times^\circ.
\end{align}
where, again, 
\begin{align}\label{equ:asym_strain}
    h^\circ_+(u,\theta,\phi) &:= \lim_{r\to\infty}r h_+ (u,r,\theta,\phi),\notag\\ 
    h^\circ_\times(u,\theta,\phi) &:= \lim_{r\to\infty}r h_\times(u,r,\theta,\phi).
\end{align}
In practice, for \gls{gw} measurements, the weakness of the perturbations allows to morally equate linearized with full shear. In this sense, the above relation between physical shear $\sigma^\circ$ and strain $h^\circ$ can be understood as being absolute, i.e., 
\begin{align}\label{equ:shear_and_strain}
    \sigma^\circ (u,z,\bar z) := -\lim_{r\rightarrow \infty} r^2(\widetilde m^\mu \widetilde m^\nu\widetilde\nabla_\mu\widetilde\ell_\nu) \equiv \frac{1}{2}( h^\circ_+ +i h^\circ_\times) = \frac{1}{2} \bar h^\circ\,,
\end{align}
 where, within the bulk, $\sigma = - \widetilde m^\mu \widetilde{m}^\nu\widetilde\nabla_\mu\widetilde\ell_\nu $ (see \cite{waveform_test_BL_III} for further details and practical applications). The above sketch of a derivation results in a fundamental relation between shear and gravitational strain. In the subsequent Sections and Chapters, this relation is heavily used to describe \gls{gw}s measured by instruments morally placed at null infinity. In this sense, in the derivations of constraint equations and other useful applications involving the shear, one can replace the latter by the gravitational strain without further explanation. This replacement becomes particularly handy when deriving constrain equations at null infinity, see Section \ref{sec:Covariant_PS}.

Before presenting a parametrization of the metric in which the shear appears as an explicit metric component in the next Section, it is instructive to relate the shear tensor, and therefore the strain, to the NPS of Section \ref{subsec:NP_scalars_and_peeling}. In fact, when studying literature on \gls{gw}s, one often finds the gravitational data, exemplarily displayed in Fig. \ref{fig:GW_waveform}, as labeled by ``$\Psi_4$''. To understand why, the relation between shear and NPS is now analyzed: The expansion in power of $1/r$ of the shear tensor (and therefore the strain) should remind the reader of the discussion regarding the peeling properties of NPS in Section \ref{subsec:NP_scalars_and_peeling}. Indeed, for $\Psi_4,\Psi_3,\Psi_2$, an explicit relation can be derived based on the contraction of the Weyl tensor \eqref{equ:pups}-\eqref{equ:double_pups} for each of the scalars. To that end, the reader is reminded of the fact that the physical NPS can be written as 
\begin{align}\label{eq:DefNPGR}
	{\Psi}_4^\circ &=K_{\mu\nu\rho\sigma}\, \bar{m}^{\mu} n^{\nu}  \bar{m}^{\rho}n^{\sigma}\notag\,,\\
	{\Psi}_3^\circ &= K_{\mu\nu\rho\sigma}\,\ell^{\mu} n^{\nu} \bar{m}^\rho n^\sigma=K_{\mu\nu\rho\sigma}\,\bar{m}^{\mu} m^{\nu} \bar{m}^\rho n^\sigma\notag\,,\\
	{\Psi}_2^\circ &=K_{\mu\nu\rho\sigma}\,\ell^{\mu} m^{\nu} \bar{m}^\rho n^\sigma= \frac{1}{2}K_{\mu\nu\rho\sigma}\,m^{\mu} \bar m^{\nu} \left(m^\rho\bar{m}^\sigma-\ell^\rho n^\sigma\right)\notag\,,\\
	{\Psi}_1^\circ &=K_{\mu\nu\rho\sigma}\,\ell^{\mu} n^{\nu} \ell^{\rho} m^{\sigma}=\notag K_{\mu\nu\rho\sigma}\,\ell^{\mu} \bar{m}^{\nu} m^{\rho} \bar m^{\sigma}\\
	{\Psi}_0^\circ &=K_{\mu\nu\rho\sigma}\, \ell^{\mu} m^{\nu}\ell^{\rho} m^{\sigma}\,.
\end{align}
where $K_{\mu\nu\rho\sigma}$ is the asymptotic Weyl tensor and $\{\ell^\mu, n^\mu ,m^\mu,\bar m^\mu\}$ the unphysical tetrad. Given that the shear is related to the Schouten tensor, and so is the Weyl tensor, a connection between these two operators seems only natural. 
\begin{proposition}
    Let $(\mathcal M,\gd)$ be the 4-dimensional conformally completed spacetime. In a divergence-free conformal frame at $\scrip$, it holds that
    \begin{align}\label{equ:proof_weyl_Schouten}
        K_{\mu\nu\rho\sigma}n^\sigma|_{\scrip} = \nabla_{[\nu}S_{\mu]\rho}|_{\scrip}\,,
    \end{align}
    where $K_{\mu\nu\rho\sigma}$ is asymptotic Weyl tensor and $S_{\mu\nu}$ the Schouten tensor corresponding to $\gd$.
\end{proposition}
\noindent \textit{Proof.}~Recall that the Weyl tensor can be written as 
\begin{align}
    C_{\mu\nu\rho\sigma}= R_{\mu\nu\rho\sigma} -g_{\rho[\mu}S_{\nu]\sigma}-S_{\rho[\mu}g_{\nu]\sigma}\,.
\end{align}
Thus, taking a covariant derivative of this expression and using the Bianchi identity 
\begin{equation}
    \nabla^\sigma R_{\mu\nu\rho\sigma}=\nabla_\nu R_{\mu\rho}-\nabla_\mu R_{\nu\rho}\,,
\end{equation}
one finds 
\begin{align}
    \nabla^\sigma C_{\mu\nu\rho\sigma}=\nabla_{[\nu} S_{\mu]\rho}\,.
\end{align}
Using the definition of the asymptotic Weyl tensor $K_{\mu\nu\rho\sigma}=\Omega^{-1}C_{\mu\nu\rho\sigma}$, one writes
\begin{equation}\label{RelWeylSchoutenI}
\nabla^{\sigma}K_{\mu\nu\rho\sigma}=\Omega^{-1}\nabla^{\sigma}C_{\mu\nu\rho\sigma}-\Omega^{-2}K_{\mu\nu\rho\sigma}\nabla^{\sigma}\Omega
\end{equation}
where $\nabla^{\mu}\Omega=n^\mu$. Moreover, from the second Bianchi identity of the Riemann tensor it follows that $\nabla^\sigma C_{\mu\nu\rho\sigma}\propto (d-3)\nabla_{[\mu}S_{\nu]\rho}$ (where $d$ is the dimensionality of the manifold) such that, on $\scrip$, it follows that
\begin{equation}
    \nabla^{\sigma}K_{\mu\nu\rho\sigma}|_{\scrip}= 0\,,
\end{equation}
Hence, on $\scrip$, the left-hand side of Eq. \eqref{RelWeylSchoutenI} vanishes such that from the right-hand side one recovers
\begin{equation}\label{RelWeylSchoutenII}
    K_{\mu\nu\rho\sigma}\,n^\sigma|_{\scrip}=\nabla_{[\nu}S\du{\mu]\rho}{}|_{\scrip}\,.
\end{equation}
Based on the definition of the Schouten tensor at $\scrip$, i.e., $\Ss\du{\mu}{\nu}:=\newpb{S\du{\mu}{\nu}}$, one finds that 
\begin{equation}\label{equ:defender2025}
     \newpb{K_{\mu\nu}}{}^{\rho \sigma}\,n_\sigma |_{\scrip}= \D_{[\nu}\Ss\du{\mu]}{\rho}|_{\scrip}\,.
\end{equation}
Using the definition of the Bondi news tensor $N_{\mu\nu}=\Ss_{\mu\nu}-\rho_{\mu\nu}$ and that
\begin{equation}
     \D_{[\nu}N\du{\mu]\rho}{}=\D_{[\nu}\Ss\du{\mu]\rho}{}-\D_{[\nu}\rho\du{\mu]\rho}{}=\D_{[\nu}\Ss\du{\mu]\rho}{}\,,
\end{equation}
the useful relation
\begin{equation}\label{RelWeylSchoutenIIII}
        \newpb{K_{\mu\nu\rho}}{}_{\sigma}\,n^\sigma |_{\scrip} =  \D_{[b}N\du{a]c}\,|_{\scrip}\,.
\end{equation}
follows.

\noindent Given this result and the Eq. \eqref{eq:DefNPGR}, for each NPS the individual relation to the shear follows by contacting \eqref{equ:proof_weyl_Schouten} with the correct mix of tetrad fields. One might notice at this point that Eq. \eqref{equ:proof_weyl_Schouten} contracts the last index of the asymptotic Weyl tensor with $n^\mu$ already when related to the shear tensor. This suggests that NPS, which cannot be written in a form in which $n^\mu$ appears contracted with at least one index, may not be related to the shear after all. Indeed, in Section \ref{subsec:NP_scalars_and_peeling} it is mentioned that outgoing radiative modes are only enclosed in selected NPS. In Appendix \ref{app:NPS_and_Shear}, it is demonstrated how to derive $\Psi^\circ_4$, $\Psi^\circ_3$, and $\Im{\Psi^\circ_2}$ in terms of shear\footnote{Note that there is a hidden subtlety in the derivation regarding the definition of the pullback of the asymptotic Weyl tensor and the vectors of the tetrad. In particular, contractions of pulled back indices with $\ell$ are ill-defined as it is not tangential to $\scrip$. For the definition of the pullback, see the ``Notations'' of this thesis.}. The remaining NPS are determined by the former ones via the Bianchi identities expressed in terms of NPS\footnote{For a detailed derivation of \eqref{equ:NPS_in_shear_0}-\eqref{equ:NPS_in_shear_II} in terms of tetrad vectors and spin-coefficients, the reader is referred to \cite{Newman_Unti_Gauge}.}. In total, one finds \cite{NPS_shear_relation}
\begin{align}\label{equ:NPS_in_shear_0}
    \dot \Psi^\circ_0 &= \eth \Psi^\circ_1 + 3\sigma^\circ \Psi^\circ_2\,,\\
    \dot \Psi^\circ_1 &= \eth \Psi^\circ_2 + 2\sigma^\circ \Psi^\circ_3\,,\\
    \dot \Psi^\circ_2 &= \eth \Psi^\circ_3 + \sigma^\circ \Psi^\circ_4\,,\\ \label{equ:NPS_in_shear_I}
    -2i\Im{\Psi_2^\circ} &= \sigma^\circ \dot{\bar{\sigma}}^\circ - \bar{\sigma}^\circ \dot{\sigma}^\circ + \eth^2\bar{\sigma}^\circ - \bar{\eth}^2 \sigma^\circ\\
    \Psi^\circ_3 &= -\eth \dot{\bar{\sigma}}^\circ\sim \dot h\,,\\
    \Psi^\circ_4 &= -\ddot{\bar{\sigma}}^\circ\sim \ddot h\,.
    \label{equ:NPS_in_shear_II}
\end{align}
The Bianchi identities also result in $\dot \Psi^\circ_3 =- \eth  \Psi^\circ_4$ which is consistent with the explicit computations. The equation for $\Im{\Psi_2^\circ}$ can be rewritten as 
\begin{align}\label{equ:ferrari_porsch}
    2i \Im{\Psi^\circ_2}= 2i\Im{-\Dot{\Bar{\sigma}}^\circ\sigma^\circ + \Bar{\eth}^2\sigma^\circ}\,.
\end{align}
For $\Psi^\circ_4$, $\Psi^\circ_3$, the previously derived relation to the \gls{gw} strain is applied to build an intuition about the relation of the scalars to actual measurement data. Note thereby that the operator $\eth$ has been previously defined only in the context of spin-weighted spherical harmonics. In the context of shear and Bondi news tensor, it makes sense to generalize its actions to arbitrary functions of spin-weight $s$: Without loss of generality, a function $f_s$ with spin-weight $s$ can be written as 
\begin{equation}
    f_s = T_{\mu_1\cdots \mu_p \nu_1\cdots \nu_{q}} m^{\mu_1}\cdots m^{\mu_p}\bar{m}^{\nu_1}\cdots \bar{m}^{\nu_q}\qquad\text{with } p-q = s\,.
\end{equation}
Introducing the abbreviation
\begin{equation}
    P^{a_1\cdots a_p b_1\cdots b_{q}} := m^{a_1}\cdots m^{a_p}\bar{m}^{b_1}\cdots \bar{m}^{b_q}\,,
\end{equation}
one can define the angular derivative of a spin-weight $s$ function as\footnote{Note that in literature one might encounter deviations regarding the prefactor.}
\begin{align}\label{EthArbitrary}
    \eth f_{s} :=&\ \frac{1}{\sqrt{2}} m^{\mu} P^{\mu_1\cdots \mu_p \nu_1\cdots \nu_{q}}\left(\D_\mu T_{\mu_1\cdots \mu_p \nu_1\cdots \nu_{q}}\right)\,,
\end{align}
as well as 
\begin{align}\label{EthArbitrary:II}
    \eth f_{s} :=&\ \frac{1}{\sqrt{2}} \bar m^{\mu} P^{\mu_1\cdots \mu_p \nu_1\cdots \nu_{q}}\left(\D_\mu T_{\mu_1\cdots \mu_p \nu_1\cdots \nu_{q}}\right)\,,
\end{align}

To conclude this Section, the most important results are summarized: Given an conformally completed spacetime $(\mathcal M, \gd)$ with a causal structure and future boundary $\scrip$ on which \gls{gr} acts as the fundamental theory of gravity, it is demonstrated that the structure of the derivative operator $\D$ at this boundary $\scrip$ encodes gravitational radiation. It is explicitly demonstrated that there exists a relation between the information contained in the derivative operator, conveniently wrapped into the shear tensor, and physical measures of curvature, represented as the Bondi news tensor. Further, in the linearized regime, e.g., in the context of real-world \gls{gw} interferometry, the shear tensor can be related to the gravitational strain, which is particularly crucial for Subsection \ref{subsec:BL_a_la_ashtekar}. It should be noted that despite outlining the symmetries of the universal structure of $\scrip$ in detail, conserved quantities associated with these symmetries have not been addressed yet. For their derivation and practical application, a relation between shear and strain is indispensable. Finally, the above discussion links between NPS and the shear tensor, underlining not only the NPS' interpretation but also highlighting the relation between peeling properties of such and the asymptotic treatment above. Note that during the constructions and proofs of this Section (and more lengthy proofs in Appendix \ref{AppendixB} belonging to this Section), it was always assumed that one acts within the Bondi frame to simplify calculations. This corresponds to choosing a specific background structure, without affecting the physics or the symmetries. For a more general presentation of some key identities analyzed above, the reader is referred to \cite{PhysRevD.110.044050}.\\
In the following Subsection, it will be demonstrated that there exists a metric decomposition such that the asymptotic shear tensor directly appears as a metric component. The latter predates the above treatment of identifying radiative degrees of freedom with derivative operators at $\scrip$ and is slightly advantageous when it comes to computing and interpreting Einstein's equations. Nonetheless, the result is completely consistent with the treatment above, including the symmetries (i.e., Killing vector field of the metric) and properties of the (physical as well as unphysical) metric.


%
%
%


\section{Bondi-Metzner-Sachs Metric and its Symmetries}
\label{sec:BMS_metric_and_Conserved_Quanties}

In a previous Section, it was demonstrated that by a specific choice of coordinates, i.e., the BMS coordinates on an asymptotically flat spacetime, at $\scrip$ one obtains the structure
\begin{align}
        n^\mu \partial_\mu= \partial_u && \text{and} && q_{\mu\nu}\dd x^\mu \dd x^\nu = \frac{4\dd z\dd \bar z}{(1+|z|^2)^2}\,,
\end{align}
where $q_{\mu\nu}$ corresponds to the metric on the $\mathcal S^2$ sphere. It can be shown that in a neighborhood of $\scrip$ and with the above choice of coordinates, the line element can be cast into a form very similar to Eq. \eqref{equ:NGC_metric}. The resulting metric morally extends the universal structure off $\scrip$, describing particularly the asymptotic behavior close to $\scrip$. 

\begin{proposition}
    Given the above structure on $\scrip$ in BMS coordinates of an asymptotically flat spacetime, there exists a unique coordinate extension $(u,\Omega=r^{-1},z,\bar z)$ in the neighborhood of $\scrip$ such that
    \begin{align}
    \label{equ:BMS_metric}
        \dd s^2 &= \widetilde{g}_{\mu\nu} \dd x^\mu\dd x^\nu \notag \\
        &=\frac{e^\beta V}{r}\dd u^2 -2 e^{2\beta} \dd u \dd r + r^2 H_{AB} (\dd y^A-U^A \dd u)(\dd y^B - U^B\dd u)
    \end{align}
    with 
    \begin{align}
        V(u,r,y^A) = \mathcal{O}(r^{3})\,,&&\beta(u,r,y^A) = \mathcal{O}(r^0)\,,&&U^A(u,r,y^A) = \mathcal{O}(r^0)\,,
    \end{align}
    as well as
    \begin{align}
    H_{AB}(u,t,y^A)= q_{AB}(z,\bar z) + \mathcal{O}(r^0)\,, && \partial_r\det (H)=0\,.
    \end{align}
    Thereby, $q_{AB}$ represents the metric of $\mathcal S^2$ with radius $1$. The last equation is a particular gauge choice.
\end{proposition}
\noindent \textit{Proof.} One starts at $\scrip$ where the coordinate choice implies a tetrad $(n|_{\scrip}=\partial_u, m|_{\scrip}=(1+|z|^2)\sqrt{2}^{-1}\partial_z, \bar m|_{\scrip}=(1+|z|^2)\sqrt{2}^{-1}\partial_{\bar z})$ which can be uniquely extended by $\ell|_{\scrip}=\partial_r$ to become a null tetrad at $\scrip$. As $\ell^\mu$ is, by construction, a null transverse to $\scrip$, it generates a \gls{ngc}. The integral curves belonging to this \gls{ngc} all solve the geodesic equation for $\gd$ at $\scrip$ and, since null geodesics are conformal invariants, they also solve the geodesic equations for $\widetilde{g}_{\mu\nu}= \Omega^{-2}\gd$. Being transverse to $\scrip$, $\ell^\mu$ allows for an extension of the \gls{ngc} into the bulk spacetime and off $\scrip$. This is done by requiring that $(u,y^A)$ remains constant along null geodesics. Further, $g_{\Omega\Omega} = 0$ because $\ell\sim \partial_\Omega$ generates the congruence, $\partial_r g_{A\Omega} = 0$ since $\ell$ is geodesic, i.e., $\nabla_\ell \ell^\mu \sim \ell^\mu$, and $g_{z\Omega}|_{\scrip} = g(\partial_z,\partial_\Omega)|_{\scrip} = m^\mu \ell^\nu \gd |_{\scrip} = 0$ by the definition of $\ell$. It also follows from the latter that $g_{z\Omega}=0$. Since $\gd|_{\scrip}$ is finite, it must further hold that $V(u,r,y^A) = \mathcal{O}(r^{3})\,,\beta(u,r,y^A) = \mathcal{O}(r^0)\,,U^A(u,r,y^A) = \mathcal{O}(r^0)\,,$ as well as $ H_{AB}(u,t,y^A)= q_{AB}(z,\bar z) + \mathcal{O}(r^0)\,$. Finally, the rescaling freedom in $\Omega \rightarrow f(x^\mu)\Omega$ is used to fix $\partial_r\text{det}(H)=0$. 

\noindent The above line element is defined on the physical spacetime and, thus, describes a physical metric. It is constructed in such a way that in the limit of $\Omega \rightarrow 0$, the adapted coordinate system at $\scrip$ outlined above is recovered. More importantly, comparing Eq. \eqref{equ:BMS_metric} with the metric defined in Subsection \ref{subsec:NGC}, one finds that Eq. \eqref{equ:BMS_metric} defines a \gls{ngc} ``starting'' at $\scrip$. The latter relation becomes evident in the above proof. It has been a key insight for the derivation of a related definition of asymptotically flat spacetimes by Newman and Unti \cite{Newman_Unti_Gauge}. Their definition and the metric \eqref{equ:BMS_metric} share the same geometric foundations. One crucial difference is the gauge\footnote{The gauge condition is not the only difference between the definitions in \cite{Newman_Unti_Gauge} and \cite{Bondi_Origin_2, Sachs_Origin_3}. The symmetry algebra of these works is fundamentally different. Thus, despite sharing the same geometric foundation, i.e., the \gls{ngc}, the \cite{Newman_Unti_Gauge} obtains a slightly larger symmetry algebra consisting of a semidirect product of the BMS algebra with an abelian algebra of infinitesimal conformal rescaling \cite{Barnich_2012}.}. While above the condition $\partial_r \text{det}(H_{AB})=0$, known as BMS gauge, is imposed, in the context of \gls{ngc} the Newman-Unti gauge (Newman-Unti coordinates) was applied, i.e., $\beta = 0$. The latter follows automatically by choosing $\Omega$ to be the affine parameter of $\ell^\mu$.\\ 
The BMS metric \eqref{equ:BMS_metric} contains multiple functions that have yet to be interpreted physically. In a straightforward but tedious manner, they could be computed by solving Einstein's equations for a given spacetime. This computation can be significantly simplified by applying the asymptotic expansion of the metric coefficients postulated by Bondi, Metzner and Sachs \cite{Bondi_Origin_2, Sachs_Origin_3}: Assuming that the metric coefficients in \eqref{equ:BMS_metric} can be written as decaying power series in $r$, i.e.,
\begin{align}
    r^{-3}V(u,r,y^A)&= \sum_{i=0}^\infty V_{(i)}(u,y^A)r^{-i}\,,&&U^A(u,r,y^A)= \sum_{i=0}^\infty U^A_{(i)}(u,y^A)r^{-i}\,,\notag\\
    \beta(u,r,y^A)&= \sum_{i=0}^\infty \beta_{(i)}(u,y^A)r^{-i}\,,&&H_{AB}(u,r,y^A)= \sum_{i=0}^\infty H_{AB}^{(i)}(u,y^A)r^{-i}\,,
\end{align}
one finds that the solution to Einstein's equation order by order in $r$ imply
\begin{align}
\label{equ:HAB}
    H_{AB} &= q_{AB} + \frac{1}{r} C_{AB}(u, z, \bar z) + \frac{1}{r^2} D_{AB}(u,z,\bar z) + r^{-3} E_{AB} + \mathcal{O}(r^{-3})\,,\\
    \label{equ:beta}
    \beta &= 0 + \frac{1}{r^2}\left(-\frac{1}{64} C^{AB}C_{AB}\right) + + \mathcal{O}(r^{-3})\,,\\
    \label{equ:V}
    r^{-3}V &= 0- \frac{1}{2}\frac{1}{r^2}+ \frac{1}{r^3} 2 M(u,z ,\bar z)  + \mathcal{O}(r^{-4}) \,,\\
    \label{equ:U}
    U^A &= 0 - \frac{1}{r^2} \left(\frac{1}{2}\nabla_C C^{CA}\right) + \frac{1}{r^3}N^A(u,z,\bar z)+ \mathcal{O}(r^{-4})\,.
\end{align}
The resulting new functions, except for one, are either fixed by Einstein's equations or constrained by evolution equations. Starting with Eq. \eqref{equ:HAB}, one finds that $C_{AB}$ is symmetric and trace-free but otherwise unconstrained while $D_{AB}$ and $E_{AB}$ are constrained by evolution equations, for instance $\partial_u D_{AB}=0$. The metric of the $\mathcal S^2$ sphere is again denoted by $q_{AB}$. Regarding Eq. \eqref{equ:beta}, all terms at all orders are fixed by Einstein's equations. The same holds true for all subleading terms in \eqref{equ:V} and \eqref{equ:U}. The functions $M,N^A$, known as \textit{mass aspect} and \textit{angular momentum aspect}, are again constrained by evolution equations. In a static, spherically symmetric case, the mass aspect reduces to the mass of the compact object, for instance, a Schwarzschild \gls{bh}\footnote{In this case, $\beta, U^A$ become zero and the metric reduces to the Eddington-Finkelstein metric.}. The angular momentum aspect $N^A$ earns is name by being derived via solving the $r,A$ component of Einstein's equation which relates, in the non-vacuum case, to the angular momentum flux of the matter fields (i.e., $r^2T_{rA}$ where $T$ is the stress-energy tensor). A detailed derivation of the above functions is provided in \cite{M_dler_2016}. 

From a \gls{gw} physics point of view, perhaps the most important metric components are the mass aspect and the tensor $C_{AB}$. The latter is often referred to as \textit{asymptotic shear}. This results from its relation to the shear of the \gls{ngc} (as defined above, right after \eqref{equ:NGC_metric}). In fact, using this prescription, one finds 
\begin{align}
    \sigma_{AB} = \partial_r H_{AB} = rC_{AB} + \mathcal{O}(r)\,.
\end{align}
The asymptotic shear, in turn, defines the \textit{Bondi news} tensor $N_{AB}:= \partial_u C_{AB}$. The Bondi news tensor captures the conformally invariant part of the Riemann tensor at $\scrip$ and is often described to encapsulate radiative information. Indeed, one can compute that in the absence of gravitational radiation, $N_{AB}=0$. Up to a conventional factor of $2$, the Bondi news tensor determines the NPS, as shown above, via
\begin{align}
\label{equ:psi_4_bondi}
    \Psi_4 = \frac{1}{r} \partial_u N_{zz} + \mathcal{O}(r^{-2})\,.
\end{align}
The Bondi news tensor also determines the evolution equation of the mass aspect. Note, however, that, to find a reasonable expression in terms of the Bondi news tensor, Einstein's equations are supplemented by the contracted Bianchi identities. Only then, it holds that
\begin{align}
\label{equ:pre_bondi_mass_loss}
    \partial_u M =&  \frac{1}{4}\nabla_A\nabla_B N^{AB}-\frac{1}{8}N_{AB}N^{AB}\,,
\end{align}
from which the Bondi mass loss formula
\begin{align}
\label{equ:bondi_mass_loss}
    \frac{\dd M_B}{\dd u}=\partial_u \int_{\mathcal{S}^2} \dd \Sigma \, M(u,z,\bar z) = -\frac{1}{4} \int_{\mathcal{S}^2}\dd \Sigma \, N_{AB}N^{AB}\,.
\end{align}
follows by integration over $\mathcal{S}^2$. Eq. \eqref{equ:bondi_mass_loss} can be loosely interpreted as a flux-energy condition. Thereby, the left-hand side acts as the variation of energy in time while the right-hand side embodies the flux flowing out of the system (thus the minus sign). This equation can be seen as a predecessor to the balance flux laws derived in Section \ref{sec:Covariant_PS}. Note, however, that the interpretation of \eqref{equ:bondi_mass_loss} is very sensitive to the chosen fall-off conditions for the metric components and the definition of asymptotic flatness. While in the above, the asymptotic behavior of the metric components has been rigorously defined, some works tend to loosen the restrictions (see, for instance, \cite{Different_Spacetimes_II} and \cite{Flanagan_2017}), resulting in an extended BMS algebra as well as slightly different constraint equations for the individual components. Moreover, in realistic scenarios such as \gls{gw} measurements, the only accessible quantity is the gravitational strain. Although it is computed based on the Bondi news tensor, the mass aspect remains unknown. Therefore, when deriving the balance flux laws below in Section \ref{sec:Covariant_PS}, one makes use of a more robust way of analyzing the theory's fundamental conserved quantities in a more practical form at null infinity. \\
It should be highlighted at this point that, as opposed to the initial value problem on space-like hypersurfaces, solving Einstein's and the evolution equations corresponds to solving partial differential equation with data from a null surface. Interestingly, the free data of the expansion above is given by the Bondi news tensor at $\scrim$ as well as an infinite tower of zero modes at $v = -\infty$,. i.e., timelike past infinity, e.g., $C_{AB}|_{v=-\infty},M|_{v=-\infty},D_{AB}|_{v=-\infty}$. The constraints of the free data result from the evolution equations of the metric components that are derived based on Einstein's equations, as well as the Bianchi identities. The News tensor thereby remains a free function. In fact, it comprises two free functions at $\scrip$ corresponding to the two helicities of the massless graviton. The latter becomes fundamentally important when discussing the scattering problem in \gls{gr} in the context of asymptotically flat spacetime, see \cite{strominger_lectures} for an exhaustive review. 

Given the metric \eqref{equ:BMS_metric}, it is natural to ask whether the asymptotic symmetries preserving the expansion above are the same that one finds by preserving the universal structure at $\scrip$, i.e., definition \eqref{equ:lie_algebra_vector}. Therefore, one is now interested in the infinitesimal symmetry generators $X$ that, given the BMS gauge (sometimes also only called \textit{Bondi gauge}), yield a metric obeying the same constraints when varied, i.e.
\begin{align}\label{equ:KillingI}
    \delta g_{rr}=0,&& \delta g_{rA}=0, && g^{AB}\delta g_{AB}=0\,,
\end{align}
where the latter prevents $g_{AB}$ from undergoing a conformal rescaling under a coordinate transformation. Further,
\begin{align}\label{equ:KillingII}
    \delta \widetilde g_{uu}=\mathcal{O}(r^{-1}), && \delta \widetilde  g_{uA}=\mathcal{O}(1), && \delta \widetilde  g_{ur}=\mathcal{O}(r^{-2}),&& \delta \widetilde  g_{AB}=\mathcal{O}(r)\,.
\end{align}
Replacing $\delta$ with the Lie derivative along a generator of a diffeomorphism, or more precisely, along a Killing vector field $X$, the Killing equations that determine the corresponding components of $X_\mu$ are obtained. For instance, the first equation in \eqref{equ:KillingI} yields
\begin{align}
    \nabla_r X_r=\partial_r X_r - \Gamma\ud{u}{rr}\xi_u - \Gamma\ud{r}{rr}X_r- \Gamma\ud{A}{rr}X_A=0\,.
\end{align}
This results in 
\begin{align}
    \partial_r X_r = 2\partial_r \beta\,,
\end{align}
which is solved by 
\begin{align}
    X_r = f(u,x^A) e^{2\beta}\,,
\end{align}
where $f$ is a differentiable function. For the other components the procedure is less trivial but follows a similar logic. All in all, it follows that
\begin{align}
    X^u &=\alpha(x^A) + \frac{u}{2}\nabla_A Y^A(x^B),\\
   X^A &=  Y^A - \frac{\nabla^A\alpha(x^A)}{r}-u\frac{\nabla^A\nabla_C Y^C(x^B)}{2r} + \mathcal{O}(r^{-2}),\\
    X^r &= -\frac{r}{2}\nabla_CY^C(x^B) + \frac{\nabla_C\nabla^C\alpha(x^A)}{2}+ u\frac{\nabla_C\nabla^C\nabla_AY^A(x^B)}{4} + \mathcal{O}(r^{-1})\,,
\end{align}
where $\nabla_A$ is the covariant derivative on the unit $\mathcal S^2$ sphere. In the limit to $\scrip$, these equations boil down to
\begin{align}\label{equ:finalKF}
    X&= \underbrace{X^u \partial_u}_{=:X_\alpha} + X^A \partial_A\notag\\
    &= \left[\alpha(x^A)+\frac{1}{2}u\nabla_AY^A(x^B) \right]\partial_u + Y^A(x^B) \partial_A\,,
\end{align}
while $X_r$ is not well-defined at null infinity and drops out. Comparing the latter to what is obtained in \eqref{equ:lie_algebra_vector}, the replacement $Y^A\partial_A = \chi \partial_z + \bar \chi \partial_{\bar z}$ leads to the conclusion that the generator $X$ is identical to the one preserving the universal structure. 
Note that $\alpha$ is an arbitrary function and $Y^A$ are the conformal Killing vectors of the metric of the unit $2$-sphere. The function $\alpha$ effectively describes the generator of the supertranslations as defined previously in Eq. \eqref{equ:lie_algebra_vector}. Again, as it is a function of the angular coordinates only, one can decompose it into spherical harmonics where $l=0,1$ represent time and spatial translations while $l>1$ are the ``proper'' supertranslations. The vector $Y^A$ encompasses six Lorentz transformations that preserve the unit-sphere metric at the boundary $\scrip$. The generators $Y^A$ form a $so(1,3)$ subalgebra of diff($S^2$) and obey, by definition, the conformal killing equation on $S^2$, i.e.,
\begin{align}\label{equ:SuperLorentz}
    \nabla_AY_B+\nabla_BY_A=h_{AB}\nabla_CY^C
\end{align}
The latter equation provides (globally) three rotations uniquely defined by $\nabla_AY^A=0$ and three boosts given by $\varepsilon^{AB}\partial_AY_B=0$, where $\varepsilon^{AB}$ is the totally antisymmetric tensor on the unit sphere. Hence, for $\alpha=0$, the Killing fields of the BMS group are isomorphic to the orthochronous Lorentz transformations \cite{Sachs_Origin_2}. Therefore, the generator $X$ again exactly results in the BMS group.

\subsection{Asymptotic Shear and its Interpretation}
\label{subsec:asym_shear}
With the metric laid out in the above format, it is instructive to elaborate on the connection between memory and asymptotic data, i.e., the asymptotic shear. As mentioned briefly in the introduction, the \gls{gw} memory can be fundamentally related to a change in vacua at $\scrip$ which are related via supertranslations \cite{IR_triangle_III}. The logic behind this statement rather intuitively follows when one relates the Bondi news tensor $\partial_u C_{AB}$ to gravitational radiation, e.g., via Eq. \eqref{equ:psi_4_bondi}, and accepts the claim that two spacetimes related by supertranslations are inequivalent \cite{Bondi_Origin_2}. Starting with the latter, one can compute the action of a supertranslation on free data on $\scrip$ by taking the Lie derivative w.r.t. the corresponding generator. Explicitly writing out the components of the involved tensors in $(u,z,\bar z)$ coordinates, one finds that 
\begin{align}\label{equ:lie_free_data_i}
    \lie _{X_\alpha} N_{zz} &= \alpha\partial_u N_{zz}\,,\\
    \lie_{X_\alpha} M &= \alpha \partial_u M + \frac{1}{4}\left(N^{zz}\nabla_z^2\alpha + 2 \nabla_z N^{zz}\nabla_z\alpha + c.c.\right)\,,\\
    \label{equ:lie_free_data_ii}
    \lie_{X_\alpha} C_{zz} &= \alpha \partial_u C_{zz}-2\nabla_z^2 \alpha\,.
\end{align}
Here, $X_\alpha$ refers to the supertranslation generator, see Eq. \eqref{equ:finalKF}. Eq. \eqref{equ:lie_free_data_i}-\eqref{equ:lie_free_data_ii} imply that, for Minkowski space, $M=C_{AB}=N_{AB}=0$ supertranslations do not change mass aspect or Bondi news. The latter is consistent with the expectation that diffeomorphism generally cannot change physical quantities like mass, or create \gls{gw}s $(N_{AB}\neq0)$. Supertranslations, however, do change the asymptotic shear $C_{AB}$. In fact, one can show that vanishing curvature (which is necessarily assumed when talking about vacuum) for the expansion above implies
\begin{align}
    \nabla_{\bar z}^2 C_{zz} - \nabla_{z}^2 C_{\bar z\bar z }=0\,.
\end{align}
The latter is solved by 
\begin{align}
\label{equ:goldstone}
    C_{zz} = -2 \nabla_z^2 C(z,\bar z)\,,
\end{align}
which, under supertranslations, transforms as $\lie_{X_\alpha} C (z.\bar z)= \alpha$, according to \eqref{equ:lie_free_data_ii} and \eqref{equ:goldstone}\footnote{Given this particular Lie derivative, one could view $C(z,\bar z)$ as a Goldstone boson resulting from a spontaneously broken supertranslation symmetry \cite{IR_triangle_III}.}. In turn, given two vacua $C_{zz}(u_2),C_{zz}(u_1)$ where $u_2<u_1$, the supertranslation relating these states can be computed by integrating \eqref{equ:pre_bondi_mass_loss}
\begin{align}
\label{equ:vac_trans}
    \nabla_z^2 \left[C_{zz}(u_2)-C_{zz}(u_1)\right] = 2 \left[M(u_2)-M(u_1)\right]\,.
\end{align}
This equation can be solved for $\Delta C(z,\bar z)$ producing $\Delta C_{zz}=C_{zz}(u_2)-C_{zz}(u_1)$ which effectively computes the corresponding supertranslation generator $X_\alpha$. Note that in case of a non-trivial energy-stress tensor, the right-hand side of \eqref{equ:vac_trans} obtains another contribution.\\
The above sketch demonstrates that BMS vacua are related by supertranslations. What is left to show is how a change of the vacuum state enforces the \gls{gw} memory, i.e., the permanent displacement of two freely floating test masses. First, note that necessarily, to translate between distinct vacuaa $C_{zz}(u_1)\neq C_{zz}(u_2)$, there needs to be a time domain for which $N_{zz}= \partial_uC_{zz} \neq 0$. Thus, the vacua can only be changed as gravitational radiation passes ``through'' $\scrip$. As presented in the introduction, this gravitational radiation necessarily entails a memory. For the perspective of the asymptotic shear changed by passing gravitational radiation, one finds that, since the shear fundamentally affects the metric, the proper distance between two points changes. Therefore, to obtain an estimate of the memory induced by $\Delta C_{zz}$ one simply computes the change in proper distance $\Delta L$ of two fixed points $(u_1,z_1,\bar z_1)$ and $(u_2,z_2,\bar z_2)$ given the changed metric component $C_{zz}$. One finds that $\Delta L$ is proportional to the previously defined $\Delta C_{zz}$ (for the explicit computation, see \cite{IR_triangle_III, strominger_lectures}). This result is of fundamental importance as it implies a direct relation between the symmetry structure of spacetime and a direct physical observable, i.e., the \gls{gw} memory. Yet to be detected, its observation would act as a smoking gun for much of the theory presented in this work. 

Combining the approaches of Section \ref{subsec:Radiative_modes} with the original works of Bondi and Sachs, one can show that the asymptotic shear $C_{AB}$ contains the radiative degrees of freedom at $\scrip$. Proving the latter corresponds to deriving a connection between the traceless part of $\Sigma_{\mu\nu}$ (i.e., $\D_\mu\ell_\nu$) and $C_{AB}$. To obtain the desired relation, one necessarily needs to extend $\widetilde{g}_{\mu\nu}$ to the unphysical metric $\gd$ as the relation to be proven is inherent to $\scrip$. The reader is reminded that the \gls{ngc} is conformally invariant so that the properties derived above are still valid for $\gd$. In particular, this implies that $\newpb{\widetilde{\nabla}_\mu \ell_\nu} = \newpb{\nabla_\mu \ell_\nu} = \mathcal{D}_\mu\ell_\nu$, where $\widetilde{\nabla}_\mu$ is the covariant derivative of the physical metric and ${\nabla}_\mu$ the one for the unphysical metric, and $\mathcal{D}$ a derivative operator on $\scrip$. As outlined in Section \ref{subsec:Radiative_modes}, one can introduce a trivial derivative operator $\Dcirc_\mu$ such that $\Sigma_{\mu\nu}=\left(\mathcal D_\mu-\Dcirc_\mu\right)\ell_\nu =\mathcal D_\mu\ell_\nu$. Now with $\Sigma_{\mu\nu}= \mathcal D_\mu\ell_\nu$ one can build a relation to the quantities in the conformally completed bulk. In the above metric \eqref{equ:BMS_metric}, one natural choice for the generator of the \gls{ngc} is $\ell_\mu \sim \partial_\Omega$. In other words $\ell_\mu = -\nabla_\mu u$ which implies that 
\begin{align}
    \nabla_\mu\ell_\nu = \Gamma\udd{\sigma}{\mu}{\nu} \nabla_\sigma u =  \Gamma\udd{u}{\mu}{\nu}\,.
\end{align}
Thus, one finds that 
\begin{align}
\label{equ:shear_metric_equality}
    \Sigma_{\mu\nu}:=\newpb{\nabla_\mu\ell_\nu} =  \Gamma\udd{u}{\mu}{\nu}|_{\scrip} =\left.\frac{C_{AB}}{2}\right|_{\scrip}\,,
\end{align}
where the asymptotic shear is, by definition, symmetric and trace-free and, thus, the expression \eqref{equ:shear_metric_equality} equals the definition of the shear above, see Section \ref{subsec:radi_degrees}. Eq. \eqref{equ:shear_metric_equality} demonstrates that the radiative degrees of freedom are inherently linked to one metric component, i.e., the shear, in the definition of Bondi and others. This insight is rather powerful as, together with the definition of the NPS, it links together large parts of the literature on \gls{gw}s, symmetries, and conserved quantities.


%
%
%


\section{Towards Conserved Quantities at Null Infinity $\scrip$}
\label{sec:Covariant_PS}

So far in this Chapter, the asymptotic spacetime structure is presented as a powerful tool for determining the presence of gravitational radiation far from the source. Two perspectives, the BMS metric and the radiative modes hidden in the derivative operator, are presented, yielding the exact same result, including the symmetries present at the boundary $\scrip$. Relations to physically observable quantities like the \gls{gw} strain and the memory are hinted at. However, so far, a discussion about the implications of the symmetries, i.e., their associated conserved quantities, is not to be found. In fact, the latter turns out to be less trivial compared to common theories, such as Maxwell's theory. A somewhat intuitive but mathematically not very sound explanation is given in the introduction \ref{sec:intro_conserved}. \\
In this Section, this shortcoming is rectified by introducing a mathematical framework developed by Wald and Zoupas \cite{Wald_Zoupas} with which conserved quantities can be obtained for \gls{gr} in asymptotically flat spacetimes. This formalism employs a different language compared to what has been presented so far. The translational steps necessary to bring the results of the latter into congruence with previous Sections, however, constitute much less effort than continuing in the language of radiative modes encoded in derivative operators \cite{Ashtekar_Streubel}. To highlight this tradeoff, the work of Ashtekar and Streubel \cite{Ashtekar_2017} is briefly presented and connected to the formalism by Wald and Zoupas in Section \ref{subsec:Ashtekar_Streubel}. In total, based on the previous endeavor, this Section is aimed at formulating the necessary tools, i.e., the balance flux laws for gravitational radiation, for practical applications of the mathematical framework discussed so far. Concretely, in this thesis, it is demonstrated how the results of this Section can be used to improve and test gravitational waveforms (see Section \ref{sec:Paper_BL}) as well as to add novel signatures (see Section \ref{sec:Paper_Mem}). Note that this treatment applies exclusively to diffeomorphism-invariant theories. 

\subsection{The Covariant Phase Space Approach}
\label{subsec:Wald_Zoupas}
When discussing conserved quantities, it is often beneficial to switch from a Lagrangian to a Hamiltonian picture. Being fundamentally connected to the phase space of a system, the latter uses the language of symplectic geometry to define crucial quantities such as the Hamiltonian conjugate. If a Hamiltonian conjugate to some (Killing) vector on a given Manifold exists, it provides a somewhat natural definition of a conserved quantity w.r.t. this vector. However, generally the existence of a Hamiltonian is not to be taken for granted and, in fact, at $\scrip$ no Hamiltonian exists in \gls{gr}. In their work \cite{Wald_Zoupas}, Wald and Zoupas argued that the reason for the absence of a Hamiltonian at $\scrip$ is the presence of a flux (of gravitational radiation). Mathematically, this flux arises in their definition of a modified Hamiltonian for diffeomorphism covariant theories by virtue of Poincar\'e's Lemma, which roughly states that a closed differential form is only determined up to its exact part. As it is elaborated in more detail below, the remaining ambiguity of a closed form relevant to the definition of the Hamiltonian \`a la Wald \& Zoupas in combination with some physical intuition yield an expression for the flux across the boundary of the bulk manifold. For a precise analysis of various mathematical subtleties encountered during the construction presented here, the reader is referred to the original work \cite{Wald_Zoupas} (see also \cite{Harlow_2020,Lehner_2016} for generalizations of the geometric setup). Throughout the demonstration, it is helpful to keep in mind that one aims at conserved quantities w.r.t. a Killing vector field $\xi^\mu$ of spacetime at $\scrip$, i.e., a generator of an asymptotic (BMS) symmetry (defined by $X^\mu$ in Eq. \eqref{equ:lie_algebra_BMS}, Section \ref{subsec:BMS_Group}).

\subsubsection{The Setup:}
One starts by outlining the general setup. Consider an $n$-dimensional manifold $ M$ \footnote{In this work, one exclusively deals with $4$-dimensional spacetime. Nonetheless, the prescription outlined in the following holds in all generality.} with a set of dynamical fields $\phi=\{\gd, \psi\}$ including the metric $\gd$ and additional tensor fields $\psi$ \footnote{Although the physical manifold is considered, until considering \gls{gr} explicitly, the tilde ontop of quantities defined on this physical manifold is dropped}. The fields admit values only within a space $\F$ which is dubbed the space of ``kinematically allowed'' field configurations. A precise definition of this term is very sensitive to the concrete system under consideration. In the context of asymptotically flat spacetimes, it particularly encompasses the asymptotic conditions on $\phi$ as well as global hyperbolicity of the metric. Given that the dynamics of the theory are encoded in the $n$-form Lagrangian density 
\begin{align}
    \mathbf L = \mathbf{L}(\gd, R_{\mu\nu\rho\sigma}, \nabla_\mu R_{\nu\rho\sigma\gamma}, ..., \psi, \nabla_\mu)\,,
\end{align}
where the derivative operators $\nabla_\mu$ are the torsion and non-metricity free covariant derivatives associated to $\gd$, the equations of motions follow upon variation, i.e., 
\begin{align}
    \delta\mathbf L = \mathbf E (\phi)\delta \phi + \dd \boldsymbol{\theta}(\phi,\delta\phi)\,.
\end{align}
The Euler-Lagrange equations result from $\mathbf E(\phi)=0$ and $\boldsymbol{\theta}$ corresponds to the boundary term resulting from partial integration of terms of the form $\nabla \delta\phi$. Since the boundary term $\boldsymbol{\theta}$ appears with the exterior derivative in the variation of $\mathbf{L}$, its influence on the equations of motion is determined up to the addition of an exact $n$-form $\mathbf{W}$
\begin{align}\label{equ:freedom_bt}
    \boldsymbol{\theta}\rightarrow\boldsymbol{\theta} + \boldsymbol{W}\,.
\end{align}
This freedom directly transfers to the presymplextic current $\boldsymbol{\omega}$ defined as
\begin{align}
    \boldsymbol{\omega}(\phi,\delta_1 \phi, \delta_2 \phi) = \delta_1 \boldsymbol{\theta}(\phi,\delta_2\phi)-\delta_2 \boldsymbol{\theta}(\phi,\delta_1\phi)\,.
\end{align}
Thereby, $\delta_1 \phi,\delta_2\phi$ are linearized perturbations off $\phi$ in field configuration space (satisfying the linearized field equations). The ambiguity in $\boldsymbol{\theta}$ results in $\boldsymbol{\omega}\rightarrow\boldsymbol{\omega} + \delta_1 \boldsymbol{W}(\phi,\delta_2\phi)-\delta_2 \boldsymbol{W}(\phi,\delta_1\phi)$. Note that technically, $\bt$ obtains another contribution upon adding another exact $n$-form $\mathbf K$ to the Lagrangian form. While the equations of motion remain unaffected, $\bt \rightarrow \bt + \delta \mathbf K$. Note, however, that this does not change the presymplectic current due to the commutation of variations. One can now define a presymplectic form $\Omega_\Sigma$ mapping field configurations and linearized perturbations into real numbers. This form is associated with a slice $\Sigma$ \footnote{For intuition, in the relevant setting of conformally compactified spacetime, a choice of $\Sigma$ would be a null slice such that the intersection of the slice with the total manifolds boundary in the future results in a cross section of $\scrip$.}, such that 
\begin{align}\label{equ:symp_pot_wald}
    \Omega_\Sigma (\phi,\delta_1\phi,\delta_2\phi)= \int_\Sigma \bw\,.
\end{align}
Note here that the ambiguity in the definition of $\bw$ also results in an ambiguity of the presymplectic form $\Omega_\Sigma$. The presymplectic form is generally degenerate and allows for the subtraction of orbits of the degeneracy subspace from $\F$ \cite{Wald_Zoupas} (see also references therein). Through this projection, $\Pi_{\Omega_\Sigma}: \F\rightarrow \Gamma$, one obtains the phase space $\Gamma$ as well as a nondegenerate symplectic form $\Omega$.  
The factoring out of orbits also yields a projection from $\F$ to $\Gamma$ which allows the definition of a Hamiltonian in the following sense: Given a vector field $\xi^\mu$ on $\mathcal M$ its Lie derivative generates a field variation $\lie_\xi \phi$ on $\F$ which may be identified with $\delta_\xi\phi = \lie_\xi\phi$ if, like the variation, $\lie_\xi \phi$ is tangent to $\F$, i.e., the corresponding diffeomorphism does not map out of $\F$. The vector field can then be seen as generating an evolution of $\phi$ within $\F$. Restricting $\F$ to the covariant phase space, or, solution submanifold $\bar \F$, if $\xi^\mu$ consistently projects to phase space, an image $\bar \Gamma$ of $\bar \F$ is generated. In other words, 
let $\bar \F$ denote the subset of $\F$ consisting of solutions to
the equations of motion. If one interprets $\Gamma$ as the space of kinematically allowed states of the system, then $\bar \Gamma$ consists
of those states that are dynamically possible. If $\bar \Gamma$ is a proper subset of $\Gamma$, then not all kinematically possible states are
dynamically possible, i.e., constraints are present (see Fig. \ref{fig:symplectic}).
If the pullback of $\Omega$ to $\bar \Gamma$ is preserved by the evolution vector field $\xi^\mu$ on $\bar \Gamma$, i.e., $\lie_\xi(\varphi^* \Omega)=0$, then it is generated by an Hamiltonian $H_\xi$ conjugated to $\xi$ \cite{Lee_Wald, Wald_Zoupas}. This rather technical construction establishes the foundation for a valid definition of the Hamiltonian: 
\begin{figure}[t!]
    \centering
    \includegraphics[width=0.35\textwidth]{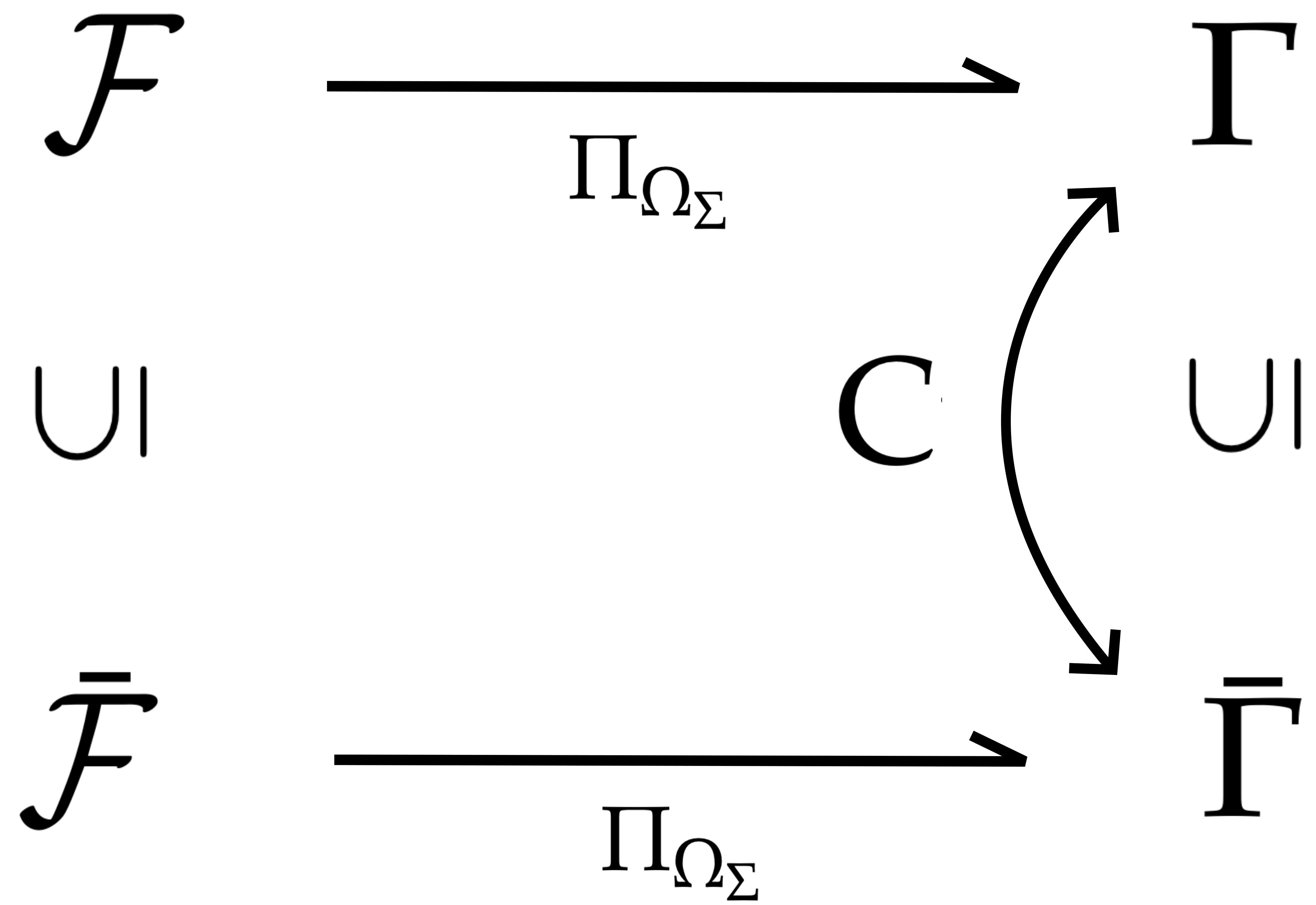}
    \caption{Relation between field configuration space $\F$ and phase space $\Gamma$ by the map $\Pi_{\Omega_\Sigma}$ described in \cite{Lee_Wald}. In the presence of constraints such as equations of motions and else collectively denoted by C, the system is restricted to the subspaces $\bar{\F}$ and $\bar{ \Gamma}$ respectively.} 
    \label{fig:symplectic}
\end{figure}
Given a particular choice for $\Omega_\Sigma$ (i.e., assuming the ambiguity has been removed) and the finiteness of $\int_\Sigma \bw(\phi,\delta\phi,\lie_\xi \phi)$ for all $\phi\in \bar \F$ and $\delta \phi$ tangent to $\F$, the function $H_\xi:\F \rightarrow \mathbb R$ defines the Hamiltonian conjugate to $\xi^\mu$ on $\Sigma$ if
\begin{align}\label{equ:hamiltonina_def}
    \delta H_\xi=\Omega_\Sigma(\phi,\delta\phi,\lie_\xi \phi)=\int_\Sigma \bw(\phi,\delta\phi,\lie_\xi \phi)\,.
\end{align}
The value given by $H_\xi$ provides a definition of a conserved quantity associated with $\xi$ on $\Sigma$.

\subsubsection{Criteria for $H_\xi$:}
The above definition of the Hamiltonian is valid only for a restricted set of cases. Most importantly, for \gls{gr} in asymptotically flat spacetimes it does not apply -- there exists no Hamiltonian in this case. The necessary criteria for the existence of $H_\xi$ can be derived by rewriting the definition \eqref{equ:hamiltonina_def} using the $(n-1)$-form Noether current associated with $\xi$,
\begin{align}
    \mathbf{j}= \bt(\phi,\lie_\xi\phi) - \xi \cdot \mathbf{L}\,,
\end{align}
where $\xi\cdot \mathbf{L}$ denotes the contraction of $\mathbf{L}$'s first index with $\xi^\mu$. It is easy to check that variation then yields
\begin{align}
    \delta \mathbf{j} = \boldsymbol{\omega}(\phi,\delta\phi,\lie_\xi\phi) + \dd(\xi\cdot \bt)\,.
\end{align}
Given that for diffeomorphism covariant theories one can rewrite \cite{Iyer_1994}
\begin{align}
    \mathbf{j}=\dd \mathbf{Q} + \xi^\mu C_\mu\,,
\end{align}
where $C_\mu$ corresponds to the constraints of the theory, and trivializes if one restricts to on-shell solutions, one finds that 
\begin{align}
    \bw(\phi,\delta\phi,\lie_\xi\phi) = \xi^\nu\delta C_\mu + \dd (\delta \boldsymbol{Q}- \xi\cdot \bt)\,.
\end{align}
Thus,
\begin{align}
    \delta H_\xi = \int_\Sigma \xi^\mu \delta C_\mu + \int_{\partial \Sigma}[\delta \boldsymbol{Q} - \xi\cdot\bt]  \,,
\end{align}
where the first term on the left-hand side drops out if one restricts $\phi$ to on-shell solutions, i.e., when confined to $\bar \F\subseteq\F$. A straightforward criterion for the existence of $H_\xi$ is then derived based on the commutation of mixed partial derivatives $\delta_1\phi,\delta_2\phi$ such that 
\begin{align}
    0=(\delta_1\delta_2-\delta_2\delta_1)H_\xi= -\int_{\partial \Sigma} \xi\cdot \bw(\phi,\delta_1\phi,\delta_2\phi)\,.
\end{align}
Thus, for the Hamiltonian to be well defined for some field configuration $\phi$ satisfying the equations of motions, $\phi\in \bar \F$, and $\delta_1\phi,\delta_2\phi$ solving the linearized solutions and being tangent to $\bar \F$, it must hold that
\begin{align}\label{equ:criteria_symplectic}
    \int_{\partial \Sigma} \xi\cdot \bw(\phi,\delta_1\phi,\delta_2\phi)=0\,.
\end{align}
The integral is thereby evaluated as a limit towards $\partial \Sigma$, i.e., one chooses a compact region $K$ of $\Sigma$ with boundary $\partial K$, integrates over $\partial K$, and then computes the limit of $K$ approaching $\Sigma$. Naturally, this involves calculating the pullback of $\xi \cdot \bt$ to $\partial K$. The criterion \eqref{equ:criteria_symplectic} is trivially satisfied if, based on the asymptotic behavior of $\phi$, $\bw$ is such that it approaches zero sufficiently fast asymptotically, such that in the limit of $K$ to $\Sigma$ the integral vanishes. Equally, if $\xi^\mu$ is always tangent to $\partial K$, the criterion \eqref{equ:criteria_symplectic} is satisfied as the pullback of $\xi \cdot \bt$ to $\partial K$ vanishes (as the pullback intrinsically is defined w.r.t. a normal vector). If neither of those two criteria holds, the Hamiltonian is generally not well defined. 

For the purpose of this Section, the above formal definition is put into the framework of \gls{gr} acting on asymptotically flat spacetimes. In this case, $\xi$ is considered to generate an infinitesimal asymptotic symmetry and a complete vector field on $\mathcal M$ with $(n-1)$-dimensional boundary $\scri$ (for physically relevant scenarios, the consideration of $\scrip$ is sufficient). It is associated with a diffeomorphism mapping $\bar \F$ into itself. The manifold $\mathcal M$ is equipped with a universal structure as described in detail in previous Sections, which enters the specification of the fall of conditions of the fields encoded in $\bar \F$. Thus, for $\xi$ to be associated with an asymptotic symmetry, it must preserve the asymptotic conditions specified in $\bar \F$, or, in other words, $\lie_\xi\phi$ is tangent to $\bar \F$. It is further assumed that $\bw(\phi,\delta_1\phi,\delta_2\phi)$ extends continuously to $\scrip$ and the slice $\Sigma$ extends smoothly to $\scrip$ such that the extended hypersurface intersects null infinity in an $(n-2)$-dimensional submanifold $\partial \scrip$, i.e., a $u=const.$ cross section of $\scrip$. Given this setup, it does in general not hold that the extension of $\bw$ has a vanishing pullback to $\scrip$, i.e., Eq. \eqref{equ:criteria_symplectic} is not satisfied. Note that it might be the case that $\xi$ is everywhere tangent to $\partial \scrip$, which enables the definition of a Hamiltonian \eqref{equ:hamiltonina_def}. This, however, does not mean that it embodies a conserved quantity as 
\begin{align}\label{equ:conserved_quant}
    \delta H_\xi |_{\delta \scrip_2}-\delta H_\xi |_{\delta \scrip_1} = -\int_{\Delta\scrip} \bw(\phi,\delta_1\phi,\delta_2\phi)
\end{align}
does not in general vanish. Here, the Section $\Delta \scrip$ denotes the portion of $\scrip$ enclosed by the cross sections $\partial \scrip_1$ and $\partial \scrip_2$ (corresponding to slices of constant $u$, i.e., $u_1$ and $u_2$ with $u_2> u_1$, at null infinity). Thus, even if all asymptotic symmetries are found to be generated by vectors everywhere tangent to $\partial \scrip$ and a Hamiltonian exists, one remains empty-handed regarding a definition of a conserved quantity\footnote{Note at this point that the failure regarding the definition of conserved quantities does not mean all hope is lost. The contrary is the case. If symmetries are present, missing conservation hints the existence of non-conservation-, or flux-laws.}.  

Interestingly, for \gls{gr} on spacetimes being asymptotically flat at spatial infinity, one finds that $\bw$ indeed has a vanishing pullback to the boundary of the manifold $\mathcal M$ which, in this case, would be spatial infinity. Thus, $H_\xi$ does exist in this case and can be shown to be independent of the cross section $\partial \Sigma_i$ such that $H_\xi$ is indeed a conserved quantity. The definition above, in fact, gives rise to the ADM mass given $\xi$ corresponds to asymptotic time translations \cite{Wald_Zoupas}.

\subsubsection{A Hamiltonian \`a la Wald and Zoupas:}
In \cite{Wald_Zoupas}, the authors circumvent the apparent issue by constructing their own version of conserved quantity (or Hamiltonian), which is conserved independent of the orientation of $\xi$ at $\scrip$. To distinguish the new Hamiltonian from the definition \eqref{equ:hamiltonina_def}, it is denoted by $\Hh_\xi$. The logic behind the definition of $\Hh_\xi$ is quite simple. Given the criteria \eqref{equ:criteria_symplectic}, one can trivially satisfy the condition by adding a term to the Hamiltonian which exactly cancels the left-hand side \eqref{equ:criteria_symplectic} after the double variation. The job is done by
\begin{align}\label{equ:why_not}
    \int_{\partial \scrip} \xi\cdot \bT\,,
\end{align}
where $\bT$ is the symplectic potential corresponding to the pullback $\newpb{\bw}$ of the extension of $\bw$ to $\scrip$, i.e., 
\begin{align}\label{equ:pullback_bw}
    \newpb{\bw}(\phi,\delta_1\phi,\delta_2\phi) = \delta_1 \bT(\phi,\delta_2\phi) - \delta_2 \bT(\phi,\delta_1\phi)\,.
\end{align}
Here, $\bT$ is conformally invariant, i.e., it depends only on the universal structure of $\scrip$ and therefore admits similar rescaling freedom. Note also that the integral \eqref{equ:why_not} is intrinsically defined on $\scrip$ and not, as before, computed in a limiting procedure involving the pullback. For more insights on potential dependencies in the construction of $\bT$, see \cite{Wald_Zoupas}.  With this term at hand, the new Hamiltonian is defined as 
\begin{align}\label{equ:new_Hamiltonian}
    \delta \Hh_\xi = \int_{\partial\scrip} [\delta \boldsymbol{Q } - \xi\cdot \bt] -\int_{\partial \scrip}  \xi \cdot\bT
\end{align}
which trivially satisfies Eq. \eqref{equ:criteria_symplectic}. Therefore, $\Hh_\xi$ establishes a conserved quantity up to an arbitrary constant which is fixed upon a suitable choice of reference solution $\phi_0\in \bar \F$. So far, so good. The involved reader, however, might notice that the modification that is adapted for the new definition of the Hamiltonian, $\Hh_\xi$, could be considered cheating as the quantities $\bt$ and $\bT$ may seem to be one and the same at first glance. This, however, is not the case as it will be demonstrated in the following, largely due to an ambiguity in the symplectic potential. Indeed, except for the additional constant, definition \eqref{equ:new_Hamiltonian} fails to be unique due to the freedom
\begin{align}\label{equ:freedom_bT}
    \bT(\phi,\delta \phi) \rightarrow \bT(\phi,\delta \phi)+\delta \boldsymbol{W}(\phi)\,,
\end{align}
where $\mathbf{W}$ is an $(n-1)$-form on $\scrip$. Note that in contrast to \eqref{equ:freedom_bt}, the freedom \eqref{equ:freedom_bT} does not result from the invariance of the equation of motion but from an ambiguity in the pullback $\newpb{\bw}$ in \eqref{equ:pullback_bw}. Assuming that $\mathbf{W}$ inherits the independence regarding the choice of the exact background structure, meaning the conformal factor, from $\bT$, it seems to be a priori completely arbitrary. Thus, further conditions have to be enforced to give physical meaning to this piece in the new Hamiltonian. Wald and Zoupas argue that additional requirements for $\bT$ can be motivated by accounting for the presence of fluxes $\bF_\xi$ on $\scrip$ in cases where the conserved quantity associated with $\xi$ is in fact not conserved. In a physical picture, this flux is linked to the presence of radiation escaping to $\scrip$ and, hence, must vanish if $\phi$ encompasses a stationary solution. That is, for all stationary solutions, $\bF_\xi$ must vanish on $\scrip$ for all $\xi$. \\
The flux across some region on $\scrip$ is generally defined by the change in the conserved quantity along cross sections of $\scrip$,
\begin{align}
    \delta \Hh_\xi |_{\partial \scrip_2} -  \delta \Hh_\xi |_{\partial \scrip_1} = -\int_{\Delta \scrip} \delta \bF_\xi\,.
\end{align}
Inserting the definition \eqref{equ:new_Hamiltonian} on the left-hand side and applying Stokes' theorem, one finds that 
\begin{align}
    \delta \bF_\xi = \newpb{\bw}(\phi,\delta\phi,\lie_\xi\phi) + \dd[\xi\cdot \bT(\phi,\delta \phi)]\,,
\end{align}
where 
\begin{align}
    \dd[\xi\cdot \bT(\phi,\delta\phi)] = \lie_\xi \bT(\phi,\delta \phi) = -\newpb{\bw}(\phi,\delta\phi,\lie_\xi \phi) + \delta \bT (\phi,\lie_\xi \phi)\,.
\end{align}
This leads 
\begin{align}
    \delta \bF _\xi = \delta \bT (\phi,\lie_\xi \phi)
\end{align}
Given that the (stationary) reference solution $\phi_0$, fixing the undetermined constant in the definition \eqref{equ:new_Hamiltonian}, is chosen such that $\Hh_\xi$ vanishes everywhere on $\scrip$, and so does $\bF_\xi$\footnote{A straightforward example for the choice of $\phi_0$ in the context of this Chapter would be Minkowski space. For additional discussions regarding $\phi_0$, see \cite{Wald_Zoupas}.}. As now both $\bT$ and $\bF_\xi$ vanish on $\phi_0$, it follows that 
\begin{align}\label{equ:def_flux}
    \bF_\xi = \bT(\phi,\lie_\xi \phi)\,.
\end{align}
Eq. \eqref{equ:def_flux} provides a valid definition of the flux that nicely follows the phenomenological intuition. It also admits the desirable property that if $\xi$ embodies an exact symmetry such that $\lie_\xi \phi=0$, the flux vanishes independent of the presence of radiation. It further establishes clarity regarding the distinction between $\bt$ and $\bT$: Namely, one finds that the difference is given by a total variation 
\begin{align}
    \newpb{\bt}-\bT = \delta \boldsymbol{W}_{\scrip}\,.
\end{align}
Here, $\delta \boldsymbol{W}$ is fixed by conditions on the flux $\bF_\xi$ that were built based on physical intuition. It follows that 
\begin{align}\label{equ:finished_song}
    \Hh_\xi = \int_{\partial\scrip} \boldsymbol{Q}-\xi \cdot\boldsymbol{W}_{\scrip}\,,
\end{align}
where $\boldsymbol{W}$ lacks a direct physical interpretation except that it is part of the flux definition. Eq. \eqref{equ:finished_song}, however, has a clear physical purpose depending on the choice of $\xi$. If, for instance, $\xi$ is chosen to be the asymptotic time translation vector $t^\mu$ at a cross section $\partial \scrip$ of $\scrip$, then \eqref{equ:finished_song} computes the Bondi mass $M_B$ (compare Eq. \eqref{equ:bondi_mass_loss} and see below for the computation of the latter in the framework of Wald and Zoupas). Finally, note that a similar argument regarding the presence of a total derivative in the integral computing a physical quantity also enables the definition of the entropy of dynamical \gls{bh}s \cite{Dynamical_Entropy}. 

In conclusion, the treatment of Wald and collaborators successfully establishes the notion of a conserved quantity and flux at null infinity for Einstein's theory of \gls{gr}. It therefore resolves one of the most pressing issues in the asymptotic spacetime formalism. The results \cite{Wald_Zoupas} thereby agree with earlier works from Ashtekar, Streubel, and Dray \cite{Ashtekar_Streubel, Dray_Streubel} as it will be briefly discussed in Section \ref{subsec:Ashtekar_Streubel}. Before doing so, the Wald-Zoupas formula in combination with the insights from previous Sections are utilized to defined a flux formula for \gls{gr} in asymptotically flat spacetime at null infinity:
Starting with the Einstein-Hilbert Lagrangian form in physical spacetime (with the physical metric $\widetilde{g}_{\mu\nu}$)
\begin{align}\label{equ:EH_wald}
    \boldsymbol{L} = \frac{1}{16 \pi} \widetilde R \,\,{}^{(4)}\boldsymbol\epsilon\,,
\end{align}
one straightforwardly obtains the presymplectic potential $\bt$ as
\begin{align}\label{equ:idk_KKKK}
    \theta_{\mu\nu\rho} = \frac{1}{16\pi}\widetilde\epsilon_{\sigma\mu\nu\rho}\widetilde g^{\mu\alpha}\widetilde g^{\beta\gamma}(\widetilde \nabla_\beta \delta \widetilde g_{\alpha\gamma} - \widetilde \nabla_\alpha \delta \widetilde g_{\beta\gamma})\,.
\end{align}
The variation of Eq. \eqref{equ:idk_KKKK} yields the current $3$-form $\bw$, i.e., 
\begin{align}\label{equ:symp_current_GR}
    \omega_{\mu\nu\rho} = \frac{1}{16\pi}\widetilde\epsilon_{\sigma\mu\nu\rho}P^{\sigma\alpha\beta\gamma\delta\tau}(\delta_2\widetilde g_{\alpha\beta}\widetilde \nabla_\gamma \delta_1 \widetilde g_{\delta \tau} - \delta_1\widetilde g_{\alpha\beta}\widetilde \nabla_\gamma \delta_2 \widetilde g_{\delta \tau} )\,,
\end{align}
where 
\begin{align}
    P^{\alpha\beta\gamma\delta\mu\nu} = \widetilde g^{\alpha\mu} \widetilde g^{\nu\beta}\widetilde g^{\gamma\delta}-\frac{1}{2}\widetilde g^{\alpha\delta}\widetilde g^{\beta\mu}\widetilde g^{\nu\gamma}-\frac{1}{2}\widetilde g^{\alpha\beta}\widetilde g^{\gamma\delta}\widetilde g^{\mu\nu}-\frac{1}{2}\widetilde g^{\beta\gamma}\widetilde g^{\alpha\mu}\widetilde g^{\nu\delta}+\frac{1}{2}\widetilde g^{\beta\gamma}\widetilde g^{\alpha\delta}\widetilde g^{\mu\nu}\,.
\end{align}
Adapting to the discussions of previous Sections, the spacetime is now conformally compactified, introducing the unphysical metric $\gd$. Further, one makes use of the tetrad adapted to the \gls{ngc} that spans the spacetime within the universal structure of asymptotically flat spacetimes. In particular, one resorts to $n_\mu$ as defined above via $n_\mu=\nabla_\mu\Omega$ where $\Omega$ is the conformal factor. The Bondi gauge is adapted throughout the remainder of this Section for which $n^\mu n_\mu = \mathcal{O}(\Omega^2)$. If $\F$ is such that it consists of metrics $\widetilde g_{\mu\nu}$ such that $\Omega^2 \widetilde g_{\mu\nu}$ extends smoothly to $\scrip$ and becomes the $2$-sphere metric at $\scrip$, then $\delta \gd \equiv \Omega^2 \delta \widetilde g_{\mu\nu}$ also extends smoothly but vanishes at $\scrip$ such that one can define 
\begin{align}
    \gamma_{\mu\nu} = \delta \gd = \Omega \tau_{\mu\nu}
\end{align}
where $\tau_{\mu\nu}$ is generally not vanishing at $\scrip$ by the virtue of the Lemma presented in Section \ref{subsec:NP_scalars_and_peeling}. Further, as $\delta n_\mu|_{\scrip}=0$, one obtains $\tau_{\mu\nu}n^\nu = \Omega\tau_\mu$. Again, $\tau_\mu$ is smooth and in general not vanishing at $\scrip$. The trace of $\tau_{\mu\nu}$ is denoted by $\tau$, without indices\footnote{Note that in subsequent computations $\tau_{\mu\nu}\tau_{\mu}$ and $\tau$ receive an numerical index to indicate the variations $\delta_1,\delta_2$ of the symplectic current}. Adjusting all quantities in Eq. \eqref{equ:symp_current_GR} by their conformal transformation as well as replacing $\delta \widetilde g_{\mu\nu}$ with $\tau_{\mu\nu}$ and $\tau_\mu$, the presymplectic current can be computed for conformally compactified, unphysical spacetime (see \cite{Wald_Zoupas} for details). To explicitly calculate $\bT$, one needs to define the pullback $\newpb{\bw}$. Doing so, first note that in conformally completed spacetime, 
\begin{align}\label{equ:some_shit_identity}
    \epsilon_{\mu\nu\rho\sigma} = 4 \epsilon_{[\mu\nu\rho}n_{\sigma]}\,,
\end{align}
which results in a positively orientated pullback ${}^{(3)}\newpb{\boldsymbol{\epsilon}}$ at $\scrip$ such that
\begin{align}
    \newpb{\bw} = -\frac{1}{16\pi} \Omega^{-4}n_\mu w^\mu{}^{(3)}\newpb{\boldsymbol{\epsilon}}\,.
\end{align}
A rather tedious calculation and evaluation at $\scrip$ further leads to
\begin{align}
    \Omega^{-4}n_\mu w^\mu|_{\scrip} = \frac{1}{2}\left([\tau_2^{\nu\rho}n^\mu\nabla_\mu \tau_{1\beta \rho} + \tau_2 n^\mu\nabla_\mu \tau_1 + \tau_2 n^\mu \tau_{1\mu}] + [1\leftrightarrow2]\right)\,.
\end{align}
Similar to Section \ref{subsec:radi_degrees}, the latter expression can be related to the Bondi new tensor capturing the radiative information by considering the Riemann and Schouten tensors. Concretely, one computes variations of those in the vacuum. The Schouten tensor is then defined by 
\begin{align}
    S_{\mu\nu }= -2\Omega^{-1} \nabla_{(\mu}n_{\nu)} + \Omega^{-2} n^\mu n_\mu \gd
\end{align}
which leads to 
\begin{align}\label{equ:pfsjidopsja}
    \newpb{\delta S_{\mu\nu}} = 2n_{(\mu}\tau_{\nu)} - n^\rho\nabla_\rho \tau_{\mu\nu} - n^\rho \tau_\rho g_{\mu\nu}\,. 
\end{align}
Note that the difference in the prefactor of the first term w.r.t. \cite{Wald_Zoupas}. This, however, does not affect the end result since the first term vanishes when multiplied with $\tau_{\mu\nu}$ as $\tau_{\mu\nu}n^\nu|_{\scrip} = 0$. Further note that on $\scrip$, one has $n^\mu\nabla_\mu \tau + 2 n^\mu\tau_\mu |_{\scrip}= 0$ which can be obtained by the direct computation of $\delta S_{\mu\nu}$ via $S_{\mu\nu} = R_{\mu\nu} - \frac{1}{6}R \gd$ with Eq. \eqref{equ:pfsjidopsja}. 
It follows that $ \Omega^{-4}n_\mu w^\mu|_{\scrip} $ can be compactly written in terms of the Schouten tensor, i.e., 
\begin{align}
     \Omega^{-4}n_\mu w^\mu|_{\scrip} = \frac{1}{2}\left(\tau_2^{\mu\nu}\newpb{\delta_1 S_{\mu\nu}} - \tau_1^{\mu\nu}\newpb{ \delta_2 S_{\mu\nu}}\right)\,.
\end{align}
With the definition of the Bondi news tensor as in Eq. \eqref{equ:Bondiiiiiii} (which is intrinsically defined at $\scrip$), one then obtains
\begin{align}
    \newpb{\bw} = -\frac{1}{32\pi}[\tau_2^{\mu\nu}\delta_1N_{\mu\nu}-\tau_1^{\mu\nu}\delta_2 N_{\mu\nu}]
\end{align}
Note that the replacement of Schouten and Bondi news tensor was enabled by $\delta \rho_{\mu\nu} = 0$ in the Bondi frame. That is due to the fixed choice of $q_{\mu\nu} = \newpb{\gd}$ and the constant Ricci curvature associated with $q_{\mu\nu}$, see Eq. \eqref{equ:rhooooo}. In conclusion, the latter result suggests that
\begin{align}\label{equ:last_step}
    \bT =  -\frac{1}{32\pi}\tau^{\mu\nu}N_{\mu\nu} {}^{(3)}\boldsymbol{\epsilon}\,.
\end{align}
Given the definition of the flux in Eq. \eqref{equ:def_flux} and the fact that the goal is to define the flux conservation law w.r.t. a vector field generating symmetries at $\scrip$, the final step towards the flux formula consists of replacing $\delta \gd \leftrightarrow \lie_\xi \gd$ in Eq. \eqref{equ:last_step}\footnote{Here, a couple of steps in the discussion about the well-posedness and uniqueness of $\bT$ as well as a suitable reference solution $\phi$ are skipped. The interested reader is referred to \cite{Wald_Zoupas} for a rigorous discussion.}. This step requires some additional caution in regards of how to replace $\tau_{\mu\nu} \sim \delta g_{\mu\nu}$ in Eq. \eqref{equ:last_step}. In previous Sections, the BMS symmetries are introduced either by considering the infinitesimal generators preserving the universal structure, see Section \ref{subsec:BMS_Group}, or by taking the limit of the BMS metric's Killing fields to $\scrip$, see Section \ref{sec:BMS_metric_and_Conserved_Quanties}. Both strategies are correct and, as demonstrated, yield to the same result. To prevent confusion about what metric, physical or unphysical, the generators of the BMS group, $\xi^\mu$, have to be Killing vectors for, the following definition of \cite{Geroch:1981ut}, commonly cited in literature, can be consolidated:
\begin{definition}
    The generators of the Bondi-Metzner-Sachs asymptotic symmetry group are given by vectors $\xi^\mu$ on conformally compactified spacetime $(\mathcal M,\gd)$ such that  
    \begin{align}\label{equ:def_bms_wald}
        \Omega^2 \lie_\xi \widetilde g_{\mu\nu} |_{\scrip}=0\,,
    \end{align}
    i.e., the $\Omega^2 \lie_\xi \widetilde g_{\mu\nu} $ is smooth and vanishes on $\scrip$.
\end{definition}
\noindent The latter expression can be computed to yield $\Omega^2 \lie_\xi \widetilde g_{\mu\nu} = \lie_\xi \gd - 2\Omega^{-1} \xi^\alpha n_\alpha \gd$ and one immediately finds that, by definition of the BMS symmetry generators (e.g., Section \ref{subsec:BMS_Group}) and the fact that $n^\mu \sim \partial_u$ on $\scrip$, $\xi^\mu n_\mu$ vanishes on $\scrip$. It is convenient to express this contraction in terms of some smooth function $K$ such that $\xi^\mu n_\mu = \Omega^{-1} K$. Eq. \eqref{equ:def_bms_wald} can then be rewritten as 
\begin{align}\label{equ:def:chi_munu}
    \Omega \chi_{\mu\nu} := \frac{1}{2}\Omega^2 \lie_\xi \widetilde g_{\mu\nu} = \nabla_{(\mu}\xi_{\nu)} - K \gd\,,
\end{align}
where $\chi_{\mu\nu}$ is a smooth tensor field on $\mathcal M$\footnote{Note that w.r.t. \cite{Wald_Zoupas} the definition of $\chi_{\mu\nu}$ differs by a factor of 2.}. Given $\chi_{\mu\nu}$ vanishes if $\xi_\mu$ is Killing, one can view the latter tensor as a measure for the extent to which $\xi$ fails to be Killing. Going back to Eq. \eqref{equ:last_step} and remembering that $\tau_{\mu\nu}= \Omega \delta \widetilde g_{\mu\nu}$ the above mentioned final replacement actually consists of switching $\delta \widetilde g_{\mu\nu}\leftrightarrow \lie_\xi \widetilde g_{\mu\nu}$ and thus $\tau_{\mu\nu} \rightarrow \Omega \lie_\xi \widetilde g_{\mu\nu} = 2 \chi_{\mu\nu}$. Thus, the flux through $\scrip$ in \gls{gr} and associated to the Killing field $\xi_\mu$ is given by 
\begin{align}\label{equ:final_flux_GR}
    \boldsymbol{F}_\xi =  -\frac{1}{16\pi}\chi^{\mu\nu}N_{\mu\nu} {}^{(3)}\boldsymbol{\epsilon}\,.
\end{align}
It is immediately clear that the flux vanishes in the absence of gravitational radiation since this implies a vanishing Bondi news tensor. If no other fields are included, Eq. \eqref{equ:final_flux_GR} therefore yields the final flux formula for the gravitational flux through $\scrip$, as desired. 

For illustration, the above treatment of pure \gls{gr} in asymptotically flat spacetimes is generalized by including additional fields. A straightforward instance of such extension is given by Einstein-Maxwell theory \cite{Bonga_2020}: Consider Lagrangian \eqref{equ:EH_wald} extended by 
\begin{align}
    \mathbf{L}_{\text{EM}} = -\frac{1}{16\pi} \widetilde F^2 \,\,{}^{(3)}\boldsymbol{\epsilon}\,.
\end{align}
The symplectic potential, then again, results from the boundary term upon variation. The tilde on the electromagnetic field strength tensor and the vector potential denote their definition of physical spacetime, i.e., indices are raised and lowered with $\widetilde g_{\mu\nu}$. In addition to Eq. \eqref{equ:idk_KKKK}, for Einstein-Maxwell, one does not find mixing terms for the boundary term $\bt_\text{Einstein-Maxwell}$ and simply complements the electromagnetic sector
\begin{align}
    \theta^{\text{Einstein-Maxwell}}_{\mu\nu\rho} = \theta^\text{GR}_{\mu\nu\rho} - \frac{1}{4\pi}\widetilde F^{\sigma\alpha}\delta \widetilde A_\alpha \epsilon_{\sigma\mu\nu\rho}
\end{align}
So far the theory can effectively be split into \gls{gr} and Maxwell, and each sector considered individually. This holds true as well for the symplectic current in the limit of asymptotically flat perturbations \cite{Bonga_2020}, resulting in
\begin{align}\label{equ:current_Einstein_Maxwell}
    \omega_{\mu\nu\rho}^\text{Einstein-Maxwell} = \epsilon_{\sigma\mu\nu\rho}\left(w^\sigma_\text{GR} + w^\sigma_\text{EM} + w^\sigma_\times\right)
\end{align}
where $w^\sigma_\text{GR}$ is extracted from \eqref{equ:symp_current_GR}, the pure Maxwell part is given by 
\begin{align}
    w^\sigma _ \text{EM} = -\frac{1}{4\pi} \widetilde g^{\sigma \rho}\widetilde g^{\nu\mu}\left(\delta_1 \widetilde F _{\rho\mu}\delta_2 \widetilde A_\nu - [1\leftrightarrow2]\right)\,,
\end{align}
and the cross term reads
\begin{align}
    w^\sigma_\times= -\frac{1}{4\pi}\left(2\widetilde g^{\rho[\sigma}\widetilde f^{\nu]\mu} + \frac{1}{2}\widetilde F^{\sigma\nu} \widetilde g^{\rho\mu} \right)\delta_2 \widetilde A_\nu \delta_1 \widetilde g_{\rho \mu}\,.
\end{align}
Although cross terms appear in $\bw$, they vanish in the limit to $\scrip$, or, more precisely, when pulling back Eq. \eqref{equ:current_Einstein_Maxwell} to $\scrip$. For the pure Maxwell sector, using Eq. \eqref{equ:some_shit_identity}, this pullback results in 
\begin{align}\label{equ:symplectic_current_em}
    \newpb{\bw}_\text{EM}= -\frac{1}{4\pi} \left( \delta_1\Upsilon^\mu \delta_2 \newpb{A_\mu}- \delta_2\Upsilon^\mu \delta_1 \newpb{A_\mu}\right) {}^{(3)}\boldsymbol{\epsilon}
\end{align}
where one defines $\Upsilon_\mu := \newpb{F_\mu}{}_\nu n^\nu = -\lie _n \newpb{A_\mu}$ and $F_{\mu\nu}$ denotes the Maxwell tensor in unphysical conformally completed spacetime. Note again that $\widetilde F_{\mu\nu}= F_{\mu\nu}$ due to the conformal invariance of the theory. A brief inspection of Eq. \eqref{equ:symplectic_current_em} yields the following symplectic potential:
\begin{align}
    \bT_\text{EM}= -\frac{1}{4\pi } \Upsilon^\mu\delta \newpb{A_\mu }\,{}^{(3)} \boldsymbol{\epsilon}\,.
\end{align}
Similar to the case of \gls{gr}, one finds for Maxwell that the symplectic current, and correspondingly the flux, are solely determined by radiative degrees of freedom encoded in $\newpb{A_\mu}$ and $\Upsilon_\mu = \newpb{F_{\mu}}{}_{\nu}n^\nu$ (compare to the NPS for Maxwell's theory, Eq. \eqref{equ:NPS_EM_Maxwell}). As the cross terms vanish, the total flux across a section of $\scrip$, $\Delta \scrip$, enclosed by two $u_1,u_2 = const.$ cross sections $\partial \scrip_1, \partial \scrip_2$ is given by \cite{Bonga_2020}
\begin{align}\label{equ:Final_Flux_EM}
    \boldsymbol F_\xi [\Delta \scrip] = \int_{\Delta \scrip} \bT_\text{GR} + \bT_\text{EM}\,.
\end{align}
The result is simple and yet powerful. It demonstrates that in the case of vanishing cross terms, the total flux of the theory results from the sum of fluxes of individual massless fields. A similar result is obtained in the context of more general applications in \cite{Janns_paper_I}, specifically regarding the gravitational memory, which is shown to be related to the flux as defined above in Section \ref{subsec:BL_a_la_ashtekar}. There, Eq. \eqref{equ:Final_Flux_EM} is further expressed in terms of the shear tensor and NPS, and a physical interpretation of Eq. \eqref{equ:Final_Flux_EM} is provided. Note further that for the electromagnetic sector, the flux is computed and discussed for two physically relevant examples in \cite{Our_Review}.

The covariant phase space approach by Wald and Zoupas is evidently very successful in computing well-defined ``conserved'' quantities for \gls{gr} and more general diffeomorphism invariant theories. Historically, as mentioned before, it was, however, not the first attempt to derive such formulas. Ashtekar and Streubel \cite{Ashtekar_Streubel} as well as Dray and Streubel \cite{Dray_Streubel} both provided equivalent expressions to Eq. \eqref{equ:final_flux_GR}. Despite their derivation being more closely related to the content of Section \ref{subsec:Radiative_modes}, it involves non-trivial mathematical tools and a lengthy derivation. Nonetheless, in the following Section, the approach of \cite{Ashtekar_Streubel} is briefly outlined and similarities and differences w.r.t. the covariant phase space approach by Wald and Zoupas are highlighted.

\subsection{The Symplectic Geometry of Radiative Modes}
\label{subsec:Ashtekar_Streubel}

Instead of treating the phase space of the full theory, i.e., including the gravitational background and additional matter fields, it is feasible to directly focus (at least morally) on the radiative modes of a theory such as \gls{gr} \cite{Ashtekar_Streubel}. This does, in fact, avoid the challenges of Section \ref{subsec:Wald_Zoupas} but is conceptually more demanding and requires extensive knowledge about the radiative mode in the first place. The construction goes, in a very simplified manner, as follows:
Constructing a phase space of radiative modes for any theory requires a kinematical background on which these modes can propagate. For most theories admitting radiative solutions, e.g., Maxwell's theory, this ``arena'' is provided by spacetime itself. For \gls{gr}, naturally, this poses a particular problem as there is no background spacetime to radiate on, but instead it is the background itself admitting propagation of radiative modes. While the radiative solution of other theories propagates on a fixed background, for \gls{gr}, there exists a collection of backgrounds admitting radiative solutions of the gravitational field. Luckily, as it is demonstrated in Section \ref{sec:Gravitational_radiation_at_Scri}, these background spacetimes share a universal structure characterized by their properties at future null infinity. Therefore, a potential arena for formulating the radiative phase space for \gls{gr} is provided by $\scrip$. In fact, Section \ref{subsec:Radiative_modes} clearly demonstrates the identification of radiative modes on $\scrip$ for the theory of gravity via the derivative operator $\D$ and associated equivalence classes $[\D]$. Based on these results, Ashtekar and Streubel \cite{Ashtekar_Streubel} proposed adapting the space of equivalence classes $[\D]$ as a suitable phase space $\Gamma$ for radiative modes in \gls{gr}\footnote{Note that mathematically speaking, one has to assure that the equivalence classes of which the phase space is build of are suitably regular. The regularity check is provided in \cite{Ashtekar_Streubel}.}. In this construction, a point in $\Gamma$ is given by a particular choice $[\D]$ and the information needed to absolve a given trajectory starting at $[\D]$ and reaching $[\D']$ is encapsulated solely in the shear tensor $\sigma_{\mu\nu}$ which has, coincidentally, two degrees of freedom. The latter follows by the virtue of Eq. \eqref{7.333333} which states that two derivative operators at $\scrip$ differ by a symmetric tensor $\Sigma_{\mu\nu}$ with $\Sigma_{\mu\nu}n^\nu = 0$. If the two operators lie within a single equivalence class, one finds $\Sigma_{\mu\nu}= \lambda q_{\mu\nu}$ for some function $\lambda$. It follows that the difference between any two equivalence classes must lay in $\sigma_{\mu\nu}$ which is defined as the trace-free part of $\Sigma_{\mu\nu}$, see Eq. \eqref{ShearTensor}. Thus, from the construction in Section \ref{subsec:Radiative_modes} it naturally follows to define the phase space of gravitational radiation as the space of all equivalence classes $[\D]$ ``coorditanized'' by the shear tensor $\sigma_{\mu\nu}$, i.e., the shear in acting morally as a tangent vector on the space of equivalence classes. 

The next natural step is to provide a symplectic structure for the phase space at hand via the symplectic potential $\Omega([\D], \sigma, \sigma')$ which is a tensor field on $\Gamma$ at any point $[\D]$ and for any tangent vectors $ \sigma_{\mu\nu}, \sigma'_{\mu\nu}$ at $[\D]$\footnote{Note that every vector field on $\Gamma$ is an assignment of tensor field $T_{\mu\nu}$ on $\scrip$ for each point $[\D]$ \cite{Ashtekar_Streubel}. Thus, the shear tensor $\sigma_{\mu\nu}$ corresponds morally to a vector field on $\Gamma$.} (compare with the definition Eq. \eqref{equ:symp_pot_wald}). One can argue that \cite{Ashtekar_Streubel} 
\begin{align}\label{equ:symp_pot_Ashtekar}
    \Omega([\D], \sigma, \sigma') = \int_{\scrip} \left(\sigma_{\mu\nu} \lie_n \sigma'_{\rho\sigma} - \sigma'_{\mu\nu}\lie_n \sigma_{\rho\sigma}\right)q^{\mu\rho}q^{\nu\sigma}\, {}^{(3)}\boldsymbol{\epsilon}\,,
\end{align}
where $n$, as usual, is the null normal to $\scrip$, provides a well defined symplectic potential. It is worth noting at this point that the integral on the right-hand side of \eqref{equ:symp_pot_Ashtekar} is independent of the point $[\D]$ and that $\sigma_{\mu\nu}$ defines constant vector fields on $\Gamma$. Moreover, the symplectic structure is conformally invariant, i.e., invariant under the rescaling $(q_{\mu\nu}, n^\mu)\rightarrow (f^2 q_{\mu\nu}, f^{-1}n^\mu)$ \cite{Ashtekar_Streubel}.\\
Similar to the approach by Wald and collaborators \cite{Wald_Zoupas}, i.e., Eq. \eqref{equ:hamiltonina_def}, a Hamiltonian conjugated to some vector field can naturally be defined using the symplectic potential \eqref{equ:symp_pot_Ashtekar}. As in the previous considerations, the relevant vector field, here denoted by $\xi$, is such that it preserves the universal structure, i.e., it is a generator of an asymptotic symmetry or, in other words, a BMS vector field. By \eqref{equ:lie_algebra_BMS}, this implies that $\lie_\xi q_{\mu\nu }= 2 k q_{\mu\nu}$ and $\lie_\xi n^\mu = -k n^ \mu$. The transformation generated by $\xi$ corresponds to a one-parameter family of diffeomorphisms, denoted as $\psi(\lambda)$. This diffeomorphisms induces an isomorphism mapping $\Gamma$ into itself while preserving the affine structure \cite{Ashtekar_Streubel}, i.e., if $\D\sim \D'$, the images under $\psi(\lambda)$ are again equivalent, $(\psi(\lambda)\circ \D)\sim (\psi(\lambda)\circ \D')$. The latter straightforwardly follows from the definition of the image acting on an $1$-form $\alpha_\mu$ intrinsic to $\scrip$, i.e., $(\psi(\lambda)\circ\D_\mu)\alpha_\nu = \psi(\lambda)\circ(\D_\mu \psi(-\lambda)\circ\alpha_\nu)$, as well as the preservation of the universal structure by the diffeomorphism $\psi(\lambda)$ by definition, $\psi(\lambda) \circ q_{\mu\nu} = q_{\mu\nu}$ and $\psi(\lambda) \circ n_{\mu} = n_{\mu}$. This characteristic is of fundamental importance as one can demontrate that, in convolution with $\Omega([\D], \sigma, \sigma')$ being a constant tensor field on $\Gamma$ (as it does not depend on $[\D]$), $\psi(\lambda)$ establishes an symplectomorphism on $\Gamma$ \cite{Ashtekar_Streubel}. In its infinitesimal version, a symplectomorphism gives rise to a symplectic vector field $\psi_\xi$ on $\Gamma$ (not to be confused with the BMS vector field $\xi$ on $\scrip$) characterized by
\begin{align}
    \lie_{\psi_\xi} \Omega([\D], \sigma, \sigma') = 0\,.
\end{align}
The Hamiltonian $H_\xi$ in this approach arises as the generator of this vector field. In particular, one expects that, for all vector fields $X$ on $\Gamma$, 
\begin{align}\label{equ:puuuh_is_dat_heis}
    \lie_X H_\xi = \Omega([\D], \psi_\xi, X)\,.
\end{align}
To compute $H_\xi$, it is imperative to determine the vector field $\psi_\xi$ on $\Gamma$ first, or rather its assigned tensor field on $\scrip$. An ansatz can be deduced by combining Eq. \eqref{eq:ShearAndDerivativeEll} and \eqref{eq:RunningOutOfNames} under the assumption that two derivative operators are related by the diffeomorphism $\psi(\lambda)$. One finds a shear tensor $\sigma_{\mu\nu}(\lambda)$ defined via $\sigma_{\mu\nu}(\lambda)\,\widetilde = \,(\psi(\lambda)\circ\D_\mu- \D_\mu)\ell_\nu$, where ``$\widetilde =$'' denotes equivalence of the trace-free part. By varying this ansatz w.r.t. $\lambda$, one finds the ansatz \cite{Ashtekar_Streubel}
\begin{align}\label{equ:not}
    (\psi_\xi)_{\mu\nu} \,\widetilde = \, (\lie_\xi \D_\mu - \D_\mu \lie_\xi)\ell_\nu \,.
\end{align}
To ensure that $(\psi_\xi)_{\mu\nu}$ belongs to the same class of vector fields on $\Gamma$ as $\sigma_{\mu\nu}$ one additionally has to check that $ (\psi_\xi)_{\mu\nu}$ is symmetric and $ (\psi_\xi)_{\mu\nu}n^\nu$ = 0. In case $\xi$ corresponds to a BMS vector field \eqref{equ:lie_algebra_BMS} with $k=0$, i.e., supertranslations, Eq. \eqref{equ:not} is the final result. Otherwise, it is straightforward to check that the latter two conditions for $ (\psi_\xi)_{\mu\nu}$ yield
\begin{align}\label{equ:not_II}
    (\psi_\xi)_{\mu\nu} \,\widetilde = \, (\lie_\xi \D_\mu - \D_\mu \lie_\xi)\ell_\nu + 2 \ell_{(\mu}\D_{\nu)} k\,,
\end{align}
where $k$ refers to the function in the definition \eqref{equ:lie_algebra_BMS}.
Given this tensor field on $\scrip$ assigned to $\psi_\xi$ on $\Gamma$, Ashtekar and collaborators showed that a Hamiltonian $H_\xi$ on $\Gamma$ that satisfies the condition \eqref{equ:puuuh_is_dat_heis} at any point $[\D]$ is
\begin{align}\label{equ:fuuu}
    H_\xi = -\frac{1}{2}\int_{\scrip} N_{\mu\nu}\left[ (\lie_\xi\D_\rho - \D_\rho \lie_\xi)\ell_\sigma + 2 \ell_{(\mu}\D_{\nu)} k\right]
    q^{\mu\rho}q^{\nu\sigma} \,{}^{(3)} \boldsymbol{\epsilon}\,,
\end{align}
where $N_{\mu\nu} \sim \lie_n \sigma_{\mu\nu}$ is the Bondi news tensor. Note that the second term in brackets vanishes for $\xi$ being associated with supertranslations as, in this case, $k=0$. 
This Hamiltonian establishes the flux of a conserved quantity associated with the asymptotic symmetry $\xi$ through null infinity $\scrip$. It therefore holds the same interpretation as the previously defined quantity \eqref{equ:def_flux}\footnote{There is no analogy for the full Hamiltonian \eqref{equ:new_Hamiltonian}; Eq. \eqref{equ:fuuu} solely recovers the flux. This is a direct consequence of focusing on the radiative phase space instead of the full phase space of the theory.}. As pointed out earlier, the main difference between this and the previous result pioneered by \cite{Wald_Zoupas} resides in the type of phase space under consideration, i.e., if one considers all allowed field configurations \cite{Wald_Zoupas} or if one explicitly focuses on radiative modes \cite{Ashtekar_Streubel}.

Since it is not immediately clear that the definitions provided in Eqs. \eqref{equ:def_flux} and \eqref{equ:fuuu} are equivalent, a comparison is indispensable. As Eq. \eqref{equ:fuuu} exclusively applies to \gls{gr}, in the following the discussion is restricted to Einstein's theory. In this case, the first step is to compute the quantity $\chi_{\mu\nu}$ in Eq. \eqref{equ:final_flux_GR} explicitly. To do so, it is helpful to consider an additional work on the symplectic structure of \gls{gr} by Ashtekar and collaborators \cite{Ashtekar_Magnon}.\\
For simplicity, one first restricts to the case of the BMS subgroup, i.e., $\xi^\mu = \alpha n^\mu $. Similar to the considerations of \cite{Ashtekar_Streubel}, starting by considering a 1-parameter family of physical spacetime metrics $\widetilde g_{\mu\nu}(\lambda)$ (compare to the discussion below Eq. \eqref{equ:Toledo}), one defines 
\begin{align}\label{equ:one_para}
    \frac{\dd}{\dd\lambda}\widetilde g_{\mu\nu}(\lambda)|_{\lambda = \lambda_0} =: \varsigma_{\mu\nu} \,.
\end{align}
Thus, the tensor $\varsigma_{\mu\nu}$ denotes the change of the metric along the path parametrized by $\lambda$ in field space. The change in the metric induces a variation in the connection induced by this metric as well as the conformally completed, unphysical metric $\gd = \Omega^2 \widetilde g_{\mu\nu}$. Using Eq. \eqref{CovDerConformalII} of Appendix \ref{App:ConformalTransformations} as well as Eq. \eqref{equ:one_para}, one can straightforwardly compute the change in the connection $\nabla$ corresponding to the unphysical metric induced by the change $\varsigma_{\mu\nu}$. Defining $\widetilde \nabla_\mu V_\nu \rightarrow \nabla_\mu V_\nu + (\delta \nabla_\mu) V_\mu$ where $V_\mu$ is any covector on conformally compactified spacetime $\mathcal M$, it follows that 
\begin{align}\label{equ:var_I}
    (\delta \nabla)_\mu V_\nu = -\frac{1}{2}\Omega^{-1}g^{\rho\sigma}\left(\nabla _\mu \Omega \,\varsigma_{\sigma\nu} + \nabla_\nu \Omega \,\varsigma_{\mu\sigma} - \nabla_\sigma \Omega \, \varsigma _{\mu\nu}\right)\,.
\end{align}
Note that the latter is directly related to the change in the Levi-Civita connection 
\begin{align}\label{equ:equ_1202}
    \delta {\Gamma^{\alpha}}_{\beta \mu}=  \frac 12 g^{\alpha \lambda} (\nabla_{\beta} \delta g_{\lambda \mu}+ \nabla_{\mu} \delta g_{\beta \lambda}-
\nabla_{\lambda} \delta g_{\beta \mu})\,,
\end{align}
when $\delta \gd = \Omega^2 \delta \widetilde g_{\mu\nu}$ and $\delta \widetilde g_{\mu\nu} \sim \varsigma_{\mu\nu}$. 
Pulling Eq. \eqref{equ:var_I} back to $\scrip$, i.e., $\newpb{\nabla_\mu V_\nu} = \D_\mu \newpb{V_\nu}$, one further finds\footnote{Note that the following equation differs by a factor of $2$ to the result in \cite{Ashtekar_Magnon}.}
\begin{align}\label{equ:var_II}
    (\delta \D)_\mu \newpb{V_\nu} = - \boldsymbol{\varsigma}_{\mu\nu} n^\rho V_\rho \,,
\end{align}
where $\boldsymbol{\varsigma}_{\mu\nu} = \lim_{\Omega \to 0} \Omega \, {\varsigma}_{\mu\nu}$ and $n^\mu$ is the null normal to $\scrip$ and part of the tetrad chosen in this Chapter. Eq. \eqref{equ:var_I} and \eqref{equ:var_II} hold for an arbitrary change of the metric along its trajectory parametrized by $\lambda$. Generally, and in particular in Section \ref{subsec:Wald_Zoupas}, one is interested in the transformation of the metric as it is Lie-dragged along a Killing vector field $\xi_\mu$ that asymptotically generates symmetries of the universal structure. As such, following definition \eqref{equ:def_bms_wald}, one sets $\varsigma_{\mu\nu} = \lie_\xi \widetilde g_{\mu\nu}$. Now, the key insight connecting Wald's work \cite{Wald_Zoupas} to its predecessors \cite{Ashtekar_Streubel, Dray_Streubel, Ashtekar_Winicour} is composed in the following Lemma (in a slightly modified manner found in \cite{Ashtekar_Magnon}):
\begin{lemma}
    Let $\varsigma_{\mu\nu}=\lie_\xi \widetilde g_{\mu\nu}$ and $\xi_\mu$ be an generator of an asymptotic supertranslation symmetry at $\scrip$. In particular, let the extension of $\xi_\mu$ to $\scrip$ be tangential to it. Then, it holds that 
    \begin{align}
        (\delta \D)_\mu \newpb{V_\nu} =  - \boldsymbol{\varsigma}_{\mu\nu} n^\rho V_\rho = (\lie_\xi \D_\mu - \D_\mu \lie_\xi) \newpb{V_\nu}\,,
    \end{align}
    for any covector on $\scrip$.
\end{lemma}
\noindent \textit{Proof.}~One starts by acknowledging that for the Levi-civita connection (w.r.t. the conformally compactified spacetime), 
\begin{align}
    \Gamma\ud{\lambda}{\mu\nu} = \frac{1}{2}g^{\lambda\rho}\left(\partial_\mu g_{\nu \rho} + \partial_\nu g_{\mu\rho} - \partial_\rho g_{\mu\nu}\right)\,,
\end{align}
the variation due to a change in the metric $\delta \gd = \lie_X \gd$ is equal to the lie derivative along the generator $X^\mu$ of the change in the metric, i.e., $\delta \Gamma\ud{\lambda}{\mu\nu}= \lie_X \Gamma\ud{\lambda}{\mu\nu}$ (one can easily convince oneself upon explicit computation). For this transformation, writing the derivative operator explicitly for the affine connection \`a la \gls{gr} yields
\begin{align}
    \nabla'_\mu V_\nu = \underbrace{\partial_\mu V_\nu - \Gamma\ud{\lambda}{\mu\nu}V_\lambda}_{= \nabla_\mu V_\nu }-  \delta\Gamma\ud{\lambda}{\mu\nu}V_\lambda\,,
\end{align}
where $\nabla'$ is the new derivative operator after the metric is changed according to $\gd \rightarrow\gd + \delta\gd$. Further note that 
\begin{align}\label{equ:dumb_money}
    \lie_X (\nabla_\mu V_\nu) = \nabla_\mu(\lie_X V_\nu)- (\lie_X \Gamma\ud{\lambda}{\mu\nu})V_\lambda\,,
\end{align}
and thus $-\lie_X\Gamma\ud{\lambda}{\mu\nu}V_\lambda = (\lie_X\nabla_\mu - \nabla_\mu \lie_X)V_\nu$. Comparing with Eq. \eqref{equ:var_I} and replacing $X$ with $\xi$ one finds that 
\begin{align}\label{equ:var_III}
    (\delta \nabla)_\mu V_\nu = (\lie_\xi\nabla_\mu - \nabla_\mu\lie_\xi)V_\nu\,.
\end{align}
Using that $\xi_\mu$ is by definition tangential to $\scrip$, one finds that the pullback of Eq. \eqref{equ:var_III} is well-defined and yields
\begin{align}
    (\delta \D)_\mu \newpb{V_\nu} = (\lie_\xi\D_\mu - \D_\mu\lie_\xi) \newpb{V_\nu}\,,
\end{align}
which holds for all covectors on $\scrip$. By virtue of Eq. \eqref{equ:var_II}, this is equal to $- \boldsymbol{\varsigma}_{\mu\nu} n^\rho V_\rho$.

\noindent Inspecting the term $- \boldsymbol{\varsigma}_{\mu\nu} n^\rho V_\rho$ further, one immediately finds that it can be simplified by choosing $V_\mu = \ell_\mu$, where $\ell_\mu$ is the generator of the \gls{ngc}. Since the cross-normalization yields $n^\mu \ell_\mu=-1$, writing out $\boldsymbol{\varsigma}_{\mu\nu} = \Omega \, \varsigma_{\mu\nu}$ with $\varsigma_{\mu\nu} = \lie_\xi \widetilde g_{\mu\nu}$, it immediately follows from Eq. \eqref{equ:def:chi_munu} that $\boldsymbol{\varsigma}_{\mu\nu} |_{\scrip} = 2 \chi_{\mu\nu}|_{\scrip}$ and, thus, 
\begin{align}\label{equ:GOD_equation}
    \chi_{\mu\nu} = \frac{1}{2}(\lie_\xi\D_\mu - \D_\mu\lie_\xi) \ell_\nu\,,
\end{align}
which is equivalent to Eq. \eqref{equ:not}. Eq. \eqref{equ:GOD_equation} holds only for generators of the supertranslation subgroup as for simplicity, the above consideration was restricted to such vector fields. The result can be generalized by means of a conformal transformation (of the left-hand side of Eq. \eqref{equ:GOD_equation}) of the form $q'_{\mu\nu}= \omega^2q_{\mu\nu}$ and $k=-\xi^\mu\D_\mu \ln(\omega)$\footnote{Given a BMS Killing vector field $\xi^\mu$ with $\mathcal L_\xi q_{\mu\nu}= 2kq_{\mu\nu}$ and $\lie_\xi n^\mu = -kn^\mu$, one can choose a Bondi frame $(q'_{\mu\nu},n'^\mu)$ such that $\mathcal L_\xi q_{\mu\nu}= 0$ and $\lie_\xi n^\mu =0$ \cite{Ashtekar_LAST_STAND}. Then, transforming back to $(q_{\mu\nu},n^\mu)$ corresponds to a conformal transformation.}, or, equally, deriving the same equation for a general BMS transformation with $k\neq 0$ in the definition \eqref{equ:lie_algebra_BMS}. This yields an additional term in Eq. \eqref{equ:GOD_equation} \cite{Ashtekar_Magnon, Ashtekar_LAST_STAND}, such that one recovers Eq. \eqref{equ:not_II}, i.e., $\chi_{\mu\nu} = \frac{1}{2}(\lie_\xi\D_\mu - \D_\mu\lie_\xi) \ell_\nu\, + \ell_{(\mu}\D_{\nu)}k$. In the computation above, when allowing for all types of symmetry generators on $\scrip$ (i.e., $k\neq0$), the extra term missing in Eq. \eqref{equ:GOD_equation} arises from the pullback of Eq. \eqref{equ:var_III} to $\scrip$. Another pathway to this result can be obtained when pursuing Eq. \eqref{equ:dumb_money} of the above proof. Starting with 
\begin{align}\label{equ:dumber_money}
    (\delta_X \Gamma\ud{\lambda}{\mu\nu})V_\lambda= \nabla_\mu(\lie_X V_\nu)-\lie_X (\nabla_\mu V_\nu)  \,,
\end{align}
explicit computation of $(\delta_X \Gamma\ud{\lambda}{\mu\nu})$ yields Eq. \eqref{equ:equ_1202} where $\delta\gd$ can be replaced with the lie derivative along $\xi$ of $\gd$. Then, using \eqref{equ:def:chi_munu}, one replaces $\lie_\xi \gd \rightarrow 2\Omega \chi_{\mu\nu}+ 2 K \gd$ and finds that 
\begin{align}\label{equ:203232}
    \delta_\xi \Gamma\ud{\lambda}{\mu\nu} = -n^\lambda \chi_{\mu\nu }- \gd \nabla^\lambda K + 2 n_{(\mu}\chi_{\nu)}{}^\lambda+2 \delta^\lambda_{(\nu}\nabla_{\mu)}K + \mathcal O(\Omega)\,.
\end{align}
Then, replacing $V_\mu\rightarrow\ell_\mu$ in Eq. \eqref{equ:dumb_money}, inserting \eqref{equ:203232}, and pulling back to $\scrip$, one can rearrange the result to obtain again
\begin{align}\label{equ:running_out_of_names_II}
    \chi_{\mu\nu} = \frac{1}{2}(\lie_\xi\D_\mu - \D_\mu\lie_\xi) \ell_\nu\, + \ell_{(\mu}\D_{\nu)}K\,.
\end{align}
Upon explicit computation of $K$ in \eqref{equ:def:chi_munu} one finds that $K=k$, where the latter appears in definition \eqref{equ:lie_algebra_BMS}, and the result equals the one above. The pullback of \eqref{equ:203232} thereby eliminates all terms $\mathcal O(\Omega)$ as well as non-contracted $n_\mu$, as $\newpb{n_\mu}=0$. The term in Eq. \eqref{equ:203232} proportional to $\gd$ would in principle survive the pullback and appear in Eq. \eqref{equ:203232}, however given that in the flux formula \eqref{equ:final_flux_GR} it is contracted with the Bondi news, which is trace-free, it can be dropped at this point. Note that we choose to ignore it in the above as well and in \cite{Ashtekar_Streubel} similar logic is applied by directly focusing at the trace-free part of the phase space vector field $(\psi_\xi)_{\mu\nu}$.
\\ 
Inserting this result into \eqref{equ:final_flux_GR} one finds an expression equivalent to Eq. \eqref{equ:fuuu}. Therefore, it is concluded that the approaches \cite{Wald_Zoupas} and \cite{Ashtekar_Streubel} are indeed identical. This equivalence reassures the correctness of the more universally applicable approach by Wald and collaborators. As it holds generally for diffeomorphism-invariant theories, one could, in principle, derive the flux for arbitrary theories. Note, however, that the discussion leading up to Eq. \eqref{equ:GOD_equation} depends on specific assumptions about the connection, the metric, and the well-posedness of the tetrad choice and may be more intricate for modifications far beyond \gls{gr}. Nonetheless, an application of Wald's formalism to selected beyond \gls{gr} theories is currently under construction by the author of this dissertation.\\
In the subsequent Section, it is explicitly demonstrated that Eq. \eqref{equ:GOD_equation} leads to an expression of the flux formulas \eqref{equ:final_flux_GR} and \eqref{equ:fuuu} in terms of the shear, combining all the interim results of this Chapter. Based on this expression, applications of the flux equation are presented in Section \ref{sec:Paper_BL}.

\subsection{Balance Flux Laws and Gravitational Strain}
\label{subsec:BL_a_la_ashtekar}

Thus far, the flux formulas do not have a particularly useful form. In order for them to be used in the context of measurement data, it is crucial to express them in terms of the linearized gravitational strain picked up by \gls{gw} interferometers. In Section \ref{subsec:shear_and_GW}, it was briefly sketched that this connection naturally appears in the linearized asymptotic shear, which allows (at least morally) to set $h^\circ = 2 \bar \sigma^\circ$. Thus, a desired form of the flux formulas would solely involve quantities related to the shear tensor or the Bondi news, as $N^\circ = -\dot{\bar \sigma}^\circ$.\\
To derive the latter, one starts by recycling Eq. \eqref{equ:some_result_for_some_equation} stating that
$ (\lie_\xi\D_\nu-\D_\nu\lie_\xi)\ell_\mu=\,\xi^\rho\left(q_{\mu[\rho}\Ss\du{\nu]}{\sigma}+\Ss_{\mu[\rho}\delta\du{\nu]}{\sigma}\right)\ell_\sigma-\ell_\rho\D_\nu\D_\mu\xi^\rho\,$. A propri, this equation cannot be simplified for general BMS vectors fields $\xi$. One can, however, consider only generators of a subgroup of the BMS group, such as supertranslations. This is a particularly interesting example as it not only includes time translations, i.e., an energy flux, but also the extensions of the translation subgroup of the Poincar\'e group for which only limited physical intuition is provided thus far in this work. Focusing on generators of supertranslations derived in Section \ref{subsec:BMS_Group}, it is demonstrated that $\xi^\mu = \alpha(z,\bar z) n^\mu$. Given the transversality of the Schouten tensor and the metric at $\scrip$, one finds that, in this case, $\lie_\xi q_{\mu\nu} = 0$ and $ (\lie_{\alpha n}\D_\nu-\D_\nu\lie_{\alpha n})\ell_\mu= \frac{1}{2}\alpha\Ss_{\mu\nu}+ \D_\nu\D_\mu \alpha$. Therefore, the flux (through $\scrip$ or a section of it $\Delta\scrip$) simply reads 
\begin{align}\label{equ:flux_BMS}
    F_{\alpha n } = -\frac{1}{32\pi} \int_{\Delta \scrip} N^{\mu\nu}\left(\frac{1}{2}\alpha \Ss_{\mu\nu} + \D_\nu\D_\mu \alpha\right)\,{}^{(3)}\boldsymbol{\epsilon}\,.
\end{align}
Given Eq. \eqref{eq:DefinitionBondi} and the fact that the Bondi news tensor is trace-free, one can replace the Schouten tensor in the result \eqref{equ:flux_BMS} with $N_{\mu\nu}$. For translations that are shared between BMS and Poincar\'e group, see discussion below Eq. \eqref{equ:shiat}, further note that the second term in Eq. \eqref{equ:flux_BMS} trivializes. This is a direct consequence of the fact that $\zeta^\mu \equiv q^{\mu\nu}\D_\nu\alpha$ is a conformal Killing vector field for all modes in $\alpha$ with $\ell<2$. Hence, it satisfies the conformal Killing equation
\begin{align}
    \D_\mu\zeta_\nu+\D_\nu\zeta_\mu=q_{\mu\nu}\D_\rho\zeta^\rho,
\end{align}
and it follows that $\D_\mu\D_\nu\alpha\propto q_{\mu\nu}$. Integrating the second term in Eq. \eqref{equ:flux_BMS} twice and using that $N_{\mu\nu}$ is trace-free, it follows that the term in the flux vanishes for these particular translations. 

In literature, e.g., \cite{waveform_test_BL_III}, the supertranslation flux \eqref{equ:flux_BMS} is commonly written as 
\begin{align}\label{equ:BL_only_Bondi}
    -\frac{1}{32\pi} \int_{\Delta \scrip} \alpha\left(\frac{1}{2}N^{\mu\nu} N_{\mu\nu} + \D_\nu\D_\mu N^{\mu\nu}\right)\,{}^{(3)}\boldsymbol{\epsilon}\,,
\end{align}
where the second term is rewritten using integration by parts. Observe that Eq. \eqref{equ:BL_only_Bondi} only contains the Bondi news tensor and the arbitrary but smooth function $\alpha(z,\bar z)$. Therefore, a version of \eqref{equ:flux_BMS} expressed solely in terms of the \gls{gw} strain is imminent. As a final step, the Bondi tensor is related to the shear tensor by virtue of $N^{\mu\nu} = 2\lie_n \sigma^{\mu\nu}$. Starting with the second term in Eq. \eqref{equ:BL_only_Bondi}, one finds
\begin{align}    \D_\mu(\D_\nu{N}^{\mu \nu})=&-2\D_\mu\left(\D_\nu\left(\dot{\Bar{\sigma}}^\circ m^\mu m^\nu+\dot{\sigma}^\circ\Bar{m}^\mu \Bar{m}^\nu\right)\right)\notag\\
    =&-2\D_\mu \left[m^\mu m^\nu\D_\nu\Dot{\Bar{\sigma}}^\circ+\Dot{\Bar{\sigma}}^\circ m^\nu\D_\nu m^\mu +\Dot{\Bar{\sigma}}^\circ m^\mu \D_\nu m^\nu \right.+\Bar{m}^\mu \Bar{m}^\nu \D_\nu \Dot{\sigma}^\circ \notag\\&+\Dot{\sigma}^\circ\Bar{m}^\nu \D_\nu \Bar{m}^\mu 
    +\Dot{\sigma}^\circ\Bar{m}^\mu \D_\nu \Bar{m}^\nu  \left.\right]\notag\\
    =&-2\D_\mu \left[m^\mu m^\nu \D_\nu \Dot{\Bar{\sigma}}^\circ+ \Bar{m}^\mu \Bar{m}^\nu \D_\nu \Dot{\sigma}^\circ +  \frac{1}{\sqrt{2}}\Dot{\Bar{\sigma}}^\circ\cot{\theta}\,m^\mu \right.\notag\\ &\left.+ \frac{1}{\sqrt{2}}\Dot{\sigma}^\circ\cot{\theta}\,\Bar{m}^\mu \right]\notag\\
    =&-2\left[m^\mu m^\nu \D_\mu \D_\nu \Dot{\Bar{\sigma}}^\circ + \Bar{m}^\mu \Bar{m}^\nu \D_\mu \D_\nu \Dot{\sigma}^\circ\right]\notag\\
    =& -2 \left[\eth ^2 \Dot{\Bar{\sigma}}^\circ + \eth^2\Dot{\sigma}^\circ\right]= -4 \Re{\eth^2\Dot{\Bar{\sigma}}^\circ}\,,
\end{align}
where it was used that $m^\mu D_\mu m^\nu =\frac{1}{\sqrt{2}}\cot{\theta}\,m^\nu $ and $D_\mu m^\mu =0$ (see Section 1.D in \cite{Our_Review}). Then, by simply inserting the definition of the Bondi news tensor \eqref{eq:bondiI}, the first term in Eq. \eqref{equ:BL_only_Bondi} is given by
\begin{align}
    N_{\mu \nu }{N}^{\mu \nu }&=4(\dot{\Bar{\sigma}}^\circ m^\mu m^\nu +\dot{\sigma}^\circ\Bar{m}^\mu \Bar{m}^\nu )(\dot{\Bar{\sigma}}^\circ m_\mu m_\nu +\dot{\sigma}^\circ\Bar{m}_\mu \Bar{m}_\nu )\notag\\ &= 4[\dot{\Bar{\sigma}}^\circ\dot{\sigma}^\circ+\dot{\Bar{\sigma}}^\circ\dot{\sigma}^\circ]=8|\dot{\sigma}^\circ|^2\,,
\end{align}
and for the total flux one finds
\begin{align}\label{equ:theUltimateBL}
    {F}_{\alpha n}=\frac{1}{8\pi G}\int_{\scrip}\alpha(\theta,\phi)\left( |\Dot{\sigma}^\circ|^2  - \Re{\eth^2\dot{\Bar{\sigma}}^\circ} \right) {}^{(3)}\boldsymbol{\epsilon}\,.
\end{align}
Note that the integral taken over $\scrip$ can be separated into a ``time'' and $2$-sphere integral, i.e., ${}^{(3)}\boldsymbol{\epsilon} \rightarrow \dd u \dd \Sigma$. Using Eq. \eqref{equ:shear_and_strain} to replace the shear tensor in the latter, one obtains a well-defined flux formula that can be readily computed with measurement data from a \gls{gw} experiment. Thereby, the value of the radius in Eq. \eqref{equ:asym_strain} is replaced with the luminosity distance $D_L$ for numerical computations. This aspect of real data applications of the flux law \eqref{equ:theUltimateBL} is elaborated in more detail in Section \ref{sec:Paper_BL}. 

In its current form, Eq. \eqref{equ:theUltimateBL} computes the flux associated with the conserved quantity of the generator $\xi^\mu = \alpha n^\mu$. However, the value alone does not yield any helpful insights towards a better understanding of gravitational radiation or gravity itself. One can, however, use previous insights of the relation between NPS and shear tensor to transform the flux \eqref{equ:theUltimateBL} into a constraint equation, gaining the name ``balance flux'' law. To that end, one considers Eq. \eqref{equ:NPS_in_shear_0} - \eqref{equ:NPS_in_shear_II}, in particular
\begin{align}
    \Psi_2^\circ - \Bar{\Psi}^\circ_2&= \Bar{\sigma}^\circ\Dot{\sigma}^\circ-\sigma\Dot{\Bar{\sigma}}^\circ+ \Bar{\eth}^2\sigma^\circ - \eth^2\Bar{\sigma}^\circ,\label{equ:Psi2ImportantI}\\
    \partial_u\Psi_2^\circ&=-\eth^2\Dot{\Bar{\sigma}}^\circ-\sigma^\circ\Ddot{\Bar{\sigma}}^\circ\label{equ:Psi2ImportantII}\,,
\end{align}
from which follows that 
\begin{align}\label{equ:ImPsi_2}
    \Im{\Psi_2^\circ} &= \Im{ \Bar{\eth}^2\sigma^\circ+\sigma^\circ N^\circ } =  \Im{ -\eth^2\sigma^\circ+\sigma^\circ N^\circ }\\
    &= \Im{ \Bar{\eth}^2\sigma^\circ-\sigma^\circ \Bar{N}^\circ } =  \Im{ -\eth^2\sigma^\circ- \sigma^\circ \Bar{N}^\circ }\,.\label{equ:the_key}
\end{align}
First, note that Eq. \eqref{equ:theUltimateBL} can be partially integrated along the time direction. While $\alpha$ remains unaffected, the remaining terms can be rewritten as total time derivatives, such that, for a time domain $(u,-\infty)$, Eq. \eqref{equ:theUltimateBL} reads
\begin{align}\label{equ:integration_flux}
    F_{\alpha n}=\frac{1}{8\pi G}\oint_{\mathcal{S}^2}\dd \Omega \,\alpha(\theta,\phi)\Re{ \Psi_2^\circ + \bar \sigma^\circ \dot \sigma^\circ }^u_{-\infty}\,,
\end{align}
where one makes direct use Eq. \eqref{equ:the_key}. The lower integration bound $-\infty$ of $u$ indicates that the integral is performed over the $2$-sphere at $i^\circ$, i.e., spatial infinity. The shear tensor vanishes\footnote{Note that $h\rightarrow 0$ for $u\rightarrow -\infty$ in the post-Newtonian Bondi frame \cite{waveform_test_BL_II}.} and the NPS $\Psi_2^\circ$ has a clear interpretation in this limit (see Section \ref{subsec:NP_scalars_and_peeling}). In particular, it holds that \cite{waveform_test_BL_III}
\begin{align}
    \lim_{u\rightarrow -\infty} \Re{\Psi^\circ_2} = -G M_{\text{ADM}}\,,
\end{align}
where $ M_{\text{ADM}}$ is the ADM mass. 
Thus, considering a section of $\scrip$, $\Delta \scrip$, limited by $i^\circ$ and some finite $u$, it holds that 
\begin{align}\label{equ:idk_now}
    F_{\alpha n}&=\frac{1}{8\pi G}\int_{\Delta\scrip}\dd u \,\dd\Sigma \,\alpha(\theta,\phi)\left( |\Dot{\sigma}^\circ|^2  - \Re{\eth^2\dot{\Bar{\sigma}}^\circ} \right) 
    \notag \\
    &=\frac{1}{8\pi G}\int_{\Delta\scrip}\dd u \,\dd\Sigma \,\alpha(\theta,\phi) |\Dot{\sigma}^\circ|^2  - \frac{1}{8\pi G}\oint_{\mathcal S^2}\dd\Sigma \,\alpha(\theta,\phi)\left.\Re{\eth^2{\Bar{\sigma}}^\circ}\right|_u
    \notag \\
    &=\frac{1}{8\pi G}\oint_{\mathcal{S}^2}\dd \Sigma \,\alpha(\theta,\phi)\left(\Re{ \Psi_2^\circ + \bar \sigma^\circ \dot \sigma^\circ }_u + G M_\text{ADM}\right)\,.
\end{align}
In the latter, $\cdot |_u$ marks the evaluation at some time $u$ on $\scrip$. Given that $\alpha(z,\bar z)$ is a smooth function of the coordinates spanning $\mathcal S^2$, one can equate Eq. \eqref{equ:idk_now} and Eq. \eqref{equ:theUltimateBL} such that 
\begin{align}\label{equ:Alex_Chillov}
    \Re{\eth^2{\Bar{\sigma}}^\circ}\bigg|_u = \int_{\Delta\scrip}\dd u \, |\Dot{\sigma}^\circ|^2 - \left(\Re{ \Psi_2^\circ + \bar \sigma^\circ \dot \sigma^\circ }_u + G M_\text{ADM}\right)\,,
\end{align}
which can be seen as ``time-series constraint'' for the strain $h$ once the shear tensor is converted using Eq. \eqref{equ:shear_and_strain}. It is important to stress at this point that the constraint Eq.  \eqref{equ:Alex_Chillov} results from the supermomentum flux only. This implies that the terms involved have a specific physical interpretation. For instance, as it is further elaborated below, they include the displacement memory of the gravitational waveform as it is demonstrated in the following Subsection. Before, note that other memory effects have been computed as well, such as the spin memory \cite{Mitman_2020, Flanagan_2017, Spin_mem}, which are not derived from the supertranslation flux conservation law but can be linked to other symmetries such as the superrotations mentioned in Section \ref{subsec:BMS_Group}.

To put Eq. \eqref{equ:Alex_Chillov} to use in the context of \gls{gw} data, one replaces $2\bar{\sigma} \rightarrow h^\circ = h^\circ_+ - i h^\circ_\times$ to find 
\begin{align}\label{equ:kristin}
    \Re{\eth^2{\Bar{h}}^\circ}\bigg|_u = \int_{\Delta\scrip}\dd u \, |\Dot{h}^\circ|^2 - 4\left(\Re{ \Psi_2^\circ + \frac{1}{4}\bar h^\circ \dot h^\circ }_u + 4 G M_\text{ADM}\right)\,.
\end{align}
For practical applications, the waveform and $\Psi_2^\circ$ are computed by complex numerical simulations (e.g., \cite{Boyle_2019, Moxon_2020, moxon2021spectrecauchycharacteristicevolutionrapid, Kidder_2017, spectrecode}). The ADM mass is estimated from the simulation templates used to extract the \gls{gw} from measurement data. For most applications, in particular the computation of the displacement memory as a time series, Eq. \eqref{equ:kristin} constitutes a benchmark. \\
In some works in literature, Eq. \eqref{equ:kristin} is slightly modified by integrating the flux \eqref{equ:integration_flux} over the entire future horizon $\scrip$. The latter is sensible as it includes the full evolution of a given \gls{gw} event, and allows for the replacement \cite{waveform_test_BL_III}
\begin{align}\label{equ:replacementtt}
    \lim_{u\rightarrow \infty}\Re{\Psi^\circ_2} = -\frac{GM_{i^+}}{\gamma^3\left(1-\frac{\mathbf{v} \cdot \hat{\mathbf{x}}}{c}\right)^3}\,,
\end{align}
where the limit $u\rightarrow \infty$ corresponds to approaching $i^+$. In this work, the limit \eqref{equ:replacementtt} assumes that gravitational radiation is produced by massive bodies far away from the observer that form a remnant object after collision, propagating with speed $|\mathbf v|$. Thus, in the latter equation, $\mathbf{v}$ stands for the kick velocity of a remnant object after merger, $\gamma \equiv \gamma(v)$ is the usual Lorentz factor from Special Relativity, and $\hat{\mathbf x} = (\sin\theta \cos\phi, \sin\theta \sin\phi,\cos\theta)$ is the unit radial vector in spherical coordinates representing the line of sight of the \gls{gw} source for an observer close to earth. With relation \eqref{equ:replacementtt}, Eq. \eqref{equ:Alex_Chillov} turns into
\begin{align}
   \oint \dd \Omega\,\alpha(\theta,\phi)\left(GM_{i^\circ} -\frac{GM_{i^+}}{\gamma^3\left(1-\frac{\mathbf{v}\cdot \hat{\mathbf x}}{c}\right)^3}\right) =\int_{\scrip}\dd \Omega\,\dd u\,\alpha(\theta,\phi) \left(-\Re{\eth^2\Dot{\Bar{\sigma}}^\circ} + |\Dot{\sigma}|^2\right)
\end{align}
or similarly, if $\alpha(\theta,\phi)$ is an arbitrary smooth function,
\begin{align}\label{equ:Hol}
   GM_{i^\circ} -\frac{GM_{i^+}}{\gamma^3\left(1-\frac{\mathbf{v}\cdot \hat{\mathbf x}}{c}\right)^3} =\int\dd u \left[-\Re{\eth^2\Dot{\Bar{\sigma}}^\circ} + |\Dot{\sigma}|^2\right]\,,
\end{align}
with the ADM mass rewritten as $M_{i^\circ}$. Eq. \eqref{equ:Hol} can be further decomposed into spherical harmonics. As each individual term is by itself a function of the angular coordinates in the sky, one can rewrite the latter equation in terms of modes as
\begin{align}\label{equ:Hol_lm}
    \left(GM_{i^\circ} -\frac{GM_{i^+}}{\gamma^3\left(1-\frac{\mathbf{v}\cdot \hat{\mathbf x}}{c}\right)^3} \right)_{l,m}=\int\dd u \left(-\Re{\eth^2\Dot{\Bar{\sigma}}^\circ}\right)_{l,m} +\int\dd u \left(|\Dot{\sigma}|^2\right)_{l,m}\,.
\end{align}
In Section \ref{sec:Paper_BL}, the decomposition into spherical harmonics is utilized to test the balance between left- and right-hand side in Eq. \eqref{equ:Hol_lm} mode by mode. 

This decomposition concludes the derivation of the flux balance laws. The entirety of Chapter \ref{chap:asymptotics} so far is aimed at the derivation of a constraint equation that can be applied to actual measurement data. In Eq. \eqref{equ:Hol_lm} this goal has been achieved. A crucial part in the derivation process is understanding the gravitational radiation from an observer's perspective, i.e., an instrument far away from the source of radiation, and its interconnection with symmetry properties of relevant spacetimes as well as \gls{gr}'s nonlinearity. Given the specification made in the previous discussions, it is emphasized that the derived constraint equations and the treatment leading up to such exclusively hold for cases where the outlined assumptions hold true and are not to be taken as a general statement for all theories of gravity of all spacetimes. Assuming that the assumptions above are correct, the balance flux laws such as Eq. \eqref{equ:Hol_lm} are put to use in the following Section.


%
%
%


\section{Applications of the Asymptotic Spacetime Formalism: Testing Numerical Waveforms}
\label{sec:Paper_BL}

The following analysis employs constraint equations derived in earlier discussions to conduct a detailed evaluation of state-of-the-art waveform models, with particular emphasis on their predictions for kick velocities and inferred \gls{gw} memory. Motivation for such investigations is drawn in light of recent advancements in \gls{gw} instrumentation that have set new standards for the precision of future measurements, and which have made it essential to advance both semi-analytical and numerical waveform models used in interferometric data analysis. Progress toward this objective can be achieved through the development and implementation of a validation pipeline that quantifies waveform model accuracy based on energy-momentum balance laws derived within the framework of full, non-linear \gls{gr}. Such numerical evaluation is carried out for both precessing and non-precessing binary simulations across models from the \EOB{}, \Phen{}, and \Surr{} families in \cite{waveform_test_BL_I} and summarized in the subsequent paragraphs. The analysis identifies statistically significant deviations, attributable to inaccuracies in the modeling of subdominant modes and to intrinsic systematic errors within the waveform families. These findings are further supported by analytical investigations into harmonic mode mixing effects on the predicted kick velocities and inferred memory signatures. The methodology, developed and validated in \cite{waveform_test_BL_I} offers a robust foundation for future waveform model evaluations and serves as a practical guide for model selection in \gls{gw} data analysis.



\subsection{Numerical Challenges in Gravitational Wave Detection}
\label{subsec:challenges}

Waveforms matched against \gls{gw} signals, such as those reported in \cite{First_LIGO_detection}, carry rich information about the source systems, including the masses and spins of the merging binaries, their distances, and the geometry of their motion. Analyzing these waveforms enables astrophysicists to probe the underlying physical mechanisms, determine the properties of exotic compact objects, and test theoretical models of gravity with remarkable precision. To date, accurate parameter estimation and rigorous tests of \gls{gr} have relied heavily on numerical modeling of gravitational waveforms across a broad range of source parameters—most notably, mass, spin, and other characteristics of the merging systems.
These modeled waveforms serve as templates that are fitted to observational data, and the effectiveness of this template matching critically depends on the accuracy of the waveform predictions. Consequently, the construction of comprehensive and precise waveform models that faithfully capture the dynamics dictated by \gls{gr} is essential for extracting reliable physical information from observed signals.
This need is further underscored by the anticipated improvements in signal-to-noise ratio provided by next-generation GW observatories such as LISA~\cite{colpi_lisa_2024}, the Einstein Telescope~\cite{ET_I}, and the Cosmic Explorer~\cite{evans_cosmic_2023}. Their enhanced sensitivity will allow for the detection of subtle phenomena such as \gls{gw} memory~\cite{Christodoulou_Mem, Henris_Mem, zeldovichRadiationGravitationalWaves1974, thorneGravitationalwaveBurstsMemory1992, PhysRevD.44.R2945, PhysRevD.80.024002, favataNONLINEARGRAVITATIONALWAVEMEMORY2009}, and may even uncover signatures of physics beyond GR~\cite{Heisenberg:2018vsk}. However, discrepancies between the template waveforms and true signals in the data can lead to systematic biases, undermining the accuracy of parameter estimation and theory testing.
To mitigate such biases, data analysis relies on a diverse library of template waveforms. Among these, those generated through NR are considered the most accurate. However, NR simulations are computationally intensive, with each waveform requiring significant resources to compute. This challenge becomes increasingly pressing as the volume of observational data grows with the deployment of multiple ground- and space-based detectors, including LISA~\cite{LISA} and LIGO/Virgo~\cite{ALIGO_2015, AVIRGO_2014}.
Moreover, as measurement precision continues to improve, the possibility of detecting deviations from GR~\cite{Heisenberg:2018vsk} becomes more tangible. Identifying such deviations requires extending the parameter space to include alternative gravitational theories, which in turn increases the number of waveform templates needed for comprehensive data analysis.

The efficiency limitations of NR waveform generation drive the development of alternative waveform models. They are however not the only motivation for other numerical methods for waveform generation. For instance, semi-analytical waveform models provide a more transparent framework for understanding the fundamental physics of \gls{gw} sources. These models enable the explicit inclusion of key physical effects and approximations across the inspiral, merger, and ringdown phases of binary \gls{bh} coalescences. Their architecture is designed to offer flexibility and adaptability, making them applicable to a wide array of astrophysical scenarios—including binary \gls{bh}s, neutron star mergers, and mixed systems—through the tuning of model parameters.
In practical settings, \gls{gw} observatories such as LIGO and Virgo rely on data analysis pipelines that require rapid generation and comparison of waveform templates. Semi-analytical models are particularly well-suited for such applications, as they support real-time or near-real-time processing of observational data.
Moreover, by leveraging the analytical Post-Newtonian (PN) framework commonly used in modeling the inspiral phase, semi-analytical waveforms can be extended to capture the early stages of binary evolution. These early phases are characterized by nearly monochromatic oscillations long before the final merger. Due to computational demands, this regime is often inaccessible to NR simulations, underscoring the complementary advantages of semi-analytical modeling in \gls{gw} astronomy.

The most prominent classes of alternative waveform models include Surrogate models (\Surr{})~\cite{Blackman:2017pcm, Blackman:2017dfb, NRSur7dq4_paper}, phenomenological models (\Phen{})~\cite{PhysRevD.93.044006, PhysRevD.93.044007, PhysRevD.103.104056, IMRPhenomTPHM_paper}, and effective-one-body models (\EOB{})~\cite{SEOBNRv1, SEOBNRv2T, SEOBNRv3, SEOBNRv4, SEOBNRv4PHM_paper, ramosbuades2023seobnrv5phm, Nagar_2018, Gamba_2022, nagar2023analytic}, with the latter two classes relying on semi-analytical techniques. To produce accurate gravitational waveforms within computationally feasible timescales, these models employ distinct strategies for simulating the inspiral, merger, and ringdown phases.
Each model targets specific physical aspects of compact binary coalescence and operates efficiently within particular regions of the source parameter space. In view of the increasing precision and scope of future \gls{gw} observations, the continued validation and refinement of waveform models is essential. This includes adapting templates to address emerging challenges in data analysis pipelines—such as incorporating gravitational memory where it is currently omitted—and establishing robust frameworks for performance evaluation.
These efforts represent active areas of research within the waveform modeling community and constitute key challenges in the broader field of \gls{gw} physics.

These challenges can be addressed by offering a comprehensive and quantitative comparison of state-of-the-art waveform models, focusing on their predictions of key physical observables such as the remnant’s kick velocity and the \gls{gw} memory (i.e., \cite{waveform_test_BL_I}). These models are benchmarked against waveform templates generated from NR simulations, and, where NR data are unavailable—i.e., beyond the current NR catalog—they are directly compared against one another. The analysis prioritizes non-precessing binary mergers for evaluating kick velocities, while the assessment of GW memory encompasses both precessing and non-precessing systems.
The primary analytical tool employed for this comparison is the set of balance flux laws expressed in Eq. \eqref{equ:Hol_lm} through which both a simplified constraint equation and explicit expressions for the kick velocity and GW memory can be obtained for any given waveform.
The utility of these balance laws as diagnostic tools for waveform validation was initially demonstrated in~\cite{Khera:2020mcz, waveform_test_BL_IV}. In turn, \cite{waveform_test_BL_I} is aimed at fully harnessing the potential of this framework for a rigorous cross-model comparison. The study extends earlier investigations that focused individually on analytical waveform models~\cite{waveform_test_BL_IV}, remnant recoil velocities~\cite{Borchers:2021vyw, waveform_test_BL_V}, or GW memory effects~\cite{Khera:2020mcz, waveform_test_BL_II}. By systematically exploring the waveform parameter space and refining the analytical techniques involved, \cite{waveform_test_BL_I} reveals recurring trends and systematic biases present in current waveform models that are comprehensively outlined in this Section.



\subsection{Waveform Models Overview}
\label{subsec:BL_overview}

Before generating waveforms using the diverse range of numerical and semi-analytical models outlined above, it is instructive to introduce the alternative waveform models considered in this study—each of which has played a significant role in the development of \gls{gw} research. Key characteristics of these models are briefly summarized, with attention given to relevant aspects of their numerical implementation. For a detailed description of the internal structure and assumptions of each model, the reader is referred to the references provided below. A comprehensive review can be found in~\cite{lisaconsortiumwaveformworkinggroupWaveformModellingLaser2023}.
To align as closely as possible with the data analysis pipelines commonly employed in GW science, the analysis focusses on a representative ensemble of waveform families accessible through the LIGO-provided \LALSuite{} software suite~\cite{ligoscientificcollaborationLALSuiteLIGOScientific2020}. Rather than implementing the models directly, \LALSuite{} offers an integrated framework developed by the LIGO Scientific Collaboration to streamline waveform-related computations within data analysis workflows.
The selected waveform families fall into four main categories, labeled \NR{}, \Surr{}, \Phen{}, and \EOB{}, based on their respective generation techniques. Additionally, a \TEOB{} model—closely related to the \EOB{} approach—is included in the analysis. The rationale for this particular selection is discussed in detail below.

For each of these five waveform model families, particular attention is given to specific representative. The models analyzed in detail include: \NR{}/\SXS{}, accessed via its \LALSuite{} interface~\cite{Schmidt:2017btt}, which is based on simulations from the SXS collaboration catalog~\cite{Boyle_2019} utilizing the Spectral Einstein Code (SpEC)\cite{kidderBlackHoleEvolution2000}; the \Surr{} model \textit{NRSur7dq4}\cite{NRSur7dq4_paper}; the \EOB{} model \textit{SEOBNRv4PHM}\cite{SEOBNRv4PHM_paper}; the phenomenological \Phen{} model \textit{IMRPhenomTPHM}\cite{IMRPhenomTPHM_paper}; and the \TEOB{} model \textit{TEOBResumS}~\cite{Nagar_2018}. For brevity, this work will frequently refer to individual models by their family names (e.g., \textit{NRSur7dq4} as \Surr{}).
These models exemplify distinct strategies in waveform generation. The \NR{} family comprises full numerical relativity simulations, offering the most complete and accurate waveforms available, and thus serving as the benchmark for other approaches. The remaining ``approximant'' models adopt different methodologies: phenomenological modeling of binary mergers (\Phen{}~\cite{PhysRevD.93.044006,PhysRevD.93.044007, PhysRevD.103.104056, IMRPhenomTPHM_paper, PhysRevLett.113.151101}), effective-one-body formalism (\EOB{}~\cite{SEOBNRv1, SEOBNRv2T, SEOBNRv3, SEOBNRv4, SEOBNRv4PHM_paper, ramosbuades2023seobnrv5phm, Nagar_2018, Gamba_2022, nagar2023analytic, Damour_2014, Buonanno_2006}) and its tuned version (\TEOB{}~~\cite{Nagar_2018, Gamba_2022, nagar2023analytic}), and surrogate modeling via interpolation of NR waveforms (\Surr{}~\cite{Blackman:2017pcm,Blackman:2017dfb, NRSur7dq4_paper}).

Table~\ref{table:1} summarizes key characteristics of the selected models, including the spin-weighted harmonic modes incorporated, their domains of validity in terms of mass ratio $q = M_1/M_2 \geq 1$, the maximal time span before merger, and the spin magnitude $|\chi_i|$. Except for the mode content, these specifications refer to the models as implemented in the \LALSuite{} framework.
For the \SXS{} family, the waveform duration—and thus the maximum time before merger—varies substantially across simulations. The median number of BBH cycles prior to merger in \SXS{} waveforms is approximately 39, with the shortest simulation consisting of 7 cycles and the longest extending up to 351.3 cycles~\cite{Boyle_2019}\footnote{The number of cycles is measured relative to the dominant strain mode $h_{\ell m}$ with $\ell=m=2$.}.
The characteristics of the \LALSuite{} implementation of \Surr{} reflect those of its underlying model, \textit{NRSur7dq4}, which is trained on \NR{} waveforms restricted to a mass-ratio range $q \leq 4$ and dimensionless spin magnitudes $|\chi_i| \leq 0.8$ (where $i=1,2$ for the two compact objects). The initial simulation time for the training \NR{} data spans from $4693$ to $5234$ $M$, which bounds the maximum time before merger in the \LALSuite{} implementation of \Surr{} to approximately $4500$ $M$~\cite{NRSur7dq4_paper}.
The \EOB{} waveform model selected in this study, \textit{SEOBNRv4PHM}, does not impose fixed limitations on waveform duration or the initial reference frequency $f_\text{ref}$, owing to its construction based on the PN ansatz for the early inspiral phase. This flexibility permits waveform generation from arbitrarily low frequencies. Nevertheless, while the model nominally supports a broad parameter space, it has been robustly validated only for $q \leq 4$ and $|\chi_i| \leq 0.8$, with typical simulations beginning at $f_\text{ref} \approx 20$ Hz for a total mass of $50\,M_\odot$, corresponding to approximately $15$–$30$ cycles before merger~\cite{SEOBNRv4PHM_paper}. Therefore, despite the wide numerical parameter range allowed by the \LALSuite{} interface, model accuracy should be considered reliable primarily within this validated region\footnote{Higher spins and mass ratios significantly increase computational cost and reduce accuracy due to the presence of disparate length scales in the binary system~\cite{SEOBNRv4PHM_paper}.}.
The same caveats regarding practical parameter limitations apply to the phenomenological model \Phen{}. The representative model \textit{IMRPhenomTPHM} is derived from its non-precessing precursor, \textit{IMRPhenomTHM} \cite{PhysRevD.105.084039}, using the so-called ``twisting-up'' technique. For additional technical details, the reader is referred to \cite{IMRPhenomTPHM_paper}. The parent model \textit{IMRPhenomTHM} has been calibrated against NR simulations up to $q = 18$ and spin magnitudes $|\chi_i|$ spanning the full physically allowed range $[0, 0.99]$. Similar to \EOB{}, the reference frequency $f_\text{ref}$ can, in principle, be chosen arbitrarily low, and the values listed in Table~\ref{table:1} reflect the maximum parameter range accessible via the \LALSuite{} interface for \Phen{}.
The second representative from the EOB family, \textit{TEOBResumS}, exhibits more stringent parameter constraints. These arise both from numerical stability considerations and prior studies that reported physically inconsistent behavior beyond the recommended parameter bounds. The model has been calibrated against NR simulations for mass ratios up to $q = 20$, but remains limited to non-precessing systems and includes only the dominant $\ell = m = 2$ mode. For the analysis outlined in this Section, \textit{TEOBResumS} serves as a reference to evaluate the impact of including additional harmonic modes, particularly as measured via the balance flux laws.
As with the other semi-analytical models, the \LALSuite{} interface for \TEOB{} permits initial frequencies extending to arbitrarily low values. However, \LALSuite{} typically imposes a maximal allowed initial frequency dependent on system parameters such as mass ratio, component masses, and spin magnitudes. This restriction ensures the waveform captures a minimal portion of the inspiral phase, thereby preserving physical fidelity in simulated gravitational waveforms.

At this point, it is important to highlight that, for the trained eye, discrepancies may appear between the intrinsic properties of waveform models—such as their mode content—and the corresponding features reported within the \LALSuite{} framework, as summarized in Table~\ref{table:1}. For instance, while \LALSuite{} allows users to load the $\ell=3, m=2$ mode for both \Phen{} and \EOB{} waveform families, these specific harmonics may not carry any physically meaningful information and should be treated as artifacts. Such artifacts arise because \LALSuite{} does not necessarily establish a one-to-one mapping between the harmonic modes provided by the original waveform models and those output within the framework. Instead, the waveform data undergoes internal processing, including time interpolation and frame corrections~\cite{ligoscientificcollaborationLALSuiteLIGOScientific2020}, which can subtly alter the strain components compared to their original forms.
Nonetheless, dedicated verification efforts have been performed for specific model versions, demonstrating that the differences introduced by the \LALSuite{} standardization procedures are negligible for practical purposes. For comprehensive information regarding these validated model versions, the reader is referred to the official \LALSuite{} documentation~\cite{ligoscientificcollaborationLALSuiteLIGOScientific2020}. Consequently, one can confidently regard the \LALSuite{} environment—which enforces consistent coordinate and time reference frames across all waveform models—as providing a highly accurate and standardized representation of the original models listed in Table~\ref{table:1}.
Finally, it is important to emphasize that none of the waveform models evaluated in this study incorporates the \gls{gw} memory effect in any form. Nonetheless, methods for including memory—either through manual addition or as built-in features of waveform models—are already available. For \SXS{} waveforms, memory can be extracted and appended using Cauchy-Characteristic-Evolution (CCE) \cite{taylorComparingGravitationalWaveform2013, mitmanComputationDisplacementSpin2020}, which, although currently implemented in only a limited number of publicly available simulations, is in principle applicable to any merger in the catalog using the \textit{Scri} software package \cite{mike_boyle_2020_4041972}. For other waveform models, comparable approaches based on flux balance laws can be employed \cite{waveform_test_BL_II}. Furthermore, recent developments in certain alternative waveform models have begun to incorporate memory effects directly into the simulations (see references pertaining to the respective model families). In the present analysis, memory-free waveforms are used, as the computation of memory serves as a diagnostic tool to assess the fidelity of individual strain modes. This methodology is explored in detail in Section \ref{subsec:BL_analysis}.

\begin{center}
\begin{table*}[t]
\fontsize{9pt}{9pt}
\begin{tabular}{
    |p{2.2cm}|p{2.8cm}|p{2.1cm}|p{2.4cm}|p{1.0cm}|p{1.0cm}|p{1.0cm}|
}
 \hline \bfseries
 Family & \bfseries Implementation & \bfseries Branch & \bfseries Mode Content & \bfseries $q$-range & \bfseries $|\chi_i|$-range & \bfseries $t_\text{init}$ 
 \\\hline\hline 
 $\text{Numerical}$ $\text{Relativity}$ & \SXS{} $\text{(NRhdf5)}$   & $\text{precessing}$    &$\{(\ell,m)| \ell \leq 8\}$  &   $\lesssim 10$ & $\lesssim 0.998$ & -
 \\\hline
   $\text{Surrogate}$ & $\text{NRSur7dq4}$&   $\text{precessing}$  & $\{(\ell,m)| \ell \leq 4\}$  &$\leq 4$&$\leq 0.8$ & $\leq 4500 M$ 
   \\\hline
  $\text{Effective}$ $\text{One Body}$& $\text{SEOBNRv4PHM}$ & $\text{precessing}$ & $\{(2,\pm2),(2,\pm1)$ $,(3,\pm3),(4,\pm4),$ $(5,\pm5)\}$&  $\leq 100$&$\leq 1$ & - 
  \\\hline
   $\text{Phenomeno-}$ $\text{logical}$ & $\text{IMRPhenomTPHM}$  & $\text{precessing}$ &$\{(2,\pm2),(2,\pm1)$ $,(3,\pm3),(4,\pm4),$ $(5,\pm5)\}$& $\leq200$& $\leq 1$&- 
 \\\hline
$\text{``Tidal'' Effective}$ $\text{One Body}$ & $\text{TEOBResumS}$ & $\text{non-precessing}$ & $\{(2,2)\}$&  $\leq30$&$\leq0.99$ & -
  \\\hline

\end{tabular}
 \caption{
    Summary of the model families under investigation in this section. Displayed are the specific implementations considered, the accessible mode content and parameter space coverage (see also \cite{waveform_test_BL_I}). Note that the maximal time before merger, $t_\text{ref}$, is given in units of total mass $M$, where $0$ marks the merger time. In practice, $t_\text{ref}$ implies a parameter-dependent lower bound for the minimum GW (reference) frequency $f_\text{ref}$ of the waveform. The blank spaces for $t_\text{ref}$ indicate that the corresponding model does not admit a uniform boundary regarding the minimal frequency valid for all simulations. All quantities, except for the mode content are displayed as listed in the \LALSuite{} documentation~\cite{ligoscientificcollaborationLALSuiteLIGOScientific2020}. The mode content follows the models' intrinsic output which may differ from the \LALSuite{} one as elaborated below. }
    \label{table:1} 
\end{table*}
\end{center}



\subsection{Preparation of Numerical Waveforms}
\label{subsec:BL_preparation}

The analysis of this Section considers multiple different points in the parameter space layed out by compact binary mergers. Each point in this parameter space corresponds to a distinct numerical waveform which is loaded for each model via \LALSuite{}. Using this framework is beneficial as it avoids potential issues arising due to waveform model-dependent choice of reference frames in each GW simulation. Since different waveform models rely on distinct numerical methods to compute the strain and its harmonic modes, their outputs may be expressed in varying reference frames. To mitigate discrepancies arising from such differences, the \LALSuite{} environment is employed, as it applies the necessary frame transformations to ensure that all output waveforms—and their harmonics—are consistently represented within a common reference frame, up to a possible global phase shift.
In the following, the waveform preparation procedure used in this Section is outlined. This step addresses residual frame alignment issues, among other considerations, and serves as the foundation for the subsequent comparative analysis of the waveform models.

\subsubsection{Alignment and residual ambiguities}
\label{subsubsec:alignment}
As noted previously, waveforms generated via \LALSuite{} are already aligned within a common reference frame, up to a rotation in the orbital plane—equivalently represented by a phase factor, denoted hereafter as $\phi_\text{ref}$. When the waveforms from different approximants are of equal duration—that is, they begin at a common reference frequency $f_\text{ref}$\footnote{Here, the reference frequency refers to the \gls{gw} frequency at the initial non-trivial time step of the waveform. This frequency can be translated into the number of orbits included prior to merger, which typically varies across waveform models, see Table \ref{table:1}.}—the orbital phase remains the sole degree of freedom to be aligned for non-precessing systems.
Challenges emerge when comparing waveform models to NR data from the \SXS{} catalog. For a BBH configuration with specific spin and mass parameters, the domain of validity for each approximant may differ due to model-dependent constraints on the reference frequency $f_\text{ref}$. Notably, $f_\text{ref}$ varies across \SXS{} simulations. In such cases, the waveform for each approximant is regenerated using a new, minimal common $f_\text{ref}$, determined by the most restrictive constraint among the models considered for the given compact binary setup. While \LALSuite{} approximants support waveform generation from different initial or reference frequencies (subject to internal limits; see Table~\ref{table:1}), \SXS{} waveforms are provided at a fixed reference frequency that cannot be modified. Consequently, it may be necessary to manually truncate the inspiral portion of an NR waveform to match the common domain, which introduces an additional phase alignment requirement w.r.t. $\phi_\text{ref}$.
Both the residual phase shift inherent to \LALSuite{} and the phase correction necessitated by waveform truncation can be simultaneously addressed using a single optimization procedure.
For precessing mergers, similar alignment considerations apply. In this case, however, the ambiguity extends beyond the orbital phase and includes the initial orientations of the individual spin vectors of the coalescing \gls{bh}s. This introduces four additional degrees of freedom—specifically, the angular components of the spin vectors—into the alignment procedure, alongside the phase factor $\phi_\text{ref}$. Potential complications arising from these additional parameters are discussed in more detail below.

The optimization procedure addresses both the adjustment of $\phi_\text{ref}$ and the alignment of the initial spin vectors by minimizing a mismatch function $\mathcal{M}(\phi_\text{ref}, \Omega_1, $ $\Omega_2)$, where $\Omega_i =$ $ (\phi_{\chi_i}, \theta_{\chi_i})$ denotes the pair of angular parameters defining the orientation of each spin vector $\chi_i$ involved in the binary constellation. Upon successful minimization, any residual discrepancies observed in the comparison of physical quantities derived from the aligned waveforms are attributed to intrinsic differences in how each model evaluates the binary configuration\footnote{This assumption is admittedly simplistic. A more rigorous treatment would require a comprehensive investigation of potential limitations in the alignment procedure, which lies beyond the scope of this investigation.}. Before addressing the removal of phase ambiguities in more detail, it is essential to ensure that the time axes of the strain and its harmonic modes are properly synchronized—i.e., all waveform time series must be defined on a common time grid. This synchronization is largely handled by the \LALSuite{} environment, which supports waveform loading with consistent time- or frequency-binning across different models\footnote{Tests performed on a small set of BBH simulations with varying time (or frequency) sampling reveal no significant effect on the results presented in Section~\ref{subsec:BL_analysis}, provided that all waveform features and oscillations are sufficiently resolved within the sampling interval.}. The origin of the time grid is automatically set to the merger time, defined as the point at which 
\begin{align}
    \text{max}\,\sqrt{\sum_{\ell,m}|h_{\ell,m}(t)|^2}\,.
\end{align}
Once uniform binning and a common time origin are established, the remaining task in time synchronization involves trimming each waveform's time series to share a common initial time $t_\text{i}$ and final time $t_\text{f}$. This step is applied after waveform loading but precedes any alignment procedures.
Turning now to the residual phase shift, which is determined through mismatch minimization, the numerically implemented mismatch function $\mathcal{M}$ requires an input related to the gravitational strain or its harmonic modes. The alignment is carried out w.r.t. the dominant $h_{2,\pm2}$ mode, and the mismatch is computed as
\begin{equation}\label{equ:mismatch}
    \mathcal{M} (\phi_\text{ref}, \Omega_1, \Omega_2) \coloneqq 1 - \frac{\langle\tilde{h}^\text{ref}_{2,2}, \tilde{h}^\text{align}_{2,2}\rangle}{\lVert\tilde{h}^\text{ref}_{2,2}\rVert \,\lVert\tilde{h}^\text{align}_{2,2}\rVert} \,,
\end{equation}
where $\tilde{h}_{2,2}$ is the Fourier transform of the $(\ell = m = 2)$-mode of the strain as expressed in the spin-weighted spherical harmonic decomposition
\begin{align}\label{equ:full_strain_decomposition}
    h = \sum_{\ell ,m} h_{\ell ,m}(u) _{-2}Y_{\ell ,m}(\theta, \phi)\, ,
\end{align}
with $_{-2}Y_{\ell ,m}$ denoting the spin-weighted spherical harmonics of weight $-2$. The inner product $\langle\cdot,\cdot\rangle$ is evaluated in the frequency domain as
\begin{equation}\label{eq:M-def}
\langle \tilde h_1, \tilde h_2 \rangle \coloneqq 4 \,\text{Re}\, \int_{f_\text{min}}^{f_\text{max}} \dd f\, \tilde h_1(f)\, \tilde h^*_2(f)\,,
\end{equation}
where $\tilde{h}^*$ is the complex conjugate of $\tilde{h}$. This inner product naturally defines a norm, denoted as $\lVert\cdot\rVert$. The integration domain is determined by the Fourier transform of the time-aligned waveform, encompassing the full frequency range computed via fast Fourier transformation of the $h_{2,2}$ time series.
The modes $\tilde h_{2,2}^\text{ref}$ and $\tilde h_{2,2}^\text{align}$ correspond to the reference and target waveform models, respectively. Throughout the analysis, \SXS{} waveforms serve as the default reference; in regions of the parameter space lacking \SXS{} data, \Surr{} is used instead.
It is worth noting that while mismatch computations such as Eq.~\eqref{equ:mismatch} can also be carried out in the time domain, the use of Fourier space offers improved comparability with the mismatch metrics employed in waveform model development (see, e.g., \cite{NRSur7dq4_paper}) and in other waveform comparison studies (e.g., \cite{waveform_test_BL_V, Borchers:2021vyw}). As such, the frequency-domain mismatch is adopted throughout this work.

A fundamental limitation of the alignment strategy defined by Eq. \eqref{equ:mismatch} is that the phase correction is restricted to the $h_{2,2}$ mode—i.e., the alignment is performed solely on the dominant harmonic. Consequently, subdominant modes with odd values of $m$ may still exhibit a residual $\pi$-phase shift, even after the $h_{2,2}$ mode has been successfully aligned. To account for this well-known issue, an additional step is introduced in the waveform preparation following the alignment of the dominant $h_{2,2}$ mode on a common time grid: for each subdominant mode with odd $m$, the mismatch relative to its reference counterpart—calculated via Eq. \eqref{equ:mismatch}—is evaluated twice: once using the original mode and once using the same mode multiplied by a phase factor $\exp(i\pi)$. If the $\pi$-shifted version yields a lower mismatch, it is adopted for that mode in the subsequent analysis. This procedure ensures consistent phase alignment across all relevant harmonic modes between the alternative waveform models and the reference waveforms provided by \NR{} or \Surr{}.

\subsubsection{Domain of alignment and precessing mergers}
\label{subsubsec:comments_few}

Given its dominant contribution to the decomposition in Eq.~\eqref{equ:full_strain_decomposition}, it is standard practice to align waveforms w.r.t. the $h_{2,2}$ harmonic mode in the frequency domain. This approach ensures adequate alignment of the full \gls{gw} strain, which can subsequently be reconstructed for a chosen orientation or line-of-sight of the binary system.
A potential limitation of this method emerges when considering the integration limits in Eq.\eqref{eq:M-def}. Specifically, minimizing the mismatch $\mathcal{M}$ using the $h_{2,2}$ mode over the full frequency interval $f \in [f_\text{min}, f_\text{max}]$—encompassing inspiral, merger, and ringdown—can mask systematic differences between waveform models. While the inspiral phase is generally described using PN techniques across models, the merger and ringdown phases are treated with varying methodologies, leading to increasingly pronounced discrepancies in these later stages. These differences are inherently included in the mismatch computed from Eq.\eqref{eq:M-def}. As such, aligning over the entire frequency domain may inadvertently smooth out or conceal meaningful model-specific features.
To preserve these differences, an alternative strategy would be to restrict the mismatch integral to frequencies below $f_{\text{end of PN-phase}}$, thereby enforcing phase alignment only during the PN regime. This ensures that any deviations in the merger and ringdown remain visible in the aligned waveforms and are not artificially minimized through overextended alignment.
Here, the only alignment-sensitive quantity examined for non-precessing BBH mergers is the remnant recoil (kick) direction. As this represents just one of several diagnostic measures, the potential alignment limitations discussed here are of secondary concern. Accordingly, the analysis adopts the standard integration limits defined in Eq.~\eqref{equ:mismatch}, while a more detailed evaluation of alternative alignment strategies is deferred to future investigations.

A few concluding remarks on precessing systems are warranted. Aligning waveforms for precessing binary \gls{bh} mergers remains a significant challenge, with no universally optimal alignment procedure currently established. In fact, improving alignment strategies for such systems continues to be an active area of research. The primary difficulty arises from the time-dependent orientation of the spin vectors $\chi_1$ and $\chi_2$ of the individual \gls{bh}s. An ideal alignment would require identifying a specific time at which these spins are consistently oriented across waveform models. However, this is computationally demanding and, in some cases, infeasible—particularly when the required synchronization point lies outside the minimal time domain provided by a given waveform model.
Even if the models offered full spin evolution data, discrepancies in the internal treatment of spin dynamics can make direct comparison unreliable. Consequently, the standard approach in the literature is to minimize the mismatch—defined as in Eq.~\eqref{equ:mismatch}—w.r.t. both $\phi_\text{ref}$ and the four spin orientation angles of the binary prior to merger. The alignment is implemented as follows: For a given BBH merger, waveforms from different models are generated using consistent physical parameters, including spin magnitudes and directions, reference frequency, and component masses. Once projected onto a common time grid, the spin magnitudes from the target simulation are held fixed, while the orientation angles serve as input parameters for the mismatch optimization.
The alignment proceeds through an iterative optimization scheme, in which the waveform models are regenerated at each step with updated spin angles, producing a reduced mismatch at each iteration. The procedure continues until the mismatch is sufficiently small—typically comparable to that achieved for non-precessing systems. This results in a set of aligned waveforms on a shared time grid, all featuring identical spin magnitudes but potentially differing spin orientations. These differences reflect the varying treatments of spin precession across waveform models.
Because the direction and magnitude of the remnant recoil (kick) are highly sensitive to the spin orientations, computing kicks for aligned precessing waveforms becomes unreliable. Therefore, in the analysis presented in Section~\ref{subsec:BL_analysis}, attention is restricted to the (nonlinear) memory contribution from precessing mergers. In contrast to the kick, the memory effect is invariant under global phase shifts and is only mildly sensitive to changes in the spin configuration. For the precessing mergers considered in the analysis (detailed list of relevant events is found in Section~\ref{subsec:BL_events}), the alignment procedure described here significantly improves the accuracy of the computed memory signal.
Both the kick and the memory are influenced by the asymmetry between the $\ell,m$ and $\ell,-m$ harmonic strain modes, which is especially pronounced in precessing systems. While both quantities depend on this asymmetry, their sensitivities differ—a fact that can be explained analytically. It should be noted that this asymmetry is explicitly incorporated only in the \Surr{} and \SXS{} models. The versions of \EOB{} and \Phen{} used in this analysis do not include this feature. For recent developments in phenomenological models that do, see, for example, \cite{New_Phen_I, New_Phen_II}.

\begin{figure}[t]
	\centering
	\includegraphics[width=0.9\linewidth]{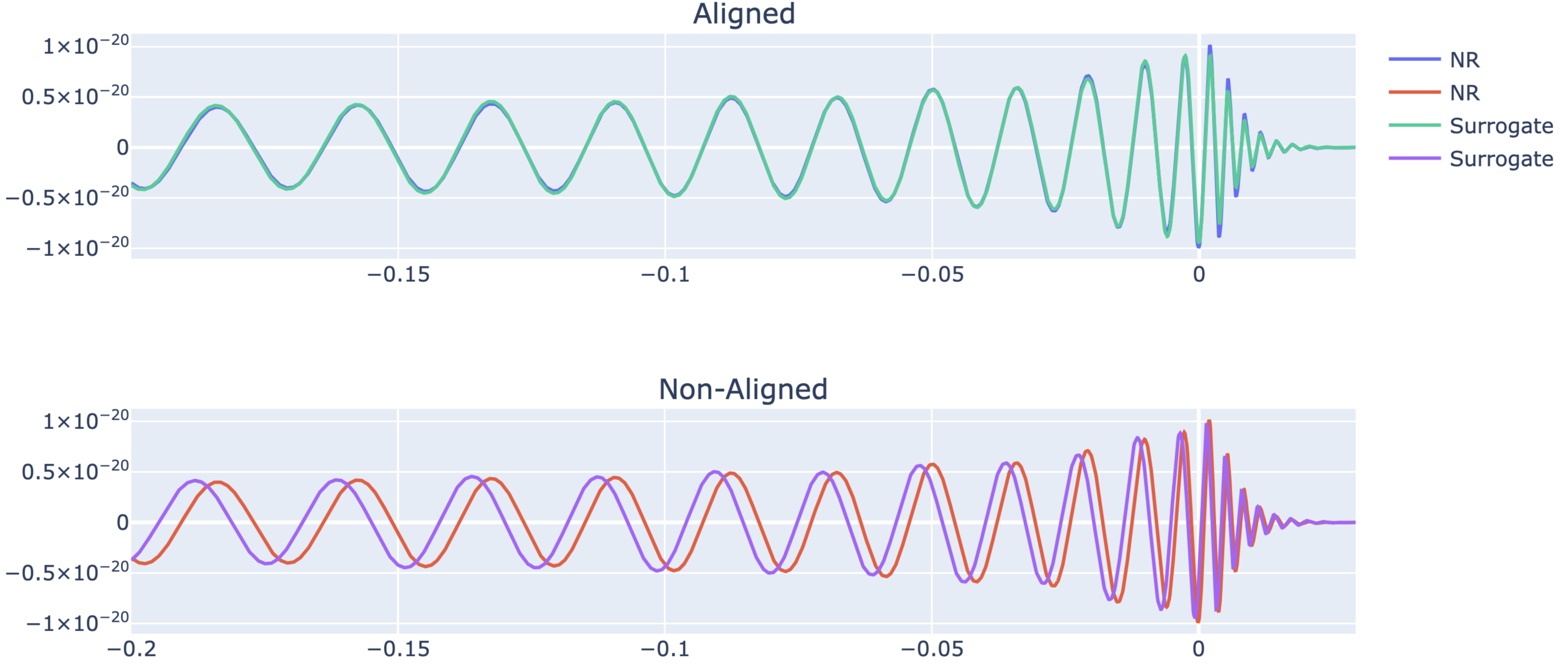}
	\caption{Aligned (top) and non-aligned (bottom) $h_{2,2}$ waveform mode for a precessing binary merger (SXS:BBH:$1011$) \cite{waveform_test_BL_I}.}
	\label{fig:alignI}
\end{figure}

\begin{figure}[t]
	\centering
	\includegraphics[width=0.9\linewidth]{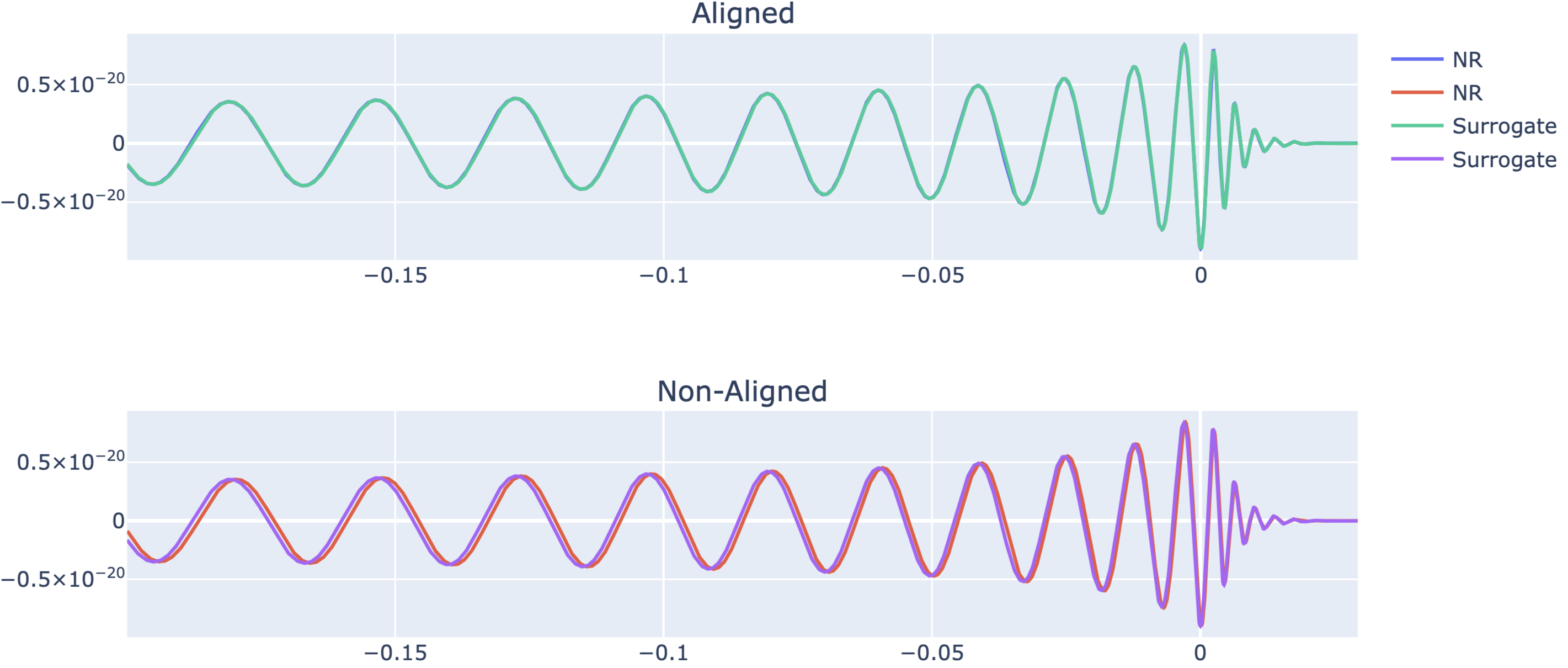}
	\caption{Aligned (top) and non-aligned (bottom) $h_{2,2}$ waveform mode for a non-precessing binary merger (SXS:BBH:$0191$) \cite{waveform_test_BL_I}.}
	\label{fig:alignII}
\end{figure}

\begin{figure}[t]
	\centering
	\includegraphics[width=0.9\linewidth]{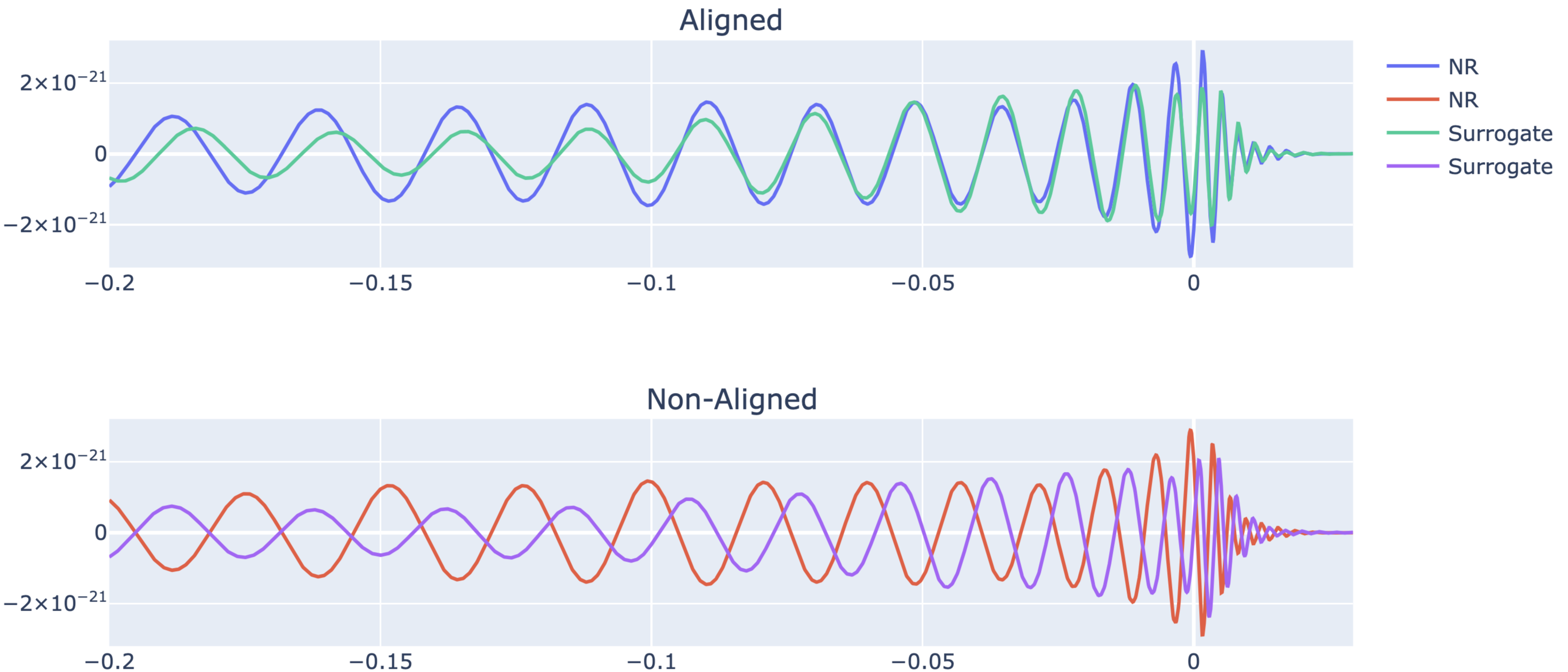}
	\caption{Aligned (top) and non-aligned (bottom) $h_{2,1}$ waveform mode for a precessing binary merger (SXS:BBH:$1011$) \cite{waveform_test_BL_I}. Here, the aligned waveform is displayed.}
	\label{fig:alignIII}
\end{figure}

\begin{figure}[t]
	\centering
	\includegraphics[width=0.9\linewidth]{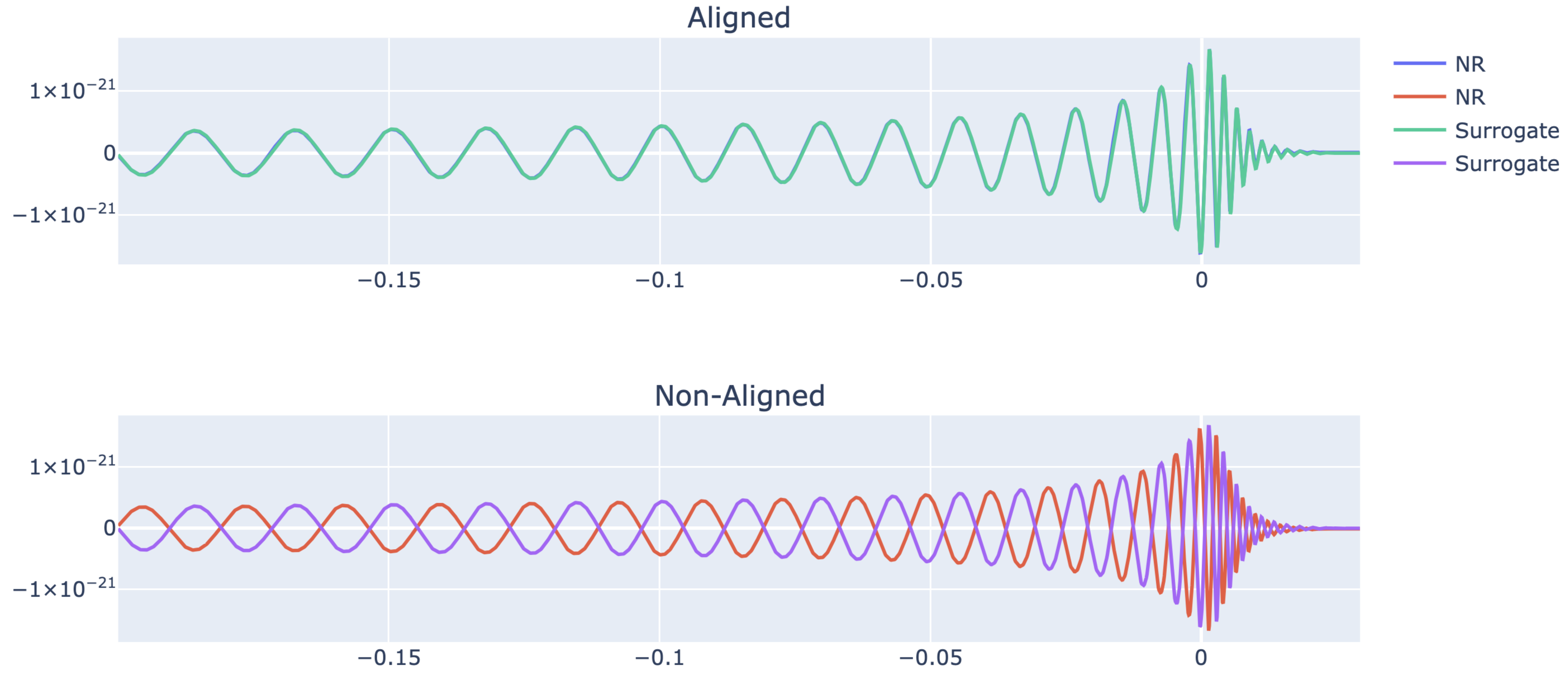}
	\caption{Aligned (top) and non-aligned (bottom) $h_{3,3}$ waveform mode for a non-precessing binary merger (SXS:BBH:$0191$) \cite{waveform_test_BL_I}.}
	\label{fig:alignIV}
\end{figure}

The effectiveness of the adapted alignment strategy is illustrated in Figures \ref{fig:alignI}-\ref{fig:alignIV}. For clarity, only results from \Surr{} and \NR{} are shown, with \NR{} serving as the reference. Figures~\ref{fig:alignI} and~\ref{fig:alignII} display the alignment of the dominant $h_{2,2}$ harmonic for precessing and non-precessing mergers, respectively, with precessing systems typically exhibiting slightly larger mismatches. Figures~\ref{fig:alignIII} and~\ref{fig:alignIV} highlight the need for applying residual $\pi$-shifts to subdominant modes with odd $m$ values. These modes are not subject to the optimization process but are corrected post-alignment via a $\pi$-shift, if doing so reduces the mismatch. The examples shown are chosen to emphasize this effect in select subdominant harmonics.




\subsection{Numerical Simulations under Consideration}
\label{subsec:BL_events}

A meaningful comparison of numerical waveform models requires a sufficient coverage of the waveform parameter space to prevent undesirable biases of the results. Thus, for the analysis of this Section, it is instructive to outline a strategy to prevent such biases, in particular given the substantial number of BBH merger simulations analyzed. Doing so enables one to draw meaningful connections between the results and underlying phenomenological trends—particularly in instances where one or more waveform models exhibit significant deviations from the reference model. Furthermore, by carefully distributing the simulated mergers across a broad parameter space, one minimizes the risk of introducing selection biases that could skew the conclusions. In Section~\ref{subsec:BL_analysis}, discrepancies in physical quantities are explicitly related to the mergers’ positions within parameter space. To ensure robust coverage, both precessing and non-precessing BBH systems are considered. For each category, one includes waveforms with existing NR counterparts and supplement the dataset with additional simulations, thereby enhancing the diversity and representativeness of the parameter space under investigation.

\subsubsection{Non-precessing binary mergers}
\begin{figure}[t]
	\centering
	\includegraphics[width=0.7\linewidth]{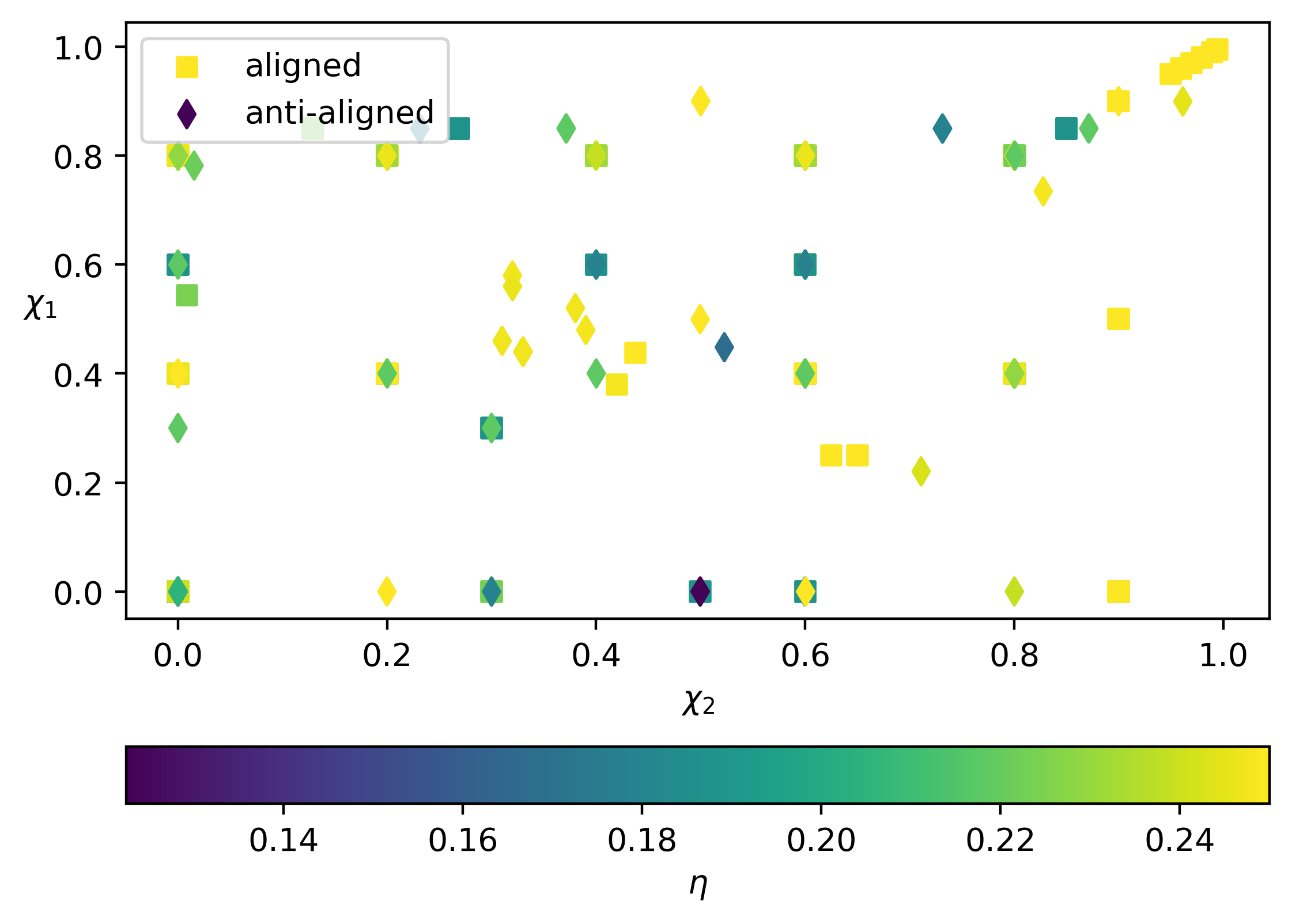}
	\caption{Parameter space for the non-precessing \SXS{} data \cite{waveform_test_BL_I}.}
	\label{fig:params_I}
\end{figure}

\begin{figure}[t]
	\centering
	\includegraphics[width=0.7\linewidth]{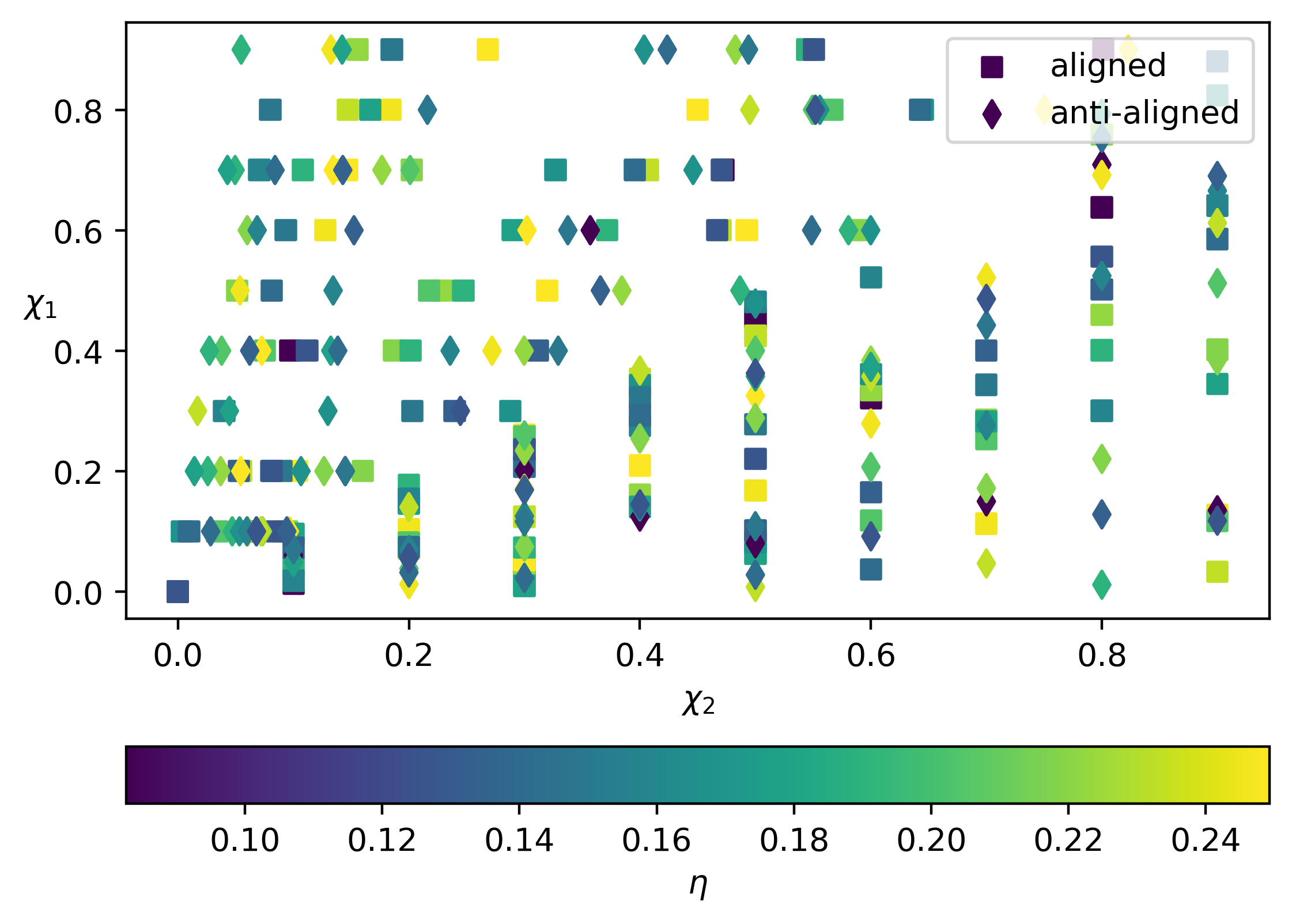}
	\caption{Parameter space for the non-precessing merger simulations without NR counterpart \cite{waveform_test_BL_I}.}
	\label{fig:params_non_prec}
\end{figure}

The primary focus of this analysis is on non-precessing BBH simulations drawn from the \SXS{} catalog. Figure~\ref{fig:params_I} illustrates the distribution of these simulations across the parameter space, distinguishing between aligned and anti-aligned spin configurations. Color coding is used to represent the symmetric mass ratio $\eta \coloneqq \frac{q}{(1+q)^2}$, where $q \coloneqq M_{1} / M_{2} \geq 1$ denotes the ratio of component masses.
A total of 175 mergers with non-negligible recoil velocities ($v > 20 , \text{km}/\text{s}$) are selected for the non-precessing case. Despite this number, Figure~\ref{fig:params_I} appears to show sparse coverage of the parameter space. This apparent sparsity arises from overlaps among simulations that differ only by mass ratio or spin alignment.
As seen in Figure~\ref{fig:params_I}, regions corresponding to low spin magnitudes $\chi_1$ and $\chi_2$ are underrepresented in the current selection from the catalog. To address these gaps, additional BBH mergers without corresponding \SXS{} counterparts are generated. These supplemental simulations include both aligned and anti-aligned spin configurations and span a wide range of symmetric mass ratios $\eta$, as shown in Figure~\ref{fig:params_non_prec}. This extended dataset contributes an additional 220 non-precessing waveform instances to the overall analysis. Combined, the cataloged and supplementary non-precessing simulations constitute a well-distributed dataset, well-suited for an unbiased and systematic investigation.

Despite considerable efforts to accurately model GW waveforms, certain non-precessing mergers reveal substantial deviations between specific approximants and both the reference model and other waveform models—particularly w.r.t. the remnant velocity or gravitational memory (anticipating results discussed in Section~\ref{subsec:BL_analysis}). A closer inspection of these cases shows that such discrepancies typically originate from a single mode—such as the $h_{2,1}$ mode—of the deviating model differing markedly from the corresponding mode in the other approximants and the reference. This behavior is most frequently observed when one of the normalized spin components, $\chi_1$ or $\chi_2$, approaches extremal values, i.e., close to 0 or 1. Representative examples include \textit{SXS:BBH:0222}, \textit{SXS:BBH:0223}, and \textit{SXS:BBH:0251}, for which \EOB{} predicts kick velocities that are significantly over- or underestimated relative to the other models. To avoid biasing the statistical analysis, such simulations are excluded from the final dataset.

\subsubsection{Precessing binary mergers}
\begin{figure}[t]
	\centering
	\includegraphics[width=0.7\linewidth]{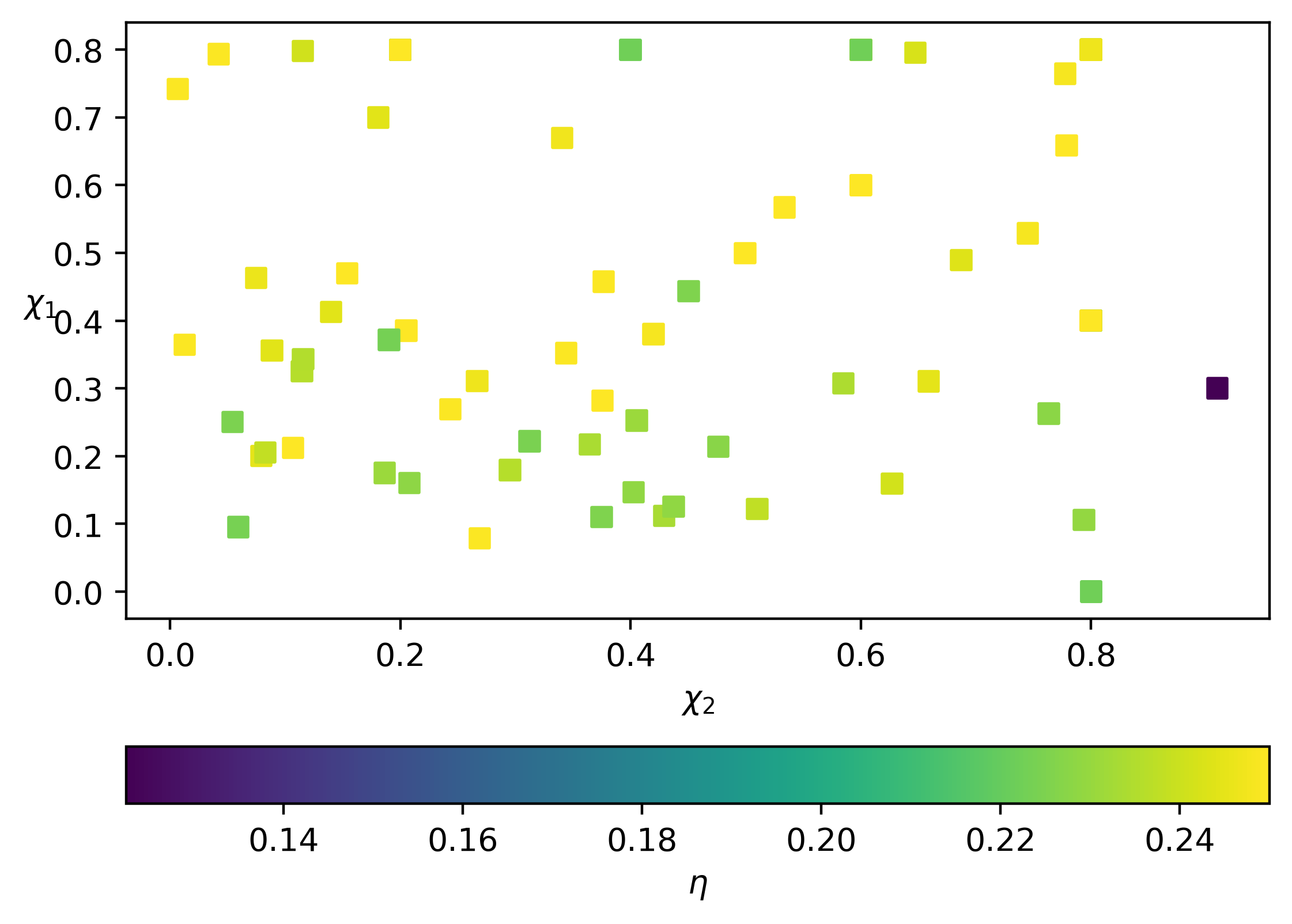}
	\caption{Parameter space for the precessing \SXS{} data \cite{waveform_test_BL_I}.}
	\label{fig:params_prec_NR}
\end{figure}

\begin{figure}[t]
	\centering
	\includegraphics[width=0.7\linewidth]{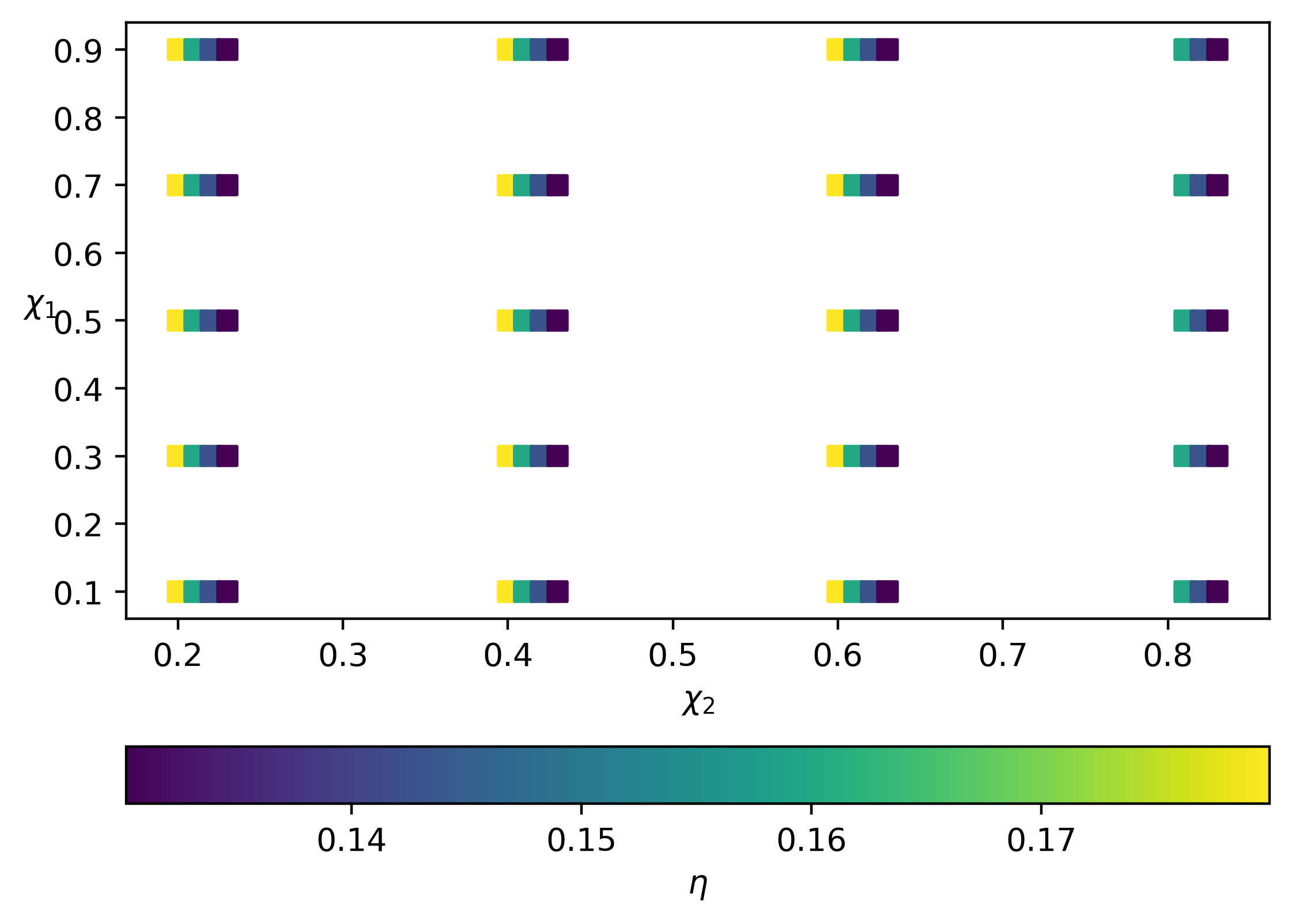}
	\caption{Parameter space for the precessing simulations without NR counterpart \cite{waveform_test_BL_I}. For every bundle of data points, the spin configuration is identical. For illustrative reasons, the plot shows a slight misplacement.}
	\label{fig:params_prec}
\end{figure}

The analysis of precessing BBH mergers is based on 130 simulations with \SXS{} counterparts, illustrated in Figure~\ref{fig:params_prec_NR}. w.r.t. the two spin vectors, the selected mergers provide a fairly uniform coverage of the accessible parameter space. However, the catalog lacks coverage at low mass ratios, with the sole exception of \textit{SXS:BBH:0165}, involving component masses of $51.4,\Msol$ and $8.6,\Msol$. This underrepresentation of systems with $\eta < 0.2$ introduces a potential bias, which is addressed by incorporating an additional 75 simulated precessing BBH mergers. These supplemental cases are shown in Figure~\ref{fig:params_prec}. The difference in both the number and distribution of added simulations between the precessing and non-precessing datasets stems from the significantly higher computational cost of aligning precessing waveforms. Alignment in these cases involves optimizing over five parameters rather than just $\phi_\text{ref}$, leading to considerably longer generation times. As a result, the added precessing simulations are fewer in number and more regularly spaced across parameter space.



\subsection{Waveform Model Assessment: Prelude}
\label{subsec:BL_analysis_pre}

To assess a waveform model's performance against a chosen reference model, an adequate measure of comparison has to be selected. The remnant's kick velocity and the \gls{gw} memory are chosen as such, which are both physical observables that can be calculated solely from the strain of a GW. To that end, Eq. \eqref{equ:Hol} is utilized. For the purpose of this Section, one thereby chooses to integrate over all of $\scrip$, i.e., from $u\rightarrow -\infty$ to $u\rightarrow \infty$. Physically, this integration of a given strain over the full range of $u$ can be interpreted as a single event in an empty spacetime starting and some far past, merging at some finite $u_0$ and ringing down towards $u\rightarrow  \infty$. In other words, the strain contains the maximal amount of information of the merger. Then, by adding suitable dimensionful parameter and replacing $h^\circ_{+,\times}$ with $D_L h^\circ_{+,\times}$ where $D_L$ is the luminosity distance (see the discussion below Eq. \eqref{equ:theUltimateBL} for details), one obtains
\begin{align}\label{equ:fullDimBL}
    c^2 \left(\frac{M_{i^+}}{\gamma^3\left(1-\frac{\mathbf {v}\cdot\hat{\mathbf x}}{c}\right)^3} - M_{i^\circ}\right) =& -\frac14 \frac{D^2_L\, c^3}{G}\int_{-\infty}^{\infty} \dd u\left(\dot{h}^2_+ + \dot{h}^2_\times\right) \\ \notag &+ \frac12 \frac{D_L\, c^4}{G} \text{Re}\left[\eth^2\left(h_+ - i\, h_\times\right)\right]\bigg|_{t=-\infty}^{t=+\infty}\,,
\end{align}
where $\mathbf{v}$ stands for the kick velocity, $\gamma \equiv \gamma(v)$ is the usual Lorentz factor from Special Relativity for $v:=|\mathbf v|$, and $\hat{\mathbf x} = (\sin\theta \cos\phi, \sin\theta \sin\phi,\cos\theta)$ is the unit radial vector in spherical coordinates. As a matter of fact, Eq. \eqref{equ:fullDimBL} does not only provide one constraint equation but infinitely many. Namely, one per point on the $2$-sphere, i.e., one so-called energy-momentum balance law per choice of $(\theta, \phi)$.

A more intuitive understanding of Eq. \eqref{equ:fullDimBL} can be obtained through a tentative interpretation of the balance laws. The left-hand side of~\eqref{equ:fullDimBL} may be viewed as the difference between two energies measured at distinct moments in time, specifically at $i^+$ and $i^\circ$: the energy of the remnant, corresponding to the system's energy after the merger, minus the energy of the binary system, measured well before the merger. The denominator, containing the Lorentz factor and the kick velocity, serves as a correction accounting for the fact that the binary’s energy is evaluated in its instantaneous rest frame, whereas the remnant generally moves with a velocity $\vec{v}$ relative to that frame. When decomposing~\eqref{equ:fullDimBL} into spherical harmonics, this correction factor for the $\ell=0$ mode reduces precisely to the familiar special relativistic expression for the kinetic energy of a moving body, $E = \gamma(v) M c^2$ \cite{waveform_test_BL_III}. Given that the binary system loses energy through the emission of \gls{gw}s, it is natural to expect the energy difference to be negative. Consistently, the right-hand side of\eqref{equ:fullDimBL} features an integral over a manifestly positive quantity, $\left(\dot{h}^2_+ + \dot{h}^2_\times\right)$, with an overall negative sign. This integral represents the energy radiated by GWs, a well-established result in linearized GR. However, it is important to note that the balance laws apply well beyond the linear regime, and no linearization is invoked in their derivation. The second term on the right-hand side captures the contribution from GW memory. Because the left-hand side of~\eqref{equ:fullDimBL} depends on $\hat{\mathbf x}$, which itself varies with the angular coordinates $(\theta, \phi)$, and since the strains $h_+$ and $h_\times$ are functions of both time and angle, it follows that there exists one balance law for each choice of $(\theta, \phi)$~\cite{waveform_test_BL_I}.

In what follows, the notation is streamlined by introducing the asymptotic shear: 
\begin{align}
    h(t, \theta, \phi) \coloneqq \frac12 \left(h_+ + i\, h_\times\right)(t, \theta, \phi)\,,
\end{align}
where $i$ denotes the imaginary unit. This definition, in particular, enables a compact rewriting of the memory term as $\eth^2\Delta\bar{h}$, with $\bar{h}$ denoting the complex conjugate of $h$, and 
\begin{align}
    \Delta \bar{h} \coloneqq \int_{-\infty}^{+\infty} \dd u \,\dot{\bar{h}}\,.
\end{align}
This compact form is made possible because $\text{Im}(\eth^2 \bar{h})$ vanishes at $t=\pm \infty$ (see, e.g., \cite{waveform_test_BL_III}). Using these definitions and expanding both sides of Eq. \eqref{equ:fullDimBL} in spherical harmonics yields a tower of constraints—one for each pair $(\ell, m)$ with $|m|\leq\ell$. Specifically, this decomposition results in 
\begin{align}\label{equ:BLinModes}
    \left(M_{i^\circ} - \frac{M_{i^+}}{\gamma^3\left(1-\frac{\mathbf {v}\cdot\hat{\mathbf x}}{c}\right)^3}\right)_{\ell,m} &= \frac{D^2_L c}{G}\int_{-\infty}^{+\infty} \dd u\left(|\dot{h}|^2\right)_{\ell,m}- \frac{D_L c^2}{G} C_\ell \Delta \bar{h}_{\ell, m}\,,
\end{align}
where the coefficients $C\ell$ are defined by 
\begin{align}
    C_\ell \coloneqq \sqrt{(\ell-1)\ell(\ell+1)(\ell+2)} \,.
\end{align}
Deriving this expression involves standard properties of spherical harmonics and the explicit form of $\eth^2\bar{h}$, as discussed in \cite{Our_Review}. Notably, $C_\ell$ vanishes for $\ell = 0$ and $\ell = 1$, implying that the memory term ($\sim \Delta \bar{h}_{\ell, m}$) contributes only for $\ell\geq2$. Separately treating the $\ell<2$ and $\ell\geq2$ modes of the balance laws enables determination of the remnant mass—if not supplied by the waveform model—as well as the kick velocity and \gls{gw} memory, using solely the GW strain. This is achieved by decomposing the left-hand side of Eq. \eqref{equ:BLinModes} into modes. For this purpose, it is convenient to align the $z$-axis of the rest frame (in which $M_{i^\circ}$ is measured) with the kick velocity $\mathbf{v}$ via a rotation in the plane spanned by $\mathbf{v}$ and $\hat{\mathbf{z}}$. The first few $(\ell,m)$ coefficients then take the form: 
\begin{align}
    (\ell = 0, m=0)&: & &2 \sqrt{\pi } (\gamma  M_{i^+} - M_{i^\circ}) ,\notag\\
    (\ell = 1, m=0)&: & &2 \sqrt{3 \pi } \gamma  M_{i^+} v,\notag \\
    (\ell = 2, m=0)&: & &\frac{\sqrt{5 \pi } \gamma M_{i^+} \left(5 v^3+3 \gamma^{-4} \tanh ^{-1}(v) -3 v\right)}{v^3} \,.\notag
\end{align}
The first term corresponds, up to a factor of $c^2$, to the difference between the rest energy $M_{i^\circ}$ of the binary and the kinetic energy $\gamma M_{i^+}$ of the remnant, while the second term describes the $z$-component of the remnant's momentum. Although the third term does not admit a simple physical interpretation, it is important to highlight that the inverse hyperbolic tangent function $\tanh^{-1}(v)$ caused numerical instabilities in the analysis code; hence, it is Taylor-expanded to sixth order in $v/c$. To recover the corresponding modes in the original frame, the modes are transformed using Wigner $D$-matrices $D^\ell_{m,m'}$ corresponding to the inverse of the rotation used to align $\mathbf{v}$ with $\hat{\mathbf{z}}$. The decomposition of the GW energy integral in Eq. \eqref{equ:BLinModes} is straightforward, as the integrand can be expressed as \begin{equation}\label{equ:strain_decompose}
    |\dot h|^2 = \sum_{\ell, m} \alpha_{\ell m} Y_{\ell m}(\theta, \phi)\,.
\end{equation}
where the coefficients $\alpha_{\ell m}$ are given by (see also \cite{Khera:2020mcz})\begin{align}\label{equ:alphass}
    \alpha_{\ell m} = \sum_{\ell_1=2}^\infty \sum_{\ell_2=2}^\infty \sum_{|m_1|\leq \ell_1}\sum_{|m_2|\leq \ell_2} (-1)^{m_2+m} \dot h_{\ell_1 m_1} \dot{\bar{h}}_{\ell_2m_2}\sqrt{\frac{(2\ell_1+1)(2\ell_2+1)(2\ell+1)}{4\pi}}  D^{\ell,\ell_1,\ell_2}_{m,m_1,m_2}
\end{align}
with 
\begin{align}
D^{\ell,\ell_1,\ell_2}_{m,m_1,m_2}:=
    \begin{pmatrix}
\ell_1 & \ell_2 & \ell\\
m_1 & -m_2 & -m
\end{pmatrix}\begin{pmatrix}
\ell_1 & \ell_2 & \ell\\
2 & -2 & 0
\end{pmatrix}\,.
\end{align}
The $\begin{psmallmatrix} \ell_1 & \ell_2 & \ell\\ m_1 & m_2 & m\end{psmallmatrix}$ denotes the Wigner-$3j$ symbol, which dictates how the strain modes $\dot h_{\ell_1 ,m_1} \dot{\bar{h}}_{\ell_2,m_2}$ couple to each other in Eq. \eqref{equ:alphass}.

For astrophysically realistic kick velocities, it is a good approximation to assume $\gamma\approx1$. Under this assumption, the $\ell=0$ mode of the balance laws reduces to an energy conservation equation: 
\begin{align}\label{equ:Energy}
    c^2 (M_{i^\circ}-M_{i^+}) = \frac{D_L^2 c^3}{16\pi G}\int_{-\infty}^{\infty }\dd u \oint \dd \Omega \,|\dot h|^2\,,
\end{align}
where $\oint\dd\Omega$ denotes integration over the unit two-sphere. As expected, the mass loss of the system is accounted for by the energy radiated away via GWs. For waveform models that provide the remnant mass, this relation offers a consistency check; for models that do not, it provides a way to infer $M_{i^+}$ from the initial total mass, luminosity distance, and GW strain.\footnote{It should be noted that the precision of analytical predictions is inherently limited by the accuracy of the strain, particularly the number of inspiral cycles captured before merger. In numerical simulations, waveforms are generally truncated, introducing a systematic uncertainty that, however, affects all waveforms equally and thus does not impact comparative studies.} Using the expansion coefficients~\eqref{equ:alphass} for $\ell=m=0$, the energy conservation equation can also be expressed as 
\begin{align}
    c^2 (M_{i^\circ}-M_{i^+}) &=\frac{D_L^2 c^3}{8\sqrt{\pi} G}\int_{-\infty}^{\infty}\dd u\,\alpha_{0,0}.
\end{align}
The $\ell=1$ mode of the balance laws encapsulates linear momentum conservation, allowing extraction of the kick velocity components: 
\begin{equation}\label{equ:kick}
    \mathbf {v}_\text{kick} = \frac{D_L^2 c^2}{16 \pi G M_{i^+}} \int_{-\infty}^\infty \dd u \oint \dd \Omega \, \hat{x}_i\, |\dot h|^2 ,
\end{equation}
in the original reference frame, where $\hat{x}_i$ are the Cartesian components of the radial unit vector $\hat{\mathbf{x}} = (\sin\theta\cos\phi,\sin\theta\sin\phi,\cos\theta)$. The kick velocity components can also be expressed in terms of the $\alpha_{\ell m}$ coefficients for $\ell=1$ as 
\begin{align}
    v_1 &= \frac{D_L^2 c^2}{16 \pi G M_{i^+}} \sqrt{\frac{2\pi}{3}} \int_{-\infty}^\infty \dd u \, (\alpha_{1,-1} - \alpha_{1,1})\label{equ:v1}\,, \\
    v_2 &= \frac{-iD_L^2 c^2}{16 \pi G M_{i^+}} \sqrt{\frac{2\pi}{3}} \int_{-\infty}^\infty \dd u \, (\alpha_{1,-1} + \alpha_{1,1})\,, \\
    v_3 &= \frac{D_L^2 c^2}{8 \pi G M_{i^+}} \sqrt{\frac{\pi}{3}} \int_{-\infty}^\infty \dd u \, \alpha_{1,0} \,.\label{equ:v3} 
\end{align}
Thus, equations~\eqref{equ:v1}–\eqref{equ:v3} determine the kick velocity from the luminosity distance, remnant mass, and strain, with $M_{i^+}$ obtainable from the energy conservation relation~\eqref{equ:Energy} if not already available.

The harmonic modes corresponding to $\ell\geq 2$ encompass contributions from the GW memory. In the limit of integrating $u$ from $-\infty$ to $\infty$ it is instructive to rewrite the memory as
\begin{equation}
\label{eq:mem_total_def}
 \Delta \bar{h}_{\ell,m} = \Delta \bar{h}^\text{lin}_{\ell,m} + \Delta \bar{h}^\text{non-lin}_{\ell,m}\,.
\end{equation}
Following conventions in the literature (including in particular \cite{waveform_test_BL_I}, i.e., \cite{waveform_test_BL_I}), the first contribution is defined as
\begin{equation}
\label{eq:lin_mem_def}
    \Delta \bar{h}^\text{lin}_{\ell,m} \coloneqq \frac{G}{C_\ell D_L c^2 }\left(\frac{M_{i^+}}{\gamma^3(1 - \frac{\vec {v}}{c}\cdot \hat{x})^3} - M_{i^\circ}\right)_{\ell,m}\,,
\end{equation}
and denoted as the \textit{linear memory}~\cite{Lin_Mem_II, zeldovichRadiationGravitationalWaves1974} in what follows. The second term, computed as
\begin{align}
\label{eq:nonlin_mem_def}
\Delta \bar{h}^\text{non-lin}_{\ell,m} \coloneqq  \frac{D_L}{4C_\ell c}\int_{-\infty}^\infty \dd t \, \alpha_{\ell m}\,,
\end{align}
in turn describes the \textit{non-linear memory}~\cite{Christodoulou_Mem, thorneGravitationalwaveBurstsMemory1992}. It is stressed at this point that other names for these two terms are commonly used, e.g., \textit{ordinary memory} for the former and \textit{null memory} for the latter~\cite{waveform_test_BL_II}. 
Most importantly, the $\ell\geq 2$ modes of the balance laws~\eqref{equ:BLinModes} are constraints on the (displacement) memory terms of a gravitational waveform\footnote{As mentioned before, other memory contributions such as the spin memory do exist and can be computed based on the strain as well. The latter is more thoroughly discussed in Section \ref{sec:Paper_Mem}}. In previous studies the same constraints are utilized to either add memory to numerical waveform models or correct numerical waveforms which do not accurately incorporate the memory effect~\cite{Khera:2020mcz, waveform_test_BL_II}. Instead, here the linear and non-linear memory inferred from~\eqref{eq:lin_mem_def} and~\eqref{eq:nonlin_mem_def}, respectively, are used as a means of comparison between different waveform models. 

Before continuing with the actual waveform assessment, it is instructive to briefly comment on the connection between the previously outlined equations and the kinematics involved in a BBH merger, the event of interest throughout this work: The utility of the balance flux laws extends beyond their mathematical elegance to their ability to accurately predict the kinematic properties of the remnant compact object formed in a BBH merger. As shown in this Section, both the final mass and recoil velocity of the remnant can be determined using only the \gls{gw} strain. Moreover, for systems exhibiting precession, the decomposition of physical quantities into spherical harmonic components, as defined in Eq. ~\eqref{equ:alphass}, reveals additional structure and subtleties in the merger dynamics that might otherwise remain obscured. To offer a brief insight, consider the selection rules—specifically, the Wigner-$3j$ symbols—that govern which harmonic strain components contribute to the relevant $\alpha_{\ell m}$ coefficients. In general, the $\alpha_{\ell m}$ terms are dominated by contributions involving at least one instance of the leading strain modes $h_{2,\pm2}$. For example, examine $\alpha_{1,0}$, which determines the out-of-plane component of the remnant’s recoil velocity, $v_3$. A detailed evaluation of the sums in Eq. \eqref{equ:alphass} for $\alpha_{1,0}$ reveals two leading contributions proportional to $h_{2,-2}\bar{h}_{2,-2}$ and $h_{2,2}\bar{h}_{2,2}$\footnote{Time derivatives are omitted here for clarity.}. In non-precessing BBH systems, these terms cancel exactly, resulting in a negligible out-of-plane kick. However, in precessing binaries, where an asymmetry between $h_{2,2}$ and $h_{2,-2}$ arises, this cancellation is no longer exact. As a result, $\alpha_{1,0}$ acquires a non-vanishing contribution, yielding a non-trivial $v_3$ component. While this asymmetry is typically small—and thus the difference between the two terms is also small—the resulting out-of-plane kick can nonetheless be comparable in magnitude to the in-plane recoil.\\
The small asymmetry between the $\ell,m$ and $\ell,-m$ modes also plays a critical role in shaping various physical observables, particularly the non-linear memory effect. This memory effect is especially sensitive to the $\alpha_{2,0}$ component, which is dominant due to its allowance for coupling between two leading strain modes, such as $h_{2,-2}\bar{h}_{2,-2}$ and $h_{2,2}\bar{h}_{2,2}$. In contrast to $\alpha_{1,0}$, these contributions enter with the same sign, resulting in constructive interference. As a consequence, even in the absence of precession, the terms proportional to $h_{2,-2}\bar{h}_{2,-2}$ and $h_{2,2}\bar{h}_{2,2}$ reinforce each other, and any asymmetry between $h_{2,2}$ and $h_{2,-2}$ leaves only a minor imprint on their sum. This constructive behavior is a key driver of the buildup of the non-linear memory signal.
While the asymmetry is responsible for the entire out-of-plane kick, its influence on the memory is confined to a fraction of its magnitude relative to the dominant integrals involving $h_{2,\pm 2}\bar{h}_{2,\pm 2}$. Therefore, unlike the recoil velocity, the non-linear memory is only mildly sensitive to asymmetries between the $\ell,m$ and $\ell,-m$ modes. This reduced sensitivity allows the memory effect to be meaningfully estimated even in waveform models that do not explicitly incorporate these asymmetries, including the \EOB{} and \Phen{} models considered in this Section. Although other mode combinations within the $\alpha_{\ell m}$ coefficients are also influenced by potential asymmetries between $\ell,m$ and $\ell,-m$, such contributions are typically subdominant and thus of lesser practical significance.

The kinematic mode analysis of the $\alpha$ components involves a range of complex features that extend well beyond the brief illustration provided above. While a more detailed exploration of these aspects is certainly valuable, presenting an exhaustive discussion without the support of empirical data would be premature at this stage. Therefore, this analysis will be revisited and expanded below, where the capabilities of state-of-the-art waveform models to fully leverage the analytical structure imposed by the balance flux laws is demonstrated. The numerical results presented, provide concrete evidence to substantiate the theoretical considerations introduced here. 

\subsection{Waveform Model Assessment: Analysis}
\label{subsec:BL_analysis}

The remnant velocity is computed using equations~\eqref{equ:v1}–\eqref{equ:v3}, while the gravitational memory is evaluated based on equations~\eqref{eq:lin_mem_def} and~\eqref{eq:nonlin_mem_def}, to assess the performance of the selected waveform approximants across the parameter space defined in Section~\ref{subsec:BL_events}. The influence of subdominant strain modes on both the kick and the inferred \gls{gw} memory is also investigated. Analysis of alignment-sensitive quantities is limited to non-precessing simulations. Prior to examining extensive datasets, the numerical pipeline is calibrated by comparing key physical quantities against metadata from the \SXS{} database. In particular, deviations in the magnitude of the kick computed via Eq. \eqref{equ:kick} are measured relative to the values reported in \SXS{}. The strong agreement observed in these comparisons supports the reliability of the balance laws as numerically implemented for the purposes of this Section.

\subsubsection{Mode Mismatch}
\begin{figure}
	\centering
 \includegraphics[width=0.7\linewidth]{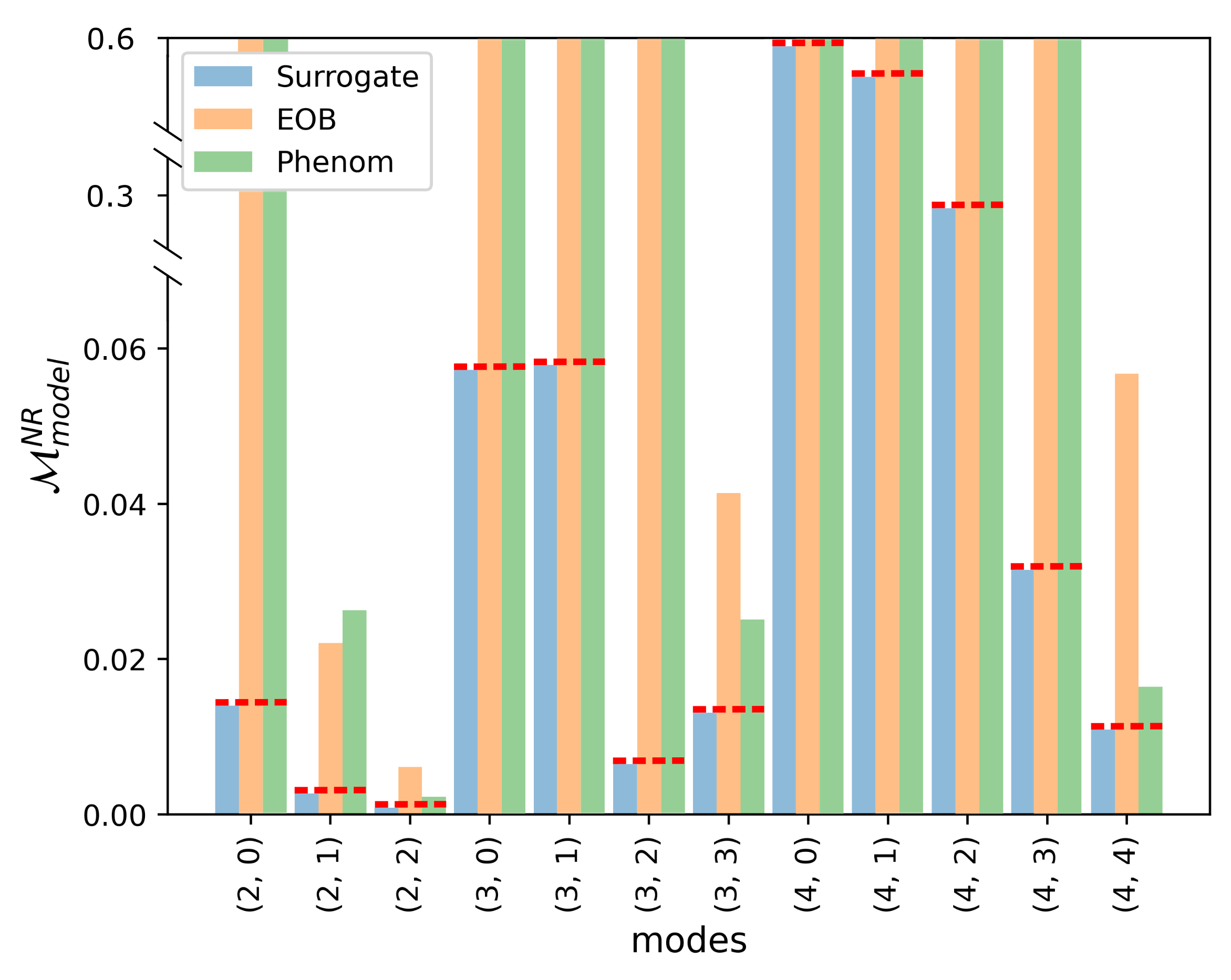}
	\caption{Mode mismatch w.r.t. NR, averaged over all non-precessing BBH simulations \cite{waveform_test_BL_I}. The mismatch axis is capped at the maximum value attained by \Surr{} among the selected modes. Modes exceeding the threshold are not produced by the algorithms of \EOB{} and \Phen{}, despite being accessible via their \LALSuite{} implementations. As such, their exact mismatch values are not meaningful. }
	\label{fig:mismatch_NR_np}
\end{figure}

The analysis begins with the non-precessing mergers described in Section~\ref{subsec:BL_events}, for which NR waveforms are available. A standard metric for assessing the similarity between waveforms (or individual modes) is the mismatch function defined in Eq. \eqref{equ:mismatch}. During the alignment procedure, the mismatch $\mathcal{M}$ is minimized for the dominant $h_{2,2}$ mode, while no direct mismatch minimization is applied to the remaining modes. Consequently, insights into the limitations of waveform models can be obtained by examining the residual mismatch in the subdominant modes and comparing it to the mismatch of the dominant $h_{2,2}$ mode. Figure~\ref{fig:mismatch_NR_np} displays the mismatch, computed via Eq. \eqref{equ:mismatch}, for the set of modes common to all waveform models\footnote{Modes are shown only up to $\ell = 4$ due to the limited mode content in \Surr{}.}, averaged across all simulations considered.
As anticipated, the lowest mismatch consistently occurs for the dominant $h_{2,2}$ mode across all approximants, with \Surr{} achieving the best performance by this metric. In contrast, the mismatch increases for subdominant $\ell$ modes. This increase is especially pronounced for the $\ell=4$ modes, as well as for $h_{2,0}$, $h_{3,2}$, $h_{3,1}$, $h_{3,0}$, and their complex conjugates, where mismatches are several orders of magnitude higher than for $h_{2,2}$.\\
The relatively poor agreement between \Surr{} and NR for subdominant modes is attributed to the model's focus on optimizing the accuracy of the dominant $h_{2,2}$ component within its interpolation scheme~\cite{Blackman:2017dfb,Blackman:2017pcm,Varma_2019:hyb,Varma:2019csw}.
As summarized in Table~\ref{table:1}, several subdominant modes are not modeled by \EOB{} and \Phen{}, despite being available through \LALSuite{}. As a result, their mismatch values are close to unity. For clarity in visualization, the $y$-axis in Figure~\ref{fig:mismatch_NR_np} is limited to the highest mismatch observed for \Surr{}. The absence of non-trivial contributions from these omitted modes carries over approximately to precessing simulations, even though internal modeling mechanisms—such as the twisting-up procedure implemented in \Phen{} models~\cite{IMRPhenomTPHM_paper}—can redistribute some mode power across subdominant components. Consequently, a similar trend in mode-by-mode mismatch, as depicted in Figure~\ref{fig:mismatch_NR_np}, is observed for precessing systems. It is important to note, however, that for precessing simulations, the mismatch calculated for a given $(\ell, m)$ mode may differ from that of its $(\ell, -m)$ counterpart, due to the inherent asymmetries present in both the \NR{} and \Surr{} waveforms. Note also that the \TEOB{} model is excluded from Figure~\ref{fig:mismatch_NR_np} since it provides only the $h_{2,2}$ mode. On average, the mismatch for this mode is only slightly higher for \TEOB{} compared to \EOB{}, confirming that \TEOB{} also serves as a highly accurate approximant for the dominant \gls{gw} strain mode. The agreement between \TEOB{} and \NR{} improves further when the analysis is restricted to the inspiral phase. This enhancement reflects the model's construction, which prioritizes high precision during the inspiral—based on NR calibration—while allowing for small deviations near the merger~\cite{Nagar_2018, Damour_2014_II}.

A similar analysis was performed for non-cataloged merger simulations, i.e., waveforms lacking corresponding NR data. In this case, \Surr{} was selected as the reference model. The results reveal a comparable trend in the mismatch behavior of \EOB{} and \Phen{} relative to \Surr{}. For the $\ell=2$ modes, both \EOB{} and \Phen{} show low mismatch values w.r.t. \Surr{}. However, for higher-order modes such as $h_{3,3}$ and $h_{4,4}$, \EOB{} waveforms display noticeably larger discrepancies from \Surr{} than those produced by \Phen{}.\\
Beyond intrinsic differences in waveform modes, the mismatch may also be affected by a broad class of nonphysical features referred to here as “numerical artifacts.” An example of such an artifact is illustrated in Appendix A of \cite{waveform_test_BL_I}. These artifacts, which are particularly pronounced in higher-order modes, warrant separate consideration. In prior studies, some of these effects have been identified and partially mitigated—for example, through the alignment of BMS charges~\cite{Mitman:2022kwt, Mitman:2024uss}. Although the present analysis does not explore the origin or detailed consequences of these artifacts, it is important to emphasize that, even with a well-defined alignment procedure, numerical factors inherent in waveform generation can influence mismatch outcomes.
Despite the elevated mismatch observed in certain modes, the complete waveforms remain sufficiently accurate approximations to NR, primarily due to the dominance of the well-modeled $h_{2,2}$ mode across most BBH simulations. At the current level of observational precision, discrepancies in individual subdominant modes are therefore of secondary concern. Nonetheless, with the advent of future GW observatories, the required precision is expected to increase substantially, potentially elevating the importance of higher-order modes~\cite{pitte_detectability_2023}. Accordingly, it is essential for next-generation GW waveform models to address the challenges highlighted in Figure~\ref{fig:mismatch_NR_np}, including incomplete mode coverage and reduced accuracy in subdominant modes. The following Section lays the foundation for tackling these limitations by analyzing alignment-independent physical observables of BBH systems, derived exclusively from the asymptotic strain and its harmonic decomposition.

\subsubsection{Computing Physical Quantities}
To assess the influence of an approximant's mode content on overall waveform fidelity, relevant physical quantities—derived from the strain—are computed for BBH mergers using two distinct sets of strain modes. The first set comprises all modes available for each individual model. The second set includes only those subdominant modes associated with low mismatch, specifically excluding modes from the set\footnote{Modes with $\ell > 4$, which are beyond the mode content of the \Surr{} model, are not excluded for the remaining models (including NR).} \begin{equation*}
H_{\text{sub}} = \{h_{2,0}, h_{3,\pm 2}, h_{3,\pm 1}, h_{3,0}, h_{4,\pm 3}, h_{4,\pm 2}, h_{4,\pm 1}, h_{4,0}\}\,. 
\end{equation*}
By excluding $H_{\text{sub}}$, all waveform approximants are placed on approximately equal footing, enabling a more consistent and quantitative comparison based solely on the modes that are robustly simulated across all models.

\subsubsection{Remnant Velocity}
\begin{figure}[!t]\centering
   \includegraphics[width=0.6\columnwidth]{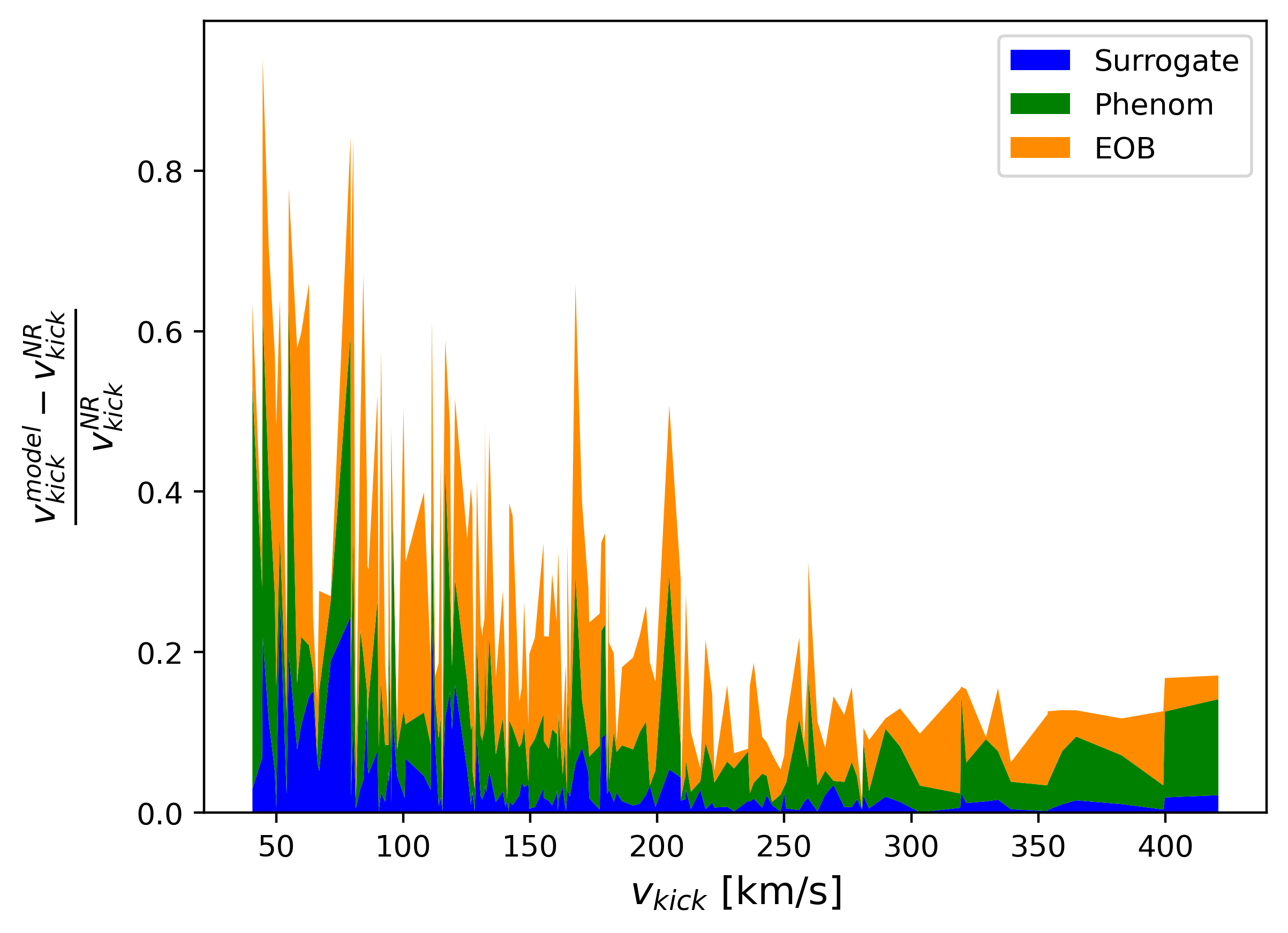}    
\includegraphics[width=0.5\columnwidth]{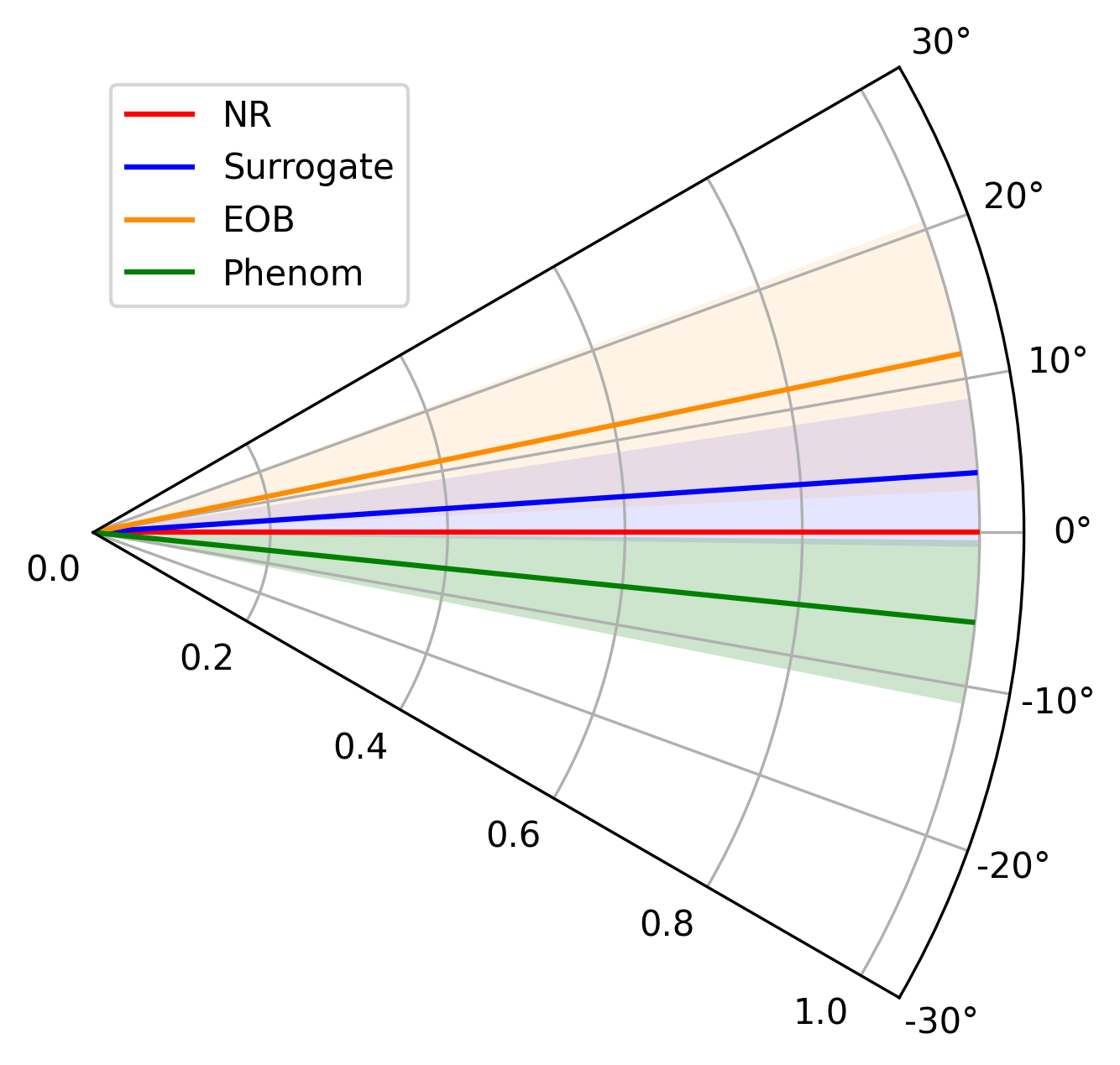}   
     \caption{Relative error of the kick magnitude (top) and absolute deviation angle (bottom) of the approximants w.r.t. NR for non-precessing merger simulations \cite{waveform_test_BL_I}. The top stack plot shows the relative errors for each individual waveform labeled by its resulting kick magnitude. The bottom plot illustrates the deviation in kick direction w.r.t. the orientation predicted by NR, averaged over all simulated waveforms. The shaded regions mark a $1\sigma$-interval above and below the mean. Despite displaying the mean absolute deviation, for better readability, the presented value for \Phen{} is equipped with an artificial negative sign.}
       \label{fig:error_kick_NR_np}
\end{figure}

For the selected set of modes, the analysis begins with the recoil velocity of the remnant BH from non-precessing simulations. The focus is restricted to mergers yielding kick velocities in the range of $20$–$400$ km/s; simulations with negligible recoil ($v < 20$ km/s) are excluded. The magnitude and direction of the remnant velocity vectors, computed via Eq. \eqref{equ:kick}, are compared by evaluating the relative error of each model w.r.t. NR. Using the full mode content, the relative error in kick magnitude and the directional deviation are presented in Figure~\ref{fig:error_kick_NR_np}. The upper panel stacks the relative errors against the corresponding kick magnitudes across all simulations. The lower panel displays the average absolute deviation in kick direction compared to NR, with negative values assigned to \Phen{} for visual clarity. The shaded regions denote the $1\sigma$ interval around the mean, where $\sigma$ is the standard deviation of the directional deviations.
Quantitatively, Figure~\ref{fig:error_kick_NR_np} confirms the strong agreement between \Surr{} and NR. In particular, for higher kick velocities, the relative errors are minimal and the directional deviation remains below $5^\circ$ on average. In contrast, \EOB{} and \Phen{} exhibit significantly larger errors and cannot reproduce the NR reference velocity vectors with comparable precision. Both models show increased relative errors at lower recoil velocities, along with strong fluctuations indicative of sensitivity to the simulation’s intrinsic parameters. Across the parameter space, \EOB{} yields the largest average errors in both kick magnitude and direction. These results are consistent with prior studies of remnant recoil (see Figures 4 and 5 in~\cite{waveform_test_BL_V}), though differences in waveform model versions and parameter coverage should be noted. For \TEOB{}, the computed kick velocities are trivial throughout the examined space, a direct consequence of its exclusive modeling of the dominant $h_{2,2}$ strain mode, as will be further elaborated below.

When computing the same statistics while excluding the high-mismatch subdominant modes $H_{\text{sub}}$, neither the magnitude nor the direction of the remnant kick shows any substantial deviation from the values obtained using the full mode content. For both NR and \Surr{}, the average change in directional deviation and relative error in kick magnitude amounts to only a few percent of the corresponding values based on the complete set of modes. Similarly, for \EOB{} and \Phen{}, no statistically significant differences are observed between the two mode selections. This conclusion also holds for non-cataloged, non-precessing merger simulations.
These findings indicate that key recoil-related quantities, such as the remnant BH's velocity, are largely insensitive to the specific subdominant strain modes included in a given GW waveform model. The same conclusion applies to the computation of the remnant \gls{bh}'s final mass (see Eq. \eqref{equ:Energy}). This insensitivity to the $H_{\text{sub}}$ mode set is consistent with the analytical structure of the $\alpha_{\ell,m}$ coefficients given in Eq. \eqref{equ:alphass}, and their role in determining the recoil components via equations~\eqref{equ:v1}–\eqref{equ:v3}:
The Wigner-$3j$ symbols in Eq. \eqref{equ:alphass} selectively isolate specific mode pairs of the form $h_{\ell_1, m_1}\bar{h}_{\ell_2, m_2}$\footnote{Time derivatives are omitted here for brevity.} that contribute to each $\alpha_{\ell m}$ coefficient. These $\alpha_{\ell m}$ coefficients, in turn, govern the computation of physical observables such as the recoil (kick) velocity. As given in equations~\eqref{equ:v1}–\eqref{equ:v3}, the relevant coefficients for the kick are $\alpha_{1,\pm1}$ and $\alpha_{1,0}$: the in-plane components of the recoil are determined by $\alpha_{1,\pm1}$, while $\alpha_{1,0}$ controls the out-of-plane contribution. As previously noted, the latter becomes significant only in the presence of precession.\\
Among all strain mode pairs contributing to the kick coefficients, the most significant terms involve at least one dominant strain mode, $h_{2,\pm2}$. Notably, terms proportional to $h_{2,2}\bar{h}_{2,2}$ appear only in $\alpha_{1,0}$, where they cancel against the corresponding terms involving $h_{2,-2}\bar{h}_{2,-2}$. On the other hand, the dominant contributions to the in-plane kick arise from interference between $h_{2,\pm2}$ and subdominant modes, such as $h_{2,\pm2}\bar{h}_{2,\pm1}$ in $\alpha_{1,\pm1}$. In contrast, $\alpha_{1,0}$ receives no non-vanishing contributions from any combination of two $\ell=2$ modes, providing an analytic explanation for the typically negligible out-of-plane kick component in non-precessing BBH mergers. Naturally, this conclusion does not extend to precessing systems.
Within $\alpha_{1,\pm1}$, the dominant $h_{2,2}$ mode also appears in products like $h_{2,\pm2}h_{3,\pm1}$ and $h_{2,\pm2}h_{3,\pm3}$, but these terms cancel out in the summation~\eqref{equ:alphass}. Furthermore, when forming the $v_x$ and $v_y$ components of the kick via $\alpha_{1,1} \pm \alpha_{1,-1}$ in Eq. \eqref{equ:kick}, the resulting directional contributions are primarily governed by terms proportional to $h_{2,\pm2}h_{2,\pm1}$. Due to the symmetry of these combinations, any phase offset applied to $h_{2,\pm2}$ enters both $\alpha_{1,\pm1}$ equally and cancels out in the magnitude of the in-plane recoil. Therefore, the in-plane kick magnitude remains invariant under phase shifts in $h_{2,\pm2}$.
Taken together, these arguments show that, across the BBH simulations considered, the dominant contributions to the $\alpha$ coefficients governing the in-plane kick are proportional to $h_{2,\pm1}$—modes not included in $H_{\text{sub}}$. Consequently, omitting the $H_{\text{sub}}$ modes has only a minimal effect on the kick’s magnitude and direction. An analogous conclusion holds for the final mass of the remnant, which depends on $\alpha_{0,0}$.
\begin{figure}
	\centering
	\includegraphics[width=0.7\linewidth]{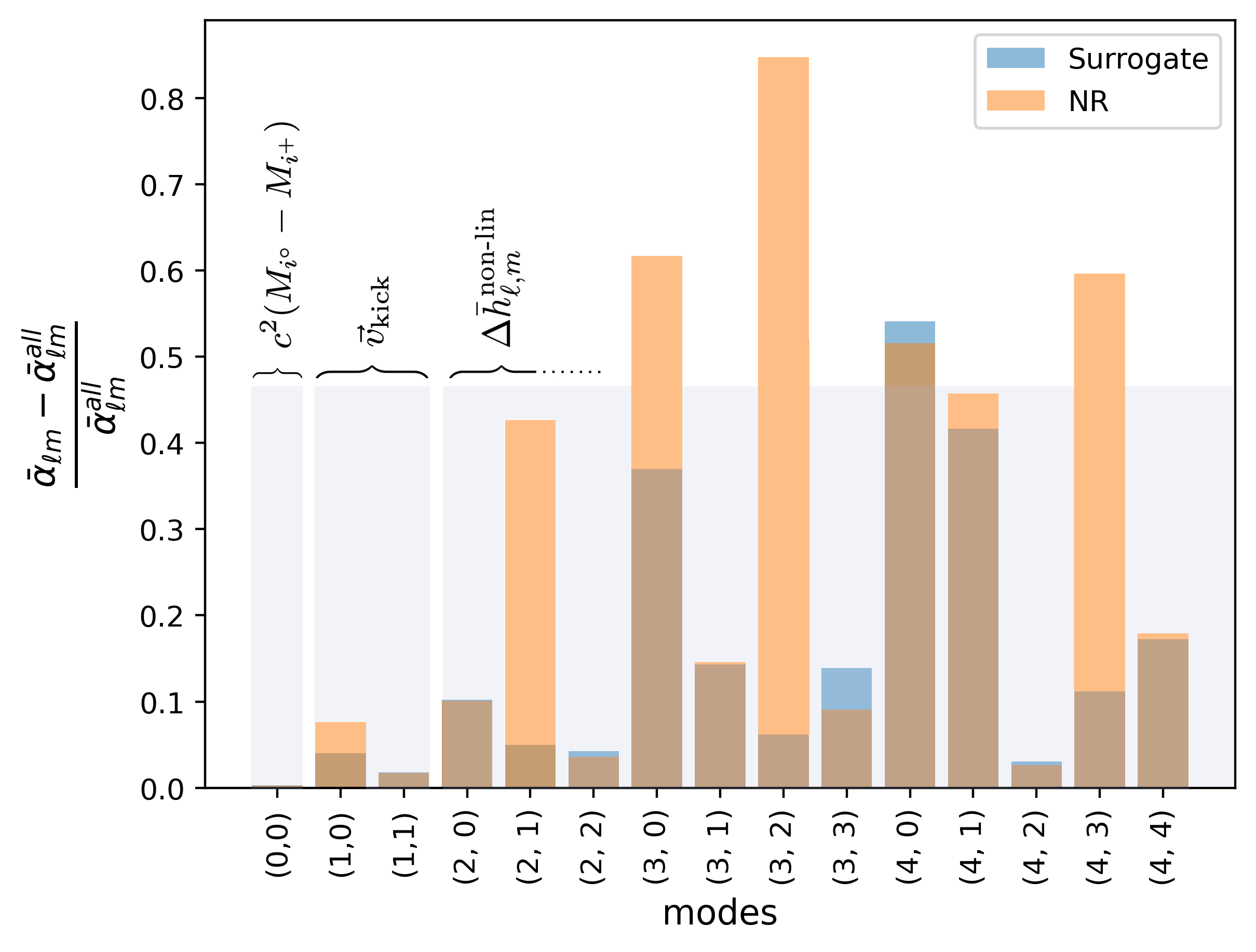}
	\caption{Time-integrated, relative error for $\alpha_{\ell m}$ \eqref{equ:alphass} \cite{waveform_test_BL_I}. The error compares the $\alpha$ computed including all strain vs excluding $H_\text{sub}$ for NR and \Surr{}. The modes $(\ell,m)$ relevant for energy, kick, and non-linear memory are indicated correspondingly.}
	\label{fig:alphas_rel}
\end{figure}

The qualitative assessment of mode contributions to the recoil is quantitatively supported by Figure~\ref{fig:alphas_rel}, which shows the relative error in the $\alpha$-coefficients for NR and \Surr{} waveforms, comparing results with and without the inclusion of modes in $H_{\text{sub}}$\footnote{\EOB{} and \Phen{} are omitted from this comparison, as they do not include the modes in $H_{\text{sub}}$ to begin with.}. The figure clearly demonstrates that the variation in the $\alpha$-coefficients relevant to the kick velocity is minimal. The relatively higher error observed for NR is attributable to its extended mode content ($\ell \leq 8$) compared to \Surr{} ($\ell \leq 4$), resulting in a substantially larger number of contributing mode-mixing terms in the summation for $\alpha_{1,\pm1}$. As a result, omitting $H_{\text{sub}}$ affects $\alpha^{\text{NR}}_{\ell m}$ more significantly than $\alpha^{\text{Surr}}_{\ell m}$.
In summary, the limited influence of subdominant modes on the kick velocity arises from the constrained set of allowed mode couplings in the $\alpha_{\ell=1,m}$ coefficients. These restrictions are imposed by the Wigner-$3j$ selection rules in Eq. \eqref{equ:alphass}, which permit mode mixing only when $|\ell_1 - \ell_2| \leq \ell$. Consequently, the dominant $h_{\ell_1 = 2,2}$ mode cannot couple to higher-order modes beyond a certain threshold, thereby limiting its contribution to the recoil. However, this restriction is no longer applicable in the context of computing \gls{gw} memory effects, where different couplings become relevant.

\subsubsection{Memory Components}
\begin{figure}\centering
   \includegraphics[width=0.58\columnwidth]{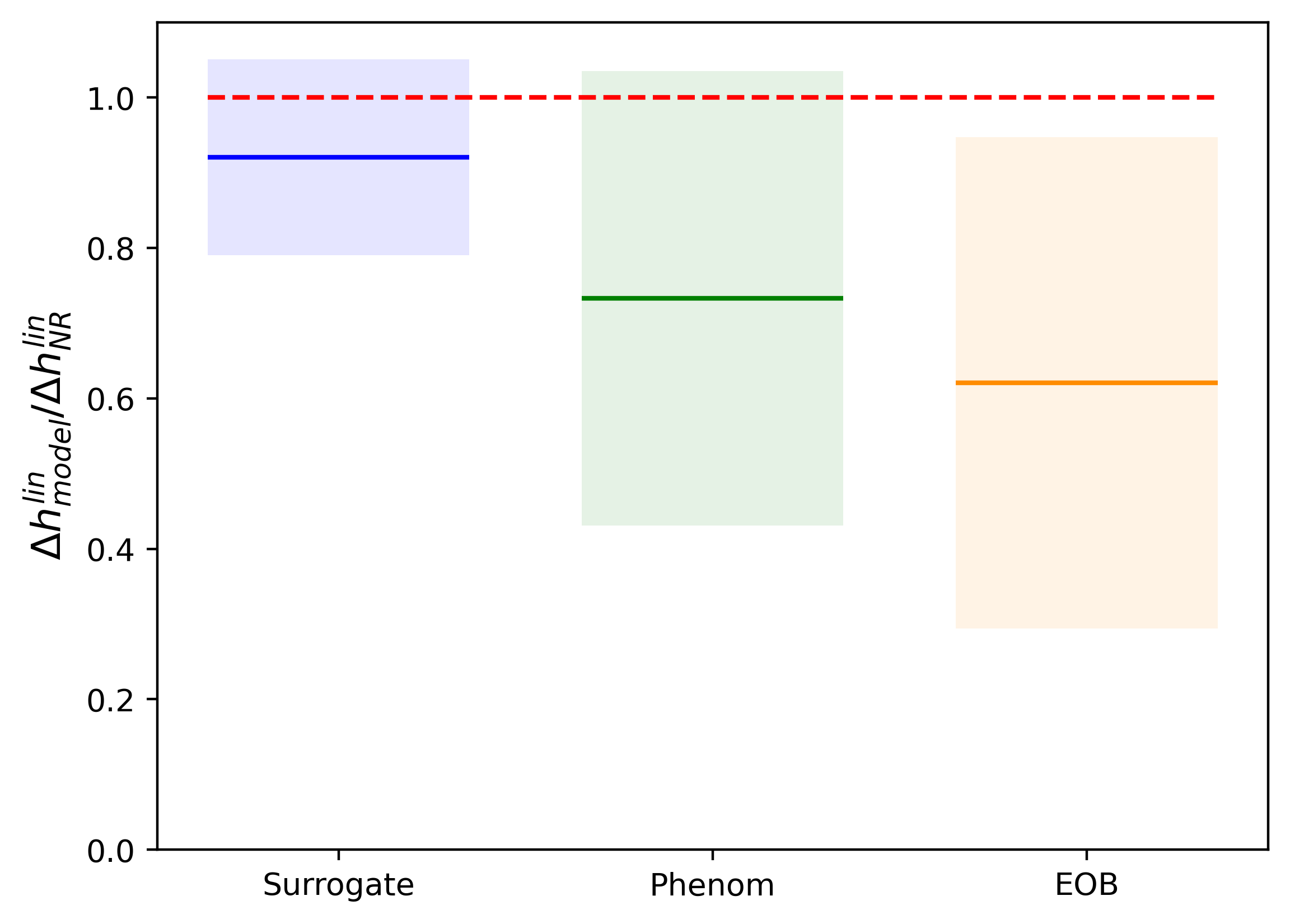}    
\includegraphics[width=0.6\columnwidth]{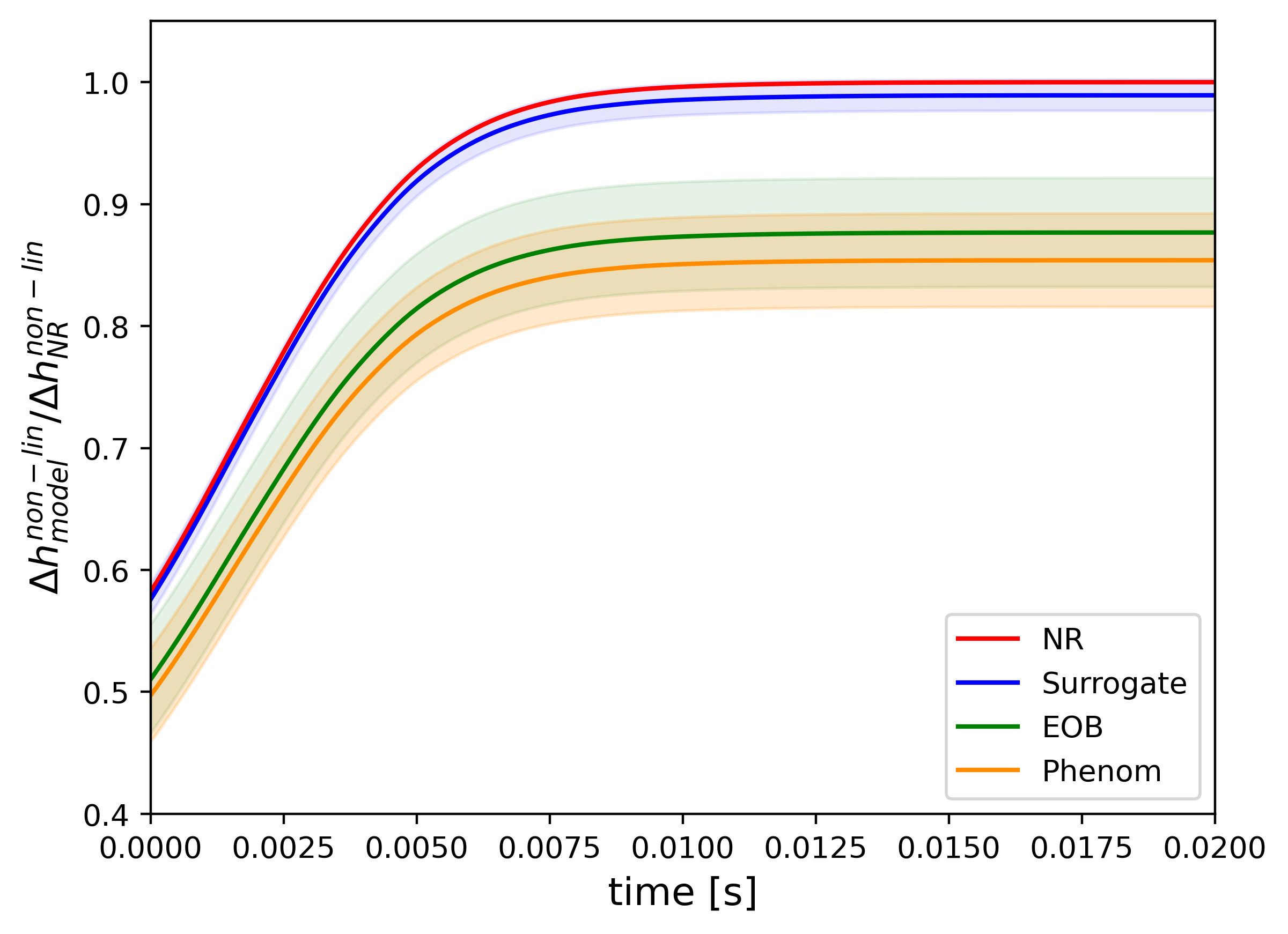} 
\includegraphics[width=0.6\columnwidth]{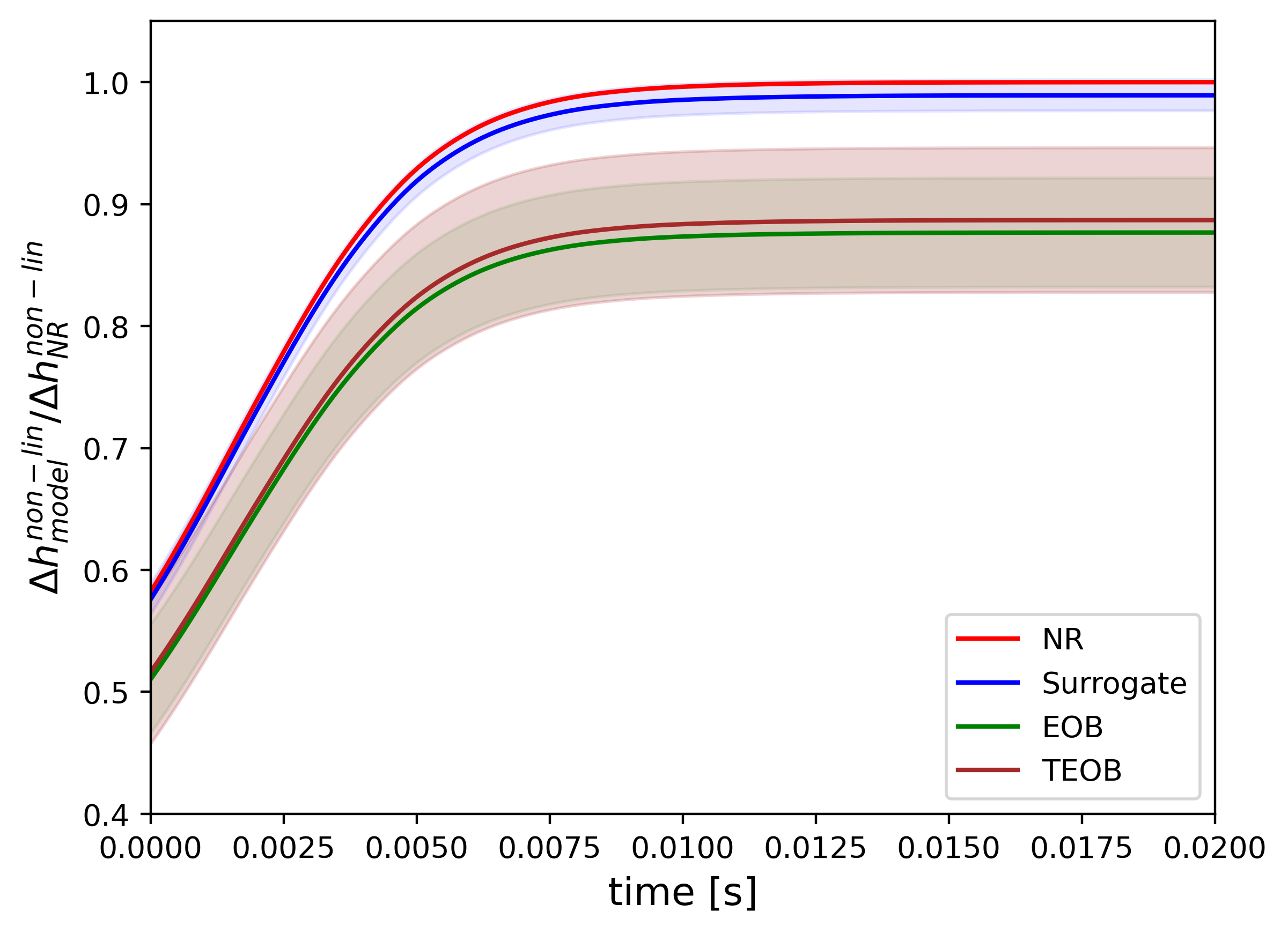}   
     \caption{Deviation of waveform models' linear (top) and non-linear (middle) memory w.r.t. NR \cite{waveform_test_BL_I}. Corresponding $1\sigma$-intervals above and below the mean are shaded with the corresponding coloring. For the non-linear memory the plot displays a time series. Both memory contributions are normalized to the NR value and averaged over all simulated mergers. The bottom plot shows again the non-linear memory when replacing \Phen{} with \TEOB{}. For better readability, two plots for the non-linear memory are provided.}
       \label{fig:memory_error}
\end{figure}

The linear and non-linear \gls{gw} memory contributions are computed and decomposed according to Eq. \eqref{eq:mem_total_def}. Figure~\ref{fig:memory_error} presents the relative errors for both memory components in non-precessing BBH mergers, using the full mode content of each waveform model. In the upper panel, the total linear memory is shown, normalized to the NR result for each model. The shaded regions denote the $1\sigma$ interval around the mean memory error across the ensemble of simulations, resulting in a total width of $2\sigma$. Solid lines indicate the average memory ratio $\Delta h_\text{model} / \Delta h_\text{NR}$. The lower panel displays the corresponding quantities for the non-linear memory, plotted as a time series to illustrate error accumulation during the gradual memory build-up. The merger occurs at $t = 0$, and normalization is performed relative to the time-integrated non-linear NR memory for each simulation.\\
In both the linear and non-linear cases, the \EOB{} and \Phen{} approximants perform significantly worse than the \Surr{} model. These two models consistently underestimate the memory contributions, with NR values lying outside their $1\sigma$ intervals, particularly in the non-linear regime. By contrast, \Surr{} yields accurate predictions, especially for the dominant non-linear memory component. In the lower panel, \Phen{} is replaced with \TEOB{} to highlight its ability to model non-linear memory despite containing only the dominant strain mode. However, similar to its kick velocity prediction, \TEOB{} produces negligible linear memory, owing to the absence of modes such as $h_{2,\pm1}$ that are essential for this contribution.
Modeling the linear memory is particularly challenging, as it requires access to additional Newman-Penrose scalars not typically provided by most waveform approximants\footnote{An exception is the SXS Collaboration’s Cauchy-Characteristic Extraction (CCE) catalog~\cite{Moxon_2020}. Similar methods have been adopted in other modeling approaches, e.g.,\cite{Surrogate_CCE}.}. This difficulty can be partially circumvented by extending the time integrals from past to future null infinity, as in Eq. \eqref{equ:fullDimBL}. While this method introduces a small systematic bias due to the finite duration of simulated waveforms, the error affects all models equally and is irrelevant in the relative error analysis. A more rigorous treatment of the linear memory can be found in~\cite{waveform_test_BL_II}. Additionally, the linear memory is several orders of magnitude smaller than its non-linear counterpart, making it inherently more sensitive to numerical noise. As a result, large fluctuations are observed in the linear memory estimates.

\begin{figure}\centering
\includegraphics[width=0.7\columnwidth]
{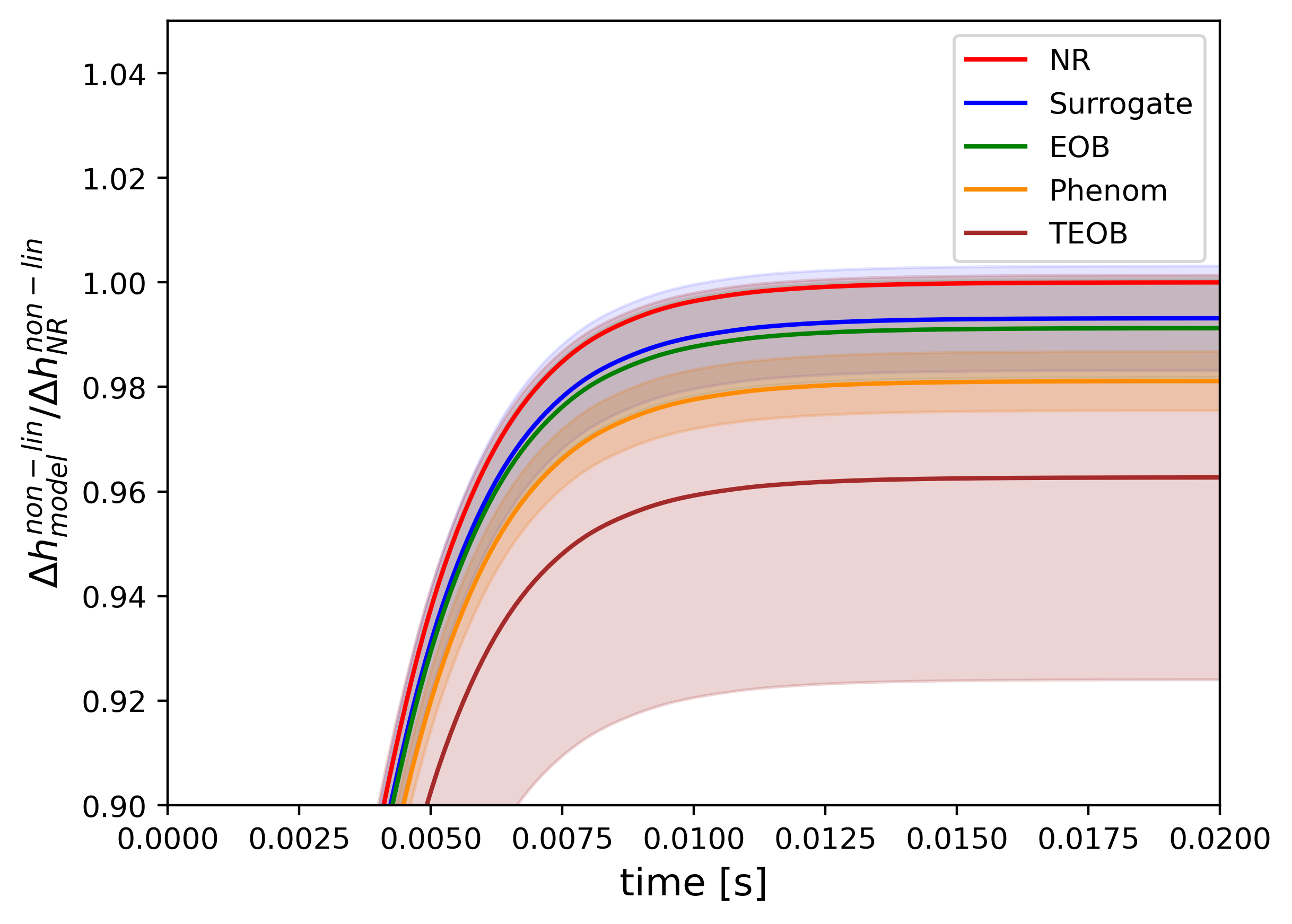}  
     \caption{Non-linear memory as depicted in Figure~\ref{fig:memory_error}, here computed excluding $H_{\text{sub}}$ \cite{waveform_test_BL_I}.}
    \label{fig:memory_error_II}
\end{figure}
As with the kick velocity, the \gls{gw} memory can also be evaluated while excluding the set of subdominant harmonic strain modes, $H_{\text{sub}}$. For the linear memory component, this exclusion leads to only a minor change in the contributions predicted by each waveform model. This result is expected, given that the linear memory—per Eq. \eqref{eq:lin_mem_def}—is entirely determined by the remnant velocity. This dependency originates from the specific approximation employed in the derivation of the balance laws used here, particularly in Eq. \eqref{equ:fullDimBL}, where the original formulation based on Newman-Penrose scalars is, under certain conditions, replaced with expressions involving the remnant velocity\footnote{For a detailed discussion of the relevant transformations and physical assumptions, see~\cite{waveform_test_BL_III}.}.
As a consequence, because the linear memory is dictated by the kick, the discussion surrounding the relevant strain modes closely mirrors that of the kick velocity presented in the previous Subsection. Notably, both the kick and the linear memory are alignment-dependent quantities. This alignment sensitivity suggests that some of the discrepancies observed across waveform models in Figure~\ref{fig:memory_error} may be partially attributed to suboptimal alignment choices rather than fundamental modeling errors.
In contrast, the non-linear memory responds quite differently when the subdominant mode set $H_{\text{sub}}$ is excluded. As illustrated in Figure~\ref{fig:memory_error_II}, the relative errors across all waveform approximants tend to converge toward the NR baseline once the mode content is limited to a common subset. While the absolute non-linear memory values for NR and \Surr{} decrease by roughly $\mathcal{O}(10)$ percent, the values for \EOB{}, \TEOB{}, and \Phen{} remain unchanged, as these models do not include the excluded modes in the first place. Consequently, upon exclusion of $H_{\text{sub}}$, the non-linear memory estimates of all models become approximately consistent. By the formulation of the balance laws, this also implies comparable energy flux predictions.
Unlike the linear component, the non-linear memory is not sensitive to the choice of reference phase and is thus independent of the alignment procedure. Furthermore, a noticeable reduction in the variance of relative errors is observed for \EOB{} and \Phen{}, stemming from the reduced number of NR modes contributing to the memory computation, which in turn limits fluctuations in comparisons with these models. This trend, however, is not evident for \TEOB{}, where notable discrepancies persist due to residual subdominant strain modes that are present in the NR data but absent from both \TEOB{} and the reduced mode set $H_{\text{sub}}$.

As anticipated by prior studies (e.g., \cite{waveform_test_BL_V}; see also \cite{Khera:2020mcz}), both the \EOB{} and \Phen{} waveform models do not capture the full information content present in \NR{} simulations, primarily due to their limited set of harmonic modes. Nonetheless, when the analysis is restricted to the strain modes these models actually generate, their non-linear memory estimates closely match the reference data. This indicates that, within the shared harmonic content, these models reliably reproduce the non-linear memory.
A direct comparison between Figures~\ref{fig:memory_error} and~\ref{fig:memory_error_II} reveals that the discrepancy in memory-related information between \EOB{} and \Phen{} on one side, and NR or \Surr{} on the other, is largely attributable to the exclusion of subdominant modes, $H_{\text{sub}}$. To make this connection explicit, the analysis revisits the mixing of strain modes in the $\alpha_{\ell m}$ coefficients defined in Eq. \eqref{equ:alphass}.\\
According to Eq. \eqref{eq:nonlin_mem_def}, the total non-linear memory consists of all $\alpha$-coefficients with $\ell > 1$, among which $\alpha_{2,0}$ is the dominant contributor. This prominence stems from the allowed mode mixing in the $\alpha_{\ell m}$ terms. As previously discussed, the dominant strain modes $h_{2,2}$ and $h_{2,-2}$ can contribute meaningfully only to coefficients with $m = 0$, such as $\alpha_{2,0}$ and $\alpha_{4,0}$, owing to the selection rules imposed by the Wigner-$3j$ symbols in Eq. \eqref{equ:alphass}. Importantly, the self-coupling terms $h_{2,2}h_{2,2}$ and $h_{2,-2}h_{2,-2}$ cancel for all values of $\ell$ except $\ell = 2$ and $\ell = 4$ in the non-precessing case. Even then, the contribution of $\alpha_{4,0}$ is heavily suppressed—by about $\mathcal{O}(10^{-2})$—making it effectively negligible.
Consequently, $\alpha_{2,0}$ alone accounts for roughly 96\% of the total non-linear memory on average\footnote{This dominance reflects the fact that the majority of \gls{gw} is radiated azimuthally symmetrically, primarily via the $h_{2,\pm2}$ modes. The resulting memory signal manifests in the $h_{2,0}$ harmonic.}.
While the largest contribution to $\alpha_{2,0}$ indeed originates from the interaction of the dominant $h_{2,2}$ and $h_{2,-2}$ modes, a substantial portion also arises from couplings between subdominant modes and $h_{2,\pm2}$. Furthermore, all other $\alpha_{\ell m}$ terms—with $\ell > 1$—likewise include contributions from such mixed-mode interactions. In contrast to the case of the kick velocity, where subdominant modes play only a marginal role, the non-linear memory incorporates a wider range of $\alpha$-coefficients that permit more extensive mode mixing. This is due to the less stringent selection rules of the Wigner-$3j$ symbols, which allow for broader coupling possibilities as the number of strain modes increases. As a result, the non-linear memory becomes a highly sensitive and robust metric for evaluating the completeness and fidelity of waveform models across all available harmonic content.
This sensitivity is clearly reflected in the disparity between Figures~\ref{fig:memory_error} and~\ref{fig:memory_error_II}, where the non-linear memory differences can largely be traced back to couplings between $h_{2,2}$ and subdominant modes within the higher-order $\alpha_{\ell m}$ terms. Removing these subdominant modes—absent in \EOB{} and \Phen{}—from the NR and \Surr{} data effectively reduces the memory to the $h_{2,\pm2}h_{2,\pm2}$ contributions. Given that the dominant modes are accurately captured by all approximants, the resulting memory estimates become consistent, with only minor mismatches—clearly illustrated in Figure~\ref{fig:memory_error_II}.
The dominant role of $h_{2,\pm2}h_{2,\pm2}$ interactions in the non-linear memory computation is further underscored by the performance of the \TEOB{} model. As shown in Figures~\ref{fig:memory_error_II} and~\ref{fig:memory_error}, \TEOB{}, which includes only the $h_{2,\pm2}$ modes, produces non-linear memory estimates comparable to those of other models with more extensive harmonic content. This outcome reinforces the critical influence of these dominant-mode couplings under mode-restricted conditions.

\begin{figure}\centering
\includegraphics[width=0.65\columnwidth]
{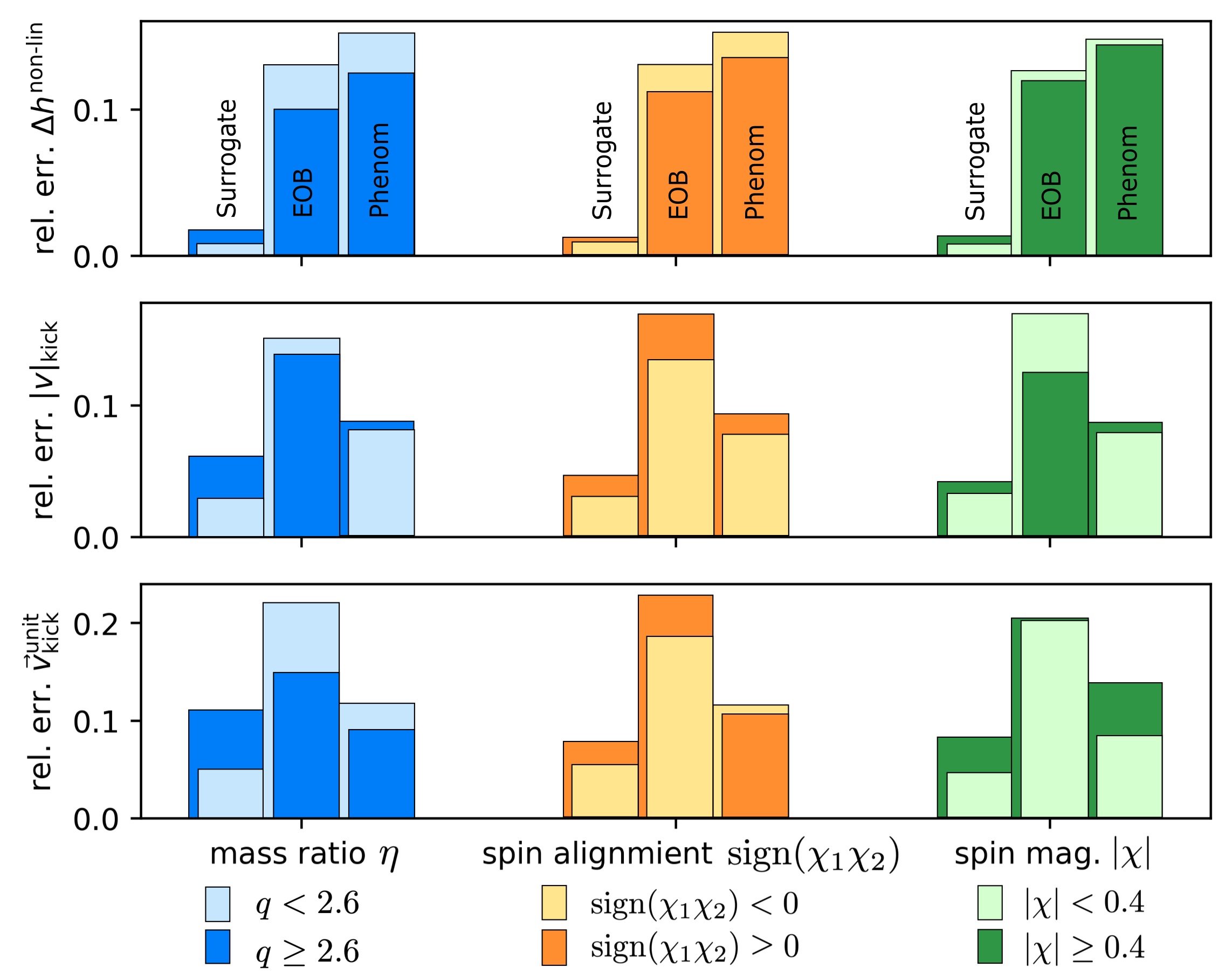}  
     \caption{Relative error for kick direction, magnitude, and non-linear memory compared to NR waveforms and averaged over all simulations selected by the displayed constraint ($x$-axis) \cite{waveform_test_BL_I}. For each value and model, two error bars corresponding to the categorizations below the $x$-axis are displayed. Smaller errors are moved to the front. To convert to the mass parameter $\eta$ used in \ref{subsec:BL_events}, one uses $\eta = q /(1+q)^2$, where $q = M_1/M_2$ and $M_1$ being the heavier of the BBH system.}
    \label{fig:param_error_dist}
\end{figure}

Prior to analyzing precessing systems, the dependence of relative errors in key physical properties of BBH mergers on intrinsic parameters of non-precessing binaries—namely, the symmetric mass ratio $\eta$, spin magnitudes $|\chi|$, and spin alignments—is examined. The results are summarized in Figure~\ref{fig:param_error_dist}, which displays relative error bars for each quantity of interest, grouped according to the relevant parameter. For each parameter, the data are divided into two subsets, separated at thresholds chosen to approximately split the simulation ensemble evenly.
In the case of mass ratio, the threshold is set at $q = M_1/M_2 = 2.6$ (corresponding to $\eta \approx 0.2$), allowing a clear comparison between low- and high-mass-ratio regimes. Among all waveform models, the \Surr{} approximant exhibits the strongest parametric dependence, with significantly larger relative errors at higher mass ratios ($q \geq 2.6$), as elaborated in Section~\ref{subsec:BL_events}. Weaker, though still discernible, trends are also observed for spin alignment and spin magnitude: \Surr{} generally performs better for anti-aligned configurations with lower spin magnitudes ($|\chi| < 0.4$), particularly in relation to the kick velocity and non-linear memory.\footnote{Linear memory is not included in this analysis, as it is directly determined by the remnant velocity; see Section~\ref{subsec:BL_analysis_pre}.}
By contrast, the \EOB{} model shows a consistent trend only with mass ratio, achieving better accuracy in the high-$q$ regime. No statistically significant dependence is found w.r.t. spin alignment or magnitude. The \Phen{} model likewise displays no clear trends with any of the examined parameters, suggesting a more uniform error distribution across the parameter space, albeit not necessarily indicative of greater accuracy.

\begin{figure}[!t]\centering
\includegraphics[width=.7\columnwidth]{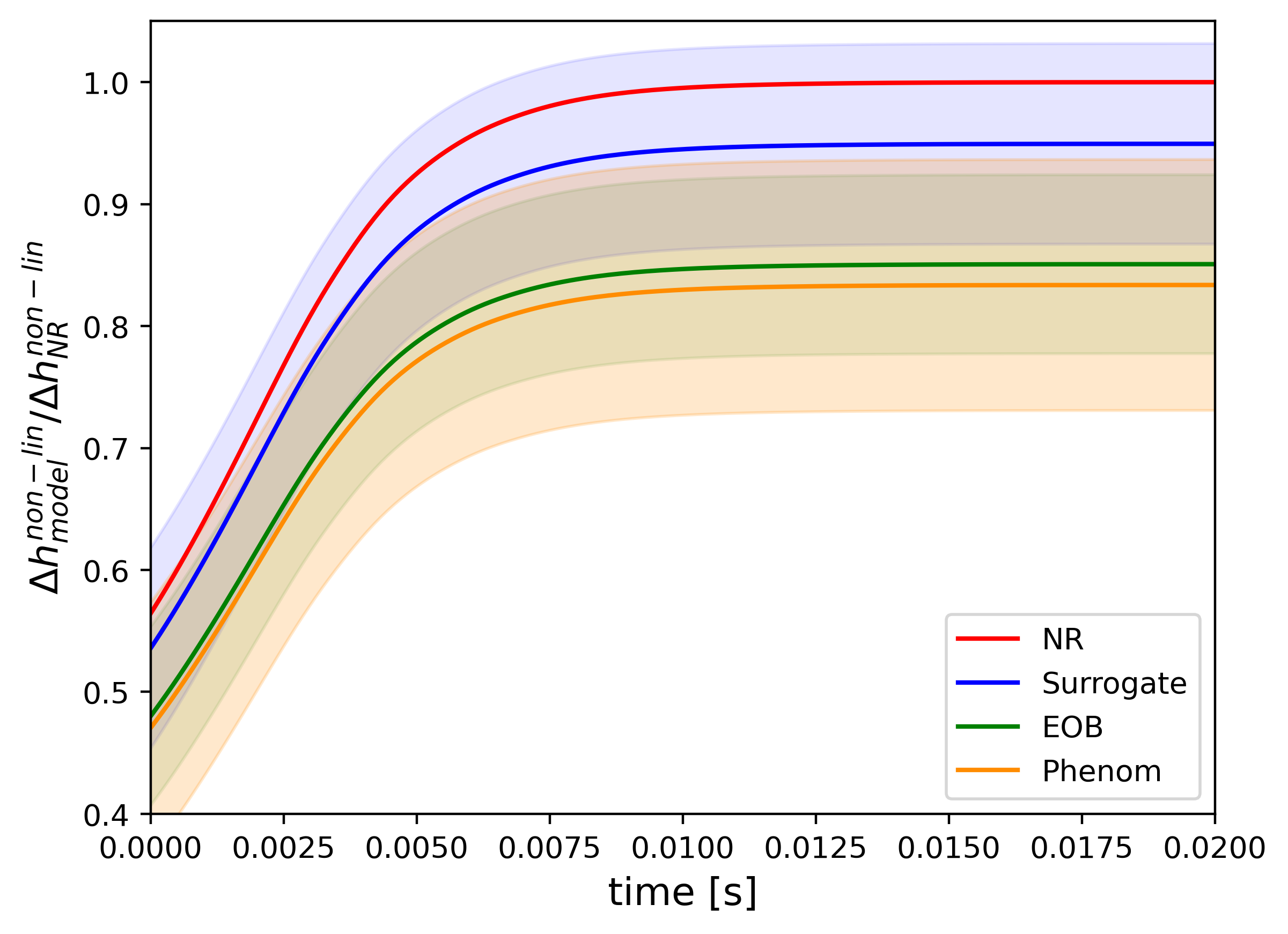}  
\includegraphics[width=.7\columnwidth]{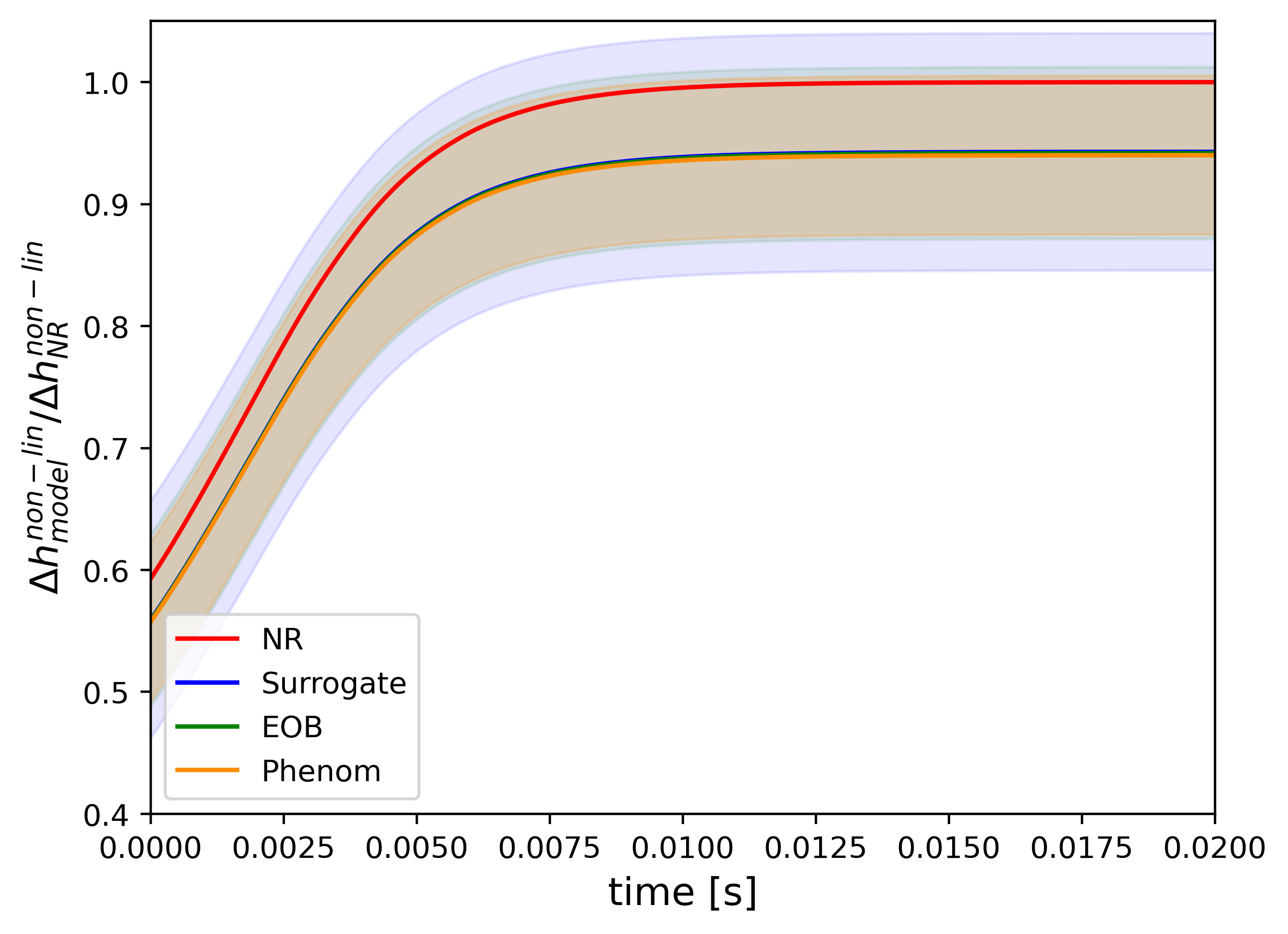}   
     \caption{Non-linear memory visualized as in Figure~\ref{fig:memory_error}, here computed excluding $H_{\text{sub}}$, for precessing mergers \cite{waveform_test_BL_I}.}
       \label{fig:memory_error_III}
\end{figure}

\subsubsection{Precessing merger simulations}
In the context of precessing binary \gls{bh} mergers, comparable results are obtained for the non-linear memory, as illustrated in Figure~\ref{fig:memory_error_III}. The top panel presents the time series of average relative errors for the full harmonic mode content, with normalization and standard deviation shading computed analogously to Figure~\ref{fig:memory_error}. Both \EOB{} and \Phen{} show deviations from NR consistent with those observed in the non-precessing case. In contrast, \Surr{} exhibits substantially greater fluctuations, resulting in a noticeably larger standard deviation relative to its performance in non-precessing simulations.
These elevated fluctuations persist even when the subdominant modes $H_{\text{sub}}$ are excluded, as seen in the bottom panel of Figure~\ref{fig:memory_error_III}. Despite this, \Surr{} continues to outperform both \EOB{} and \Phen{} in terms of overall accuracy in non-linear memory predictions. When $H_{\text{sub}}$ is disabled, the non-linear memory contributions from \Surr{}, \EOB{}, and \Phen{} become nearly identical. Notably, \Surr{} exhibits the widest standard deviation interval in this configuration.
Although the non-linear memory estimates from all approximants become comparable, no convergence to the NR baseline is observed, in contrast to the non-precessing scenario depicted in Figure~\ref{fig:memory_error_II}. Instead, all models settle around 90\% of the total NR non-linear memory on average. This suggests that, in the precessing case, differences between \Surr{}, \EOB{}, and \Phen{} are primarily driven by the presence or absence of modes in $H_{\text{sub}}$. Nevertheless, even though \Surr{} uniquely captures the asymmetry between $(\ell, m)$ and $(\ell, -m)$ modes, all models appear to lack information beyond the content of $H_{\text{sub}}$, as implied by the residual gap with NR in the lower plot of Figure~\ref{fig:memory_error_III}.
Consistent with prior discussions on waveform alignment for precessing systems, it is important to note that the contribution of alignment choices to the observed deviations in Figure~\ref{fig:memory_error_III} remains unquantified. However, the roughly uniform offset exhibited by all models argues against alignment procedures as the primary source of the discrepancy. Since each waveform model is aligned independently, alignment-related inaccuracies would be expected to manifest unevenly. An alternative explanation for the collective deviation from \SXS{}'s non-linear memory is its significantly richer harmonic mode content (see Table~\ref{table:1}). Nevertheless, truncating the \SXS{} strain data at $\ell \leq 4$ in the statistical analysis underlying Figure~\ref{fig:memory_error_III} does not significantly narrow the gap between it and the other waveform models.

The results presented in Figure~\ref{fig:param_error_dist} are further evaluated in the context of precessing binary \gls{bh} mergers. As in the non-precessing case, this analysis focuses exclusively on the non-linear memory component. The spin alignment marker is omitted, as it lacks physical meaning for precessing systems. To maintain a balanced division of the simulation dataset, the parameter thresholds are slightly adjusted: the mass ratio boundary is set at $q = 1.8$ (corresponding to $\eta = 0.23$), and the spin magnitude threshold is set at $|\chi| = 0.45$.
Under these conditions, a consistent trend emerges for all approximants except \Surr{}: lower mass ratios (i.e., larger $\eta$) are associated with increased relative errors in the non-linear memory when compared to NR. In contrast, \Surr{} exhibits the inverse behavior, with higher relative errors occurring at higher mass ratios (smaller $\eta$). For spin magnitude, no statistically significant trend is observed across any of the models.
When compared to the non-precessing results in Figure~\ref{fig:param_error_dist}, the performance of the waveform models shows a qualitatively similar dependence on system parameters, indicating that these trends are robust across spin configurations. Moreover, these parameter-dependent error distributions remain largely unaffected by the inclusion or exclusion of the subdominant mode set $H_{\text{sub}}$: while removing $H_{\text{sub}}$ reduces the overall amplitude of the relative errors, it does not alter the underlying trends.

Utilizing the full available parameter space, the analysis assesses whether conclusions regarding the non-linear memory are sensitive to the specific selection of NR binary mergers. By incorporating additional simulations beyond the \SXS{} catalog, the consistency of previously reported trends is confirmed for both precessing and non-precessing waveforms. The comparative behavior of \Surr{}, \EOB{}, and \Phen{} in reproducing the memory signal remains unchanged, supporting the robustness of the findings and indicating that they are not driven by selection biases in the underlying source parameters.

\subsection{Conclusion}
\label{sec:discussion}
An extensive comparison of gravitational waveform models was carried out, focusing on the remnant kick velocity and gravitational memory associated with BBH mergers. The study evaluated four state-of-the-art waveform models (see Table~\ref{table:1}), reviewing their numerical implementations and addressing limitations inherent to standard waveform alignment procedures. In particular, attention was drawn to the challenge of disentangling alignment residuals from intrinsic model discrepancies. Following the application of a robust alignment method, the balance laws were employed to compute both the kick velocity and the memory contributions. To mitigate potential selection bias, the analysis extended beyond the cataloged \SXS{} simulations to include randomly sampled binary mergers from the broader parameter space. Kick and memory estimates were compared to reference models: NR waveforms served as the baseline for cataloged simulations, while \Surr{} was used as a reference for cases outside the \SXS{} dataset.

The results offer a fresh perspective on longstanding discrepancies between the selected waveform approximants and the reference model in the context of both kick velocity and gravitational memory, in agreement with previous studies~\cite{waveform_test_BL_V}. By tracing the origin of these inconsistencies, it becomes evident that physical observables derived from the balance flux laws serve as effective diagnostics for two key waveform features: the accuracy of the dominant $h_{2,\pm2}$ mode and the distribution of information across the full harmonic mode content of the approximants. The former can be reliably assessed through the remnant velocity (or the linear memory, as evaluated here), while the non-linear memory proves particularly sensitive to the latter, offering insight into the models’ ability to reproduce subdominant mode contributions.\\
Using the remnant kick velocity as a diagnostic, substantial agreement was found across all waveform approximants when compared to the reference model, with \Surr{} exhibiting the highest accuracy, followed by \Phen{}, and then \EOB{}. This agreement holds for both the magnitude and direction of the kick. The ranking of approximants mirrors the post-alignment mismatch values of the dominant strain modes, underscoring the central role of $h_{2,2}$ in determining the kick. While analytical arguments identify $h_{2,\pm1}$ and $h_{2,\pm2}$ as the primary contributors to the kick, the majority of observed discrepancies are attributable to subdominant modes—particularly $h_{2,\pm1}$—as indicated by the remnant velocity errors exceeding the $h_{2,2}$ mismatches by roughly an order of magnitude across all models.
The kick velocity is shown to be highly sensitive to the waveform alignment procedure and, for non-precessing systems, to subtle characteristics of the harmonic mode structure. Accurate computation of the out-of-plane kick in precessing binaries, in particular, requires the faithful modeling of asymmetries between the $\ell,m$ and $\ell,-m$ modes—features that are intrinsic to precessing waveforms. Relaxing the assumption of perfect alignment, a portion of the approximants’ relative errors in remnant velocity may be attributed to residual misalignment. While such artifacts are expected to be subdominant, further refinement of waveform alignment—especially for precessing mergers—remains an essential direction for future research, particularly in the pursuit of high-precision \gls{gw} measurements.

The known limitations in the harmonic mode content of \Phen{} and \EOB{} were clearly exposed through the calculation of the non-linear \gls{gw} memory. Since the non-linear memory accumulates contributions from numerous combinations of strain modes—as expressed in the summation over $\alpha_{\ell,m}$ coefficients in Eq. \eqref{equ:alphass}—it is highly sensitive to any reduction in mode content. This sensitivity is evident in the convergence of NR and \Surr{} estimates toward those of \Phen{} and \EOB{} once the subdominant mode set $H_\text{sub}$ is excluded.
Interestingly, \EOB{} generally produces smaller non-linear memory errors than \Phen{}, although both are, on average, surpassed in accuracy by \TEOB{}. This observation underscores that for practical purposes, reliable non-linear memory predictions can be achieved through accurate modeling of only the dominant harmonic mode. Furthermore, it is demonstrated that asymmetries in the harmonic structure of precessing systems have only a limited impact on the memory calculation, as shown by evaluating the non-linear memory of precessing waveforms with $H_\text{sub}$ removed.

Overall, the combined analysis of kick velocity and gravitational memory offers a robust and complementary diagnostic for evaluating waveform models, and can be readily applied to any waveform model with numerical access. For the models examined here, this dual approach reveals that \Phen{} waveforms tend to provide more accurate estimates of the dominant $h_{2,2}$ mode, whereas the subdominant mode content in \EOB{} yields comparatively better non-linear memory predictions. Despite this, both are outperformed by the \Surr{} model, which benefits from a richer set of harmonic modes, including non-negligible contributions up to $\ell < 5$.
Interestingly, only \Surr{} shows statistically meaningful performance variations across different regions of parameter space. In particular, systems with mass ratios $q < 2.6$ (or equivalently, $\eta > 0.2$) exhibit markedly improved agreement with NR data, potentially indicating a parameter-space bias introduced by the interpolation techniques employed in the \Surr{} construction.
These findings are further substantiated by analytical insights derived from mode mixing in the spin-weighted spherical harmonic decomposition of both the kick and the non-linear memory. The role of the $\alpha$-coefficients in Eq. \eqref{equ:alphass}, which encode allowed combinations of strain modes based on spin-weight $-2$ selection rules, is central to this interpretation. The associated Wigner-$3j$ symbols govern the mode coupling structure, and their appearance in the $\alpha$ terms reflects the underlying symmetry constraints. This formalism, via Eq. \eqref{equ:BLinModes}, connects the harmonic content of the waveform directly to observable physical quantities such as radiated energy, recoil velocity, and gravitational memory, explaining, for instance, why the dominant memory contribution arises from the azimuthally symmetric $h_{2,0}$ mode due to the primarily symmetric nature of the energy flux.

By subtracting the dominant memory contribution (proportional to $\alpha_{2,0}$), the analytical framework reveals a pronounced dependence of the residual non-linear memory on subdominant harmonic modes. This sensitivity offers a promising avenue for evaluating the fidelity of subdominant modes in future gravitational waveform models. Section~\ref{subsec:BL_analysis} demonstrated a practical application of this approach by comparing non-linear memory contributions from various harmonic mode sets. From a physical perspective, probing individual $\alpha$-coefficients with $\ell \geq 2$ and $m \neq 0$ amounts to investigating anisotropic components of the non-linear memory — equivalently, the anisotropic energy flux emitted by the binary to null infinity. While this flux is inherently configuration-dependent and typically subdominant, it could serve as a valuable diagnostic in precision studies of waveform features, especially for precessing BBH systems. A comprehensive exploration of this potential is reserved for future work.
More broadly, assessing waveform quality through the lens of physically derived observables—rather than relying solely on mode-by-mode mismatch comparisons such as those shown in Figure~\ref{fig:mismatch_NR_np}—presents several advantages. This physically motivated strategy circumvents many of the ambiguities inherent in mismatch analyses, particularly those stemming from waveform alignment uncertainties. In contrast, observables like kick velocity and gravitational memory are invariantly defined and derived directly from general relativity, making them robust indicators of waveform fidelity. This is especially beneficial for precessing binaries, where precise alignment remains a challenge, and reinforces the value of physical diagnostics as phenomenologically grounded tools for evaluating waveform models.

In conclusion, the study presented in this Section \ref{sec:Paper_BL} establishes a robust framework for evaluating gravitational waveform models, offering clear guidance for improving waveform precision in anticipation of data from next-generation detectors. By leveraging the balance flux laws, the analysis reveals model-specific strengths and limitations in capturing key physical observables and highlights how these vary across different regions of parameter space. These insights not only inform targeted model selection for applications such as \gls{gw} memory extraction and metadata inference but also lay the groundwork for systematic benchmarking of future waveform model iterations. Extending this methodology to updated or next-generation models presents a natural direction for subsequent investigations.


\chapter{Quantum Signatures in Gravitational Waves} 

\label{chap:quantum} 



\textit{Based on the findings of Chapter II, in this Chapter, the balance flux laws are explored as a tool for propagating quantum corrections throughout all portions of the gravitational waveform. The Chapter thereby presents the findings of the joint works [B, C] listed under ``Publications'' below and referenced as \cite{Maibach2025} and \cite{Other_features_VIII}, respectively, in the following. More precise references are given in the main text where fitting.}

\noindent The previous Chapter provides an exhaustive derivation of the asymptotic spacetime formalism in which the flux equations at future null infinity $\scrip$ can be defined. As it is demonstrated in Section \ref{sec:Paper_BL}, a natural application of such laws is found in the context of numerical waveforms. It is, however, crucial to highlight that the power of these exact (in the sense of validity for full, non-linear \gls{gr}) constraint equations reaches way beyond the pure test of waveforms. Instead, they allow for the computation of higher-order (quantum) effects that manifest in the \gls{gw} strain. One such instance is provided by the addition of the \gls{gw} (displacement \cite{Flanagan_2017} and spin \cite{Spin_mem}) memory to numerical waveforms \cite{waveform_test_BL_II,Khera:2020mcz, mitmanComputationDisplacementSpin2020,Liu_2021}. The (displacement) memory being a fundamental contribution to the flux equation \eqref{equ:BLinModes} is equipped with an elegant interpretation closely tied to the energy flux across $\scrip$. It is, however, regarded as an intrinsic feature of the waveform and a prediction of plain \gls{gr}. Modification of the underlying theory (or others) naturally requires a corresponding adaptation of the procedure outlined in Chapter \ref{chap:asymptotics}, potentially yielding new terms in Eq. \eqref{equ:BLinModes}, see for instance \cite{Janns_paper_I,Spin_mem, Different_Spacetimes_II}. However, if the waveform is intrinsically altered within the regime of validity of \gls{gr}, i.e., by quantum corrections or other novel effects, Eq. \eqref{equ:BLinModes} still holds and allows for propagating the corrections through all (linear and non-linear) regimes of the gravitational strain. This key feature of the balance flux laws is only poorly reflected by literature so far. In this chapter, the shortcoming is rectified by setting an example in the context of the \gls{gw} echo effect by which the gravitational strain resulting from a BBH merger obtains additional features caused by the reflective properties of the BHs' horizons. A detailed overview of crucial notations, production mechanisms, and numerical computations of the \gls{gw} echo is provided in Section \ref{sec:quantum_BH}. Detectability prospects with the LISA instrument, as well as corrections to the non-linear sector of the \gls{gw} strain picked up by an interferometer far away from the merging BBH are discussed in Section \ref{sec:quantum_detectibility}.

\section{Echo Effect, Quantumness and Gravitational Radiation}
\label{sec:quantum_BH}

Classically, BHs represent a well-established solution to Einstein’s equations and serve as a theoretical framework for exploring the coexistence of strong gravitational fields and macroscopic quantum phenomena. Their three classical parameters—mass, charge, and angular momentum—offer a relatively simple (particularly compared to NSs) means of computing the gravitational strain emitted during the merger of two such objects. In most astrophysical scenarios, charge is negligible, allowing standard predictions for classical BHs to depend primarily on two parameters: mass and spin. As a result, for instance, the \gls{qnm}s of a spinning BH are determined solely by these two properties of the Kerr BH (see, e.g., \cite{Kokkotas_1999}). \\
Potential deviations from the no-hair theorem—whether due to classical extensions or quantum modifications of \gls{gr}—are therefore expected to manifest in the QNM content measured, for instance, as the ringdown phase of gravitational waveforms \cite{Ringdown_VIII}. The predictive nature of BH QNMs makes them, along with signatures that inherently rely on them, a widely utilized tool for probing physics beyond GR. One of those intriguing signature examined in this section is the GW echo \cite{Cardoso_2016,Cardoso_2016:I,Cardoso_2017} (see also earlier works \cite{kokkotas1996pulsatingrelativisticstars, Tominaga_1999,Ferrari_2000, Seb_Voe_I, Seb_Voe_II}), which is considered a potential smoking gun for modifications to the near-horizon structure predicted by GR.
Such modifications often stem from proposed resolutions to the BH information paradox. Notable examples include wormholes \cite{Cardoso_2016}, firewalls \cite{Almheiri_2013}, gravastars \cite{Other_features_IV}, and other exotic compact objects (ECOs) \cite{Mark_2017}, such as fuzzballs \cite{Lunin_2002}, among others \cite{Quantum_BH_in_the_Sky}. Despite their theoretical diversity, these scenarios share a common feature: a deviation from the classically all-absorbing event horizon, localized near the BH surface. In the absence of detailed knowledge about the precise nature of this modification, the echo phenomenon—characterized by a repeated ringing in the GW signal after the ringdown phase—is typically modeled by assuming partial reflectivity at the surface of the merging compact object.

The echo effect was proposed as a potentially detectable phenomenon shortly after the first observation of a GW signal. Despite extensive investigations conducted by LIGO and other collaborations \cite{Afshordi_I, Uchikata_2019, Abedi_2019, Holdom_2020, Westerweck_2018, PhysRevD.99.104012, Salemi_2019}, the statistical significance of reported echo signals remains actively debated \cite{Westerweck_2018, Salemi_2019, ashton2016commentsonechoesabyss, abedi2018commentonlowsignificance, Abedi:2017isz}, and a definitive confirmation of the existence of GW echoes is still outstanding. Note that there exists also some research questioning the echo's conceptional foundataions, e.g., \cite{Chen_2019, zimmerman2023rogueechoesexoticcompact, shit_1,shit_2,shit_3,shit_4,shit_5}. Due to their ambiguous theoretical support and inconclusive observational evidence, GW echoes have become a focal point of interest within the quantum gravity community, especially in light of the enhanced precision expected from upcoming GW detectors.\\
Accordingly, the subsequent (sub)sections concentrate on the connection between GW echoes and quantum gravitational phenomenology, as well as their detectability with future instruments. For a more comprehensive review covering the theoretical background, current status, and future prospects of GW echoes beyond the scope of this chapter, the reader is referred to \cite{Quantum_BH_in_the_Sky}.

\subsection{Gravitational Wave Echo Effect}
\label{sec:quantum_echo}

Arguably, the most significant feature of a BH is its horizon, which delineates the surface beyond which all light cones tilt toward the singularity. In this framework, any matter or radiation crossing the horizon becomes causally disconnected from the exterior universe, unless it exceeds the speed of light, thereby rendering the classical BH effectively ``all-absorbing.'' As the causal boundary of a BH in \gls{gr}, the event horizon is determined solely by the BH's classical hair. Its structure, particularly in the near-horizon regime, is central to discussions involving the \gls{gw} echo.\\
When evaluating the prospects of detecting echoes in observational data, it is important to recognize that the concept of an (event) horizon is inherently idealized. By definition, the event horizon demarcates the boundary from which no signal can reach future null infinity. However, determining its precise location requires complete knowledge of the entire future evolution of spacetime, which is impractical in dynamical scenarios. As a result, the literature often substitutes the event horizon with the apparent horizon in such contexts. Apparent horizons can be identified at any given time without information about the future and are defined as the outermost surfaces for which both null expansions (of both null vectors) are non-positive (see Section \ref{sec:Gravitational_radiation_at_Scri} for the formal definition of null expansions).
An intuitive distinction between event and apparent horizons can be illustrated by considering a spherical null shell collapsing into a BH \cite{Quantum_BH_in_the_Sky}. The apparent horizon expands precisely when the shell crosses it, whereas the event horizon begins to grow even before the shell arrives. This anticipatory behavior arises from the increasing gravitational field that influences the global causal structure prior to the shell's arrival, causing the event horizon to extend further outward.\\
In the following discussion, references to the BH horizon during dynamical phases —such as the ringdown—implicitly pertain to the apparent horizon. As the system evolves toward future timelike infinity ($i^+$), the apparent and event horizons asymptotically approach each other. Additional explanatory footnotes are included for clarity where needed.

With a well-defined notion of the BH horizon in place, a consistent framework for constructing \gls{gw} echoes can be established. It is worth noting that several methodologies for echo construction exist in the literature, many of which deliberately avoid explicit modeling of the merger dynamics. Alternative approaches incorporate frameworks such as the EOB formalism \cite{EOB_I,EOB_II} or the Close-Limit Approximation (CLA) \cite{CLA_I,CLA_II,CLA_III,CLA_IV} (for a comprehensive overview, see \cite{Ma_2022}). Among the most prominent echo models are those that eschew merger dynamics entirely. They can generally be classified as either \textit{inside} (e.g., \cite{Wang_2020, Maggio_2019_II}) or \textit{outside} (e.g., \cite{Afshordi_II}) formulations\footnote{Note that in both approaches, the main \gls{gw} signal is computed within the framework of unmodified \gls{gr}, without invoking quantum gravitational corrections.}.
In the outside approach, the primary \gls{gw} emitted from a BBH merger is conceptualized as originating from a pulse reflected at the light-ring potential after having been sent in from past null infinity. Part of this pulse undergoes multiple reflections between the peak of the BH potential and a near-horizon structure—whose nature remains unspecified—producing a sequence of echoes that trail the main signal in detectors effectively located at future null infinity.\\
Conversely, the inside approach models the primary \gls{gw} as the transmitted portion of a wave that emerges from the past (BH) horizon. Due to the partial transmissivity of the BH potential, this results in an echo pattern identical to that produced by the outside formulation. Both methods enable a direct connection between the principal BBH \gls{gw} and the subsequent echoes, thereby circumventing the need for detailed modeling of the merger process.
A unifying aspect across most of the echo models lies in the phenomenological mechanism responsible for the echo generation: \textbf{gravitational radiation is repeatedly reflected between the BH potential barrier and a semi-reflective structure near the BH horizon, emitting a fraction of its energy toward $\scrip$ during each cycle (illustrations can be found in Fig. 1 of \cite{Wang_2020} and \cite{Afshordi_II})}.

\begin{figure}[!t]
	\centering
	\includegraphics[width=0.5\linewidth]{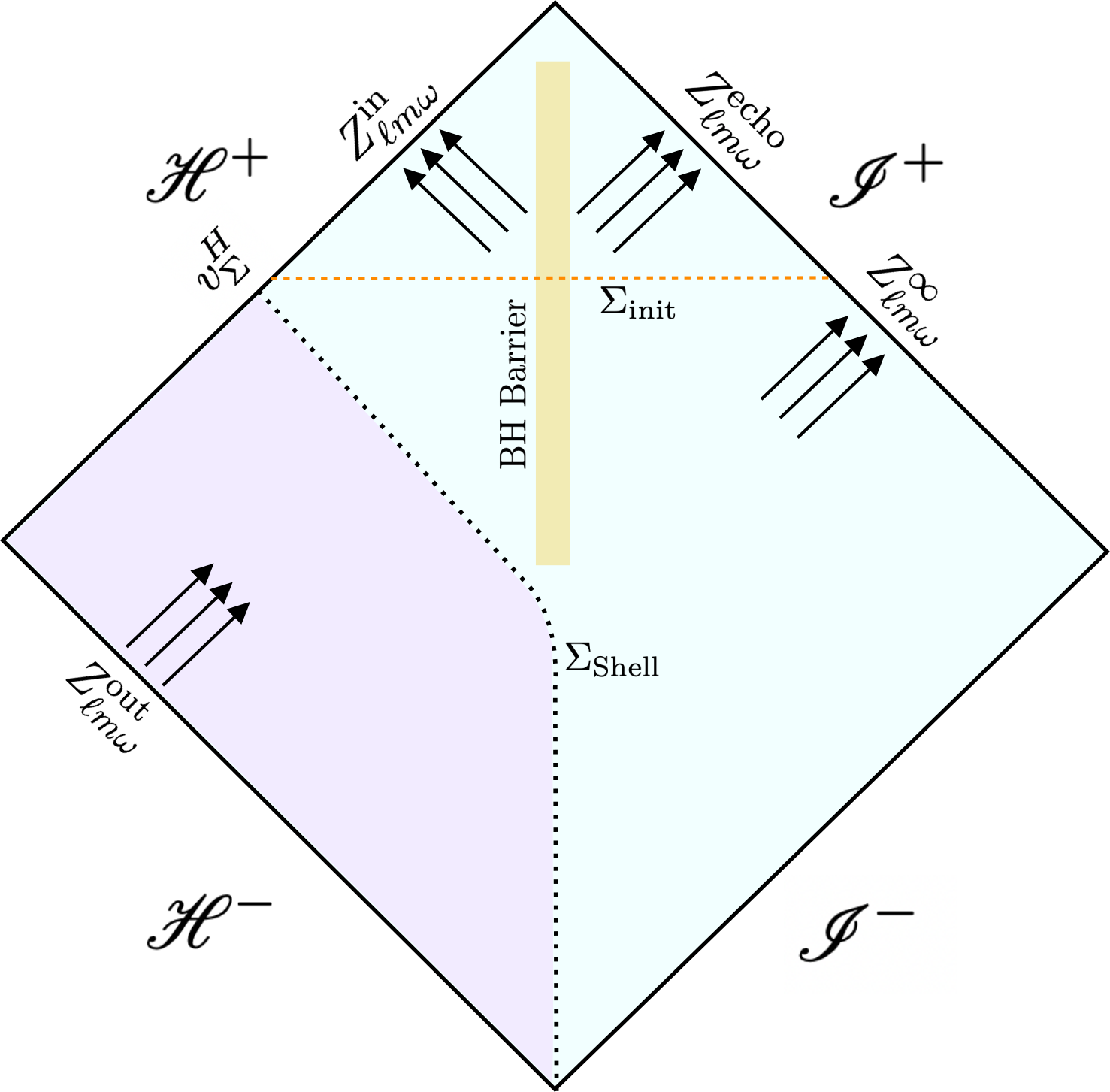}
	\caption{Spacetime diagram of the BBH merger \cite{Maibach2025}. The amplitudes received or emitted by each individual horizon are denoted by three arrows. Here, $\Zinf$ describes the principal waveform including the ringdown while $\Zecho$ denotes all subsequent echoes. The diagram holds analogously also for all $Y_{\ell m \omega}$. The construction outlined in \cite{Ma_2022} considers the blue-shaded regions only. The time at which $\Sigma_\text{shell}$ intersects the future horizon marks the onset of the ringdown phase of the main GW.}
	\label{fig:STR}
\end{figure}

In this Chapter, the modeling of echoes adopts a novel methodology \cite{Ma_2022} based on the so-called \textit{hybrid approach} \cite{Nichols_2010,Nichols_2012}, which integrates Post-Newtonian (PN) theory with Black Hole Perturbation (BHP) theory to model \gls{gw}s from comparable-mass BBH mergers. Within this framework, the construction of the relevant spacetime geometry is partitioned into two distinct regions, as illustrated in Fig.~\ref{fig:STR}: a blue-shaded region where BHP theory is applicable, and a purple-shaded region governed by the PN regime, separated by a three-dimensional timelike hypersurface denoted as $\Sigma_\text{shell}$. This hybrid method enables the modeling of \gls{gw} signals, including the determination of the recoil velocity.\\
The echo construction employed here (see \cite{Ma_2022} for further details) builds upon this framework by similarly dividing spacetime into corresponding regions. Rather than solving for the complete spacetime geometry, however, the method under consideration reconstructs the geometry in the BHP regime using the NR waveform at future null infinity $\scrip$, with particular attention to \gls{gw}s propagating toward the future BH horizon $\Hp$. These NR waveforms are extracted in the form of $\Psi^\circ_4$ and $\Psi^\circ_0$ at $\scrip$ via the CCE (Cauchy Characteristic Extraction) simulation pipeline \cite{Moxon_2020, moxon2021spectrecauchycharacteristicevolutionrapid}, which is implemented in the NR code \texttt{SpECTRE} \cite{Kidder_2017, spectrecode}\footnote{Note that the same numerical framework is utilized in Section \ref{sec:Paper_BL}.}.

Given the separation of the spacetime geometries and the availability of numerical data at $\scrip$, the echo construction in \cite{Ma_2022} centers on the region governed by BHP theory. By enforcing a no-incoming radiation condition at $\scrim$ alongside outgoing waveforms at $\scrip$—obtained via the CCE NR pipeline—the spacetime corresponding to the blue-shaded region in Fig.~\ref{fig:STR} can be reconstructed through a superposition of solutions to the homogeneous Teukolsky equation \cite{Teukolsky_1974} that satisfy the specified boundary conditions\footnote{In the literature, this class of solutions to the Teukolsky equation is commonly referred to as the \textit{up-solution}.}.
Rewriting the homogeneous solutions for $\Psi^\circ_4, \Psi^\circ_0$ as 
\begin{subequations}
\begin{align}
    \Psi_4(t,r,\theta,\phi) &= \frac{1}{r^4}\sum_{\ell,m} \int \dd \omega \,_{-2}R_{\ell m\omega} \,_{-2}Y_{\ell m }(\theta,\phi)e^{-i\omega t}\,,\\
    \Psi_0(t,r,\theta,\phi) &= \sum_{\ell,m} \int \dd \omega \,_{+2}R_{\ell m\omega} \,_{+2}Y_{\ell m }(\theta,\phi)e^{-i\omega t}\,,
\end{align}
\end{subequations}
the radial functions$\,_sR_{\ell m \omega}(r)$ (with spin-weight $s$) satisfy the radial Teukolsky equations
\begin{align}
    \Delta^{-s}\frac{\dd}{\dd r}\left(\Delta^{s+1} \frac{\dd}{\dd r}\,_sR_{\ell m\omega}\right) + V_\ell\,_sR_{\ell m \omega}=0,
\end{align}
with $\Delta = r^2 - 2r$ and 
\begin{align}\label{equ:BH_pot}
    V_\ell = 4is\omega r -\ell (\ell +1 ) + \frac{r^4 \omega^2-2is(r-M)r^2\omega}{\Delta}
\end{align}
for a BH with mass $M$. The solution of the Teukolsky equation is known to, asymptotically, converge to the form
\begin{subequations}
\label{equ:up_teukoslky}
    \begin{align}
       \,_{-2}R^\T{up}_{\ell m \omega} 
       \sim\begin{cases} 
            r^3 e^{i\omega r^*}, & r^* \rightarrow +\infty, \\
             D^{\T{out}}_{\ell m \omega} e^{i\omega r^*} + \Delta^2 D^{\T{in}}_{\ell m \omega}e^{-i\omega r^*}, & r^*\rightarrow - \infty,
            \end{cases}\\
        \,_{+2}R^\T{up}_{\ell m \omega} 
       \sim\begin{cases} 
            r^{-5} e^{i\omega r^*}, & r^* \rightarrow +\infty, \\
             C^{\T{out}}_{\ell m \omega} e^{i\omega r^*} + \Delta^{-2} C^{\T{in}}_{\ell m \omega}e^{-i\omega r^*}, & r^*\rightarrow - \infty,
            \end{cases}
    \end{align}
\end{subequations}
where $r^* = r + 2\ln{(\frac{r}{2}-1)}$ is the tortoise coordinate and numerical values of the coefficients $\Din, \Dout, \Cin, \Cout$ are computed using the Black-Hole Perturbation Toolkit \cite{BHPToolkit}. They encapsulate the physics of the BH potential barrier: the transitivity for radiation traveling from $\Hm$ to $\scrip$ is given by $1/\Dout$ ($1/\Cout$), the reflectivity of the potential barrier displayed in Fig. \ref{fig:STR} by $\Din/\Dout$ ($\Cin/\Cout$). The key insight of the construction presented in \cite{Ma_2022} now resides in fact that for a BBH merger, in the regime where BHP theory is valid, the asymptotic solutions for $\Psi^\circ_4, \Psi^\circ_0$ can be rewritten in similar form to Eq. \eqref{equ:up_teukoslky} as
\begin{subequations}
\label{equ:BH_teukolsky}
    \begin{align}
       \,_{-2}R^{\T{BBH}}_{\ell m \omega} 
       \sim\begin{cases} 
            r^3 Z^\infty_{\ell m \omega}e^{i\omega r^*}, & r^* \rightarrow +\infty, \\
             Z^{\T{out}}_{\ell m \omega} e^{i\omega r^*} + \Delta^2 Z^{\T{in}}_{\ell m \omega}e^{-i\omega r^*}, & r^*\rightarrow - \infty,
            \end{cases}\\
        \,_{+2}R^{\T{BBH}}_{\ell m \omega} 
       \sim\begin{cases} 
            r^{-5}Y^\infty_{\ell m \omega} e^{i\omega r^*}, & r^* \rightarrow +\infty, \\
             Y^{\T{out}}_{\ell m \omega} e^{i\omega r^*} + \Delta^{-2} Y^{\T{in}}_{\ell m \omega}e^{-i\omega r^*}, & r^*\rightarrow - \infty,
            \end{cases}
    \end{align}
\end{subequations}
where, again, $r^* = r + 2\ln{(\frac{r}{2}-1)}$ and the waves escaping to infinity, $\Zinf$ and $\Yinf$, are determined by the numerical values for $\Psi^\circ_4, \Psi^\circ_0$ via
\begin{align}
    r M _f\Psi_4^\circ |_{\scrip} &= \sum_{\ell,m}\, {}_{-2}Y_{\ell m} (\theta,\phi)\Zinf \,,\\
    r^5 M_f^{-3} \Psi_0^\circ |_{\scrip} &= \sum_{\ell,m}\, {}_{+2}Y_{\ell m} (\theta,\phi)\Yinf\,,\label{equ:Psi_0_NR}
\end{align}
where $M_f$ is the final mass of the remnant object.
Note that because of the direct relation between $\Psi_4^\circ$ and the shear/strain (see Section \ref{subsec:Radiative_modes}), one has 
\begin{align}
\label{equ:strain_to_Z}
    h^{\infty}_{\ell m}(\omega) = \frac{1}{\omega^2}\Zinf\,,
\end{align}
with 
\begin{align}
    r M_f^{-1} h^\circ |_{\scrip} &= \sum_{\ell,m}\, {}_{-2}Y_{\ell m} (\theta,\phi)h^{\infty}_{\ell m}\,.
\end{align}
The remaining amplitudes, i.e., $\Zin,\Zout$ ($\Yin, \Yout$), denote the amplitudes traveling towards and away from the horizon $\Hp$, see Fig. \ref{fig:STR}.

\begin{figure}
	\centering
	\includegraphics[width=0.6\linewidth]{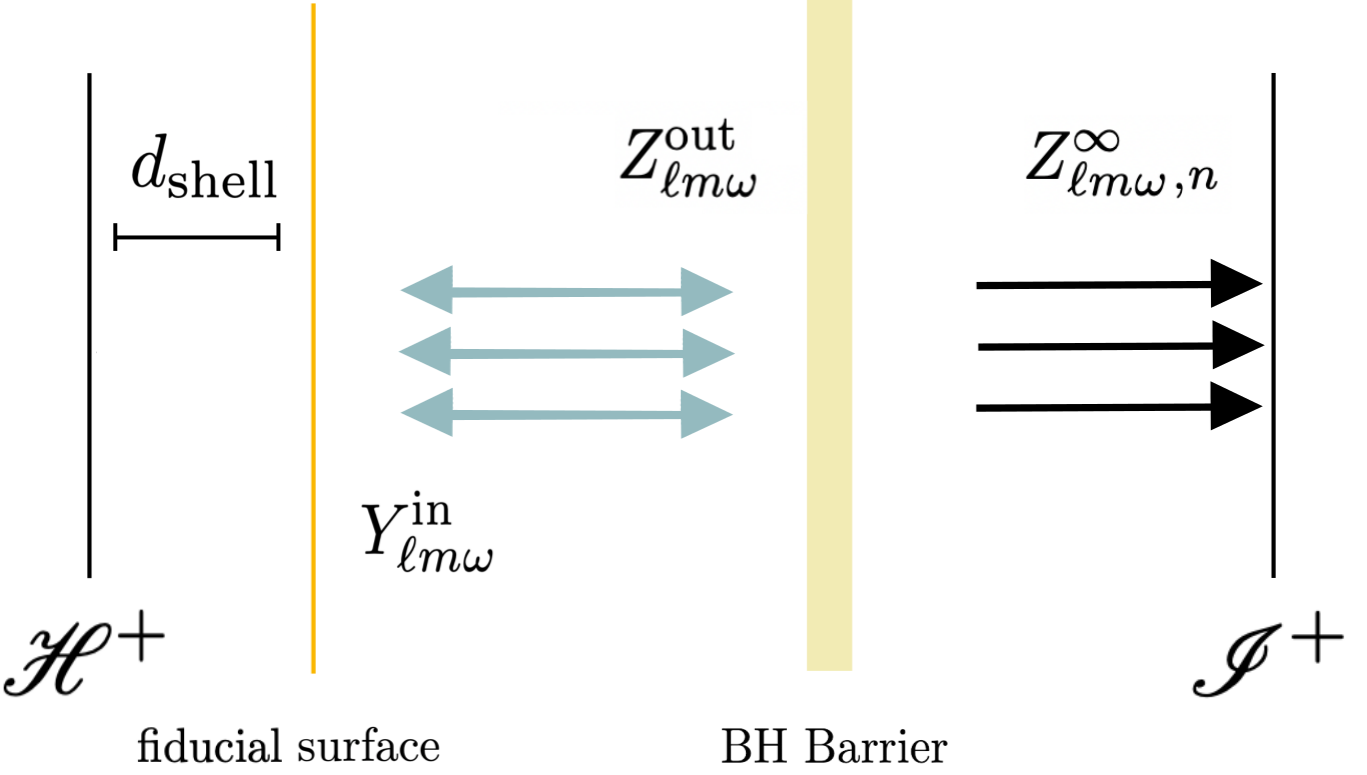}
	\caption{Illustrative sketch of the cavity formed by the BH potential barrier and the near-horizon structure separated by $d_\T{Shell}$ from the future BH horizon $\Hp$ \cite{Maibach2025}. The (radial) amplitudes of gravitational radiation sourced by the ringdown of the perturbed BH are denoted by $\mathcal Y_{\ell m \omega},Z_{\ell m \omega}$. A observer (GW detector) is located at future null infinity $\scrip$.}
	\label{fig:Sketch_intuition}
\end{figure}

For both Eqs. \eqref{equ:up_teukoslky} and \eqref{equ:BH_teukolsky}, the solution corresponding to $r^*\rightarrow - \infty$ describes the limit towards the future BH horizon, $\Hp$, and $r^*\rightarrow \infty$ towards null infinity $\scrip$. Given the structure of the solutions, a relation connecting the waves escaping to infinity, $\Zinf$ and $\Yinf$, with the ingoing waves at the future horizon, $\Hp$, $\Zin$ and $\Yin$, can be derived. One finds that \cite{Ma_2022}
\begin{subequations}
\label{equ:coefficients_interpolate}
    \begin{align}
        \Zout  &= \Dout \Zinf\,,\\
        \Zin  &= \Din \Zinf\,,\\
        \Yout  &= \Cout \Yinf\,,\\
        \Yin  &= \Cin \Yinf\,,
    \end{align}
\end{subequations}
such that the knowledge of $\Zinf,\Yinf$ from numerical simulations in combination with $\Din, \Cin$ allows for the reconstruction of the ingoing waves $\Zin,\Yin$ at $\Hp$. Assuming the presence of a semi-reflective structure surrounding the BH (event) horizon, ingoing \gls{gw}s are partially reflected near $\Hp$—at least conceptually, as the waves never strictly reach the horizon—resulting in outgoing wave components that generate a sequence of echoes following the primary \gls{gw} signal from the merger. Crucially, this process produces not just a single echo but a series of them over time. This behavior arises from the semi-reflective nature of the BH potential barrier, which, together with the near-horizon structure, forms an effective cavity. This cavity traps a fraction of the gravitational radiation, subsequently releasing an echo during each oscillation of the trapped radiation between its boundaries. An illustration of this mechanism is presented in Fig.~\ref{fig:Sketch_intuition}.

The resulting strain measured by a \gls{gw} detector, morally situated at $\scrip$, can be decomposed into two distinct components: the principal waveform originating from the BBH merger, with amplitude denoted by $\Zinf$, and the subsequent echo contributions, labeled as $\Zecho$. The transformation from the radial amplitude solution of the Teukolsky equation to the \gls{gw} strain at $\scrip$ is straightforwardly expressed in Fourier space, as given by Eq.~\eqref{equ:strain_to_Z}. For the transformation from ingoing to outgoing solution, one relies on the Teukolsky-Starobinsky (TS) relations 
\begin{subequations}
\label{equ:CD}
\begin{align}
    \frac{4\omega^4}{C^*}\Yinf = \Zinf,\\
    \Yin = \frac{D}{C}\Zin,
\end{align}
\end{subequations}
where
\begin{subequations}
    \begin{align}
        C &= (\ell-1)(\ell +1 )(\ell +2)\ell +12 i\omega,\\
        D &= 64i\omega(128\omega^2 + 8)(1-2i\omega).
    \end{align}
\end{subequations}
An overview of the relationships between the asymptotic information at $\scrip$ and at the horizon $\Hp$ for both ingoing and outgoing radiation is provided in Fig.~3 of \cite{Ma_2022}. To illustrate the physical interpretation of the asymptotic components of the radial functions $\Psi_0$ and $\Psi_4$, consider Fig.~\ref{fig:STR}. Beginning with $\Zout$ (and equivalently $\Yout$), which appear to be emitted from the past horizon, these waves can be thought of as corresponding to linear extrapolations of the asymptotic information at $\scrip$. In this sense, the \gls{gw}s at future null infinity can be understood as being generated by ``image waves'' $\Zout$ ($\Yout$) emitted from the past horizon $\Hm$ \cite{Ma_2022}. Upon encountering the BH potential barrier, part of the wave is reflected towards $\Hp$, while the other part is transmitted towards $\scrip$. These correspond to $\Zin$ and $\Zinf$ ($\Zecho$), respectively\footnote{Note that the information transmitted towards $\scrip$ is referred to as $\Zinf$ only for the first imprint of the image waves. For each subsequent cycle within the cavity, the transmitted information is captured by $\Zecho$.}.
Then, with the transitivity proportional to $1/\Dout$ ($1/\Cout$) and the reflectivity given by $\Din/\Dout$ ($\Cin/\Cout$), it follows that
\begin{subequations}
\label{equ:ref_trans}
    \begin{align}
        \Zinf = \frac{1}{\Dout}\Zout\,,\\
        \Zin = \frac{\Din}{\Dout}\Zout\,,\\
        \Yinf = \frac{1}{\Cout}\Yout\,,\\
        \Yin = \frac{\Cin}{\Cout}\Yout\,.
    \end{align}
\end{subequations}
At this point, it is important to acknowledge that, based on the arguments presented above, the outlined reconstruction using the hybrid method corresponds to the same reconstruction as the inside prescription within the regime of validity of BHP theory. For a detailed comparison between the two methods, the reader is referred to \cite{Ma_2022}. However, it is essential to highlight that the two methods generally apply linear BHP theory in different regimes. In this context, the region of spacetime is constrained by $\Sigma_\text{shell}$. The intersection of $\Sigma_\text{shell}$ with $\Hp$ marks the time $v_\Sigma$ (see Fig.~\ref{fig:STR}), which corresponds to the onset of the ringdown phase of the main \gls{gw}. It is only at this point that the Teukolsky equation accurately describes the gravitational information propagating through spacetime. Therefore, the validity of Eq.~\eqref{equ:up_teukoslky} and \eqref{equ:BH_teukolsky} is ensured as the merger exits the strong-field regime. Thus, when converting the information at null infinity $\scrip$ into in- and outgoing radiation, one imperatively converts the relevant information into time-domain waveforms, e.g.,
\begin{align}
    \mathcal{Y}^\text{in}_{\ell m}(v)=\int \dd \omega\Yin e^{-i\omega v}\,,
\end{align}
and truncates the waveform at $v_\Sigma$ such that it only contains \gls{qnm} information. Numerically, the waveform section corresponding to the \gls{qnm}s is computed via the fitting procedure outlined in \cite{Ma_2022}, where a minimal mismatch between the NR waveform for $\Yin$ and an analytical expression for the ringdown modes (here exemplarily for the $\ell=2=m$ mode),
\begin{align}
\label{equ:overtones}
    \mathcal Y^\text{in}_{22}(v>v_\Sigma) = \sum_{n=0}^{n_\text{max}}\left(\mathcal{A}_ne^{-i\omega_nv}+\mathcal{B}_ne^{i\overline\omega_nv}\right)\,,
\end{align}
is computed with the time marking the onset of the ringdown, $v_\Sigma$, as a free parameter. The extracted time parameter then determines the Planck Filter $\mathcal{F}(v)$,
\begin{align}\label{equ:taper}
  \mathcal{F}(v,\Delta v, v_\Sigma) =  
  \begin{dcases*} 
  0, & $v < v_\Sigma - \Delta v$ \\ 
  \left(\exp{\chi+1}\right)^{-1} &  $v_\Sigma -\Delta v <v <v_\Sigma$ \\ 
  1, & $v>v_\Sigma $
  \end{dcases*} \,,
\end{align}
such that the ringdown information in $\Yin$ is contained in
\begin{align}
\label{equ:filter_Y}
    \mathcal Y^\T{in QNM}_{\ell m} (v) =\mathcal Y^\T{in}_{\ell m} (v) \mathcal{F}(v) + \T{Const.} \cdot \big(1-\mathcal{F}(v)\big)\,.
\end{align}
The function $\chi$ is defined as $\chi(v,\Delta v, v_\Sigma) = \left(\frac{\Delta v}{v-v_\Sigma} + \frac{\Delta v}{v-v_\Sigma + \Delta v}\right)$. The Planck-taper window function \eqref{equ:taper} is applied to mitigate spectral leakage associated with the abrupt onset of the ringdown phase of the binary merger. The precise choice for $\delta v$ is provided below. The mismatch function used to compare the numerical data with the overtone decomposition \eqref{equ:overtones} is identical to Eq. \eqref{equ:mismatch}. It is important to note, however, that the norm in this chapter is computed in the time domain, as opposed to the frequency domain integral in Eq. \eqref{eq:M-def}. Additionally, the mismatch function depends solely on the starting time $v_\Sigma$ and is evaluated for varying numbers of overtones included in the series \eqref{equ:overtones}.

For the computation of the echo, the relevant information is encapsulated in the ingoing radiation at the future horizon $\Hp$, represented here by $\Yin$ (see Fig. \ref{fig:STR}). As discussed in Section \ref{sec:Gravitational_radiation_at_Scri}, the ingoing gravitational radiation is associated with $\Psi_0$, which, in principle, constitutes the sole numerical information required for the echo computation. This justifies the choice of $\mathcal Y^\text{in}$ for illustrating the \gls{qnm} isolation procedure. Physically, the ingoing radiation travels toward the (partially) reflective surface, where part of it becomes trapped in the cavity. To determine how each frequency mode interacts with the BH's near-horizon structure, both a reflectivity parameter and an appropriate boundary condition are necessary. For the cavity wall corresponding to the light-ring potential, as depicted in Fig. \ref{fig:Sketch_intuition}, these are given by numerical functions dependent on the mass and spin of the perturbed BH. However, for the boundary replacing the would-be horizon, the computation of a meaningful reflectivity requires further specifications of the boundary's properties. The cases considered in this chapter, based on \cite{Maibach2025, Other_features_VIII}, are outlined in Section \ref{sec:quantum_BH}. A specific interpretation of the echo as originating from the quantum phenomenology of the BH horizon is adopted. Before elaborating further on this perspective, it is useful to formalize the reflectivity of the BH potential barrier. Specifically, as one links between $\Zout$ and $\Yin$ within the cavity, the relevant computations can be simplified by absorbing the corresponding conversion factors into the reflectivity function. For instance, as previously mentioned, the reflectivity of the BH potential barrier is given by $\Din/\Dout$. However, the outgoing radiation after reflection is typically expressed in terms of $\Zout$ (further details are provided in the next section).
Thus, to obtain a ``new'' $\Yin$ after the first reflection off the light-ring potential, one applies the identities above to find that 
\begin{align}\label{equ:application_RBH}
\mathcal Y^{\text{in, } n+1}_{\ell m} = \frac{\Din}{\Dout} \frac{D}{C} Z^{\text{out}, n}_{\ell m } = \RBH Z^{\text{out}, n}_{\ell m }\,,
\end{align}
where the index $n$ indicates the cycle in the cavity as depicted in Fig. \ref{fig:Sketch_intuition}.

\subsection{Black Hole Reflectivity}
\label{subsec:quantum_BH}

With the discussion above, the final piece required to complete the evaluation of the \gls{gw} echo lies in the contribution from the near-horizon structure of the BH. As highlighted in the introduction of this Section, there exists a wealth of phenomenology that could, in principle, lead to echo production. However, the subsequent focus will be on the aspects of quantumness and the membrane paradigm. Echo production in this context has been examined in prior studies \cite{Cardoso_2016,Cardoso_2016:I,Cardoso_2017,Cardoso_2019,Agullo_2021}, with additional insights provided in \cite{Afshordi_I,Afshordi_II,Damico_2020,Manikandan_2022, Chakraborty_2022}, and a more generalized treatment of quantum modifications such as echoes can be found in \cite{fransen2024gravitationalwavesignaturesdepartures}. In the following, a novel perspective on the subject is presented, with a focus on the detectability prospects through the advanced space-based LISA instrument in Section \ref{sec:quantum_detectibility}.

\subsubsection{Quantum Black Holes}

While a limited body of works has attempted to treat gravitational radiation from a quantum mechanical perspective, such as \cite{Parikh_2021, Guerreiro_2022}, quantum effects in \gls{gw} physics are predominantly explored in the context of the \gls{gw} echo phenomenon. Notably, the \gls{gw} echo does not require a complete quantum gravity theory but can be derived from first principles within the framework of a semi-classical BH treatment: In classical \gls{gr}, the BH horizon is typically viewed as an unremarkable region of spacetime, with no extraordinary effects expected locally. However, in the quantum regime, this picture can change dramatically, as demonstrated by Hawking’s seminal discovery of thermal radiation \cite{Hawking:1975vcx}. According to this theory, BHs emit thermal radiation at the Hawking temperature, $T_\T{H} = \kappa/2\pi$, where $\kappa$ is the surface gravity, leading, combined with earlier work linking the BH entropy to its horizon area, to the famous entropy-area relation $S_\text{BH} = A/4$ \cite{Area_I,Area_II}, where $A$ represents the area of the event horizon. While this formula has been derived within quantum gravity frameworks, it has only been achieved under specific assumptions and idealizations. A significant challenge lies in understanding the nature of the BH’s microstates, which are responsible for enumerating this entropy, but which remain unknown. In String Theory, these microstates are linked to higher-dimensional fuzzball solutions \cite{Quantum_BH_in_the_Sky}, while Loop Quantum Gravity associates quantum geometries of the horizon with the microstates \cite{Rovelli_1996}. \\
Notably, it is possible to not only derive an entropy formula for a BH but also a complete set of thermodynamic relations \cite{BH_Thermodynamics}. In a broader sense, they can be formulated as follows:

1. Zeroth law: A stationary BH has constant surface gravity, i.e., constant Hawking temperature in thermal equilibrium;

2. First law: A chande of mass is related to changes in horizon area, angular momentum and electric charge via $\dd M = \frac{\kappa}{8\pi} \dd A + \Omega \dd J + \Phi\dd Q$ where $\Omega$ is the angular velocity and $\Phi$ the electrostatic potential;

3. Second law: The entropy and thus the area of the BH never decreases;\footnote{Note that arguably, this law is in contradiction with the emission of thermal radiation. Hawking postulated the law in its original form in 1971, before discovering that BHs must radiate in 1974. Taking Hawking radiation into account, an updated version of the second law states that the total entropy, i.e., the Beckenstein-Hawking entropy of the BH plus the von Neumann entropy of quantum fields outside the horizon should never decrease over time.}

4. Third law: A BH with zero surface gravity $\kappa$ cannot exist as no matter can reach zero temperature.

When attempting to integrate the above reasoning with the principles of quantum gravity, one is immediately confronted with the challenge of ensuring compatibility with the quantization framework. While not all quantum gravity approaches utilize area quantization, the majority of contemporary theories incorporate such (thermodynamic) principles, albeit in various forms. The combination of quantization with the BH area and entropy laws was first explored by Mukhanov and Bekenstein \cite{BH_quant_I, BH_quant_II} (see also \cite{BH_quant_III, BH_quant_IIII}), leading to the area quantization formula $A_N = \alpha \ell_\T{Pl}^2 N$, where $\ell_\T{Pl}$ is the Planck length and $N$ is a positive integer. The constant $\alpha$ is a phenomenological parameter determined by the underlying quantum gravity theory ($\alpha \in \mathbb{R}$)\footnote{In Bekenstein's original works, the value $\alpha = 8\pi$ was chosen. Interestingly, this same value is recovered in various unrelated derivations (see, for example, \cite{Maggiore_2008}).}. This framework results in a discrete mass spectrum for a given BH, which in turn quantizes its emission and absorption processes \cite{Agullo_2021}. In the subsequent analysis, this argument forms the cornerstone of the reflective properties associated with a semi-classical BH.
Concretely, only frequencies matching the mass gaps
\begin{align}
\label{equ:char_freq_init}
    \omega = \frac{|\Delta M|}{\hbar} = \frac{\alpha \Delta N}{32\pi M}\,,
\end{align}
can be absorbed or emitted. $M$ thereby refers to the mass of the BH and $\Delta N$ to the number of microstates. As these frequencies scale in $1/M$, Planck-scale effects are magnified, pushing them into the realm of detectability for \gls{gw} interferometers (e.g., \cite{Other_features_VIII}). For spinning macroscopic BHs, i.e., $N\gg1$, one can compute the characteristic frequencies of the BH as a function of the phenomenological constant, $\alpha$, and dimensionless spin, $a$, as \cite{Agullo_2021}
\begin{align}
\label{equ:characteristic_frequency}
    \omega_N(\alpha,a)= \frac{\kappa \alpha N}{8\pi} + 2\Omega_H + \mathcal{O}(N^{-1})\,,
\end{align}
where $\kappa = \sqrt{1-a^2}/[2M(1+\sqrt{1-a^2})]$ and $\Omega_H = a/[2M(1+\sqrt{1-a^2})]$. Here, $\kappa$ and $\Omega_H$ denote the surface gravity and the angular momentum, respectively. With the frequency $\omega_N$ scaling as $1/M$, Planck-scale effects are magnified and elevated into the frequency regime relevant for present-day GW interferometers. 

In principle, the characteristic frequencies serve as the only narrow pathway for GWs to enter a \gls{bh} by crossing its horizon. However, if the \gls{bh} is spinning sufficiently fast, the width of the quantized energy levels, denoted as $\Gamma$, becomes significant. Treating the semiclassical BH, in the following also refered as quantum BH (QBH), as an atom-like macroscopic object, this width is inversely proportional to the decay rate associated with the spontaneous emission of Hawking radiation, which leads to the de-excitation of the \gls{bh} \cite{Agullo_2021}. If the energy states were to overlap, BHs would behave as true absorbers, allowing any frequency $\omega$ to cross the horizon. However, this scenario does not hold, even for highly spinning remnant BHs, as long as the phenomenological constant $\alpha$ exceeds a critical value $\alpha > \alpha_\T{crit}$. Notably, even for rapidly spinning BHs, this critical value $\alpha_\T{crit}$ is much smaller than the lowest phenomenological constant typically considered in the literature \cite{Agullo_2021}, i.e., $\alpha = 4\log 2$. Consequently, the overlap of energy levels is generally not anticipated, and even for $\alpha = 4\log 2$, they remain quite narrow. This implies that a considerable portion of the GW ringdown's mode content cannot be absorbed by a \gls{qbh}. As argued in \cite{Cardoso_2019}, the remaining modes could be reflected, leading to a late-time echo in the gravitational waveform. Modeling the reflectivity of a QBH is highly non-trivial and requires careful consideration of various (quantum) effects. In this chapter, the simplified, phenomenologically motivated toy model described in \cite{Other_features_VIII} is employed. In outlining its construction, it is crucial to highlight the underlying assumptions: i) GR is the effective theory describing the propagation of GWs throughout spacetime, ii) in the transition from classical to QBH, the ``quantumness'' manifests by discretizing the mass spectrum of BHs, iii) the radiation directed towards the BH is sourced during the ringdown phase of a BBH merger, iv) of such radiation, the unabsorbed portion is reflected off the BH.

A BH reflectivity function derived from area and, by extension, frequency quantization (cf. Eq.~\eqref{equ:char_freq_init}) can be constructed by adhering to guiding principles i)--iv) \cite{Other_features_VIII}. The model assumes that the characteristic frequencies $\omega_N(\alpha, a)$ manifest as sharp absorption lines in the QBH reflectivity spectrum, resembling an atom-like structure—a direct consequence of the Bekenstein-Mukhanov proposal \cite{Bekenstein1997QuantumBH} (see also \cite{Cardoso_2019, Foit_2019} for further motivation). In this framework, the BH is treated as an excited multilevel quantum system. However, spectral line broadening can arise from various physical mechanisms. As demonstrated in \cite{Agullo_2021}, for instance, the linewidth of these absorption features increases with the spin of the QBH. Additionally, quantum fluctuations and uncertainty near the horizon may smear the otherwise sharp roots of the reflectivity coefficient into broader cusps centered at the discrete frequencies $\omega_N$. Such quantum-induced modifications are also explored in the context of altered \gls{qnm} spectra \cite{Berti_2009, Konoplya_2011} and Hawking radiation corrections \cite{Amelino_Camelia_2001, York_1986, Giddings_2016}. Based on these assumptions, a first ansatz for the reflectivity reads
\begin{align}
    \RQBH \sim \begin{cases} 
            1, & \omega\leq\omega_1 \,, \\
             \left|\sin{\left( \frac{\pi\omega}{\omega_N(\alpha)} \right)}\right|^\delta\,, & \omega > \omega_1 \,,
            \end{cases}
\end{align}
where $\delta$ parameterizes the ``sharpness'' of the spectral lines associated to $\omega_N$. Assuming an equal spacing between the characteristic frequencies allows for the use of a periodic function\footnote{Exceptions to this assumptions are provided in some quantum gravity theories, for details see \cite{Agullo_2021} and references therein.}. As the subsequent analysis below focuses primarily on the identification of first characteristic frequency $\omega_1$, the exact spacing does not affect the considerations besides the concrete functional description of $\RQBH$. 

On a quantum level, the reflection of ingoing GWs at the horizon is not the only emission channel available to a black hole. Even at the semi-classical level, Hawking radiation contributes an additional channel, which is similarly influenced by the arguments for frequency quantization \cite{BH_quant_III, Hod_2015}. In this context, the characteristic frequency associated with the Hawking temperature $T_H$ often serves as a cutoff, beyond which the reflectivity is exponentially suppressed \cite{Abedi_2017, Oshita_2019, Oshita_2020, Chen_2021, Chakraborty_2022}. This assumption is widely adopted in studies of ECOs and aligns with the concept of BH microstates \cite{Mathur_2014, Chowdhury_2013}. Further support for a suppression mechanism arises from astrophysical observations \cite{Broderick_2009, Broderick_2015} and theoretical investigations addressing the BH information paradox \cite{Afshordi_2016, Pen_2014}. Accordingly, a conservative exponential suppression factor $\exp{[-|\omega|/(2T_{H})]}$ is introduced into the reflectivity ansatz. To remain agnostic regarding possible modifications to the cutoff temperature, in the exponent adapted in this model, the Hawking temperature $T_H$ is replaced by $T_{QH} = \epsilon T_H$, where $T_H = 1/(8\pi)$. In the limit $T_{QH} \rightarrow 0$, the reflectivity vanishes, and classical \gls{gr} is recovered.

The phase information encoded in the reflectivity function $\mathcal{R}^\text{QBH}(\omega)$ governs the time separation between successive echoes. This relationship can be understood by analyzing the echo production mechanism illustrated in Fig.~\ref{fig:Sketch_intuition}: Consider gravitational radiation emitted during the ringdown phase of a post-merger BH entering a cavity formed between the black hole potential barrier and a reflective shell at radius $r_\text{Shell}$. This shell represents the near-horizon structure confined to a narrow region near the would-be horizon. During each cycle of reflection within this cavity, a fraction of the radiation escapes, producing a distinct echo detectable at future null infinity, $\mathscr{I}^+$. While the precise physical nature of the shell remains unspecified, it is modeled here as a fiducial surface (see also the discussion in \cite{Giddings_2016}). A simple time-delay argument implies $r_\text{Shell} > r_H$, with $r_H = 2M$ denoting the Schwarzschild radius, to ensure that reflected radiation can reach an asymptotic observer within finite time.
Slightly generalizing the notion of the BH radius to account for spinning BHs, $r_H := r_\pm = M(1\pm \sqrt{1-a^2})$, where $a$ is the dimensionless spin parameter, the time a GW takes to complete a single cycle within the cavity is given by \cite{Abedi_2017, Wang_2018}
\begin{align}
\label{equ:t_echo_0}
    \Delta t_\T{echo} = 2 r^* \big|_{r_\T{Shell}}^{r_\T{Barrier}}\,,
\end{align}
where $r^*$ is the tortoise coordinate and $r_\T{Barrier}$ marks the location of the BH potential barrier, see Eq. \eqref{equ:BH_pot}. The latter implies that \cite{Wang_2018}
\begin{align}
\label{equ:t_echo_II}
    \Delta t_\T{echo} 
    \approx 2 \frac{r_+^2 +a^2M^2}{r_+-r_-}\ln{\left(\frac{M}{r_\T{Shell}-r_+}\right)} + M f(a)\,,
\end{align}
where ${r_\T{Shell}}-r_+ =: d_\T{Shell}^2\sqrt{1-a^2}/4M(1+\sqrt{1-a^2})$ and $f(a)\approx 0.335/(a^2-1) + 4.77 + 7.42(a^2-1) + 4.69(a^2-1)^2$. The free parameter in this context is denoted by $d_\T{Shell}$ and represents the proper distance between the fiducial shell at $r_\T{Shell}$ and the BH horizon at $r_H$. In the limit $d_\T{Shell} \rightarrow 0$, the resulting time separation between successive echoes diverges. Importantly, the temporal offset between an echo and the associated classical waveform ($\Delta t_\T{echo}$) is not subject to any fundamental physical constraint. As a result, echoes may emerge in interferometer data without a clear temporal correlation to a preceding classical merger signal \cite{zimmerman2023rogueechoesexoticcompact}.

Including exponential suppression and phase information, one obtains, using the convention $M=1$ and setting $a=0$ \footnote{For simplicity, the subsequent analysis is restricted to non-rotating remnant objects. A comment on the generalization is found below.},
\begin{align}
\label{equ:reflectivity_QBH}
    \RQBH =e^{-i\omega8\ln{\left[\beta(a=0)\right]}}
     e^{-\frac{|\omega|}{2T_{QH}}}\begin{cases} 
            1, & \omega\leq\frac{\omega_1}{2}, \\
             \left|\sin{\left( \frac{\pi\omega}{\omega_N(\alpha, a=0)} \right)}\right|^\delta, & \omega > \frac{\omega_1}{2},
            \end{cases}
\end{align}
where constant phase contributions have been absorbed by a new model parameter $\beta \sim d_\T{Shell}$. In the limit $\delta \rightarrow 0$ and with a suitable choice for $\beta$, Eq. \eqref{equ:reflectivity_QBH} recovers the ECO reflectivity outlined in \cite{Wang_2020, Oshita_2020} and subject of the next subsection. Throughout Sections \ref{sec:Paper_LISA} and \ref{sec:Paper_Mem}, the parameter space for the QBH reflectivity $\RQBH$ is spanned by the reflectivity parameter $\alpha, \beta, \epsilon$ and $\delta$. \\
The reflectivity function \eqref{equ:reflectivity_QBH} encapsulates the quantum-induced semi-reflective properties of a QBH. It can be applied in similar fashion as $\mathcal R^\T{BH}$ as Eq. \eqref{equ:application_RBH} when choosing a suitable boundary condition. From a phenomenological standpoint, there is no compelling reason to assume significant alterations to the ingoing wave beyond the possibility of a phase shift. All other effects are encapsulated in the model-dependent reflectivity coefficient $\RQBH$. Therefore, inspired by the arguments that lead to Eq. (24) in \cite{Ma_2022}, the boundary condition for the QBH is assumed to be \footnote{It must be acknowledged that the boundary condition presented in (24) \cite{Ma_2022} is intrinsically linked to the tidal deformability of ECOs. Since classical BHs do not possess tidal deformability, this analogy does not strictly apply. As a result, it remains plausible that BHs could exhibit entirely different response characteristics. Nevertheless, for the purposes of this Chapter, a mildly generalized ECO-like boundary condition is employed, while the compelling question of the appropriate QBH boundary condition is deferred to future studies.}
\begin{align}
\label{equ:boundary_QBH}
    \Zoutqbh{}^{,1} = \frac{(-1)^{\ell + m +1 }}{\zeta} \RQBH \Yinqbh{}^{,1} \,,
\end{align}
where $\zeta$ is an additional model parameter. In the latter equation, $\Yinqbh{}^{,1}$ corresponds to the relevant part \gls{gw} (i.e., $\mathcal Y _{\ell m \omega}^{\text{in QNM}}$) hitting the potential barrier during the initial cavity cycle of the gravitational radiation originating from the merger. While a part of the reflected information escapes to $\scrip$ establishing the first echo's imprint in the detector data, i.e., 
\begin{align}
     \Zecho{}^{,1} = \frac{1}{ \Dout} \Zoutqbh{}^{,1}  \,.
\end{align}
the remaining radiation repeats the cycle within the cavity. Using Eq. \eqref{equ:application_RBH}, it follows that 
\begin{align}\label{equ:FCSII}
    \Yinqbh{}^{,2} = \frac{(-1)^{\ell + m +1 }}{\zeta} \RQBH \RBH \Yinqbh{}^{,1}\,.
\end{align}
Note at this point that in literature, the prefactor $ \frac{(-1)^{\ell + m +1 }}{\zeta}$ is commonly absorbed into $\RBH$. Here, for clarity, the notation is kept as in the above definition for $\RBH$, Eq. \eqref{equ:application_RBH}. To provide more compact equations below, it is, however, helpful to rewrite $\widetilde{\mathcal{R}}^\text{BH}:= \frac{(-1)^{\ell + m +1 }}{\zeta} \RBH$. Combining Eq. \eqref{equ:boundary_QBH}-\eqref{equ:FCSII} the total echo can be described as a series of the form 
\begin{align}
\label{equ:sum_echo}
    \Zecho     &= \frac{C}{D \Din}\sum_{n=1} \left(\frac{(-1)^{\ell + m +1 }}{\zeta}\Reco\RBH\right)^n \Yinqbh\notag\\
    &= \frac{(-1)^{\ell +m+1}\Reco}{1-\Reco\widetilde{\mathcal{R}}^\text{BH}}\frac{1}{\zeta\Dout}\Yinqbh\notag\\
    &=\sum_n\Zecho{}^{,n}\,.
\end{align}
In the latter equation, the total echo is expressed as a sum of individual contributions, each corresponding to $n$ cycles within the cavity bounded by the potential barrier and the reflective surface. In a final step, the amplitude of the outgoing echo can be translated into the gravitational strain via Eq. \eqref{equ:strain_to_Z}. To prevent instabilities in the model, it is essential to ensure that $|\RQBH\widetilde{\mathcal{R}}^\text{BH}|<1$ holds at all times. An exception to this condition arises at the QNMs, $\omega_n$, of the QBH, where $\RQBH(\omega_n)\widetilde{\mathcal{R}}^\text{BH}(\omega_n)=1$.
The QNMs therefore (roughly) appear as poles of the corresponding transfer function 
\begin{align}\label{equ:trans_func_sum}
    \mathcal{K}(\omega) := \frac{C}{D \Din}\sum_{n=1}\left(\frac{(-1)^{\ell + m +1 }}{\zeta}\RQBH\RBH\right)^n \,.
\end{align}
Note at this point that $|\RQBH|^2$ represents the corresponding energy reflectivity of the ECO and QBH, respectively. Similarly, $|\RBH|^2$ represents the energy reflectivity of the potential barrier \cite{Xin_2021}. Based on this interpretation, one can equally define a coefficient of transmission for the \gls{gw} amplitude of ingoing radiation at $\Hp$. One writes
\begin{align}
\label{equ:energy_transmittivity}
    |\TQBH|^2 := 1- | \RQBH|^2 \,,
\end{align}
such that, for the initial amplitude penetrating the QBH's reflective shell and propagating towards the horizon $\Hp$, it holds
\begin{align}
    \ZBH{}^{,1} = \TQBH\Yinqbh\,.
\end{align}
As for the reflected radiation sourcing the echo in Eq. \eqref{equ:sum_echo}, the transmitted portion receives a contribution for every cycle. Adding all subsequently transmitted radiation, one finds obtain
\begin{align}
\label{equ:sum_echo_BH}
    \ZBH &=  \left(\TQBH+ \frac{\TQBH\RQBH\widetilde{\mathcal{R}}^\text{BH}}{1-\RQBH\widetilde{\mathcal{R}}^\text{BH}}\right)\Yinqbh\notag\\
    &= \TQBH  \sum_{n=1} \left(\frac{(-1)^{\ell + m +1 }}{\zeta}\RQBH\RBH\right)^{n-1}\Yinqbh\notag\\
    &=\sum_n \ZBH{}^{,n} \,.
\end{align}
Eqs. \eqref{equ:sum_echo} and \eqref{equ:sum_echo_BH} are employed in the following sections to evaluate the energy and angular momentum fluxes across both $\scrip$ and $\Hp$ \cite{Other_features_VIII}. It should be emphasized that Eq. \eqref{equ:energy_transmittivity} does not completely specify $\Teco$ due to an undetermined complex phase. Resolving this ambiguity requires the introduction of additional model-specific assumptions. Nonetheless, the current section focuses solely on the energy and angular momentum fluxes. Since Eq. \eqref{equ:energy_transmittivity} accurately characterizes the energy transmittivity, its definition suffices for the forthcoming analysis.

\subsubsection{Membrane Paradigm and Exotic Compact Objects}
 
Shifting the focus of the discussion now towards a more prominent source of echoes, one starts be acknowledging the ``ignorance'' of an outside observer: As aforementioned, in classical \gls{gr}, freely falling observers encounter no extraordinary phenomena upon crossing the BH event horizon. However, from the perspective of a distant, static observer, any in-falling object appears to asymptotically freeze at the horizon due to the extreme gravitational blue-shift. Consequently, the BH interior becomes physically irrelevant to such outside observers. This complementary viewpoint near the horizon forms the foundation of the Membrane Paradigm \cite{Membrane_Paradigm}, in which the BH horizon is effectively modeled as a classically radiating membrane. The paradigm has proven to be a powerful tool for analyzing exterior physics while sidestepping the ambiguities associated with the BH interior. For a concise overview of membrane modeling in \gls{gr}, the reader is referred to \cite{Quantum_BH_in_the_Sky}. Replacing the classical BH horizon with a (quantum) membrane fundamentally modifies the behavior of linear perturbations in the BH spacetime. It was proposed in \cite{Oshita_2020} that quantum effects near the horizon—responsible for thermal emission at the Hawking temperature $T_\T{H}$—lead to corrections in the governing equations of these linear perturbations.
Concretely, the Membrane Paradigm and other effects in convolution with the fluctuation-dissipation theorem lead to\footnote{Eq. \eqref{equ:fluctuation_dissipation} is to be compared to the classical version $\left[\frac{\dd^2}{\dd x^2}+\widetilde{\omega}+ V_\ell\right]\psi_{\widetilde{\omega}}=0$ with asymptotic solution $\lim_{x\rightarrow\pm \infty} = e^{\pm i\widetilde{\omega}x}$.}
\begin{align}
\label{equ:fluctuation_dissipation}
    \left(-i\frac{\gamma \Omega(x)}{E_\T{Pl}}\frac{\dd^2}{\dd x^2} + \frac{\dd^2}{\dd x^2} + \omega^2 -V(x)\right)\psi_\omega(x)=\xi_\omega(x)\,,
\end{align}
representing the evolution equation for QNM functions, $\psi_\omega$, of GWs under a stochastic fluctuation field $\xi_\omega$. In the above expression, $\Omega(x) := \omega/\sqrt{g_{00}}$ denotes the blueshifted frequency, $E_\T{Pl}$ is the Planck energy, $\gamma$ represents a dimensionless dissipation parameter, and $V(x)$ is the \gls{bh} potential. The dissipation term mimics the viscous dissipation observed in sound wave propagation but is suppressed by the gravitational coupling factor $\Omega/E_{Pl}$. This suppression aligns well with phenomenological expectations from quantum gravity theories, where constraints on spacetime viscosity can be incorporated through effective viscous terms \cite{Liberati_2014}.

According to Eq. \eqref{equ:fluctuation_dissipation}, the fluctuation-dissipation theorem implies a thermal spectrum for the mode functions $\psi_\omega$ if dissipation and fluctuations are in balance. Otherwise, classical solutions are recovered (see footnote 9). In the near-horizon region, the analytical solution for the QNM functions can thus be obtained by assuming a constant surface gravity $\kappa = 2\pi T_\T{H}$ and approximating the exterior with the Rindler metric $\dd s^2 = e^{2\kappa x}(-\dd t^2 + \dd x^2) + \dd y^2 + \dd z^2$. The resulting mode function is constant in the limit $x\rightarrow -\infty$. From a physical perspective, this corresponds to the scenario in which the energy flux of ingoing GWs is unable to penetrate the BH horizon and is either absorbed or reflected. This interpretation aligns with models in which the BH interior is effectively inaccessible, such as the Membrane Paradigm and the String Theory–motivated fuzzball complementarity conjecture \cite{Mathur_2014,Chowdhury_2013}.\\
In the near horizon limit, $-\log(E_{Pl}/\gamma|\omega|)\ll \kappa x\ll-1$, one can thus rewrite the function $\psi_{\omega}$ as a superposition of in- and outgoing modes and finds the Boltzmann (flux) reflectivity \cite{Oshita_2020}
\begin{align}
\label{equ:boltzmann_ref}
    |\mathcal{R}| = \left| \frac{e^{-\pi|\omega|/(2\kappa)}\bar Y}{e^{\pi|\omega|(2\kappa)} Y} \right|^2 = e^{-|\omega|/(2T_\T{H})}\,,
\end{align}
It is important to note that this equation is independent of the dissipation parameter $\gamma$ that appears in the dissipation term of Eq. \eqref{equ:fluctuation_dissipation}. This independence arises from taking the absolute value. The parameter $\gamma$ determines the time separation between the reflected echoes \cite{Oshita_2020}, entering as a complex phase. Specifically, when dissipation effects are included, the reflection point is defined by the distance $x_0$, where $\gamma\Omega(x)\sim E_\T{Pl}$.
 Thus, in this model,
\begin{align}
\label{equ:t_echo_I}
    \Delta t_\T{echo} = 2|x_0| = -\frac{\ln{(\gamma|\omega|)}}{\pi T_\T{H}}\,.
\end{align}
For a more exhaustive derivation of the latter, the reader is refer to \cite{Wang_2020}. The time delay can be rewritten as a complex phase and incorporated into the reflectivity function $\mathcal{R}$ such that
\begin{align}
\label{equ:reflectivity}
    \mathcal{R} = \exp{\left(-i\omega\frac{\ln{(\gamma|\omega|)}}{\pi T_\T{H}}\right)}\exp{\left(-\frac{|\omega|}{2T_\T{H}}\right)}\,.
\end{align}
It should be noted that, while the derivation of the Boltzmann reflectivity is relatively model-independent, the phase of the reflectivity arises solely from Eq.~\eqref{equ:fluctuation_dissipation}, particularly from the structure of the dissipation term. As a result, the time delay between echoes exhibits a strong sensitivity to the specifics of the underlying (quantum) theory\footnote{For example, \cite{Chakraborty_2022} presents a more rigorous treatment within the Membrane Paradigm, replacing the BH horizon with a hypothetical quantum membrane. This membrane represents an ensemble of microscopic degrees of freedom in the ground state, governed by a Gaussian wave function. Consequently, a frequency-dependent reflectivity can be defined (see Fig. 1 in \cite{Chakraborty_2022}), along with a corresponding echo time separation given by $\Delta t = 2M[(1-\sigma/\sqrt{\pi}-2\log{(\sigma/\sqrt{\pi})}]$. Both the reflectivity and $\Delta t$ depend significantly on the variance $\sigma^2$ of the quantum state describing these microscopic degrees of freedom.}. Comparable results to those in \cite{Oshita_2020,Wang_2020,Chakraborty_2022} have also emerged from frameworks with distinct physical motivations, such as \cite{Burgess_2018}.\\
In this work, the reflectivity defined in Eq.~\eqref{equ:reflectivity} is adopted as a toy model to describe the reflective characteristics of ECOs. For clarity, Eq.~\eqref{equ:reflectivity} is henceforth denoted as $\Reco$. A slight generalization is introduced by replacing the standard Hawking temperature $T_\T{H}=1/8\pi$ (as for the QBH model) with an effective horizon temperature $T_\T{QH}$, enabling an investigation into how variations in horizon temperature affect the features of GW echoes. This provides a means to probe the influence of quantum horizon effects on the observable signals.\\
Physically, the first exponential factor in Eq.~\eqref{equ:reflectivity} corresponds to the location of the membrane near the horizon from which the GW is reflected. In radial coordinates, this reflective surface is approximately situated at $r_\T{echo} \sim \frac{\ln{\gamma}}{2\pi T_\T{QH}}$. The second exponential factor in Eq.~\eqref{equ:reflectivity} dictates the frequency range over which the reflectivity remains appreciable. Accordingly, while the dissipation parameter $\gamma$ determines the temporal separation between successive echoes, the effective horizon temperature $T_\T{QH}$ predominantly controls the bandwidth of the reflected signal.\\
From a quantum mechanical standpoint, the exponential suppression at high frequencies is supported by modeling an isolated BH as a multilevel quantum system \cite{Oshita_2020}. In this so-called ``giant atom'' framework, Hawking radiation facilitates the spontaneous de-excitation of the BH, whereas reflectivity can be interpreted as a form of stimulated emission. As a result, the frequency range $\omega \lesssim T_\T{H}$ naturally emerges as the domain in which stimulated emission occurs, while for $\omega \gg T_\T{H}$, the BH effectively behaves as a fully absorbing object.

\begin{figure}
	\centering
	\includegraphics[width=0.7\linewidth]{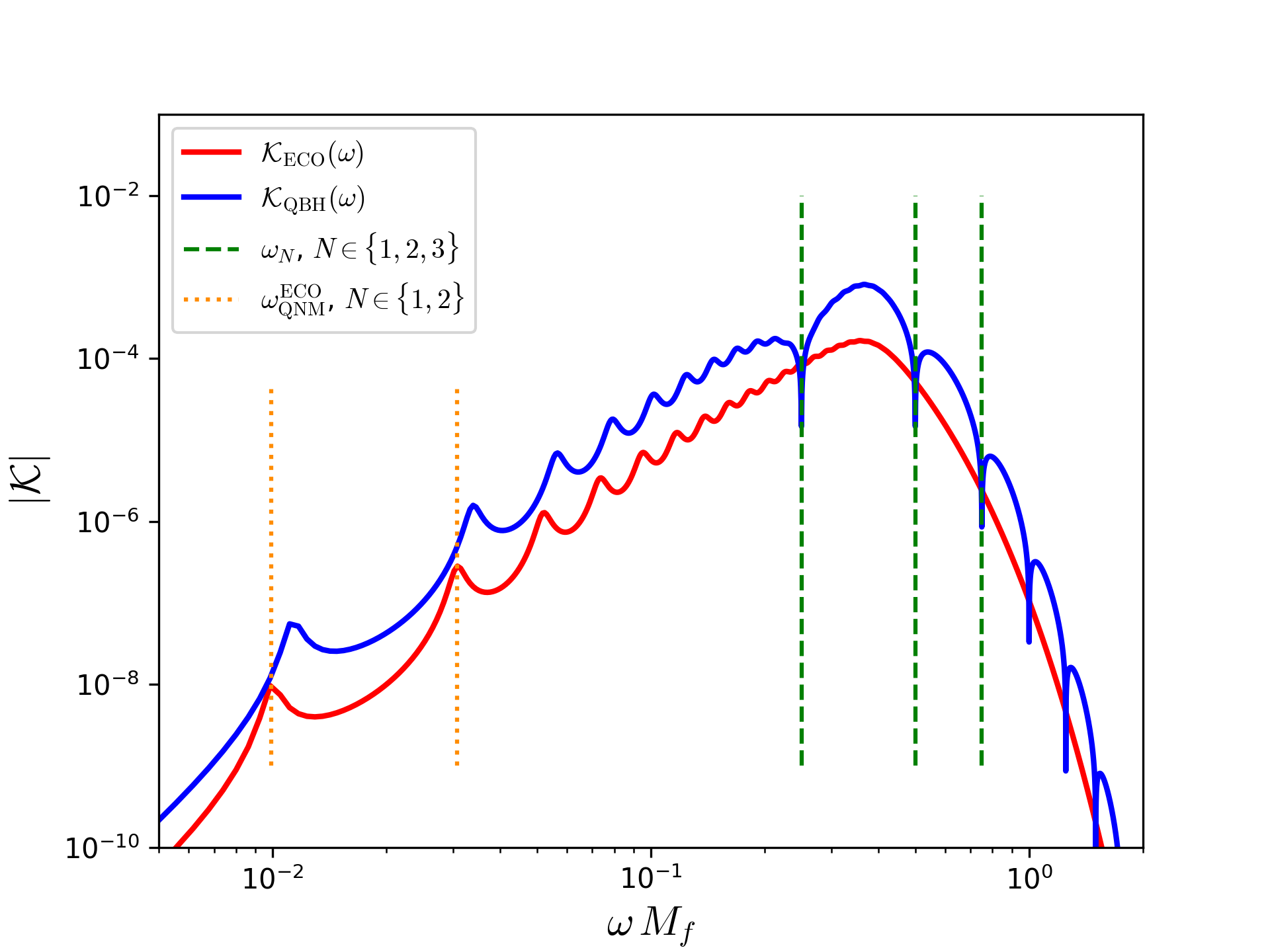}
	\caption{Transfer functions for $\Reco$ and $\RQBH$ \cite{Maibach2025}. The parameters (see Table \ref{table:2}) are chosen as $T_\T{QH}= 1/8\pi, \gamma = 10^{-15}$ and $\alpha = 8\pi, \delta = 0.5, \epsilon = 1, \beta = 10^{-15}$, $\zeta=4$. The green dashed lines mark the characteristic absorption frequencies for the QBH. The dotted orange lines mark the QNM of the ECO. Similarly, the peaks of $\mathcal{K}_\T{QBH}$ mark the QNMs of the QBH. For better readability, the transfer function of the QBH, $\mathcal{K}_\T{QBH}$, is multiplied with an overall factor of 5.}
	\label{fig:Transfer_functs}
\end{figure}

The considerations following Eq.~\eqref{equ:reflectivity_QBH} apply analogously to the reflectivity of ECOs. An important exception lies in the boundary condition, which can be rigorously derived within the framework of the Membrane Paradigm \cite{ECO_boundary}. Specifically, by analyzing the tidal tensor field $\mathcal{E}_{ij}$ as experienced by fiducial observers with zero angular momentum in their own rest frame, one can compute the transverse components that characterize the tidal deformation induced by \gls{gw}s. Following the tidal response of a neutron star, the ECO is proposed to react linearly to tidal stress resulting in 
\begin{align}
\label{equ:FIDO}
    -\frac{r^2}{\Delta}\Bar\Psi_4  = \frac{\Reco}{\Reco-1}\mathcal{E}^\T{trans}\,,
\end{align}
at the surface of the ECO, i.e., at the location of the membrane. Given for the transverse part of the tidal tensor it holds that $\mathcal{E}^\T{trans}\sim -\frac{\Delta}{4r^2}\Psi_0 - \frac{r^2}{\Delta}\Bar{\Psi}_4$, for an ECO, one obtains
\begin{align}
\label{equ:boundary_ECO}
    \Zouteco{}^{,1} = \frac{(-1)^{\ell + m +1 }}{4} \Reco \Yineco{}^{,1} \,.
\end{align}
As previously noted, the boundary condition is recovered from Eq.~\eqref{equ:boundary_QBH} by setting $\zeta = 4$. The subsequent discussion following the definition of $\RQBH$ applies in a similar manner. This includes, in particular, the aspects concerning the stability and the QNMs of the transfer function.\\
The transfer functions corresponding to $\Reco$ and $\RQBH$ (Eq. \eqref{equ:trans_func_sum} for the latter) are illustrated in Fig.~\ref{fig:Transfer_functs}. A distinct difference in the QNMs of ECOs and QBHs (being closely aligned with the poles of the respective transfer functions) is clearly visible. It is important to highlight that the poles shown do not precisely correspond to the QNMs, but are located very close to them. Instead, these poles of the transfer function indicate the positions of the resonances in frequency space associated with the effective cavity depicted in Fig.~\ref{fig:Sketch_intuition} resulting from the corresponding phenomenologies.
The global maxima of both transfer functions coincide, as they correspond to the fundamental QNM frequency of a Schwarzschild BH. The roots of the QBH reflectivity are marked by dashed green lines. At high frequencies, both transfer functions are exponentially suppressed. At low frequencies, the $1/\Dout$ factor in Eq.~\eqref{equ:sum_echo} dominates the behavior of both transfer functions. For ECOs, both $T_\T{QH}$ and $\gamma$ influence the location of the QNMs, while $T_\T{QH}$ additionally sets the exponential suppression scale. For QBHs, the parameters $\beta$ and $\delta$ determine the position of the QNMs, $\omega_\T{QNM}^\T{QBH}$, whereas $\alpha$ controls the root structure and $\epsilon$ regulates the exponential decay. The overall amplitude is modulated by $\zeta$.

\begin{figure}
	\centering
	\includegraphics[width=0.70\linewidth]{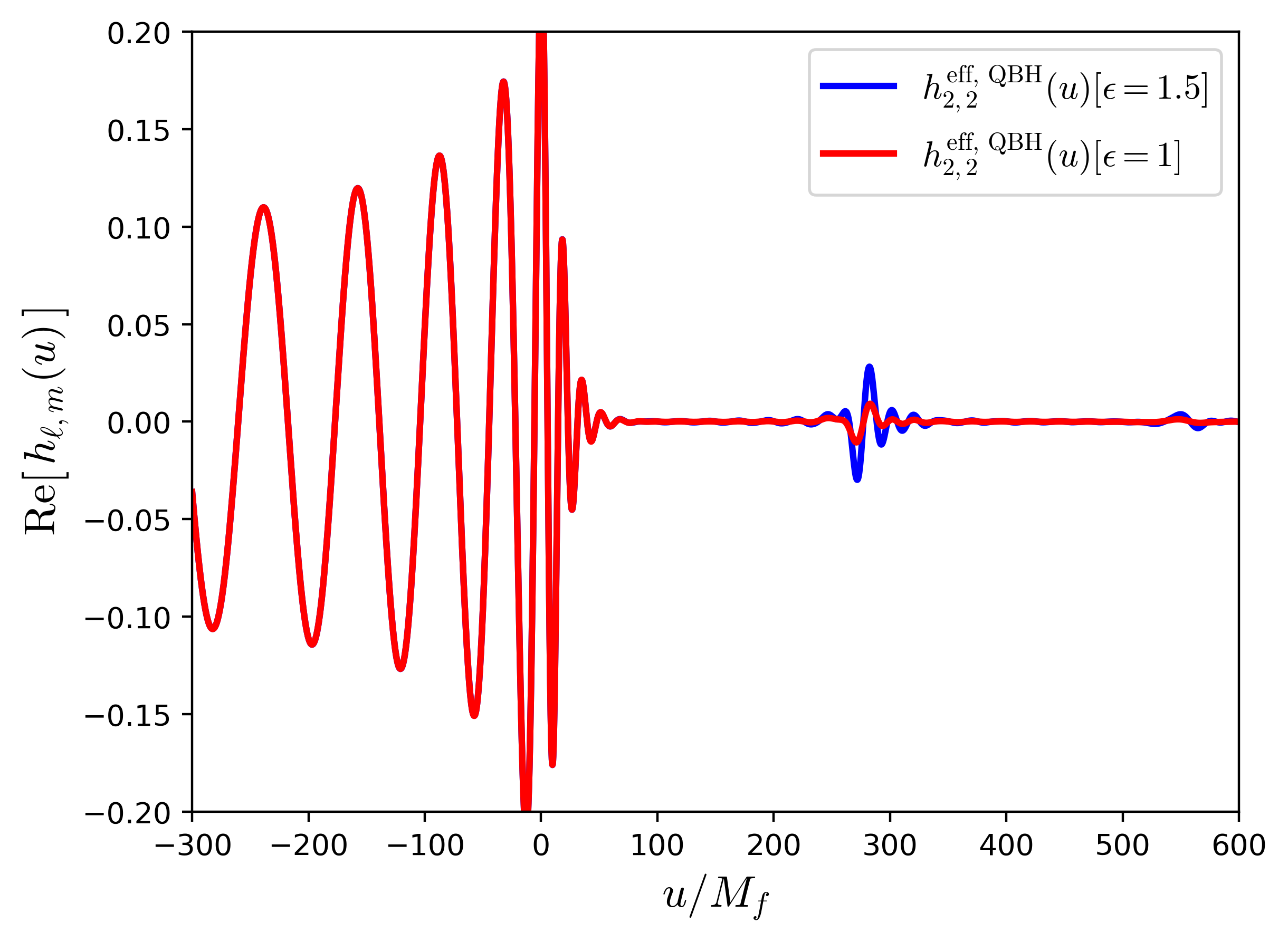}
	\caption{Exemplary waveform with echo computed for \textit{SXS:BBH:1936} with the baseline reflectivity parameter $\alpha = 8 \pi,\beta = 10^{-15}, \gamma = 4, \delta = 0.2$ (see Table \ref{table:2}) \cite{Maibach2025}. The exponential suppression parameter $\epsilon$ indicated in the plot.}
	\label{fig:ECHO}
\end{figure}

With the reflectivity function and boundary conditions established, the \gls{gw} echo can be directly computed for an arbitrary NR strain time series. For both models examined in this chapter, the primary \gls{gw} signal can be modified by superimposing additional $n$ echoes onto the time series, where $n$ is an arbitrary integer. It is important to note, however, that by construction $|\RQBH|<1$ ($|\Reco|<1$), leading to a progressive attenuation of the echo amplitudes. Consequently, only the first few echoes are expected to be significant in realistic detection scenarios.\\
An illustrative example of a computed echo is shown in Fig.~\ref{fig:ECHO}, based on the NR event \textit{SXS:BBH:1936} \cite{Boyle_2019} and give set of reflectivity parameters selected for illustrative purposes. Additional visualizations are provided in Fig.~\ref{fig:waveform_QBH} of Appendix~\ref{app:ECHO}. The dependence of QBH echoes on model parameters is presented in Fig.~\ref{fig:echo_QBH} of Appendix~\ref{app:ECHO} (see also Table~\ref{table:2}). A comparable plot for ECOs can be found in \cite{Ma_2022}. As anticipated, the morphological differences between the two types of echoes are minor.
To offer a comprehensive overview of the relevant model parameters—especially w.r.t. Sections~\ref{sec:Paper_LISA} and~\ref{sec:Paper_Mem}—they are summarized in Table~\ref{table:2}. 

To reconstruct the complete strain shown in Fig.~\ref{fig:ECHO}, one can leverage the separability between the main \gls{gw} signal and the subsequent echoes.
Thus, one can simply compute the complete waveform as
\begin{align}
\label{equ:strain_decomposition}
    \hfull = \hecho + \hnorm\,,
\end{align}
where $\hnorm$ results from $\Zinf$, the \textit{raw} asymptotic waveform at $\scrip$. A depiction of $\hfull$ is provided in Fig. \ref{fig:ECHO}. Thereby, $\hecho$ can be explicitly computed as a convolution of all aforementioned sub-steps, using equations \eqref{equ:coefficients_interpolate}, \eqref{equ:strain_to_Z}, \eqref{equ:sum_echo}, \eqref{equ:filter_Y}, and \eqref{equ:Psi_0_NR} as
\begin{align}
\label{equ:hecho}
    \hecho(\omega) &= \sum_n \sum_{\ell, m} \,_{-2}Y_{\ell m}(\theta, \phi) \hecho_{\ell m, n}(\omega)\,,
\end{align} 
where (for the QBH)
\begin{align}
\label{equ:some_label}
\hecho_{\ell m, n}(\omega)=\frac{1}{\omega^2}\frac{C}{D \Din}\left(\frac{(-1)^{\ell + m +1 }}{\zeta}\RQBH\RBH\right)^n \notag\cdot\mathfrak F\left\{C^\T{in}_{\ell m} \Psi_{0, \ell m}(v)\mathcal{F}(v) \right\}\,,
\end{align}
and $\Psi_{0, \ell m }$ is defined as in Eq. \eqref{equ:Psi_0_NR}, i.e., 
\begin{align}
    r^5M_f^{-3}\Psi_0^\circ = \sum_{\ell,m} \,_{+2}Y_{\ell m}(\theta, \phi) \Psi_{0,\ell m}\,.
\end{align}
In the latter equation, one uses $\mathfrak F (\Psi_{0,\ell m}) =: \Yinf$ and $\mathfrak F (\cdot)$ denotes the Fourier transform. Note also that $C^\T{in}_{\ell m}$ represents the Fourier transform of $\Cin$. The statement in Eq. \eqref{equ:strain_decomposition} is independent of the time separation that is determined by the corresponding parameters of a given reflectivity model, i.e., independent of the complex phase factors in \eqref{equ:reflectivity} and \eqref{equ:reflectivity_QBH}. Therefore, the echo can easily be added to (numerical) waveforms by adding it on top of the main \gls{gw} strain time series in a post-processing step similar to the memory, see \cite{waveform_test_BL_II}. Crucially, the echo, as well as its memory, can be treated as individual ``events''\footnote{This approach aligns with recent proposals suggesting that echoes should be treated as independent signals rather than as extensions of the BBH merger waveform due to the potentially substantial time delays involved \cite{zimmerman2023rogueechoesexoticcompact}.}.

Before proceeding with explicit computations involving the \gls{gw} echo, it is important to recognize that the phenomenological models of both QBHs and ECOs give rise to more than just echo signals. In particular, semi-reflective properties of BHs or other coalescing compact objects also induce a secondary effect on the \gls{gw} strain. For completeness, this additional contribution is briefly discussed in the following paragraph.

\begin{center}
\begin{table*}[t]
\centering
\begin{tabular}{
    |c|c|c|
}
 \hline \bfseries
 $\mathcal{K}(\omega)$ attribute & \bfseries $\,\,\,\,$Parameter ECO$\,\,\,\,$ & \bfseries $\,\,\,\,$Parameter QBH $\,\,\,\,$
 \\\hline\hline 
 Time separation echo & $\gamma, T_\T{QH}$   & $\beta$   
 \\\hline
  $\,\,\,\,$ Exponential supp. for large $\omega$ $\,\,\,\,$& $T_\T{QH}$&   $\epsilon$  
   \\\hline
  Location QNMs & $\gamma, T_\T{QH}$  & $\beta,\epsilon$ 
  \\\hline
Separation of roots & - & $\alpha$ 
  \\\hline
 ``Sharpness'' of roots & -  & $\delta$ 
 \\\hline
 Boundary suppression & 1/4  & 1/$\zeta$ 
 \\\hline
\end{tabular}
 \caption{Model parameters for the reflectivity functions of ECOs and QBHs, i.e., $\Reco$ and $\RQBH$ respectively \cite{Maibach2025}. The corresponding transfer functions $\mathcal{K}(\omega)$ are displayed in Fig \ref{fig:Transfer_functs}. As the ECO's transfer function does not obtain roots, a correspondence of $\alpha, \delta$ for ECOs is absent.
    }
    \label{table:2} 
\end{table*}
\end{center}

\subsubsection{Tidal Heating}
\label{sec:Tidal_heat}

During the ringdown phase, echoes are generated by gravitational radiation that propagates toward the BH horizon and is subsequently reflected by a near-horizon structure. However, a similar mechanism can be identified during the earlier inspiral phase of a BBH merger. In the classical framework, each BH absorbs a portion of the incoming \gls{gw}s during the inspiral, although this absorbed fraction is small relative to the radiation escaping to $\scrip$ \cite{Poisson_1995}. Nonetheless, the incident waves induce deformations in the (event) horizons of the BHs \cite{goswami2020tidalforcesgravitationalwaves}. As the BHs emit \gls{gw}s and lose rotational energy, these horizon distortions in turn influence the orbital dynamics of the binary \cite{PhysRevD.8.1010}. For NSs \cite{Tidal_Heat_I,Tidal_Heat_II}, or BHs modeled via the Membrane Paradigm \cite{Tidal_Heat_III,Tidal_Heat_IV}, such tidal deformations provide an additional channel for dissipating orbital energy, which ultimately results in a characteristic phase evolution in the \gls{gw} signal. This process is known as \textit{tidal heating}.\\
Introducing an absorption energy gap, as predicted by the Bekenstein-Mukhanov spectrum in Eq. \eqref{equ:char_freq_init}, significantly suppresses the absorption of \gls{gw}s by the compact objects. This leads to a distinct inspiral evolution compared to the fully classical GR predictions \cite{Hughes_2001, PhysRevLett.120.081101}. Specifically, the \gls{gw} phase deviates from the classical expectation, where tidal heating introduces a correction at 2.5 PN order, scaling with the logarithm of the orbital velocity. A detailed parametrization of the impact of a discretized energy spectrum on tidal heating can be found in \cite{PhysRevLett.120.081101}.\\
In the context of BH area quantization, the influence of tidal heating has been analyzed in \cite{Datta:2021row}. Additional studies focusing on the associated phase shifts for ECOs \cite{PhysRevLett.120.081101, Datta_2019, Datta_2020, Datta_2021} highlight the strong potential of advanced interferometers such as LISA and ET in probing these effects. Despite the promising outlook regarding detectable imprints of modified tidal heating in future \gls{gw} observations (see also \cite{PhysRevLett.120.081101}), this topic is not pursued further in the current Chapter. Nevertheless, it is noted that the sensitivity of LISA to potential tidal heating signatures—using a fully operational and state-of-the-art analysis pipeline—is currently under active investigation.


%
%
%


\section{Detectability and Waveform Corrections}
\label{sec:quantum_detectibility}

With the preceding section having established the fundamental characteristics of the \gls{gw} echo, introduced two representative phenomenological origins, and demonstrated the inclusion of echoes in NR waveforms, the subsequent analysis—based on \cite{Other_features_VIII} and \cite{Maibach2025}—makes use of these tools to achieve two main objectives: First, it is shown that LISA possesses the sensitivity required to detect \gls{gw} echoes and to distinguish between the different phenomenological models discussed above by identifying the characteristic frequency predicted by the Bekenstein-Mukhanov spectrum \eqref{equ:char_freq_init}. Details follow in Section \ref{sec:Paper_LISA}; Second, it is demonstrated that the quantum effects responsible for generating the \gls{gw} echo also induce memory corrections in the waveform, which can become significant within certain regions of the reflectivity models' parameter space, as discussed in Section \ref{sec:Paper_Mem}.

\subsection{Gravitational Wave Echoes in LISA Data}
\label{sec:Paper_LISA}

\begin{figure}
	\centering
	\includegraphics[width=0.7\linewidth]{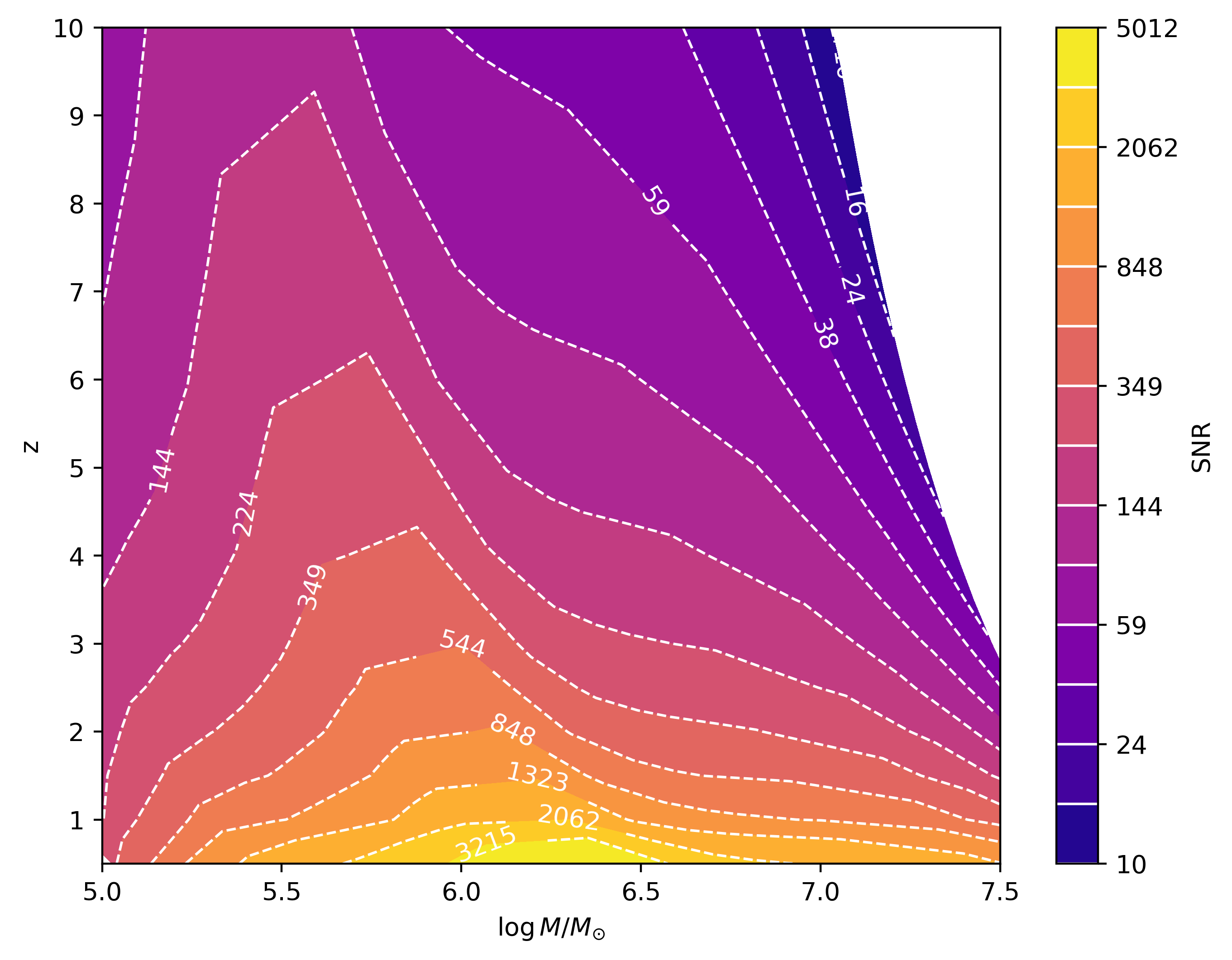}
	\caption{LISA-SNR of the \textit{SXS:BBH:1936} waveform for different redshifts and total (redshifted) masses \cite{Other_features_VIII}. The SNR is computed following \cite{Henris_Mem}.}
	\label{fig:SNR}
\end{figure}

The detectability of \gls{gw} echoes in data from the future space-based LISA observatory is evaluated using the simulation and data analysis pipeline developed in \cite{Henris_Mem}. The analysis is conducted on signals recorded in the \gls{tdi} X channel of the LISA instrument\footnote{Readers unfamiliar with the LISA instrument and the concept of TDI channels may consult Section \ref{sec:SGWB_detection_Paper_HI} and references therein for a brief overview.}. It should be emphasized that the X channel is not inherently preferred; other TDI channels with adequate signal \gls{snr}, as well as the full strain data, may equally be used depending on the orientation of the GW source relative to the detector. Concerning sky location and other orientation-sensitive parameters affecting the \gls{snr}, the conservative baseline listed in Table I of \cite{Henris_Mem} is adopted. For reference, Fig. \ref{fig:SNR} presents the \gls{snr} of the full waveform corresponding to event \textit{SXS:BBH:1936}, as measured by LISA, plotted against redshift and redshifted mass. These results may be qualitatively compared with Fig. 6 of \cite{Henris_Mem}, noting that slight discrepancies arise due to differing simulation parameters such as binary spins and mass ratio in the merger events under consideration. In Sections \ref{sec:Paper_LISA} and \ref{sec:Paper_LISA_II}, the echo analysis is confined to the QBH scenario, since its reflectivity model encompasses that of the ECO and can be interpreted as a generalization thereof. As such, ECO-based estimates may be straightforwardly derived by restricting to parameter values $\zeta = 4$ and $\delta = 0$ in the QBH framework.

The \gls{snr} of the echo is numerically determined for a set of 11 NR simulations of massive binary BH mergers (with total mass $M_\text{tot} > 10^5 \Msol$) extracted from \cite{Boyle_2019}. These include \textit{SXS:BBH:1936}, \textit{0207}, \textit{0334}, \textit{1155}, \textit{1424}, \textit{1448}, \textit{1449}, \textit{1455}, \textit{1936}, and \textit{2108}. The events are selected primarily based on their vanishing remnant spin. Exceptions are \textit{SXS:BBH:0334}, \textit{1155}, and \textit{2108}, which exhibit remnant spin amplitudes $|\vec{\chi}|$ between $0.28$ and $0.68$. These simulations are included to investigate potential systematic differences in echo-related features when $a \neq 0$. While the reflectivity function Eq. \eqref{equ:reflectivity_QBH} holds in general, the echo reconstruction method used in \cite{Ma_2022}, and applied in \cite{Maibach2025, Other_features_VIII}, is specifically adapted for non-spinning remnants due to computational challenges in the presence of spin. Thus, extending the analysis to Kerr remnants introduces a small systematic error. However, with the acceptance of these minor corrections, the analysis is not restricted to a specific region of parameter space and can be applied to any arbitrary waveform, provided the knowledge of $\Psi_0^\circ$ and $\Psi_4^\circ$. Extracting the Newmann-Penrose scalars beyond $\Psi_4^\circ$ typically requires additional simulation efforts. For \textit{SXS} events, the CCE scheme offers the necessary tools.

\begin{figure}
	\centering
	\includegraphics[width=0.7\linewidth]{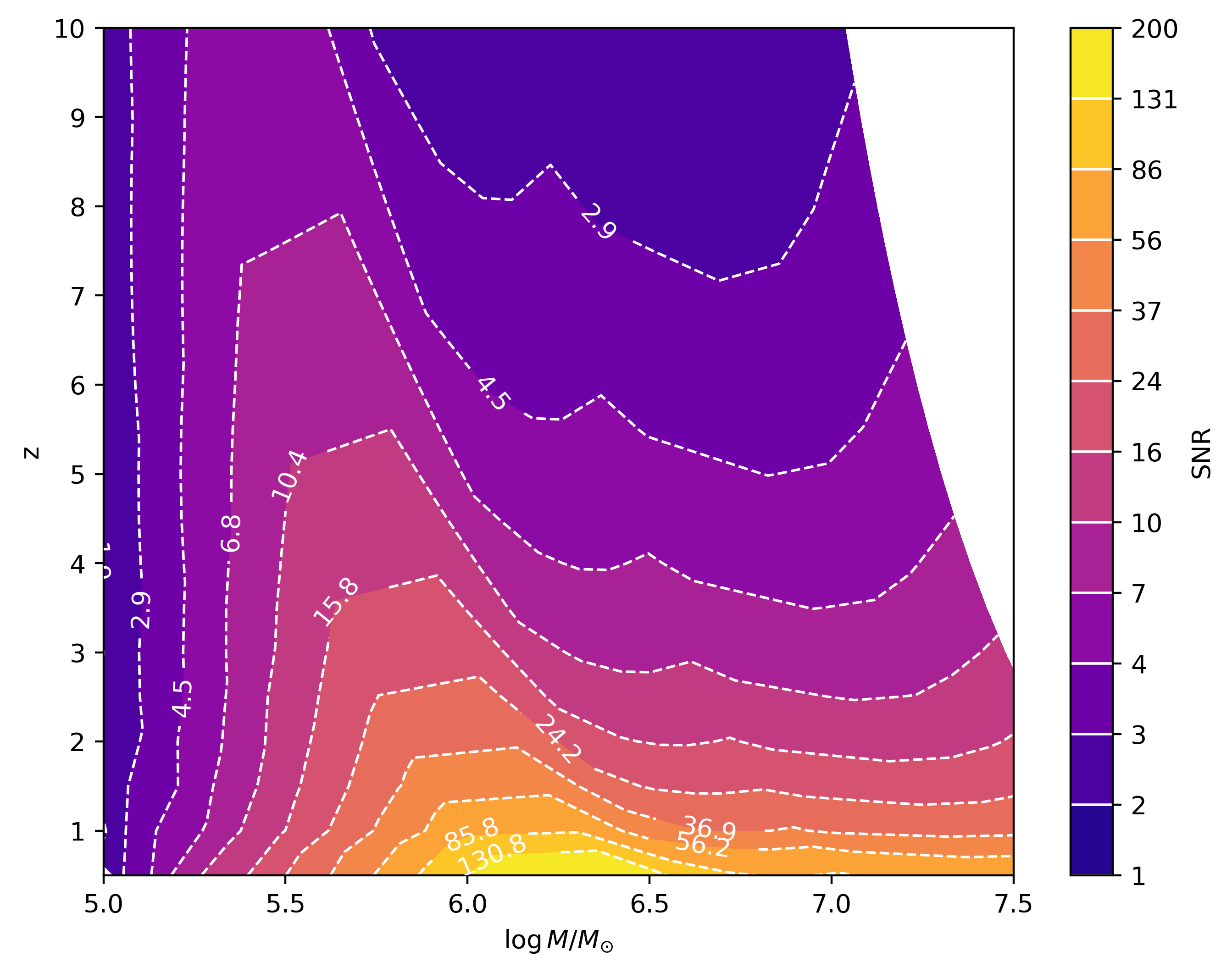}
	\caption{LISA-SNR of the echo produced by the \textit{SXS:BBH:1936} waveform over redshift and total (red-shifted) mass \cite{Other_features_VIII}. One fixes $ \log M_\text{tot} / \Msol = 6, z=1, \alpha = 8\pi, \delta = 0.2, \epsilon =1 ,\gamma=4$.}
	\label{fig:SNR_ECHO}
\end{figure}

\begin{figure}
	\centering
	\includegraphics[width=0.7\linewidth]{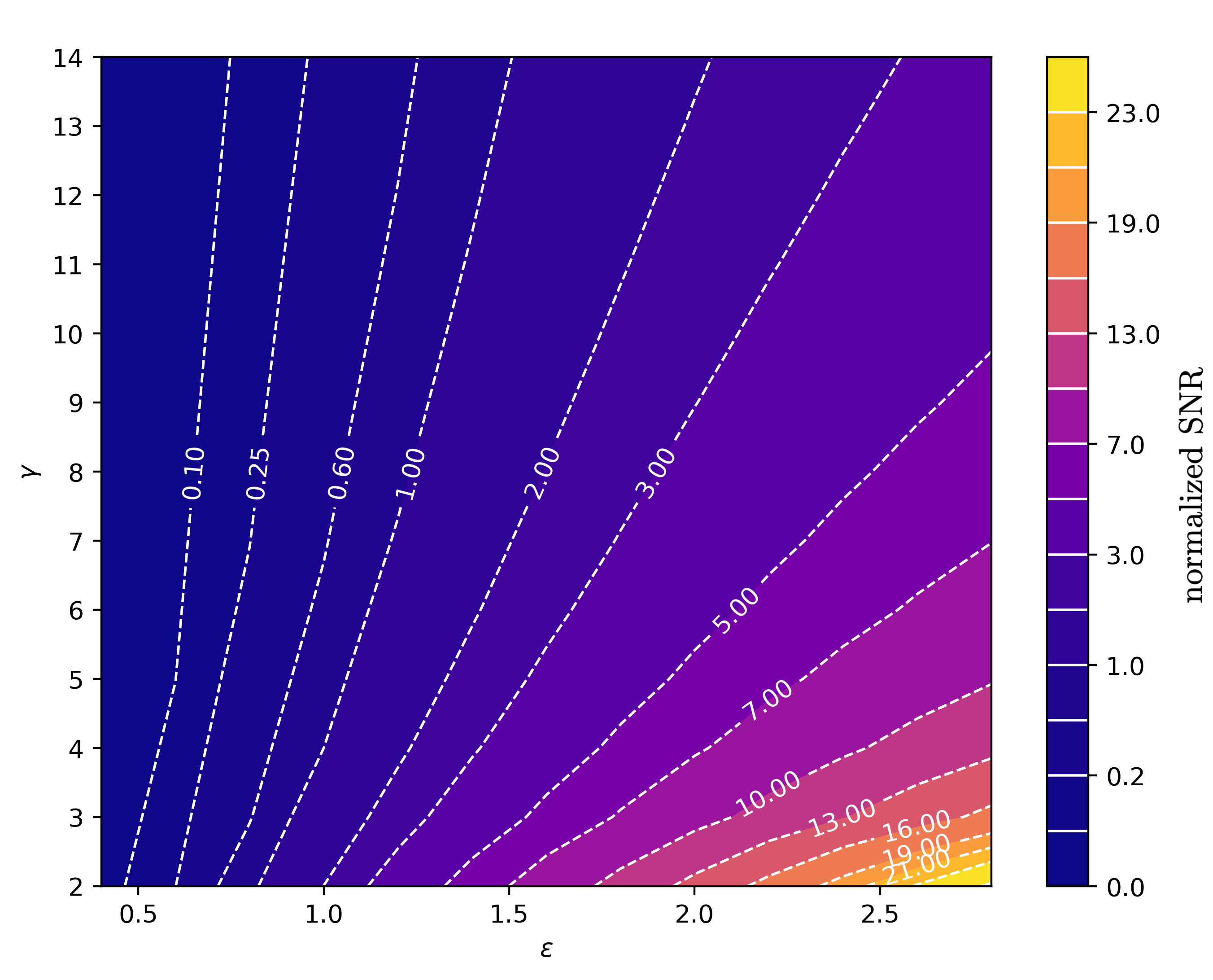}
	\caption{Normalized SNR for the echo of \textit{SXS:BBH:1936} \cite{Other_features_VIII}. Relevant parameters are fixed as $ \log M / \Msol = 6, z=1, \alpha = 8\pi, \delta = 0.2$. Further, $\epsilon,\gamma$ are varied. The resulting SNR is normalized to the SNR resulting from the choice $\epsilon=1,\gamma=4$, which is $\approx 86$.}
	\label{fig:ECHO_PARAM}
\end{figure}

Computing the SNR of the waveform echo requires fixing the reflectivity function $\RQBH$. The choice of the parameters involved can significantly alter the shape of the echo in interferometer data. The phenomenological constant $\alpha$ and the cusp-parameter $\delta$ influence the location and depth of the features corresponding to the characteristic frequencies (see Fig. \ref{fig:SNR_TDI}). The temperature coefficient $T_{QH}$ (or rather $\epsilon$) and boundary suppressor $\gamma$ directly impact the amplitude of the echo. The time dilation parameter $\beta$ is irrelevant for the SNR as it only shifts the echo along the time axis, i.e., it regulates the time separation between the echo and waveform time series as well as among echoes. Generally, one assumes that the time separation is large enough ($\beta$ is small enough) such that the echo does not interfere with the ringdown. \\
The first estimate of echo SNR applies the baseline parameter set $\alpha = 8\pi$, $\beta = 10^{-15}$, $\gamma = 4$, $\delta = 0.2$, and $\epsilon = 1$. The echo SNR is computed as a function of mass and redshift. The result for \textit{SXS:BBH:1936} is displayed in Fig. \ref{fig:SNR_ECHO}. If the mass and redshift ($M_\text{tot} = 10^{6}M_\odot$ and $z=1$) are fixed, but $\gamma$ and $\epsilon$ are varied, the echo SNR can deviate significantly from its baseline profile. It can be observed that lower $\gamma$ and larger $\epsilon$ yield a larger SNR. This is to be expected as these parameters drive the overall amplitude of the echo. A visualization of the functional dependence of the SNR on $\gamma,\epsilon$ is given by Fig. \ref{fig:ECHO_PARAM}. The latter exemplarily displays the echo SNR for the NR simulation \textit{SXS:BBH:1936} for fixed mass and redshift. Thereby, the SNR is normalized w.r.t. the baseline reflectivity parameters ($\gamma=4, \epsilon=1$). For a given merger determined by redshift and mass, the parameter-dependent echo SNR is given by the product of the SNR value in Fig. \ref{fig:SNR_ECHO} and the corresponding factor in Fig. \ref{fig:ECHO_PARAM}, depending on the choice of reflectivity. For instance, for a merger at redshifted mass $\log M / \Msol = 6$ and redshift $z=1$, the SNR for $\epsilon = 1$ and $\gamma = 6$ is given by $\T{SNR} \approx 86 \cdot 0.6 \approx 52$.

The results indicate that, generally, there is a vast regime in the reflectivity parameter space that suggests a potential echo detection with LISA. Given the expected extremely ``loud'' signals from very massive events, even very weak echoes that are strongly damped either by the boundary condition or the exponential decay of the reflectivity exhibit high SNR $\gtrsim 10$ in the conservative baseline of \cite{Henris_Mem} for LISA. As demonstrated in \cite{Henris_Mem}, depending on the population, the number of such ultra-loud events in the regime $M_\text{tot} = 10^{6}M_\odot$ and $z=1$ can reach up to $\mathcal{O}(100)$ over an observation period of $4$ years with LISA. \\
For favorable reflectivity parameters, the echo SNR can be boosted considerably. Note, however, that the absence of any echo detection by the LIGO collaboration so far (e.g., \cite{Westerweck_2018, Other_features_VII}) imposes constraints on the choice of the reflectivity parameters $\gamma,\epsilon$. Due to data analysis-related challenges, however, such constraints are rather vague, such that for the simulated regime $\gamma \in [1,14]$ and $\epsilon \in [0.2,2.7]$, the echo amplitude remains within an unconstrained regime ($\epsilon = 2, \gamma = 4$ roughly corresponds to a damping factor $\gamma = 0.4$ in \cite{Westerweck_2018}). \\
Finally, note that the above results are obtained independently of the time separation between the initial waveform and gravitational echo. Moreover, the cusp parameter $\delta$ and the location of the characteristic frequencies $\alpha$ play only marginal roles in the SNR computation.

\subsection{Measuring Characteristic Frequencies - a Smoking Gun for Black Hole Physics.}
\label{sec:Paper_LISA_II}
\begin{figure}[t]
	\centering
	\includegraphics[width=0.7\linewidth]{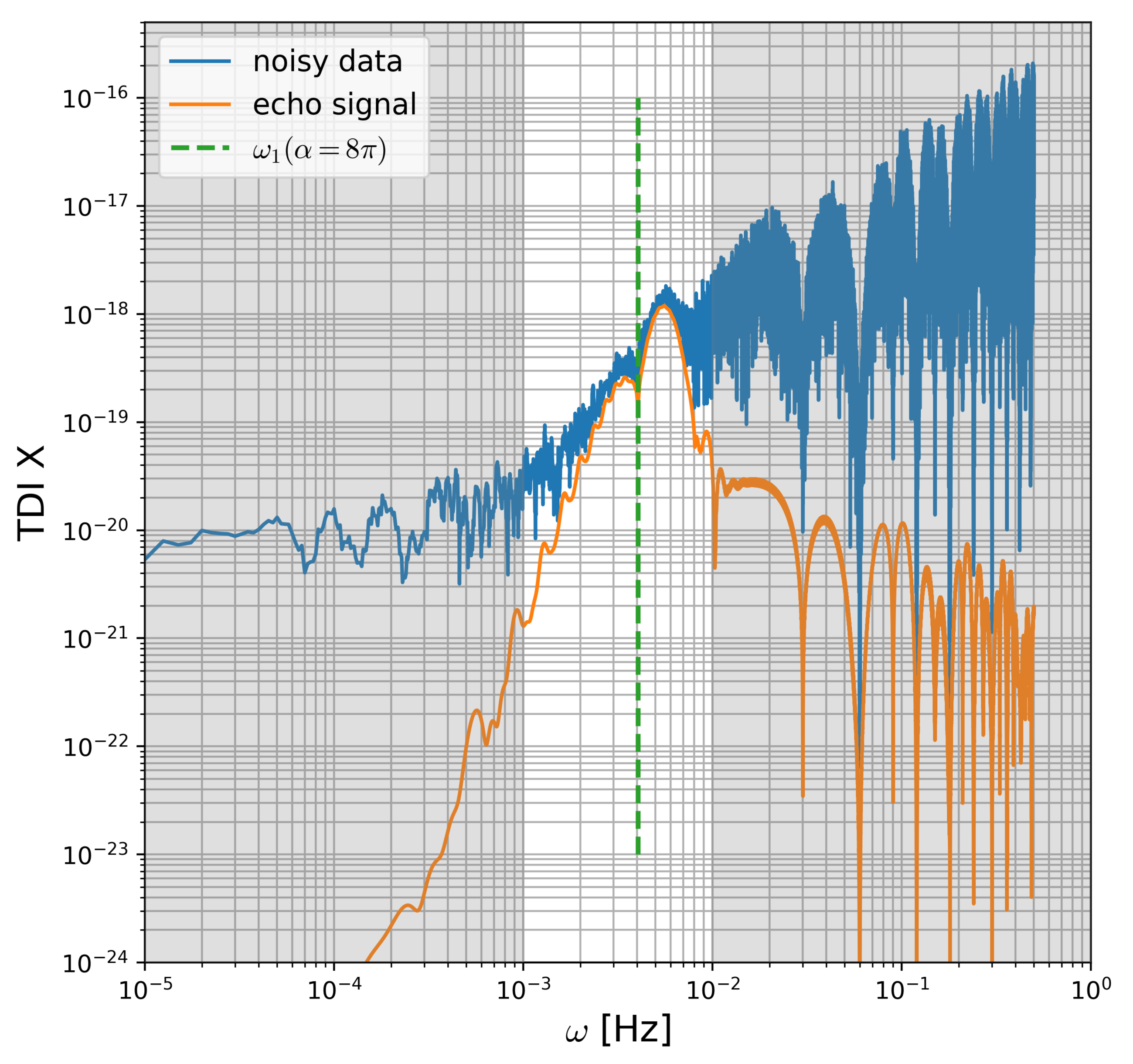}
	\caption{LISA TDI X channel data for the echo of \textit{SXS:BBH:1936} (including noise) with the baseline parameters for the reflectivity and $\delta = 0.5$ at redshift $z=1$ and $\log M_\text{tot}/M_\odot = 6$, including noise. The orange graph represents the signal without noise. The dashed line indicates the location of the first characteristic frequency of the QBH for the given mass and $\alpha$. The gray-shaded frequency domains are excluded from the fitting procedure. }
	\label{fig:SNR_TDI}
\end{figure}

Assuming the detection of an echo with LISA, valuable information can be extracted from its distinct features in the interferometer data. For example, by searching for the characteristic frequencies of the Bekenstein-Mukhanov spectrum, Eq. \eqref{equ:char_freq_init}, the origin of echo production can be narrowed down among the various potential phenomenologies. Consequently, this analysis focuses on features within the data collected in the \gls{tdi} channels associated with the characteristic frequency $\omega_N$. Detecting these characteristic frequencies serves as a definitive signature for BH physics, as their measurement would directly probe the area quantization of the BH event horizon. These frequencies correspond to the roots of the echo's strain in frequency space, which manifest as cusps (due to $\delta > 0$) in the TDI data, as shown in Fig. \ref{fig:SNR_TDI}. \\
To obtain a preliminary estimate for the detectability of the characteristic frequency, determined by the fundamental constant $\alpha$ (and spin $a$), with LISA, the uncertainty of the corresponding feature within the TDI X signal is computed. Specifically, the echo's transfer function, Eq. \eqref{equ:trans_func_sum}, is fitted to the simulated TDI X data in frequency space using a non-linear weighted least-squares method and a Markov-Chain-Monte-Carlo (MCMC) scheme. The MCMC thereby is applied on the Fourier transform of the echo strain time series convoluted the noise-contaminated data simulated for a realistic LISA measurement. To reduce the computational complexity, instead of simulating the TDI X data for each point in parameter space and fitting it against the latter, the essence of the signal, in particular the signature associated with the characteristic frequency, is isolated and represented by the transfer function. This can be done due to the shape of the echo strain in frequency space being accurately captured through the transfer function's morphology, \eqref{equ:trans_func_sum}. Based on this property, a rescaled version is selected as the fitting function in this procedure, with the rescaling parameter serving as an additional fit parameter. \\
The TDI data used for fitting includes simulated LISA noise based on a conservative noise model, the same model employed in \cite{Henris_Mem}. The fit function is informed by the noise's statistics via its Power Spectral Density (PSD). Fig. \ref{fig:SNR_TDI} displays an instance of noisy data (including the TDI X features of an echo) alongside the pure echo signal, absent of the corresponding merger waveform. The uncertainty associated with the fitting scheme is denoted by $\sigma_\T{fit}$, representing the standard deviation error of the fit parameter $\omega_1$.

To emphasize the robustness of the analysis, it is performed over 20 distinct noise realizations. Empirically, it is found that the scatter of the 20 recovered frequencies $\omega_1$ is statistically consistent with the theoretical errors $\sigma_\T{fit}$. Furthermore, to obtain reliable estimates for detectability, the uncertainty $\sigma_\T{fit}$ is calculated using a weighted least squares method, complemented by a more comprehensive Bayesian MCMC analysis. Both the MCMC and fitting procedures are carried out without prior knowledge of the location of the characteristic frequencies. The MCMC analysis is initiated based on a no-cusp scenario, i.e., $\alpha=0=\delta$. The parameter ranges for $\alpha$ and $\delta$ in both methods are set to span $\alpha \in [4\log 2, 8\pi]$ and $\delta \in (0.05, 0.6]$. The chosen range for $\alpha$ includes all phenomenological constants found in the literature, while the range for $\delta$ is sufficiently large to cover a wide variety of phenomenologies. The uncertainties derived from the MCMC analysis are consistent with an expected likelihood approach, where the likelihood function is weighted by the noise PSD, and the noise realization in the data converges to zero.\\
Additionally, the analysis considers an event with $\log M/M_\odot = 6$ and $z=1$ to ensure a sufficient echo SNR. This selection sets the peak of the echo's frequency signal within the frequency domain (as shown in Fig. \ref{fig:SNR_TDI}). In general, the remnant mass can significantly impact the uncertainty. Increasing (or decreasing) the mass shifts the characteristic frequency to lower (or higher) values, as $\omega_N \sim 1/M$. For approximately $\log M/M_\odot \gtrsim 7.5$ ($\log M/M_\odot \lesssim 5.5$), the frequency moves outside the LISA sensitivity band. Therefore, the mass range where echoes are strong and where the characteristic frequency can potentially be identified aligns well with LISA's sensitivity.

\begin{figure}
	\centering
	\includegraphics[width=0.7\linewidth]{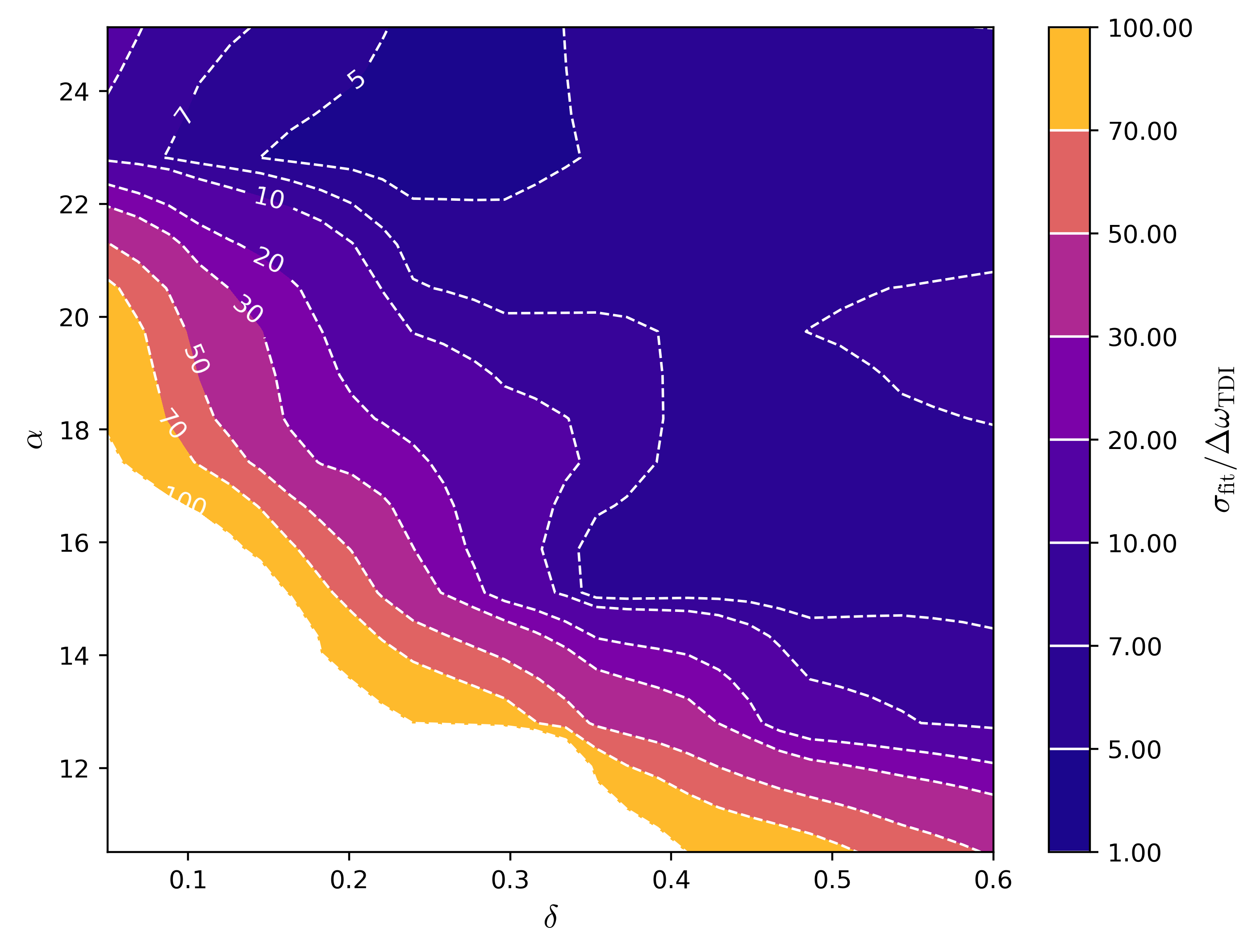}
	\caption{Uncertainty of the characteristic frequency normalized by the TDI frequency resolution ($\approx 2.5$ $\mu$Hz) and extracted from a fit to the TDI data of echo of the simulated waveform \textit{SXS:BBH:1936} \cite{Other_features_VIII}. One fixes $ M/M_\odot = 10^6, z=1, \epsilon=1, \gamma=4$ and varies $\alpha,\delta$.}
	\label{fig:SNR_PEAK_PARAMS}
\end{figure}

The characteristic frequency fit is performed w.r.t. $\omega_1$ and $\delta$, where the former can be converted to $\alpha$ using Eq. \eqref{equ:characteristic_frequency}. Apart from $\epsilon$ and $\gamma$, the other reflectivity parameters do not affect the amplitude of the signal and are thus irrelevant to the fitting procedure. For the analysis, $\epsilon = 1$ is selected as it corresponds to the phenomenologically preferred value (where $T_\T{QH} = T_\T{H}$), as explained earlier. To recover a boundary condition similar to those of ECOs, $\gamma = 4$ is chosen. Since $\epsilon$ and $\gamma$ directly influence the echo's amplitude, larger values of $\epsilon$ and smaller values of $\gamma$ generally result in smaller uncertainties for the identification of $\omega_1$, as long as the fitting precision is limited by noise, i.e., for $\epsilon \lesssim 1.5$ and $\gamma \gtrsim 5$. For $\epsilon > 1.5$ and $\gamma < 5$, the echo's signal dominates the conservative noise model, and the estimate of $\omega_1$ is constrained only by the frequency resolution of the TDI channel. \\
The uncertainties are computed for $\alpha \in [4 \log 2, 8\pi]$ and $\delta \in [0.05, 0.6]$. The results show that, for most of the parameter space considered, the amplitude of the echo's signal in the TDI X channel is sufficiently large to resolve the characteristic frequency within $5\sigma_\T{fit}$ of the true value. In the tested regime, where $M/M_\odot = 10^6$ and $z = 1$, the characteristic frequencies lie between $1$ and $4$ mHz. Given a frequency resolution $\Delta \omega_\T{TDI}$ of approximately $2.5$ $\mu$Hz, the uncertainties range from 5 to 100 $\Delta \omega_\T{TDI}$. It is specifically found that for large values of $\alpha$ (around $8\pi$), or equivalently $\omega_1 \approx 4$ mHz, the uncertainties remain small across the tested values of $\delta$. In the parameter regime where the frequency resolution becomes the limiting factor, uncertainties (i.e., $\sigma_\text{fit}$) could theoretically be reduced by further improving the frequency resolution of the corresponding TDI channel, $\Delta \omega_\T{TDI}$. However, the TDI frequency resolution is limited by the duration of the signal, i.e., the duration of the echo. \\
Additionally, a trade-off w.r.t. the reflectivity parameters is observed when computing the uncertainty: smaller values of $\alpha$ shift the characteristic frequencies in the TDI data toward the low-frequency noise-dominated regime, resulting in larger uncertainties unless the features are more pronounced (i.e., $\delta$ is larger). The presence of higher-order $\omega_N$ can only marginally offset the noise domination of the feature associated with $\omega_1$, as it becomes impossible to distinguish $\omega_2$ from $\omega_1$ when the latter is hidden behind the noise. Therefore, it is expected that the small-$\alpha$ sector of the quantum gravity theory space will generally be less tightly constrained by echo searches. Finally, it is worth noting that these results hold for all the tested waveforms listed above. For \textit{SXS:BBH:1936}, they are displayed in Fig. \ref{fig:SNR_PEAK_PARAMS}.

Summarizing the results from the preceding two Subsections \ref{sec:Paper_LISA} and \ref{sec:Paper_LISA_II}, it is shown that, even with a conservative baseline and moderate echo amplitude, LISA will be capable of detecting \gls{gw} echoes originating from non-trivial QBH reflectivity. Importantly, the chosen range of reflectivity parameters also includes values typically used to describe ECOs, thereby extending the results to both scenarios. Consequently, the absence of echo detection would impose significant constraints on both BH and ECO phenomenology. The forecast for echo detectability is independent of the specific QNM content and has been evaluated across 11 NR simulations. For the mergers tested which involve non-negligible remnant spin, no significant modifications to the echo are observed when compared to spin-less remnants. Based on this detection potential, it is demonstrated that, across a broad spectrum of masses, redshifts, and binary BH mergers, the features in the LISA TDI data corresponding to characteristic frequencies are likely to be detectable with high precision. This outcome suggests that future \gls{gw} instruments such as LISA will provide unprecedented probes of spacetime's quantum structure, magnified by BHs. A direct measurement of the characteristic absorption frequency of BHs would offer a unique test of the BH mass-area relation and provide stringent constraints on QG theories. The non-detection of absorption frequencies in echo data would imply a more membrane-dominant QBH phenomenology strongly hinting towards a favorable interpretation of the BH information paradox. In general, the detection of echo signals is essential for gaining crucial insights into BH physics and may provide direct evidence for QG effects, highlighting the necessity of including GW echo searches in the scientific agenda of future GW experiments.

At this stage, it is essential to recognize that several critical assumptions in this work (specifically, the validity of the chosen boundary condition and the reflectivity function $\RQBH$) necessitate further refinement, supported by both experimental and theoretical considerations. Nevertheless, for the purposes of this investigation, these assumptions are deemed adequate, and a more rigorous derivation of the mechanisms underlying $\RQBH$, which may include quantum information theoretical aspects, is deferred to future studies. Finally, although the analysis presented has demonstrated robustness against significant variations in certain model parameters, the issue of model dependence in echo searches is discussed in the following section.

%
%
%


\subsection{Quantum Corrections to the Gravitational Wave Memory}
\label{sec:Paper_Mem}

Following the introduction to this Chapter, this section aims to demonstrate how the balance flux laws can be utilized to correct the gravitational waveform derived from NR simulations by incorporating quantum features. Specifically, the feature of interest in this section is the \gls{gw} echo. In Section \ref{sec:quantum_BH}, a procedure for adding the echo to the \gls{gr} waveform is outlined. This procedure addresses the linear order of the gravitational strain, while neglecting corrections arising in the non-linear regime. The latter is further complicated by the fact that most numerical waveforms do not fully include non-linear information, even in the classical regime; for instance, the displacement memory is typically added post-waveform generation \cite{waveform_test_BL_II}. The goal of the ensuing discussion is to incorporate echo corrections into the non-linear regime of the numerical waveform, or, more precisely, to add quantum corrections to the \gls{gw} memory.

To this end, it is instructive to present a new perspective on the memory. A variety of approaches to the GW memory have been explored in the literature \cite{First_hint_Mem, Christodoulou_Mem, Memory_paper_I, Memory_paper_VI, Memory_paper_V, Memory_paper_II}, with the equivalence of the resulting expressions for memory being well established. In the context of numerical waveforms, a particular formulation of the gravitational memory based on energy-momentum or balance laws \cite{waveform_test_BL_III} (see also \cite{waveform_test_BL_IV, Our_Review} for further technical details) has become widely adopted \cite{Mitman_2020, Khera:2020mcz, waveform_test_BL_II}. A thorough derivation of this formulation can be found in Chapter \ref{chap:asymptotics}. Notably, these formulations include expressions for both the linear and non-linear GW memory, as shown in Eq. \eqref{eq:mem_total_def} and below. The flux balance laws have been applied in previous studies for waveform analysis and NR applications, among others \cite{waveform_test_BL_I, waveform_test_BL_V, Borchers:2021vyw}. \\
For the numerical analysis of this section, the memory components are derived from the BMS balance laws, as implemented in Eq. \eqref{equ:kristin}. This choice is motivated by the applicability of these equations to NR waveforms and the convenient presentation format, namely in terms of the strain time series. Specifically, the memory components, first outlined in \cite{waveform_test_BL_II}, allow for the direct computation of time-dependent memory corrections for each individual harmonic strain mode. Given that the BMS laws are derived from full, non-linear GR, it is expected that these equations will be equally applicable to any strain (quantum) corrections that emerge within the regime where GR serves as an effective theory.\\
The desired memory equation, given by (9) in \cite{waveform_test_BL_II}, is directly derived from the balance flux laws Eq. \eqref{equ:kristin} by using
\begin{align}
\Im{\eth^2h^\circ} = 4\Im{\Psi_2^\circ + \frac{1}{4}\dot h^\circ \bar h^\circ}  \,,  
\end{align}
which can be directly obtained from Eq. \eqref{equ:ImPsi_2}.
The time series containing only the memory-induced contributions to the strain, after a few simplifications, can be written as
\begin{align}\label{equ:strain_mem}
    \hmem = \frac{1}{2} \bar{\eth}^2 \mathfrak{D}^{-1}\left[\frac{1}{4}\int_{u_i}^u \dd u |\dot h^\circ|^2 - \Delta\left(\Psi_2 + \frac{1}{4}\dot h ^\circ\bar h^\circ\right)\right]\,,
\end{align}
where $\mathfrak{D}^{-1}$ is defined via
\begin{align}
    \mathfrak D = \frac{1}{8} D^2(D^2 + 2)\,,\\
    D^2 = \bar \eth \eth\,,
\end{align}
and $\eth$ defines the spin-weighted derivative operator whose precise definition can be found in~\cite{waveform_test_BL_III, waveform_test_BL_II, Our_Review}. The ``$\Delta$'' for a tensor or scalar thereby denotes the change in time, i.e., $\Delta f \equiv f(u_2)-f(u_1)$. The quantity $h^\circ$ appearing in Eq. \eqref{equ:strain_mem} represents the full GW strain. It is fundamentally dependent on the line-of-sight, which spans from the detector to the binary frame. This frame can be described by two angles, $\theta$ and $\phi$, that specify a point on the celestial sphere. The initial time $u_i$ in Eq. \eqref{equ:strain_mem} corresponds to the beginning of the gravitational waveform. In this section, numerical waveforms are considered; by construction, the precise definition of Eq. \eqref{equ:strain_mem}, which would require taking the limit $u_i \rightarrow -\infty$, is not used. Instead, $u_i$ is identified with the starting time of the provided NR waveforms. This truncation is compensated by an angular-dependent constant $\alpha$ \cite{waveform_test_BL_II}\footnote{Additionally, frame-related issues might contribute to constant angular-dependent terms \cite{waveform_test_BL_II}. Here, both contributions are summarized into a single term.}.\\
The full memory given by Eq. \eqref{equ:strain_mem} can be separated into a dominating term associated with the energy flux and an oscillatory term, 
\begin{subequations}
\label{equ:mem_gross}
\begin{align}
    h_{\mathcal E} = \frac{1}{2} \bar \eth^2 \mathfrak D^{-1} \left[ \frac{1}{4} \int_{u_i}^u \dd u |\dot h^\circ|^2 \right] + \alpha\,,\\
    h_{\T{osc.}} = \frac{1}{2} \bar \eth^2 \mathfrak D^{-1} \Delta\left[- \left(\Psi_2^\circ + \frac{1}{4} \dot h^\circ\bar h^\circ\right) \right]\,,
\end{align}
\end{subequations}
respectively. 
While the latter equations both describe the displacement memory that can be directly related to the supertranslation part of the asymptotic symmetry group at $\scrip$ (e.g., \cite{Flanagan_2017, IR_triangle_III}), one may also consider spin memory contributions \cite{Flanagan_2017, Spin_mem, Mem_outlook_I} for completeness. The spin memory in turn, is connected to the superrotations subgroup. Both superrotations and supertranslations each generalize the know translation and rotation symmetries included in the Poincar\'e group respectively by an infinite tower of ``new'' generators. For details, we refer to \cite{Flanagan_2017}\footnote{Note that there exist recent works \cite{Mem_outlook_V} proposing to experimental infer relevant spacetime symmetries with future \gls{gw} instruments by measuring different types of memory.}. 

While the latter equations describe the displacement memory (e.g., \cite{Flanagan_2017}) one may also consider spin memory contributions \cite{Spin_mem} for completeness. Rewriting both types of memories in similar form to Eq. \eqref{equ:strain_mem} and rearranging individual contributions, one can write them as \cite{waveform_test_BL_II, Mitman_2020}
\begin{subequations}
\label{equ:mem_net}
    \begin{align}
        h_m &= \frac{1}{2} \bar \eth^2 \mathfrak D^{-1} \Delta m\,,\\
        h_\mathcal{E} &= \frac{1}{2} \bar \eth^2 \mathfrak D^{-1} \left[ \frac{1}{4} \int_{u_i}^u \dd u |\dot h^\circ|^2 \right] + \alpha\,,\\
        h_{N} &=\frac{1}{2} i\bar \eth^2 \mathfrak D^{-1} D^{-2} \Delta \T{Im}\left[\bar \eth (\partial_u N)\right]\,,\\
        h_\mathcal{J} &=\frac{1}{2} i\bar \eth^2 \mathfrak D^{-1} D^{-2} \Delta\Im{ \frac{1}{8}\eth(3h^\circ\bar \eth \dot{\bar{h}}^\circ - 3 \dot h^\circ \bar \eth \bar h^\circ + \dot{\bar{h}}^\circ\bar \eth h^\circ- \bar h ^\circ\bar \eth \dot h^\circ)} \,.
    \end{align}
\end{subequations}
Here, $m$ and $N$ correspond to the Bondi mass aspect and the angular momentum aspect\footnote{Not to be confused with the Bondi News tensor, Eq. \eqref{eq:bondiNewsexpansion}.} respectively, i.e., in terms of strain and Newman Penrose scalar they read
\begin{subequations} 
    \begin{align}
        m &= -\Re{\Psi_2^\circ + \frac{1}{4}\dot h^\circ\bar h^\circ}\,\\
        N &= 2 \Psi_1^\circ - \frac{1}{4} \bar h^\circ \eth h^\circ - u \eth m - \frac{1}{8}\eth(h^\circ\bar h^\circ   )\,.
    \end{align}
\end{subequations}
While, in principle, deviations induced by echoes in the strain time series can be computed for all components appearing in Eq. \eqref{equ:mem_net}, the primary focus in the following is on the components associated with the energy and angular momentum flux across the horizon $\scrip$, namely $h_\mathcal{E}$ and $h_\mathcal{J}$. Both components are related to the \textit{null memory} (also referred to as \textit{non-linear memory}, compare to previous discussions, Chapter \ref{chap:asymptotics}), but exhibit different parity: $h_\mathcal{E}$ corresponds to the electric part, while $h_\mathcal{J}$ corresponds to the magnetic part. For practical applications, it is generally sufficient to consider only $h_\mathcal{E}$, as it constitutes the dominant contribution to the non-linear memory. However, given the interest in the kinematical properties of the binary under consideration, the memory component associated with the angular momentum flux is also included in the analysis.

Applying the notation of Section \ref{sec:quantum_BH}, one can identify and separate contributions in $h_\mathcal{E}$ and $h_\mathcal{J}$. Start by considering $h_\mathcal{E}$, where Eq. \eqref{equ:strain_decomposition} leads to mixed terms of the form
\begin{align}
    \dot h ^{\T{eff}} \dot{\bar{h}}^\T{eff} = |\dot{h}^\infty|^2 + \dot{h}^\infty\dot{\bar{h}}^\T{echo} + \dot{\bar{h}}^\infty\dot{h}^\T{echo} + |\dot{h}^\T{echo}|^2.
\end{align}
Given that both $\hecho$ and $\hnorm$ are time series waveforms, it is reasonable to assume that the echo—and consequently its time derivative—is temporally well separated from the principal waveform $\hnorm$, as illustrated, for example, in Fig.~\ref{fig:ECHO} and Fig.~\ref{fig:waveform_QBH} of Appendix \ref{app:ECHO}. Under this assumption, the relations $\dot{h}^\infty\dot{\bar{h}}^\T{echo} = \dot{\bar{h}}^\infty\dot{h}^\T{echo} = 0$ hold, and the memory $h_\mathcal{E}$ naturally decomposes into a waveform and an echo contribution. The assumption of temporal separability requires a sufficiently large phase shift encoded in the reflectivities $\Reco$ and $\RQBH$. The corresponding parameter constraints can be estimated as $\gamma \lesssim 10^{-4}$ (for $T_\T{QH} = T_\T{H}$, and $T_\T{QH} \lesssim 5 T_\T{H}$ for $\gamma = 10^{-15}$) in the case of an echo from an ECO. For echoes generated by QBHs, one finds $\beta \lesssim 10^{-7}$. For the remainder of this section, the analysis is restricted to a range in parameter space where this no-overlap assumption holds valid, such that
\begin{align}
\label{equ:mem_electic_echo}
    \hfull_{\mathcal{E}} &= \frac{1}{2} \bar \eth^2 \mathfrak D^{-1} \left[ \frac{1}{4} \int_{u_i}^u \dd u |\dot h^\T{echo}|^2  + \frac{1}{4} \int_{u_i}^u \dd u |\dot h^\infty|^2 \right]\notag\\
    &=: \hecho_\mathcal{E} + \hnorm_\mathcal{E}\,,
\end{align}
and where the constant $\alpha$ is omitted.

\begin{figure}[!t]\centering
\includegraphics[width=0.6\columnwidth]{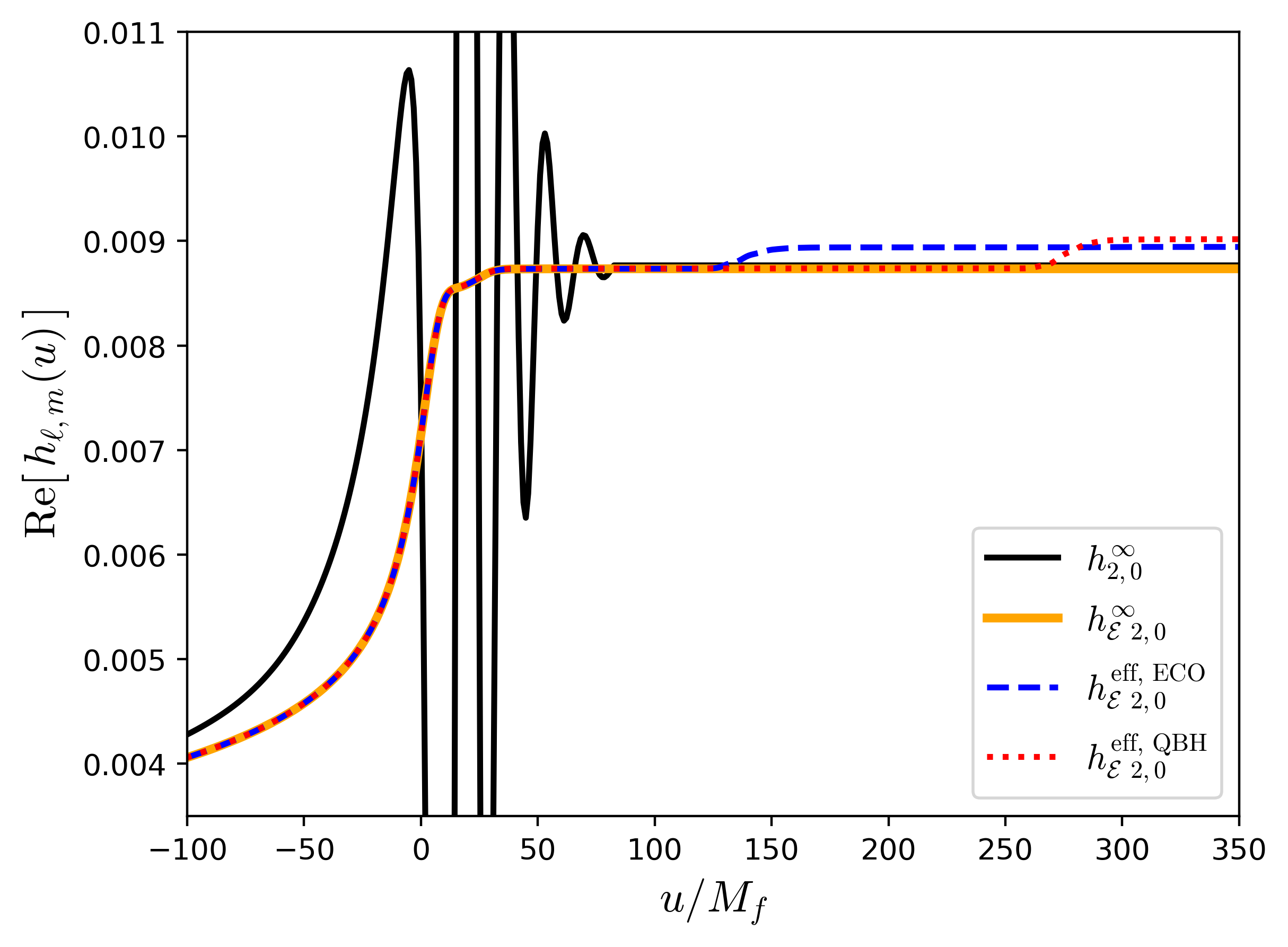} 
\includegraphics[width=0.6\columnwidth]{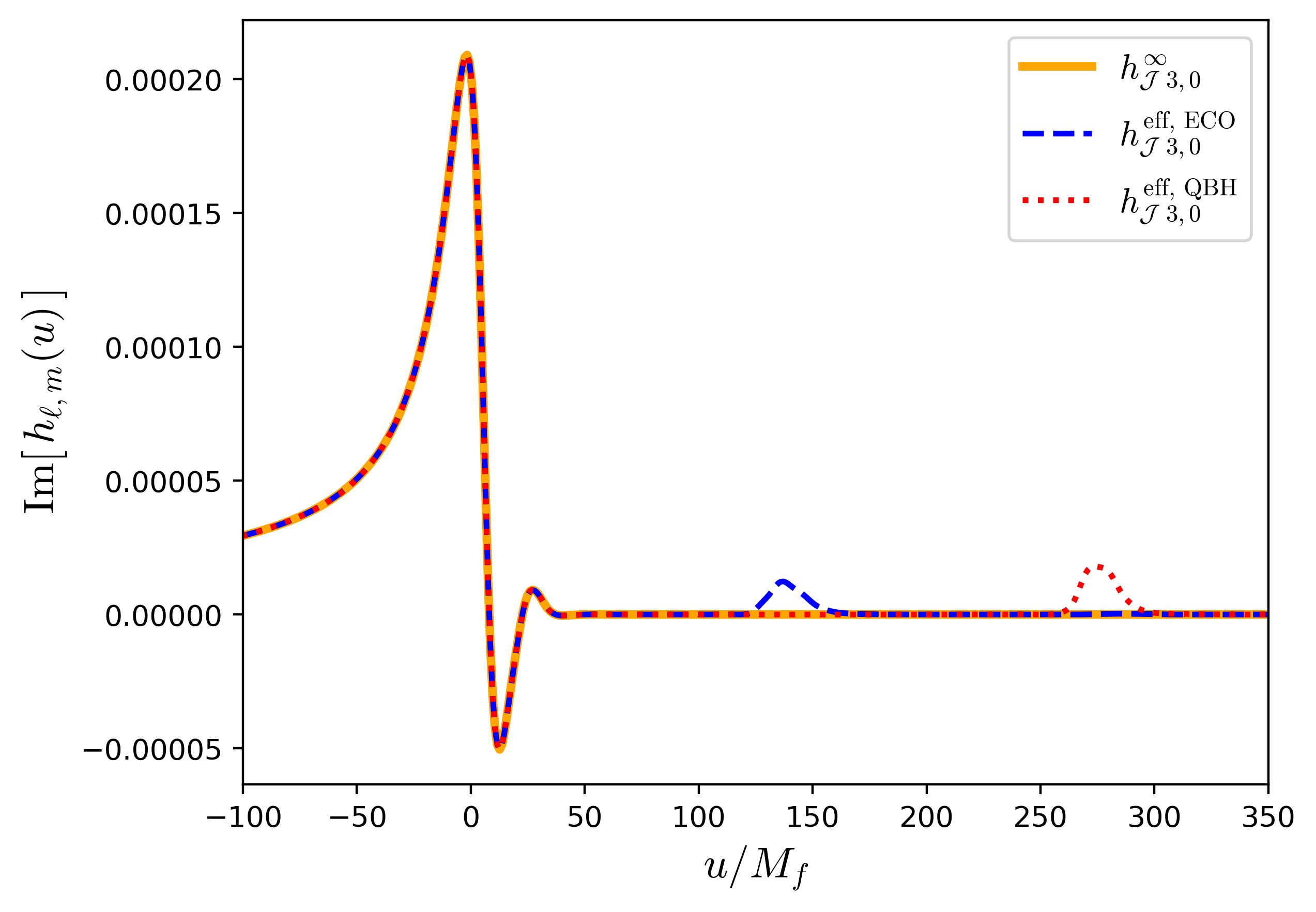}   
     \caption{Selected modes of the memory corrected waveform \cite{Maibach2025}. The dashed and dotted lines mark the memory contributions $h_\mathcal{E}$ and $h_\mathcal{J}$ for echoes from the ECO and QBH reflectivity models. The ``no-echo'' counterpart is displayed in both the bottom and top plots as the solid orange line. The parameter choice and the event are adopted from Fig. \ref{fig:waveform_QBH}.}
       \label{fig:mem_echo}
\end{figure}

Next, consider $h_\mathcal{J}$. Structurally, the main difference w.r.t. $h_\mathcal{E}$ are the angular derivative $\eth h$. One can argue that the derivative operator only acts on the angular dependent part of the strain, i.e., it can be decomposed
\begin{align}
    h(u,\theta,\phi) = \sum_{\ell,m } h_{\ell m}(u) \,_{-2}Y_{\ell m}(\theta, \phi)\,,
\end{align}
with
\begin{align}
    \eth\,_sY_{\ell m} = \sqrt{(\ell -s)(\ell +s+1)}\,_{s+1}Y_{\ell m }\,\\
    \bar \eth\,_sY_{\ell m} = - \sqrt{(\ell +s)(\ell -s+1)}\,_{s-1}Y_{\ell m }\,,
\end{align}
such that the strain modes, and thus $\Zin$ ($\Yin$), as functions of retarded (advanced) time, are not affected. Consequently, even with the angular derivative operator $\eth$ applied on the strain, the separation argument still holds, and terms like $\eth \hnorm \bar \eth \hecho$ vanish due to the absence of non-trivial time series overlap. Thus, inserting \eqref{equ:strain_decomposition} in $h_\mathcal{J}$ it is found that $\hecho$ does not mix with $\hnorm$ and thus 
\begin{align}
    \hfull_\mathcal{J} = \hecho_\mathcal{J} + \hnorm_\mathcal{J}\,,
\end{align}
where $\hecho_\mathcal{J}$ is given by
\begin{align}
\label{equ:mem_angular_echo}
    \hecho_\mathcal{J} &=\frac{1}{16} i\bar \eth^2 \mathfrak D^{-1} D^{-2} \T{Im}\left\{\eth(3h^\T{echo}\bar \eth \dot{\bar{h}}^\T{echo}\right.\notag\\
    &\left.- 3 \dot h^\T{echo} \bar \eth \bar h^\T{echo}+ \dot{\bar{h}}^\T{echo}\bar \eth h ^\T{echo}- \bar h^\T{echo} \bar \eth \dot h^\T{echo})\right\}\,.
\end{align}
A comparison between $\hecho_\mathcal{J}$, $\hecho_\mathcal{E}$ and $\hnorm_\mathcal{J}$, $\hnorm_\mathcal{E}$ is presented in Fig.~\ref{fig:mem_echo}, where the temporal separation of the effects is clearly visible. It is observed that, for $h_\mathcal{E}$, the waveform and echo contributions share an identical shape up to a rescaling in amplitude. In contrast, the corresponding contributions to $h_\mathcal{J}$ exhibit fundamentally different structures. Furthermore, if the no-overlap assumption is violated, the step-like behavior characteristic of the non-linear memory undergoes non-negligible modifications due to the mixing terms $\dot{\bar{h}}^\infty\dot{h}^\T{echo} + \text{c.c.}$, which are generally oscillatory in nature. This feature constitutes a fundamental manifestation of the memory's non-linear character. Consequently, an echo may substantially modify the conventional memory contribution associated with BBH waveforms.

For the computation of the memory, Eq.~\eqref{equ:mem_angular_echo} completes the derivation of all required expressions for the numerical evaluation of echo-induced memory corrections. Before proceeding to the main analysis, it is worth emphasizing that the memory, as described above, possesses a clear physical interpretation that proves valuable when examining the energy balance within the cavity formed between the BH potential barrier and the near-horizon structure. A particularly relevant question in this context is the following: How much energy is reflected by the BH or ECO? Naturally, the answer depends on the reflectivity function specific to the underlying model. To establish a numerical reference, the energy and angular momentum fluxes associated with the echo are also computed in this section. In this framework, Eq.~\eqref{equ:mem_net} enables the extraction of both energy and angular momentum fluxes carried by the GWs. Since these equations decompose into components associated with the initial strain and the echo contributions, the corresponding flux expressions follow the same structure.
Concretely, for a generic strain, the (dimensionless) energy and momentum flux per unit time and angle read \cite{Mitman_2020}
\begin{subequations}
\label{equ:fluxes}
\begin{align}
    \frac{\dd E}{\dd \Omega \dd u} &= \frac{1}{16\pi} {|\dot h|^2} \,,\\
    \frac{\dd J}{\dd \Omega \dd u} &= \frac{1}{16\pi} \left(3\bar h \eth \dot{h} - 3 \dot{\bar{h}} \eth  h + \dot{h} \eth \bar h -  h \eth \dot{\bar{h}}\right)\,.
\end{align}
\end{subequations}
The flux formulas in Eq.~\eqref{equ:fluxes} are defined in terms of asymptotic strains evaluated at $\scrip$, thereby justifying the use of the retarded time coordinate $u$. Strictly speaking, in Eq.~\eqref{equ:fluxes}, the strain $h$ should be interpreted as the asymptotic strain $h^\circ$, which is defined exclusively at $\scrip$ and extended into the physical spacetime through Eq.~\eqref{equ:asym_strain}. This distinction becomes particularly relevant when computing the dimensionful versions of Eq.~\eqref{equ:fluxes}, in which case the asymptotic strain must be replaced with its physical counterpart. For practical applications, such as observational measurements of \gls{bbh} mergers, the radial coordinate is substituted with the estimated luminosity distance $D_L$ of the event, as previously discussed in Section~\ref{sec:Paper_BL}. Naturally, to ensure dimensional consistency, appropriate powers of $c$ and $G$ must be introduced alongside the distance rescaling.

When computing \eqref{equ:fluxes}, it may be helpful to make use of strain's decomposition into spherical harmonics outlined, for instance, in \cite{waveform_test_BL_I} (see also Appendix \ref{app:decomp}). For instance, for the memory flux associated with the echo one computes
\begin{align}
    \frac{\dd E^\T{echo}}{\dd \Omega \dd u} &= \frac{1}{16\pi} |\dot{h}^\T{echo}|^2\,,
\end{align}
which can be decomposed into spherical harmonics of spin-weight zero,
\begin{align}
     |\dot{h}^\T{echo}|^2 = \sum_{\ell, m} \alpha^\T{echo}_{\ell m} Y_{\ell m}(\theta, \phi)\,.
\end{align}
The resulting frequency-dependent expansion coefficients, up to a prefactor depending on $\ell$ and $m$ (see Appendix \ref{app:decomp} for details), read 
   \begin{align}
     \alpha^\T{echo}_{\ell m \omega} \sim \sum_n\sum_m&\frac{1}{\omega^4}\left(\frac{C}{D}\right)^2(\frac{(-1)^{\ell + m +1 }}{\zeta}\RQBH\RBH)^{n+m}\frac{1}{D^\T{in}_{\ell_1 m_1 \omega}D^\T{in}_{\ell_2 m_2 \omega}}\notag\\
     &\cdot\partial_v\mathfrak F\left[C^\T{in}_{\ell_1 m_1} \Psi_{0, \ell_1 m_1}(v)\mathcal{F}(v)\right]\partial_v\mathfrak F\left[\overline{C^\T{in}_{\ell_2 m_2} \Psi_{0, \ell_2 m_2}(v)\mathcal{F}(v)}\right]\,.
\end{align} 
A similar expression holds for the angular momentum flux ${J}^\T{echo}$, which is formulated in terms of the coefficients $\beta^\T{echo}_{\ell m \omega}$. The coefficients $\alpha^\T{echo}$ and $\beta^\T{echo}$ differ only by a mode-dependent factor, the explicit form of which is provided in Appendix~\ref{app:decomp}. It is important to emphasize that this decomposition is equally applicable to the memory expressions given in Eqs.~\eqref{equ:mem_electic_echo} and~\eqref{equ:mem_angular_echo}.

In the context of echo generation within the cavity formed by the BH potential barrier and the surface of the ECO (or QBH), a natural question arises as to whether the associated GWs adhere to a form of (energy) flux conservation for gravitational radiation. This can be directly addressed by referring to Fig.~\ref{fig:Sketch_intuition}. Consistent with the notation used in the previous section, the perspective of the QBH is adopted, with analogous considerations applying to the ECO case. The process begins with the ingoing gravitational radiation $\Yinqbh{}^{,1}$ interacting with the reflective surface. This interaction results in partial reflection, producing the outgoing wave $\Zoutqbh{}^{,1}$, and partial transmission, which generates $\ZBH{}^{,1}$ that propagates towards the BH horizon $\Hp$. The outgoing wave $\Zoutqbh{}^{,1}$ then encounters the BH potential barrier, where it may either be reflected or transmitted. If transmitted, the radiation escapes to future null infinity, where it is detected as the first echo $\Zinf{}^{,1}$. The reflected portion, however, forms a new ingoing wave $\Yinqbh{}^{,2}$, beginning the next cycle within the cavity. This iterative process underlies the production of successive echoes, characterized by the continuous exchange of radiative energy between the reflective surface and the BH potential barrier.
Propagating this configuration infinitely far into the future, one concludes that the energy flux carried by the initial strain directed toward the QBH's surface must equal the sum of the energy flux passing through both horizons, $\Hp$ and $\scrip$, i.e., 
\begin{align}
\label{equ:energy_balance}
    \frac{\dd E^\T{out QBH}}{\dd t}+ \frac{\dd E^\T{horizon}}{\dd t} = \frac{\dd E^\T{in QBH}}{\dd t} \,.
\end{align}
Naturally, the energy flux conservation equation can be expressed in terms of the amplitudes at each stage of the GW's interaction with the barriers or horizons. For instance, combining the configuration depicted in Fig. \ref{fig:Sketch_intuition} with the radial solutions close to the ECO, one could also write
\begin{align}
\label{equ:some_equ_}
    \frac{\dd E^\T{in QBH}}{\dd t}- \frac{\dd E^\T{out QBH}}{\dd t} =\frac{\dd E^\infty{}^\T{\,in}}{\dd t}- \frac{\dd E^\infty{}^\T{\,out}}{\dd t}\,,
\end{align}
which is similar to the formulations found in \cite{Teukolsky_1974, Chen_2021}. Note that in Eq. \eqref{equ:some_equ_}, the \textit{up solution} was supplemented by a set of ingoing waves at infinity, $\Yinf{}^\T{in}$. Thus, $E^\infty{}^\T{\,in}$ is proportional to the absolute value of $(\Yinf{}^\T{in})^2$, while $E^\infty{}^\T{\,out} \propto |\Zinf|^2$. It is important to highlight that the notation for radial solutions can vary across the literature, particularly when the Sasaki-Nakamura (SN) formalism \cite{Sasaki_2003} is employed instead of the Teukolsky framework. For more detailed discussions on echo computations using the SN formalism, the reader is directed to references such as \cite{Teukolsky_1974,Chen_2021,Xin_2021,Hughes_2000}.

For the fluxes reaching $\scrip$, the definitions provided in Eq. \eqref{equ:fluxes}, in conjunction with the asymptotic strain, yield the correct result. In cases where the asymptotic strain is not directly available, Eq. \eqref{equ:strain_to_Z} can be used to extract it from $\Psi_4^\circ$. However, defining the flux that falls into the BH or ECO, or the initial flux directed toward the surface, is more complex. Specifically, to compute the energy carried by $\Yinqbh$, one must convert $\Psi_4$ to $\Psi_0$ (since it pertains to the perturbations of the shear on the horizon) using the TS identities. Afterward, the energy flux can be derived from the change in area according to Hartle's formula for BH area increase. After a lengthy computation, one finds \cite{Teukolsky_1974, Chen_2021}
\begin{align}
\label{equ:energy_dE_in_ECO}
    \frac{\dd E^\T{in QBH}}{\dd \omega} = \sum_{\ell,m} \frac{\omega}{64\pi k (k^2+4\epsilon^2)(2r_+)^3}|\Yinqbh|^2\,,
\end{align}
where $k = \omega - m\frac{a}{2Mr_+}$,  $r_+ = M + \sqrt{M^2+a^2}$ and $\epsilon = \frac{\sqrt{M^2-a^2}}{4Mr_+}$. For the events relevant to this article, the dimensionless spin $a$ of the ECO/BH is (mostly) negligible, thus $k\approx \omega$ and $r_+ \approx 2M$. A similar result can be obtained for the reflected quantity $\Zoutqbh$ (see Fig. \ref{fig:Sketch_intuition}), i.e.,
\begin{align}
    \frac{\dd E^\T{out QBH}}{\dd \omega} = \sum_{\ell,m} \frac{\omega}{4\pi k (k^2+4\epsilon^2)(2r_+)^3}|\Zoutqbh|^2\,.
\end{align}
With the defined reflectivities $\RQBH$ and $\RBH$, the above energies can be related as follows:
\begin{align}
\label{equ:first_relevant}
    \frac{\dd E^\T{out QBH}}{\dd \omega} =|\RQBH|^2\frac{\dd E^\T{in QBH}}{\dd \omega}\,,
\end{align}
and thus 
\begin{align}
\label{equ:some_stuff_echo}
    \frac{\dd E^\T{echo}}{\dd \omega} &= (1-|\RBH|^2)\frac{\dd E^\T{out QBH}}{\dd \omega}\notag \\&= (1-|\RBH|^2)|\RQBH|^2 \frac{\dd E^\T{in QBH}}{\dd \omega}\,.
\end{align}
Finally, the energy flux across $\Hp$ is expressed in terms of the same quantity at $\scrip$, namely,
\begin{align}
\label{equ:energy_law_the_best}
     \frac{\dd E^\T{horizon}}{\dd \omega} &= |\TQBH|^2 \frac{\dd E^\T{in QBH}}{\dd \omega} \,,
\end{align}
where $E^\T{in QBH}$ can be replaced using equation \eqref{equ:some_stuff_echo}.
The energy loss per cycle in the cavity, as a function of frequency, then corresponds to
\begin{align}
    \label{equ:energy_loss_per_cycle}
     &\Delta E^\T{loss} (\omega)=  \int\dd \omega \frac{\dd E^\T{horizon}}{\dd \omega} + \int \dd \omega \frac{\dd E^\T{echo}}{\dd \omega} \notag\\
    &=\int \dd \omega \left(\underbrace{|\TQBH|^2}_{=:\Gamma} + \underbrace{|\RQBH|^2 |T^\T{BH}|^2}_{=:\Theta} \right) \frac{\dd E^\T{in QBH}}{\dd \omega}\,,
\end{align}
where $|T^\T{BH}|^2:= 1 - |\RBH|^2$. The coefficients $\Gamma$ and $\Theta$ denote the fraction of energy loss due to gravitational radiation crossing the ECO/QBH horizon and the potential barrier, respectively. Equation \eqref{equ:energy_loss_per_cycle} allows for the explicit calculation (and comparison) of the energy crossing the horizon $\Hp$ for a given event. This becomes particularly significant in the context of ECOs, as there may be instances where they collapse to form a BH after the first echo (the first cycle) \cite{Chen_2019}. If the BH is considered within a classical framework, no echoes would be detected at $\scrip$. However, if the resulting BH is modeled as a QBH, the transfer function of subsequent echoes is altered, leading to a distinct echo morphology.

\subsubsection{Numerical Evaluation - Echo-induced Memory}

With the analytical frameworks for the GW echo and its associated memory established, the next step is to quantify their influence on the waveform detected by relevant GW interferometers. In particular, attention is directed toward \gls{lisa}, which is highly suited for detecting potential memory effects due to its exceptional sensitivity in the low-frequency range, where such effects are typically more pronounced. To assess the \gls{snr} of the echoes and the corresponding memory effects, the pipeline developed in \cite{Henris_Mem} is employed once more. A realistic forecast for the detection of echo memory requires taking into account various event-specific factors, such as the orientation and sky position of the merger relative to the LISA frame. For the sky position and all other orientation-dependent characteristics relevant to the SNR, the conservative baseline parameters presented in Table I of \cite{Henris_Mem} are adopted.

In this section, several \textit{SXS} simulations with negligible or vanishing spin are employed, including \textit{SXS:BBH:0205}, \textit{0206}, \textit{0207}, \textit{1424}, \textit{1448}, \textit{1449}, \textit{1455}, and \textit{1936}. Exceptions to the low-spin condition include \textit{SXS:BBH:0334}, \textit{1155}, and \textit{2108}, which exhibit remnant spin amplitudes $|\vec{\chi}|$ ranging from $0.28$ to $0.68$. Although the algorithm used for echo computation may introduce systematic errors for these events, their impact on the analysis is expected to be minimal, as corrections for non-trivial spin amplitudes are typically small. With regard to the reflectivity models, the presence of a non-negligible spin causes line broadening, as discussed in \cite{Agullo_2021}. However, for the events of interest, the effective broadening remains close to the frequency resolution imposed numerically by the pipeline, $\mathcal{O}(1)$ $\mu$Hz. Consequently, these events are included in the analysis to serve as consistency checks for any potential parameter biases associated with the remnant spin. For each event, the echo is computed numerically for the harmonic strain modes $h_{2,\pm2}$, which dominate the overall signal. It is important to note, however, that all modes are included in the flux computation. As a result, not all strain modes involved in the calculations below contribute to the echo. The impact of restricting the echo to the $h_{2,\pm2}$ mode will be discussed in further detail below.

\begin{figure}[!t]\centering
\includegraphics[width=0.6\columnwidth]{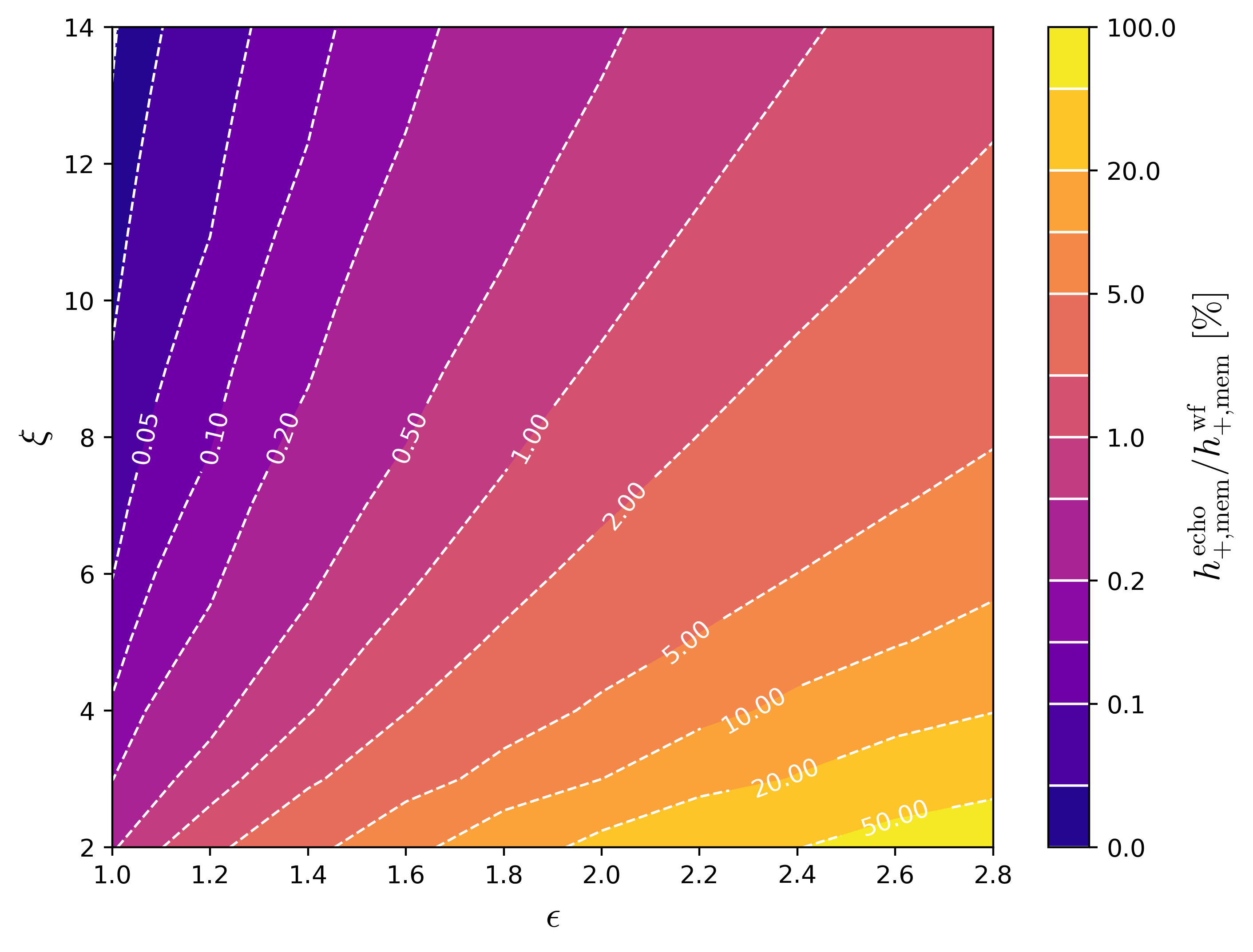} 
\includegraphics[width=0.6\columnwidth]{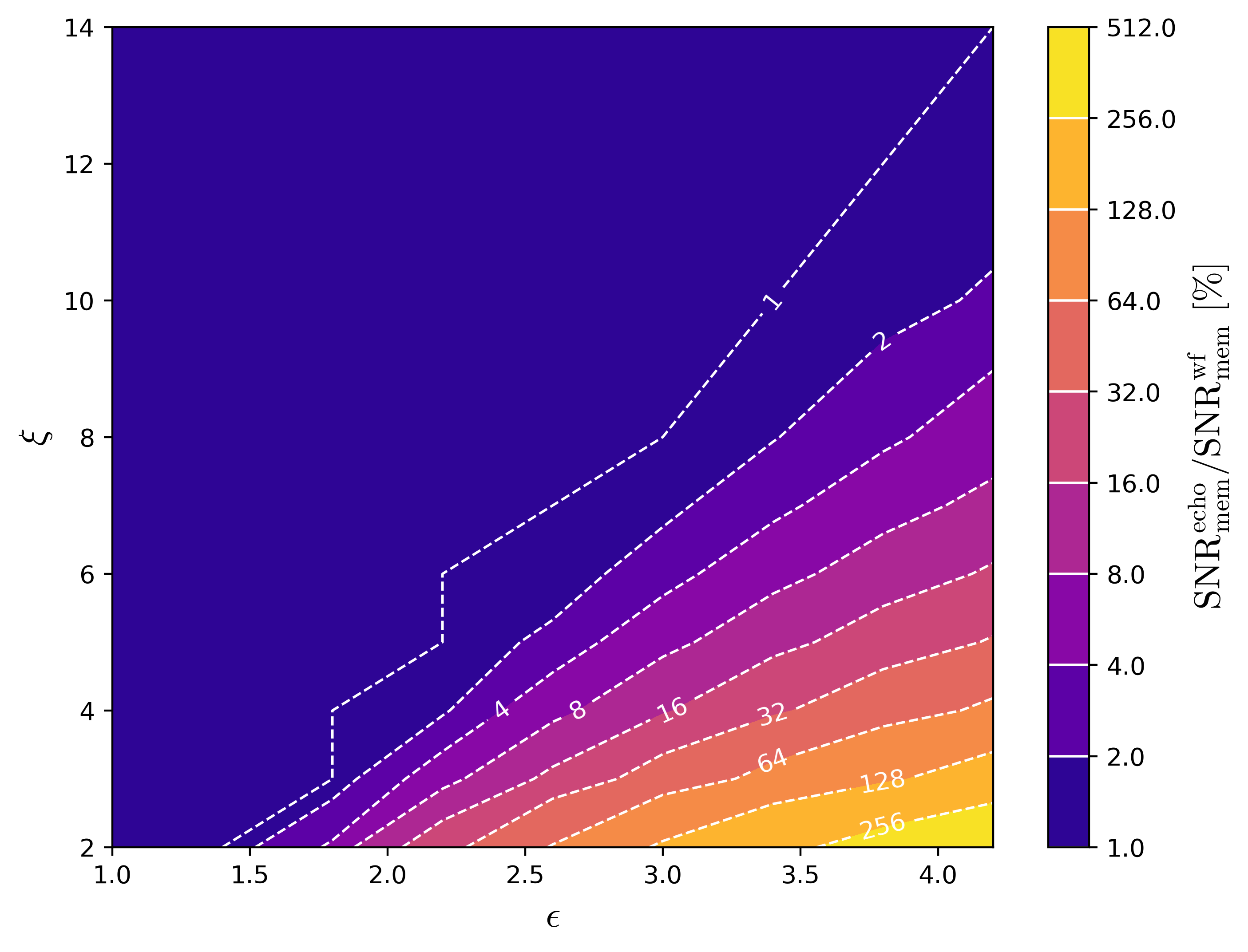}
     \caption{Memory amplitude and SNR gain for event \textit{SXS:BBH:1936} under the reflectivity model of a QBH with $\alpha=8\pi$, $\beta = 10^{-15}$ and $\delta=0.1$ \cite{Maibach:2021qqf}. The top plot shows the memory amplitude purely due to the echo as a fraction of the classical waveform memory (without echo). The bottom plot shows the gain in overall memory SNR relative to the SNR of the waveform memory without echo (at redshift $z=1$ and mass $M=10^6 M_\odot$ the total value for the SNR for the chosen baseline is roughly 10). Note that for better readability, the $x$-axis of the bottom plot is extended up to $\epsilon=4$.}
       \label{fig:numerics_I}
\end{figure}

To gain insight into the significance of the echo-induced memory based on the QBH model, different reflectivity functions are simulated, and the memory is computed for each waveform listed. The results are exemplarily shown for \textit{SXS:BBH:1936} in Fig. \ref{fig:numerics_I}. This event is selected to align with the results presented in the previous section, which discusses the general prospects for echo detectability. The contour plots in Fig. \ref{fig:numerics_I} display the fraction of memory attributed solely to the echo in comparison to the classical waveform's memory (top panel), and the corresponding \gls{snr} associated with the additional memory contribution (bottom panel)\footnote{Note the differing scales of the $\epsilon$-axis for better readability of the bottom panel.}. The \gls{snr} shown pertains specifically to the detectability of the memory and is independent of the waveform itself. To simplify the interpretation, the relative increase in memory is presented, effectively factoring out the redshift and mass dependencies. For a comprehensive exploration of the detectability of memory concerning these remnant-specific parameters, the reader is referred to \cite{Henris_Mem}. As expected, Fig. \ref{fig:numerics_I} reveals a clear trend indicating that lower suppression factors, $\xi$ and $\epsilon$, are associated with higher echo memory contributions. Hence, the echo's memory contribution is proportional to its amplitude. Furthermore, the contour plot shows that over a wide range of parameter space, the echo significantly contributes to the overall memory, with \gls{snr} levels comparable to those of the initial waveform's memory in extreme cases. \\
While the \gls{snr} for the echo memory alone is unlikely to be sufficient for individual detection, it is plausible that the echo memory enhances the overall memory \gls{snr}. This enhancement strongly depends on the specific event and the temporal separation between the echo and the waveform's ringdown. The sensitivity to time separation arises primarily from \gls{lisa}'s response to the step-like increase characteristic of the memory. The closer the two memories (sourced from the waveform and the echo) are in time, the more pronounced the low-frequency features appear in \gls{lisa} data (see \cite{Henris_Mem} for more details). Moreover, as discussed earlier, terms like $\dot{\bar{h}}^\infty\dot{h}^\T{echo} + c.c.$ now contribute to the non-linear memory, leading to a non-linear increase. On the other hand, if the time separation becomes too large, the synergistic effects between the waveform’s and echo’s memories diminish. This effect is particularly evident in the analysis of the ECO model, which will be addressed further below.

It is important to emphasize that, in parameter regions where the echo memory contributes more than $\mathcal{O}(10)$ percent of the total waveform memory, the oscillatory component of the echo’s time series itself becomes sufficiently large to fall within the detectable range of LISA. The detectability of this echo is strongly dependent on factors such as redshift and mass, as discussed in the previous section (see also \cite{Maibach2025} and \cite{Other_features_VIII}). For a reference on the pure echo SNR for a parameter region similar to that displayed in Fig. \ref{fig:numerics_I}, the reader is referred to Fig. 5 therein. In this context, it is observed that for larger values of $\epsilon$ and smaller values of $\xi$, the echo’s magnitude increases much faster than the memory contribution. Once the echo amplitude surpasses approximately $\gtrsim 5\%$ of the waveform amplitude, LISA’s SNR becomes sufficiently high to allow for individual echo detection, assuming the appropriate redshift and mass range for the merger events. This makes the echo-induced memory a less critical tool for detecting the echo in such cases. Nonetheless, it remains surprising that an event of such a short duration, compared to the original waveform, can contribute so significantly to the memory.

\begin{figure}\centering
\includegraphics[width=0.6\columnwidth]{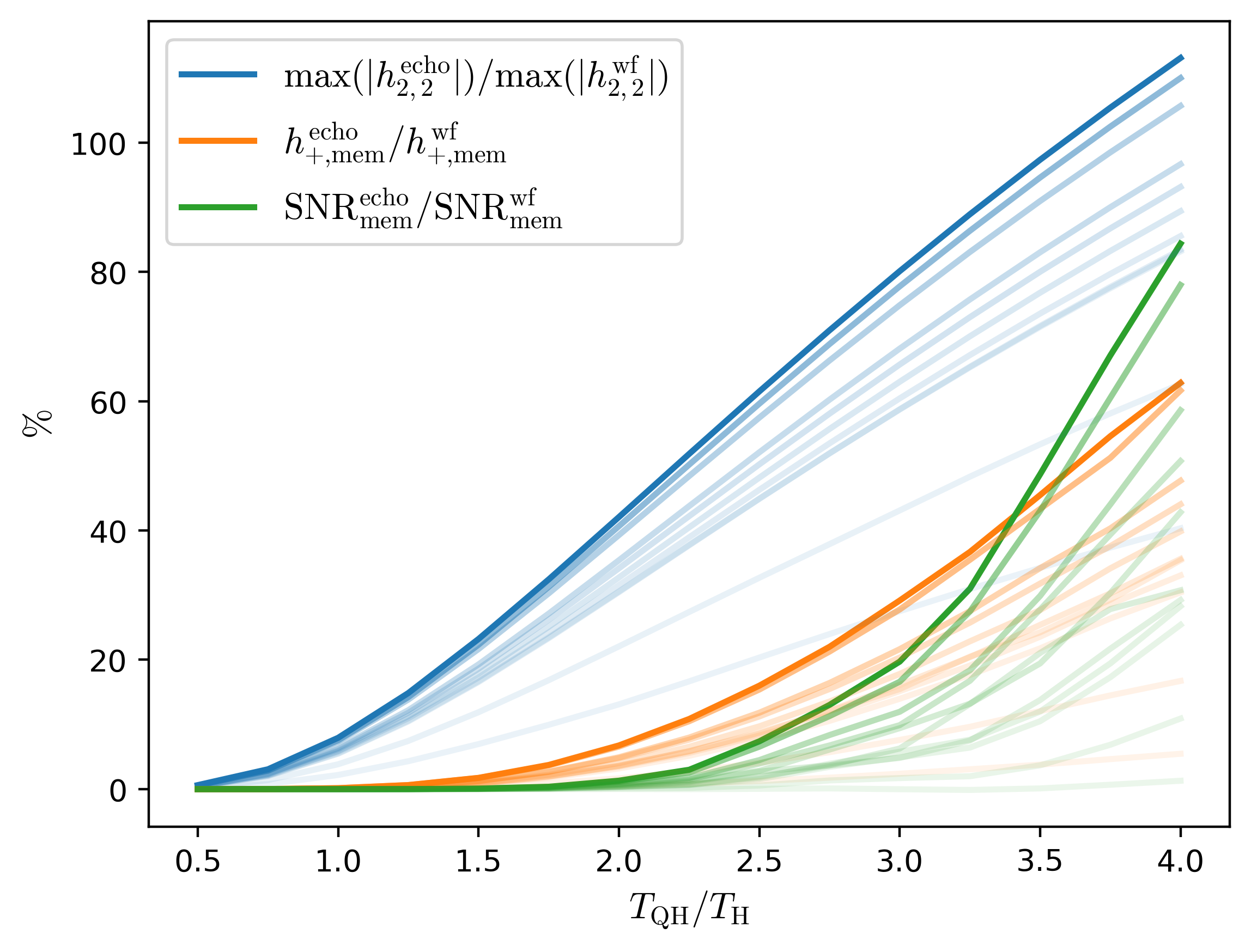}   
     \caption{Comparison of the echoes maximal magnitude, memory, and memory SNR w.r.t. corresponding quantities for the classical waveform (without memory) for the ECO model \cite{Maibach2025}. Each line corresponds to a given event of the list above, where the color intensity is arranged such that the lines fade according their value for the relevant fraction. The graphs represent the ECO analog to Fig. \ref{fig:numerics_I}.}
       \label{fig:numerics_III}
\end{figure}

Transitioning to the ECO model, the parameter dependence of the echo memory contribution, its SNR, and the echo amplitude is presented in Fig. \ref{fig:numerics_III}. The maximum echo amplitude serves as a benchmark for comparing the echo memory and pure echo SNR. Due to the simplified parameter dependence, instead of focusing on a single event, all the SXS simulations listed above are incorporated into the plot. It is apparent that while all events follow a trend similar to that of \textit{SXS:BBH:1936} for the QBH, there are notable event-dependent variations in the ratios of echo and echo memory amplitudes. These variations arise from differences in simulation parameters, such as mass ratios and initial spins, which lead to a diverse range of QNM content and, as a result, different echo characteristics. Fig. \ref{fig:numerics_III} further highlights important differences between the reflectivity models and their associated parametrizations. Specifically, while the response and transfer functions for the ECO and QBH are relatively similar (as shown in Fig. \ref{fig:Transfer_functs}), an increase in $T_\T{QH}$ for the ECO not only impacts the exponential suppression but also compresses the time interval between the waveform ringdown and the echo. This leads to a non-linear response in LISA’s memory SNR, as opposed to the linear increase in the amplitude of the echo memory contribution when $T_\T{QH}$ is varied. A similar effect is observed for the QBH when $\beta$ is adjusted accordingly.

Most importantly, when comparing the results for the QBH model and the ECO, no significant differences are found in the memory contributions, provided that the parameters are chosen to ensure comparable amplitudes of the transfer functions. This suggests that, despite the distinct underlying physics of the two models, their effect on \gls{gw} memory is nearly identical. To further validate this hypothesis, the remaining parameters, namely $\gamma$ in $\Reco$ and $\alpha$, $\beta$, and $\delta$ in $\RQBH$, are examined. While $\gamma$ and $\beta$ primarily influence the QNMs of the ECO and the QBH, respectively (distinct from the QNMs of the ringdown that feed the echo), $\alpha$ and $\delta$ affect the positioning of the characteristic frequencies and the sharpness of their cusps. The results confirm that, as expected, none of these parameters has a significant impact on the echo memory, except for $\delta$. When $\delta$ is sufficiently large, it can reduce the amplitude of the echo memory due to increased absorption by the BH horizon. Although this feature could allow for a distinction between reflectivity models based solely on the amplitude of the resulting echo memory, current detection estimates suggest that this amplitude is unlikely to be determined with sufficient accuracy.

For the tests of different reflectivity model parameters, the following parameter ranges were considered: $\gamma \in [10^{-15}, 10^{-4}]$, $\alpha \in [4 \log 2, 8\pi]$, and $\delta \in [0, 1]$. The upper bound for $\gamma$ was selected to ensure that the separability condition remains intact, while the lower bound is arbitrary, as it would only delay the arrival of the echoes. For $\alpha$, the chosen interval reflects relevant values from the literature \cite{Agullo_2021}. Lastly, for $\delta$, the lower bound marks the point at which the ECO and QBH transfer functions become indistinguishable at high frequencies. The upper bound represents a critical value, beyond which the cusps of the characteristic frequencies become sufficiently pronounced to influence the overall amplitude of the transfer function.

\subsubsection{Numerical Evaluation - Fluxes across $\Hp$ and $\scrip$}

\begin{figure}[!t]\centering
\includegraphics[width=0.6\columnwidth]{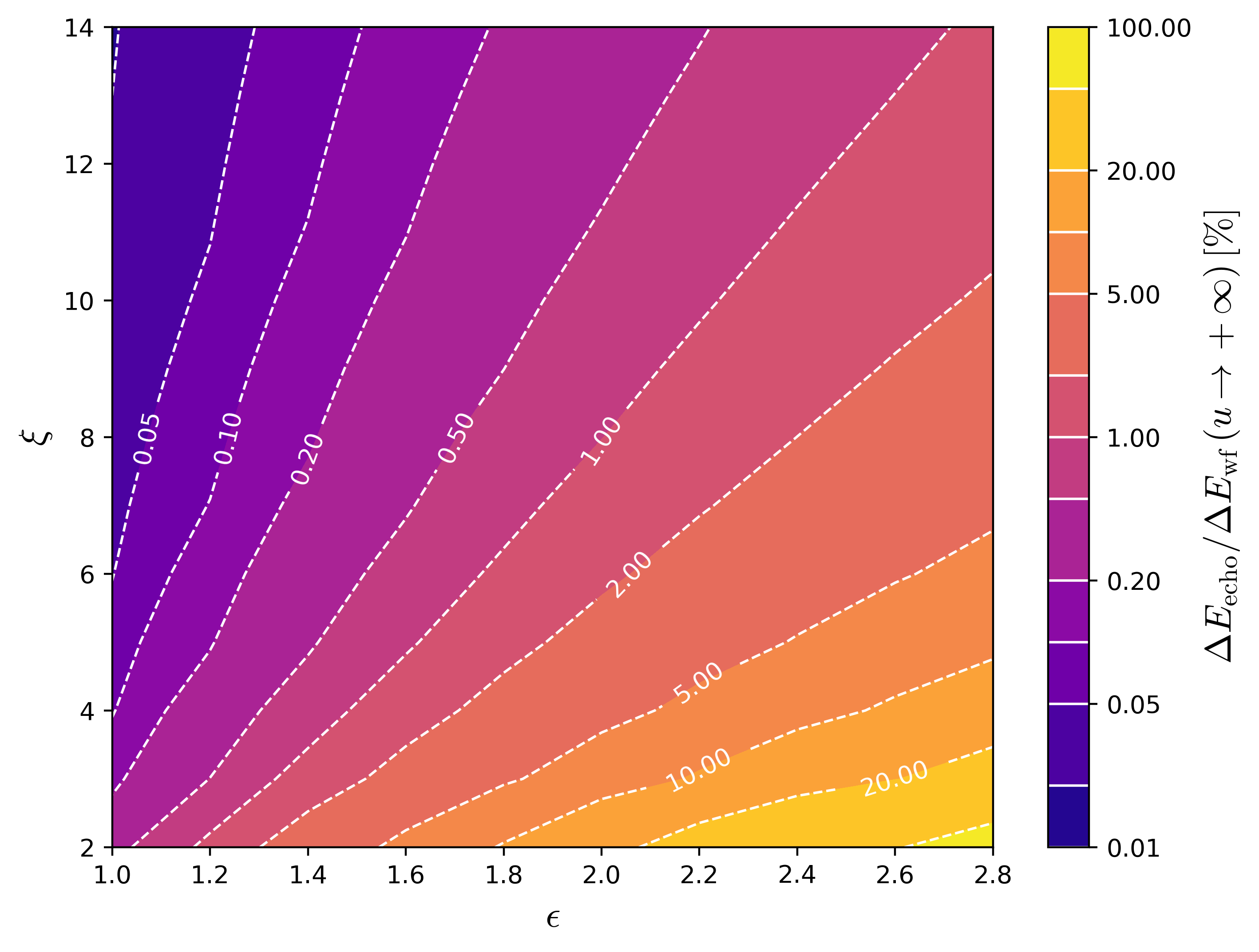} 
\includegraphics[width=0.6\columnwidth]{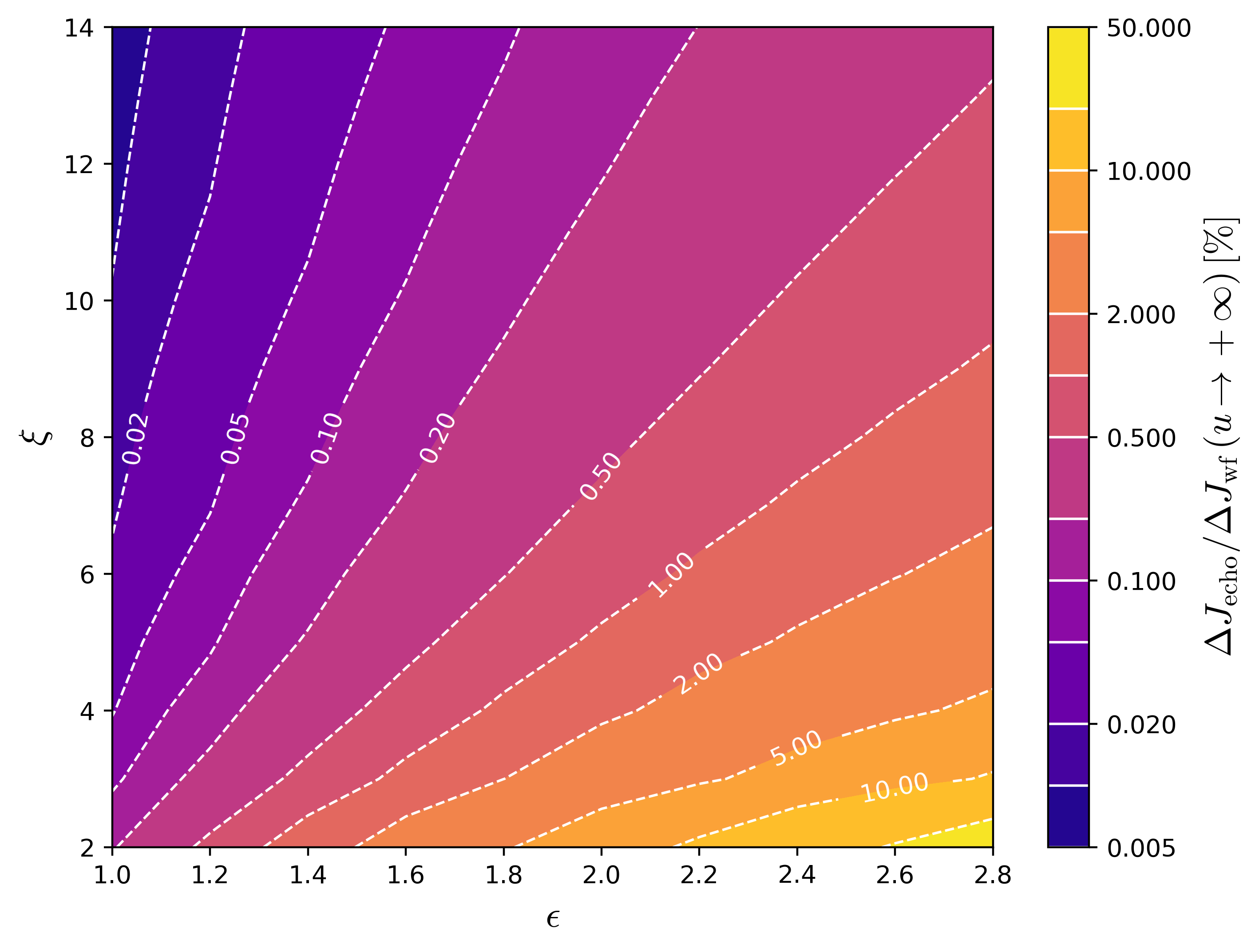}
     \caption{Integrated energy (top) and momentum (bottom) flux for \textit{SXS:BBH:1936} under the reflectivity model of a \gls{qbh} \cite{Maibach2025}. For the parameters not displayed in the plots, the same values as in Fig. \ref{fig:numerics_I} have been chosen. }
       \label{fig:numerics_IV}
\end{figure}

\begin{figure}[!t] \centering
\includegraphics[width=0.6\columnwidth]{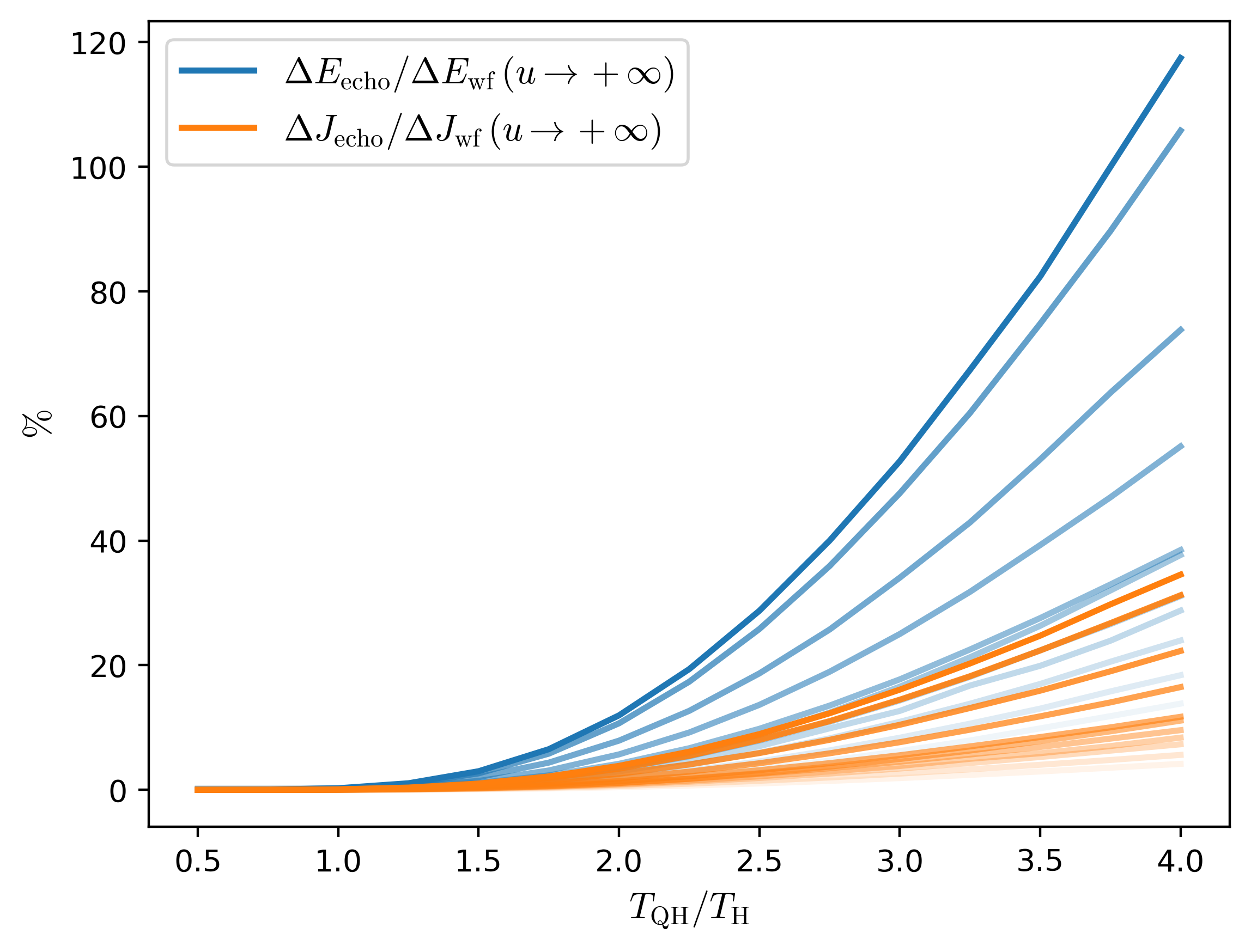}   
     \caption{Integrated energy and momentum flux for the listed events under the reflectivity model of an ECO \cite{Maibach2025}. Again, each line corresponds to a given event of the list above, where the opacities are arranged such that the lines fade according to the largest value of the relevant fraction. The graphs represent the ECO analogue to Fig. \ref{fig:numerics_IV}.}
       \label{fig:numerics_V}
\end{figure}

To gain an initial characterization of the echo in terms of fluxes, the time-integrated energy and momentum fluxes of the echo for \gls{eco} and \gls{qbh} are compared against the corresponding fluxes of the waveform at $\scrip$. The relevant parameters are varied in the same manner as in the previous subsection. For the \gls{qbh}, Fig. \ref{fig:numerics_IV} shows the integrated energy and momentum fluxes measured \textit{morally} at ($u \rightarrow +\infty$). Numerically, this boundary is approximated by the maximum time extent of the simulated \textit{SXS} waveform. The values displayed in Fig. \ref{fig:numerics_IV} suggest that, within the parameter region where the memory contributions are non-negligible, the fractional integrated energy flux behaves similarly to the fractional memory. Specifically, the energy flux associated with the echo can constitute a substantial fraction of the total energy carried by the full waveform. A similar pattern is observed for the momentum flux; however, it is important to note that, for the tested events, the echo carries less momentum than energy. A similar energy-momentum flux relationship is also observed for events with higher remnant spins, as shown in Fig. \ref{fig:numerics_V}. In these cases, waveforms with larger energy fluxes also exhibit larger angular momentum fluxes, although the latter is generally much smaller in magnitude compared to the full waveform's flux. This behavior is found to be uncorrelated with the remnant's spin. Physically, the observation that the echo carries less angular momentum flux than the initial waveform is consistent with the process generating the respective strain signals. While a coalescing binary typically radiates away large amounts of angular momentum, the perturbed Schwarzschild \gls{bh} does not.

In Fig. \ref{fig:numerics_V}, the analysis of the \gls{eco} reflectivity model is extended to more extreme ratios of $T_\T{QH}/T_\T{H}$, demonstrating that the energy flux due to the echo can indeed surpass that of the waveform. It is important to note that, in this scenario, the individual echoes computed for the ECO model transition into a continuous signal due to the decreasing time separation between successive echoes as $T_\T{QH}$ increases. The strain associated with this situation is illustrated in Fig. 12 of \cite{Ma_2022}.
Similar to the results obtained for the memory, the findings show that the parameters $\gamma$, $\alpha$, and $\beta$ do not significantly affect the energy or angular momentum flux when varied within the previously defined ranges. Instead, Fig. \ref{fig:numerics_V} demonstrates that the fluxes carried by the echo exhibit substantial variation across the simulated waveforms. This leads to the conclusion that the direct and indirect features of \gls{gw} echoes are fundamentally determined by the classical QNM content of the remnant, as the reflectivity models presented here sufficiently capture the physical characteristics of these systems. This conclusion is supported by the equations above, where $\Psi_{0, \ell m}(v)\mathcal{F}(v)$ encapsulates the information contained in the ingoing QNMs. This observation implies that, in general, the echo, its memory, and associated fluxes are primarily determined by the physical properties of the remnant compact object, with the reflectivity primarily influencing the amplitude. Further support for this statement is provided in the Appendix of \cite{Maibach2025}. Thus, the selected models can be regarded as robust with respect to potential errors due to additional phenomenological effects. 
This result holds true when the echo is computed for all numerically accessible strain modes, rather than just for $h_{2,\pm2}$. Naturally, the fluxes increase slightly when the mode content of the echo is extended, which is to be expected given the sum over strain modes in the flux-determining factors $\alpha^\T{echo}$ ($\beta^\T{echo}$) in Eq. \eqref{equ:alphas} (Eqs. \eqref{equ:alphas_0} and \eqref{equ:alphas_1}).

Turning to the computation of an energy balance given by Eq. \eqref{equ:energy_loss_per_cycle}, the focus is particularly on how the energy measured by the GW detector compares to the energy absorbed by the compact object or the BH at $\Hp$. To establish a flux balance, Eq. \eqref{equ:energy_dE_in_ECO} is computed, and in combination with Eqs. \eqref{equ:first_relevant}-\eqref{equ:energy_loss_per_cycle}, the energy consumed by the \gls{qbh}/\gls{eco} is estimated in comparison to the energy radiated to $\scrip$. The findings indicate that, even when adjusting the reflectivity of the remnant object to enhance the emission of large echoes, the overwhelming majority of energy—over $99\%$ of the total energy loss per cycle—is absorbed by the celestial body itself. In other words, for a stable cavity between the event horizon and the surrounding potential barrier, nearly all of the energy $E^\T{in ECO}$ is returned to the compact object, whether it is a \gls{qbh} or an \gls{eco}.

To assess the impact of characteristic absorption frequencies as a primary distinguishing factor between the reflectivity models investigated in this study, the reflectivity parameter $\delta$ is varied. The remaining parameters are held constant to ensure that the \gls{eco} and \gls{qbh} models produce identical transfer functions up to the cusps associated with $\omega_N$. It is observed that the flux exhibits a strong sensitivity to $\delta$. For $\delta = 0.01$, the flux of the \gls{qbh} decreases by $\mathcal{O}(1)$ percent relative to the flux of the \gls{eco}. This trend continues approximately linearly, affecting both the energy and momentum flux in a similar manner. As $\delta$ increases, the \gls{qbh} reflectivity diminishes significantly, and for $\delta \gg 1$, the remnant behaves like a classical \gls{bh}, absorbing all energy via $\Hp$. The sensitivity of the flux to $\delta$ implies that, in principle, the flux measurement, or alternatively, the amplitude of the echo-related memory contribution, could serve to distinguish the origin of echo-like features in the gravitational waveform, provided the strain time series of the echo achieves a sufficiently high SNR. However, it is unlikely that the precision of the echo memory amplitude will surpass the SNR of the characteristic frequency features in LISA's \gls{tdi} data, as discussed in Section \ref{sec:Paper_LISA_II}. An investigation into the synergy between echo strain and memory in LISA data, along with the detection prospects for instruments more focused on memory, is left for future research.

Finally, the impact of the cavity's ``walls'' on confining the energy traversing them is tested. This confinement can be quantified by the factors $\Gamma$ and $\Theta$ in equation \eqref{equ:energy_loss_per_cycle}, which represent the energy contributions to the BH horizon, $\Hp$, and to future null infinity, $\scrip$, respectively. The results show that the signal entering the cavity as $\Yineco$ encounters both the horizon of the \gls{eco} and \gls{qbh} as well as the potential barrier, with integrated transmissivities that are roughly equal\footnote{As a reference, the \gls{eco}'s transitivity with $\gamma=10^{-15}$ and $T_\T{QH}=T_\T{H}$ was chosen.}. However, when integrated over the frequency band, the reflectivity is less than $\mathcal{O}(1)$ percent, leaving the echo with only a minimal amount of energy after escaping the cavity toward $\scrip$, as depicted in Fig. \ref{fig:Sketch_intuition}. This significant energy absorption suggests that, for a broad range of events, the \gls{eco} may transform into a \gls{qbh} within the first cycle of radiation traversing the cavity, as noted in \cite{Chen_2019}. As a result, even if features corresponding to characteristic frequencies are detected, either as cusps in the \gls{tdi} data (see Section \ref{sec:Paper_LISA_II}) or in diminished echo memories, it is not immediately apparent whether the remnant object was originally a \gls{qbh} or an \gls{eco}. Thus, the detection of echoes with characteristic frequencies cannot definitively rule out the presence of \gls{eco}s.

\subsection{Discussion}

The \gls{gw} memory effect, along with other subtle features of the gravitational waveform, is expected to enter the detection range of future space-based detectors such as LISA \cite{LISA} and next-generation ground-based detectors like Cosmic Explorer \cite{evans_cosmic_2023} and the Einstein Telescope \cite{ET_I,ET_II}. In addition to testing the fundamental non-linearity of \gls{gr}, this provides a pathway for testing deviations beyond \gls{gr} \cite{Heisenberg:2018vsk} and other phenomena that could influence the permanent displacement of freely floating test masses. The primary challenge in detecting these new and subtle features lies in the need to develop highly accurate \gls{gw} templates to extract signals from data that are often contaminated with noise. Each additional feature incorporated into a template significantly expands the parameter space of the simulated waveform, leading to a substantial increase in computational costs. \\
In Section \ref{sec:quantum_detectibility}, a comprehensive test of the \gls{gw} echo feature is conducted. This includes testing the detectability of echo-induced features with the LISA instrument, incorporating a detailed Fourier space analysis of the TDI data to pinpoint the phenomenology responsible for the echo. The investigation is extended by considering \gls{eco}s, and the resulting signature for both phenomenologies can be added to the waveform templates without substantial changes. The quantum approach is based on fundamental area quantization arguments, which facilitated the development of a reflectivity model for \gls{qbh}s. This model is then compared to established models for ECOs. For both models, corresponding \gls{gw} echoes are constructed following the methodology described in \cite{Ma_2022}. Building on the investigation from Section \ref{sec:Paper_LISA}, Section \ref{sec:Paper_LISA_II} computes the memory and flux contributions of these echoes. Semi-analytical expressions for the gravitational memory effect induced by the echo, along with the corresponding fluxes, are provided. Additionally, an energy balance is formulated for the ingoing radiation towards $\Hp$ during and after the merger's ringdown phase. Numerical investigations of both the echo-induced memory, focusing on its amplitude and \gls{snr}, as well as the flux balance, are carried out. Various scenarios within a reasonable parameter space are considered, and their phenomenological consequences are thoroughly explored.

The results highlight that for a significant fraction of events, the \gls{gw} echo effect is expected to be detectable by the LISA instruments throughout the measurement period. Additionally, there are promising prospects for accurately predicting signatures uniquely associated with BH area quantization, specifically the Bekenstein-Mukhanov spectrum, based on the collected measurement data. This leads to the conclusion that LISA could, at the very least, rule out quantization models similar to Bekenstein's by detecting an echo with sufficient SNR. Regarding the echo's impact on non-linearities, it is found that memory corrections can contribute substantially to the overall memory of the principal waveform. While in a large portion of the parameter space the echo memory is too faint to be detected individually, it has the potential to significantly enhance the total memory's \gls{snr}. Furthermore, when the time gap between the echo and the merger waveform is small, additional synergies emerge. This is interpreted as the reflective shell being closer to the BH potential barrier. Therefore, any deviation from the expected memory could point to the presence of echoes, offering an indirect detection approach that does not necessitate extensive adjustments to waveform models. Based on the findings in Section \ref{sec:Paper_LISA}, a combined search incorporating both memory-related and strain-related time series features of echoes is likely to yield the most robust detection prospects, given that the time series SNR dominates over the induced memory SNR for a given echo. \\
In general, the time separation between echoes is highly dependent on the model, and there is no consensus on its precise value. Some studies even suggest searching for \textit{rogue} echoes, which cannot be definitively linked to a particular merger event \cite{zimmerman2023rogueechoesexoticcompact}. Similar approaches, targeting primarily outlier events or instrumental glitches, have also been proposed in different contexts, such as in investigations of topological Dark Matter (see for example \cite{Heisenberg:2023urf} and references therein).

With respect to the flux analysis in Section \ref{sec:Paper_LISA_II}, the investigation demonstrates that the majority of the energy and angular momentum contained in the ingoing $\Psi_0$ towards $\Hp$ will ultimately be absorbed by the BH or \gls{eco}. Therefore, no significant phenomenological or astrophysical consequences are expected from the reflective shells considered in this study, aside from the subtle echoes observed in the interferometer data\footnote{Naturally, slight modifications to the Tidal Heating effect are also expected, resulting in minor phase discrepancies in the inspiral compared to \gls{gr} waveforms, as discussed towards the end of Section \ref{sec:quantum_BH}.}. Despite transferring only a negligible fraction of the energy and momentum flux of the ingoing $\Psi_0$ to future null infinity $\scrip$, the echo's contribution to the total energy and angular momentum of the waveform can still be considerable. Additionally, it is noted that the echo consistently carries less angular momentum than energy in comparison to the classical waveform across all tested events. This behavior is largely independent of the specific reflectivity parameters of the system and aligns with expectations based on the dynamics that generate the echo.

It is important to note that the manifestation of the echo memory within the gravitational waveform is largely model-independent, with the primary variation being in its amplitude, which depends on the choice of the reflectivity model. The main distinction between models is captured by the reflectivity parameter $\delta$, meaning that, in principle, echo-induced memory corrections provide a simple yet powerful addition to waveform templates for future \gls{gw} detections. In this context, the findings suggest that the reflectivity models play a secondary role in determining the exact shape of the echo time series and the associated memory. Instead, it is the dynamics of the binary configuration and the resulting QNM spectrum of the remnant body that govern the fluxes and $h^\text{echo}$. Thus, measuring an echo, in principle, offers an alternative avenue for studying QNMs in the context of quantum \gls{bh}s. The distinguishing feature of the echo, for the models considered here, lies within the Fourier space data as detailed in Section \ref{sec:Paper_LISA_II}. However, it is to be emphasized that the model-independence of the memory comes with certain challenges. A key difficulty in identifying the echo-induced memory is differentiating it from other memory corrections, such as those arising from deviations in \gls{gr} \cite{Heisenberg:2018vsk} or quantum effects unrelated to echoes \cite{Guerreiro_2022, Parikh_2021}. Further theoretical and analytical investigations of these effects will be necessary to pinpoint distinct signatures of these features in both the memory and the oscillatory components of the strain time series.

Another theoretical aspect, which has been only partially addressed in Section \ref{sec:quantum_BH} and warrants further exploration, is the BH reflectivity model itself. While the arguments presented in Section \ref{sec:quantum_BH}, primarily based on \cite{Other_features_VIII}, align with phenomenological and observational constraints, the investigation of \gls{gw} echoes necessitates reflectivity models grounded in both quantum information theory and astrophysics. Specifically, it is crucial to ensure that these models do not violate the core principles of BH information theory. The boundary conditions for \gls{qbh}s, in particular, require further scrutiny. It is important to acknowledge that the prescription provided in Section \ref{sec:quantum_BH} is an oversimplification. The ingoing radiation toward the BH horizon could, in fact, introduce a feedback term in the Teukolsky equation. Alternatively, GW scattering, rather than reflection off the horizon, might be a more appropriate consideration. These key aspects of QBH models must be thoroughly investigated in future research to achieve a more precise understanding of their physical behavior and the corresponding GW signatures.

In conclusion, it is found that echoes and their associated signatures serve as a smoking gun for quantum corrections to the \gls{bh}'s horizon, as well as the existence of \gls{eco}s, assuming sufficient measurement precision. Since these features manifest in the non-linear GW memory, potentially alongside numerous other phenomenological signatures, the detection of this memory holds the potential to become a pivotal milestone in gravitational physics, fundamentally altering the current understanding of \gls{bh}s and other compact stellar objects.


\chapter{Gravitational Wave Backgrounds} 

\label{chap:analysis} 




\textit{In the following Chapter, the discussion moves away from resolved \gls{gw}s and towards unresolved sources of gravitational radiation. In particular, cosmological contributions to the SGWB are in focus, culminating in an analysis for detection prospects in Section \ref{sec:SGWB_detection_Paper_HI}. The latter reproduces the results of the joint work [E], which is subsequently cited as \cite{SGWB_analysis_I}. For more exhaustive reviews and cutting-edge literature on the topic, relevant references are provided.}

\noindent The continuous enhancement of \gls{gw} measurements has ushered in a new era in GW astronomy. While the initial detection efforts focused on resolved waveforms, such as the first measurement by the LIGO/Virgo collaboration in 2015 \cite{First_LIGO_detection}, we are now approaching the capability of detecting unresolved \gls{gw} sources with upcoming probes \cite{SGWB_1, baghi_uncovering_2023, hartwig_characterization_2022}. On astrophysical scales, sheer number of unresolved sources is so vast that they collectively create a stochastic signal. Additionally, a wide array of unresolved cosmological sources mixes with the astrophysical background. Combined, the form what is known as the \gls{sgwb} (see \cite{SGWB_1} and \cite{SGWB_1} for comprehensive reviews on cosmological and astrophysical sources, respectively). Rich in its phenomenology, the \gls{sgwb} can serve as compelling evidence for new physics on cosmological scales, offering a distinct avenue of exploration beyond high-energy physics at the TeV scale and probing the early Universe prior to the \gls{cmb}. Very recently, PTAs \cite{Principles_PTA} (see \cite{PTA_parkes, PTA_europe, PTA_CPTA, PTA_ipta, PTA_meerkat, PTA_nanograv} for relevant experiments) have managed to collect initial evidence of a stochastic background \cite{Nanograv_SGWB_I, Nanograv_SGWB_II} flooding the Universe. Despite the origins of this background remaining uncertain to this day, the results of PTA represent the initial step in identifying the astrophysical contributions to the \gls{sgwb}. With forthcoming space-based instruments like the \gls{lisa} \cite{LISA} or TianQin \cite{TianQin_I,TianQin_II}, insights gained from PTA can be supported by additional data, potentially enhancing the current upper limits on the \gls{sgwb} from ground-based instruments \cite{LIGO_limit, LV_2009, renzini2022stochastic}.
In addition, future ground-based instruments, such as the Einstein Telescope \cite{ET_I, ET_II, Branchesi_2023} and Cosmic Explorer \cite{evans_cosmic_2023, reitze2019cosmic}, will play a pivotal role in extending the frequency range over which both resolved and unresolved \gls{gw} sources can be detected. Jointly, space- and ground-based instruments possess the potential to guide prospective detection efforts regarding the \gls{sgwb}, offering exciting prospects for future \gls{gw} research.

After the intensive investigations around resolved gravitational waveforms and their subtle features, this Chapter aims to switch gears by studying the prospects for the detection of unresolved sources of gravitational radiation. To this end, the following Sections \ref{sec:SGWB_intro}, \ref{sec:SGWB_astro}, and \ref{sec:SGWB_cosmo} provide a thorough introduction to theoretical and phenomenological aspects of the SGWB. Section \ref{sec:SGWB_detection} outlines recent results of PTAs in more detail and provides an exhaustive insight into current efforts with the LISA instrument \cite{SGWB_analysis_I}. 

\section{From Gravitational Wave Sources to Backgrounds}
\label{sec:SGWB_intro}

Gravitational radiation reaches Earth from all over the Universe. Its sources are not at all limited to mergers of binary compact objects but also result from cosmological phenomena. Naturally, the larger the distance between the detector and the source of gravitational radiation, the more likely it is that the \gls{gw} is affected by the expansion of the Universe. Moreover, for cosmological sources forming a \gls{gw} background, as introduced in Section \ref{sec:Intro_SGWB}, the radiation passes the late time structure in the course of its journey towards Earth. Naively, one would therefore expect non-trivial interactions of \gls{gw} with matter and themselves, by means of Einstein's equations. In practice, however, the weakness of the gravitational interaction decouples \gls{gw} from other contributions to the stress-energy tensor already since the Planck scale. As it will be demonstrated below, it is therefore sufficient to remain at linear order in perturbation theory and neglect any interaction with matter or self-interactions. Thus, \gls{gw}s propagate freely (within the Hubble radius).
To properly understand the latter statements, this Chapter is initiated by reviewing relevant basic concepts of \gls{gw} propagation in cosmologically relevant spacetimes before exploring potential contributions to the \gls{sgwb}, of both astrophysical and cosmological nature.

%

\subsection{Gravitational Waves in realistic Spacetimes}
\label{subsec:real_st_GW}
In previous Chapters, it has mostly been assumed that \gls{gw}s are encapsulated in linearized perturbations $h_{\mu\nu}$ around a background metric $\bar g_{\mu\nu}$ that is asymptotically flat, i.e., $\gd = \bar{g}_{\mu\nu} + h_{\mu\nu}$ with $ g_{\mu\nu} \approx \eta_{\mu\nu} + \mathcal O(1/r)$ for large $r$. Already at this stage, technically, the assumptions for the extraction of TT gauge part of the metric perturbation break down as it can only be chosen in the true (non-asymptotic) vacuum. Thus, the question arises whether the corresponding two polarization states of $h_{\mu\nu}^{TT}$ remain intact when deviating from the vacuum assumption. Naturally, this must be true for all kinds of spacetimes as the presence of matter should not affect the propagating degrees of freedom of the underlying theory of gravity. Showing the latter explicitly, however, requires the use of the \textit{scalar-vector-tensor decomposition} (SVT). To that end, the following derivation is based on \cite{SGWB_intro_I, SGWB_intro_II, Carroll_Book_2019} as well as \cite{SGWB_1}:
Adapting the previous notation, in the case of non-vanishing stress-energy tensor $T_{\mu\nu}$, the metric is written as $\gd = \bar{g}_{\mu\nu} + \delta g_{\mu\nu}$ with $\bar g_{\mu\nu} = \eta_{\mu\nu}$\footnote{Note that it is implicitly assumed that $T_{\mu\nu}$ as well as $\delta \gd$ depend on $x^\mu$.}. For cosmological spacetimes, the latter equation technically fails to capture the expansion of the Universe\footnote{Note that the flat background assumption implies that $T_{\mu\nu}$ vanishes at the background level.}. However, for the purpose of this demonstration, the background is assumed to be flat. Propagation of \gls{gw}s in curved and expanding backgrounds is discussed below.\\
At the core of the SVT decomposition stands the splitting of metric perturbations and energy-stress tensor into irreducible parts w.r.t. rotations. By standard textbook treatments, this results in 
\begin{align}
\label{equ:1}
    \partial g_{00} &= -2\phi\,,\\
    \partial g_{0i} &= \partial_i B + S_i\,,\\
    \partial g_{ij} &= -2\psi\delta_{ij} + (\partial_i\partial_j - \frac{1}{3}\delta{ij}\nabla^2)E + \partial_i F_j +\partial_jF_i + h_{ij}\,,
\end{align}
as well as 
\begin{align}
    T_{00}&= \rho\,,\\
    T_{0i}&= \partial_i u + u_i\,,\\
    T_{ij}&= p \delta_{ij} + (\partial_i\partial_j -\frac{1}{3}\delta_{ij}\nabla^2)\sigma + \partial_iv_j+\partial_j+v_i + \Pi_{ij}\,.\label{equ:2}
\end{align}
Note that both $\delta g_{\mu\nu}$ and $T_{\mu\nu}$ are symmetric. The above decomposition naturally picks out four scalars, two vectors, and one tensor for each $T_{\mu\nu}$ and $\delta \gd$ under spatial ($3$-dimensional) rotations. Thereby, it holds that $\partial_i S_i=0, \partial_iF_i = 0, \partial_i h_{ij}=0, h_{ii}=0 $. The latter constraints are inherent to the decomposition and reduce the propagating degrees of freedom of the metric to $10$ (as expected for a symmetric $4\times4$ tensor). Similar constraint equations analogously hold for the decomposition of the stress-energy tensor. Assuming asymptotic flatness, the components of $T_{\mu\nu}$ have to vanish far away from a source. The unique decompositions \eqref{equ:1}-\eqref{equ:2} are further constraint by the conservation equations $\partial_\mu T_{\mu\nu}=0$, eliminating each $4$ extra degrees of freedom. For the metric perturbation $\delta \gd$ one can do a similar reduction by choosing an adequate gauge, or equally, apply a suitable diffeomorphism $\delta \gd \rightarrow \delta \gd -\partial_\mu\xi_\nu - \partial_\nu\xi_\mu$. The latter leaves the tensor part of the metric decomposition invariant (for small metric perturbation), while one can define new scalars and a vector as 
\begin{align}
    \Phi &= \phi + \dot B -\frac{1}{2}\Ddot E\\
     \Theta &= -2\psi -\frac{1}{3}\nabla^2E\\
    \Sigma_i &= S_i -\dot F_i \,,
\end{align}
where $\partial_i \Sigma_i=0$. The above definition reduces the degrees of freedom in the metric perturbations down to $4$ as well. Note that the newly defined scalars and the vector $\Sigma_i$ are, in fact, diffeomorphism invariant. Therefore, no more degrees of freedom can be removed by coordinate transformations\footnote{Note that the stress-energy tensor is automatically gauge invariant as it was chosen to vanish at the background level, i.e., strictly speaking $T_{\mu\nu}\equiv \delta T_{\mu\nu}$. For tensors with vanishing background contribution, the Stewart-Walker Lemma predicts gauge invariance \cite{Stewart_1991}.}. Given the new components of the metric perturbations $\delta \gd$, one can rewrite Einstein's equations, fully expressed by $\Phi,\Theta,\Sigma_i$ and $h_{ij}$. Directly inserting the stress-energy tensor's decomposition, Einstein's equations read
\begin{align}\label{equ:31}
    -\nabla ^2 \Theta &= 8\pi \rho\,,\\
    -\frac{1}{2}\nabla^2\Sigma_i - \partial_i \dot \Theta&= 8\pi(\partial_i u + u_i)\,,\\
    -\frac{1}{2}\Box h_{ij} -\partial_{(i}\Sigma_{j)} -\frac{1}{2}\partial_i\partial_j (2\Phi + \Theta) &+ \delta_{ij}\left[\frac{1}{2}\nabla^2(2\Phi + \Theta)- \Ddot \Theta\right]\notag\\
    = 8\pi(p \delta_{ij} + (\partial_i\partial_j -\frac{1}{3}\delta_{ij}\nabla^2)\sigma &+ \partial_iv_j+\partial_j+v_i + \Pi_{ij})\,.\label{equ:41}
\end{align}
Eqs. \eqref{equ:31}-\eqref{equ:41} can be simplified into 
\begin{align}\label{equ:dof_with_matter}
  \nabla ^2 \Theta = -8\pi \rho\,, && \nabla^2 \Phi = 8\pi(\rho +3p+3\dot u)\,,&& \nabla^2 \Sigma_i = -8\pi S_i\,,&&  \Box h_{ij}=-8\pi \Pi_{ij}\,.
\end{align}
Evidently, solely the dynamics of the tensor part of the decomposition of the metric perturbations is determined by a wave equation. The remaining components, in contrast, obey variations of the Poisson equation. This key result indicates that only the tensor component and its degrees of freedom can be assigned to represent radiative degrees of freedom. In vacuum ($\Pi_{ij}=0$), these propagate freely. The remaining components of the metric perturbation are equally physical, but non-radiative. Despite choosing a specific gauge for the above derivation, the statements hold without restriction of the applied gauge as long as the perturbations are small. This is to be expected, as physical degrees of freedom cannot be altered by gauge choices. In conclusion, the result of Eq. \eqref{equ:dof_with_matter} demonstrates that one can describe the dynamics of gravity's radiative degrees of freedom in the presence of matter, alleviating any initial concerns. For gravitational radiation travelling on cosmological scales, however, it is crucial to also include expanding spacetime into the above consideration. 

For many reasons, including curvature into the (background) picture is generally non-trivial. Most importantly, it necessitates considerations beyond leading order in $\delta g_{\mu\nu}$ as the presence of \gls{gw}s themselves causes a non-trivial curvature, as explained in the introduction \ref{sec:intro_memory}. Further, in particular for gravitational radiation sourced during the very early stages of the Universe, the background, previously assumed to be Minkowski, is now given by the \gls{flrw} metric. This introduces a time-dependent expansion coefficient in $\bar g_{\mu\nu}$ with $|\delta g_{\mu\nu}|\ll |\bar g_{\mu\nu}|$ where, despite $\delta \gd\ll1$ non-linear terms in $\delta \gd$ now become relevant. Although the background is much larger in amplitude than the fluctuations by definition, its time-dependence makes the distinction between fluctuations and background dynamics non-trivial. The solution to the latter issue unfolds when considering the concrete scales on which the background fluctuations manifest, both in time and space. In the context of FLRW for instance, the expected background fluctuations appear on much larger wavelengths ($\bar\lambda \sim 1/H_0$ where $H_0$ is the Hubble factor) then \gls{gw}s and at much smaller frequencies (for earths gravitational field fluctuations are visible at roughly $\bar f \lesssim 0.1$ Hz \cite{SGWB_1}). Thus, in cosmologically relevant cases, one can approach the problem of defining gravitational radiation on a curved background via separation of scales, similar to Isaacson's definition of the \gls{gw} memory introduced in \ref{sec:intro_memory}. In practice, the separation of scales becomes evident when averaging over length scales $L$ with $\lambda_\text{GW}\ll L \ll \bar \lambda $ and frequencies $f$ where $\bar f\ll f\ll f_\text{GW}$ \cite{averaging}. Note that averaging necessarily requires the consideration of higher-order terms in $\delta\gd$ as linear contributions average to zero. 
As demonstrated in the introduction \ref{sec:intro_memory}, perhaps the most important result at second-order in the metric perturbation is the non-trivial stress-energy tensor contribution associated with the presence of \gls{gw}s. This can be computed by expanding the $G_{\mu\nu}$ to second-order in $\delta \gd$. Setting $h_{\mu\nu}= \delta \gd$ morally, for the trace reversed metric $\bar h_{\mu\nu} = h_{\mu\nu} -\frac{1}{2}\bar g_{\mu\nu} \bar g^{\mu\nu} h_{\mu\nu}$ in the Lorentz gauge $\nabla^\mu \bar h_{\mu\nu}=0$, one finds a contribution 
\begin{align}
\label{equ:quadratic_stuff}
    T^\text{GW}_{\mu\nu} = \frac{1}{32\pi}\braket{\nabla_\mu \bar h_{\alpha\beta}\nabla_\nu \bar h^{\alpha\beta}}\,.
\end{align}
Naturally, because now the background metric is considered in full generality, $\nabla$ corresponds to the covariant derivative w.r.t. $\bar g_{\mu\nu}$. This means also that all spurious gauge modes have to be removed before recovering \eqref{equ:quadratic_stuff}. The average over $L,f$ in the latter equation singles out the radiative contributions only. Remarkably, this result is a direct generalization of Eq. \eqref{equ:GW_stress_energy} with $\partial_\mu\rightarrow\nabla_\mu$. The term \eqref{equ:quadratic_stuff} contributes to the background spacetime curvature as it is of order $\mathcal{O}(h^2/\lambda_\text{GW}^2)$. It becomes immediately evident that it carries the \gls{gw} energy in the $00$ components, i.e, $T^\text{GW}_{00}\sim \rho_\text{GW}$. 

For the propagation of \gls{gw}s in a non-trivial background, going to first order in $\delta \gd$ suffices. Note however, that terms such as \eqref{equ:quadratic_stuff} enter via the background equations determining $\bar g_{\mu\nu}$, which now also enters the first-order equation via the Riemann and Ricci tensor. In the case of Minkowski backgrounds, these have been trivialized. All in all, Einstein's equations to first order give
\begin{align}
\label{equ:kein_plan}
    -\frac{1}{2}\Box \bar h_{\mu\nu} &+ R^\lambda_{\mu\nu}{}^\sigma \bar h_{\lambda\sigma} + \nabla_{(\nu}\nabla^\sigma\bar h_{\mu)\sigma} -\frac{1}{2} \bar g_{\mu\nu}\nabla^\alpha \nabla^\beta \bar h_{\alpha\beta}\notag\\
    &+R^{\alpha\beta}\left[\frac{1}{2}\bar g_{\mu\nu} \bar h_{\alpha\beta} -\frac{1}{2}\bar h _{\mu\nu} \bar g_{\alpha\beta} + \bar g_{\beta (\mu}\bar h_{\nu)\alpha}\right] = 8\pi \delta T_{\mu\nu}
\end{align}
Here, $\Box,\nabla, R_{\mu\nu\alpha\beta}, R_{\mu\nu}$ are computed w.r.t. $\bar g_{\mu\nu}$. The stress-energy tensor has been split into background contributions $\bar T_{\mu\nu}$ and first-order perturbations $\delta T_{\mu\nu}$ that act as \gls{gw} sources. By scaling arguments it holds that $T^\text{GW}_{\mu\nu}\subset \bar T_{\mu\nu}$. Note that in Eq. \eqref{equ:kein_plan}, no gauge choice was made. Applying the Lorentz gauge, for instance, simplifies the equation further by dropping the third and fourth terms on the left-hand side. Independent of the gauge, however, $\bar h_{\mu\nu}$ manifestly couples to the background, even when one assumes vacuum $(R_{\mu\nu}=0)$. This is due to the direct coupling to the background Riemann tensor as well as the implicit background coupling via $\Box \bar h_{\mu\nu}$. Practically, this coupling may lead to changes in the \gls{gw} amplitude and polarization. With Eq. \eqref{equ:kein_plan} at hand, the propagation equation for \gls{gw}s in FLRW spacetimes can be derived. To simplify the calculation, it is instructive to be reminded of FLRW's symmetries. Namely, a constant time hypersurface is homogeneous and isotropic. This implies that an identification of contributions pertaining to the background or the metric perturbations, in particular $h_{\mu\nu}$, is possible even when $\lambda_\text{GW}\simeq \bar \lambda$. Generally, on these hypersurfaces, because of the symmetries present, one can apply a SVT decomposition of the metric perturbations $\delta g_{\mu\nu}$ as above. Similarly to the case above, on FLRW, the tensor part of the metric perturbations carries the two radiative degrees of freedom, while scalar and vector modes cannot propagate (in vacuum). The decomposed parts of $\delta g_{\mu\nu}$ do not couple at linear order, such that one can write for the background
\begin{align}\label{equ:fuck_shit}
    \dd s^2 = -\dd t^2 +a^2(t)(\delta_{ij}+h_{ij})\dd x^i \dd x^j\,.
\end{align}
By the Stewart-Walker Lemma, $h_{ij}$ is gauge invariant as it appears only at the first order in perturbation theory. In Eq. \eqref{equ:fuck_shit}, $h_{ij}$ is to satisfy $h_{ii}=0=\partial_ih_{ij}$. Being symmetric, this leaves only the two \gls{gw} polarizations as independent components. The background \eqref{equ:fuck_shit} can now be inserted into \eqref{equ:kein_plan}. By keeping only the TT component of the tensor perturbations and computing the relevant curvature tensors for the background \eqref{equ:fuck_shit}, one finds
\begin{align}
\label{equ:fuck_shit_ii}
    \Ddot h_{ij}(\mathbf{x},t) + 3H \dot h_{ij}(\mathbf{x},t)-\frac{1}{a^2}\partial_i\partial_i h_{ij}(\mathbf{x},t) = 16 \pi \Pi_{ij}^{TT}(\mathbf{x},t)\,.
\end{align}
The Hubble rate is given by $H=\dot a/ a$ and $\Pi^{TT}_{ij}$ is the TT part of the anisotropic stress tensor, defined as $\Pi_{ij}=a^{-2}(T_{ij}-pa^2(\delta_{ij}-h_{ij}))$. In principle, Eq. \eqref{equ:fuck_shit_ii} describes the relevant dynamics of the propagating degrees of freedom sourced by $\Pi_{ij}$. In practice, in particular in the context of early Universe physics, it is advantageous to switch into the Fourier domain as one eventually separates between perturbations smaller or larger than the Hubble radius. The latter distinction highly affects the dynamics of the corresponding modes. For most \gls{gw} sources in the early Universe, it further holds true that their propagation time (up to detection) is much larger than the time in which they are sourced. Consequently, interested primarily in propagation effects for now, one can simply put $\Pi^{TT}_{ij}=0$. Working in conformal time $\dd \eta = \dd t/a$, Eq. \eqref{equ:fuck_shit_ii} turns into
\begin{align}
\label{equ:fuck_shit_k}
    H_{+,\times}''(\mathbf{k},\eta) + \left(k^2 - \frac{a''}{a}\right)H_{+,\times}(\mathbf{k},\eta) = 0\,.
\end{align}
In the latter equation, the prime denotes the derivative w.r.t. conformal time. The comoving wave vector is denoted by $\mathbf{k}$ with $k=|\mathbf{k}|$. The components $H_{+,\times}$ result from the Fourier transform of the TT metric perturbations $h_{ij}$, i.e., 
\begin{align}
\label{equ:shot}
    h_{ij}(\mathbf{x},t) = \sum_{p=+,\times} \int \frac{\dd^3\mathbf k}{(2\pi)^3}h_p(\mathbf{k}, t) e^{-i\mathbf{k}\mathbf{x}} \epsilon^p_{ij}(\hat{\mathbf{k}})\,,
\end{align}
with $h_{ij}(\mathbf{k},\eta)= a^{-1}H_{ij}(\mathbf{k},\eta)$. The polarization tensors $\epsilon^{+,\times}_{ij}$ are symmetric, real and $\epsilon^{+,\times}_{ij}(\hat{\mathbf{k}})=\epsilon^{+,\times}_{ij}(-\hat{\mathbf{k}})$. They are dependent only on the direction of $\mathbf{k}$, transverse to it and are traceless, $\epsilon^{+,\times}_{ii}=0$. For $h_{ij}$, it holds that $h_{ij}^*(\mathbf{k},\eta)=h_{ij}(-\mathbf{k},\eta)$. 

To describe a particular era of the early Universe, one simply inserts the corresponding scale factor $a(\eta)$. The general solution of Eq. \eqref{equ:fuck_shit_k} is given by 
\begin{align}
\label{equ:solution_fuck_shit}
    h_{+,\times}(\mathbf{k},\eta) = \frac{A_{+,\times}(\mathbf{k})}{a(\eta)}\eta j_{n-1}(k\eta) + \frac{B_{+,\times}(\mathbf{k})}{a(\eta)} \eta y_{n-1}(k\eta)\,,
\end{align}
where $j_n,y_n$ are the spherical Bessel functions. The functions $A_{+,\times},B_{+,\times}$ are dimensional constants. The solution \eqref{equ:solution_fuck_shit} is more insightful when computed in the limits of super- ($k\ll\mathcal H$) or sub-Hubble scales ($k\gg \mathcal H$), where $\mathcal H = a'/a$ \footnote{Note that these horizon limits correspond to $k\eta \ll1$ and $k\eta\gg 1$ respectively.}. Using that the early Universe is described by a power law scale factor $a\sim \eta^n$, the Bessel function in \eqref{equ:solution_fuck_shit} simplifies as $\eta j_{n-1}\rightarrow e^{ik\eta}$ and $\eta y_{n-1}\rightarrow e^{-ik\eta}$ in the sub-Hubble limit $k\gg \mathcal H$. Then, one finds decaying plane wave solutions for the spacetime metric perturbation given by 
\begin{align}\label{equ:sub_hub}
    h_{ij}(\mathbf{x},\eta ) = \frac{1}{a(\eta)} \sum_{p=+,\times} \int \frac{\dd^3 \mathbf{k}}{(2\pi)^3}\epsilon^p_{ij}(\hat{\mathbf{k}})(A_p(\mathbf{k})e^{ik\eta-i\mathbf{k}\mathbf{x}}+c.c.)\,, &&\text{for}&& k\gg \mathcal H\,,
\end{align}
In the latter, the amplitude decay is caused by the factor $a^{-1}$. Note that the solution does not include $B_{+,\times}$ as, by the reality contition for $h_{ij}(\mathbf{x},\eta )$ it holds that $A_p(-\mathbf{k})=B^*_p(\mathbf{k})$ and $A^*_p(-\mathbf{k})=B_p(-\mathbf{k})$.\\
On super-Hubble scales, a similar strategy yields 
\begin{align}
\label{equ:out_of_names}
    h_{ij}(\mathbf{x},\eta ) = \sum_{p=+,\times} \int \frac{\dd^3 \mathbf{k}}{(2\pi)^3} \left(A_p(\mathbf{k}) + B_p(\mathbf{k})\int \frac{\dd \eta'}{a^2(\eta')}\right) \,, &&\text{for}&& k\ll \mathcal H\,.
\end{align}
Here, the solution is composed of a term constant in time and another contribution decaying with the growth of the scale factor. Solution \eqref{equ:out_of_names} plays a crucial role when it comes to \gls{gw} emission in the early Universe. During the inflationary period, all types of perturbation evolve, and due to the exponential expansion, a considerable amount will be on super-Hubble scales. There, the exponential scale factor effectively renders the term proportional to $B_p$ in \eqref{equ:out_of_names} irrelevant and the modes stay constant for $k\ll \mathcal H$. Eventually, they reenter the Hubble radius after inflation and form the so-called \textit{irreducible \gls{gw} background}. A more detailed explanation is provided in Section \ref{sec:SGWB_irreducible}.

%

\subsection{Stochasticity }
\label{subsec:cosmo_bkg_stochasticity}
The example of tensor perturbations created during the inflationary period begs the question of how the corresponding metric perturbations of early Universe \gls{gw} sources obtain their stochastic properties. Before answering this question, one needs to define what it means for tensor perturbations $h_{ij}$ to be stochastic. Generally, when speaking of the \gls{sgwb}, one refers to unresolved tensor perturbations $h_{ij}$ acting as a random variable that is characterized statistically via the ensemble average. Naturally, taking the ensemble average becomes impossible in practice as one only has one set of instances, i.e., the observable Universe, to probe this random variable. Instead, one adopts the \textit{ergodic hypothesis}, which effectively equates the ensemble average with a spatial or temporal average. Adapting this hypothesis, observations of large patches of the Universe for long times replace the ensemble, yielding many instances of the same system. These instances are what is averaged over, for example, in Eq. \eqref{equ:quadratic_stuff}. For the ergodic hypothesis to be approximately true, two conditions have to be met.\\
First, the Universe must be homogeneous and isotropic, guaranteeing the same initial conditions for all (even causally disconnected) \gls{gw} sources. This is guaranteed by the definition of the FRWL background, which, by definition, is isotropic and homogeneous. Note that this property is passed on to the distribution of \gls{gw}s, such that one can write the $2$-point function of the tensor perturbations defined above as
\begin{align}
    \braket{h_{ij}(\mathbf{x}_1,\eta_1),h_{lm}(\mathbf{x}_2,\eta_2)} = \xi_{ijlm}(|\mathbf{x}_1-\mathbf{x_2}|,\eta_1,\eta_2)\,.
\end{align}
Being homogeneous and isotropic ensures that temperature and particle density do not vary significantly, even beyond the causal horizon. Therefore, processes like \gls{pt}s happen everywhere in the Universe at the same time and with the same outcome. The same is true also during inflation as the (quantum) fluctuations evolve on an FLRW background. \\
Second, the \gls{gw} sources must not violate causality. The latter is of fundamental importance to ensure statistical independence of the above-mentioned instances, or ensemble, and to be able to do meaningful statistics with such. In physical terms, the causality constraint implies that \gls{gw} sources cannot produce a signal correlated beyond the length/time of the cosmological horizon at this time. Marking the time of $\gls{gw}$ production with an index $p$, this condition is given by $\ell_p\leq H_p^{-1}$. Here, one takes the Hubble radius as the cosmological horizon, which holds for all cosmological epochs except for inflation\footnote{The production of \gls{gw}s during inflation is commented on below.}. Today, our observable Universe is much larger than the redshifted scale at production, $H_0^{-1}\gg (1+z_p)H_p^{-1}$ with $z_p$ being the redshift at production. Therefore, the \gls{gw} signals in the sky is composed of many uncorrelated (in space and time) signals. 

To form a reasonable statistical ensemle, the number of uncorrelated instances of \gls{gw} signals must be large. Determining this number, one estimates the fraction of redshifted correlation length $\ell_p^0$ against the Hubble radius today,
\begin{align}
\label{equ:stat_i}
    \frac{\ell^0_p}{H_0^{-1}}\leq \frac{H_p^{-1}}{H_0^{-1}}(1+z_p)= \frac{(1+z_p)}{\sqrt{\Omega_\text{m}(z_p)+\Omega_\gamma(z_p) + \Omega_\Lambda(z_p)}}\,.
\end{align}
Here, $H_0=100\mathfrak{h} \text{km}\text{s}^{-1} \text{Mpc}^{-1}$, $H(z)=\sqrt{\Omega_\text{m}(z)+\Omega_\gamma(z) + \Omega_\Lambda(z)}$, and $\Omega_{\text{m},\gamma,\Lambda}(z)= \rho_{\text{m},\gamma,\Lambda}/\rho_\text{crit}$ with $\rho_\text{crit}= 3H_0^2/(8\pi)$. For \gls{gw} sources active during the early phases of the Universe, the radiation density $\Omega_\gamma$ dominates the square root in the denominator of \eqref{equ:stat_i}. Assuming an adiabatic expansion (conservation of entropy per comoving volume) and some early Universe cosmology (see \cite{SGWB_1} and references therein for details), one finds that
\begin{align}
    (1+z_p) \simeq 1.25 \cdot 10^{13}\left(\frac{g_s(T_p)}{100}\right)^{1/3}\left(\frac{T_p}{\text{GeV}}\right)\,,
\end{align}
where $g_s(T)$ defines the Standard Model degrees of freedom with 3 light neutrinos\footnote{In this calculation, one must pretend that the neutrinos were still relativistic today, see \cite{SGWB_1}.}. Together with 
\begin{align}
    \rho_\gamma = \frac{\pi^2}{30}g_*(T)T^4\,,
\end{align}
where $g_*(T)$ is the effective number of relativistic degrees of freedom at temperature $T$, it then follows that 
\begin{align}
    \frac{\ell_p^0}{H_0^{-1}}\simeq 1.3\cdot 10^{-11} \left(\frac{100}{g_s(T_p)}\right)^{1/6}\left(\frac{\text{GeV}}{T_p}\right)\,.
\end{align}
Therefore, the correlation scale of \gls{gw} signals sourced in the early Universe is small compared to the Hubble radius today\footnote{Note that the latter was obtained by equating the correlation length with the Hubble scale at production, $\ell_p\equiv H_p^{-1}$. For inflation, this no longer holds as the causal horizon grows exponentially. The stochasticity of the tensor perturbations sourced during inflation, therefore does potentially violate the causality constraint in this sense. Instead, their stochasticity follows from the quantum nature of the perturbations, rendering them intrinsically stochastic.}. Computing today's angular scale in the sky corresponding to $\ell_p$, $\Theta = \ell_p/d_A(z_p)$, one uses the \text{angular diameter distance} defined as
\begin{align}
    d_A=\frac{1}{H_0 (1+z_p)} \int_0^{z_p} \frac{\dd z'}{\sqrt{\Omega_\text{m}(z')+\Omega_\gamma(z') + \Omega_\Lambda(z')}}\,.
\end{align}
This results in $\Theta_p^{-2}$ uncorrelated regions forming the ensemble to average over. As an example, consider \gls{gw}s emitted during the electroweak \gls{pt}. One finds that $T_\text{EWPT}\sim \mathcal{O}(10^2)$ GeV and $g_s(T_\text{EWPT} \sim 100)$. It therefore follows that $\Theta \simeq2\cdot 10^{-11}$ degrees which, in turn, implies a superposition of at least $10^{24}$ uncorrelated patches\footnote{Here, this establishes a lower bound because of equating the correlation lengths with the Hubble radius at production, when, generally, it holds $\ell_p\leq H_p^{-1}$.}. For this particular example, the signal can only be described statistically. For explorations beyond its stochastic nature, it would require an instrument of angular resolution of approximately $\Theta_p$.

The assumption that the Universe is isotropic and homogeneous guarantees that the \gls{gw}s of a certain production mechanism are sourced everywhere at roughly the same time but without causal contact. This same physical origin justifies the ergodic hypothesis above and indicates that the replacement of the ensemble average with the spatial average is well-motivated. Indeed, one finds that this statement holds even for much smaller energies compared to the electroweak \gls{pt}. For instance, at photon decoupling, $z_\text{dec}\simeq1090$, one finds $\Theta_\text{dec}\simeq 0.9$ degrees. The latter still remains unresolved and yields a large enough quantity of patches such that the stochasticity of the \gls{gw} signal remains valid. In fact, it does so well into the matter-dominated era of the Universe. Note that so far, the consideration has remained purely spatial, which raises the question of the time resolution. With similar considerations, one can characterize the signal in terms of a characteristic time scale which, for now, we assume to be $\Delta t_p\simeq H_p^{-1}$. Redshifting this time interval to today's value, one finds that the scale of time correlation for \gls{gw}s sourced by the electroweak \gls{pt} $\Delta t_\text{EWPT}^0\simeq 8$ h, while for the QCD \gls{pt}, it is around 9 months. On first glance, the latter result sounds like a reasonable time span for a resolved observation. However, in order to promote this type of \gls{gw}s from a stochastic ensemble to a resolved source, one would need to observe a patch $\Theta_p$ for the duration $\Delta t_p^0$, which remains impossible for today's and (near-)future instruments. 
Potentially spoiling subsequent Sections, it needs to be emphasized that there exist sources which continuously produce \gls{gw} signals not violating causality, i.e., signals which are causal but not localized in time. It will be demonstrated below that, in contrast to time-wise localized sources, they extend over many frequencies due to the source being ``active'' during multiple Hubble times. An instance of such are topological defects, in particular, \textit{Cosmic Strings} (CS). For these types of sources, the stochastic nature naturally results from the superposition of multiple horizons that, at each time $t_i$, fit within the Hubble radius of today $H_0^{-1}$.

Besides homogeneity and isotropy, the above properties also render the \gls{sgwb} to be statistically Gaussian and unpolarized. The former is a direct consequence of the central limit theorem. The large ensemble of uncorrelated patches of \gls{gw} signal originating from the same production mechanism at time $t_p$ can be viewed as independent samples. By the central limit theorem, their distribution converges towards a Gaussian. The unpolarized nature of the \gls{sgwb} follows from the absence of any significant source of parity violation in the Universe. The majority of primordial \gls{gw}s is sourced by interactions that are symmetric under parity. Therefore, their polarizations must be uncorrelated such that
\begin{align}\label{equ:stats_ii}
    \braket{h_+(\mathbf{k},\eta),h_\times(\mathbf{k},\eta)}=0\,.
\end{align}
The criteria can equally be formulated in a helicity basis. Then, Eq. \eqref{equ:stats_ii} can be translated to the two helicity $\pm2$-modes having, on average, the same amplitude, or, equally, the same expectation value. 

The above features of the \gls{sgwb}, i.e., being homogeneous, isotropic, unpolarized, and Gaussian, are all rooted in solid motivation and approximately hold true for the majority of cosmological sources. Yet, there can be exceptions to the norm. Besides tensor perturbations from the inflationary period, these instances will not be discussed in this work. We refer to \cite{SGWB_1} for an exhaustive treatment in this regard.

%

\subsection{Background Characterization}
\label{subsec:bkg_char}
As discussed above, for the majority of primordial \gls{gw} sources, the random variable $h_{+,\times}(\mathbf{k},\eta)$ follows a Gaussian distribution. This implies that all the information about the random variable is encoded in the $2$-point function
\begin{align}\label{equ:2_point}
    \braket{h_p(\mathbf{k},\eta), h_{p'}^*(\mathbf{q},\eta)} = \frac{8\pi^5}{k^3}\delta^{(3)}(\mathbf{k}-\mathbf{q})\delta_{p,p'}h^2(k,\eta)\,.
\end{align}
while all odd-point functions vanish. Non-vanishing higher-point functions can be expressed in terms of $h$. The latter is dimensionless, real, and depends only on the absolute of the wave vector $\mathbf{k}$. The delta functions in wave vectors and polarization are a direct consequence of homogeneity, isotropy, and uncorrelated polarizations. Note that here a particular normalization was chosen, hence the factor of $8\pi^5$. This factor can vary across literature and needs to be carefully considered in relevant computations. Applying the Fourier transform to Eq. \eqref{equ:2_point}, i.e., 
\begin{align}
    \braket{h_{ij}(\mathbf{x},\eta),h_{ij}(\mathbf{x},\eta)} \sim \int_0^{\infty} \frac{\dd k}{k}h^2(k,\eta)
\end{align}
one finds that $h$ can be interpreted as the \gls{gw} amplitude per logarithmic wave vector and polarization at time $\eta$. \\
In literature, it has further established common practice to characterize a given contribution to the \gls{sgwb} by its energy density spectrum per logarithmic wave vector, $\dd \rho_\text{GW}/\dd \log k$. Applying similar logic as above, it holds that 
\begin{align}\label{equ:energy_density_gw}
    \rho_\text{GW} = \frac{\braket{ h'_{ij}(\mathbf{x},\eta),h'_{ij}(\mathbf{x},\eta)}}{32 \pi a^2(\eta)} = \int_0^{\infty}\frac{\dd k}{k}\frac{\dd \rho_\text{GW}}{\dd \log k}\,.
\end{align}
Eq. \eqref{equ:energy_density_gw} is obtained by direct computation of $\delta T_{00}$ as above, where the average is to be understood as an ensemble average under the ergodic hypothesis. For practical calculations based on observational data, the wave vector in Eqs. \eqref{equ:2_point} and \eqref{equ:energy_density_gw} has to be converted into the frequency measured by the observing instrument. One finds $f=k/(2\pi a(\eta_0))$, where $a(\eta_0)$ denotes the scale factor of today. Given the amplitude is converted into redshifted frequency, $h(f)=h(k,\eta)$, one can further define the spectral density as
\begin{align}
    S_h(f) =\frac{h^2(f)}{2f}\,.
\end{align}
In particular, in instrumental considerations, the spectral density plays a significant role and is often used to characterize the \gls{gw} signal, see Section \ref{sec:SGWB_detection}. It has unit $\text{Hz}^{-1}$ and can be related to the \gls{gw} energy density spectrum as
\begin{align}
    \Omega_\text{GW}(f,\eta_0):=\frac{1}{\rho_c}\frac{\dd \rho_\text{GW}}{\dd \log f}(k,\eta_0) = \frac{4\pi^2}{3H_0^2}f^3 S_h(f)\,.
\end{align}
In literature, the frequency dependency of the tensor perturbation $h_{ij}$ is often introduced already at the level of the Fourier transform. Then, Eq. \eqref{equ:shot} is replaced by 
\begin{align}
\label{equ:shotot}
    h_{ij}(\mathbf{x},t) \sim \sum_{p=+,\times}\int_{-\infty}^{\infty}\dd f \int \dd^2 \hat{\mathbf{k}} \,h_p(\hat{\mathbf{k}}, f) e^{2\pi if(t-\hat{\mathbf{k}}\mathbf{x})} \epsilon^p_{ij}(\hat{\mathbf{k}})\,.
\end{align}
Note that in \eqref{equ:shotot} the integration over negative frequencies ensures that $h_{ij}$ is real by virtue of $h_p(\hat{\mathbf{k}}, -f)=h^*_p(\hat{\mathbf{k}}, f)$. The notation \eqref{equ:shotot} can be utilized to demonstrate a crucial fact about \gls{gw} energy density spectrum. Using the solution for tensor perturbations on sub-Hubble scales, compare Eq. \eqref{equ:sub_hub}, one can demonstrate that for this free wave solution, the energy density spectrum scales as $\rho_\text{GW}\sim a^{-4}$, while the shape of the spectrum remains unaffected by the propagation. The evolution of the wave number, $k\sim a^{-1}$, yields a shift of the spectrum in frequency. It holds that 
\begin{align}
   \Omega_\text{GW}(k) = \rho_p\left(\frac{a_p}{a_0}\right)^4\left(\frac{1}{\rho_c}\frac{\dd \rho_\text{GW}}{\dd \log k}\right)_p=\rho_p\left(1+z_p\right)^{-4}\left(\frac{1}{\rho_c}\frac{\dd \rho_\text{GW}}{\dd \log k}\right)_p\,,
\end{align}
where $\rho_p$ denotes the total energy density of the Universe at the production time $t_p$ of the \gls{gw} source of interest. 

\begin{figure}[t!]
    \centering
    \includegraphics[width=0.75\textwidth]{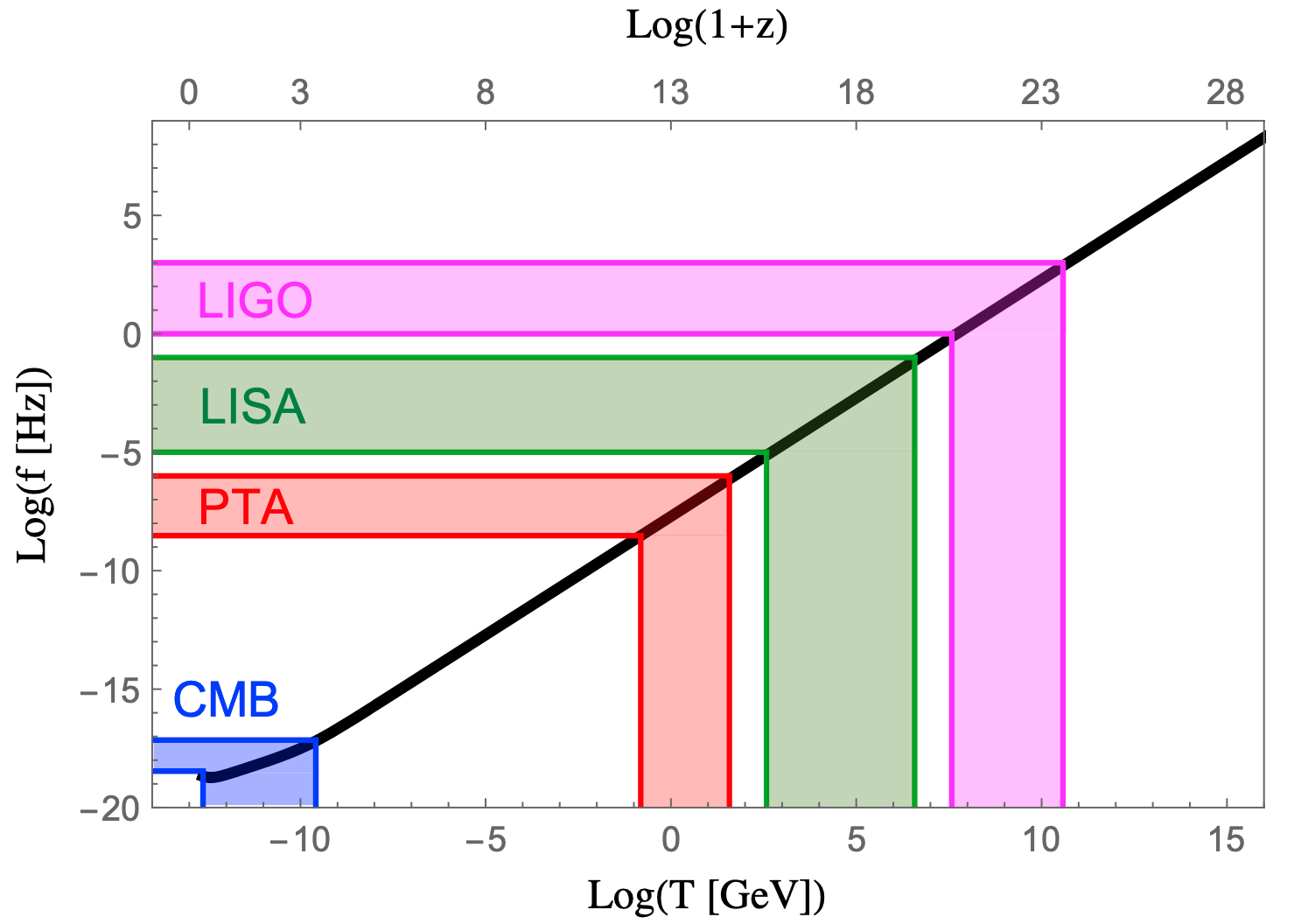}
    \caption{Frequency-Epoch relation for LIGO, LISA, PTA and the CMB \cite{SGWB_1}. The colorized regions mark the corresponding sensitivity domains of the corresponding experiments, the black line denotes the characteristic frequency $f\sim (1+z_p)^{-1}H_p$ or, equally, as a function of temperature $f\sim T_p$.} 
    \label{fig:SGWB_Epoch}
\end{figure}

When describing a particular source of contribution to the \gls{sgwb}, one usually resorts to $\Omega_\text{GW}$ as a function of frequency. The resulting shape of $\Omega_\text{GW}$ is unique among most of the sources. A common systematic of the spectrum is given by the correlation length of the \gls{gw} source at the time of production $\ell_p$ (given the source is localized in time). For the majority of sources, $\ell_p$ can be associated with a peak or another feature within $\Omega_\text{GW}$. The location of such a feature in frequency space is constrained from above by virtue of $\ell_p\leq H_p^{-1}$. Thus, one expects that $f_\text{feature}\geq (1+z_p)^{-1}H_p/(2\pi)$. Similar logic applies to signals not localized in time, such as \gls{gw}s from CS. Here, the fact that the source radiates over an extended period of $H_p/(1+z_p)$ leads to an almost plateau-like extent of the spectrum over frequency. A detailed depiction of the energy density spectrum of CS follows in Section \ref{sec:SGWB_beyond_irreducible}. The relation between correlation length and spectral features of \gls{gw}s enables an important prediction for \gls{gw} experiments. Namely, one can relate a certain epoch of the early Universe to a frequency range of \gls{gw}s as it is recorded today. For a small selection of experiments, the relation is displayed in Fig. \ref{fig:SGWB_Epoch}. The plot highlights two crucial insights. First, each \gls{gw} experiment covers a specific epoch of the early Universe in terms of sensitivity w.r.t. the \gls{gw}s produced during this period. Second, the detection of \gls{gw}s allows for insights into the early Universe well before photons decouple. As gravitational radiation is practically not affected by any form of matter, there is no opacity, in contrast to the photons case, which prevents looking into the earliest stages of our Universe.

%

\subsection{Example: A generic source}
So far, most discussions have been focused on the propagation and characterization of tensor perturbations identified with the radiative degrees of freedom of the gravity sector. In this subsection, the sourcing of such \gls{gw}s via the tensor anisotropic stress $\Pi_{ij}$, see Eq. \eqref{equ:fuck_shit_ii}, is outlined. To that end, one starts by decomposing $\Pi_{ij}$ as before, similar to the metric perturbation $h_{ij}$, such that
\begin{align}
    \Pi_{ij}(\mathbf{x},t) = \sum_{p=+,\times}\int \frac{\dd ^3 \mathbf{k}}{(2\pi)^3}\Pi_p(\mathbf{k},t)e^{-i\mathbf{k}\mathbf{x}}e^p_{ij}(\hat{\mathbf{k}})\,.
\end{align}
If one assumes that the stochastic properties of $h_{ij}$ are already inherent at the stage of source generation, the $\Pi_{ij}$ as well is isotropic, homogeneous, Gaussian distributed, and unpolarized. Therefore, it is equally well-described by the $2$-point function
\begin{align}\label{equ:idk}
    \braket{\Pi_p(\mathbf{k},\eta), \Pi_{p'}^*(\mathbf{q},\eta')} = \frac{(2\pi)^3}{4}\delta^{(3)}(\mathbf{k}-\mathbf{q})\delta_{p,p'}\Pi^2(k,\eta,\eta')\,,
\end{align}
where a different normalization in this case was chosen for convenience. The propagation Eq. \eqref{equ:fuck_shit_k} under the activation of a source $\Pi_{ij}$ becomes
\begin{align}
\label{equ:fuck_shit_kk}
    H_{+,\times}''(\mathbf{k},\eta) + \left(k^2 - \frac{a''}{a}\right)H_{+,\times}(\mathbf{k},\eta) = 16\pi a^3 \Pi_{+,\times}(\mathbf{k},\eta)\,.
\end{align}
Consider now a source during the radiation-dominated era, i.e., $a(\eta) = a_\gamma\eta$, and introduce a dimensionless variable $\zeta=k\eta$. Then, \eqref{equ:fuck_shit_kk} reduces to
\begin{align}
    \frac{\dd^2}{\dd \zeta^2} H_{+,\times}^\gamma(\mathbf{k},\zeta) + H_{+,\times}^\gamma(\mathbf{k},\zeta) = \frac{16\pi a_\gamma^3 \zeta^3}{k^5} \Pi_{+,\times}(\mathbf{k},\zeta)\,.
\end{align}
Defining a start and endpoint of \gls{gw} production (again, assuming the source acts localized in time) as $\zeta_i=k\eta_i$ and $\zeta_f=k\eta_f$, one finds two branches of the solutions, for $\zeta<\zeta_f$ and $\zeta>\zeta_f$,
\begin{align}
    H_{+,\times}^\gamma  (\mathbf{k},\zeta) = \begin{cases}
\frac{16\pi a_\gamma^3}{k^5}\int_{\zeta_i}^\zeta \dd \zeta' \zeta'^3 \sin(\zeta-\zeta')\Pi_{+,\times}(\mathbf{k},\zeta')&, \zeta <\zeta_f \\
A_{+,\times}^\gamma(\mathbf{k}) \cos \zeta + B^\gamma_{+,\times}(\mathbf{k})\sin \zeta &,\zeta>\zeta_f
\end{cases}\,.
\end{align}
Matching these branches at $\zeta_f$ one finds a solution for the amplitides $A^\gamma_{+,\times},B^\gamma_{+,\times}$. From the solution for $H_{+,\times}^\gamma$, one finds a solution for $h_{+,\times}(\mathbf{k},\zeta)$ which can be plugged into the $2$-point function \eqref{equ:2_point} to obtain, in combination with \eqref{equ:idk}
\begin{align}
    h^2_\gamma(k,\eta_0) = 64 \frac{a_\gamma^6}{a_0^2k^7} \int_{\zeta_i}^{\zeta_f} \dd \zeta' \zeta'^3\int_{\zeta_i}^{\zeta_f} \dd \zeta'' \zeta''^3 \cos{(\zeta'-\zeta'')} \Pi(k,\zeta',\zeta'')\,,
\end{align}
and thus
\begin{align}
    \frac{\dd \rho_\text{GW}^\gamma}{\dd \log k}(k,\eta_0) = \frac{4}{\pi a_0^4} k^3 \int_{\eta_{i}}^{\eta_f}\dd \eta a^3(\eta)\int_{\eta_{i}}^{\eta_f}\dd \eta' a^3(\eta')\cos{(k(\eta-\eta'))\Pi(k,\eta,\eta')}\,.
\end{align}
The index $\gamma$ indicates the source being active during the radiation-dominated phase. 

The above procedure holds in all generality and can be repeated for any source of stochastic \gls{gw} given that the anisotropic stress tensor $\Pi_{ij}$ is known. Naturally, depending on the era in which the source produces \gls{gw}s, the scale factor has to be adapted, which may change the solutions of \eqref{equ:fuck_shit_kk}. Note, however, that the change in scale factor predominantly affects the scaling of the spectral density. In the following two Sections, a series of sources are presented for which the derivation above is applied. As the main interest in this Chapter is the resulting energy density spectrum, for any concrete derivations of $\Pi_{ij}$ for such sources, the reader is referred to \cite{SGWB_1} and references therein.

%

\section{Astrophysical Sources}
\label{sec:SGWB_astro}
The cosmological contribution to the \gls{sgwb} encompass a plentiful of unprecedented information about the early Universe. However, it is to be expected that these contributions are overshadowed by much ``brighter'' astrophysical background contributions. These signals, at the same time, hold important insights into structure formation and physics of compact objects, and act as noise, masking the primordial stochastic background contributions. Because of the latter, astrophysical contributions are sometimes called \textit{foregrounds}. Generally, they result from a superposition of a large number of unresolved sources, which are either too faint or overlap in time, preventing a resolution of individual contributions. As they are connected to astrophysical compact objects, their signal cannot be older than the dawn of stellar activity. Therefore, the detection of the astrophysical background may be insightful, for instance, regarding the star formation history.

As it was concluded before, the cosmological background is homogeneous, isotropic, unpolarized, and Gaussian. The nature of the astrophysical contributions may not necessarily include the same attributes. First and foremost, the signals result from large-scale structure in the relative vicinity to Earth ($\sim 100 $ Mpc). On such scales, galaxies are not distributed isotropically. Further, for astrophysical sources, it is not given that they form a continuous signal. As they are sourced by individual mergers of compact objects which are too far to be detected but too ``loud'' to be ignored, the time interval between events could be much larger than the duration of a single \gls{gw} event. Depending on this ratio, a given source may exhibit distinct statistical properties when forming a stochastic background. To characterize the radio between a signal's duration and the time passing between successive events, one computes the so-called \textit{duty cycle} \cite{SGWB_2}
\begin{align}
\label{equ:duty}
    \Delta(z) = \int_0^z\dd z' \tau (1+z')\frac{\dd R^\circ(z')}{\dd z'}\,.
\end{align}
Here, $\tau$ denotes the average time scale of an event actively producing \gls{gw}s, and $\dd R^\circ/\dd z$ is the number of sources per unit time and redshift interval\footnote{The function $R^\circ$ is naturally very much orientation dependent because of the anisotropy in astrophysical sources. In the latter equation, this dependence was integrated out.}. Depending on Eq. \eqref{equ:duty}, which can also be understood as describing the average number of events present at the detector at a given observation time, \gls{gw} sources can be characterized into continuous signals, shot noise, or ``Popcorn''. The first describes frequently appearing signals with small time delays between events compared to the duration of an individual one. As an ensemble, for such sources, because they are so numerous, the central limit theorem can be applied and the resulting distribution is Gaussian. The second category, shot noise, includes sources which are less numerous such that the time interval between events is long compared to the production time of \gls{gw}s of a single event. The incoming signals are separated by extended periods of silence. The popcorn-type sources describe an intermediate case between shot noise and continuous signals. For those sources, the time duration of an individual event is of similar magnitude to the interval between events. Even if some waveforms overlap, the statistical distribution cannot be assumed to be Gaussian anymore. 
In the following, it is assumed that the sources mentioned form a continuous signal. Shot noise and popcorn-like sources are treated as confusion noise from now on.

Before diving into individual types of contributions to the astrophysical background, one should point out that there is a clear difference regarding the detection of such sources w.r.t. electromagnetic signals. While radio telescopes have to scan large regions of the sky to capture the full signal, \gls{gw} interferometer receives the signals of events no matter their direction in the sky. Although by construction, they have a favorable orientation, physically, sources from every direction can reach the detector. The latter is particularly important when thinking in terms of a stochastic background. As signals from all directions can penetrate the detector at any given time, it is much harder to disentangle even very localized sources if they are numerous enough, just as it is the case for the continuous-type signals mentioned above. 
Instead of an individual resolution, therefore, treating such a conglomerate of signals as a background can be beneficial. Note, however, that in this case the best way of detecting the signal is via cross-correlation of measurements of multiple detectors. A prime example of this instance are PTAs. A more detailed description of the recent findings is provided in Section \ref{sec:SGWB_detection}.

In principle, there is an abundance of possible astrophysical backgrounds in literature. For the purpose of this Chapter, the most significant contributions are singled out. Note that, as for the cosmological contributions, different astrophysical contributions occupy different regimes in frequency space. Therefore, not all sources are equally relevant for all instruments, as the latter have comparably narrow sensitivity curves. In the following, for each source, relevant instruments are indicated. For detailed studies of astrophysical contributions to the \gls{sgwb}, the reader is referred to \cite{SGWB_2, Astro_Bkg_I, Astro_Bkg_II, Astro_Bkg_III}.

\subsection{Stellar Mass Binaries and Extreme Mass Ratio Inspirals}
\begin{figure}[t!]
    \centering
    \includegraphics[width=1\textwidth]{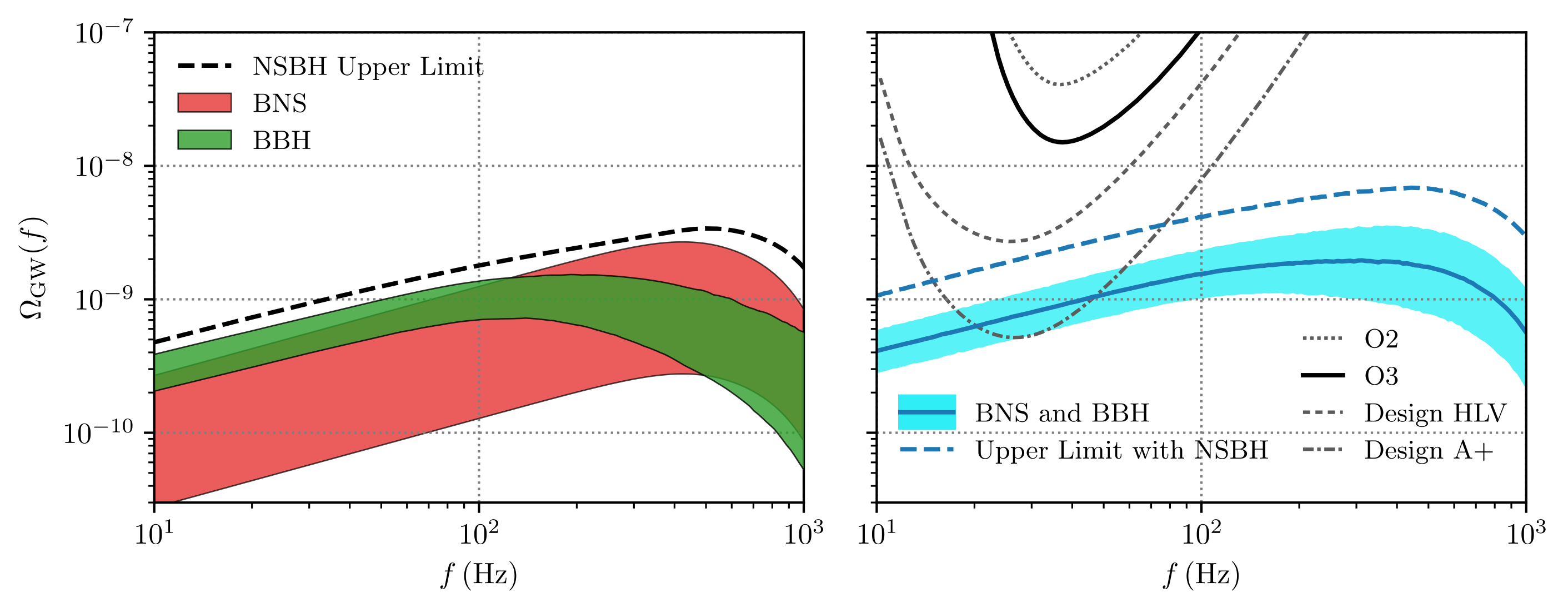}
    \caption{Model predictions for the astrophysical background from BBHs, BNS, and NS-BH binaries along with sensitivity curves of LIGO detectors \cite{LIGO_limit}. The left panel shows the 90 \% confidence interval for backgrounds sourced by BBH and BNS mergers. Uncertainties are due to the merger rates and mass distributions. The right panel compares background energy densities with LIGO sensitivities of observation runs O2 and O3. In addition, predictions for new designs of the LIGO instruments are shown in grey dashed lines. Again, the blue shaded region marks the 90\% confidence interval for the joint contributions from BHNs and BBHs.} 
    \label{fig:BNS_BBH}
\end{figure}

Stellar-mass binaries emit \gls{gw}s primarily during their inspiral and merger phases. While individual events from these binaries have been detected by ground-based observatories like LIGO, the collective signal from numerous such binaries across cosmic history contributes to the astrophysical background. In the higher frequency bands, particularly above the mHz range, the GW background from stellar-mass binaries becomes more significant. Generally, one differentiates between binary BHs and binary NS (BNS)\footnote{Note that also BH-NS binaries are a valid option here. However, there is some uncertainty in their identification of measurements so far. Further, the signal from stellar mass binaries is expected to be dominated by BH and NS binaries (see, for instance, \cite{Astro_Bkg_III}). Thus, the BH-NS binaries are not considered here.}. The former are expected to form a popcorn-like noise, the latter a continuous signal due to a large overlap in time. The energy density spectrum for these sources is described by \cite{Astro_Bkg_II}
\begin{align}
    \Omega_\text{GW}(f) \sim \frac{f}{\rho_c H_0} \int_{\mathcal S^2} \dd \hat\Omega \int_0^\infty \frac{\dd R^\circ/\dd V(z,\hat \Omega)}{(1+z)\sqrt{\Omega_\text{m}(z)+ \Omega_\Lambda(z)}}\frac{\dd E_\text{GW}}{\dd f}(f(1+z),\hat \Omega)\,,
\end{align} 
where $\dd R^\circ/\dd V$ is the source formation rate per unit comoving volume as a function of redshift and $\dd E_\text{GW}/\dd f$ is the energy emitted per frequency. Both quantities are direction-dependent, hence the integral over the sky. The expected frequency range of this background component peaks around 10 Hz to 1 kHz. Therefore, it lies within the regime of sensitivity of ground-based detectors such as LIGO and the third-generation Einstein Telescope (ET) \cite{ET_I, ET_II}\footnote{The sensitivity in frequency range is roughly inversely proportional to the arm length of \gls{gw} interferometers. Thus, ground-based instruments having the shortest arm lengths are sensitive to the largest frequencies. On the other end of the spectrum are PTAs, where the effective arm length is given by the distance between involved pulsars, making them sensitive to the nHz regime. LISA, with arm length of $\mathcal{O}(10^6)$ km is sensitive to $\mathcal{O}(10^{-3}-10^{-1})$ Hz.}. Leading up to the peak, the spectral energy density roughly follows a power law $\Omega_\text{GW} (f) = \Omega_\text{ref}(f/f_\text{ref})^\alpha$. The corresponding components $\Omega_\text{ref},\alpha,f_\text{ref}$ are determined by simulations. So far, the LIGO collaboration has only been able to establish upper limits on the astrophysical background from BNSs and BBHs, i.e., $\Omega_\text{GW}\lesssim 10^{-8}$. The latter are expected to strongly improve within the upcoming years \cite{LIGO_limit}. The current status is depicted in Fig. \ref{fig:BNS_BBH}. For the ET, all types of binaries, NS, BH, and NS-BH, are expected to lie within the sensitivity of the instrument \cite{Astro_Bkg_III}.

Despite the background from stellar-mass binaries peaking around slightly below the kHz regime, this contribution propagates all the way into LISA's sensitivity band of around $\mathcal{O}(10^{-3})$ to $\mathcal{O}(10^{-2})$ Hz. In there, it appears as a constant exponent power law \cite{Astro_Bkg_II}. The background is expected to constitute a very loud contribution w.r.t. astrophysical sources \cite{Dvorkin_2016}. Note, however, that with the upcoming local measurements of LIGO and other detectors, it is expected that precise modeling of this background will allow for an efficient handling in LISA. The latter explicitly refers to the treatment of astrophysical contributions as confusion foregrounds in the context of searches for cosmological contributions to the \gls{sgwb}.
In addition to stellar-mass binaries, the intermediate frequency regime ($\mathcal{O}(10^{-4})-\mathcal{O}(10^{-1})$ Hz) covered by LISA's sensitivity band is victim to various other compact binary \gls{gw} sources. The signal expected to be largely dominant amongst such contributions results from white dwarf-white dwarf binaries within the galactic vicinity to LISA. It is expected that up to $\mathcal{O}(10^8)$ white dwarfs in the Milky Way contribute to a monochromatic background in the LISA band, dominating even the instrumental noise of the apparatus \cite{Lamberts_2019}. The density spectrum thereby has negative power over the relevant frequency regime. For optimistic rates of white dwarf-white dwarf \gls{gw} sources, the background is found to increase LISA's overall noise level by a factor of $\sim 2$ in the domain of 1-10 mHz \cite{SGWB_2}. 

Another highly relevant contribution in a similar bandwidth as white dwarf binaries are extreme mass ratio inspirals (EMRIs). EMRIs usually form when a massive BH of $M\sim 10^5-10^7 \Msol$ captures a less massive stellar-mass BH $m\sim 10-50 \Msol$. Such events are generally expected to arise in dense galactic centers. However, the precise astrophysical processes leading to the formation of such systems still remain poorly understood, resulting in large uncertainties in the prediction of the cosmic rate of these sources, spanning at least three orders of magnitude \cite{Babak_2017, Bonetti_2020}. It is estimated that between 1 and $10^4$ events will contribute to the background per year. Thereby, one has to clearly differentiate between resolved and unresolved events. For redshift $z\leq 1$, EMRIs can be individually resolved. 
Due to their extreme mass ratio, these systems inspiral incredibly slowly. Completing up to $10^5$ cycles in LISA's sensitivity band, EMRIs source the ideal background to construct maps for massive BHs, perform tests of \gls{gr}, and perform tests of matter present around the central massive BH. For $z>1$, the signals of EMRIs fall below the detection threshold, adding up to an unresolved confusion noise contributing to the astrophysical \gls{sgwb}. In \cite{Bonetti_2020} it has been shown that for most astrophysical models, this background is easily detectable with LISA with an \gls{snr} of $\mathcal{O}(100)$. In fiducial EMRI models, the background reaches magnitudes comparable to the LISA noise around $3$ mHz. Throughout literature, the spectral density power law followed by EMRIs varies between $\alpha = 0$ and $\alpha=2/3$ with $\Omega_\text{GW}\sim f^\alpha$ in the LISA band due to the uncertainty in the models.

\subsection{Collapsing and Rotating Stars}
After a star has burnt all its combustible material, it may explode via a type II supernova, shedding large amounts of mass and leaving behind a massive core that can collapse to form either a NS or BH, depending on the initial mass. In this process, large amounts of \gls{gw}s are emitted. The estimation of these types of backgrounds is rather difficult due to the large uncertainty in the waveforms and usually relies on complex numerical simulations. The \gls{gw} content thereby strongly depends on the initial properties of the progenitor, such as mass and spin. While for core-collapse into \gls{ns}, the \gls{gw}s are emitted due to the rapid compactification of large amounts of stellar material, for the core of the star collapsing into a BH, the \gls{gw}s result from the ringdown of the newly formed perturbed BH. Depending on the exact shape of the waveform, conservative estimates predict a spectral energy density peak of $\Omega_\text{GW}\sim \mathcal{O}(10^{-12})$ for the collapse into NS and $\Omega_\text{GW}\sim \mathcal{O}(10^{-9})$ for collapse into black holes \cite{SGWB_2} (and references therein). The spectra peak between 100 Hz and 1 kHz, which again falls into the sensitivity of ground-based instruments. For a recent study on potential detection prospects with (advanced) LIGO and ET, the reader is referred to \cite{Radice_2019}. To obtain these results, one usually relies on numerical simulations and the input of the most relevant properties of the progenitor. In literature, simulations commonly use masses between $1M_\odot$ and $\mathcal{O}(100)M_\odot$ for simulating both collapse to NSs and collapse to BHs. 

\begin{figure}[t!]
    \centering
    \includegraphics[width=0.7\textwidth]{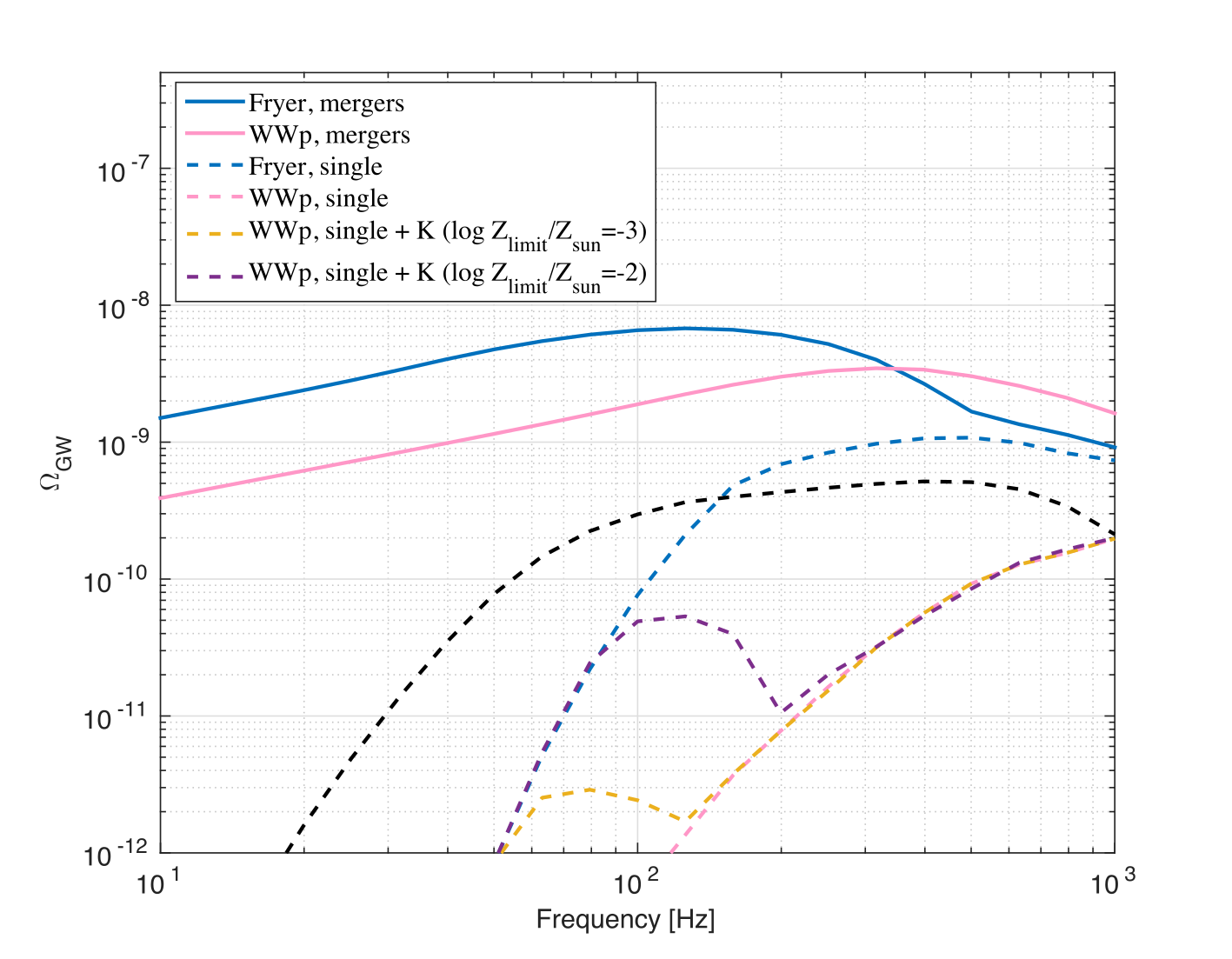}
    \caption{Spectral energy density of the SGWB from single star collapse (dashed) and from merger events (solid) \cite{Dvorkin_2016_i}. The signals from collapse primarily result from the ringdown of the newly formed BH. The colors of the respective category describe a specific model for star collapse and merger. For details, see \cite{Dvorkin_2016_i}.} 
    \label{fig:Star_Collapse}
\end{figure}

Besides \gls{gw} emission from collapse, the rapid rotation of massive NS with triaxial shape induces a quadrupole moment, leading to the \gls{gw} emission at two times the rotational frequency $\nu$. While radio pulsars are less prone to contribute significantly to a \gls{gw} background, magnetars can form strong contributions within the sensitivity band of Earth-based detectors, enhanced by their magnetic torque \cite{SGWB_2}. Such a magnetar-background can be shown to peak at roughly 760 Hz and with a spectral energy density of $\Omega_\text{GW} \sim 2\cdot 10^{-8}$, strongly depending on the star's ellipticity and magnetic field \cite{Stella_2005}. Reaching a maximum in this frequency regime, the magnetar-induced background contributions are likely to be observable with the ET. Besides the emission of \gls{gw}s purely due to rotational motion, in some instances of core collapse supernovae, an enhanced gravitational radiation can be caused by post-collapse processes involving (dynamic) instabilities such as \textit{bar modes}, \textit{r modes}, or \textit{collapse to quark matter}. As these phenomena reach beyond the scope of this Chapter, the interested reader is referred to \cite{SGWB_2}.

\subsection{(Super)-massive Black Hole Binaries}
\label{sec:SMBHB}
Just as EMRIs, super massive BBHs (SMBBH) can be both resolved individually and contribute to backgrounds in a given frequency regime. While resolved events lay within the LISA sensitivity band for BH masses in the range of $10^5-10^7 \Msol$, leading to very loud signals, SMBBH with masses above this range ($10^7-10^9 \Msol$) typically lead to an unresolved background in the nHz regime, i.e., within the sensitivity band of \gls{pta}. While the key processes behind the formation of SMBBHs are still unknown, it is most likely that they form during the merger of galaxies, with the massive BHs at the core of each galaxy being the protagonists. As for EMRIs, the still somehow mysterious conditions under which SMBBHs form lead to large uncertainties in the expected rates for resolved events ranging from $0.5$ to 100 events per year (e.g., \cite{Barausse_2020}). Yet, if detected, the signal duration can reach up to months, establishing excellent tests of \gls{gr} as well as galaxy evolution and high redshift cosmology, $z>6$. The background contributions form SMBBHs, on the other hand, are generally less sensitive to the differences between their formation models and fall within the detection threshold of PTAs (see Section \ref{sec:SGWB_detection} for more details on recent PTA results) \cite{Barausse_2020, Sesana_2008, Sesana_2018}. The spectral energy density of the SMBBH background is expected to follow $\Omega_\text{GW}\sim f^{2/3}$ between $10^{-9}-10^{-7}$ Hz.

\begin{figure}[t!]
    \centering
    \includegraphics[width=0.8\textwidth]{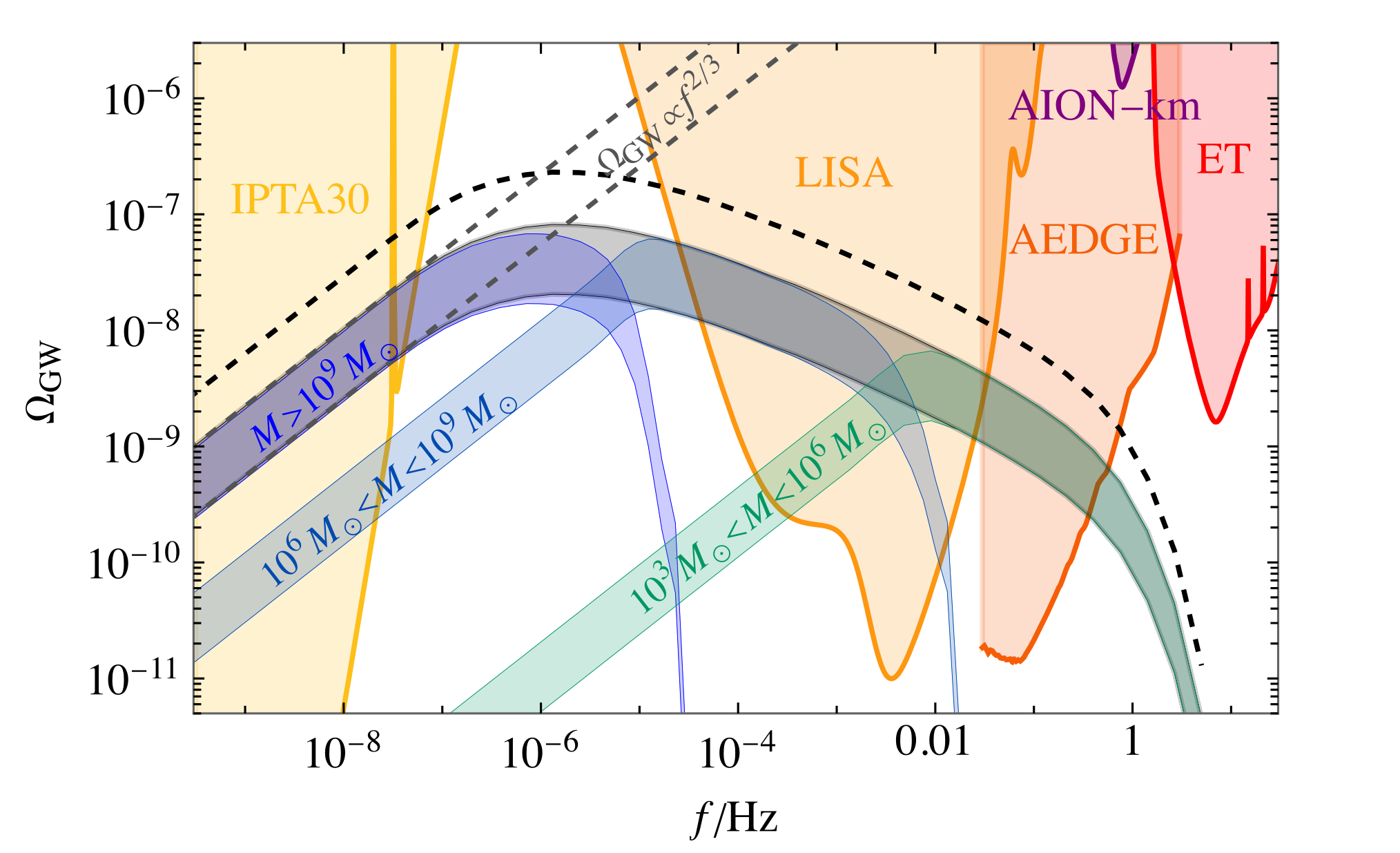}
    \caption{Mean spectral energy density from massive and supermassive \gls{bh} binaries in comparison to different experiment sensitivities \cite{Ellis_2023}. The dashed line demonstrates the accumulation of all GW signals of galactic mergers with BHs heavier than $10^3\Msol$. The shaded regions mark the sensitivities of the respective experiments. For individual references on such, see \cite{Ellis_2023}.} 
    \label{fig:SMBH}
\end{figure}
%

\section{Cosmological Sources}
\label{sec:SGWB_cosmo}

Hidden behind the dominant astrophysical background lies the cosmological \gls{sgwb}. Encompassing unprecedented information about the very early stages of the Universe, in particular pre-\gls{cmb} signals, a detection of the cosmological background has the potential to shed light on the remaining dark spots of modern-day cosmology. The ensemble forming the cosmological background consists of various distinct sources producing gravitational radiation since the inflationary period up until today. Thereby, one differentiates between the \textit{irreducible} \gls{gw} background and reducible background components. As suggested by the name, while the former establishes a fundamental (almost featureless) background radiation, in other words, a noise floor, reducible components can, in principle, be subtracted from the total measurement data due to particular delineating features such as highly localized peaks in the spectral energy density. In the following, the irreducible as well as multiple contributions of the reducible primordial \gls{gw} background are explored.
Thereby, the discussion predominantly relies on \cite{SGWB_1}. Where fitting, additional references are provided. In principle, there is a large number of potential sources of gravitational radiation in the early Universe in the common literature. However, for the purpose of this work and being mainly focused on a potential detection of \gls{gw} backgrounds with \gls{lisa}, in particular in Section \ref{sec:SGWB_detection}, the discussion is narrowed down to what is considered most helpful in understanding the remaining Sections.

\subsection{The irreducible Gravitational Wave Background}
\label{sec:SGWB_irreducible}

It is common ground in cosmology that the Big Bang was followed by a prolonged phase of rapid expansion, called \textit{Inflation}, resolving many shortcomings of the Big Bang framework \cite{Guth_1981,Linde_1982,Albrecht_1982}. Its major success can be attributed to explaining the origin of primordial density fluctuations by stretching out quantum fluctuations into the classical regime. These primordial perturbations act as seeds for the formation of structures in the Universe, leaving imprints, among other probes, in \gls{cmb} data. The very same (quantum) perturbations that stretched out during the inflationary period, however, not only imprint electromagnetic backgrounds, but also produce an abundance of unresolved gravitational radiation. Generally speaking, for any field with mass smaller than the Hubble rate during inflation $m^2\ll H^2$, quantum fluctuations are expected to occur on sub-Hubble scales, $k>aH$. Independent of their nature, these fluctuations are stretched to super-Hubble scales $k<aH$ during inflation and eventually reenter the Hubble radius at a later time, during radiation domination. This in particular also holds for massless tensor perturbations of the metric, which, post-reentry, turn into the classical irreducible \gls{gw} background with a quasi scale-invariant spectrum (thus dubbed ``irreducible'').

To characterize the irreducible background, it is instructive to review some trademarks of the inflationary period: In the simplest models, inflation is triggered by a single inflaton scalar field $\phi$ slowly rolling down a potential $V(\phi)$ and minimally coupling to gravity. The action describing such \textit{slow-roll inflation} model reads
\begin{align}\label{equ:infla_action}
    S = \int \dd ^4 x \sqrt{-g} \left(\frac{1}{16\pi}R-\frac{1}{2}\partial^\mu\phi\partial_\mu\phi-V(\phi)\right)\,.
\end{align}
To yield an inflating Universe, the kinematic energy of the scalar field needs to be dominated by the potential energy, $\dot\phi^2\ll V(\phi)$. For this period to last sufficiently long\footnote{Here, the criteria for sufficiency are mostly derived from the challenges inflation is constructed to solve. In literature, its duration is often quoted to lay between 50-60 $e$-folds corresponding to an expansion by a factor of $e^{50}\approx 10^{22}$.}, the acceleration must be suppressed w.r.t. the fields velocity per Hubble time, $|\ddot{\phi}|\ll|\dot\phi|H$. Together, the latter conditions define the slow-roll parameters
\begin{align}
    \epsilon_\phi = \frac{3}{2}\frac{\dot \phi^2}{V}\,, &&\eta_\phi = -\frac{\ddot \phi}{h\dot \phi}.
\end{align}
For inflation to be present, both $\epsilon_\phi\ll1$ and $\eta_\phi\ll1$. If one were to solve the Friedman equations, including the inflaton field and its potential, a similar slow-roll parameter can also be defined in terms of only the potential and its derivative. Either way, the $\epsilon$ can be understood as controlling the deviation from the pure de Sitter Universe (where $H$ is constant), i.e.
\begin{align}
    \epsilon \simeq -\frac{\dot H }{H^2}\,,
\end{align}
which is very small. This implies that instead of entering a regime of perfect exponential growth $a(t)\sim e^{Ht}$, the Universe during inflation grows quasi-exponentially, i.e., the Hubble rate decreases very slowly with time $\Delta H/H \simeq \epsilon \Delta N$ where $N$ is the number of $e$-folds. To deduce the behavior of metric perturbations within the inflationary regime, one can expand the pure gravitational part of the action \eqref{equ:infla_action} to second-order in $h_{ij}$ and on an expanding background. With a few standard textbook quantization procedures \cite{SGWB_1}, one finds that while tensor fluctuations oscillate on sub-Hubble scales, they freeze out and are constant in time for $k\ll a H$. For the \gls{gw} field $h_{ij}$, written as
\begin{align}
    h_{ij}(\mathbf{x},\eta)=\sum_{p=+,\times}\int \frac{\dd^3 k}{(2\pi)^{3/2}}\left(h_k(\eta)e^{i\mathbf{k}\mathbf{x}}\hat{a}_{\mathbf{k},p}+h^*_k(\eta)e^{-i\mathbf{k}\mathbf{x}}\hat{a}_{\mathbf{k},p}^\dagger\right)\epsilon_{ij}^p(\mathbf{k}/|\mathbf{k}|)\,,
\end{align}
where $\hat{a}_{\mathbf{k},p},\hat{a}^\dagger_{\mathbf{k},p}$ are creation and annihilation operators, the distinction between sub- and super-Hubble propagation is mostly captures by the amplitudes $h_k$. For instance, for modes leaving the horizon during inflation, the amplitude remains, to good approximation, constant in time, i.e.,
\begin{align}\label{equ:init_val}
  |h_k(\eta)|\simeq \frac{H}{k^{3/2}} &&\text{for} &&k\ll aH\,,
\end{align}
which is in agreement with the findings at the end of Section \ref{subsec:real_st_GW}.
As the modes re-enter the Hubble radius at a later time, $k=a_kH_k$, the latter evaluated at the instant of Hubble radius crossing will provide the initial condition for the evolution of the modes post-re-entry.

For the computation of the spectral energy density for the re-entered \gls{gw} modes, it is convenient to define the tensor power spectrum first. Similar to the first Sections (in particular \ref{subsec:cosmo_bkg_stochasticity}) of this Chapter, one defines
\begin{align}\label{equ:fs}
    \braket{0|\hat{h}_{ij}(\mathbf{k},\eta)\hat{h}^*_{ij}(\mathbf{k}',\eta)|0}=\frac{2\pi^2}{k^3}\mathcal{P}_h(k)\delta^{(3)}(\mathbf{k}-\mathbf{k'})\,,
\end{align}
such that 
\begin{align}\label{equ:fss}
    \braket{0|\hat{h}_{ij}(\mathbf{x},\eta)\hat{h}^*_{ij}(\mathbf{
    x},\eta)|0} = \int\frac{\dd k}{k}\mathcal{P}_h(k)\,,
\end{align}
and $\mathcal{P}_h$ denotes the tensor power spectrum. At horizon crossing, $k=a_kH_k$, the power spectrum is proportional to the Hubble scale at that time, i.e., $\mathcal{P}_h(k)\sim H_k^2$. As it has been demonstrated in previous Sections, the super-Hubble modes behave like a classical random field. Therefore, the average in \eqref{equ:fs}, again, can be understood as an ensemble average over a stochastic field. Thus, the power spectrum $\mathcal{P}_h$ relates to $h^2$ in Eq. \eqref{equ:2_point}. At this point, it is worth mentioning that similar procedures can equally be applied to scalar perturbations of the metric generated from inflation. These play a fundamental role in structure formation in the early Universe as well as the generation of \textit{primordial black holes} (PBHs). Equally to the \gls{gw}s, these curvature perturbations $\mathcal{R}$ are conserved on super-Hubble scales \cite{Liddle_Lyth_2000} and their power spectrum, following a similar definition as \eqref{equ:fs} and \eqref{equ:fss}, is proportional to the Hubble radius at crossing, i.e., $\mathcal{P}_\mathcal{R}(k) \sim H_k^2$ at $k=a_kH_k$. The ratio between powers in scalar and tensor perturbations from the inflationary period is commonly expressed as 
\begin{align}\label{equ:rrr}
    r = \frac{\mathcal{P}_h(k)}{\mathcal{P}_\mathcal{R}(k)}\,,
\end{align}
and assumed to be small in slow-roll inflation models, i.e., $r_k\sim \epsilon$ at $k=a_kH_k$. In single-field slow-roll inflation models, the ratio $r_k$ can be related to the scaling in $k$ of the \gls{gw} power spectrum (see \cite{SGWB_1} for details). In fact, one finds that, depending on the mode, $\mathcal{P}_h(k)\sim k^{n_T(k)}$ with $n_T(k)=-r_k/8$. This prediction is remarkably resistant to micro-physical changes in the potential that drives the inflaton field. It is, however, rather sensitive w.r.t. the scalar fields involved or the vacua prescriptions deep within the inflationary period (i.e., deviations from the standard Bunch-Davies vacuum). Therefore, the relation between the $n_t(k)$ and the ratio \eqref{equ:rrr} establishes a unique verification of a simple inflationary model and a definite proof of the concept of inflation itself. 

The ratio $r$ in \eqref{equ:rrr} is very tightly constrained by \gls{cmb} data, in particular for modes crossing during radiation domination. These constraints allow for translation into bounds on the spectral energy density of primordial \gls{gw}s, making use of some additional assumptions about regarding, for instance, relativistic particle species \cite{SGWB_1}. It is found that an upper bound is given by 
\begin{align}\label{equ:bound}
    \Omega_\text{GW}\simeq 5\cdot 10^{-16}\left(\frac{H_k}{H_{max}}\right)^2\,,
\end{align}
where $H_k$ is the Hubble rate at the time when CMB scales leave the horizon and $H_{max}\approx 9\cdot 10^{13}$ GeV, the maximum Hubble rate validating the CMB constraints, i.e., converting the maximum tensor-to-scalar ratio \eqref{equ:rrr} into a Hubble radius via the Friedman equations. Being the upper bound, \eqref{equ:bound} demonstrates that even in the most favorable case, there is no instrument near the required sensitivity to pick up the scale invariant tensor spectrum today. In reality, this spectral energy density may even fall way below the bound \eqref{equ:bound}, being further reduced by the sub-inflationary evolution of the Universe.

\begin{figure}[t!]
    \centering
    \includegraphics[width=1\textwidth]{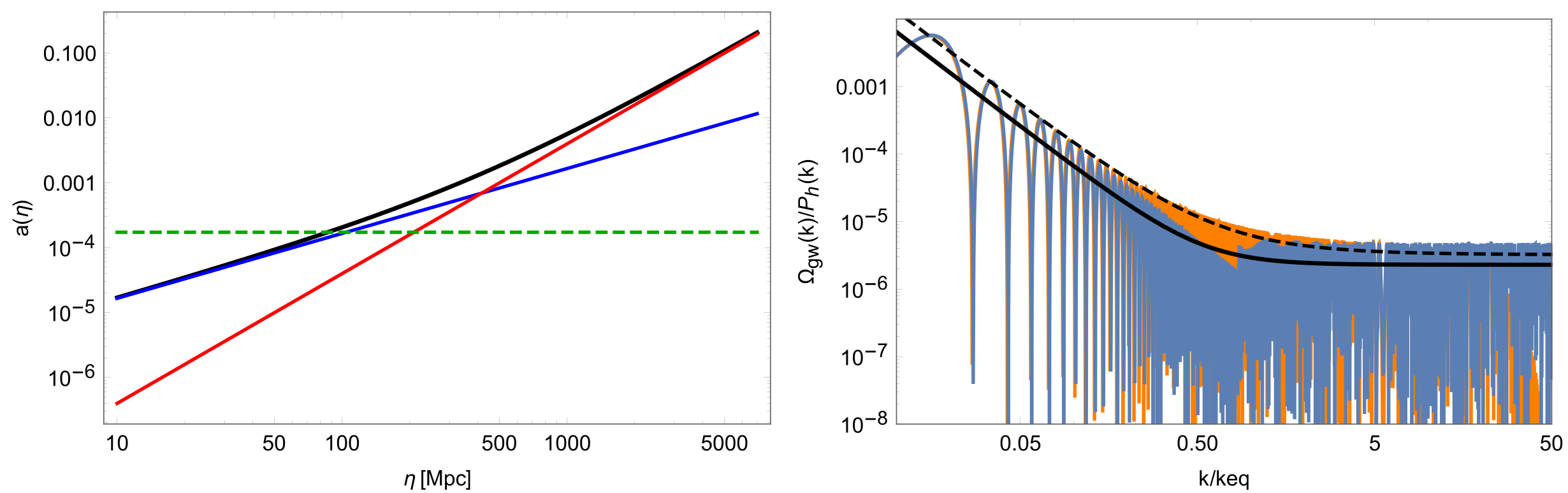}
    \caption{Left panel: scale factor for radiation and mapper dominated Universe (blue and red curves respectively). The dashed line represents the scale factor at equality, the black line depicts the scale factor of a Universe transitioning between radiation and matter domination. Right panel: spectral energy density of tenor perturbations resulting in GWs, normalized by $\mathcal{P}_h(k)$. The orange curve represents a numerical solution, the blue curve an analytical approximation. The black curve is an approximation, averaging out the oscillations, while the dashed curve depicts a result from \cite{Turner_1993}. Note in particular the oscillatory nature of this background contribution. The graphic is found in \cite{SGWB_1}.} 
    \label{fig:irreducible}
\end{figure}

As it was demonstrated in Section \ref{subsec:real_st_GW}, upon re-entry inside the Hubble radius post-inflation, the amplitudes of the modes start oscillating and decay as $1/a(\eta)$, starting from the initial value at $k=a_kH_k$ given by \eqref{equ:init_val}. The exact nature if this oscillation is thereby influenced by the period in which the modes re-enter the horizon, i.e., during radiation or matter domination. To accommodate for both cases, the full solutions incorporates a transfer function $T(k,\eta_0)$ differentiates between scales $k>k_*$ and $k<k_*$ where $k_*$ approximates the scales entering the horizon at which $a_\text{m}(\eta)$ and $a_\gamma(\eta)$ cross \cite{SGWB_1}. In the spectral energy density, the transfer function then appears quadratically as it multiplies the amplitude of the tensor perturbations $h_{ij}$. One finds
\begin{align}\label{equ:sed_des}
    \Omega_\text{GW}(k) = \frac{1}{12 H_0^2a_0^2}T'(k,\eta_0)^2\mathcal{P}_h(k)\,.
\end{align}
Note here that one usually approximates $T'(k,\eta_0)^2\simeq k^2T(k,\eta_0)$ due to the oscillatory behavior of the tensor modes on sub-Hubble scales. By performing an oscillation-averaging procedure, on sub-horizon scales $k\eta_0\gg1$ (where $\eta_0$ corresponds to today), one finds \cite{Boyle_2008, Turner_1993}
\begin{align}\label{equ:transfer_mat_rad}
    T'(k,\eta_0)^2 \rightarrow \begin{cases}
       \eta_*^2/(2\eta_0)\,, & k>k_* \\
       9/(2\eta_0^4k^2)\,, & k<k_*
    \end{cases}\,,
\end{align}
where $\eta_*= 1/k_*$. Computing the energy density \eqref{equ:sed_des} using the latter transfer function in the sub-horizon limit \eqref{equ:transfer_mat_rad}, one finds that the energy density spectrum today scales as $k^{-2}$ for modes that entered the horizon during matter domination, while for modes entering earlier, during radiation domination, the spectrum appears to be almost perfectly flat. Fig. \ref{fig:irreducible} illustrates this transition and the associated features. It should be emphasized that these results have neglected multiple effects that may or may not yield substantial corrections to this behavior. In particular, it is derived without going beyond the Standard Model of particle physics and cosmology. A more thorough treatment would discard, for instance, the assumption that the transition between radiation and matter domination happens instantaneously; late-time expansion would need to be addressed, and free-streaming neutrino species have to be acknowledged. Perhaps most importantly, the period of reheating immediately following inflation in the cosmic history needs to be modeled in detail to obtain a precise spectrum for the irreducible background. All of the mentioned issues have been adressed in literature and can be found summarized in \cite{SGWB_1}. In particular, the period of reheating is well-known to also leave other potentially more feature-full 
contributions to the cosmological \gls{gw} background. In this context, various types of corrections to the results presented above have been presented to literature, e.g., \cite{Turner_1993,Kuroyanagi:2012yp,Nakayama_2008,Kuroyanagi_2009,Kuroyanagi_2015}.

%

\subsection{Beyond the irreducible Background}
\label{sec:SGWB_beyond_irreducible}

In Section \ref{sec:SGWB_irreducible}, it was demonstrated that according to the simplest cosmological models for inflation, a ``fundamental'' background of \gls{gw}s is established early on in the cosmic history. Naturally, over the subsequent evolution of the Universe, many more layers will be added to the cosmological \gls{gw} background. In this Section, the contributions of particular interest arise due to PBHs, topological defects, and first-order \gls{pt}s. The treatment will remain superficial, and more thorough calculations can be found in \cite{SGWB_1}. However, before shifting the focus away from the inflationary period, it is instructive to mention that significant deviations from the above demonstrated scale invariance of the irreducible background from inflation can arise if new particle species or symmetries are present during this period of rapid expansion. In contrast to the irreducible part, these contributions strongly depend on the underlying modifications to the standard scenario of Section \ref{sec:SGWB_irreducible}. The main drivers of modifications to inflation's background contributions are the presence of additional fields leading to particle production, new symmetry patterns allowing, for instance, for short-term graviton mass, or generally alternative theories of gravity. While some modify the tensor perturbation described above (for instance, through the presence of spectator fields), other sources add gravitational radiation on top via separate production schemes. For the latter to emit \gls{gw}s during inflation, tensor anisotropic stress must be present. As \gls{gw}s are diluted on sub-Hubble scales, contributions of such kind are only meaningful if generated sufficiently close to the Hubble scale. A candidate checking all boxes in this regard is, for instance, the particle production during inflation.

\subsubsection{Particle Production During Inflation}
\begin{figure}[t!]
    \centering
    \includegraphics[width=0.7\textwidth]{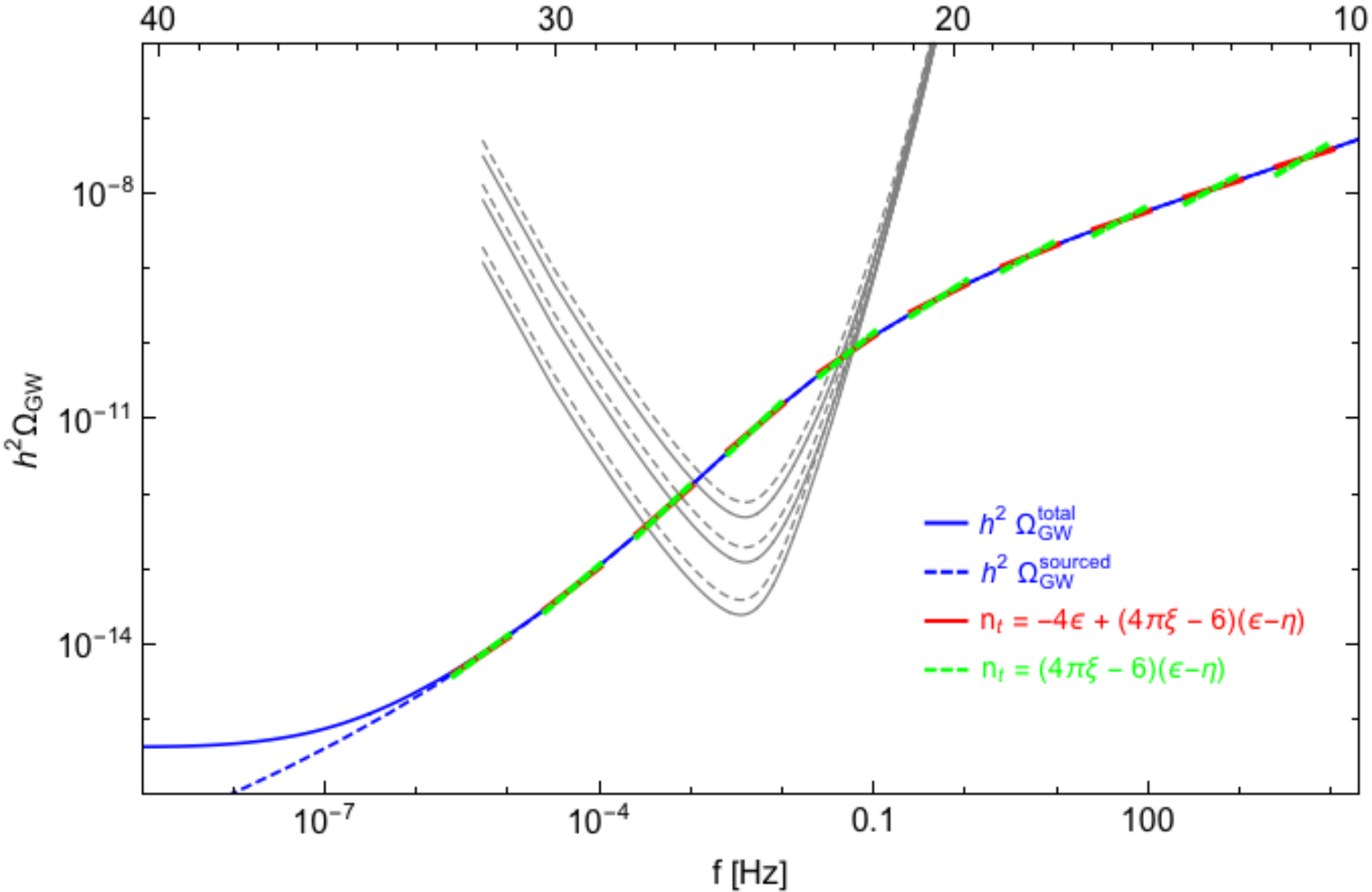}
    \caption{Sustained particle production \cite{Bartolo_2016}: numerical spectrum of GWs today $h^2\Omega_\T{GW}$ for a model of quadratic inflaton potential $V(\phi) = \frac{1}{2}m^2\phi^2$, with inflaton-gauge field coupling $\Lambda = M_\T{Pl}/35$ (continuous line) where $M_\T{Pl}$ is the Planck mass. For comparison, a local parametrization $h^2\Omega_\T{GW}\propto (f/f_*)^{n_T}$, evaluated at various pivot frequencies $f_*$ and with spectral tilt obtained from successive approximations to $n_T$, is displayed. Further displayed are Power Law-Integrated Curves of six LISA configurations \cite{Bartolo_2016}.} 
    \label{fig:smack_that}
\end{figure}

The emission of \gls{gw}s by particle production during inflation is made possible by the time-dependent background established as the inflaton rolls down its potential. This background carries sufficient energy to produce light species of particles. Independent of the particle species, i.e., scalar, vector, or fermionic, the relevant field necessarily couples to the inflaton with its mass vanishing at some time $t_0$, marking the time of particle production. The latter is achieved by selecting mass terms of the form $g(\phi-\phi_0)\bar\psi\psi$ where $\phi(t_0)=\phi$ and $\phi$ can but not necessarily has to be the inflaton. The \gls{gw} features corresponding to particle production added on top of the irreducible background spectral energy density are expected to arise around the frequency today corresponding to time $t_0$. For such mass term couplings, particle production usually happens quite rapidly around the time $t_0$. A continuous or sustained particle production can be obtained by coupling the inflaton to the derivative of, for instance, a gauge field. An exemplary interaction term would be $\mathcal{L}_{int}\sim \phi F_{\mu\nu}F^{\mu\nu}$. The factor of proportionality is a dimension-full constant $\Lambda$, subsequently denoted as the inflaton-gauge field coupling. These types of models have gained significant attention in literature, see \cite{SGWB_1} and references therein. In these models, the standard GWs from vacuum fluctuations and the additional
sourced GWs created by the excited gauge fields are statistically uncorrelated. For illustration, an instance of a GW background caused by such an interaction is displayed in Fig. \ref{fig:smack_that}. Instead of contributing a localized feature, as for the explosive particle production, sustained particle production results in a continuous increase of $\Omega_\text{GW}$ over frequency. The onset of particle production thereby marks the deviation from the roughly scale-invariant irreducible background into a continuous ascent. Therefore, their power spectra simply add together. The latter causes the characteristic shape of the total spectrum displayed in Fig. \ref{fig:smack_that} which can be separated into three distinct regimes: First, a regime where vacuum fluctuations contribution at large scales dominate ($f\lesssim10^{-5}$ Hz); second, dominance of the sourced GW contributions at intermediate scales with negligible back reaction of the coupled gauge field\footnote{Note that in this regime, the evolution of the inflaton $\dot \phi$ and the Hubble radius $H$ are still determined by slow-roll equations.} ($10^{-5} \T{ Hz } \lesssim f \lesssim 1$ Hz); third, the complete dominance of sourced GW contributions including strong back reactions ($f\gtrsim 1$ Hz). Interestingly, depending on the exact model specifications and coupling strength, the resulting GW background exhibits an almost constant slope over the sensitivity band of the LISA instrument, making it a promising candidate for further analysis. The shape of the spectrum from particle production is in stark contrast to what one expects to see for enhanced scalar perturbations. It is postulated that the latter results in sharper features within the frequency regime probed by the LISA mission \cite{Kohri_2018,PBH}. The enhanced perturbation thereby leads to predictions rich in phenomenologically interesting features such as PBHs.

\subsubsection{Primordial Black Holes}
It is commonly assumed that the amplitude of primordial density perturbations —measured with high precision by the Planck satellite at cosmological scales—remains approximately constant across all scales. This expectation is supported by the standard class of single-field slow-roll inflationary models, which predict near scale invariance of the scalar perturbation spectrum, provided the spectral index exhibits negligible running. However, such an extrapolation to smaller scales is substantial, given the lack of a definitive underlying model for inflation. Consequently, it remains entirely plausible that the scalar perturbation spectrum departs significantly from scale invariance at small scales.
In the following, a deviation from quasi-scale invariance at small scales is assumed. Several inflationary scenarios readily permit the generation of large-amplitude scalar fluctuations at small scales while remaining consistent with CMB observations on cosmological scales (see, e.g., \cite{Shit_I,Shit_II}). Notably, substantial density perturbations on small scales may lead to the formation of primordial black holes, which have been proposed as potential contributors to the dark matter content of the Universe (e.g., \cite{DM_SHIT_I,DM_SHIT_II,DM_SHIT_III,DM_SHIT_IIII}).
Instead of choosing a specific model, it is to be emphasized that it is a generic and model-independent consequence that first-order scalar perturbations—regardless of their amplitude or statistical distribution—inevitably induce \gls{gw}s at second and higher orders, see, e.g., \cite{Fuck_Shit_I,Fuck_Shit_II,Fuck_Shit_III,Fuck_Shit_IIII}. In particular, second-order contributions to the perturbed Einstein equations, constructed from products of first-order scalar modes, act as effective sources for tensor perturbations. The resulting stochastic \gls{gw} background thus encodes valuable information about the amplitude and statistical properties of scalar perturbations at small scales.

The effect can easily be quantified by considering the solution to the perturbed Einstein equations using the perturbed FRW metric 
\begin{align}
    \dd s^2 = a^2 (\eta) [-(1+2\Phi)\dd^2\eta + [(1-2\Psi)\delta_{ij}+h_{ij}]\dd x^i \dd x ^j]\,,
\end{align}
where $\Psi,\Phi$ are scalar metric perturbations and $h_{ij}$ the TT tensor perturbations. To second order in perturbations, the resulting field equations can be brought to the form 
\begin{align}
    h_{ij}'' + 2\mathcal H h_{ij}' + k^2h_{ij}= S_{ij}^{TT}\,,
\end{align}
where $S_{ij}^{TT}$ acts as a source term for tensor perturbations but consists of scalar perturbations, i.e., \cite{Fuck_Shit_II}
\begin{align}
    &S_{ij} = 2\Phi\partial_i\partial_j \Phi - 2\Psi \partial_i\partial_j\Phi + 4\Psi \partial_i\partial_j \psi + \partial_i\phi\partial_j\Phi - \partial_i\Psi\partial_j\Phi + 3\partial_i\partial_i \Psi\partial_j\Psi \notag \\
    &-\frac{4}{3(1+w)\mathcal{H}^2}\partial_i(\Psi' + \mathcal H \Phi)\partial_j(\Psi' + \mathcal H \Phi)- \frac{2c^2_s}{2w\mathcal H}[3\mathcal H(\mathcal H\Phi-\Psi') + \nabla^2\Psi]\partial_i\partial_j(\Phi-\Psi)\,.
\end{align}
Observationally, in particular second-order induced GWs are interesting, which are produced during the radiation era, as only modes entering the horizon during that early stage of the universe are probed by the frequency range of GW detectors. For second-order induced GWs during the radiation-dominated era, i.e., $w = 1/3$, it can be shown \cite{Fuck_Shit_II} that the GW background (corresponding to $f>10^{-17}$ Hz) is characterized by
\begin{align}\label{equ:iphone16}
    \Omega_\T{GW} (k)= F_\T{rad} \Omega_\T{rad} \mathcal P_{\mathcal R}^2(k)\,,
\end{align}
where $\mathcal P_{\mathcal R}(k)$ is the mostly Gaussian and smooth power-spectrum of scalar perturbations $\propto (k/k_*)^{n_s-1}$ and 
\begin{align}
    F_\T{rad}= \frac{8}{3}\left(\frac{216^2}{\pi^3}\right) 8.3\cdot 10^{-3} f(n_s)\,,
\end{align}
and $f(n_s)$ is a weekly-dependent function on the spectral tilt $n_s$ \cite{SGWB_1}. For bounds on the primordial scalar perturbation by means of the constraints on a stochastic background of GWs, the reader is referred to \cite{Assadullahi_2010}. For a discussion regarding the working assumptions of Eq. \eqref{equ:iphone16}, see \cite{SGWB_1}. 

\begin{figure}[t!]
    \centering
    \includegraphics[width=0.7\textwidth]{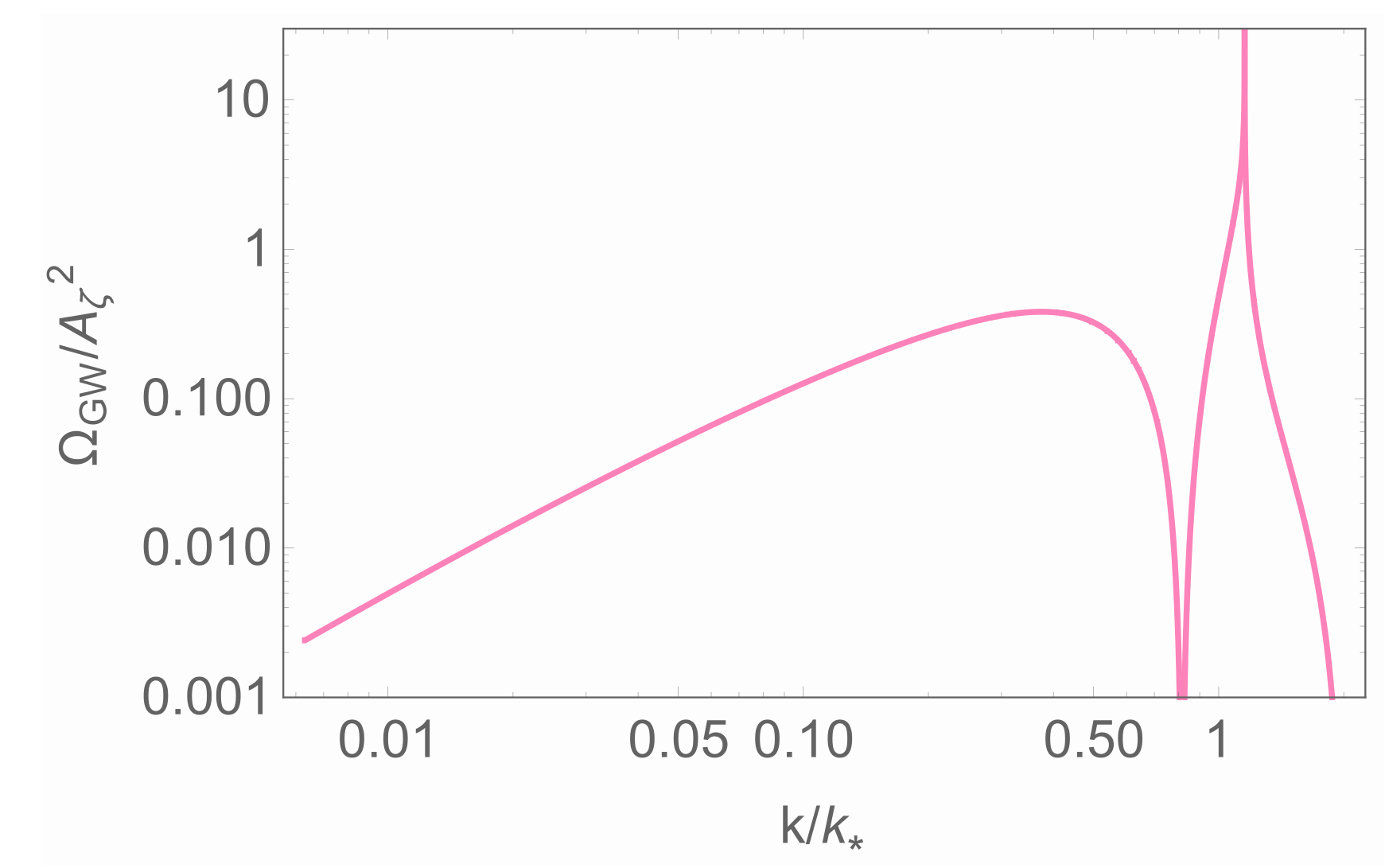}
    \caption{Energy density fraction $\Omega_\T{GW}$ of GWs produced in the radiation- dominated era for monochromatic sources according to \cite{Kohri_2018}. The factor $A_\zeta$ is a normalization constant. } 
    \label{fig:fuckthisshitimout}
\end{figure}

Large enhancements in the curvature power spectrum at specific scales can lead to the formation of PBHs when those scales re-enter the horizon during the post-inflationary evolution of the Universe. This production mechanism generally results in a population of isolated PBHs that may serve as cold dark matter candidates \cite{Garc_a_Bellido_1996}. For PBHs with masses in the range $M_\T{PBH} \sim 10^2-10^4 \Msol$, the corresponding peak frequency of the associated \gls{gw} background could fall within the sensitivity band of LISA. Given their potential role as dark matter and their influence on early galaxy formation, \gls{gw}s emitted by merging PBHs represent a compelling observational window into the early structure formation of the Universe. Indeed, it has been suggested that such PBHs may already have been observed by aLIGO (e.g., \cite{DM_SHIT_II}). Looking ahead, future observations with both ground- and space-based interferometers are expected to enable precise measurements of the broad PBH mass spectrum \cite{Raidal_2017, clesse2018detectinggravitationalwavebackground}. The computation of the spectrum induced by PBHs, being part of the reducible contribution, is sensitive to the chosen model of inflation. While \cite{SGWB_1} provides a more general discussion, in this work, in particular Section \ref{sec:SGWB_detection_Paper_HI}, inspiration is drawn from \cite{Kohri_2018}. The latter computes a semi-analytical spectrum during the radiation-dominant era. Assuming monochromatic curvature perturbations, $\mathcal P \sim \delta(\log k/k_*)$, the spectrum obtains distinctive features as displayed in Fig. \ref{fig:fuckthisshitimout}. Note that this depiction provides merely an instant of the PBH spectrum for which there is generally no consent in literature. What is, however, widely accepted is the exhibition of sharp features within or close to the frequency regime to which LISA is sensitive. This establishes large benefits for its detection, see Section \ref{sec:SGWB_detection_Paper_HI}. 

\subsection{Phase Transitions}
\label{sec:SGWB_Phase_Transitions}

During its adiabatic expansion, the Universe may have undergone several \gls{pt}s as a consequence of the decreasing temperature. A wide range of phenomena associated with primordial PTs can result in the generation of a SGWB. In many cases, such a relic SGWB is the only observable remnant of the transition, offering valuable insight into its underlying nature.
First-order PTs are distinguished by the emergence of a potential barrier separating a metastable, symmetric (false) vacuum from a more energetically favorable, symmetry-breaking (true) vacuum. As the Universe cools, this barrier prevents the field from transitioning smoothly to the true vacuum; instead, the transition proceeds via quantum tunneling or thermal fluctuations. In physical space, this process manifests as the nucleation of bubbles of the true vacuum within the surrounding false vacuum. These bubbles expand due to the pressure differential between the two vacua, releasing the latent energy stored in the false vacuum.
In the idealized scenario of a PT occurring in empty space, this energy would be entirely converted into gradient energy, causing the bubble walls to accelerate relativistically. In the early Universe, however, the presence of a hot, dense plasma changes the dynamics. The majority of the released energy is transferred into thermal energy, heating the plasma, while the remainder is split between the gradient energy of the bubble walls and the kinetic energy imparted to the plasma through bulk fluid motion.
Both the gradient energy of the scalar field and the kinetic energy of the plasma correspond to energy-momentum tensors with non-zero anisotropic stress—one of the key conditions for sourcing a \gls{gw} background. If a tensor component is present in these stress distributions, it can act as a \gls{gw} source (see Eq.~\eqref{equ:fuck_shit_ii}).
\gls{gw} production becomes especially efficient towards the end of a first-order PT, when bubbles of true vacuum collide and the entire Universe transitions to the symmetry-broken phase. These collisions inherently break the spherical symmetry of both the bubble walls and the surrounding plasma flows, generating a non-zero tensor anisotropic stress that actively sources \gls{gw}s (see \cite{PhysRevD.30.272} for early indications and \cite{SGWB_1} for a comprehensive review).

The GW signal produced by first-order PTs depends on a limited set of parameters that characterize the dynamics of the broken-phase bubble evolution—such as the typical bubble size at the time of collision and the velocity of the bubble walls—as well as the energy available to source the GWs, which is governed by the tensor anisotropic stresses. These stresses are themselves determined by the strength of the PT and the interaction between the transitioning field and the surrounding plasma constituents.
Although the specific values of these parameters are model-dependent and rooted in the particle physics details of the PT, the resulting GW signal can be expressed in a phenomenological and largely model-independent framework. One particularly critical parameter is $T_*$, the temperature of the thermal bath at the time $t_*$  when the \gls{gw}s are generated—typically coinciding with the final stages of the transition when bubble collisions occur. For transitions without substantial supercooling or reheating, $T_*$ approximately corresponds to the nucleation temperature, which is set by the nucleation rate of the true vacuum bubbles. The strength of the PT is often quantified by the dimensionless parameter $\alpha = \rho_\T{vac}/\rho_\T{rad}^*$ representing the ratio of vacuum energy density released during the transition to the radiation energy density of the Universe at the time of the transition. This parameter plays a key role in determining the amplitude of the resulting GW background.

\begin{figure}[t!]
    \centering
    \includegraphics[width=1.0\textwidth]{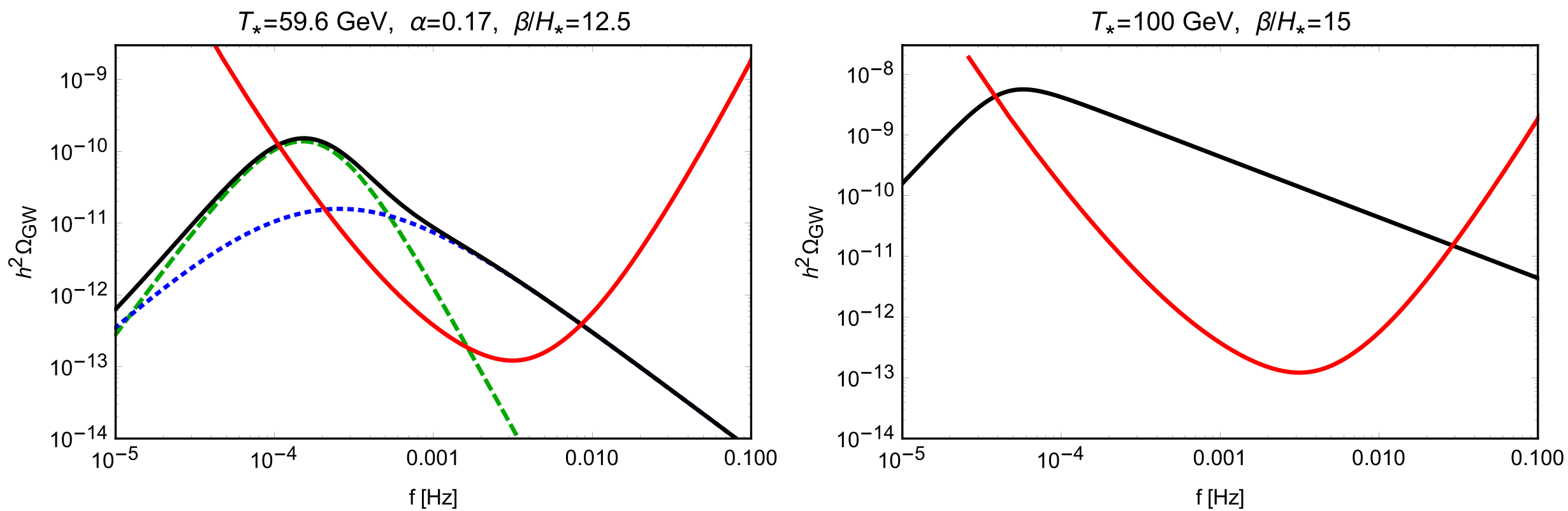}
    \caption{SGWB spectra in two examples of first order PT, compared with the estimated sensitivity curve of the interferometer LISA calculated from \cite{LISA} (red curve). This depiction is extracted from \cite{SGWB_1}. The black solid lines in the left and right plot provide instances of PT-sourced GW backgrounds.} 
    \label{fig:surfingggg}
\end{figure}

A rough estimate of the GW amplitude reveals its scaling behavior with the duration of the source and the magnitude of the tensor anisotropic stress(e.g. \cite{Caprini_2008}). Assuming the process responsible for generating the tensor anisotropic stress has a characteristic timescale equal to the PT duration, $1/\beta$, and that this timescale is shorter than the Hubble time at the corresponding epoch, $H_*$, the relation $\beta^2 h \sim 16\pi G \Pi$ can be applied, where $h$ denotes the tensor perturbation amplitude and $\Pi$ the tensor part of the energy momentum source tensor. This leads to an expression for the present-day SGWB as \cite{SGWB_1}
\begin{align}
    h^2\Omega_\T{GW} \sim 1.6\cdot 10^{-5} \left(\frac{100}{g_*(T_p)}\right)^{1/3}\left(\frac{H_*}{\beta}\right)^2\left(\frac{\kappa \alpha}{1+\alpha}\right)^2\,,
\end{align}
where $\kappa\sim \Pi/\rho_\T{vac}$ and $T_p$ is the temperature of production. The characteristic frequency today accordingly follows as \cite{SGWB_1}
\begin{align}\label{jsadjlkgbaksbg}
    f\sim 1.6 \cdot 10^{-5} \T{ Hz } \frac{\beta}{H_*}\left(\frac{100}{g_*(T_p)}\right)^{-1/6} \frac{T_*}{100 \T{ GeV}}\,.
\end{align}
Since at the end of the PT one expects the entire Universe to be converted to the broken phase, in general, the PT must complete faster than a Hubble time, so that $\beta/H_*>1$. Then Eq. \eqref{jsadjlkgbaksbg} suggests that the characteristic frequency of GW emitted around the EW symmetry breaking at 100 GeV falls in the frequency range of LISA for $1\lesssim \beta/H_*\lesssim10^5$, for instance. The slope of the GW spectrum at wave-numbers smaller than the Hubble radius at the time of production can be determined on general grounds, valid for any transient stochastic source after inflation. It is a consequence of the fact that the causal process (the PT) generating the GW signal cannot operate on time/length-scales larger than $H_*^{-1}$. From this it follows \cite{SGWB_1} that the infrared tail of the present-day GW spectrum behaves as $h^2\Omega_\T{GW} \propto f^3$ for $k<k_*$ and some $k_*$, and decays with a slope sensitive to the details of the sourcing PT for $k>k_*$. Examples of \gls{gw} backgrounds from PTs are sketched in Fig. \ref{fig:surfingggg}. The details regarding their computations can be found in \cite{SGWB_1}. Note that from an observational perspective, one is particularly interested in models that peak in the LISA band, as the peak establishes a distinguishing feature (see also the discussion of Section \ref{sec:SGWB_detection_Paper_HI}).

\subsection{Cosmic Defects}
\label{sec:SGWB_Cosmic_defects}

A PT in the early Universe corresponds to a process of spontaneous symmetry breaking, transitioning from a symmetric phase (false vacuum) to a broken phase (true vacuum). This process is typically driven by one or more scalar fields acquiring a non-zero vacuum expectation value within a vacuum manifold $\mathcal{M}_\T{vac}$. If $\mathcal{M}_\T{vac}$ satisfies certain topological conditions, cosmic defects may form as a consequence of the phase transition \cite{Vilenkin:2000jqa, CS_II}. Specifically, when the vacuum manifold is topologically non-trivial—i.e., it possesses a non-trivial homotopy group, $\pi_n(\mathcal{M}_\T{vac}) \neq \mathcal{I}$—topological field configurations emerge, giving rise to defects such as strings $(n = 1)$, monopoles $(n = 2)$, or textures $(n = 3)$. For higher values of $n$, there is no topological obstruction to the symmetry-breaking field reaching the vacuum manifold across space-time, resulting instead in non-topological field configurations. Depending on whether the broken symmetry is global or gauged, the resulting defects are classified as global or local, respectively. In all such cases—topological or not, local or global—the resulting objects are generically referred to as cosmic defects.
Cosmic strings (CSs), regardless of whether they are global or local, as well as all types of global defects, display a scaling behavior once a sufficient amount of time has elapsed since their formation \cite{CS_II}. This scaling regime is characterized by a self-similar evolution of the defect number density within a causal volume throughout cosmic history. As the cosmic defect network evolves, its energy-momentum tensor adjusts to maintain the scaling behavior. Consequently, the time evolution of the transverse-traceless component of the energy-momentum tensor associated with the defect network inevitably generates GWs \cite{KRAUSS1992229}.
Among scaling defects, cosmic strings arguably represent the best-motivated example from the perspective of particle physics. Their tendency to form loops during evolution results in the emission of a particularly large amount of GWs, making them a highly promising target for detection.\\
Cosmic strings are one-dimensional topological defects that emerge during a phase transition in the early Universe, provided the fundamental homotopy group of the corresponding vacuum manifold $\mathcal{M}_\T{vac}$ is non-trivial, i.e., $\pi_1(\mathcal{M}_\T{vac}) \neq \mathcal{I}$. These objects arise naturally in well-motivated inflationary frameworks. For example, local strings are generically produced at the end of inflation in supersymmetric GUT models of Hybrid inflation, assuming certain reasonable conditions are satisfied \cite{Jeannerot_2003}. Additionally, cosmic strings can also correspond to fundamental superstrings, rather than field-theoretic configurations, as encountered in scenarios such as brane inflation \cite{Sarangi_2002}.
Both field-theoretic strings and fundamental superstrings are characterized by a linear energy density $\mu$, which, in the Nambu-Goto approximation (describing infinitely thin strings), is identified with the string tension. A typical network of cosmic (super-)strings at any epoch consists of a mixture of ``large'' loops and ``small'' loops. Small loops have diameters smaller than the causal horizon, while large loops extend beyond the horizon scale, with only a portion residing within the observable volume.
A key aspect governing the evolution of such networks is the process of \textit{intercommutation}, wherein strings intersect (or self-intersect), exchange segments, and generate new loops. A primary distinction between field-theoretic strings and superstrings lies in the intercommutation probability: while field strings intercommute with probability $p = 1$, superstrings may have significantly lower values of $p$, which impacts the network’s evolution and, consequently, the resulting GW signal. Once formed, loops—owing to their substantial tension—undergo relativistic oscillations and lose energy predominantly through the emission of GWs.

\begin{figure}[t]
    \centering
    \includegraphics[width=1.0\textwidth]{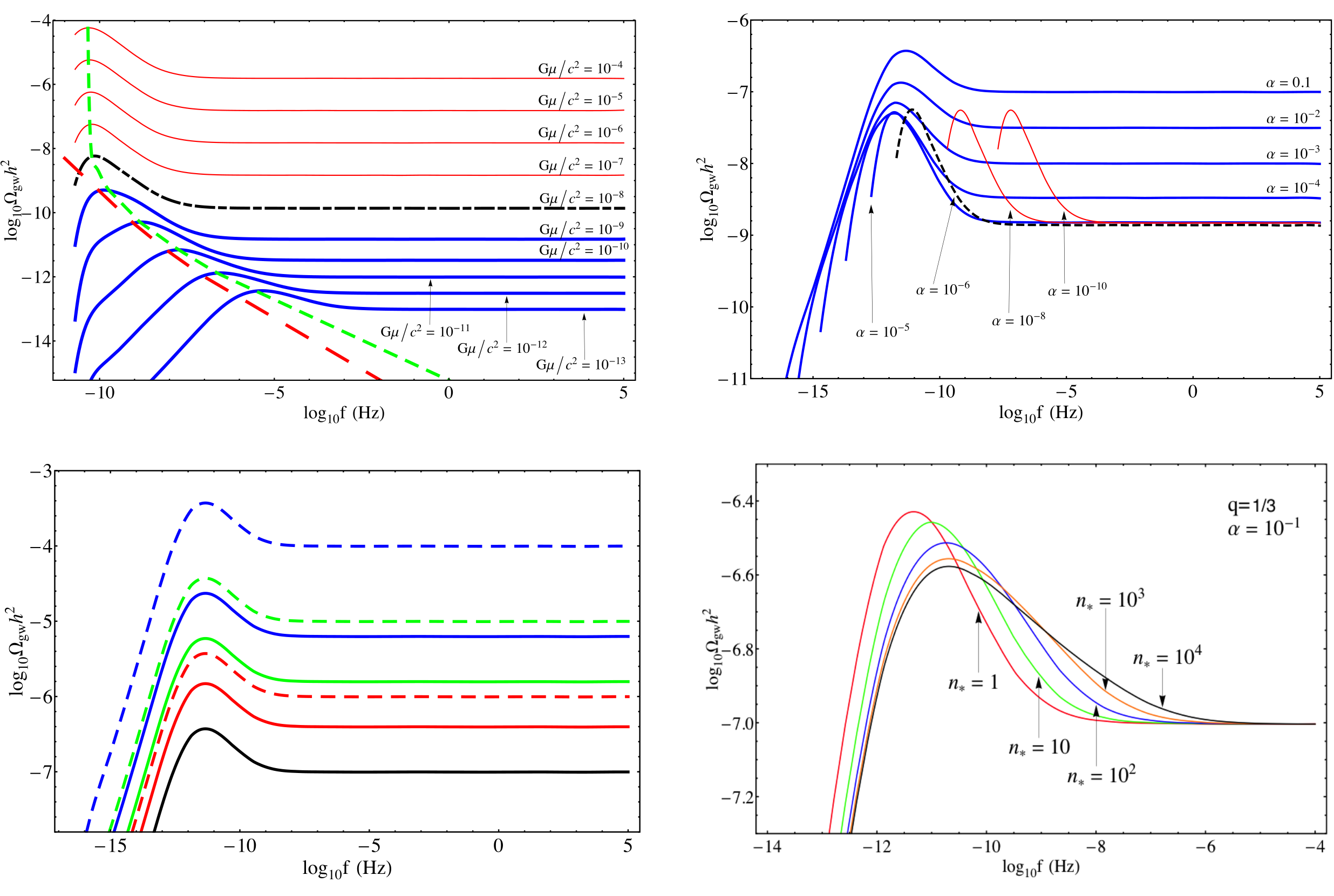}
    \caption{Amplitude of the energy density spectrum today of the GW background emitted by the decay of loops chopped oﬀ from a string network all through cosmic history \cite{Sanidas:2012tf}. The set of fiducial parameter is chosen as $G\mu = 10^{-7}$ [$c=1$], $\alpha = 10^{-7}$, $q=4/3$, $n_*=1$, and $p=1$. In each panel, one parameter is varied.} 
    \label{fig:nose_dive}
\end{figure}

The details of GW emission from cosmic string loops are highly sensitive to several intrinsic properties of the string network. The resulting spectrum is influenced in particular by: (i) the string tension $G\mu$, (ii) the initial loop size relative to the horizon $\alpha$, (iii) the spectral index $q$ of the emission spectrum, (iv) the high-frequency cut-off $n_*$ (modeling radiation backreaction), and (v) the intercommutation probability $p$. Given the number density of loops per unit volume and loop length, $n(l,t)$ (as given in Eq. (361) of \cite{SGWB_1}), the spectral power emitted by a loop of length $l$ is given by
\begin{align} 
\dd P_\T{GW}(f) = \Gamma G\mu^2 l \mathcal{P}(f)\dd f\,,
\end{align}
where $\mathcal{P}(f) \equiv 2^q q ,(f\cdot l)^{-1-q}$, and $\alpha$ characterizes the loop size at formation. The present-day energy density of GWs produced by the cosmic string network, emitted at time $t_*$ and observed at $t_0$, is then expressed as \cite{SGWB_1}
\begin{align} 
\dd \rho /\dd f \equiv \Gamma G \mu^2 \int_{t_*}^{t_0} \dd t \left(\frac{a(t)}{a_0}\right) \int_0^{a/H(t)} \dd l \, l\, n(l,t) \mathcal{P} \left(\frac{a_0}{a(t)} f, l \right)\,. \end{align}
Representative spectra, illustrating the dependence on the aforementioned model parameters, are shown in Fig. \ref{fig:nose_dive}. Observational constraints on such spectra have been derived, for example, in \cite{Sanidas:2012tf}, which reports a $95\%$ confidence upper limit on the string tension of $G\mu < 5.3 \cdot 10^{-7}$ for $\alpha \approx 10^{-5}$ and $n_* = 1$. For larger initial loop sizes, i.e., higher values of $\alpha$, bounds from the EPTA \cite{PTA_europe}, as analyzed in \cite{Sanidas:2012tf}, constrain $G\mu \lesssim 10^{-10}$. These limits are comparable to the tighter constraint $G\mu < 1.5 \cdot 10^{-11}$ obtained in \cite{Blanco_Pillado_2017}, among others. It is worth noting that looser constraints have been derived from $21$cm observations of cosmic strings, such as in \cite{Maibach:2021qqf}. Despite being less stringent, such searches involve fewer model parameters, making them an attractive avenue for cosmic string detection from a data analysis standpoint.

%

\section{Detecting the Stochastic Gravitational Wave Background}
\label{sec:SGWB_detection}

In previous Sections, a detailed overview of the theoretical and phenomenological origins of the SGWB is provided. Arguably, this background is rich in information about both astrophysics and cosmology. Its detection may thus be regarded as the next milestone for \gls{gw} physics or even cosmology in its entirety. So far, constraints on the SGWB have been established through various approaches \cite{PhysRevLett.97.021301} (see also \cite{PhysRevD.85.123002, Pagano_2016} for more a more recent analysis of similar data) for the reducible as well as the irreducible background of GW \cite{Cabass_2016, 2016, Bartolo_2016, Constraint_I, Constraint_II}\footnote{The major constraint on the irreducible \gls{gw} background is formally known as ``COBE-bound'' due to being derived using CMB quadrupole measurements by the COBE satelite \cite{COBE}.} (see also \cite{LIGO_limit, LV_2009, renzini2022stochastic}). The latter is constrained much better in comparison to the reducible part. In fact, their constraints differ by many orders of magnitude. In the following two Sections, two instruments actively participating in the search for a GW background are highlighted. In Section \ref{sec:SGWB_detection_astro}, given recent results, PTAs' efforts towards the detection of an astrophysical background are discussed. Section \ref{sec:SGWB_detection_Paper_HI} on the other hand, focuses on the potential detection of cosmological SGWB with the LISA instrument, outlining (in much more detail) a promising measurement strategy studied in \cite{SGWB_analysis_I}.

\subsection{Astrophysical Background Detection Efforts}
\label{sec:SGWB_detection_astro}

The pursuit of detecting the astrophysical GW background has been a central objective in \gls{gw} astronomy. Among many predictions and upper detection boundaries (e.g., \cite{LIGO_limit, LV_2009, renzini2022stochastic}), PTAs have emerged as a pivotal methodology in this endeavor, leveraging the exceptional rotational stability of millisecond pulsars to detect perturbations in spacetime induced by passing \gls{gw}s. Thus, in this Section, emphasis lies on the latest results of the North American Nanograv collaboration, which has made significant strides, particularly with its 15-year data set \cite{Nanograv_SGWB_I, Nanograv_SGWB_II, Tan_2025}. Nanograv's approach involves meticulous monitoring of the pulse arrival times from an array of millisecond pulsars distributed across the sky \cite{choi2024stochasticgravitationalwavebackground}. \gls{gw}s traversing the space between these pulsars and Earth induce minute variations in the pulse arrival times by perturbing the spacetime metric along the path from a pulsar to Earth. This causes a slight variation in the proper time of arrival of the pulsar’s signals as measured by Earth-based observers. In the case of a SGWB—arising, for instance, from a cosmological population of SMBBHs—these variations manifest as a spatially correlated signal across a pulsar array, superimposed upon intrinsic pulsar noise and other uncorrelated sources.\\
To isolate the GW signal, PTAs analyze the cross-correlations of timing residuals between pulsar pairs. The key theoretical prediction used to identify the stochastic GWB signal is the Hellings-Downs curve \cite{1983ApJ...265L..39H}—a quadrupolar angular correlation pattern predicted by \gls{gr} for an isotropic background of \gls{gw}s. 
The Hellings-Downs curve provides the expected angular correlation between the timing residuals of two pulsars as a function of their angular separation on the sky $\theta$, assuming the background of GWs is isotropic, unpolarized, and Gaussian. The correlation function is given by \cite{1983ApJ...265L..39H}
\begin{align}
    \zeta(\theta) = \frac{3}{2}x \ln x - \frac{1}{4} x + \frac{1}{2} - \frac{1}{2}\delta(x)\,,
\end{align}
where $x:=\frac{1-\cos \theta}{2}$ and $\delta(x)$ encapsulates the autocorrelation term. This correlation is quadrupolar in nature, peaking for co-aligned pulsars $\theta=0$ and becoming anti-correlated for pulsars separated by approximately $90$ degrees. The presence of this angular signature in the data is regarded as a ``smoking gun'' for the stochastic \gls{gw} background, as no other known source of correlated noise exhibits this precise spatial dependence.
Therefore, detecting this pattern is crucial for confirming the presence of a GW background.\\
Recently, Nanograv released its 15-year data set, encompassing observations from 67 pulsars \cite{Nanograv_SGWB_I, Nanograv_SGWB_II} and revealing multiple lines of first evidence for a stochastic signal consistent with the Hellings-Downs correlation. This result can be seen as providing compelling hints towards the existence of a nanohertz-frequency GW background \cite{Nanograv_SGWB_II}. The suspected GW background is characterized by a power-law spectrum, indicative of a population of inspiraling SMBBHs distributed throughout Earth's cosmological vicinity. The cumulative effect of numerous such systems contributes to the stochastic background observed, as outlined in Section \ref{sec:SMBHB}. A confirmation of this GWB would have profound implications for our understanding of galaxy evolution and the population statistics of SMBBHs. Moreover, it would open new avenues for probing fundamental physics, including potential contributions from early Universe phenomena such as cosmic strings or \gls{pt}s.\\
While the current data align well with the SMBBH interpretation, ongoing and future observations aim to refine the spectral characterization of the GW background, potentially disentangling contributions from various sources. Nanograv's achievement thereby marks a significant milestone in \gls{gw} astronomy, demonstrating the efficacy of PTAs in exploring the low-frequency \gls{gw} spectrum and enriching our comprehension of the dynamic Universe. For details regarding Nanograv's detection strategy, data analysis pipeline and (cosmological) working assumptions, the interested reader is referred to \cite{Nanograv_SGWB_I, Nanograv_SGWB_II}.
%
%

%

%

%

%

\subsection{Cosmological Background Detection Efforts}
\label{sec:SGWB_detection_Paper_HI}

Due to its rich phenomenology, the cosmological component of the \gls{sgwb} represents a smocking gun for new physics. However, directly detecting this component poses a significant challenge for data analysts. The \gls{sgwb} is predominantly shaped by astrophysical sources—such as white dwarf binaries, stellar-mass black hole inspirals, and extra-galactic mergers—which obscure and complicate the identification of primordial features in the spectrum. 
One of the primary challenges and priorities in \gls{sgwb} measurements is the disentanglement of instrumental noise, galactic foregrounds, and extra-galactic (including cosmological) components of the stochastic signal, without compromising the integrity of any individual contribution \cite{romano_detection_2017, lisa_cosmology_working_group_maximum_2020, domcke_measuring_2020, SGWB_analysis_III, cusin_doppler_2022, SGWB_analysis_II, dallarmi_dipole_2022} (see also \cite{wmap_initial_doppler, planck_initial_doppler, aghanim_planck_2014, Bonvin_2006, Maartens_2018, dalang_kinematic_2021, nadolny_new_2021} for analogous analyses in the electromagnetic spectrum).\\
A promising strategy to address this issue involves the exploitation of observer-dependent features that enhance the detectability of cosmological signals over foreground and instrumental noise. One such feature is the kinematic signature introduced by the motion of the local group of galaxies \cite{aghanim_planck_2014, Bonvin_2006, cusin_doppler_2022, dallarmi_dipole_2022}. While galactic sources remain unaffected, the motion of (both ground- and space-based) detectors relative to the rest frame of primordial stochastic \gls{gw} sources induces Doppler anisotropies, affecting only the cosmological contributions to the \gls{sgwb}. This motion is expected to enhance power in the lower-order multipoles of the cosmological component, mirroring effects observed in the \gls{cmb} several decades ago. 
As a result, the study of kinematic anisotropies in the \gls{sgwb} holds significant promise and has already sparked considerable interest in the scientific community \cite{romano_detection_2017, baghi_uncovering_2023, cusin_doppler_2022, SGWB_analysis_II, lisa_cosmology_working_group_maximum_2020, Contaldi_with_variance}.

In this section, the goal is to extend previous investigations by developing a diagnostic map-making framework aimed primarily—though not exclusively—at identifying extra-galactic sources in \gls{sgwb} data observed by LISA. A full four-year time-domain simulation of the anisotropic \gls{gw} sky is carried out using the LISA Simulation Suite (\texttt{LISAGWResponse} \cite{bayle_lisa_2023, bayle_lisa_2022}, \texttt{LISAInstrument} \cite{bayle_unified_2023, lisa_instrument}) along with the post-processing package \texttt{PyTDI} \cite{staab_pytdi_2023}, which performs the \gls{tdi} combination of optical measurements and constructs the final interferometric observables \cite{hartwig_time-delay_2022}. \\
Instead of relying on the commonly used Fisher matrix approach, the analysis employs a map-making strategy based on a Markov Chain Monte Carlo (MCMC) framework, similar to the methodology outlined in Section \ref{sec:Paper_LISA_II}, to assess the detectability of kinematic anisotropies in simulated \gls{sgwb} data. For related investigations focused on the galactic background, see \cite{banagiri_mapping_2021}.

\subsubsection{Doppler-boosted anisotropies as smoking-gun for extra-galactic origin}

This investigation begins with a concise review of the fundamental principles underlying stochastic kinematic anisotropies, following the discussion in \cite{SGWB_analysis_II, cusin_doppler_2022}. As outlined in the introduction, the \gls{sgwb} comprises contributions from both galactic and extra-galactic sources. The extra-galactic component is subject to Doppler shifts relative to the source frame, owing to the motion of our galaxy with respect to the \gls{sgwb} rest frame.\\
In particular, extra-galactic contributions to the \gls{sgwb} exhibit a directional modulation of their apparent, frequency-dependent energy density, expressed as $\Omega_\T{GW}(f, \mathbf{\hat{k}})$, where both frequency and angular dependencies are made explicit. To formalize this effect, consider two inertial frames: the source frame $\mathcal{S}'$, co-moving with the \gls{sgwb} source, and the observer frame $\mathcal{S}$, moving with constant velocity $\mathbf{v}$ relative to $\mathcal{S}'$. The fractional energy density of the \gls{sgwb} in the source frame, $\Omega'_\T{GW}(f)$, is assumed to be perfectly isotropic and solely frequency-dependent. This assumption holds under the condition that intrinsic anisotropies from the source are subdominant\footnote{Such anisotropies may arise from the production mechanism of the \glspl{sgwb}, Sachs-Wolfe (SW) and integrated SW effects, and from propagation through a perturbed Universe. A comprehensive review of anisotropy sources is provided in \cite{SGWB_analysis_II}.}.\\
Under these assumptions, a Lorentz boost is applied to transform the isotropic density spectrum $\Omega'_\T{GW}(f)$ from the rest frame $\mathcal{S}'$ into the observer frame $\mathcal{S}$, resulting in a modulated spectrum $\Omega_\T{GW}(f, \mathbf{\hat{k}})$. Here, the boost velocity is defined as $\mathbf{v} = \beta \mathbf{\hat{v}}$, with the convention $c = 1$, so that $\beta = |\mathbf{v}| =: v$\footnote{Throughout this work, natural units are adopted with $c=1$.}.
For the energy density spectrum, the boost mapping $\mathcal S '$ to $\mathcal S$ yields
\begin{align}
    \Omega(f, \mathbf{\hat{k}}) = \left(\frac{f}{f'}\right)^4\Omega(f').
\end{align}
The latter equation is generally defined for any Lorentz-boost where $0\leq \beta < 1$. Following the assumption of an isotropic spectrum in the source frame, this  result can be expanded up to second-order in $\beta$, using the relations between $f,f'$ outlined in \cite{SGWB_analysis_II}. One finds 
\begin{align}\label{equ:-1}
    \frac{f}{f'} = \mathcal{D} = \frac{\sqrt{1-\beta^2}}{1-\beta\  \mathbf{\hat{k}}\cdot\mathbf{\hat{v}}}
\end{align}
such that
\begin{align}\label{equ:0}
    \Omega_\T{GW}(f, \mathbf{\hat{k}}) = \mathcal{D}^4 \ \Omega_\T{GW}'\left( \mathcal{D}^{-1} f\right),
\end{align}
which expands to
\begin{align}\label{equ:1111}
    \Omega_\T{GW}(f,\mathbf{\hat{k}})=\Omega'_\T{GW}(f)\bigg(1+M(f)+\mathbf{\hat{k}\cdot\hat{v}}\, D(f) +
    \left[\left((\mathbf{\hat{k}\cdot\hat{v}})^2-\frac{1}{3}\right)Q(f)\right]\bigg).
\end{align}
Here, the functions $M(f), D(f), Q(f)$ correspond to the monopole, dipole, and quadrupole contributions respectively. They are given by 
\begin{align}\label{equ:M}
    M(f)&=\frac{\beta^2}{6}\left(8+n_{\Omega}(n_{\Omega}-6)+\alpha_{\Omega}\right),
\end{align}
\begin{align}\label{equ:D}
    D(f)&=\beta(4-n_\Omega),
\end{align}
\begin{align}\label{equ:Q}
    Q(f)&=\beta^2\left(10-\frac{9n_\Omega}{2}+\frac{n^2_\Omega}{2}+\frac{\alpha_\Omega}{2}\right),
\end{align}
and
\begin{align}\label{equ:alpha_n}
    n_\Omega(f)=\frac{\dd \ln(\Omega'_\T{GW}(f))}{\dd \ln{f}},&&\alpha_{\Omega}(f)=\frac{\dd n_{\Omega}(f)}{\dd \ln{f}}.
\end{align}
It is important to emphasize that the functions $M(f)$, $D(f)$, and $Q(f)$ introduced herein are inherently model-dependent and encode specific characteristics associated with the nature of the sources contributing to the extra-galactic component of the \gls{sgwb}. Their dependence on $\mathbf{\hat{k}} \cdot \mathbf{\hat{v}}$ arises naturally from an expansion in the boost parameter $\beta$. It must be noted, however, that $M(f)$, $D(f)$, and $Q(f)$ can receive additional contributions beyond those defined in Eqs. \eqref{equ:M}--\eqref{equ:Q} in the presence of intrinsic anisotropies in the source-frame spectrum. Consequently, the use of Eq.~\eqref{equ:1111} in combination with the specified forms of $M(f)$, $D(f)$, and $Q(f)$ is strictly valid only under the assumption of an isotropic source-frame background.
\\
As can be verified computationally, the functions discussed above directly determine the harmonic expansion coefficients $a_{lm}$, up to constant prefactors. These coefficients are instrumental in defining the angular power spectrum $C_{l}^{\text{GW}}$: By rewriting Eq. \eqref{equ:1111} as
\begin{align}
    \Omega_\T{GW}(f,\mathbf{\hat{k}})=\Omega'_\T{GW}(f)\left(1+\delta_\T{GW}^{\text{kin}}(f,\theta,\phi)\right),
\end{align}
where $\delta_\T{GW}^{\text{kin}}$ now describes the anisotropic part of the measured spectral energy density $\Omega_\T{GW}$ in the observer frame $\mathcal S$, one can exploit that $\delta_\T{GW}^{\text{kin}}$ has a smooth angular dependence, and expand
 \begin{align}
     \delta_\T{GW}^{\text{kin}}(f,\mathbf{\hat{k}})=\sum_\ell\sum_m\delta_{GW,\ell m}^{\text{kin}}(f)Y_{\ell m}(\mathbf{\hat{k}}).
 \end{align}
 The angular power of the anisotropies is then given as an ensemble average
 \begin{align}\label{equ:3}
\langle\delta_{GW,\ell m}^{\text{kin}},\delta_{GW,\ell'm'}^{\text{kin}}\rangle=:C_\ell^{GW}(f)\delta_{\ell \ell'}\delta_{mm'}.
 \end{align}
 The coefficient $C_\ell^{GW}$ can be derived straightforwardly by inserting the anisotropies $\delta_\T{GW}^{\text{kin}}(f,\mathbf{\hat{k}})$ in Eq. \eqref{equ:3}. Then, by orthogonality of the spherical harmonics one finds\footnote{Note that here the convention $\int_0^\pi\dd \theta\int_{\varphi=0}^{2\pi}Y_{\ell m} \, Y_{\ell'm'}^*d\Omega = \frac{4 \pi}{(2 \ell + 1)} \delta_{\ell\ell'}\, \delta_{mm'}$ is used.}
 \begin{align}
     \frac{1}{4\pi} \int_{\Omega}\left|\delta_\T{GW}^{\text{kin}}\right|^2\dd \Omega
     =\sum_\ell\frac{1}{2\ell+1}\sum_m\left|\delta_{GW,\ell m}^{\text{kin}}\right|^2=:\sum_\ell C_\ell^{GW},
 \end{align}
 and therefore,
 \begin{align}\label{equ:4}
     C_\ell^{GW}=\frac{1}{2\ell+1}\sum_{m=-\ell}^{\ell}\left|\delta_{GW,\ell m}^{\text{kin}}(f)\right|^2.
 \end{align}
Combining this result with Eq. \eqref{equ:1111}, one can assign each mode $\delta_{GW,lm}^{\text{kin}}$ to the corresponding kinematic mode, that is\footnote{Here, the frame is chosen in which $\mathbf{\hat{k}}\cdot \mathbf{\hat{v}}\approx \cos \theta$. In subsequent Sections, this is transitioned to the International Celestial Reference System (ICRS) frame and the attention lays mostly on Eq. \eqref{equ:0}, albeit it is expressed in terms of monopole, dipole, and quadrupole language utilizing the expansion coefficients $M(f)$, $D(f)$, and $Q(f)$.}
\begin{align}
    \delta_{GW,00}^{\text{kin}}(f)=2\sqrt{\pi}M(f),&&\delta_{GW,10}^{\text{kin}}(f)=2\sqrt{\frac{\pi}{3}}D(f) ,&& \delta_{GW,20}^{\text{kin}}(f)=\frac{4}{3}\sqrt{\frac{\pi}{5}}Q(f).
\end{align}
It follows trivially that
\begin{align}
    C_0^{GW}\sim|M(f)|^2, && C_1^{GW}\sim|D(f)|^2, && C_2^{GW}\sim|Q(f)|^2.
\end{align}
It is important to emphasize that the proposed decomposition provides several notable advantages. For example, Eqs. \eqref{equ:M}--\eqref{equ:Q} are governed by the gradient and higher-order derivatives of the spectrum. It can be readily demonstrated that, in scenarios involving rapid spectral variations—such as sharp peaks or discontinuities—the individual components $M(f)$, $D(f)$, and $Q(f)$ are significantly amplified due to the influence of $n_\Omega$ and $\alpha_\Omega$. In certain regimes, this amplification can result in the kinematic quadrupole exceeding the kinematic dipole in magnitude. The direct implications of such effects are further discussed in Section~\ref{sec:discussion}.

\begin{figure}[t]
    \centering
    \includegraphics[width=0.6\columnwidth]{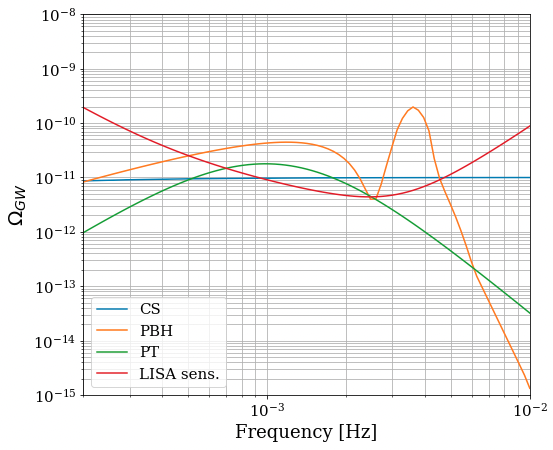}
    \caption{LISA sensitivity curve \cite{Scird} compared to the expected stochastic GW spectrum for cosmic strings (blue) modeled as a scale-free contribution at a reference power of $\Omega_\T{GW}=10^{-11}$, primordial black holes (orange) modeled according to \cite{Kohri_2018,PBH}, and first order PTs (green) as implemented in the \texttt{PTPlot} package \cite{caprini_detecting_2020}. For the LISA sensitivity curve (red) we use the implementation within the \texttt{PTPlot} package \cite{SGWB_analysis_I}.}
    \label{fig:spectrum}
\end{figure}

To investigate the impact of kinematic anisotropies, we consider specific instances of the spectrum $\Omega_\T{GW}$, focusing on three prominent early-Universe cosmological sources contributing to the \gls{sgwb}. These sources are of particular interest due to their expected amplitudes and frequency ranges, which fall within the sensitivity band of LISA \cite{Scird}, as illustrated in Fig.~\ref{fig:spectrum}. The primary spectrum under examination exhibits an approximately scale-free form and is modeled after the energy density spectrum associated with \gls{cs}, which is predicted to appear nearly flat within the frequency range relevant to LISA \cite{PhysRevLett.98.111101, CHANG2020100604, PhysRevD.98.063509}\footnote{For a given model $i$, the relevant frequency regime is defined as the region where the energy density $\Omega_\T{GW}^i$ exceeds the LISA sensitivity threshold, as shown in Fig.~\ref{fig:spectrum}. See also Section~\ref{sec:SGWB_Cosmic_defects}.}. 
Despite its conceptual simplicity, the study of \gls{cs}-induced features in the \gls{sgwb} remains compelling, as cosmic strings are regarded as promising probes of physics beyond the Standard Model and are of significant relevance in string-theoretic contexts \cite{CS_I, CS_II}\footnote{It is worth noting that scale-free spectra are not exclusive to \gls{cs}. See \cite{SGWB_1} for alternative scale-free sources.}. \\
The second spectrum considered originates from a first-order \gls{pt} in the early Universe, typically characterized by a broken power-law behavior \cite{caprini_detecting_2020} (see also Section~\ref{sec:SGWB_Phase_Transitions}) and illustrated in Fig.~\ref{fig:spectrum}\footnote{The precise details of the spectral shape—such as the peak frequency $f_{\text{peak}}$ or the slope—are not essential for the present analysis. What matters is the emergence of nontrivial values for $n_\Omega$ and $\alpha_\Omega$.}. Phase transitions provide valuable phenomenological insights, particularly in identifying scales associated with symmetry breaking in the early Universe. \\
Finally, we consider the contribution to the \gls{sgwb} from primordial black holes (PBHs), as discussed in \cite{Kohri_2018,PBH} (see also Section~\ref{sec:SGWB_beyond_irreducible}). It is important to acknowledge that there is currently no consensus in the literature regarding the exact frequency profile of PBH-induced \gls{sgwb} spectra, although the infrared behavior is relatively well understood \cite{Pi_2020} and references therein. Independent of their exact spectral shape, PBHs remain phenomenologically rich, particularly in models where PBHs constitute a partial or complete component of Dark Matter—a hypothesis that has attracted considerable recent interest \cite{Carr_2016, Carr_2020}.\\
In the main part of this Section, the focus is placed on the simplest featureless model—the CS-like signal—to study kinematic anisotropies via a realistic full time-domain simulation. The impact of more complex, feature-rich spectra such as those from PTs and PBHs on the presented analysis pipeline is addressed toward the end of this Section.

\subsubsection{Anisotropic GW stochastic sky simulation and instrument response}

Section~\ref{sec:SGWB_detection_Paper_HI} aims to introduce a comprehensive, time-domain simulation framework designed to produce realistic LISA data, integrated with a dedicated analysis pipeline optimized for detecting $l=0,1,2$ (kinematic) anisotropies in a \gls{sgwb} signal.
To provide context, the following outlines the interaction between the detector response and the \gls{sgwb} signal. \\
The \gls{sgwb} can be represented as a random, direction-dependent strain time series $h_{P}(t, \mathbf{\hat{k}})$ for each polarization state $P$. Consistent with the methodology presented in \cite{lisa_cosmology_working_group_maximum_2020}, the simulation assumes Gaussian statistical properties for the signal\footnote{A wide range of anticipated signals are known to deviate from Gaussianity \cite{bartolo_characterizing_2020}. Nonetheless, the detection strategies proposed here remain applicable with only minor modifications in such cases (see, for example, \cite{Drasco_2003, Buscicchio_2023}, and the discussion in \cite{lisa_cosmology_working_group_maximum_2020}).}. This time series is fully characterized by the second-order moments of its Fourier components, denoted as $\langle h_{P}(f, \mathbf{\hat{k}}) h^*_{P'}(f, \mathbf{\hat{k}}) \rangle$, which define the cross-power spectra $S_{PP'}(f,\mathbf{\hat{k}})$ corresponding to the stochastic process $h_{P}(t, \mathbf{\hat{k}})$.
Under the assumption of statistical homogeneity, $\langle {h}_{P}(f, \mathbf{\hat{k}}) {h}^*_{P'}(f, \mathbf{\hat{k}}) \rangle$ corresponds to 
\begin{align}
S_{PP'}(f,\mathbf{\hat{k}}) & =
\begin{bmatrix}
    \langle {h}_{+}(f, \mathbf{\hat{k}}) {h}^*_{+}(f, \mathbf{\hat{k}}) \rangle &  \langle {h}_{+}(f, \mathbf{\hat{k}}) {h}^*_{\times}(f, \mathbf{\hat{k}}) \rangle \\
    \langle {h}_{\times}(f, \mathbf{\hat{k}}) {h}^*_{+}(f, \mathbf{\hat{k}}) \rangle & \langle {h}_{\times}(f, \mathbf{\hat{k}}) {h}^*_{\times}(f, \mathbf{\hat{k}}) \rangle
\end{bmatrix}\notag
\\
& = \frac{1}{2} \delta^2 (\mathbf{\hat{k}}-\mathbf{\hat{k}}')\delta(f,f')
\begin{bmatrix}
    I + Q &&  U + i V  \\
    U - i V &&  I - Q  \\
\end{bmatrix},\notag
\end{align}
where one can introduce the Stokes parameters $I, Q, U$ and $V$, well-known in CMB physics and encoding intensity, linear polarization and circular polarization respectively. As in \cite{lisa_cosmology_working_group_maximum_2020}, this analysis is restricted to the intensity $I(f, \mathbf{\hat{k}})$ with 
\begin{equation}\label{equ:I_def}
    I(f, {\mathbf{\hat{k}}}) = \langle {h}_{+}(f, \mathbf{\hat{k}}) {h}^*_{+}(f, \mathbf{\hat{k}}) \rangle + \langle {h}_{\times}(f, \mathbf{\hat{k}}) {h}^*_{\times}(f, \mathbf{\hat{k}}) \rangle
\end{equation}
which can be related to the normalized logarithmic energy density $\Omega_\T{GW}(f,\mathbf{\hat{k}})$ as \cite{allen_detection_1997}
\begin{align}\label{equ:energydensity}
    \Omega_\T{GW}(f,\mathbf{\hat{k}}) = \frac{32\pi^3 f^3}{3H_0^2}I(f,\mathbf{\hat{k}}),
\end{align}
where $H_0$ is the Hubble constant.
Note here that the direction dependence of the latter two equations is commonly dropped by the assumption that
\begin{align}\label{equ:energydensity_II}
    \Omega_\T{GW}(f, \mathbf{\hat{k}}) = \Omega_\T{GW}(f) \mathcal{E}(\mathbf{\hat{k}}).
\end{align}
The first factor on the right-hand side of Eq. \eqref{equ:energydensity_II} remains applicable, while the second factor encapsulates the angular distribution of the background. It requires the choice of a normalization, here selected to be 
\begin{align}
    \int_{\mathcal{S}^2}\dd^2 \mathbf{\hat{k}}\, \mathcal{E}(\mathbf{\hat{k}})=1.
\end{align}
With the stochastic strain ${h}_P$ and the associated intensity $I(f, \mathbf{\hat{k}})$ at hand, one can now characterize the incoming signal as a time-frequency series. Following Eq. (12) in \cite{lisa_cosmology_working_group_maximum_2020}, the signal component of the time stream $s^\tau_I$ measured by a single \gls{tdi} channel $C$ is defined as a Fourier expansion between $t$ and $t+\Delta t$, such that it reads
\begin{align}\label{equ:defresponce}
s^\tau_C(f)=\sum_P\int_{S^2}\dd\mathbf{\hat{k}}\,R^{\tau,P}_C(f,\mathbf{\hat{k}}) h_P(f,\mathbf{\hat{k}}).
\end{align}
The superscript $\tau$ indicates that $s^\tau_C(f)$ denotes a potentially time-dependent frequency series. In this context, $R^P_C$ refers to the LISA response function, which varies based on the chosen channel $C \in [X, Y, Z]$ and the polarization state $P \in [+, \times]$ of the strain. The components $R^P_C$ and the underlying concept of channels will be elaborated upon in the following Subsection. Based on the definitions above, it holds that 
\begin{align}
     \langle{h_P(f,\mathbf{\hat{k}}),h^*_{P'}(f',\mathbf{\hat{k}}')\rangle}
    =\frac{1}{2}\frac{1}{4\pi}\delta_{f,f'}\delta^2(\mathbf{\hat{k}},\mathbf{\hat{k}}')\delta_{P,P'}S_{PP'}(f,\mathbf{\hat{k}}).
\end{align}
Under the Gaussian assumption, the power spectrum $S_{P}(f,\mathbf{\hat{k}})$ becomes the primary measurable quantity. The factor of $4\pi$ in the denominator arises from an integral over the unit sphere. While intensity is a quantity defined per pixel, the power spectrum density is an integral over the entire sky; hence, they differ by a factor of $4\pi$. One can further assume that the \gls{sgwb} is not polarized, so that $S_+(f,\mathbf{\hat{k}})=S_\times(f,\mathbf{\hat{k}})=\frac{1}{2}S_\T{GW}(f,\mathbf{\hat{k}})$ where the latter can be conveniently characterized by $\Omega_\T{GW}(f,\mathbf{\hat{k}})$ via \cite{SGWB_1}
\begin{align}\label{equ:sim_two}
    S_\T{GW}(f,\mathbf{\hat{k}})=\frac{3H_0^2}{4\pi^2f^3}\Omega_\T{GW}(f,\mathbf{\hat{k}}).
\end{align}
Note at this point the similarities between Eqs. \eqref{equ:energydensity} and \eqref{equ:sim_two}. 

\begin{figure}[t!]
    \centering
    \includegraphics[width=0.5\columnwidth]{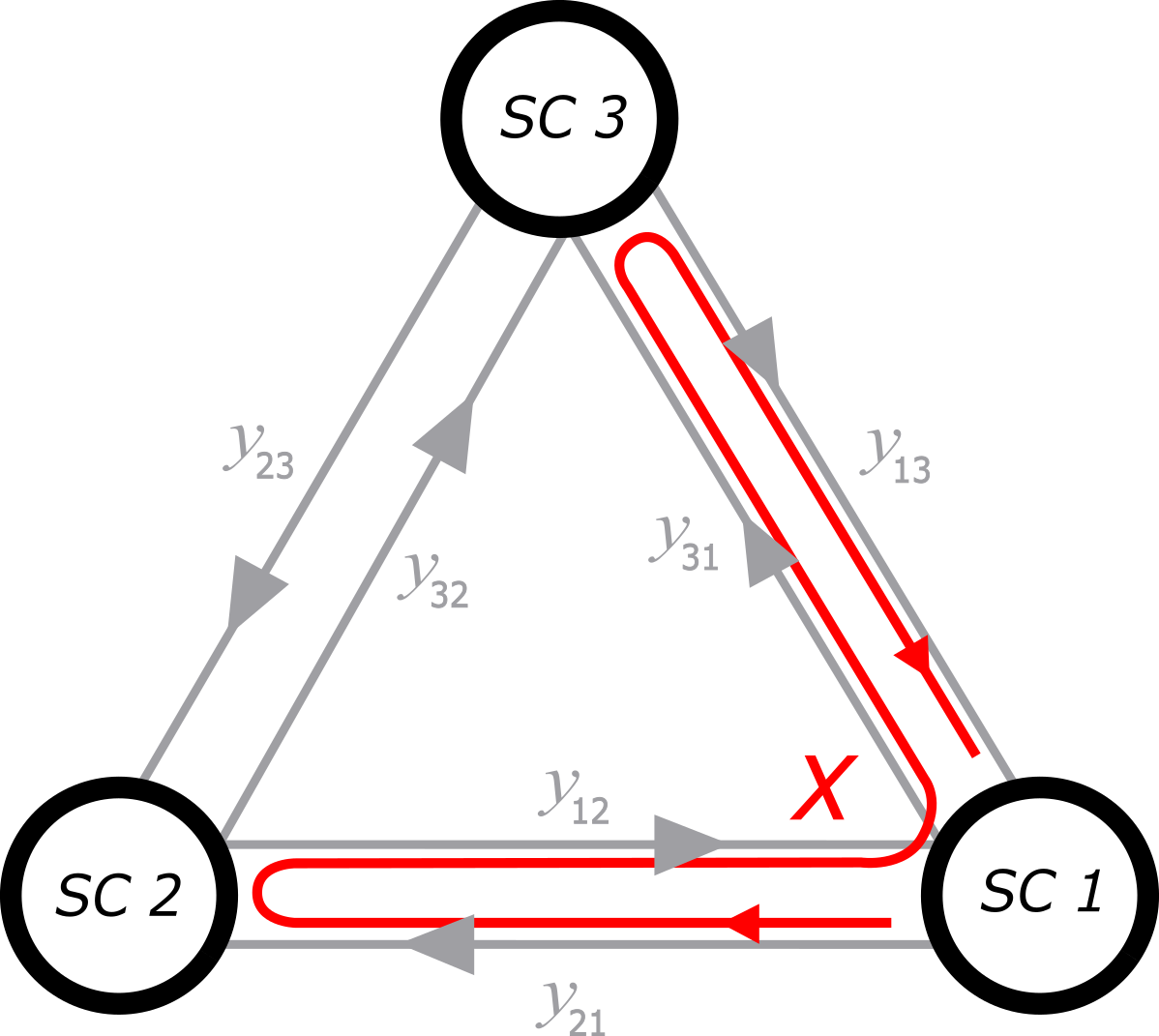}
    \caption{Illustration of three LISA spacecrafts in triangular formation, connected via six links (gray). Displayed in red is the 1.5 TDI $X$ channel as a linear combination of links \cite{SGWB_analysis_I}. }
    \label{fig:LISA_TDI}
\end{figure}

With the spectrum $\Omega_\T{GW}$ specified, the proposed analysis requires the definition of the response function associated with the chosen instrument. In the case of LISA, deriving this response function demands careful consideration of the instrument's links and the resulting \gls{tdi} channels. The LISA constellation consists of three spacecraft connected by six distinct optical links, as depicted in Fig.~\ref{fig:LISA_TDI}, each of which is perturbed in a time-dependent fashion by incident gravitational radiation.
Due to the linearity of the response function, the overall response of link $ij\in\{12,21,13,31,23,32\}$ is given by the sum of individual responses to a source allocated in pixel $p$,
\begin{align}\label{equ:indiv_resp}
    \mathbf{\tilde{y}}^{\tau}(f) = y^\tau_{ij}(f) = \sum_p y_{ij,p}^\tau(f).
\end{align}
As before, time dependence is indicated by the superscript $\tau$. Expressing the total link response in terms of the response per pixel is essential for numerical modeling and is further justified by the finite resolution of the instrument. At each time step $\Delta t$, the quantity $y_{ij,p}$ represents the frequency shift imparted to the laser beam along the link $ij$ by the gravitational strain originating from pixel $p$.
To obtain an explicit expression for $y_{ij,p}$, it is thus imperative to project the strain from point source $p$ onto the unit vector pointing along the link $ij$. Under the approximation of immobile spacecrafts during the light propagation along a single link, one finds \cite{baghi_uncovering_2023}
\begin{align}
    y_{ij,p}(t) \approx \frac{1}{2\left(1-\mathbf{\hat k}_p\cdot \mathbf{\hat n}_{ij}(t)\right)}\left[H_{ij,p}\left(t-\frac{L_{ij}(t)}{c}-\frac{\mathbf{\hat k}_p\cdot \mathbf{n}_j(t)}{c}\right)-H_{ij,p}\left(t-\frac{\mathbf{\hat k}_p\cdot \mathbf{n}_i(t)}{c}\right)\right].
\end{align}
In this context, $L_{ij}(t)$ denotes the time-dependent distance between the two spacecraft, defined by their positions $\mathbf{x}_i = \mathbf{x}_j + L_{ij} \mathbf{\hat{n}}_{ij}$, where $\mathbf{\hat{n}}_{ij}$ is the unit vector along the link. Assuming $c=1$, the quantity $L_{ij}(t)$ corresponds to the signal delay time along the link $ij$ measured at the reception time $t$. The vector $\mathbf{\hat{k}}_p$ denotes the wave vector of the gravitational waves originating from pixel $p$, and can equivalently be interpreted as the sky direction associated with pixel $p$ under the plane wave approximation.
The projection function $H_{ij,p}$ is given by
\begin{align}
    H_{ij,p}(t)  = h_+(t,\mathbf{\hat{k}}_p)\ \xi_+(\mathbf{\hat{u}}_p,\mathbf{\hat{v}}_p,\mathbf{\hat n}_{ij}) 
     + h_\times(t,\mathbf{\hat{k}}_p)\ \xi_\times(\mathbf{\hat{u}}_p,\mathbf{\hat{v}}_p,\mathbf{\hat n}_{ij}),
\end{align}
where the functions $\xi_{+}$ and $\xi_{\times}$ are the \textit{antenna pattern functions} and $\mathbf{\hat{u}}_p,\mathbf{\hat{v}}_p$ the polarization vectors associated to the propagation vector $\mathbf{\hat k}_p$. For details, see appendix A in \cite{baghi_uncovering_2023} and references therein.

To compute TDI observables, LISA combines the six link-signals resulting in three correlated channels commonly referred to as $X,Y,Z$, illustrated in Fig. \ref{fig:LISA_TDI}. The explicit linear combination of links leading to the designated channels can be encapsulated in one matrix, $\mathbf M_{TDI}$ \cite{baghi_uncovering_2023}. For instance, the second generation \gls{tdi} $X_2$ channel as it is sketched in Fig. \ref{fig:LISA_TDI} can be constructed using links $y_{12},y_{21},y_{13},y_{31}$, 
\begin{align} X_2 =&\, X_1 + \mathbf{D}_{1321}y_{12}+\mathbf{D}_{131212}y_{21}+\mathbf{D}_{1312121}y_{13}+\mathbf{D}_{13121213}y_{31}\notag\\&- (\mathbf{D}_{12131}y_{13}+\mathbf{D}_{121313}y_{31}+\mathbf{D}_{1213131}y_{12}+\mathbf{D}_{12131312}y_{21}),\label{equ:X2}
\end{align}
with
\begin{align}
    X_1 = \,y_{13} + \mathbf{D}_{13}y_{31} + \mathbf{D}_{131}y_{12}+ \mathbf{D}_{1312}y_{21}- ( y_{12} + \mathbf{D}_{12}y_{21}+\mathbf{D}_{121}y_{13}+\mathbf{D}_{1213}y_{31}).\label{equ:X1}
\end{align}
The delay operator for a single link is defined as 
\begin{align}
    \mathbf{D}_{ij}x(t)= x(t-L_{ij}(t)),
\end{align}
and can be accumulated forming the chained delay operator, $\mathbf{D}_{i_1,i_2,..}$, which, applied on a function $f(t)$, induces a chained delay in the reception, i.e.
\begin{align}
    \mathbf{D}_{i_1,i_2,..}f(t) = f\left(t- \sum_{l=1}^{n-1} L_{i_l,i_l+1}(t)\right).
\label{eq: total_delay}
\end{align}
Analogous construction can be done for the $Y,Z$ channels by permuting $ij$ in Eq. \eqref{equ:X1} and \eqref{equ:X2} correspondingly.

In the frequency domain, the delay operators appearing in Eq.~\eqref{eq: total_delay} reduce to simple phase operators. Utilizing the analytical formulation provided in Eq.~(B.5) of \cite{baghi_uncovering_2023}, the single-link frequency-domain signal arising from a given unit sky-direction $\mathbf{\hat{k}}$ is expressed in Eq.~\eqref{eq: kernel} as the projection of the two polarization modes onto the $2 \times 6$ (two polarizations, six links) frequency-domain response kernel $G_{ij,P}^\tau(f, \mathbf{\hat{k}})$.
\begin{equation}
\mathbf{\tilde{y}}^{\tau}_{p}(f) = \sum_{P=(+,\times)} G^\tau_{ij,P}(f, \mathbf{\hat k}_p)\ h_{P}^{\tau}(f,\mathbf{\hat k}_p)
\label{eq: kernel}
\end{equation}
The analysis adopts the approximation that the transfer functions $G^\tau_{ij,P}(f, \mathbf{\hat{k}}_p)$ remain stationary over each time segment labeled by $\tau$, which requires selecting time windows significantly shorter than the LISA orbital timescale ($\ll 1\,\text{year}$). Under this assumption, the delay operators introduced in Eq.~\eqref{eq: total_delay} are simplified in the frequency domain as complex phasing operators, as in \cite{baghi_uncovering_2023}:
\begin{align}
    \mathbf{\tilde{D}}_{ij} \mathbf{\tilde{y}}^{\tau}_{p}(f)\ \approx \ \mathbf{\tilde{y}}^{\tau}_{p}(f) \exp\left(-2 \pi i f L_{ij}^\tau\right)
\label{eq: approx tdi}
\end{align}
These are then combined according to Eq. \eqref{equ:X2} and assembled to build a $3 \times 6$ single-link to \gls{tdi} channel operator matrix $\mathbf M^\tau_{TDI}(f)$ \cite{baghi_uncovering_2023}. The time dependence ($\tau$ upper script) accounts for the the annual orbital breathing of LISA constellation. The measured data within a discrete time and frequency interval, $\mathbf{\tilde{d}}^{\tau}(f) \equiv (\tilde X,\tilde Y,\tilde Z)^T$, can be expressed as a vector of \gls{tdi} channels:
\begin{align}\label{equ:data_vecI}
    \mathbf{\tilde{d}}^{\tau}(f,\mathbf{\hat k}) = \mathbf M_{TDI}^\tau (f)\ \mathbf{\tilde{y}}^{\tau}(f,\mathbf{\hat k})
\end{align}
Here, $\mathbf{\tilde{y}} = (\tilde{y}_{12},\tilde{y}_{23},\tilde{y}_{31},\tilde{y}_{13},\tilde{y}_{32},\tilde{y}_{21})^T$. Note also that the time dependence was made explicit for each tensor. \\
Using definition \eqref{equ:defresponce} and the explicit formula for the individual link response, one can construct a three-vector (see appendix B in \cite{baghi_uncovering_2023})
\begin{align}
    \mathbf{R}_{P}^\tau(f,\mathbf{\hat k}) = \mathbf M^\tau_{TDI}(f) \ G^\tau_{ij,P}( f, \mathbf{\hat k})\ \mathbf M^\tau_{TDI}(f)^{\dag},
\end{align}
that contains the individual response components for each 1.5 TDI channel $X,Y,Z$. The subscript indicates the polarization dependence induced by the strain appearing in \eqref{equ:defresponce}\footnote{Compare also to \cite{lisa_cosmology_working_group_maximum_2020}.}. The linear response TDI vectors $\mathbf{R}^P=(R_X,R_Y,R_Z)^P$ can be in turn merged into the quadratic response $\mathbf{A}^\tau$ ($N_f \times N_\tau \times 3 \times 3 \times N_{\text{pix}}$):
\begin{align}
\mathbf{A}^\tau(f,\mathbf{\hat k})=\mathbf{R}^+\otimes \mathbf{R}^{+*}+\mathbf{R}^\times \otimes \mathbf{R}^{\times*}, 
\label{eq: quadratic response}
\end{align}
Thus, the covariance matrix for a measured intensity $\Tilde{I}$ and noise matrix $\mathbf{N}$ reads
\begin{align}
    \langle\mathbf{S}^\tau(f)\rangle = \mathbf{C}^\tau_f \approx \sum_p \mathbf{A}_p\Tilde{I}^p + \mathbf{N},
\label{eq: covariance_raw}
\end{align}
defined based on the quadratic strain tensor
\begin{align}
    \langle\mathbf{S}^\tau(f)\rangle = \int_{S^2}\dd\mathbf{\hat k}\, \mathbf{s}^\tau(f) \otimes \left(\mathbf{s}^\tau(f)\right)^*  = \int_{S^2} \dd\mathbf{\hat k}\,\mathbf{A}^\tau(f, \mathbf{\hat k}) I(f,\mathbf{\hat k}).
    \label{equ: quadratic strain}
\end{align}
In this context, a vector with components defined in Eq.~\eqref{equ:defresponce} is used to construct a matrix in channel space. Specifically, Eq.~\eqref{equ: quadratic strain}, and consequently the covariance matrix in Eq.~\eqref{eq: covariance_raw}, represent frequency-dependent time series of $3 \times 3$ matrices, with each entry corresponding to a specific channel. Analogously, the TDI data vector measured by LISA comprises three components, each associated with one of the three channels discussed above. Eq. \eqref{equ:data_vecI} can be decomposed into response, signal and noise via 
\begin{align}
    \mathbf{\tilde{d}} = \mathbf{R}h + \mathbf{n},
\label{eq: data - linear}
\end{align}
where the TDI noise $\mathbf{n}$ is assumed to be Gaussian with zero mean and a covariance $\mathbf{N}_{d} = \mathbf{n}\otimes \mathbf{n}$. The data components are modeled as Gaussian-distributed, with their covariance specified by Eq.~\eqref{eq: covariance_raw}. It is important to note that the data vector $\mathbf{\tilde{d}}$ remains dependent on both time and frequency. The individual link response described in Eq.~\eqref{equ:indiv_resp} exhibits directional sensitivity through its dependence on sky pixels, and as a result, the response matrix $\mathbf{R}^P$ inherits this directional dependence. Although analytic formulations for $\mathbf{R}^P$ exist in the low-frequency regime of LISA’s sensitivity band—see, for example, \cite{lisa_cosmology_working_group_maximum_2020}—the present analysis favors the use of numerical tools to more accurately capture the complexities of a realistic detection scenario. Further implementation details are provided in the following subsection.\\ 
In Eq.~\eqref{equ: quadratic strain}, the product $\mathbf{A}\Tilde{I}$ involves an integration over the entire sky. However, under realistic conditions with finite angular resolution, this integral is discretized into a sum over sky pixels, i.e., $\sum_p \mathbf{A}_p\Tilde{I}^p$. This pixelization converts angular dependence into pixel-based dependence, rendering the response matrix a three-dimensional object of shape $3 \times 3 \times N_\text{pix}$, which varies with both frequency and time. Here, $N_\text{pix}$ denotes the total number of pixels used in the discretization. Importantly, since the covariance matrix in Eq.~\eqref{eq: covariance_raw} results from an integration over all directions, it becomes independent of specific sky directions.

For the purposes of this Section, summation over sky pixels is adopted as the preferred representation of full-sky integration, reflecting the inherently numerical nature of the analyses conducted. However, from an analytical standpoint, it is advantageous to employ a continuous basis of spherical harmonics to decompose the covariance matrix into its constituent angular modes. This yields an analogous decomposition of the intensity field $I$ and, consequently, the power spectrum $\Omega_\T{GW}$, in a form reminiscent of Eq.~\eqref{equ:1111}. It is important to emphasize, however, that unlike Eq.~\eqref{equ:1111}, the mode decomposition introduced here does not rely on any assumptions regarding the (kinematic) origin of the signal. As such, this framework maintains a higher degree of model independence.\\
Although both $\mathbf{A}$ and $I$ are functions of sky direction, Eq.~\eqref{equ: quadratic strain} benefits from the use of a Mollweide projection, which establishes a bijective mapping $\mathbf{A}(\hat{\mathbf{k}}) \rightarrow \mathbf{A}^p$ and vice versa. Since any continuous map on the sphere admits a decomposition in spherical harmonics, the Mollweide projection allows for the translation between pixel-space and mode-space representations. This flexibility ensures that the covariance matrix can be expressed in terms of angular modes at any stage of the analysis.
In practice the mapping can be achieved as follows: Writing the sum over pixels explicitly, one can replace the latter by the approximation
\begin{align}\label{equ:pix_vs_smooth}
    \sum_p \mathbf{A}_p\Tilde{I}^p \approx \frac{1}{\Delta_{\text{pixel}}} \int_{S^2} \dd \mathbf{\hat k}\,\mathbf{A}(\mathbf{\hat k})\Tilde{I}(\mathbf{\hat k}).
\end{align}
Here, $\Delta_{\text{pixel}}$ corresponds to the area per pixel \footnote{Note that for most commonly used mappings of the Riemann sphere $\mathcal S^2$ this area per pixel measure is not constant, however, for numerical applications there exist suitable python packages taking care of this transformation.}. Frequency and time dependence of both response matrix and intensity are omitted in Eq. \eqref{equ:pix_vs_smooth}. Naturally, the approximation improves with the number of pixels. One can further simplified the latter equation by rewriting $\mathbf{A}$ and $I$ as
\begin{align}\label{equ:mode_decomp}
    \mathbf{A}(\mathbf{\hat k})=\sum_{\ell, m}a_{\ell m}Y_{\ell m}(\mathbf{\hat k})&&\text{and}&&I(\mathbf{\hat k})=\sum_{\ell,m}i_{\ell m}Y_{\ell m}(\mathbf{\hat k})
\end{align}
where
\begin{align}
     \langle i_{\ell m},i^{*}_{\ell'm'} \rangle =: C_\ell^{GW}\delta_{ll'}\delta_{mm'}, &&
     \langle a_{\ell m},a^{*}_{\ell'm'} \rangle =: \mathbf{A}_\ell\delta_{\ell \ell'}\delta_{mm'},
\end{align}
with
\begin{align}
    C_\ell^{GW}= \frac{1}{2\ell+1}\sum_{m=-\ell}^\ell\left|\int_{S^2}\frac{\dd\mathbf{\hat k}}{4\pi}Y_{\ell m}(\mathbf{\hat k})I(\mathbf{\hat k})\right|^2,
\end{align}
\begin{align}
\mathbf{A}_\ell= \frac{1}{2\ell+1}\sum_{m=-\ell}^\ell\left|\int_{S^2}\frac{\dd\mathbf{\hat k}}{4\pi}Y_{\ell m}(\mathbf{\hat k})\mathbf{A}(\mathbf{\hat k})\right|^2,
\end{align}
so that 
\begin{align}
    \mathbf{A}_pI^p\rightarrow \int_{S^2}\dd\mathbf{\hat k}\sum_{\ell',m'}\sum_{\ell,m}a_{\ell m}i_{\ell'm'}Y_{\ell' m'}(\mathbf{\hat k})Y_{\ell m}(\mathbf{\hat k}).
\end{align}
Note, however, that a pixel-area normalization factor is omitted in the above expression; mathematically, equality holds only in the limit of a large number of pixels, where the discrete sum approaches the continuous integral. Nevertheless, the right-hand side can be simplified by utilizing the identities $Y_{\ell,-m} = (-1)^m Y^*_{\ell m}(\mathbf{\hat{k}})$ and $i_{\ell,-m} = (-1)^m i^*_{\ell m}$\footnote{This identity applies specifically to the mode components of $I(\mathbf{\hat{n}})$, as the intensity is a real-valued function, in contrast to the potentially complex-valued response.}, together with the orthogonality relation of the spherical harmonics.
Then one finds
\begin{align}
&\int_{S^2}\dd\mathbf{\hat k}\sum_{\ell',m'}\sum_{\ell,m}a_{\ell m}i_{\ell'm'}Y_{\ell' m'}(\mathbf{\hat k})Y_{\ell m}(\mathbf{\hat k})
=\int_{S^2}\dd\mathbf{\hat k}\sum_{\ell,m}\sum_{\ell',m'}a_{\ell m}i^{*}_{\ell',-m'}Y_{\ell m}(\mathbf{\hat k})Y^*_{\ell',-m'}(\mathbf{\hat k})\notag\\
    =&\int_{S^2}\dd\mathbf{\hat k}\sum_{l,m}\sum_{\ell' m'}a_{\ell m}i^{*}_{\ell'm'}Y_{\ell m}(\mathbf{\hat k})Y^*_{\ell',m'}(\mathbf{\hat k})=\sum_{\ell,m}a_{\ell m}i^{*}_{\ell m},
\label{bla^3}
\end{align}
and thus
\begin{align}\label{equ:1.234}
    \mathbf{A}_pI^p \approx \frac{1}{\Delta_{\text{pixel}}}\sum_{\ell<3}\sum_{m=-\ell}^{\ell}a_{\ell m}i^{*}_{\ell m}.
\end{align}
The summation is performed only over $\ell = 0, 1, 2$, with higher modes being truncated, under the assumption that the signal intensity $I$ is only significant up to $\ell = 2$, i.e., up to the quadrupole contributions, as outlined in \eqref{equ:1111}. It is important to note that this is equivalent to setting $i_{\ell m} = 0$ for $\ell > 2$.

\subsubsection{Data Generation}

\begin{figure}
\centering
\frame{\includegraphics[width=\linewidth, trim={0.0cm 0.8cm 0.0cm 2.5cm}, clip]{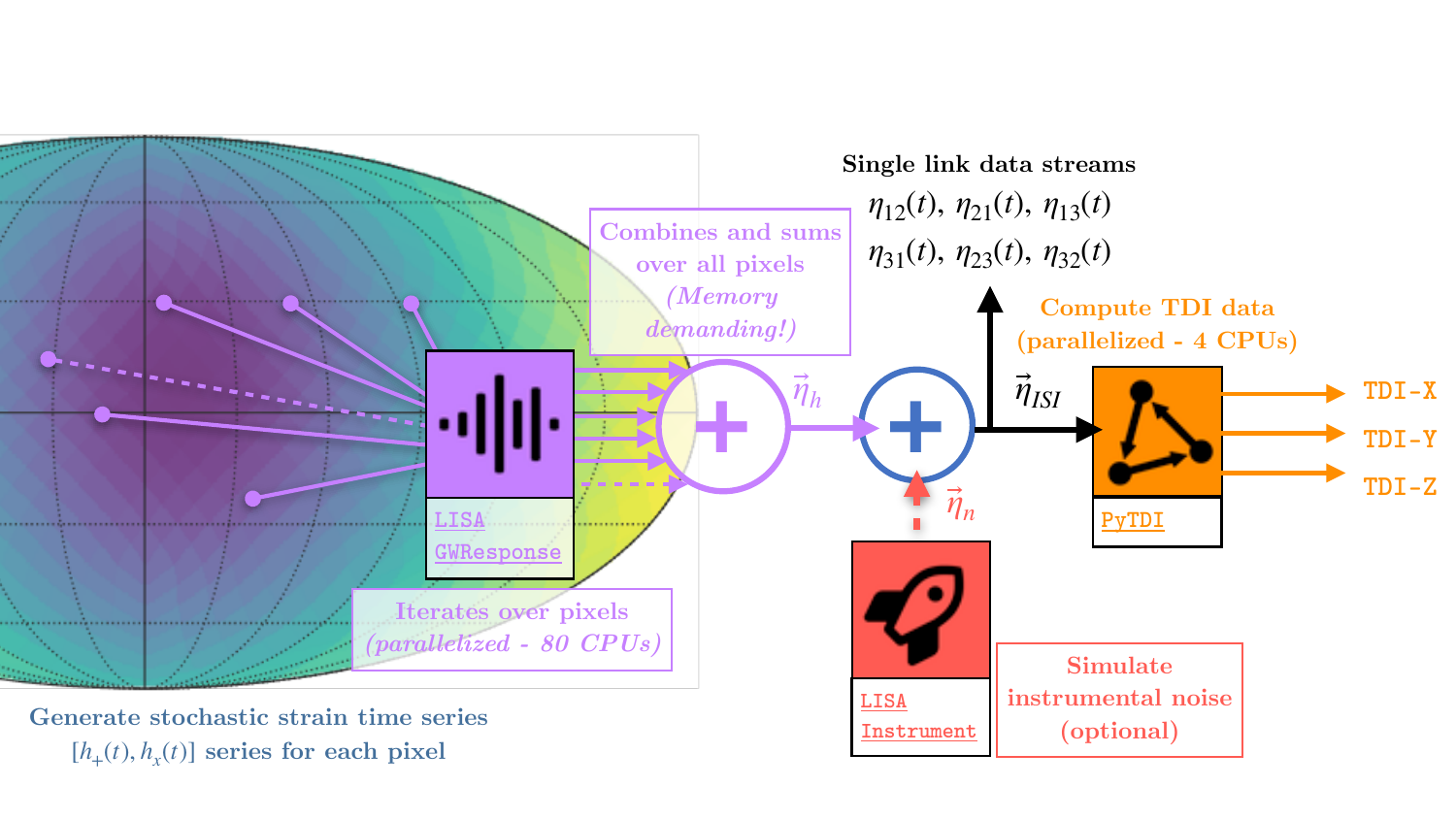}}
\caption{E2E simulation flow of LISA response to an anisotropic, stochastic GW sky \cite{SGWB_analysis_I}.}
\label{fig: sim flow}
\end{figure}
The proposed approach entails simulating the LISA detector's response to a Doppler-boosted, anisotropic \gls{sgwb} sky. A synthetic dataset spanning four years is generated as described by Eq.~\eqref{eq: data - linear}, with the covariance characterized by Eq.~\eqref{eq: covariance_raw}. For the data generation, attention is restricted to a simplified scenario in which the \gls{sgwb} exhibits a flat energy density spectrum, representative of possible CS signals within LISA’s sensitivity band \cite{SGWB_1}. For details, see Section \ref{sec:SGWB_cosmo}.\\
Sky discretization is performed via angular pixelation using the \href{https://healpy.readthedocs.io/en/latest/}{\texttt{healpy}} package, which is employed extensively throughout the analysis for map generation and transformations between pixel and spherical harmonic domains. Each of the $N_{\text{pix}} = 12 \times (N_{\text{side}})^2$ sky pixels is modeled as an independent stochastic strain time series with power spectral density $S_h$, as defined in Eq.~\eqref{equ:sim_two}. At this stage, instrumental noise and astrophysical confusion foregrounds are omitted to isolate LISA’s sensitivity to purely stochastic signals. The central computational challenge lies in the randomness of the strain time series assigned to each pixel direction (see Fig.~\ref{fig: sim flow}).\\
The individual LISA single-link response to each pixel is computed using the cutting-edge \href{https://pypi.org/project/lisagwresponse/}{\texttt{LISAGWResponse}} software \cite{bayle_lisa_2023}, which is part of the LISA Consortium simulation suite and implements a time-domain projection of the $h_+$ and $h_\times$ polarizations onto the detector response with minimal approximations\footnote{Assumptions include static spacecraft over the photon travel time and a first-order expansion in GW propagation time.}. The simulation adopts a simplified orbital model featuring an equilateral, equal-arm configuration \cite{bayle_lisa_2022}\footnote{For analyses involving non-equilateral, unequal-noise LISA configurations, see \cite{SGWB_analysis_IV}.}. Due to the linear nature of the response, the net detector response to the full anisotropic \gls{gw} sky is obtained by summing over the $N_{\text{pix}}$ individual pixel contributions. This operation is parallelized across up to $>80$ CPUs to optimize the balance between computational speed and memory consumption. For an $N_{\text{side}} = 32$ resolution, generating a four-year dataset sampled at $0.2$~\si{\hertz} requires up to 128 CPUs and approximately $3$~\si{\tera\byte} of RAM, with a total runtime of about 15 hours on a dedicated computing node.\\
The selected sampling frequency represents a trade-off between computational tract-ability and optimal signal-to-noise ratio (\gls{snr}) retention. Higher sampling rates rapidly increase memory demands, particularly when modeling hundreds of pixels\footnote{These requirements are memory-bound; for example, the runs described here employed 80 CPUs and roughly $1.5$~\si{\tera\byte} of memory per four-year simulation.}. Various rates up to $0.4$~\si{\hertz} were evaluated, with $f_s = 0.2$~\si{\hertz} emerging as the most practical compromise. At higher frequencies, instrumental noise becomes dominant (see Fig.~\ref{fig:spectrum}), leading to diminished \gls{snr}. Accordingly, increasing the sampling frequency beyond $0.2$~\si{\hertz} yields only marginal improvements in the precision of dipole measurements. Although this limitation arises from simulation constraints, the analysis framework remains compatible with full-resolution $f_s = 4$~\si{\hertz} LISA data, implying that the results presented herein may be conservative.\\
Special care is taken to ensure statistical independence across pixel-wise stochastic sources during parallel computation. Each subprocess uses a distinct \emph{local seed} to initialize its random number generator, while a shared \emph{global seed} governs the overall sky configuration to guarantee independence across repeated simulation runs.

The simulated single-link data are subsequently processed using the LISA Consortium’s \texttt{PyTDI} software \cite{staab_pytdi_2023} to generate the second-generation \gls{tdi} time-domain Michelson-like interferometric data streams $X_{2}(t)$, $Y_{2}(t)$, and $Z_{2}(t)$. Instrumental noise can optionally be added at the single-link level using the \href{https://pypi.org/project/lisainstrument/}{\texttt{LISAInstrument}} software from the LISA simulation suite. For simplicity and computational efficiency, only secondary noise sources are enabled, as the \texttt{PyTDI} pipeline is specifically designed to suppress primary noise contributions. The complete simulation workflow is outlined in Fig.~\ref{fig: sim flow}.

It is important to emphasize that the choice of clock frame with respect to which phase measurements are recorded and time-stamped plays a critical role when beat note data streams are combined to construct the \gls{tdi} time-series. Although forming the $X$, $Y$, and $Z$ channels using local spacecraft reference times is physically consistent—since each can be interpreted as measurements made from a single spacecraft situated at the vertex of a Michelson-like interferometer \cite{hartwig_time-delay_2022}—in-consistencies arise when these time-series are mathematically compared without synchronization to a common time reference. A consistent reference frame is required for coherent analysis. For the associated transformation, see \cite{pireaux_time_2007}. Failure to account for relativistic timing corrections introduces significant biases in the resulting sky maps, particularly affecting the dipole component. Aligning the $X$, $Y$, and $Z$ data streams to a common time coordinate, specifically the \gls{tcb} referenced to the \gls{bcrs}, effectively mitigates this systematic bias and enables the accurate recovery of the kinematic dipole, as discussed in the following Subsection.
\begin{figure}[t]
    \centering
    \includegraphics[width=0.8\columnwidth, trim={0.61cm 0.7cm 0.0cm 0.0cm}, clip]{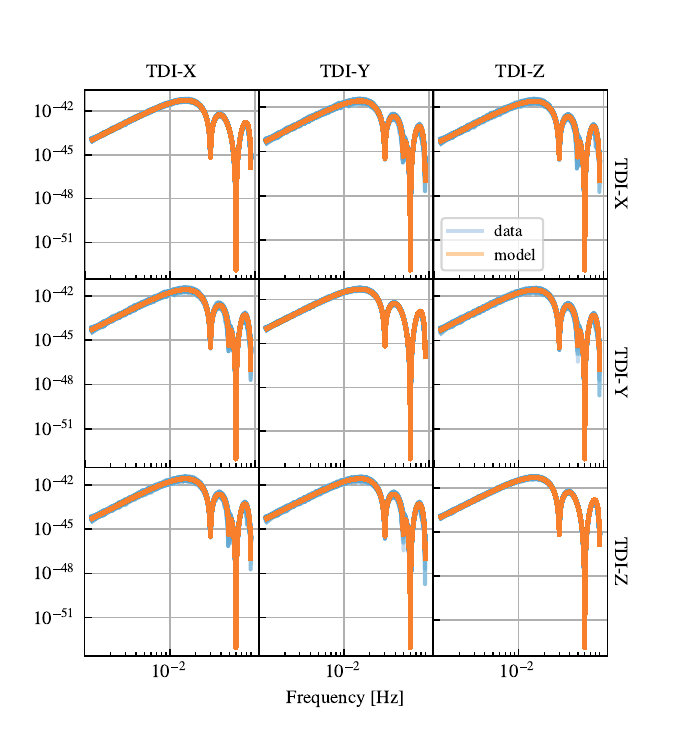}
    \caption{Averaged spectra of the $4$-years-long data set compared to the space interferometer response model to an anisotropic GW sky \cite{SGWB_analysis_I}. There are $N_t=384$ spectra over plotted for each in the figure, in shades of blue for the data, and in shades of red for the model.}
    \label{fig: data_vs_model_spectrum}
\end{figure}

\subsubsection{Data Analysis}

The analysis utilizes 4 years of \gls{tdi} 2.0 data streams sampled at $0.2\,\si{\hertz}$. As a first step, the data undergo a pre-processing procedure that includes both frequency and time compression. The continuous 4-year TDI data stream is segmented into $N_t$ discrete time intervals, which effectively sets the angular resolution for the subsequent analysis. Additional data compression is performed by averaging the spectral content over frequency bins of width $n_j$.
Based on the $X,Y,Z$ TDI channel data vector $\mathbf{d}$, one can determine the averaged data matrix as in \cite{baghi_uncovering_2023}
\begin{align}\label{equ:data_matrix}
    \mathbf{\bar{D}}(\tau_{i}, f_{j}) \equiv \frac{1}{n_{j}} \sum_{k=j-\frac{n_j}{2}}^{j+\frac{n_j}{2}} \mathbf{\tilde{d}}(\tau_{i}, f_k) \otimes \mathbf{\tilde{d}}(\tau_{i}, f_k)^{\dag}.
\end{align}
Here, the $N_t \times N_f \times 3$ data vectors $\mathbf{\tilde{d}}$ are the Fourier transforms of the time-split, simulated $X$, $Y$ and $Z$ time series. The matrix $\mathbf{\bar{D}}(\tau_{i}, f_{j})$ is the tensor product across \gls{tdi} channels of the data vectors $\mathbf{\tilde{d}}$, before averaging over the spectral window of width $n_j$ for data compression. It measures the cross-spectral density of the \gls{tdi} data streams. The statistical expectation of $\mathbf{\bar{D}}$ is the theoretical covariance $\mathbf C_d:= \mathbf C_{[X,Y,Z]}$ of the \gls{tdi} time series, which the Bayesian map-making method utilized in the analysis is ultimately solving for. The theoretical covariance $\mathbf C_d$ matrix can be computed, for each time and frequency bin, as
\begin{align}
    \mathbf{C}_{d}(\tau_{i},f_j) = \mathbf{A}_{d}(\tau_{i},f_j,p)\ I(f,p)\ + \mathbf{N}_{d}(\tau_{i},f_j),
\label{eq: covariance}
\end{align}
where $\mathbf{A}$ is the $N_t \times N_f \times 3 \times 3 \times N_{\text{pix}}$ matrix encoding the quadratic response function of the instrument \cite{lisa_cosmology_working_group_maximum_2020}, $I(f, p)$ is the \gls{gw} intensity sky map $N_{\text{pix}}$-vector, both showing explicit frequency and angular (pixel) dependence. As in previous expressions, the product $\mathbf{A}_d I$ is understood as a summation over sky pixels. The quantity $\mathbf{N}_{d}$ denotes the $N_t \times N_f \times 3 \times 3$ covariance matrix of \gls{tdi} noise, which may be optionally included in the analysis. The pixelized intensity map $I(p)$ contains the free parameters that constitute the primary target of the inference procedure.

To obtain a best-fit estimate of the intensity map relative to the observed data, a suitable probabilistic framework must be selected, guided by the structure and statistical properties of the data vector $\mathbf{\tilde{d}}$. In this analysis, a likelihood-based approach is adopted, utilizing the statistical behavior of the matrices $\mathbf{\bar{D}}(\tau_{i}, f_{j})$. As introduced in Eq.~\eqref{equ:data_matrix}, for each time segment $\tau$, the matrix $\mathbf{\bar{D}}$ is a $(3 \times 3)$ random matrix formed by averaging the outer products of $n_j$ Gaussian-distributed frequency series within the frequency window $\left[ j - \frac{n_j}{2}, j + \frac{n_j}{2} \right]$. Provided that the frequencies $f_k$ used to compute the average $\mathbf{\tilde{d}}(\tau_{i}, f_k) \otimes \mathbf{\tilde{d}}(\tau_{i}, f_k)^{\dag}$ are uncorrelated, the resulting matrix $\mathbf{\tilde{D}}(\tau_i, f_j)$ follows a multivariate Wishart distribution $W_{q=3}(n_j, \mathbf{C})$ \cite{SINHARAY201098}, where $\mathbf{C}$ denotes the covariance matrix of the underlying Gaussian process $\mathbf{\tilde{d}}$.
For $n_j \geq 3$ the probability density function\footnote{The density function is valid w.r.t. Lebesque measure on the cone of symmetric positive definite matrices.} of $\mathbf{\bar{D}}$ reads \cite{SINHARAY201098}
\begin{align}\label{equ: wishart_density}
    f(\mathbf{\bar{D}}) = \frac{1}{2^{n_j q/2} \Gamma_q(\frac{n_j}{2})|\mathbf{C}|^{n_j/2}}|\mathbf{\bar{D}}|^{(n_j-q-1)/2}\exp{(-\frac{1}{2}\text{Tr}(\mathbf{C}^{-1}\mathbf{\bar{D}}))}.
\end{align}
Here, $\Gamma_q(\alpha)$ is the multivariate gamma function, and the dimension parameter $q = 3$ accounts for the three \gls{tdi} channels\footnote{Regarding the notation for matrices, taking the absolute, $|\cdot|$, here accounts for computing the determinant.}. Based on the preceding equation, the likelihood function $\mathcal{L}$ to be maximized can now be formulated. It is important to note that any constant prefactors become irrelevant in the computation of the logarithmic likelihood ratio used in the MCMC sampling. Consequently, only the terms dependent on the fitting parameters contribute to the inference process.
For the purposes of this investigation, the log-likelihood function corresponding to a data sample $\mathbf{\bar{D}}$ is defined as in Eq. \eqref{equ:data_matrix}, i.e.
\begin{align}
    \log \mathcal{L} = \sum_{\tau_{i}}{\sum_{f_j}{ \bigg[ -\operatorname {Tr} (\mathbf{C}_{d}^{-1}\mathbf {\bar D}(\tau_{i},f_j) ) - \nu \log |\mathbf{C}_{d}(\tau_{i},f_j)| \bigg] }}\,,
\label{eq: logL}
\end{align}
where the trace is taken over the \gls{tdi} channels. The effective number of degrees of freedom is introduced as $\nu = \frac{n_j}{N_{bw}}$, where the reduction factor $N_{bw}$—referred to as the normalized equivalent noise bandwidth—accounts for the overlap of time segments $\tau_i$ and the specific window functions employed during the time segmentation of the data. For additional details, the reader is referred to the comprehensive review in \cite{heinzel_spectrum_2002}.

Evidently, the fitting parameters enter explicitly through the covariance matrix $\mathbf{C}_d$, and specifically via the intensity map embedded within it. A Bayesian map-making approach requires numerical evaluation of the covariance matrix $\mathbf{C}_d$ at each step of the MCMC sampling process, utilizing the expression on the left-hand side of Eq.~\eqref{equ:pix_vs_smooth}. Although the evaluation of the $\log \mathcal{L}$ function necessitates the full intensity pixel map $I_p$, various strategies can be adopted to extract the relevant information from $I_p$ using a reduced set of fitting parameters. The present analysis considers two such approaches.\\
The first method assumes that the anisotropies in the signal originate from a Doppler boost. For a scale-invariant energy density spectrum as described in Eq.~\eqref{equ:1111}, a suitable reference frame can be selected without loss of generality. This leaves four parameters that define the pixel map $I_p$: the monopole amplitude in the observer frame, $\Omega_\T{GW}'(f)$, and the three components of the velocity vector $\vec{\beta} = [\beta_x, \beta_y, \beta_z]$, as introduced in Eq.~\eqref{equ:1111}.
These components represent the boost of the solar system relative to the \gls{cmb} frame, expressed in the \gls{bcrs}. For a solar motion relative to the \gls{cmb} with a speed of approximately $|\vec{\beta}| \approx 369\,\text{km/s}$ \cite{planck_initial_doppler}, or $|\vec{\beta}| \approx 1.23 \cdot 10^{-3}$ when normalized to the speed of light, the corresponding components in the \gls{bcrs} are $\vec{\beta} = 10^{-3} \cdot [-1.23, 0.25, -0.18]$. These values are taken as the \textit{true} boost parameters targeted for recovery through the MCMC inference procedure.
By combining Eqs.~\eqref{equ:0} and \eqref{equ:energydensity}, the full intensity map $I_p$ is constructed, thereby defining $\mathbf{C}_d$. This approach offers the advantage of a minimal parameter space, which significantly alleviates computational demands. However, it is inherently model-dependent, applying exclusively to scenarios where Doppler-induced anisotropies from a scale-invariant source dominate the cosmological \gls{sgwb}.\\
Another method for extracting the intensity map from a given model involves its decomposition into spherical harmonic modes. In this approach, the pixel map $I_p$ is expressed as a sum of 6 independent modes, considering only up to quadrupolar anisotropies (i.e., $\ell = 2$). The $\ell, \pm m$ modes are interrelated, as $I_p$ must remain real for each pixel. These 6 modes correspond to 9 independent parameters, since 3 of the relevant modes are complex. This decomposition is model-independent and, in principle, captures any anisotropy present in the data, including intrinsic anisotropies within the source frame.\\
It is important to note that both parametrizations described above are valid only for a scale-invariant spectrum. For more complex spectra, such as those with non-trivial spectral dependencies, knowledge of the relevant fit parameters is required for each frequency bin individually.

In this study of the two parameter spaces, the optimal fit for the intensity map, denoted as $I_p$, is sought using simulated \gls{lisa} data through MCMC sampling of the likelihood function outlined in Eq. \eqref{eq: logL}. 
While exploring the parameter space via MCMC sampling incurs significant computational costs, this method provides a key advantage by avoiding the numerically challenging inversion of the Fisher matrix. The latter can introduce systematic errors that undermine the overall robustness of the analysis \cite{lisa_cosmology_working_group_maximum_2020, SGWB_analysis_II}. Moreover, the Fisher matrix approach assumes that the likelihood function can be well approximated by a Gaussian distribution around its peak \cite{bond_estimating_1998}, a condition that does not necessarily hold in this analysis. If this assumption is violated, the algorithm may converge to an undesirable local maximum of the likelihood function. While this limitation may not be as critical in signal forecasting, it becomes problematic when developing tools for real measurement data. For these reasons, the Fisher matrix approach is less suited to our needs.
In contrast, the MCMC method does not rely on any analytical or numerical approximations of the log-likelihood function, as defined in Eq. \eqref{eq: logL}. Given that our focus is primarily on low modes when studying kinematic anisotropies, an MCMC map-making strategy, based on the methodology presented in \cite{lisa_cosmology_working_group_maximum_2020}, is especially relevant. Furthermore, the inherently low angular resolution of \gls{lisa}, which limits the resolution of higher modes, makes the MCMC map-making approach a practical and effective solution for many scenarios encountered in \gls{lisa} analyses.

The mapping scheme is structured as follows: at each iteration of the algorithm, the likelihood in Eq. \eqref{equ:0} is evaluated based on the current location in the parameter space. In principle, the covariance matrix can be computed by inserting Eq. \eqref{equ:1111} into Eq. \eqref{eq: logL}; however, numerically, there is no need to expand in small velocities. Instead, the full expression in Eq. \eqref{equ:0} is used to compute both the covariance matrix $\mathbf{C}_d$ and the corresponding likelihood. As discussed previously, for a flat and isotropic energy density spectrum $\Omega_\T{GW}(f)$ in the \gls{cmb}-frame, the Doppler shift resulting from the observer's velocity relative to the stochastic emission causes a sky modulation of the amplitude that is independent of the frequency, i.e., $\alpha_\Omega = n_\Omega = 0$.
This implies that changing the velocity vector $\vec{\beta}$ will have an overall effect on the sky map at all frequencies, and the treatment of frequency and angular dependence remains separate,
\begin{align}\label{equ: sim cov}
    \mathbf{C}(\tau,f)  = \mathbf{A}(\tau,f,p) \ I(f,p) 
                = \mathbf{A}(\tau,f,p) \ \frac{E(f)}{E(f_0)} \ I(f_{0}, p)
\end{align}
where the spectral dependency of the energy density in the \gls{cmb}-frame is extracted via the function $E(f)$ as in \cite{lisa_cosmology_working_group_maximum_2020}, and the (arbitrary) reference frequency of $f_0 = 1\ \si{\milli\hertz}$ is introduced to define the intensity map $I(p) = I(f_{0}, p)$ that is fitted for. Thus, for each iteration $i$, the evaluation of Eq. \eqref{equ:0} boils down to computing the prefactor \eqref{equ:-1} for the new velocity vector $\vec \beta^i$ when applying the model-dependent fit ansatz. Given the fit parameters $\{\Omega^i_\T{GW},\beta^i_x, \beta^i_y, \beta^i_z\}$ of iteration $i$ selected by an arbitrary walker of the MCMC the algorithm proceeds by calculating 
\begin{align}
     I^i(f_0,p) &= \frac{3 H_0^2}{32 \pi^3 f_0^3} \mathcal{D}^{4}(p) \ \Omega_\T{GW}^i\left( f_0 \right) \label{equ: HI and the BW ghost}
\end{align}
with
\begin{align}
     \mathcal{D}(p) &= \frac{\sqrt{1-|\vec\beta^i|^2}}{1 - |\vec\beta^i| \  \left[ \mathbf{\hat k}\cdot\hat{\beta}^i \right]_{p}}. 
\label{equ: sim omega}
\end{align}
The angular or pixel dependence of $I(f_0,p)$ is introduced implicitly through the Doppler boosting of $\Omega_\T{GW}^i$ from the source to the observer frame via $\mathcal{D}$. It is important to note that, for a slope-free spectrum such as for \gls{cs}, $\Omega_\T{GW}(f) = \Omega_\T{GW}$ remains constant. The resulting discrete $I^i(f_0, p)$ is then summed over pixels, as shown in Eq. \eqref{equ: sim cov}, yielding the covariance matrix $\mathbf{C}^i$ for the $i$-th iteration. With this newly computed covariance matrix, the likelihood function \eqref{eq: logL} is evaluated, and the results are compared to those of the other walkers within the same iteration. According to the scheme's definition, the walkers tend to favor regions in parameter space around high-likelihood points. Over time, they converge toward the maximum of \eqref{eq: logL} for the given data. \\
The MCMC can, in principle, be initialized in any arbitrary state. However, incorporating prior knowledge about the expected signal is likely to speed up the convergence toward the correct maximum likelihood. In this analysis, an agnostic approach is chosen, initializing the MCMC with an isotropic map configuration $\Tilde{I}_p = \text{constant}$. For a detailed description of the specific MCMC scheme used in this work, the reader is referred to \cite{foreman-mackey_emcee_2013, goodman_ensemble_2010}.
\begin{figure}[t]
    \centering
    \includegraphics[width=0.8\textwidth]{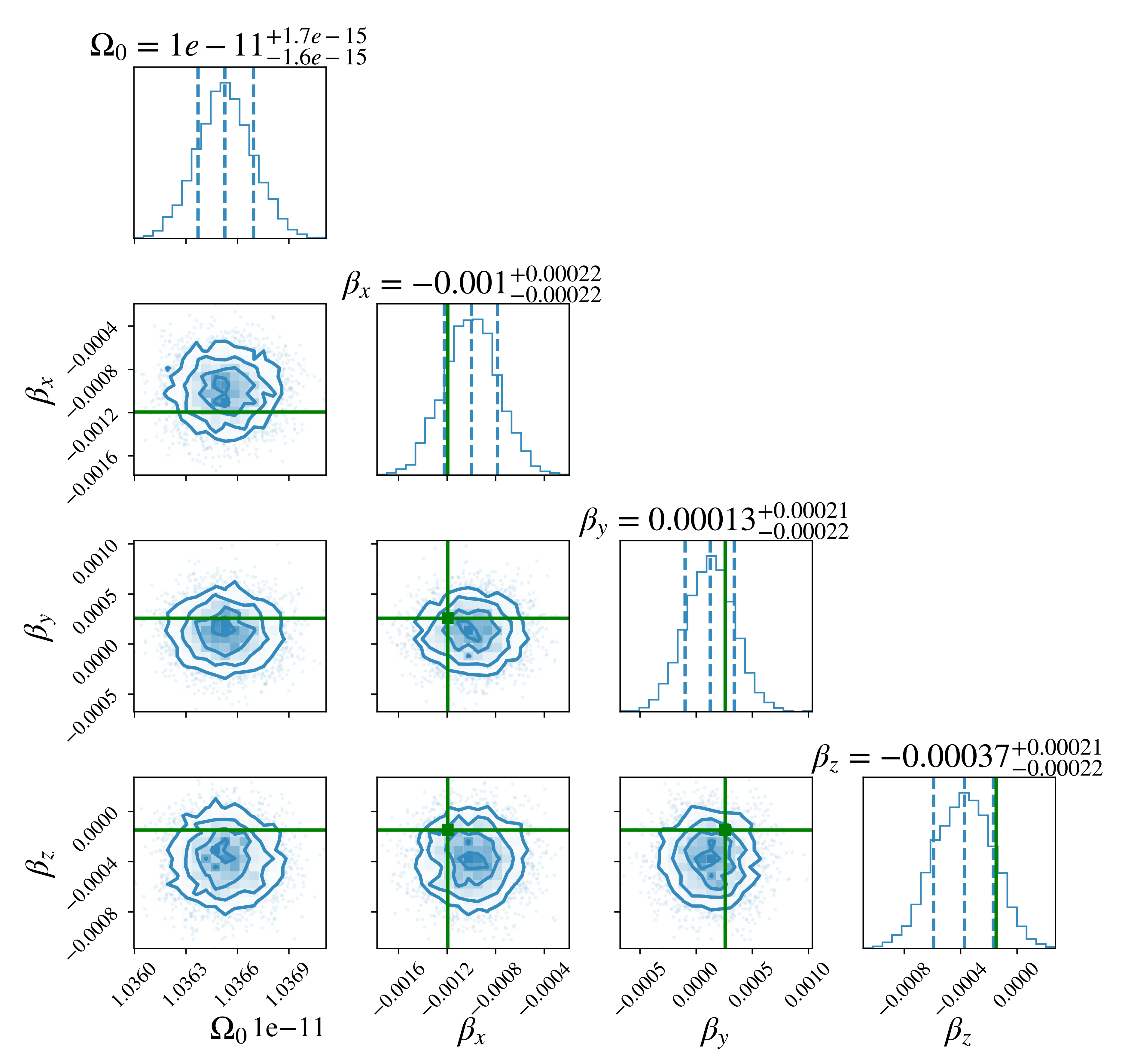}
    \caption{Samples histogram of the $(\Omega_\T{GW},\Vec{\beta})$ space for a single $4$-years-long GW sky realization after convergence of the MCMC \cite{SGWB_analysis_I}. The {\it true} boost components in the BCRS frame  targeted by the MCMC inference are drawn in green plain lines and given by $\vec{\beta} = 10^{-3}\cdot[-1.23,$ $0.25,-0.18]$.}
    \label{fig: cornerplot}
\end{figure}
\begin{figure}[t]
    \centering
    \includegraphics[width=\textwidth, trim={0.0cm 40.0cm 0.0cm 0.0cm}, clip]{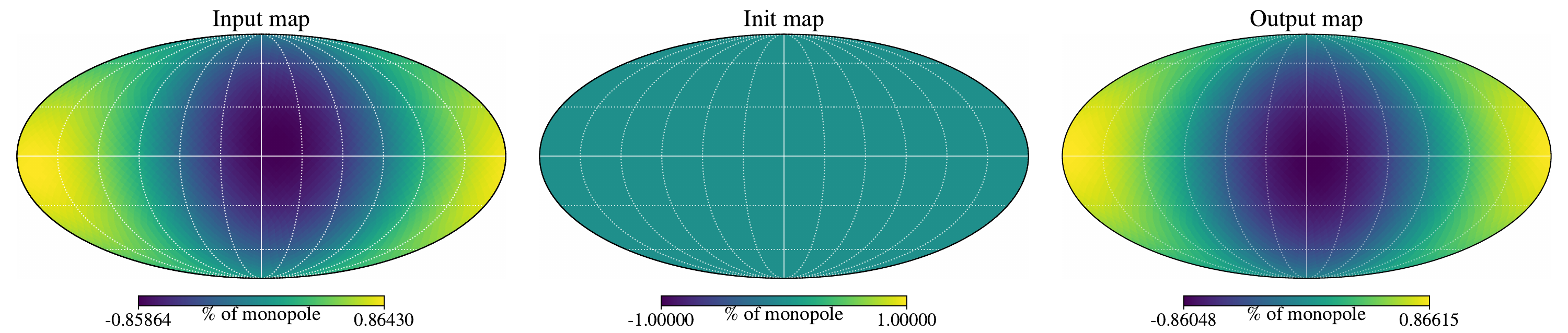}
    \caption{Intensity GW sky maps (in percentage of the monopole $\Omega_0$) \cite{SGWB_analysis_I}. From left to right: the input intensity map injected, the a priori map used to initialize the MCMC chains, and the map recovered by the map-making algorithm. The monopole is removed from the intensity maps, to ease the visualization the dipolar anisotropy. These projection plots have been made from \texttt{healpy} python package.}
    \label{fig: skymaps}
\end{figure}
In Fig. \ref{fig: cornerplot}, an exemplary histogram and correlation diagram of the MCMC chains after convergence are presented, focusing on a single realization of the \gls{sgwb} sky. The MCMC chains show convergence towards the true velocity $\vec{\beta}^{\text{true}}$. The statistical properties of the MCMC samples, which represent the posterior distribution, provide a theoretical estimate of the measurement precision. For this particular sky realization, $\vec{\beta}^{\text{true}}$ is found to be more than $4$-$\sigma$ away from zero, indicating a clear detection of the observer's motion relative to the \gls{cmb} rest frame. Fig. \ref{fig: skymaps} displays the reconstructed skymap derived from the $\vec{\beta}^{\text{meas}}$ measurement shown in Fig. \ref{fig: cornerplot}. A high degree of agreement between the input and output intensity maps is observed, especially considering that the initial sky map, prior to measurement, was perfectly isotropic.

\subsubsection{Numerical Results}

Before presenting the results of the numerical investigation outlined above, it is important to highlight a statistically significant bias in the amplitude monopole $\Omega_0$, as shown in Fig. \ref{fig: cornerplot}. The inferred value is $\Omega_0 = 1.0365 \pm 0.0002 \times 10^{-11}$, revealing a systematic error that is two orders of magnitude larger than the statistical error. This systematic bias has been traced to numerical inaccuracies in the fitting quadratic frequency-domain response model given by Eq. \eqref{eq: quadratic response}, which becomes pronounced near zero-response frequencies (see Fig. \ref{fig: data_vs_model_spectrum}) and close to the Nyquist frequency $f_s$. At zero-response frequencies, \gls{lisa} becomes insensitive to the \gls{gw} strain because the constellation arm lengths are integer multiples of the \gls{gw} wavelength, leading to no effective variation in optical path length. Numerical issues can arise near these response dips, and, as observed in this case, relative errors between the model and data can diverge as the signal approaches zero. The MCMC sampling attempts to compensate for these inaccuracies when adapting the inferred value of $\Omega_0$. It has been verified that masking the data around these dips and near the Nyquist frequency reduces the systematic error by an order of magnitude, without affecting the precision of the $\vec{\beta}$ inference. This phenomenon is particularly significant in signal-dominated regimes, where the analysis is most susceptible to model inaccuracies. \\
A second potential source of systematic error arises from the assumption that the statistics of the random matrices $\mathbf{\bar{D}}(\tau_{i}, f_{j})$ (Eq. \eqref{equ:data_matrix}) follow multivariate Wishart statistics, Eq. \eqref{equ: wishart_density}, which is only exact when averaging over uncorrelated frequency bins $k$ to construct $\mathbf{\bar{D}}(\tau_{i}, f_{j})$. In this work, however, the application of the Hanning apodization window function to compute the frequency series $\mathbf{\tilde{d}}(\tau_{i}, f_k)$ from the time-domain \gls{tdi} data streams induces correlations between adjacent frequency bins $k$. A more general statistical model, which accounts for these frequency correlations, is required to replace Eq. \eqref{equ: wishart_density} when deriving the log-likelihood function to be sampled. It is believed that this approximation contributes to the residual bias, and ongoing investigations, including those by the authors of \cite{baghi_uncovering_2023}, are exploring the derivation of the exact likelihood function. This topic merits further study, beyond the scope of the present document.

Nonetheless, the results shown in Fig. \ref{fig: cornerplot} and \ref{fig: skymaps} indicate that resolving kinematic anisotropies using the MCMC map-making strategy is indeed feasible. However, it is important to recognize that, at this stage, the analysis remains sensitive to potential statistical fluctuations in the input sky map realization, which could lead to a fortuitous amplification of the dipolar signal, deviating from its expected value. This intrinsic statistical error in the $\ell=1,2$ kinematic modes is referred to as ``sample variance''. Such variance will limit the precision of anisotropy characterization even in an idealized scenario with a signal-dominated instrument, due to the finite duration of the observations (4 years), as thoroughly discussed in \cite{Contaldi_with_variance}. 
Moreover, as noted earlier, it is essential to confirm that the inferred anisotropy and the measurement of the velocity $\vec{\beta}$ are robust with respect to any potential systematic errors in the analysis, including those affecting the monopole amplitude $\Omega_0$ inference. Our approach will be empirical and agnostic, focusing on ensuring that the successful inference of $\vec{\beta}^{\text{meas}}$ is independent of any specific data realization or injection map, and is statistically consistent with the properties of the stochastic signal.

\begin{figure}[h!]
    \centering
    \includegraphics[width=0.7\columnwidth, trim={0.55cm 0.3cm 0.05cm 0.0cm}, clip]{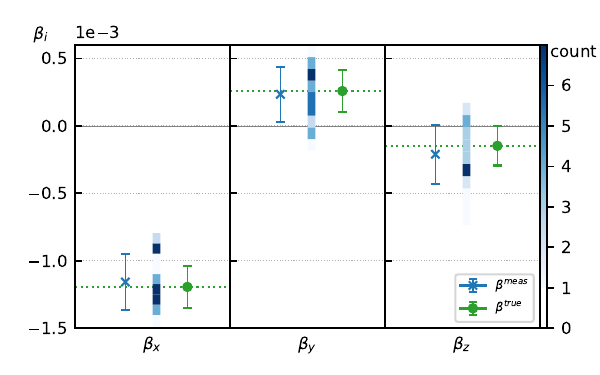}
    \caption{Variance of the measured signal over the $30$ realizations of the GW sky (blue histograms), compared to the resulting averaged values and standard deviations (in blue), and the theoretical values and MCMC error bars (in green) \cite{SGWB_analysis_I}. Measured signals, their statistics and the true values are plotted for $\beta_x$, $\beta_y$ and $\beta_z$, from left to right. The dispersion of $\vec{\beta}^{\text{meas}}$ realizations is statistically consistent with the true values $\vec{\beta}^{\text{true}}$ and the theoretical error bars from the MCMC sampling.  $\beta_x$ is resolved $> 4 \sigma$ away from $0.0$ value.}
    \label{fig: stats_beta}
\end{figure}

\begin{figure}[t]
    \centering
    \includegraphics[width=.7\columnwidth, trim={0.8cm 0.5cm 2.0cm 0.0cm}, clip]{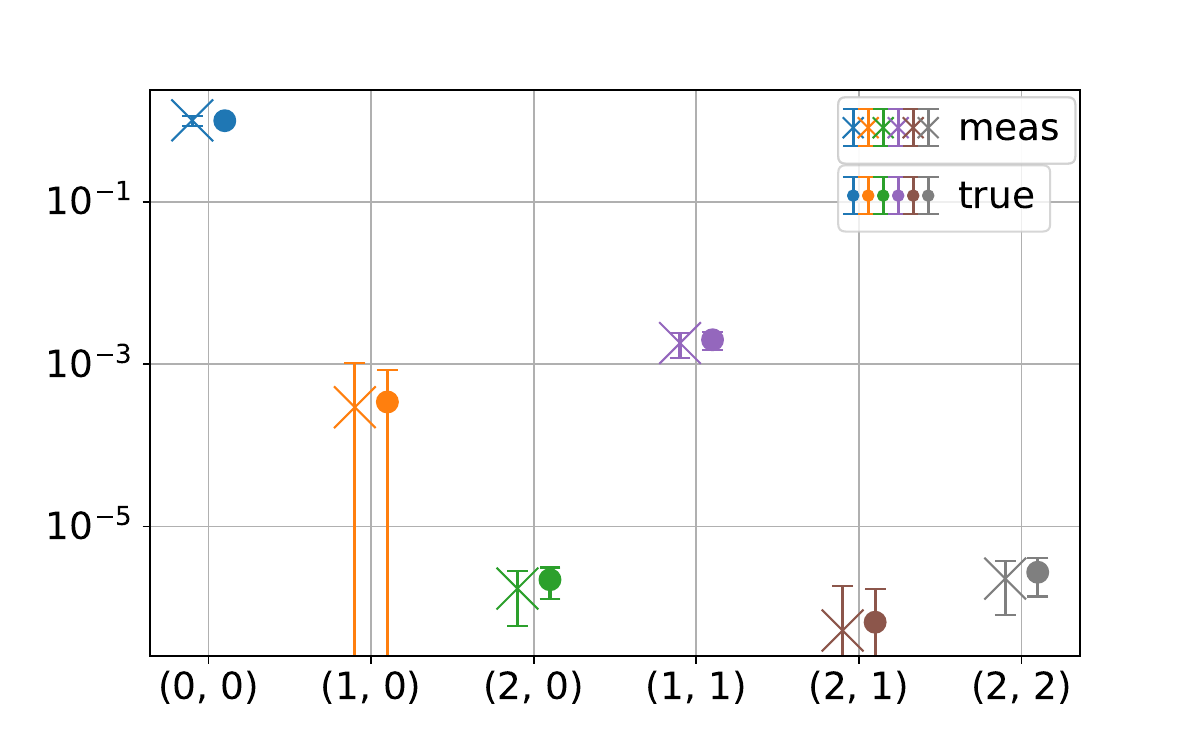}
    \caption{Variance of the measured signal against the theoretical error bars for the mode components of the SGWB in the observer frame up to $\ell=2$ based on the statistics for $\beta_i$, Fig. \ref{fig: stats_beta} \cite{SGWB_analysis_I}.}
    \label{fig: stats_alms}
\end{figure}

The potential artifacts discussed previously can be shown to be excluded as plausible sources contributing to the observed signals in Fig. \ref{fig: cornerplot} and \ref{fig: skymaps}. To demonstrate this, the statistical robustness of the signal shown in Fig. \ref{fig: cornerplot} and \ref{fig: skymaps} is evaluated by comparing 30 randomly generated realizations of a stochastic \gls{gw} sky. Figure \ref{fig: stats_beta} illustrates the dispersion of the measured velocity $\vec{\beta}^{\text{meas}}$ in the \gls{sgwb} rest frame relative to each distinct realization of the \gls{sgwb}, where the injected boost vectors remain fixed for each realization. The figure also presents a comparison between the mean value and standard deviation across the 30 realizations, alongside the true velocity values, determined by the choice of frame (see the discussion below Eq. \eqref{eq: logL}), as well as the theoretical uncertainty estimates from the MCMC sampling. The analysis in Fig. \ref{fig: stats_beta} reveals that the statistics of the 30 measurements are consistent with the theoretical values. Specifically, the average $\overline{\beta}^{\text{meas}}$ converges towards $\vec{\beta}^{\text{true}}$, and the measured standard deviation is in agreement with the theoretical error bars. Notably, the average measured $\overline{\beta}_x^{\text{meas}}$ is well resolved and found to be more than $4 \sigma$ away from $0$. Therefore, across all random realizations, the algorithm consistently recovers the \textit{true} dominant boost component responsible for inducing the kinematic anisotropy of the \gls{sgwb}, as determined earlier. The result shown in Fig. \ref{fig: cornerplot} corresponds to one of the 30 realizations chosen at random and is thus included in the histogram in Fig. \ref{fig: stats_beta}. This numerical demonstration confirms that, even in the challenging scenario of a boosted, scale-free \gls{cs}-like stochastic signal, the inherent sample variance is effectively mitigated.
\\
Based on the $30$ realizations, the corresponding intensity maps and their $a_{\ell m}$ components in the spherical harmonics decomposition are also computed. As shown in Fig. \ref{fig: stats_alms}, statistical analyses are conducted to assess the robustness of the resolution of kinematic spherical harmonics modes. A clear detection of the dipolar $(1,1)$ mode is observed, which represents the dominant component of the boosted \gls{sgwb} data with a flat energy density spectrum. As expected, the $\ell=1$ modes are significantly brighter than the $\ell=2$ modes, since their intensity scales as $\beta^{\ell}$. Notably, the MCMC method begins to exhibit sensitivity to quadrupolar components, such as the $(2,0)$ and $(2,2)$ modes, which are detected with approximately $2 \sigma$ confidence. This sensitivity stems from the fact that \gls{lisa} is much more responsive to quadrupolar signals than dipolar ones due to the pronounced parity of its response \cite{SGWB_analysis_II}. This enhanced response compensates for the inherent weakness of the quadrupolar component, making it observable by \gls{lisa}. In some instances, cosmological signals with more complex spectral signatures can further amplify the $\ell=2$ component, potentially making the $\ell=2$ modes comparable to or even dominant over the kinematic signatures in the observed signal. It is crucial to note that the specific distribution of power across the $(\ell, m)$ modes, as depicted in Fig. \ref{fig: stats_alms}, is influenced by the choice of reference frame for the boost vector. In the \gls{bcrs} frame, where the dominant component of the boost vector is aligned with the $x$-axis, the primary contribution to the dipole $D(f)$, as described in Eq. \eqref{equ:4} (and subsequent equations), is found in the $(1,1)$ mode. Likewise, in this reference frame, the quadrupole $Q(f)$ is predominantly determined by the $(2,2)$ and $(2,0)$ modes. Furthermore, the variance of the monopole arises from the variability in the measured $\beta_i$ and the dependence of $C_0^\text{GW}\sim |M(f)|^2$ on these components.

Having established the statistical significance of the result, the next step is to assess whether the observed signal could be attributed to biases or systematic errors in the instrument response or the analysis. To investigate this possibility, a Doppler-induced anisotropic sky map is introduced by injecting a rotated velocity vector $\vec{\beta}^{\text{rot}}$. The ability to recover this modified injection is then tested. The new velocity vector to be recovered is given by $\vec{\beta}^{\text{rot}} = R_{ZYX}(\phi, \eta, \theta)\ \vec{\beta}^{\text{true}}$, where the rotation matrix $R_{X,Y,Z}$ incorporates the Euler rotation angles $\phi = 100^\circ$, $\eta = 0^\circ$, and $\theta = 60^\circ$. These angles correspond to rotations around the $Z$, $Y$, and $X$ axes of the \gls{ssb} frame, respectively.
\begin{figure}[t]
    \centering
    \includegraphics[width=.7\columnwidth, trim={0.45cm 0.3cm 0.85cm 0.0cm}, clip]{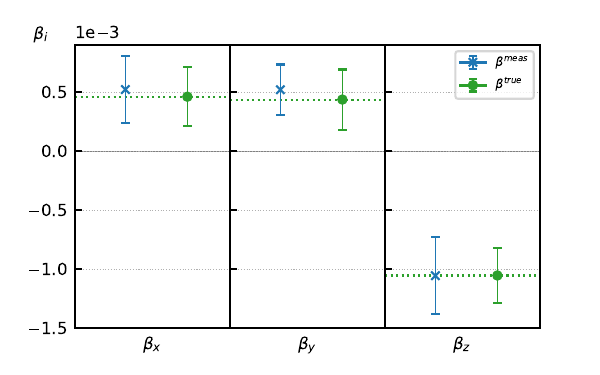}
    \caption{Variance of the measured signal over the $10$ realizations of the rotated GW sky (blue histograms), compared to the resulting averaged values and standard deviations (in blue), and  the theoretical values and MCMC error bars (in green) \cite{SGWB_analysis_I}. Are plotted the data for $\beta_x^{\text{rot}}$, $\beta_y^{\text{rot}}$ and $\beta_z^{\text{rot}}$, from left to right. We do verify that new, modified input velocity $\beta^{\text{rot}}$ is recovered with identical precision that in Fig \ref{fig: stats_beta}}
    \label{fig: stats_beta_rotated}
\end{figure}\\
Fig. \ref{fig: stats_beta_rotated} presents the injected values $\vec{\beta}^{\text{rot}}$ and the recovered values for $10$ realizations of the sky. As observed in Fig. \ref{fig: stats_beta}, the analysis effectively recovers this newly injected signal. This result demonstrates that the analysis pipeline is capable of accurately reconstructing the injected sky map, providing further evidence that the measured intensity map in Fig. \ref{fig: skymaps} is not due to a systematic error.

Thus far, the analysis has been centered on the model-dependent approach for selecting the fit parameters, specifically fitting with respect to $\{\Omega_\T{GW}, \vec \beta\}$. Now, the focus shifts to a fully agnostic approach, where the fit is performed using the spherical modes of the signal, parameterizing the intensity map $I_p$ via its mode decomposition $i_{\ell m}$, as given in \eqref{equ:mode_decomp}, up to $\ell \leq 2$. It is important to note that, assuming the signal is purely kinematic, i.e., the result of the Doppler boost of an isotropic background, the set $\{\Omega_\T{GW}, \vec \beta\}$ represents the minimum set of parameters needed to characterize the signal. However, in a more general context, a model-independent approach increases the flexibility of the outlined pipeline. Unlike the model-dependent approach, utilizing the modes $i_{\ell m}$ allows for the resolution of any angular dependence in the signal $I_p$, capturing both intrinsic and Doppler-induced sky anisotropies.

\begin{figure}
    \centering
    \includegraphics[width=.6\columnwidth, trim={0.8cm 0.5cm 2.0cm 0.0cm}, clip]{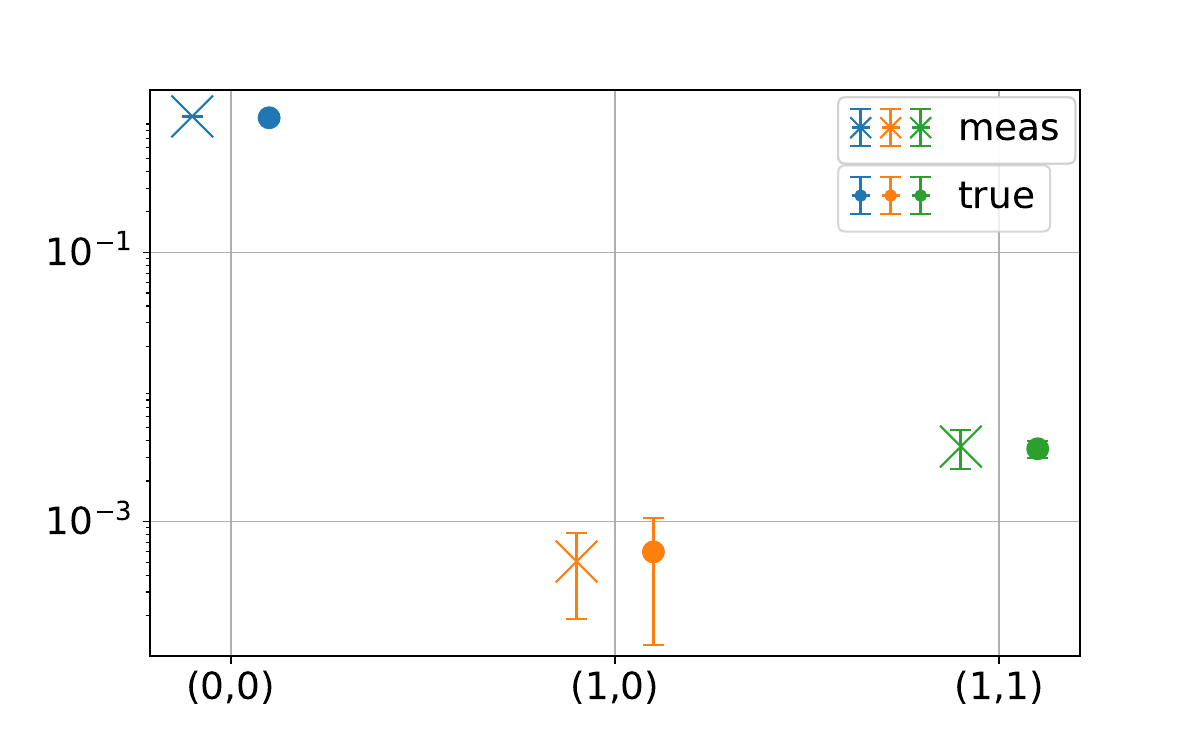}
    \caption{Variance of the measured signal against the theoretical error bars for the mode components of the SGWB in the observer frame up to $\ell=1$ based on a MCMC-sampling using $a_{\ell m}$ \cite{SGWB_analysis_I}.}
    \label{fig: stats_alms_alms}
\end{figure}

As in the previous analysis, one investigates the model-independent choice of fit parameters for $10$ random realizations of a \gls{cs}-like \gls{sgwb} signal. For each iteration $k$, the intensity $I^k_p$ is computed using the modes $i_{\ell m}^k$, which are selected by the MCMC based on the likelihood function \eqref{eq: logL}. The results are presented in Figure \ref{fig: stats_alms_alms}. For $\ell = 0, 1$, the model-independent approach achieves accuracy comparable to the parametrization $\{\Omega_\T{GW}, \vec \beta\}$. The measured mean values $\bar i_{\ell m}$ are statistically significant and exhibit variance in agreement with the theoretical error bars. When $\ell = 2$ is included, the MCMC converges to an undesirable maximum, significantly overestimating the monopolar and quadrupolar contributions. This issue arises partly due to the correlation between the monopole and quadrupole, which manifests in equal power being carried by both, as shown in Eq. \eqref{equ:M} and \eqref{equ:Q}, and confirmed numerically. Therefore, it can be concluded that, due to the correlation between $M$ and $Q$, the model-independent approach allows for statistical uncertainties of the monopole to leak into the quadrupole. This undesirable effect can be mitigated by fine-tuning analysis parameters such as frequency and time binning, or by disentangling the monopole and quadrupole through different parametrizations. One possible approach is to use a recursive scheme, where $\ell = 0, 1$ is first fixed via the MCMC before fitting for $\ell = 2$ exclusively. 
However, it should be noted that the correlations between the monopole and anisotropies are also influenced by \gls{lisa}'s complex and non-compact sky response functions \cite{Contaldi_with_variance} (see also \cite{mentasti_probing_2023, mentasti2023prospects}). Accounting for this factor introduces additional complexity to the analysis. At this stage, a more detailed investigation of these issues is deferred to future work. It is important to highlight that such complications do not arise in the model-dependent fit, which, due to the minimal set of parameters, eliminates redundancies and potential correlations between parameters. Consequently, the model-dependent approach results in a modified likelihood space that, in comparison to the model-independent approach, mitigates the peaks of the likelihood function that would otherwise favor the leakage between the monopole and quadrupole.

\begin{figure}[t]
    \centering
    \includegraphics[width=\columnwidth, trim={0.0cm 0.0cm 0.0cm 0.0cm}, clip]{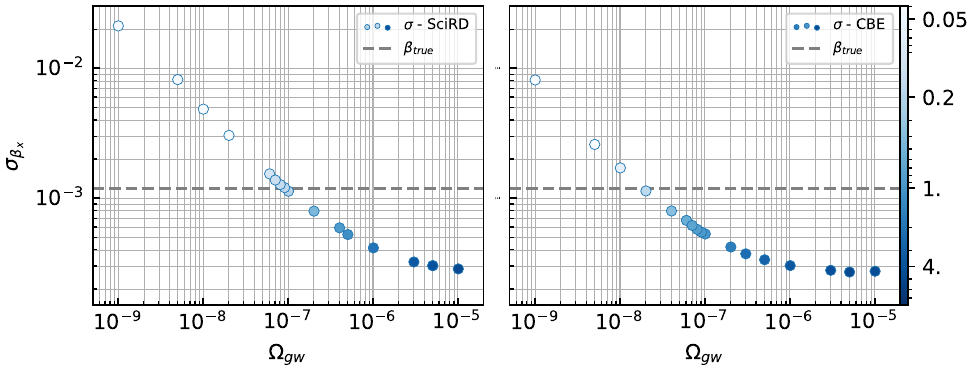}
    \caption{Analysis of the evolution of $\sigma_{\beta_{x}}$ as the signal amplitude $\Omega_\T{GW}$ varies. The blue color bar provides the estimated standard deviation $\frac{\sigma_{\beta_{x}}}{\mid \beta_{x}^{\text{true}} \mid}$ in $|\beta_{x}^{\text{true}}|$ unit \cite{SGWB_analysis_I}. Two instrumental noise models are considered: SciRD specifications (left) and current best estimate (right), differing essentially by the level of OMS noise (displacement noise floor at $1.5 \times 10^{-11}$ and $6.35 \times 10^{-12} \, \T{m}/\T{Hz}^{1/2}$ resp. \cite{Scird, bayle_unified_2023}). Both cases confirm that the kinematic dipole is reachable for $\Omega_\T{GW}$ values typically above $10^{-7}$. We observe that for $\Omega_\T{GW} > 10^{-7}$ we converge to the sample variance limit displayed in Fig. \ref{fig: stats_beta}.}
    \label{fig: noise_analysis}
\end{figure}

At this stage, instrumental noise has not been considered. The only sources of randomness and uncertainty under consideration have been the inherent stochastic characteristics of the signal, specifically the intrinsic variance of its spectral density, which competes with its annual temporal fluctuations. As shown in Fig. \ref{fig: stats_beta}, the precision of the map-making process exceeds the anticipated sample variance for a data span of $4$ years. Importantly, this outcome remains unaffected by the signal amplitude $\Omega_\T{GW}$ in the noise-free scenario, which has not been discussed until now. 
To explore the detectability of Doppler boost-induced anisotropies in a scale-free extra-galactic \gls{sgwb}, including instrumental noise, one now incorporates the noise model. Here, a zero-noise likelihood MCMC map-making approach is employed. In this method, the covariance matrix \eqref{eq: covariance} is computed, accounting for an expected value for the noise matrix $\mathbf{N}_d$ derived from the noise model detailed in \cite{baghi_uncovering_2023}. While the data matrix $\mathbf{D}$ in the log-likelihood definition \eqref{eq: logL} remains unchanged, the modeled covariance matrix $\mathbf{C}_d$ now includes instrumental noise, which affects the uncertainties of $\vec{\beta}^{\text{meas}}$ as assessed through the MCMC analysis. The results of this analysis are presented in Fig. \ref{fig: noise_analysis}\footnote{Note that only the dominant velocity direction, $\beta_x$, given our choice of reference frame are displayed. This is sufficient to identify the kinematic anisotropy, as in the chosen frame $\beta_x$ accounts for over $90$\% of the total boost's magnitude. If another reference system is chosen, the magnitude is distributed across all components depending on the transformation applied, as demonstrated in Fig. \ref{fig: stats_beta_rotated}.}. 
In contrast to the approach in \cite{baghi_uncovering_2023}, here, one considers two noise performance scenarios: one that is conservatively aligned with the LISA \gls{scird} \cite{Scird}, and a second, more optimistic scenario based on the current best understanding of the instrument's performance (see Appendix B and Eq. (B5) in \cite{bayle_unified_2023}). The figure illustrates the estimated $\sigma^{mcmc}_{\beta_x}$ as a function of $\Omega_\T{GW}$, with the color scale indicating, for each data point, the distance to the target $\beta_x^{\text{true}}$ in units of $\sigma^{mcmc}_{\beta_x}$.

Fig. \ref{fig: noise_analysis} confirms the results presented in \cite{SGWB_analysis_II}. In the top subplot, under a conservative noise configuration, the detectability of the kinematic dipole becomes significant only for relatively large values of $\Omega_\T{GW} \gtrsim 10^{-7}$. The more optimistic scenario, based on the latest LISA performance model, suggests detectability at $\Omega_\T{GW} \gtrsim 5 \times 10^{-8}$, potentially reaching $10^{-8}$. This range aligns with expectations, given LISA's suboptimal sensitivity to odd-$\ell$ modes \cite{SGWB_analysis_II}. 
In contrast to the approach outlined in \cite{SGWB_analysis_II}, it is important to note that their analysis relies on the assumption of a noise-dominated covariance matrix, leading to the simplified Fisher matrix expression in Equation (6.5). This assumption treats the intrinsic variance of the stochastic cosmological signal as negligible, thereby simplifying the analysis by removing a layer of complexity. The limitations of \gls{sgwb} detection due to its intrinsic variance, often referred to as sample variance, have been discussed in \cite{Contaldi_with_variance}. In contrast, the results presented in Fig. \ref{fig: noise_analysis} take into account both sample variance and instrumental noise.

In summary, Section \ref{sec:SGWB_detection_Paper_HI} presents a comprehensive time-domain map-making approach based on a maximum-likelihood MCMC tailored for \gls{lisa}, with the aim of detecting kinematic anisotropies in the \gls{sgwb}. Stochastic strain time series are generated for each sky pixel and projected onto the simulated full instrument response function. For the resulting data streams, which consist of random data matrices convolved with the instrumental response kernel, a suitable model-based likelihood function is chosen to test the data against a physical model. A novel MCMC mapping scheme is then developed based on this likelihood function. The effectiveness of the proposed pipeline is demonstrated by successfully recovering the injected map of Doppler-boosted primordial anisotropic data, thereby showcasing the utility of the MCMC-based mapping scheme for \gls{lisa}. The results show that the pipeline outperforms sample variance limitations in \gls{lisa} data and remains robust against both systematic and statistical errors. Recovery of the injected data is achieved using both model-dependent and independent sets of fit parameters. When instrumental noise is incorporated, the pipeline converges to a statistically meaningful recovery of the kinematic dipole for $\Omega_\T{GW}\gtrsim 10^{-7}$ under conservative noise scenarios, with slightly better performance under more favorable instrumental noise conditions. These results are consistent with and complement those reported in \cite{SGWB_analysis_II}. \\
Despite the promising outcomes of the analysis, several areas for improvement are acknowledged, which will be addressed in future work. First, the systematic bias observed in the monopole amplitude measurements has highlighted inaccuracies in the fitting response model at the zero-response frequencies, where the relative inaccuracy is amplified due to the very low amplitude response. While this issue does not affect the dipole measurements, it will require a dedicated treatment (such as masking or model improvement), along with an enhancement of the statistical model for the data matrix in Eq. \eqref{equ:data_matrix} for Bayesian inference. These modifications will help mitigate the monopole bias and potentially make the overall analysis more robust with respect to the time-frequency analysis settings (e.g., time window length and spectral averaging).

In relation to the above, several opportunities have been identified to optimize the data sampling process by transitioning to a time-frequency domain analysis instead of relying on a full time-domain approach. Replacing the short-Fourier transform with a wavelet-based analysis during the preprocessing phase could improve the likelihood of recovering kinematic anisotropy from the data. Additionally, it was noted that orbital dynamics have a significant influence on the output of the map-making scheme, suggesting that simulating more realistic orbital dynamics—rather than the static equal-arm-length assumption used here—would enhance the kinematic signatures in the data. However, the use of more realistic Keplerian orbits has introduced notable biases in the map-making analysis, which are currently under investigation. These biases appear to be strongly linked to the initial preprocessing step, where the 4-year \gls{tdi} data streams are divided using 50\% overlapping Hanning windows. It is believed that moving to a time-frequency representation of the \gls{tdi} data will resolve this issue, allowing the complex temporal features arising from the full orbital dynamics to be exploited. More broadly, the sensitivity of the proposed map-making scheme to key analysis parameters, particularly the number of time and frequency bins ($n_{\text{samples}}$ and $n_{\text{avg}}$), has been noted. These parameters determine how the input data is divided \footnote{See Eq. \eqref{equ:data_matrix} and the subsequent discussions.}. It was found that the convergence of the MCMC is sensitive to the choice of $n_{\text{samples}}$ and $n_{\text{avg}}$. In particular, suboptimal choices for bin size can significantly shift the monopole. For the results presented in the previous section, the parameters $n_{\text{samples}}$ and $n_{\text{avg}}$ were optimized to achieve convergence of the random walkers in parameter space. While the full implications of a shifted time (or frequency) grid within the MCMC context are not yet completely understood, they are under investigation and will be addressed in future work. Again, it is expected that wavelet transforms will help facilitate these investigations.

\subsection{Conclusion and Outlook}

The detection of the SGWB could represent a major breakthrough for numerous branches of physics. As a result, the scientific community is pushing the limits of existing and upcoming datasets to extract as much information as possible. Section \ref{sec:SGWB_detection} provides a brief overview of the current state of active research in this field. While astrophysical contributions to the SGWB are on the verge of being detectable, as discussed in Subsection \ref{sec:SGWB_detection_astro}, cosmological contributions are likely to be uncovered only with future instruments, as demonstrated in detail in Section \ref{sec:SGWB_detection_Paper_HI}. In this section, considerable effort was made to use the LISA instrument as a detector for the simplest possible signature of the SGWB, i.e., a scale-free spectrum. Future efforts will, however, need to focus on more complex primordial contributions to the SGWB, which exhibit non-trivial spectral dependencies, as illustrated in Fig. \ref{fig:spectrum}. Extending the analysis presented here to encompass more intricate signatures could significantly improve the lower bound on $\Omega_\T{GW}$ for which the MCMC method recovers the desired contributions. This improvement is primarily due to the dependence of the components $D(f)$ and $Q(f)$ on the spectral slope, as discussed in Eq. \eqref{equ:D} and \eqref{equ:Q}. These components, which represent the dipole and quadrupole modes of the signal, experience significant amplification when the slope of the spectrum $\Omega_\T{GW}$ varies with frequency. This variation is quantified by the functions $n_\Omega$ and $\alpha_\Omega$. In some cases, this can result in the quadrupolar power surpassing that of the dipole. As a result, spectra arising from phenomena such as, for example, PTs or PBHs strongly amplify the power of kinematic anisotropies for $\ell>0$, with a notable emphasis on enhancing the quadrupole. The amplification of signal power for $\ell = 2$ is particularly significant for LISA, as the instrument is particularly sensitive to the quadrupolar mode of sky anisotropies \cite{SGWB_analysis_II}.

In addition to identifying cosmic origins of signatures in LISA data, the pipeline developed in Section \ref{sec:SGWB_detection_Paper_HI} can also be utilized to analyze galactic contributions to the SGWB, aiming for results comparable to those in Section \ref{sec:SGWB_detection_astro} using the LISA instrument. In the previous section, a framework was established under the assumption that an effective foreground removal scheme has been applied to the raw data. Although this step is highly non-trivial and falls outside the scope of this article, our map-making algorithm can be effectively employed to map the galactic confusion noise. The potential of the MCMC mapping scheme to map the remaining Milky Way GW sources in LISA Data Challenge data, following the extraction of resolved binaries, will be explored in a follow-up project.

In conclusion, GW) astronomy has only begun to explore the vast wealth of information provided by the SGWB. While collaborations related to PTAs are likely to publish enhanced data in the coming years, potentially strengthening their evidence for the astrophysical background, new instruments like LISA have the potential to complement this data by providing insights into the cosmological background contributions. Investigations such as those presented in Section \ref{sec:SGWB_detection_Paper_HI} are therefore indispensable, as they establish ready-to-use toolkits for testing suitably processed LISA data of SGWB nature for signatures of cosmic origin. Specifically, given that the LISA sensitivity band is dominated by an SGWB signal, spectral-rich signatures may be recovered for $\Omega_\T{GW}\gtrsim 10^{-8}$ with a less conservative instrumental noise realization. This capability becomes particularly important in the context of potential PTA signals of a stochastic background \cite{Nanograv_SGWB_I} extending into the LISA band. If such a signal is (partially) of cosmic origin, significant contributions from the same source would be expected in the LISA data band. The cosmic nature of these hypothetical signatures can be confirmed using the analysis pipeline outlined in Section \ref{sec:SGWB_detection_Paper_HI}, i.e., \cite{SGWB_analysis_I}.

 \setcounter{chapter}{7}

\chapter{Concluding Remarks} 

\label{chap:discussion} 



The detection of \gls{gw}s represents one of the most significant achievements in 21st-century physics. However, only a small fraction of the extensive cosmological, astrophysical, and gravitational information encoded in \gls{gw} signals has yet been revealed. It remains to be an exciting journey where, even for the most experienced members of the research community, anticipating the discoveries enabled by future observations remains challenging—particularly regarding the potential emergence of new physics within the measurement data. \\
Yet, despite the remarkable advances in \gls{gw} physics to date, numerous theoretical and experimental challenges remain. On the theoretical front, the accurate modeling of highly precessing, eccentric, and high mass-ratio systems continues to demand more sophisticated analytical and numerical tools. The inclusion of matter effects in NS mergers, especially in strong magnetic fields and finite temperature regimes, introduces additional complexities requiring multi-physics simulations. From a foundational perspective, quantum gravitational effects (including imprints of a theory of quantum gravity) and potential GW imprints from early-Universe phase transitions remain speculative but profoundly important areas of inquiry. \\
Experimentally, the next generation of detectors—such as LISA, the Einstein Telescope, and Cosmic Explorer—will extend the GW observational band, enhancing sensitivity to lower-frequency signals and enabling the detection of intermediate-mass BHs, extreme mass-ratio inspirals, and primordial \gls{gw} backgrounds. However, these instruments face formidable engineering and data analysis challenges, including stringent requirements on noise suppression, calibration, and waveform extraction in high-dimensional parameter spaces. Moreover, the development of coherent multi-messenger frameworks, integrating GWs, electromagnetic, and neutrino data, will be crucial for maximizing the scientific return of future detections.\\
In sum, \gls{gw} physics stands at the threshold of a new era in fundamental physics and astrophysics. Continued progress will require a synergistic effort across theoretical modeling, computational techniques, and experimental innovation. The path forward is challenging but holds the promise of addressing some of the most profound questions in modern science, from the nature of spacetime to the origin of the Universe.

In this thesis, each Chapters \ref{chap:asymptotics}, \ref{chap:quantum} and \ref{chap:analysis} can be associated with one of three major research branches in \gls{gw} physics, respectively: The development of analytical tools to improve numerical evaluation of \gls{gw} data, the theoretical and numerical derivation of high-precision features of gravitational waveforms, and, finally, the instrument-specific development of data analysis pipelines ready to be applied on measurement data. As such, this thesis establishes a comprehensive framework that bridges foundational theoretical insights with practical methodologies applicable to ongoing and future \gls{gw} observations. By systematically addressing challenges across these three pillars of \gls{gw} research, it contributes both to the refinement of gravitational waveform modeling and the enhancement of data interpretation strategies. The analytical tools assembled in Chapter \ref{chap:asymptotics} lay the groundwork for more efficient numerical simulations by exploiting asymptotic structures and symmetries of spacetime, thereby reducing computational complexity without sacrificing precision. In Chapter \ref{chap:quantum}, the investigation of quantum corrections and high-order relativistic effects informs the subtle features of gravitational waveforms, which may become detectable as detector sensitivity improves. Finally, Chapter \ref{chap:analysis} translates various theoretical developments into actionable tools for data analysis, integrating them into pipelines optimized for the future space-based LISA detector.\\
In unifying these directions, this dissertation not only advances each domain independently but also elucidates the synergies between them—demonstrating, for instance, how asymptotic analyses inform waveform reconstruction, or how quantum-induced corrections may be encoded in detectable signatures. This integrated approach reflects the evolving nature of \gls{gw} physics, where theory, computation, and instrumentation converge to enable precision tests of \gls{gr} and to explore physics beyond the Standard Model.


\newpage

\phantomsection
\addcontentsline{toc}{chapter}{Publications}
\chapter*{Publications}

\textit{The following list includes all released or to-be-released publications of the author of this thesis (subsequently abbreviated by his initials ``\textbf{DM}'') at the date of submission. Publications that are part of this thesis are marked with ``${}^*$''.}

\noindent
[A] Superfluid Dark Matter through Cold Atoms near Tricriticality, (to be published), 2025.

\noindent
[B]${}^*$ Nils Deppe, Lavinia Heisenberg, Henri Inchausp\'e, \textbf{DM}, et al. \textit{Signatures of quantum gravity in the gravitational wave}, (recently accepted by PRD), arXiv:2502.20584, 2025.

\noindent 
[C]${}^*$ Nils Deppe, Lavinia Heisenberg, Henri Inchausp\'e, \textbf{DM}, et al. \textit{Echoes from beyond: Detecting gravitational wave quantum imprints with Lisa}, (recently accepted by PRD), arXiv:2411.05645, 2024.

\noindent
[D]${}^*$ Fabio D’Ambrosio, Francesco Gozzini, Lavinia Heisenberg, \textbf{DM}, et al. \textit{Testing gravitational waveforms in full General Relativity}. JCAP, 02:060, 2025.

\noindent
[E]${}^*$ Lavinia Heisenberg, Henri Inchauspé, and \textbf{DM},. \textit{Observing kinematic anisotropies of the stochastic background with LISA}. JCAP, 01:044, 2025.

\noindent
[F] Lavinia Heisenberg, \textbf{DM}, and Do\u{g}a Veske. \textit{Searching for topological dark matter in LIGO data}. Phys. Rev. D, 110(5):055037, 2024.

\noindent
[G]${}^*$ Fabio D’Ambrosio, Shaun D. B. Fell, Lavinia Heisenberg, \textbf{DM} et al. \textit{Gravitational waves in full, non-linear general relativity}, arXiv:2201.11634, 2022.

\noindent
[H] \textbf{DM}, Robert Brandenberger, Devin Crichton, and Alexandre Re-
fregier. \textit{Extracting the signal of cosmic string wakes from 21-cm observations}.
Phys. Rev. D, 104(12):123535, 2021.
\bibliographystyle{natbib_up}


\bibliography{biblatex-examples}


\appendix 
\counterwithin{figure}{chapter}
\numberwithin{equation}{chapter}


\chapter{Computational Tools and Techniques} 

\label{AppendixA} 

\section{Levi-Civita Connection for the Geodesic Null Congruence Metric}
\label{app_sec:Levi_Civita}
The Levi-Civita connection for the metric 
\begin{align}
    \dd s^2 &= V\dd u ^2 + 2B\dd u \dd r + A\dd r^2 + H_{AB}(\dd y^A - U^A \dd u)(\dd y^B -U^B \dd u)+ D_A \dd y^A \dd r\notag\\
    &=:\gd^\text{NGC}\dd x^\mu\dd x^\nu\,,
\end{align}
is computed. One finds
\begin{align}
\Gamma_{0,0,0}\to & \frac{1}{2} \bigg(\partial_{u} H_{11}{U_1}^2+2 \left({U_2}\partial_{u} H_{12}+{H_{11}}
  \partial_{u} U_1+{H_{12}}\partial_{u} U_2\right){U_1}+{U_2}^2\partial_{u} H_{22}\notag \\ 
   & + 2{U_2} \left({H_{12}}
  \partial_{u} U_1 + {H_{22}}\partial_{u} U_2\right)+\partial_u V\bigg),\\ 
\Gamma_{0,0,1}\to & \frac{1}{2} \bigg(\partial_{r} H_{11}
  {U_1}^2+2 \left({U_2}\partial_{r} H_{12}+{H_{11}}
  \partial_{r} U_1+{H_{12}}\partial_{r} U_2\right){U_1}+{U_2}^2\partial_{r} H_{22}
  \notag \\ 
   &+2{U_2} \left({H_{12}}
  \partial_{r} U_1+{H_{22}}\partial_{r} U_2\right)+\partial_r V\bigg),\\ 
  \Gamma_{0,0,2}\to & \frac{1}{2} \bigg(\partial_{y^1} H_{11}
  {U_1}^2+2 \left({U_2}\partial_{y^1} H_{12}+{H_{11}}
  \partial_{y^1} U_1+{H_{12}}\partial_{y^1} U_2\right){U_1}+{U_2}^2\partial_{y^1} H_{22}\notag \\ 
   &+2{U_2} \left({H_{12}}
  \partial_{y^1} U_1+{H_{22}}\partial_{y^1} U_2\right)+\partial_{y^1} V\bigg),\\
  \Gamma_{0,0,3}\to & \frac{1}{2} \bigg(\partial_{y^2} H_{11}
  {U_1}^2+2 \left({U_2}\partial_{y^2} H_{12}+{H_{11}}
  \partial_{y^2} U_1+{H_{12}}\partial_{y^2} U_2\right){U_1}+{U_2}^2\partial_{y^2} H_{22}\notag \\ 
   &+2{U_2} \left({H_{12}}
  \partial_{y^2} U_1+{H_{22}}\partial_{y^2} U_2\right)+\partial_{y^2} V\bigg),
  \end{align}
\begin{align}
  \Gamma_{0,1,0}\to & \frac{1}{2} \bigg(\partial_{r} H_{11}
  {U_1}^2+2 \left({U_2}\partial_{r} H_{12}+{H_{11}}
  \partial_{r} U_1+{H_{12}}\partial_{r} U_2\right){U_1}+{U_2}^2\partial_{r} H_{22}\notag \\ 
   &+2{U_2} \left({H_{12}}
  \partial_{r} U_1+{H_{22}}\partial_{r} U_2\right)+\partial_r V\bigg),\\
  \Gamma_{0,1,1}\to & \partial_r B-\frac{1}{2}
   \partial_{u} A,\\ 
   \Gamma_{0,1,2}\to & 0.5 \left(\partial_{y^1} B-1.
  {U_1}\partial_{r} H_{11}-1.{U_2}\partial_{r} H_{12}-1.{H_{11}}\partial_{r} U_1-1.{H_{12}}\partial_{r} U_2-0.5\partial_{u} D_1 \right),\\ 
  \Gamma_{0,1,3}\to & 0.5
   \left(\partial_{y^2} B-1.{U_1}\partial_{r} H_{12}-1.{U_2}
  \partial_{r} H_{22}-1.{H_{12}}\partial_{r} U_1-1.{H_{22}}\partial_{r} U_2-0.5\partial_{u} D_2\right),\\ 
  \Gamma_{0,2,0}\to & \frac{1}{2} \bigg(\partial_{y^1} H_{11}
  {U_1}^2+2 \left({U_2}\partial_{y^1} H_{12}+{H_{11}}
  \partial_{y^1} U_1+{H_{12}}\partial_{y^1} U_2\right){U_1}+{U_2}^2\partial_{y^1} H_{22}\notag \\ 
   &+2{U_2} \left({H_{12}}
  \partial_{y^1} U_1+{H_{22}}\partial_{y^1} U_2\right)+\partial_{y^1} V\bigg),\\ 
  \Gamma_{0,2,1}\to & 0.5 \left(\partial_{y^1} B-1.
  {U_1}\partial_{r} H_{11}-1.{U_2}\partial_{r} H_{12}-1.{H_{11}}\partial_{r} U_1-1.{H_{12}}\partial_{r} U_2-0.5\partial_{u} D_1 \right),\\ 
  \Gamma_{0,2,2}\to &
   -{U_1}\partial_{y^1} H_{11}-{U_2}\partial_{y^1} H_{12}-{H_{11}}\partial_{y^1} U_1-{H_{12}}\partial_{y^1} U_2-\frac{1}{2}\partial_{u} H_{11},\\ 
   \Gamma_{0,2,3}\to & \frac{1}{2}
   \bigg(-{H_{11}}\partial_{y^2} U_1-{H_{12}}\partial_{y^2} U_2-{U_1} \left(\partial_{y^2} H_{11}+\partial_{y^1} H_{12}\right)-{U_2} \left(\partial_{y^2} H_{12}+\partial_{y^1} H_{22}\right)\notag \\ 
   &-{H_{12}}\partial_{y^1} U_1-{H_{22}}\partial_{y^1} U_2-\partial_{u} H_{12}\bigg),\\
   \Gamma_{0,3,0}\to & \frac{1}{2}
   \bigg(\partial_{y^2} H_{11}{U_1}^2+2 \left({U_2}
  \partial_{y^2} H_{12}+{H_{11}}\partial_{y^2} U_1+{H_{12}}
  \partial_{y^2} U_2\right){U_1}+{U_2}^2\partial_{y^2} H_{22}\notag \\ 
   &+2{U_2} \left({H_{12}}\partial_{y^2} U_1+{H_{22}}\partial_{y^2} U_2\right)+\partial_{y^2} V\bigg),\\ 
  \Gamma_{0,3,1}\to & 0.5 \left(\partial_{y^2} B-1.{U_1}\partial_{r} H_{12}-1.{U_2}\partial_{r} H_{22}-1.
  {H_{12}}\partial_{r} U_1-1.{H_{22}}\partial_{r} U_2-0.5\partial_{u} D_2\right),\\
  \Gamma_{0,3,2}\to & \frac{1}{2}
   \bigg(-{H_{11}}\partial_{y^2} U_1-{H_{12}}\partial_{y^2} U_2-{U_1} \left(\partial_{y^2} H_{11}+\partial_{y^1} H_{12}\right)-{U_2} \left(\partial_{y^2} H_{12}+\partial_{y^1} H_{22}\right)-\notag \\ 
   &{H_{12}}\partial_{y^1} U_1-{H_{22}}\partial_{y^1} U_2-\partial_{u} H_{12}\bigg),\\ 
   \Gamma_{0,3,3}\to & -{U_1}\partial_{y^2} H_{12}-{U_2}\partial_{y^2} H_{22}-{H_{12}}\partial_{y^2} U_1-{H_{22}}\partial_{y^2} U_2-\frac{1}{2}
  \partial_{u} H_{22},\\
  \Gamma_{1,0,0}\to & \frac{1}{2}
   \bigg(-\partial_{r} H_{11}{U_1}^2-2 \left({U_2}
  \partial_{r} H_{12}+{H_{11}}\partial_{r} U_1+{H_{12}}
  \partial_{r} U_2\right){U_1}-{U_2}^2\partial_{r} H_{22}\notag \\ 
   &-2{U_2} \left({H_{12}}\partial_{r} U_1+{H_{22}}\partial_{r} U_2\right)-\partial_r V+2 \partial_u B\bigg),
   \end{align}
\begin{align}
  \Gamma_{1,0,1}\to & \frac{1}{2} \partial_{u} A,\\ 
  \Gamma_{1,0,2}\to & 0.5 \left(\partial_{y^1} B+{U_1}
  \partial_{r} H_{11}+{U_2}\partial_{r} H_{12}+{H_{11}}
  \partial_{r} U_1+{H_{12}}\partial_{r} U_2+0.5\partial_{u} D_1 \right),\\ \Gamma_{1,0,3}\to & 0.5 \left(\partial_{y^2} B+{U_1}\partial_{r} H_{12}+{U_2}\partial_{r} H_{22}+{H_{12}}\partial_{r} U_1+{H_{22}}\partial_{r} U_2+0.5
  \partial_{u} D_2\right),\\ \Gamma_{1,1,0}\to & \frac{1}{2} \partial_{u} A,\\ \Gamma_{1,1,1}\to & \frac{1}{2} \partial_{r} A,\\ \Gamma_{1,1,2}\to & \frac{1}{2} \partial_{y^1} A,\\ \Gamma_{1,1,3}\to & \frac{1}{2} \partial_{y^2} A,\\ \Gamma_{1,2,0}\to & 0.5 \left(\partial_{y^1} B+{U_1}
  \partial_{r} H_{11}+{U_2}\partial_{r} H_{12}+{H_{11}}
  \partial_{r} U_1+{H_{12}}\partial_{r} U_2+0.5\partial_{u} D_1 \right),\\ \Gamma_{1,2,1}\to & \frac{1}{2} \partial_{y^1} A,\\ \Gamma_{1,2,2}\to & 0.5\partial_{y^1} D_1-0.5
  \partial_{r} H_{11},\\ \Gamma_{1,2,3}\to & 0.25\partial_{y^2} D_1+0.25\partial_{y^1} D_2-0.5\partial_{r} H_{12},\\ \Gamma_{1,3,0}\to & 0.5 \left(\partial_{y^2} B+{U_1}
  \partial_{r} H_{12}+{U_2}\partial_{r} H_{22}+{H_{12}}
  \partial_{r} U_1+{H_{22}}\partial_{r} U_2+0.5\partial_{u} D_2\right),\\ \Gamma_{1,3,1}\to & \frac{1}{2} \partial_{y^2} A,\\ \Gamma_{1,3,2}\to & 0.25\partial_{y^2} D_1+0.25
  \partial_{y^1} D_2-0.5\partial_{r} H_{12},\\ \Gamma_{1,3,3}\to & 0.5\partial_{y^2} D_2-0.5
  \partial_{r} H_{22},\\
\Gamma_{2,0,0}\to & \frac{1}{2}
   \bigg(-\partial_{y^1} H_{11}{U_1}^2-2 \left({U_2}
  \partial_{y^1} H_{12}+{H_{11}}\partial_{y^1} U_1+{H_{12}}
  \partial_{y^1} U_2+\partial_{u} H_{11}\right){U_1}\notag \\ 
   &-{U_2}^2\partial_{y^1} H_{22}-\partial_{y^1} V-2{U_2} \left({H_{12}}\partial_{y^1} U_1+{H_{22}}\partial_{y^1} U_2+\partial_{u} H_{12}\right)\notag \\ 
   &-2{H_{11}}\partial_{u} U_1-2
  {H_{12}}\partial_{u} U_2\bigg),\\
  \Gamma_{2,0,1}\to &
   -0.5 \partial_{y^1} B-0.5{U_1}\partial_{r} H_{11}-0.5{U_2}\partial_{r} H_{12}-0.5{H_{11}}\partial_{r} U_1-0.5{H_{12}}\partial_{r} U_2+0.25\partial_{u} D_1 ,\\ \Gamma_{2,0,2}\to & \frac{1}{2}\partial_{u} H_{11},\\ \Gamma_{2,0,3}\to & \frac{1}{2} \bigg(-{H_{11}}
  \partial_{y^2} U_1-{H_{12}}\partial_{y^2} U_2+{U_1}
   \left(\partial_{y^1} H_{12}-\partial_{y^2} H_{11}\right)+{U_2}
   \left(\partial_{y^1} H_{22}-\partial_{y^2} H_{12}\right)\notag \\ 
   &+{H_{12}}
  \partial_{y^1} U_1+{H_{22}}\partial_{y^1} U_2+\partial_{u} H_{12}\bigg),\\ 
  \Gamma_{2,1,0}\to & -0.5 \partial_{y^1} B-0.5{U_1}\partial_{r} H_{11}-0.5{U_2}\partial_{r} H_{12}-0.5
  {H_{11}}\partial_{r} U_1-0.5{H_{12}}\partial_{r} U_2+0.25\partial_{u} D_1 ,
  \end{align}
\begin{align}
\Gamma_{2,1,1}\to & 0.5
  \partial_{r} D_1-0.5 \partial_{y^1} A,\\ \Gamma_{2,1,2}\to &
   \frac{1}{2}\partial_{r} H_{11},\\ \Gamma_{2,1,3}\to & \frac{1}{2} \left(0.5
  \partial_{y^2} D_1-0.5\partial_{y^1} D_2+\partial_{r} H_{12}\right),\\
\Gamma_{2,2,0}\to & \frac{1}{2}\partial_{u} H_{11},\\ \Gamma_{2,2,1}\to & \frac{1}{2}\partial_{r} H_{11},\\ \Gamma_{2,2,2}\to & \frac{1}{2}\partial_{y^1} H_{11},\\ \Gamma_{2,2,3}\to & \frac{1}{2}\partial_{y^2} H_{11},\\
  \Gamma_{2,3,0}\to & \frac{1}{2} \bigg(-{H_{11}}
  \partial_{y^2} U_1-{H_{12}}\partial_{y^2} U_2+{U_1}
   \left(\partial_{y^1} H_{12}-\partial_{y^2} H_{11}\right)+{U_2}
   \left(\partial_{y^1} H_{22}-\partial_{y^2} H_{12}\right)\notag \\ 
   &+{H_{12}}
  \partial_{y^1} U_1+{H_{22}}\partial_{y^1} U_2+\partial_{u} H_{12}\bigg),\\ 
  \Gamma_{2,3,1}\to & \frac{1}{2} \left(0.5\partial_{y^2} D_1-0.5
  \partial_{y^1} D_2+\partial_{r} H_{12}\right),\\ 
  \Gamma_{2,3,2}\to & \frac{1}{2}\partial_{y^2} H_{11},\\ \Gamma_{2,3,3}\to &\partial_{y^2} H_{12}-\frac{1}{2}
  \partial_{y^1} H_{22},\\ 
  \Gamma_{3,0,0}\to & \frac{1}{2}
   \bigg(-\partial_{y^2} H_{11}{U_1}^2-2 \left({U_2}
  \partial_{y^2} H_{12}+{H_{11}}\partial_{y^2} U_1+{H_{12}}
  \partial_{y^2} U_2+\partial_{u} H_{12}\right){U_1}\notag \\ 
   &-{U_2}^2\partial_{y^2} H_{22}-\partial_{y^2} V-2{U_2} \left({H_{12}}\partial_{y^2} U_1+{H_{22}}\partial_{y^2} U_2+\partial_{u} H_{22}\right)-2{H_{12}}\partial_{u} U_1\notag \\ 
   &-2
  {H_{22}}\partial_{u} U_2\bigg),\\
  \Gamma_{3,0,1}\to &
   -0.5 \partial_{y^2} B-0.5{U_1}\partial_{r} H_{12}-0.5{U_2}\partial_{r} H_{22}-0.5{H_{12}}\partial_{r} U_1-0.5{H_{22}}\partial_{r} U_2+0.25\partial_{u} D_2,\\ \Gamma_{3,0,2}\to & \frac{1}{2} \bigg({H_{11}}\partial_{y^2} U_1+{H_{12}}\partial_{y^2} U_2+{U_1}
   \left(\partial_{y^2} H_{11}-\partial_{y^1} H_{12}\right)+{U_2}
   \left(\partial_{y^2} H_{12}-\partial_{y^1} H_{22}\right)\notag \\ 
   &-{H_{12}}
  \partial_{y^1} U_1-{H_{22}}\partial_{y^1} U_2+\partial_{u} H_{12}\bigg),\\
  \Gamma_{3,0,3}\to & \frac{1}{2}\partial_{u} H_{22},\\ \Gamma_{3,1,0}\to & -0.5 \partial_{y^2} B-0.5{U_1}
  \partial_{r} H_{12}-0.5{U_2}\partial_{r} H_{22}-0.5{H_{12}}\partial_{r} U_1-0.5{H_{22}}\partial_{r} U_2+0.25
  \partial_{u} D_2,\\ 
  \Gamma_{3,1,1}\to & 0.5\partial_{r} D_2-0.5 \partial_{y^2} A,\\ \Gamma_{3,1,2}\to & \frac{1}{2} \left(-0.5
  \partial_{y^2} D_1+0.5\partial_{y^1} D_2+\partial_{r} H_{12}\right),\\ \Gamma_{3,1,3}\to & \frac{1}{2}\partial_{r} H_{22},\\ 
  \end{align}
  \begin{align}
  \Gamma_{3,2,0}\to & \frac{1}{2} \bigg({H_{11}}\partial_{y^2} U_1+{H_{12}}\partial_{y^2} U_2+{U_1}
   \left(\partial_{y^2} H_{11}-\partial_{y^1} H_{12}\right)+{U_2}
   \left(\partial_{y^2} H_{12}-\partial_{y^1} H_{22}\right)\notag \\ 
   &-{H_{12}}
  \partial_{y^1} U_1-{H_{22}}\partial_{y^1} U_2+\partial_{u} H_{12}\bigg),\\
  \Gamma_{3,2,1}\to & \frac{1}{2} \left(-0.5\partial_{y^2} D_1+0.5
  \partial_{y^1} D_2+\partial_{r} H_{12}\right),\\ 
  \Gamma_{3,2,2}\to &\partial_{y^1} H_{12}-\frac{1}{2}
  \partial_{y^2} H_{11},\\ \Gamma_{3,2,3}\to & \frac{1}{2}
  \partial_{y^1} H_{22},\\ \Gamma_{3,3,0}\to & \frac{1}{2}
  \partial_{u} H_{22},\\ \Gamma_{3,3,1}\to & \frac{1}{2}
  \partial_{r} H_{22},\\ \Gamma_{3,3,2}\to & \frac{1}{2}
  \partial_{y^1} H_{22},\\ \Gamma_{3,3,3}\to & \frac{1}{2}
  \partial_{y^2} H_{22}\,.
\end{align}


%
%
%


\section{Conformal Transformations}
\label{App:ConformalTransformations}
The concept of conformal transformations is summarized and a collection of formulas for the transformation properties of the most important tensors of this work is gathered. For more details the reader is referred to the appendices in \cite{Geroch_1977} and \cite{Wald_book}.

Consider a $d$-dimensional manifold $\mathcal M$ with metric $g_{\mu\nu}$ with a Levi-Civita connection and associated covariant derivative $\nabla_{\mu}$. In a covariant theory, a Weyl or conformal transformation is the introduction of a new metric 
\begin{equation}
    g'_{\mu\nu}(x)=\omega^2(x) g_{\mu\nu}(x)\,,
\end{equation}
or in other words a point-dependent rescaling of the metric, where $\omega$ is a smooth, strictly positive scalar field on $\mathcal M$. Note that in this sense a conformal transformation is not a priori associated to a diffeomorphism, hence a coordinate change\footnote{It is important to stress that in field theory the term conformal transformation is used differently: In this context with a flat background metric, a conformal transformation denotes an actual coordinate transformation which changes the metric up to a conformal factor, such that for instance Poincar\'{e} transformations are a subset with trivial conformal factor (leaving the metric invariant). Thus, in field theory conformal invariance is a statement about a combined conformal diffeomorphism and a Weyl rescaling where this last transformation is defined as in the main text. On the other hand, in a covariant theory such as GR it does not make much sense to talk about conformal diffeomorphisms, as the theory is already invariant under such transformations anyway and the concept of conformal invariance is equivalent to what is called Weyl invariance from a field theory perspective.}. Moreover, with two metrics at hand, there is an ambiguity to which metric is used to raise or lower indices, which is important to keep in mind. It is ensured, however, that the manifold with the metric $g_{ab}$ or $g'_{ab}$ have the same causal structure, such that for instance a vector retains its timelike, null or spacelike property with respect to each metric.
The determinant of the metric transforms as
\begin{equation}\label{DetMetricConformal}
    \det g' = \omega^{2d}\,\det g \,,
\end{equation}
while from the requirement $g'^{\mu\nu}g'_{\nu\rho}=g^{\mu\nu}g_{\nu\rho}=\delta\ud{\mu}{\rho}$ one finds for the inverse metric
\begin{equation}
    g'^{\mu\nu} = \frac{1}{\omega^2}\,g^{\mu\nu}\,.
\end{equation}
Note that \eqref{DetMetricConformal} implies that the Levi-Civita tensor transforms as
\begin{equation}
    \varepsilon'_{\mu_1\mu_2..\mu_d}=\sqrt{- \det g'}\,\tilde\varepsilon_{\mu_1\mu_2..\mu_d} = \omega^d\,\sqrt{- \det g}\,\tilde\varepsilon_{\mu_1\mu_2..\mu_d} = \omega^d\,\varepsilon_{\mu_1\mu_2..\mu_d}\,.
\end{equation}
Furthermore, the two covariant derivatives $\nabla_\mu$ and $\nabla'_\mu$ are related to each other for $\alpha$ a vector field on the manifold via
\begin{equation}\label{CovDerConformalI}
    \nabla'_\mu\alpha_{\nu}=\nabla_\mu\alpha_{\nu}+C\du{\mu\nu}{\sigma}\alpha_{\sigma}\quad\text{and}\quad \nabla'_\mu\alpha^{\nu}=\nabla_\mu\alpha^{\nu}-C\du{\mu\sigma}{\nu}\alpha^{\sigma}
\end{equation}
where using the metric compatibility of the convariant derivatives it can be shown that
\begin{equation}\label{CovDerConformalII}
    C\du{\mu\nu}{\rho}=-\frac{1}{\omega}g^{\rho\sigma}\left(g_{\nu\sigma}\nabla_\mu\omega+g_{\mu\sigma}\nabla_\nu\omega-g_{\mu\nu}\nabla_\sigma\omega\right)\,.
\end{equation}
It is important to emphasize, that this formula is independent of the dimension $d$ of the manifold. Moreover, it can be shown that while geodesics with respect $\nabla_\mu$ are in general not geodesics with respect to $\nabla'_\mu$ anymore, except for null geodesics (see \cite{Wald_book}).

Next, consider the relations between the curvature tensors associated to the two covariant derivatives. From \eqref{CovDerConformalI} and \eqref{CovDerConformalII} it follows that
\begin{align} \label{equ:help_connection}
        \Gamma'{}\ud{\rho}{\mu\nu} = \Gamma\ud{\rho}{\mu\nu} + \frac{1}{\omega}\left(2\delta\ud{\rho}{(\mu}\nabla_{\nu)}\omega-g_{\mu\nu}\nabla^\rho\omega\right)\,,
\end{align}
which can simply be plugged into the definition of the Riemann tensor
    \begin{equation}
        R'\du{\nu\rho\sigma}{\mu} =  \partial_{\rho}\Gamma'^{\mu}{}_{\sigma\nu} -\partial_{\sigma}\Gamma'^{\mu}{}_{\rho\nu} +\Gamma'^{\mu}{}_{\rho\tau}\Gamma'^{\tau}{}_{\sigma\nu} -\Gamma'^{\mu}{}_{\sigma\tau}\Gamma'^{\tau}{}_{\rho\nu}\,.
    \end{equation}
to obtain
 \begin{align}
    R'\du{\mu\nu\rho}{\sigma}=R\du{\mu\nu\rho}{\sigma}&-\frac{4}{\omega^2}\delta\ud{\sigma}{[\mu}\nabla_{\nu]}\omega\nabla_\rho\omega+\frac{4}{\omega^2}g_{\rho[\mu}\nabla_{\nu]}\omega\nabla^\sigma\omega+\frac{2}{\omega^2}\delta\ud{\sigma}{[\mu}\,g_{\nu]\rho}g^{\alpha\beta}\nabla_{\alpha}\omega\nabla_\beta\omega\notag\\
    &-\frac{2}{\omega}\nabla_\rho\nabla_{[\mu}\omega\,\delta\du{\nu]}{\sigma}+\frac{2}{\omega}g^{\sigma\alpha}\nabla_\alpha\nabla_{[\mu}\omega \,g_{\nu]\rho}\,.
\end{align}
Notice again that the result is independend of the dimension $d$. Contracting this transformation rule by defining $R'_{\mu\rho}=R'\du{\mu\sigma\rho}{\sigma}$ and $R'=g'^{\mu\nu}R'_{\mu\nu}$ and using $g^{\mu\nu}g_{\mu\nu}=d$ we obtain
\begin{align}
    R'_{\mu\nu}=&\,R_{\mu\nu}-\frac{(d-2)}{\omega}\nabla_\mu\nabla_\nu\omega-\frac{g_{\mu\nu}}{\omega}g^{\rho\sigma}\nabla_\rho\nabla_\sigma\omega+\frac{2(d-2)}{\omega^2}\nabla_\mu\omega\nabla_\nu\omega\notag\\&-\frac{(d-3)g_{\mu\nu}}{\omega^2}g^{\rho\sigma}\nabla_\rho\omega\nabla_\sigma\omega  \\
    R'=&\,\frac{1}{\omega^2}\left(R-\frac{2(d-1)}{\omega}g^{\mu\nu}\nabla_\mu\nabla_\nu\omega-\frac{(d-1)(d-4)}{\omega^2}g^{\mu\nu}\nabla_\mu\omega\nabla_\nu\omega\right)\,.
\end{align}
This means that the Shouten tensor defined for $d>2$ as
\begin{equation}\label{ShoutenArbitraryD}
    S_{\mu\nu}=\frac{2}{d-2}\left(R_{\mu\nu}-\frac{1}{2(d-1)}g_{\mu\nu}R\right)\,,
\end{equation}
transforms as
\begin{align}
    S'_{\mu\nu}&=S_{\mu\nu}-\frac{2}{\omega}\nabla_\mu\nabla_\nu\omega+\frac{4}{\omega^2}\nabla_\mu\omega\nabla_\nu\omega-\frac{g_{\mu\nu}}{\omega^2}g^{\rho\sigma}\nabla_\rho\omega\nabla_\sigma\omega  \,,
\end{align}
which is again independent of the dimension $d$.
On the other hand, the Weyl tensor is conformally invariant
\begin{equation}
    C'\du{\mu\nu\rho}{\sigma}=C\du{\mu\nu\sigma}{\sigma}\,,
\end{equation}
which can be checked by plugging the above transformation rules into the definition of the Weyl tensor. However, this invariance is of course dependent on the index position, since it is easily seen for example that
\begin{equation}
     C'\du{\mu\nu\rho\sigma}{}=g'_{\sigma\tau}C'\du{\mu\nu\rho}{\tau}=\omega^2g_{\sigma\tau}C\du{\mu\nu\rho}{\tau}=\omega^2C\du{\mu\nu\rho\sigma}{}\,.
\end{equation}
The conformal transformations outlined so far are limited to objects which are completely determined by the metric and hence their conformal transformation is entirely fixed. In order to talk about conformal transformations of arbitrary fields $\Phi$, the concept of conformal weight or dimension $s\in \mathbb{R}$ of the field has to be introduced, i.e., a field is said to possess conformal weight $s$ if
\begin{equation}
    \Phi'=\omega^s\, \Phi\,.
\end{equation}
An equation for the field $\Phi$ is then said to be conformally invariant if there exists a number $s$ such that $\Phi$ is a solution with respect to metric $g_{\mu\nu}$ if and only if $\Phi'$ is a solution with metric $g'_{\mu\nu}$. Note, however, that for a general tensor field the overall conformal weight depends on the index position, such that one defines
\begin{equation}
    \Phi'\phantom{}\ud{a\dots b}{c\dots d}=\omega^{s-N_u+N_l}\,\Phi\ud{a\dots b}{c\dots d}\,,
\end{equation}
where $N_u$ and $N_l$ are the number of upper and lower indices of the tensor component respectively. In this way, the conformal weight $s$ is fixed and in general corresponds to the physical dimension of the field. With these conventions, the metric, as well as the Levi-Civita tensor have dimension zero, such that indeed raising or lowering indices does not change the dimension of a field. For instance, in this language, the Weyl tensor has conformal weight $s=-2$.
Furthermore, Maxwell's equations
\begin{equation}
    g^{\mu\rho}\nabla_\rho F_{\mu\nu}=0\;,\quad\nabla_{[\mu}F_{\nu\rho]}=0\,,
\end{equation}
are conformally invariant in four dimensions (and only in $d=4$) for a conformal weight $s=-2$ of the field strength, while the conservation equation 
\begin{equation}
    \nabla_\mu T^{\mu\nu}\,,
\end{equation}
for a symmetric tensor $T^{\mu\nu}$ is conformally invariant if and only if $s=-d$ and its trace vanishes.

In the context of conformal compactifying the chosen spacetime, i.e., transforming $g_{\mu\nu}=\Omega^2\widetilde g_{\mu\nu}$, the inverse of the above transformations becomse relevant. Given the conformal factor $\Omega$ which in this sense is the inverse of $\omega$ used above, the transformation laws read
\begin{align}
    \widetilde R_{\mu\nu}=&R_{\mu\nu}+\frac{(d-2)}{\Omega}\nabla_\mu\nabla_\nu\Omega+\frac{g_{\mu\nu}}{\Omega}g^{\rho\sigma}\nabla_\rho\nabla_\sigma\Omega-\frac{(d-1)g_{\mu\nu}}{\Omega^2}g^{\rho\sigma}\nabla_\rho\Omega\nabla_\sigma\Omega \label{eq:RicciConformalAppI} \\
    \widetilde R=&\frac{1}{\Omega^2}\left(R+\frac{2(d-1)}{\Omega}g^{\mu\nu}\nabla_\mu\nabla_\nu\Omega-\frac{d(d-1)}{\Omega^2}g^{\mu\nu}\nabla_\mu\Omega\nabla_\nu\Omega\right) \label{eq:RicciConformalAppII}\\
    \widetilde S_{\mu\nu}=&S_{\mu\nu}+\frac{2}{\Omega}\nabla_\mu\nabla_\nu\Omega-\frac{g_{\mu\nu}}{\Omega^2}g^{\rho\sigma}\nabla_\rho\Omega\nabla_\sigma\Omega\,.\label{subeq:ShoutenTransfConformal}
\end{align}


%
%
%


\section{Decomposition of Strain Functions}
\label{app:decomp}
Energy and angular momentum flux as stated in Eq. \eqref{equ:fluxes} are both angle-dependent and thus possess a non-trivial decomposition in the bases of spin-weighted spherical harmonics. This choice of basis has been proven beneficial in many applications regarding gravitational waveforms, e.g., \cite{waveform_test_BL_I} and, in general, allows for a more compact denotation. In this appendix, an explicit decomposition of the terms appearing in the flux formulas \eqref{equ:fluxes} is provided. One starts by considering the energy flux, which contains contributions such as
\label{app:decomp}
\begin{align}
     |\dot h|^2 = \sum_{\ell, m} \alpha_{\ell m} Y_{\ell m}(\theta, \phi)\,,
\end{align}
which has spin-weight zero. It follows that 
    \begin{align}\label{equ:alphas}
        \alpha_{\ell m} = \sum_{\ell_1=2}^\infty \sum_{\ell_2=2}^\infty \sum_{|m_1|\leq \ell_1}\sum_{|m_2|\leq \ell_2} &(-1)^{m_2+m} \dot h_{\ell_1 m_1} \dot{\bar{h}}_{\ell_2m_2}\sqrt{\frac{(2\ell_1+1)(2\ell_2+1)(2\ell+1)}{4\pi }}
        \notag\\
        &\cdot\begin{pmatrix}
\ell_1 & \ell_2 & \ell\\
m_1 & -m_2 & -m
\end{pmatrix}\begin{pmatrix}
\ell_1 & \ell_2 & \ell\\
2 & -2 & 0
\end{pmatrix}\,,
    \end{align}
as previously also pointed out in Eq. \eqref{equ:alphas} of Section \ref{subsec:BL_analysis_pre}. 
For the decomposition of the angular momentum flux, two types of mixings have to be considered where one neglects the time derivative for simplicity as it does affect the spin-weighted basis. First, one has
\begin{align}
    \bar h \eth h = \sum_{\ell, m} \beta_{\ell, m} \,_{1}Y_{\ell m}(\theta, \phi)
\end{align}
which is of spin-weight one and where we find 
    \begin{align}\label{equ:alphas_0}
        \beta_{\ell m} = \sum_{\ell_1=2}^\infty \sum_{\ell_2=2}^\infty \sum_{|m_1|\leq \ell_1}\sum_{|m_2|\leq \ell_2} &(-1)^{m_1+m-1} \bar h_{\ell_1 m_1} h_{\ell_2m_2}\sqrt{\frac{(2\ell_1+1)(2\ell_2+1)(2\ell+1)}{4\pi(\ell_2+2)^{-1}(\ell_2-1)^{-1}}}\notag\\
        &\cdot\begin{pmatrix}
\ell_1 & \ell_2 & \ell\\
-m_1 & m_2 & -m
\end{pmatrix}\begin{pmatrix}
\ell_1 & \ell_2 & \ell\\
-2 & 1 & 1
\end{pmatrix},
    \end{align}
and similarly 
\begin{align}
     h \eth \bar h = \sum_{\ell, m} \Tilde{\beta}_{\ell, m} \,_{1}Y_{\ell m}(\theta, \phi)
\end{align}
with spin weight $1$ as well. Further,
    \begin{align}\label{equ:alphas_1}
        \Tilde{\beta}_{\ell m} = \sum_{\ell_1=2}^\infty \sum_{\ell_2=2}^\infty \sum_{|m_1|\leq \ell_1}\sum_{|m_2|\leq \ell_2} &(-1)^{m_2+m-1}  h_{\ell_1 m_1} \bar h_{\ell_2m_2}\sqrt{\frac{(2\ell_1+1)(2\ell_2+1)(2\ell+1)}{4\pi(\ell_2-2)^{-1}(\ell_2+3)^{-1}}}
        \notag\\
        &\cdot\begin{pmatrix}
\ell_1 & \ell_2 & \ell\\
m_1 & -m_2 & -m
\end{pmatrix}\begin{pmatrix}
\ell_1 & \ell_2 & \ell\\
2 & -3 & 1
\end{pmatrix}.
    \end{align}
The latter two coefficients cover all terms within the angular momentum flux \eqref{equ:fluxes}, as the four terms present in their differ only by a time derivative of $h_{\ell_1 m_1}$ or $h_{\ell_2 m_2}$.
\chapter{Supplementary Proofs} 
\label{AppendixB} 

\section{Ambiguity in the Definition of Newman-Penrose Scalars}
\label{app_sec:amb_NPS}
In the following, it is demonstrated that distinct contractions of the Weyl tensor with members of the tetrad lead to the same result, i.e., the same Newman-Penrose scalar. The Weyl curvature tensor in $4$ dimensions is a real traceless tensor with $10$ independent components. The NPS in which the tensor is decomposed are complex however. This implies that one needs to come up with an complexified version of the Weyl curvature that we then decompose into scalar contributions. The complexification can be easily achieved via
\begin{align}
    \widehat{C}_{\mu \nu \rho\sigma}= C_{\mu \nu \rho\sigma} - i C^\star_{\mu \nu \rho\sigma},&&\text{with} &&C^\star_{\mu \nu \rho\sigma} = \frac{1}{2}C\du{\mu \nu }{\alpha \beta}\varepsilon_{\rho\sigma \alpha \beta}\,,
\end{align}
where  $C^\star_{\mu \nu \rho\sigma} = \frac{1}{2}C\du{\mu \nu }{\alpha \beta}\varepsilon_{\rho\sigma \alpha \beta}$. The latter two tensors are decomposed in terms of a base of complex bivectors, $U_{\mu \nu },V_{\mu \nu }, W_{\mu \nu }$ defined as
\begin{align}
        U_{\mu \nu }=-2n_{[\mu}\Bar{m}_{\nu]},&&V_{\mu \nu } = 2\ell_{[\mu}m_{\nu]},&&W_{\mu \nu }=2n_{[\mu}\ell_{\nu]} + 2m_{[\mu}\Bar{m}_{\nu]}\,.
\end{align}
Naturally, the decomposition must preserve the tracelessness of the Weyl curvature tensor which equally implies the tracelessness of the complex version. The only admissible combinations of the base of complex bivectors must thus be traceless as well.

One starts with the simplest double combination and immediately finds that only $U_{\mu \nu }U_{\rho\sigma}$ and $ V_{\mu \nu }V_{\rho\sigma}$ are trace free. On the other hand, $W\ud{\mu}{\nu} W_{\mu\rho} = \ell_\nu n_\rho + \ell_\rho n_\nu - m_\nu\Bar{m}_\rho - m_\rho \Bar{m}_\nu$. This, however, can be compensated by $U\ud{a}{b}V_{ac}$ and $V\ud{a}{b}U_{ac}$ so that $U_{\mu \nu }V_{\rho\sigma}+V_{\mu \nu }U_{\rho\sigma}+W_{\mu \nu }W_{\rho\sigma}$ is trace free as well. Thus, one only needs two more independent combinations which are given by $U_{\mu \nu }W_{\rho\sigma}$ and $V_{\mu \nu }W_{\rho\sigma}$. Indeed, it is found that $ U_{\mu \nu }W_{\rho\sigma}+ W_{\mu \nu }U_{\rho\sigma}$ and $V_{\mu \nu }W_{\rho\sigma}+ W_{\mu \nu }V_{\rho\sigma}$ are both traceless as well. Thus, the complex Weyl tensor in four dimensions can be decomposed using the following basis:
$U_{\mu \nu }U_{\rho\sigma} \,, V_{\mu \nu }V_{\rho\sigma} \,, U_{\mu \nu }V_{\rho\sigma}+V_{\mu \nu }U_{\rho\sigma}+W_{\mu \nu }W_{\rho\sigma} \,, U_{\mu \nu }W_{\rho\sigma}+ W_{\mu \nu }U_{\rho\sigma} \,,  V_{\mu \nu }W_{\rho\sigma}+ W_{\mu \nu }V_{\rho\sigma} \,.$
Then, as they are independent and trace-free, it can be immediately conclude that the complex Weyl tensor can be written as
\begin{align}\label{equ:Ambig_Weyl}
    \frac{1}{2}\widehat{C}_{\mu \nu \rho\sigma} &= \lambda_0\left( U_{\mu \nu }U_{\rho\sigma}\right) + \lambda_1\left( V_{\mu \nu }V_{\rho\sigma}\right)+ \lambda_2\left(U_{\mu \nu }W_{\rho\sigma}+ W_{\mu \nu }U_{\rho\sigma} \right) \notag\\
    & + \lambda_3\left( V_{\mu \nu }W_{\rho\sigma} + W_{\mu \nu }V_{\rho\sigma}\right) + \lambda_4\left(U_{\mu \nu }V_{\rho\sigma} + V_{\mu \nu }U_{\rho\sigma} + W_{\mu \nu }W_{\rho\sigma} \right) \,.
\end{align}
Here, the constants $\lambda_i\in \mathbb{C}$ are a priori arbitrary and describe the 10 independent components of the complex Weyl tensor. Following the convention chosen by Newman and Penrose \cite{Newman_Penrose}, i.e., parameterizing the 10 independent components using 5 complex scalars, these can be identified with the Newman-Penrose scalars introduced in chapter~\ref{chap:asymptotics}.

Based on equation~\eqref{equ:Ambig_Weyl}, the individual scalars are projected out by contracting $\frac{1}{2}\widehat{C}_{\mu \nu \rho\sigma}$ with a suitable combination of tetrad vectors. Starting with $\lambda_0$, one finds that $U_{\mu \nu }V^{\mu \nu }=2$, i.e., $\lambda_0$ can be projected out using $V^{\mu \nu }V^{\rho\sigma}$. Explicit calculations yield
\begin{align}\label{equ:lambda0}
    \lambda_0 = \frac{1}{8} \widehat{C}_{\mu \nu \rho\sigma}V^{\mu \nu }V^{\rho\sigma} = C_{\mu \nu \rho\sigma}\ell^\mu m^\nu\ell^\rho m^\sigma \,.
\end{align}
To convert $\widehat {C}_{\mu \nu \rho\sigma}$ to $C_{\mu \nu \rho\sigma}$ it is helpful to write $\varepsilon_{\mu \nu \rho\sigma}= 4! i \ell_{[\mu}n_\nu m_\rho \Bar{m}_{\sigma]}$. Then it holds that
\begin{align}
    &\ \frac{1}{2} \widehat{C}_{\mu \nu \rho\sigma}V^{\mu \nu }V^{\rho\sigma} = \frac{1}{2}(C_{\mu \nu \rho\sigma} - \frac{i}{2}\varepsilon_{\rho\sigma \alpha\beta} C\du{\mu \nu }{\alpha \beta}) 2\ell^{[\mu}m^{\nu]} 2\ell^{[\rho}m^{\sigma]} \notag\\
    =&\ \frac{1}{2}C_{\mu \nu \rho\sigma} 2\ell^{[\mu}m^{\nu]} 2\ell^{[\rho}m^{\sigma]} + \frac{4!}{4} C\du{\mu \nu }{\alpha\beta} \ell_{[\rho}n_\sigma m_\alpha \Bar{m}_{\beta]} 2\ell^{[\mu}m^{\nu]} 2\ell^{[\rho}m^{\sigma]}\notag\\
    =&\ 2C_{\mu \nu \rho\sigma} \ell^\mu m^\nu  \ell^\rho m^\sigma + \frac{1}{2} C\du{\mu \nu }{\alpha \beta} 2\ell^{[\mu}m^{\nu]} 2\ell_{[\alpha}m_{\beta]} = 4 C_{\mu \nu \rho\sigma} \ell^\mu m^\nu \ell^\rho m^\sigma \,,
\end{align}
where the usual (cross-) normalization prescription as defined in chapter~\ref{chap:asymptotics} applies. Using the decomposition in bivectors on the other side of \eqref{equ:Ambig_Weyl} yields
\begin{align}
    \frac{1}{2} \widehat{C}_{\mu \nu \rho\sigma}V^{\mu \nu }V^{\rho\sigma}= \lambda_0 U_{\mu \nu }U_{\rho\sigma}V^{\mu \nu }V^{\rho\sigma} = 4 \lambda_0,
\end{align}
hence one finds \eqref{equ:lambda0}. Comparin equation \eqref{equ:lambda0} with the results of \cite{Newman_Penrose} implies that one can associate $\lambda_0\equiv \Psi_0$. Equally, one can project out $\lambda_1$ by using $U_{\mu \nu }U_{bc}$ such that 
\begin{align}
    \lambda_1 = \frac{1}{8} \widehat{C}_{\mu \nu \rho\sigma}U^{\mu \nu }U^{\rho\sigma} = C_{\mu \nu \rho\sigma}n^\mu\Bar{m}^\nu n^\rho \Bar{m}^\sigma,
\end{align}
and by comparing again to the convention of Newmann-Penrose, one finds $\lambda_1 \equiv \Psi_4$. The calculation is analogous in every step.

For $\lambda_2$ the procedure is less trivil as the right combination of bivectors picking out only terms proportional to $\lambda_2$ has to be determined first. Using again $U_{\mu \nu }V^{\mu \nu }=2$ and $W_{\mu \nu }W^{\mu \nu }=-4$, it is found that 
\begin{align}
    \frac{1}{2}\widehat{C}_{\mu \nu \rho\sigma}(V^{\mu \nu }W^{\rho\sigma}+ W^{\mu \nu }V^{\rho\sigma}) &= \lambda_2 (U_{\mu \nu }W_{\rho\sigma} +U_{\rho\sigma}W_{\mu \nu })(V^{\mu \nu }W^{\rho\sigma}+V^{\rho\sigma}W^{\mu \nu }) = -16 \lambda_2\notag\\
    &=\widehat{C}_{\mu \nu \rho\sigma}V^{\mu \nu }W^{\rho\sigma}
\end{align}
which can be use to show, by explicit contraction of $C^\star_{\mu \nu \rho\sigma}$ with the tetrad vectors, to result in
\begin{align}
    4 \lambda_2 &= -\frac{1}{2}\widehat{C}_{\mu \nu \rho\sigma}V^{\mu \nu }W^{\rho\sigma} = \frac{1}{2} (C_{\mu \nu \rho\sigma} - \frac{i}{2}\varepsilon_{\rho\sigma \alpha\beta}C\du{\mu \nu }{\alpha\beta}) 2\ell^{[\mu}m^{\nu]} (2n^{[\rho}\ell^{\sigma]} + 2m^{[\rho}\Bar{m}^{\sigma]}) \notag\\
    &= -\frac{1}{2} C_{\mu \nu \rho\sigma} 2\ell^{[\mu}m^{\nu]} (2n^{[\rho}\ell^{\sigma]} + 2m^{[\rho}\Bar{m}^{\sigma]}) + \frac{4!}{4} \ell_{[\rho}n_\sigma m_\alpha\Bar{m}_{\beta]} C\du{\mu \nu }{\alpha \beta} 2\ell^{[\mu}m^{\nu]} (2n^{[\rho}\ell^{\sigma]} + 2m^{[\rho}\Bar{m}^{\sigma]}) \notag\\
    &= \frac{1}{2} \left(4C_{\mu \nu \rho\sigma}(\ell^\mu m^\nu\ell^\rho n^\sigma + \ell^\mu m^\nu\Bar{m}^\rho m^\sigma) + 4C_{\mu \nu \rho\sigma}(\ell^\mu m^\nu \ell^\rho n^\sigma + \ell^\mu m^\nu \Bar{m}^\rho m^\sigma)\right)\notag\\
    &= 4C_{\mu \nu \rho\sigma}(\ell^\mu m^\nu \ell^\rho n^\sigma + \ell^\mu m^\nu \Bar{m}^\rho m^\sigma).
\end{align}
Hence,
\begin{align}\label{equ:theAbove}
    \lambda_2 = \frac{1}{2}C_{\mu \nu \rho\sigma}(\ell^\mu m^\nu\ell^\rho n^\sigma + \ell^\mu m^\nu\Bar{m}^\rho m^\sigma).
\end{align}
In this form $\lambda_2$ cannot be immediately be identified with a Newmann-Pensorse scalar by simple comparison to \cite{Newman_Penrose}. According to \cite{Newman_Penrose}, the first term resembles $\Psi_1$. Rephrasing the tracelessness condition as 
\begin{align}
    g^{\mu \rho}C_{\mu \nu \rho\sigma} = (-2n^{(\mu }\ell^{\rho)}+2m^{(\mu}\Bar{m}^{\rho)}) C_{\mu \nu \rho\sigma}=0 \,,
\end{align}
where 
\begin{align}
    g_{\mu \nu } = -2\ell_{(\mu} n_{\nu)} + 2m_{(\mu} \bar{m}_{\nu)}
\end{align}
is used, one further finds, exploiting the symmetries of the Weyl tensor,
\begin{align}\label{equ:important2.0}
    2n^{(\mu}\ell^{\rho)}C_{\mu \nu \rho\sigma} &= 2m^{(\mu}\Bar{m}^{\rho)}C_{\mu \nu \rho\sigma}\notag\\
    \xRightarrow\quad  n^\mu\ell^\rho(C_{\mu \nu \rho\sigma}+ C_{\rho \nu \mu \sigma}) &=\Bar{m}^\mu m^\rho(C_{\mu \nu \rho\sigma}+ C_{\rho \nu \mu \sigma}) \,.
\end{align}
The latter equation can be multiplied by $l^\nu m^\sigma$ and, in combination with $n^\mu\ell^\nu\ell^\rho m^\sigma C_{cbad}=\Bar{m}^\mu m^\rho \ell^\nu m^\sigma C_{\mu \nu \rho\sigma}=0$ (which can be shown via explicit calculation) this yields
\begin{align}
    n^\mu \ell^\rho \ell^\nu m^\sigma (C_{\mu \nu \rho\sigma} + C_{\rho \nu \mu \sigma}) &= \Bar{m}^\mu m^\nu \ell^\rho m^\sigma (C_{\mu \nu \rho\sigma} + C_{\rho \nu \mu \sigma})\notag\\
    \xRightarrow\quad n^\mu \ell^\rho \ell^\nu m^\sigma C_{\mu \nu \rho\sigma} &= \Bar{m}^\mu m^\rho \ell^\nu m^\sigma C_{\rho \nu \mu \sigma} \,.
\end{align}
Finally, using again the symmetries of the Weyl curvature tensor, 
\begin{align}
    \ell^\mu m^\nu \ell^\rho n^\sigma C_{\mu \nu \rho\sigma}= \ell^\mu m^\nu\Bar{m}^\rho m^\sigma C_{\mu \nu \rho\sigma}.
\end{align}
This means that 
\begin{align}
    \lambda_2 = C_{\mu \nu \rho\sigma}\ell^\mu n^\nu \ell^\rho m^\sigma , &&\text{or}&& \lambda_2 = C_{\mu \nu \rho\sigma}\ell^\mu m^\nu \Bar{m}^\rho m^\sigma  \,.
\end{align}
Based on the first equation in the latter, $\lambda_2\equiv \Psi_1$ \cite{Newman_Penrose}.

For $\lambda_3$ the calculations can be repeated step by step as above for $\lambda_2$. By contracting \eqref{equ:Ambig_Weyl} with $U^{\mu\nu }W^{\rho\sigma}+W^{\mu\nu }U^{\rho\sigma}$ one finds that
\begin{align}
    \lambda_3 = -\frac{1}{16}\widehat{C}_{\mu\nu \rho\sigma}U^{\mu\nu }W^{\rho\sigma} \,,
\end{align}
with $U_{\mu\nu }V^{\mu\nu }=0$, $W_{\mu\nu }W^{\mu\nu }=-4$. When calculating explicitly, $\lambda_3$ reads
\begin{align}\label{equ:Psi3lambda3}
    \lambda_3 = -\frac{1}{2}C_{\mu\nu \rho\sigma}(n^\mu \Bar{m}^\nu \ell^\rho n^\sigma  + n^\mu \Bar{m}^\nu \Bar{m}^\rho m^\sigma ) \,.
\end{align}
Based on the result for $\lambda_2$, the first intuition now would be to test, again, whether the two terms in \eqref{equ:Psi3lambda3} are equivalent. And indeed, returning to equation~\eqref{equ:important2.0} and multiplying by $n^\nu \Bar{m}^\sigma $, one obtains
\begin{align}
    n^\mu \Bar{m}^\nu \ell^\rho n^\sigma  C_{\mu\nu \rho\sigma} =  n^\mu \Bar{m}^\nu \Bar{m}^\rho m^\sigma C_{\mu\nu \rho\sigma} \,,
\end{align}
such that
\begin{align}
    \lambda_3 = C_{\mu\nu \rho\sigma}n^\mu \Bar{m}^\nu n^\rho \ell^\sigma , &&\text{or}&& \lambda_3 = C_{\mu\nu \rho\sigma}n^\mu \Bar{m}^\nu m^\rho  \Bar{m}^\sigma  \,.
\end{align}
The first equation identifies with $\Psi_3$, thus $\lambda_3 \equiv \Psi_3$.

Lastly, to compute $\lambda_4$ one contracts \eqref{equ:Psi3lambda3} with $U^{\mu\nu }V^{\rho\sigma}+ V^{\mu\nu }U^{\rho\sigma} + W^{\mu\nu }W^{\rho\sigma}$ and finds
\begin{align}
    24\lambda_4= \frac{1}{2} \widehat{C}_{\mu\nu \rho\sigma}\left(U^{\mu\nu }V^{\rho\sigma}+ V^{\mu\nu }U^{\rho\sigma} + W^{\mu\nu }W^{\rho\sigma}\right) \,.
\end{align}
Based on the latter, an explicit calculation results in 
\begin{align}\label{equ:important3.0}
    \lambda_4 = \frac{1}{6}\left(-2C_{\mu\nu \rho\sigma}n^\mu \Bar{m}^\nu \ell^\rho m^\sigma  + 2 C_{\mu\nu \rho\sigma}n^\mu \ell^\nu m^\rho \Bar{m}^\sigma  + C_{\mu\nu \rho\sigma}n^\mu \ell^\nu n^\rho \ell^\sigma  + C_{\mu\nu \rho\sigma}m^\mu \Bar{m}^\nu  m^\rho \Bar{m}^\sigma  \right) \,.
\end{align}
With the help of \eqref{equ:important2.0}, it can now be shown that
\begin{align}
    C_{\mu\nu \rho\sigma}n^\mu \ell^\rho \Bar{m}bm^\sigma  + C_{cbad}n^\mu \ell^\rho \Bar{m}^\nu m^\sigma  &= -C_{\mu\nu \rho\sigma}m^\mu \Bar{m}^\nu m^\rho \Bar{m}^\sigma  \,,\\
     C_{\mu\nu \rho\sigma}n^\mu \ell^\rho \Bar{m}bm^\sigma  + C_{cbad}n^\mu \Bar{m}^\nu \ell^\rho m^\sigma  &= -C_{\mu\nu \rho\sigma}n^\mu \ell^\nu n^\rho k^\sigma  \,.
\end{align}
The latter combine to $C_{\mu\nu \rho\sigma}n^\mu \ell^\nu n^\rho \ell^\sigma  = C_{\mu\nu \rho\sigma}m^\mu \Bar{m}^\nu m^\rho \Bar{m}^\sigma $. Inserting these relations into equation~\eqref{equ:important3.0} yields
\begin{align}\label{equ:important4.0}
    \lambda_4 = \frac{1}{3} \left( -C_{\mu\nu \rho\sigma}n^\mu \Bar{m}^\nu  \ell^\rho  m^\sigma  +C_{\mu\nu \rho\sigma}n^\mu \ell^\nu  m^\rho \Bar{m}^\sigma  +C_{\mu\nu \rho\sigma}n^\mu \ell^\nu n^\rho \ell^\sigma  \right) \,,
\end{align}
or 
\begin{align}\label{equ:important5.0}
    \lambda_4 = \frac{1}{3} \left( -C_{\mu\nu \rho\sigma}n^\mu \Bar{m}^\nu \ell^\rho m^\sigma  +C_{\mu\nu \rho\sigma}n^\mu \ell^\nu  m^\rho \Bar{m}d +C_{\mu\nu \rho\sigma}m^\mu \Bar{m}^\nu m^\rho \Bar{m}^\sigma  \right) \,.
\end{align}
With the use of the Bianchi identity $C_{\mu\nu \rho\sigma} + C_{\mu\rho\sigma \nu} + C_{\mu \sigma \nu \rho} = 0$ and the derived equivalences between contractions of the Weyl curvature tensor one can rewrite~\eqref{equ:important4.0} as 
\begin{align}
    \lambda_4 = -\frac{1}{2}(C_{\mu\nu \rho\sigma}\ell^\mu n^\nu m^\rho \Bar{m}^\sigma  + C_{\mu\nu \rho\sigma}\ell^\mu n^\nu n^\rho \ell^\sigma ) = \frac{1}{2}C_{\mu\nu \rho\sigma}\ell^\mu n^\nu (\ell^\rho n^\sigma -m^\rho \Bar{m}^\sigma ) \,.
\end{align}
Given $C_{\mu\nu \rho\sigma}n^\mu \ell^\nu n^\rho \ell^\sigma  = C_{\mu\nu \rho\sigma}m^\mu \Bar{m}^\nu m^\rho \Bar{m}^\sigma $, this is equivalent to 
\begin{align}
\lambda_4=\frac{1}{2}C_{\mu\nu \rho\sigma}m^\mu \Bar{m}^\nu (m^\rho \Bar{m}^\sigma - \ell^\rho n^\sigma ) \,.
\end{align}
Finally, replacing $C_{\mu\nu \rho\sigma}n^\mu \ell^\nu m^\rho \Bar{m}^\sigma  = -2 C_{\mu\nu \rho\sigma}n^\mu  \Bar{m}^\nu  \ell^\rho  m^\sigma  - C_{\mu\nu \rho\sigma}n^\mu \ell^\nu n^\rho \ell^\sigma $, one finds
\begin{align}
    \lambda_4 = C_{\mu\nu \rho\sigma}m^\nu  \ell^\mu n^\rho \Bar{m}^\sigma  \,,
\end{align}
which identifies $\lambda_4 = \Psi_2$ \cite{Newman_Penrose}.

In summary, writing the Weyl curvature tensor in terms of $5$ independent complex scalars, the ambiguity in the definition of the NPS boils down to
\begin{align}
    \Psi_0 &= C_{\mu\nu \rho\sigma}n^\mu \Bar{m}^\nu  n^\rho  \Bar{m}^\sigma  \\
    \Psi_1 &= C_{\mu\nu \rho\sigma}\ell^\mu m^\nu \Bar{m}^\rho m^\sigma  = C_{\mu\nu \rho\sigma}\ell^\mu n^\nu \ell^\rho m^\sigma  \,,\\
    \Psi_2 &= \frac{1}{2}C_{\mu\nu \rho\sigma}\ell^\mu n^\nu (\ell^\rho n^\sigma -m^\rho \Bar{m}^\sigma )= \frac{1}{2}C_{\mu\nu \rho\sigma}m^\mu \Bar{m}^\nu (m^\rho \Bar{m}^\sigma - \ell^\rho n^\sigma ) \notag\\ &= C_{\mu\nu \rho\sigma}m^\nu  \ell^\mu n^\rho \Bar{m}^\sigma ,\\
    \Psi_3 &= C_{\mu\nu \rho\sigma}n^\mu \Bar{m}^\nu n^\rho \ell^\sigma  = C_{\mu\nu \rho\sigma}n^\mu \Bar{m}^\nu m^\rho  \Bar{m}^\sigma ,\\
    \Psi_4 &= C_{\mu\nu \rho\sigma}n^\mu \Bar{m}^\nu  n^\rho  \Bar{m}^\sigma  \,,
\end{align} 
which concludes this demonstration.

%
%
%


\section{Transformation Behavior of the Symmetric Tensor $\rho_{\mu\nu}$}
\label{app:rho_conformal}

Consider a Bondi frame $(q_{\mu\nu},n^\mu)$ with constant $\R$ as well as the solution $\rho_{\mu\nu}=\frac{1}{2}q_{\mu\nu}\R$. When applying a conformal transformation of $\rho$ to a generic frame $(q'_{\mu\nu}=\omega q_{\mu\nu},n'^\mu=\omega^{-1}n^\mu, \lie_n\omega=0)$, one needs to ensure that $\rho'$ still satisfies the properties
\begin{equation}\label{PropertiesPrime}
    \rho'_{\mu\nu}n'^b=0\,,\quad \rho'_{\mu\nu}q'^{\mu\nu}=\R'\,,\quad \D'_{[\mu}\rho'_{\nu]\rho}=0\,.
\end{equation}
From $\rho'_{\mu\nu}q'^{\mu\nu}=\R'$ and the general transformation property of the Ricci-scalar $\R=q^{\mu\rho}q^{\nu\sigma}\R_{\mu\nu\rho\sigma}$ (see also Appendix~\ref{App:ConformalTransformations}), it follows that
\begin{equation}
    \rho'_{\mu\nu}q'^{\mu\nu}=\rho'_{\mu\nu}q^{\mu\nu}\frac{1}{\omega^2}=\frac{1}{\omega^2}\left(\R-\frac{1}{\omega}q^{\mu\nu}\D_\mu\D_\nu\omega+\frac{2}{\omega^2}q^{\mu\nu}\D_\mu\omega\D_\nu\omega\right)\,,
\end{equation}
which restricts $\rho'_{\mu\nu}$ to be of the form
\begin{equation}\label{rhoTransformationInter}
    \rho'_{\mu\nu}=\rho_{\mu\nu}+\frac{A}{\omega}q_{\mu\nu}q^{\rho\sigma}\D_\rho\D_\sigma\omega+\frac{B}{\omega}\D_\mu\D_\nu\omega+\frac{C}{\omega^2}q_{\mu\nu}q^{\rho\sigma}\D_\rho\omega\D_\sigma\omega+\frac{D}{\omega^2}\D_\mu\omega\D_\nu\omega\,,
\end{equation}
with
\begin{equation}\label{rhoCoeffTransformationInter}
    2A+B=-2\quad \text{and}\quad 2C+D=2\,,
\end{equation}
where $A$, $B$, $C$ and $D$ are real coefficients. Note that at this stage, the first primed equation $q'_{\mu\nu}n'^\nu=0$ is already satisfied, because of $q_{\mu\nu}n^\nu=0$ and the requirement to remain in a divergence free conformal frame $\lie_n\omega=n^\mu\D_\mu\omega=0$ which also implies that $n^\mu\D_\mu\D_\nu\omega=n^\nu\D_\mu\D_\nu\omega=0$. Hence, we have to use the last property $\D_{[\mu}\rho_{\nu]c}=0$ in order to determine the values of the coefficients. Writing out the primed version of this equation $\D'_{[\mu}\rho'_{\nu]\rho}\overset{!}{=}0$ while using \eqref{rhoTransformationInter} and \eqref{rhoCoeffTransformationInter} as well as the general transformation property of covariant derivatives 
\begin{equation}
    \D'_\mu\alpha_{\nu\rho}=\D_\mu\alpha_{\nu\rho}+C\du{\mu\nu}{\sigma}\alpha_{\sigma\rho}+C\du{\mu\rho}{\sigma}\alpha_{\nu\sigma}\,,
\end{equation}
where
\begin{equation}
    C\du{\mu\nu}{\rho}=-\frac{1}{\omega}q^{\rho\sigma}\left(q_{\nu\sigma}\D_\mu\omega+q_{\mu\sigma}\D_\nu\omega-q_{\mu\nu}\D_\sigma\omega\right)\,,
\end{equation}
one finds
\begin{align}
   \D'_{[\mu}\rho'_{\nu]\rho}=&\,\D_{[\mu}\rho'_{\nu]\rho}+\frac{1}{\omega}\left(\rho'_{\rho[\mu}\,\D_{\nu]}\omega+q_{\rho[\mu}\,\rho'_{\nu]\sigma}\,\D^\sigma\omega\right)\\
   =&\,\D_{[\mu}\rho_{\nu]\rho}+\frac{1}{\omega}\left(\rho_{\rho[\mu}\,\D_{\nu]}\omega+q_{\rho[\mu}\,\rho_{\nu]\sigma}\,\D^\sigma\omega\right)-\frac{A}{\omega}q^{\sigma\alpha}q_{\rho[\mu}\D_{\nu]}\D_\sigma\D_\alpha\omega \notag \\&-\frac{2(1+A)}{\omega} \D_{[\mu}\D_{\nu]}\D_\rho\omega
   -\frac{2(1+2A+C)}{\omega^2}q^{\sigma\alpha}q_{\rho[\mu}\D_{\nu]}\D_\sigma\omega\D_\alpha\omega\notag\\&+\frac{2(1+2A+C)}{\omega^2}\D_{[\mu}\omega\,\D_{\nu]}\D_\rho\omega\nonumber
   +\frac{3A}{\omega^2}q^{\sigma\alpha}q_{\rho[\alpha}\D_{\nu]}\omega\D_{\sigma}\D_{\alpha}\omega\notag\\&+\frac{2(1-C)}{\omega^2}\D_{[\mu}\D_{\nu]}\omega\D_\rho\omega+\frac{2(1+C)}{\omega^3}q^{\sigma\alpha}q_{\rho[\mu}D_{\nu]}\omega\D_\sigma\omega\D_\alpha\omega\,.
\end{align}
The last term results in the condition $C=-1$, while the term $-\frac{A}{\omega}q^{\sigma\alpha}q_{\rho[\mu}\D_{\nu]}\D_\sigma\D_\alpha\omega$ requires $A=0$. As concerns the remaining terms, note that $\D_{[\mu}\rho_{\nu]\rho}=0$. Moreover, the second to last term also vanishes 
\begin{equation*}
    \frac{4}{\omega^2}\D_{[\mu}\D_{\nu]}\omega\D_\rho\omega=0\,,
\end{equation*}
since $\D_{[\mu}\D_{\nu]}\omega=0$ as the covariant derivative $\D$ is torsion free. Moreover, using the definition of the Riemann tensor $\D_{[\mu}D_{\nu]}\alpha_\rho=\frac{1}{2}\R\du{\mu\nu\rho}{\sigma}\alpha_\sigma$ one can write
\begin{equation*}
    -\frac{2}{\omega} \D_{[\mu}\D_{\nu]}\D_\rho\omega=-\frac{1}{\omega}q^{\sigma\alpha}\R_{\mu\nu\rho\sigma}\D_\alpha\omega\,.
\end{equation*}
Gathering everything, one is therefore left with
    \begin{align}
        \D'_{[\mu}\rho'_{\nu]\rho}=&\,\frac{1}{\omega}\left(\rho_{\rho[\mu}\,\D_{\nu]}\omega+q_{\rho[\mu}\,\rho_{\nu]\sigma}\,\D^\sigma\omega\right)-\frac{1}{\omega}q^{\sigma\alpha}\R_{\mu\nu\rho\sigma}\D_\alpha\omega\,.
    \end{align}
At this stage, the reader is reminded that since the Riemann tensor $\R_{\mu\nu\rho\sigma}$ on $\scrip$ lives in two dimensions orthogonal to $n^\mu$ it is completely determined by the scalar curvature and can be written as\footnote{This follows from the Biancci identities which dictate, that the Riemann tensor in two dimensions only has one independent component.}
\begin{equation}
    \R_{\mu\nu\rho\sigma}=\R \,q_{\mu[\rho}\,q_{\sigma]\nu}\,,
\end{equation}
while using the solution in the Bondi frame $\rho_{\mu\nu}=\frac{1}{2}q_{\mu\nu}\R$ gives 
\begin{align}
    \D'_{[\mu}\rho'_{\nu]\rho}=q_{\rho[\mu}\,\D_{\nu]}\omega\,\R-q_{\rho[\mu}\,\D_{\nu]}\omega\,\R=0\,.
\end{align}
Thus, plugging in $A=0$, $B=-2$, $C=-1$, $D=4$ into \eqref{rhoTransformationInter}, required conformal transformation property is retrieved, i.e., 
\begin{equation}\label{rhoTransformation}
    \rho'_{ab}=\rho_{ab}-\frac{2}{\omega}\D_a\D_b\omega-\frac{1}{\omega^2}q_{ab}q^{cd}\D_c\omega\D_d\omega+\frac{4}{\omega^2}\D_a\omega\D_b\omega\,,
\end{equation}
which is needed in order to satisfy \eqref{PropertiesPrime}.

%
%
%


\section{Derivation of the Newman-Penrose Scalars in Terms of the Shear Tensor}
\label{app:NPS_and_Shear}
The relation between physical NPS and shear tensor (and derivatives thereof), i.e., equations \eqref{equ:NPS_in_shear_I}-\eqref{equ:NPS_in_shear_II}, are derived. One start by acknowledging that, as derived in Eq. \eqref{equ:defender2025}\footnote{Note that here only two indices are pulled back to $\scrip$.},
 \begin{equation}\label{RelWeylShouten75}
        \newpb{K_{\mu\nu}}{}^{\rho\sigma}\,n_\sigma |_{\scrip}=\D_{[\nu}\Ss\du{\mu]}{\rho}|_{\scrip}\,,
    \end{equation}
or, equivalently, written in terms of Bondi news tensor
\begin{equation}\label{RelWeylShoutenII75}
        \newpb{K_{\mu\nu\rho}}{}_{\sigma}\,n^\sigma|_{\scrip}=\D_{[\nu}N\du{\mu]\rho} \,|_{\scrip}\,.
\end{equation}
If contracted with another $n^\mu$, this yields
\begin{equation}\label{RelWeylShoutenIII75}
    \newpb{K_{\mu\nu\rho\sigma}}\,n^\nu n^\sigma|_{\scrip} = \frac{1}{2}\lie_n N\du{\mu\rho}{}|_{\scrip} = \frac{1}{2}n^\nu\D_{\nu}N\du{\mu\rho}{}|_{\scrip}\,,
\end{equation}
within a divergence free conformal frame. Using the NPS as defined in \eqref{eq:DefNPGR}\footnote{In the definition of the NPS, all indices are contracted and thus a pullback does not act here. Note however that when defining the scalars at null infinity, the bulk Weyl tensor needs to be pulled back as, at $\scrip$, it is only contracted with quantities tangent to this boundary. At least in most cases, as is elaborated below. In section \ref{subsec:NP_scalars_and_peeling}, the NPS are defined in the bulk and their limit to $\scrip$ is computed. Because of the peeling, the results at $\scrip$ agree and can be related to the shear.} as well as the definition of the Bondi news tensor, $N_{\mu\nu}=2\left(N^\circ m^\mu m^\nu+\bar{N}^\circ \bar{m}^\mu\bar{m}^\nu\right)$, relating to the shear as \eqref{eq:bondiII}
\begin{equation}\label{eq:bondiIIApp}
    N^\circ=-\partial_u\bar \sigma^\circ:= - \dot{\bar{\sigma}}^\circ\,,
\end{equation}
the derivation of the scalars is straightforward.

\paragraph{$\Psi_4^\circ$:} 
From the definition~\eqref{eq:DefNPGR}, $\Psi_4^\circ$ is obtained by contracting the asymptotic Wely tensor by transverse vectors to $\scrip$ only, such that in this case one can simply use the pulled back equation \eqref{RelWeylShoutenII75} and contract it with $\bar{m}^{\mu} n^{\nu}  \bar{m}^{\rho}$ in order to obtain $\Psi_4^\circ$. The desired result then follows upon expanding the right-hand side of \eqref{RelWeylShoutenIII75} and expressing it in terms of the asymptotic shear, i.e., 
\begin{align}\label{eq:IntermediatePsi4Rel}
   \Psi^\circ_4= \bar{m}^{\mu} n^{\nu}  \bar{m}^{\rho}\,\D_{[\nu}N\du{\mu]\rho}{}=&\,\bar{m}^{\mu} n^{\nu}  \bar{m}^{\rho}\,\D_\nu\left(N^\circ m_\mu m_\rho+\bar{N}^\circ \bar{m}_\mu\bar{m}_\rho\right)\notag\\
    =&\,\bar{m}^{\mu} n^{\nu}  \bar{m}^{\rho}\left(m_\mu m_\rho\D_\nu N^\circ+N^\circ m_\mu\D_\nu m_\rho+N^\circ m_\rho\D_\nu m_\mu\right)\notag\\
    =&\, n^{\nu} \D_\nu N^\circ+2N^\circ\bar{m}^{\rho}n^{\nu}\D_\nu m_\rho\,,
\end{align}
where the (cross-)normalization relations between the tetrad vectors are used. Moreover, note that in a divergence free conformal frame it holds that
\begin{align}
    \lie_nm^\mu =0= n^\nu\D_\nu m^\mu-m^\mu\D_\nu n^\nu=n^\nu\D_\nu m^\mu\,.
\end{align}
Hence, the last term in \eqref{eq:IntermediatePsi4Rel} vanishes and one is left with
\begin{align}\label{eq:IntermediatePsi4RelII}
    \bar{m}^{\mu} n^{\nu}  \bar{m}^{\rho}\,\D_{[\nu}N\du{\mu]\rho}{}= n^{\nu} \D_\nu N^\circ=\dot{N}^\circ=-\ddot{\bar{\sigma}}^\circ\,.
\end{align}
Thus, 
\begin{equation}
    {\Psi}^\circ_4=-\ddot{\bar{\sigma}}^\circ\,.
\end{equation}
Note that this relation holds in any divergence free conformal frame on $\scrip$ for a tetrad basis constructed by Lie-dragging the vector $m^a$ along the null normal $n^\mu$. For the rest of this exercise, however, the treatment is restrict to a Bondi frame for simplicity. For more general expressions of the NP scalars, the reader is referred to \cite{GomezLopez:2017}.

\paragraph{$\Psi_3^\circ$:} In this case the first equation in \eqref{eq:DefNPGR}, $\Psi_3^\circ$ involves the non-transverse tetrad vector $\ell^a$ such that one can no longer simply use \eqref{RelWeylShoutenII75}, as a contraction of $\ell^a$ with a pulled back index is not well-defined. One should therefore either use \eqref{RelWeylShouten75}\footnote{The pullback is defined as a map between co-tangent spaces, hence only well-defined for covariant indices. For mixed tensors, the upper indices are pushed forward.} or alternatively use the second definition of $\Psi_3^\circ$ in \eqref{eq:DefNPGR} together with \eqref{RelWeylShoutenII75}. The latter option is easier and immediately yields
\begin{align}\label{eq:IntermediatePsi3Rel}
    \Psi_3^\circ=\bar{m}^{\mu} m^{\nu}  \bar{m}^{\rho}\,\D_{[\nu}N\du{\mu]\rho}{}=&\,\frac{1}{2}\bar{m}^{\mu} m^{\nu}  \bar{m}^{\rho}\D_{\nu}N\du{\mu\rho}{}-\frac{1}{2}\bar{m}^{\mu} m^{\nu}  \bar{m}^{\rho}\D_{\mu}N\du{\nu\rho}{}\,.
\end{align}
At this point it is instructive to remind oneself of the action of the angular derivative in a Bondi frame, $\eth$, on a function of spin-weight $s$,
\begin{equation}
f_s=T_{\mu_1...\mu_p\nu_1..\nu_q}m^{\mu_1}...m^{\mu_p}\bar{m}^{\nu_1}...\bar{m}^{\nu_q}\,,\quad\text{with}\quad p-q=s\,,
\end{equation}
, i.e., 
\begin{equation}
    \eth{f}_s=m^\mu P^{\mu_1...\mu_p \nu_1..\nu_q}\left(\D_\mu T_{\mu_1...\mu_p \nu_1..\nu_q}\right)\,.
\end{equation}
where
\begin{equation}
    P^{\mu_1\cdots \mu_p \nu_1\cdots \nu_{q}} := m^{\mu_1}\cdots m^{\mu_p}\bar{m}^{\nu_1}\cdots \bar{m}^{\nu_q}.
\end{equation}
This means in particular that for $2N^\circ=N_{\mu\nu}\bar{m}^\mu\bar{m}^\nu$ one finds $p=0$, while $q=2$. Thus $N^\circ$ is of spin-weight $s=-2$ and
\begin{equation}
   \eth N^\circ=\frac{1}{2}m^\mu\bar{m}^\nu\bar{m}^\rho\D_\mu N_{\nu\rho}\,,
\end{equation}
which corresponds precisely to the first term in \eqref{eq:IntermediatePsi3Rel}. The second term on the other hand vanishes. To see this, expand the Bondi News tensor once more
\begin{align}\label{eq:IntermediatePsi3Rel2}
    -\frac{1}{2}\bar{m}^{\mu} m^{\nu}  \bar{m}^{\rho}\D_{\mu}N\du{\nu\rho}{}=&-\bar{m}^{\mu} m^{\nu}  \bar{m}^{\rho}\D_{\mu}\left(N^\circ m_\nu m_\rho+\bar{N}^\circ\bar{m}_\nu\bar{m}_\rho\right)\notag\\
    =&-\bar{m}^{\mu}\left(N^\circ m^\nu \D_\mu m_\nu+\bar{N}^\circ\bar{m}^{\rho}\D_\mu\bar{m}_\rho\right)\,.
\end{align}
Now from equation (1.54) in \cite{Our_Review} it follows that
\begin{equation}
    \bar{m}^\mu\D_\mu m_\nu\propto m_\nu\quad\text{and}\quad \bar{m}^\mu\D_\mu\bar m_\nu\propto \bar m_\nu\,.
\end{equation}
Applying the standard (cross-)normalization relations of the tetrad vectors, it thus follows that \eqref{eq:IntermediatePsi3Rel2} vanishes and thus
\begin{equation}\label{equ:result_Psi_3}
    \Psi^\circ_3=\eth N^\circ=-\eth \dot{\bar{\sigma}}^\circ\,.
\end{equation}

\paragraph{$\Psi_2^\circ$:} Looking at the second equation for $\Psi^\circ_2$ in \eqref{eq:DefNPGR}, one finds that only the second term contains a null tetrad vector $n^\mu$ and therefore one can only draw a connection with the shear with respect to this term, as the fundamental connecting equation \eqref{RelWeylShouten75} necessarily requires at least one contraction with $n^\mu$. It turns out that this second term is the imaginary part of $\Psi^\circ_2$ which is easily verified
\begin{equation}
    2i\Im{\Psi^\circ_2}=\Psi^\circ_2-\Bar\Psi^\circ_2=-\lim_{\Omega\rightarrow0}\frac{1}{2}K_{\mu\nu\rho\sigma}\,m^{\mu} \bar m^{\nu}\ell^\rho n^\sigma\,.
\end{equation}
Now, there is still a contraction of the asymptotic Weyl tensor with the null tetrad vector $\ell^\mu$ such that $K_{\mu\nu\rho\sigma}n^\sigma$ cannot just be pulled back as before. Hence, one inevitably has to work with \eqref{RelWeylShouten75}, i.e., starting with 
\begin{equation}\label{equ:unstoppable}
    \frac{1}{2}\newpb{K_{\mu\nu}}{}^{\rho\sigma}\,m^{\mu} \bar m^{\nu}\ell_\rho n_\sigma|_{\scrip} =  m^{\mu} \bar m^{\nu}\ell_\rho\D_{[\nu}\Ss\du{\mu]}{\rho}|_{\scrip}\,.
\end{equation}
To relate the right-hand side of Eq. \eqref{equ:unstoppable} to the shear tensor, one makes use of Eq. \eqref{equ:kack_equation}. Taking the derivative and contracting both sides with $m^\mu, \bar m^\nu$ twice, one finds an equation for $m^{\mu} \bar m^{\nu}\ell_\rho\D_{[\nu}\Ss\du{\mu]}{\rho}|_{\scrip}$ in terms of $\sigma_{\mu\nu}$ and contractions with $m^\mu, \bar m^\nu$. 
By means of tedious calculations that would surpass the intended scope of even this thesis, in the end, one obtains
\begin{equation}
    2i\Im{\Psi^\circ_2}=\dot \sigma^\circ\bar\sigma^\circ-\sigma^\circ\dot{\bar{\sigma}}^\circ+\bar{\eth}^2\sigma^\circ-\eth^2\bar\sigma^\circ\,,
\end{equation}
implying that the imaginary part of the second NPS can be expressed fully in terms of the shear tensor. Note that this final expression is again only valid in a Bondi frame. Moreover, it is no surprise that the real part $\Re{\Psi^\circ_2}$ has no similar connection to the shear as it contains the Coulombic information about the mass of the space-time which is not radiative in nature. Finally, note that, as an exercise $\Psi^\circ_3$ can be computed computed using the ansatz \eqref{RelWeylShouten75} as well. Here, starting from 
\begin{align}
 \Psi_3^\circ = \newpb{K_{\mu\nu}}^{\rho\sigma}\bar m^\mu n^\nu \ell_\rho n_\sigma = \bar m^\mu n^\nu \ell_\rho \D_{[\nu}\Ss_{\mu]}{}^\rho\,,
\end{align}
one applies a similar strategy, contracting the derivative of Eq. \eqref{equ:kack_equation} with $n^\mu, m^\nu$ and twice with $\bar m ^\mu$. After an equally tedious computation, one arrives at the result \eqref{equ:result_Psi_3}.

\chapter{Additional Plots} 

\label{AppendixC} 

\section{Gravitational Wave Echo from Quantized Black Holes and Exotic Compact Objects}
\label{app:ECHO}

\begin{figure}
	\centering
	\includegraphics[width=0.8\linewidth]{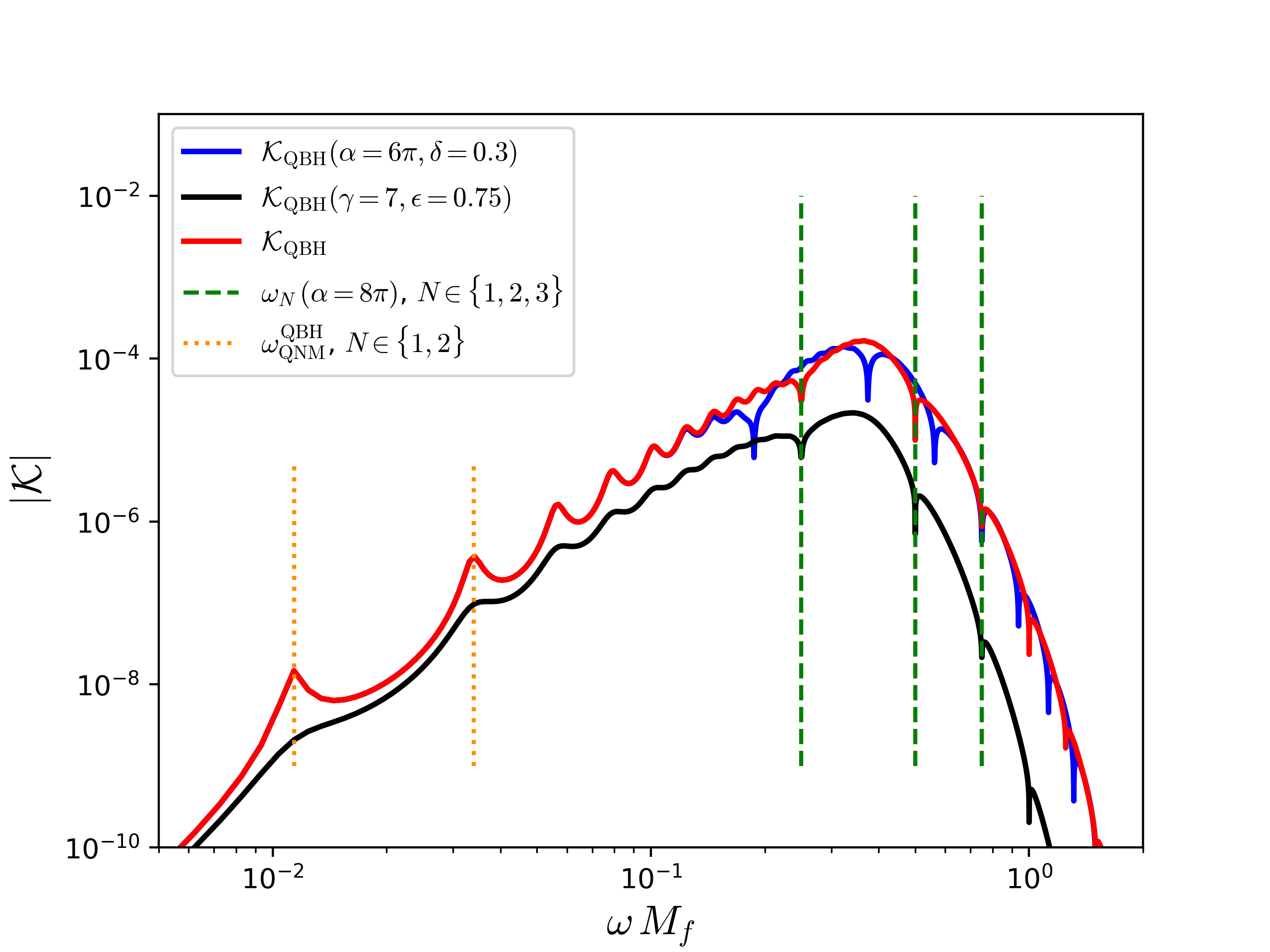}
	\caption{Transfer function for the baseline reflectivity parameter $\alpha = 8 \pi,\beta = 10^{-15}, \gamma = 4, \delta = 0.2, \epsilon=1$ \cite{Other_features_VIII}. Variations of the these parameters are indicated.}
	\label{fig:TRANSFER}
\end{figure}

An additional example of the transfer function for the QBH is depicted in Fig. \ref{fig:TRANSFER}. In the latter, the QBH's QNMs as well as the characteristic BH frequencies $\omega_N$ are marked. The latter are dictated by $\RQBH$ in the numerator of \eqref{equ:sum_echo} (which is equal to Eq. \ref{equ:trans_func_sum}), appearing as zeros. Note that the poles in Fig. \ref{fig:TRANSFER} do not correspond exactly to the QNMs but to the resonances of the system described by Fig. \ref{fig:Sketch_intuition} of Section \ref{sec:quantum_BH}. The QNMs reside very close to these resonances in frequency space.

\begin{figure}\centering
\includegraphics[width=0.7\columnwidth]{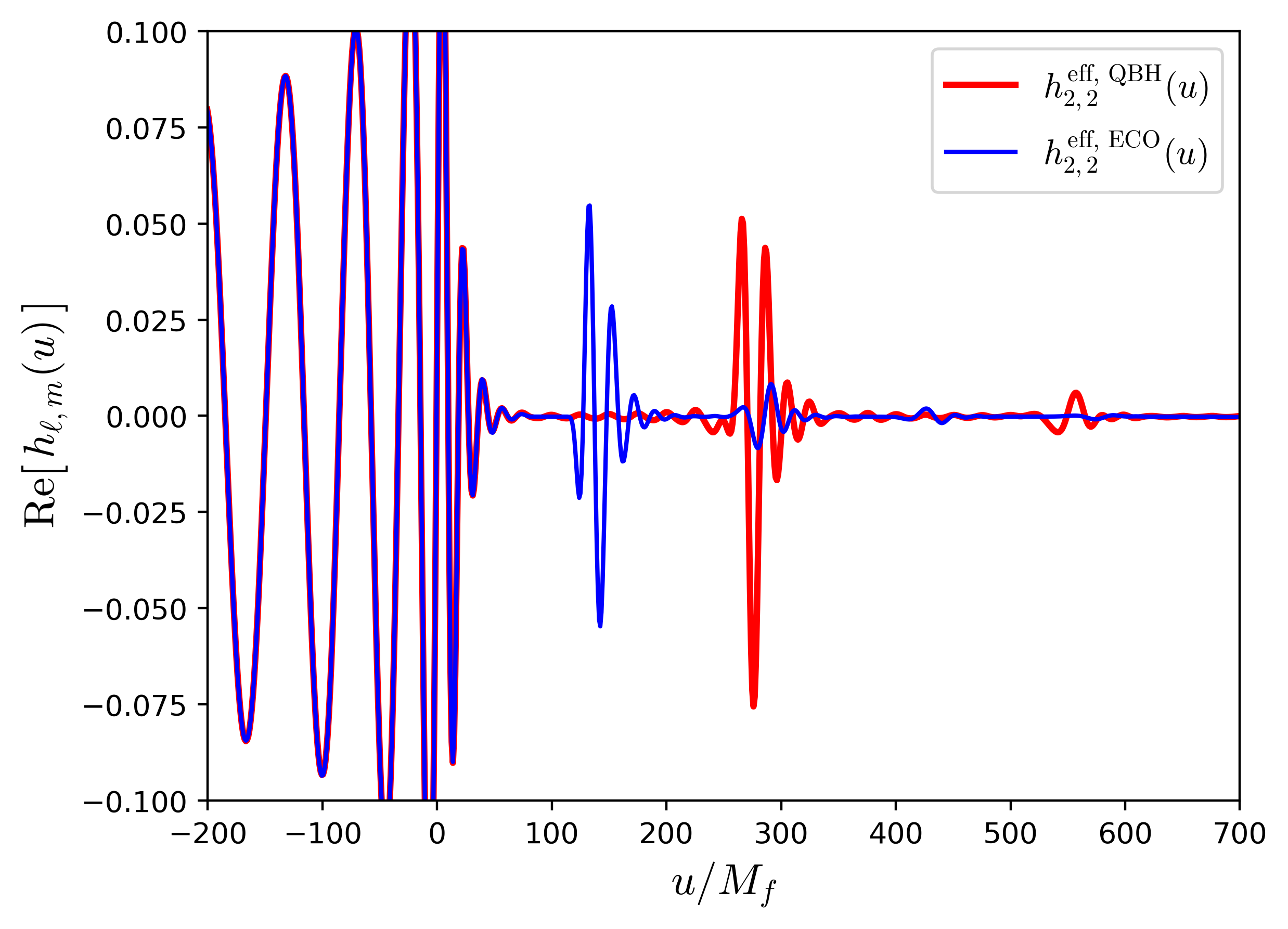} 
\includegraphics[width=0.7\columnwidth]{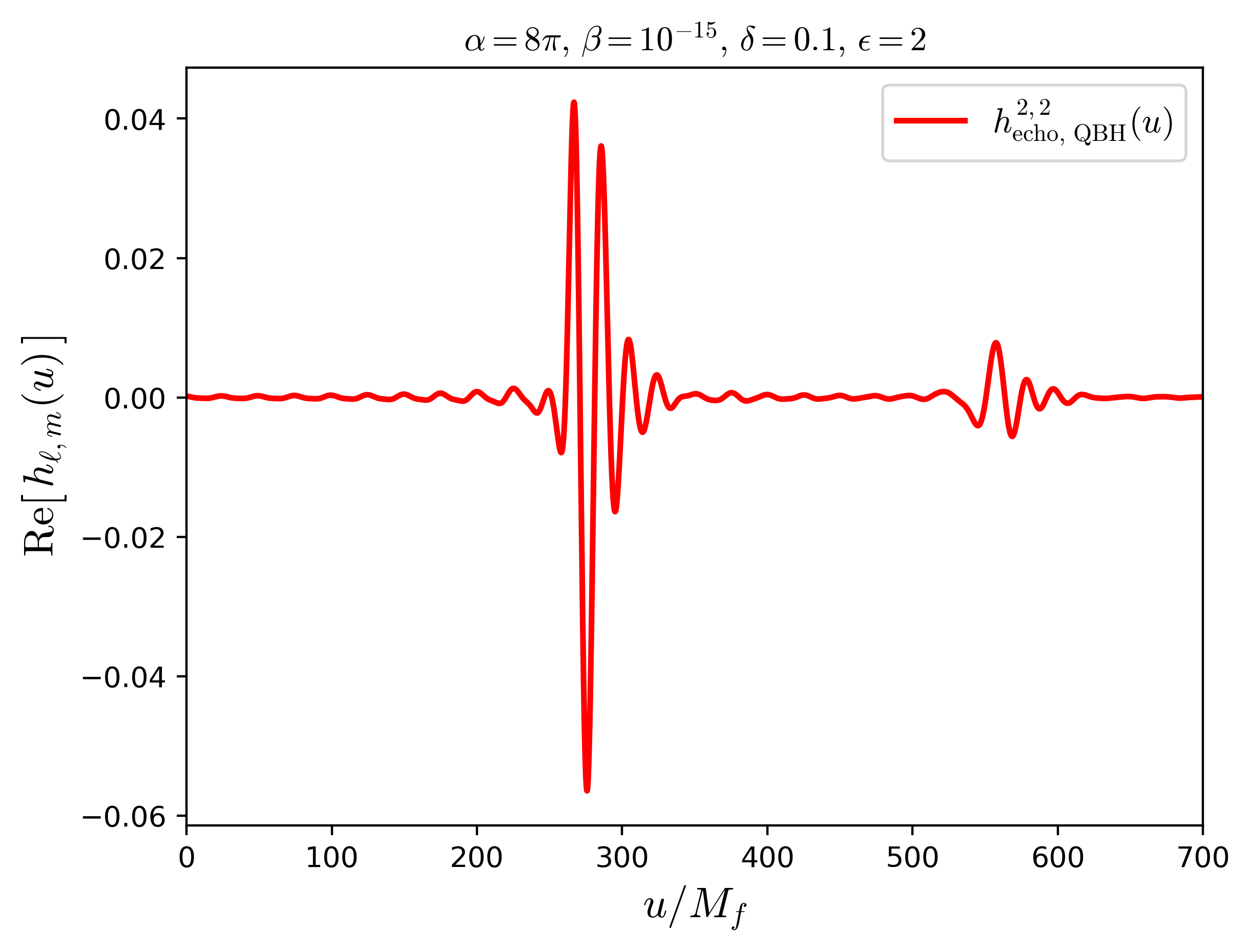}   
     \caption{Full waveform $\hfull=\hecho+\hnorm$ (top) for ECO and QBH simulated for event \textit{SXS:BBH:0207}. For the ECO, we choose $T_\T{QH}=2T_\T{H}$ and $\gamma=10^{-15}$. For the QBH, the parameter choice is displayed on top of the bottom plot. The latter displays the isolated $\hecho$ from the upper plot.}
       \label{fig:waveform_QBH}
\end{figure}

The QBH echo is exemplarily shown for different parameter configurations in Fig. \ref{fig:echo_QBH} below. The small noise-like oscillations are due to numerical issues with handling the transfer functions poles, representing the resonances of the QBH. For the results of Section \ref{sec:Paper_Mem}, these oscillations are irrelevant. To compare the echo against the full waveform consider Fig. \ref{fig:waveform_QBH}.

\begin{figure}[!t]
  \centering
  \begin{subfigure}[b]{0.45\textwidth}
    \centering
    \includegraphics[width=\linewidth]{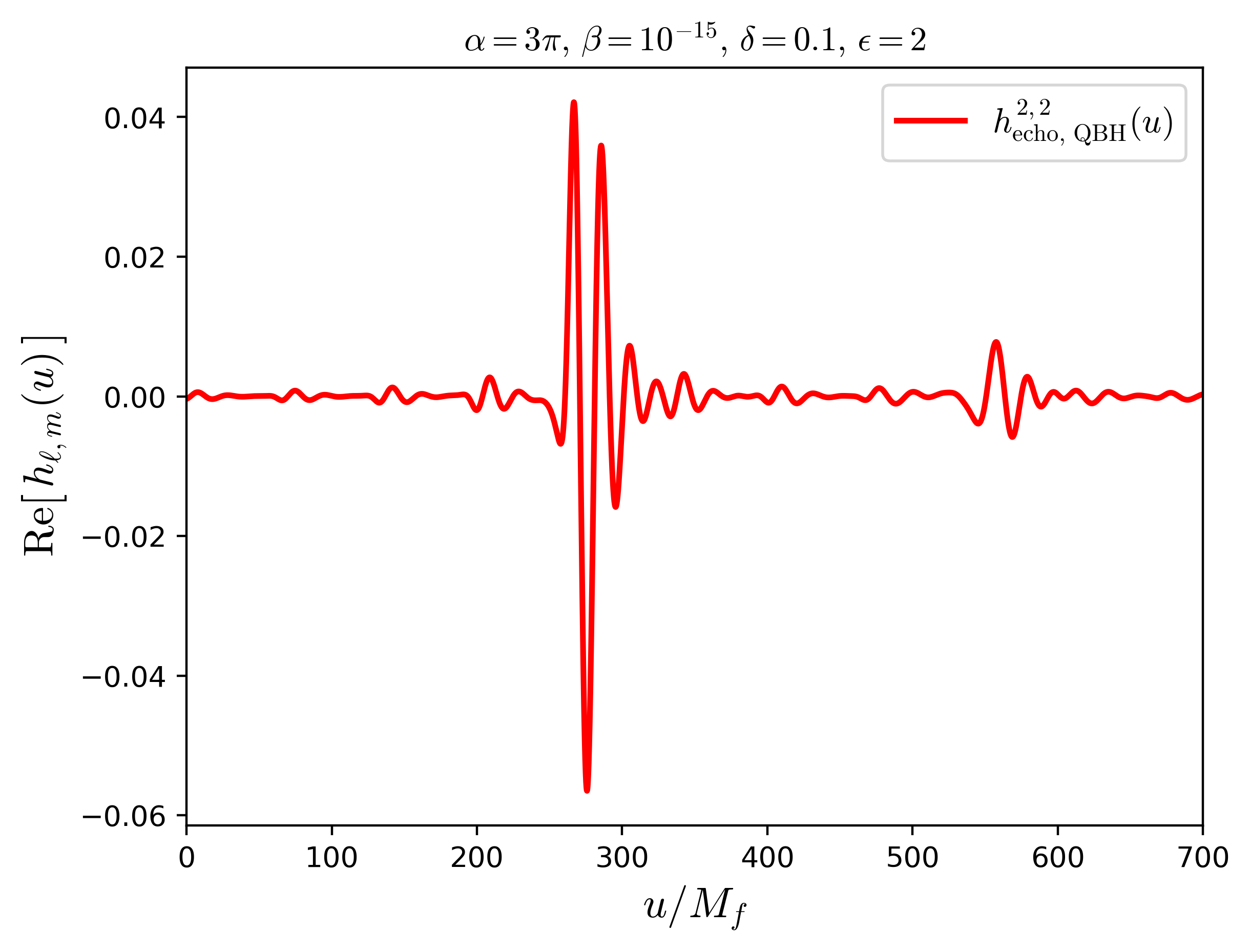}

    \label{fig:sub1}
  \end{subfigure}
  \hfill
  \begin{subfigure}[b]{0.45\textwidth}
    \centering
    \includegraphics[width=\linewidth]{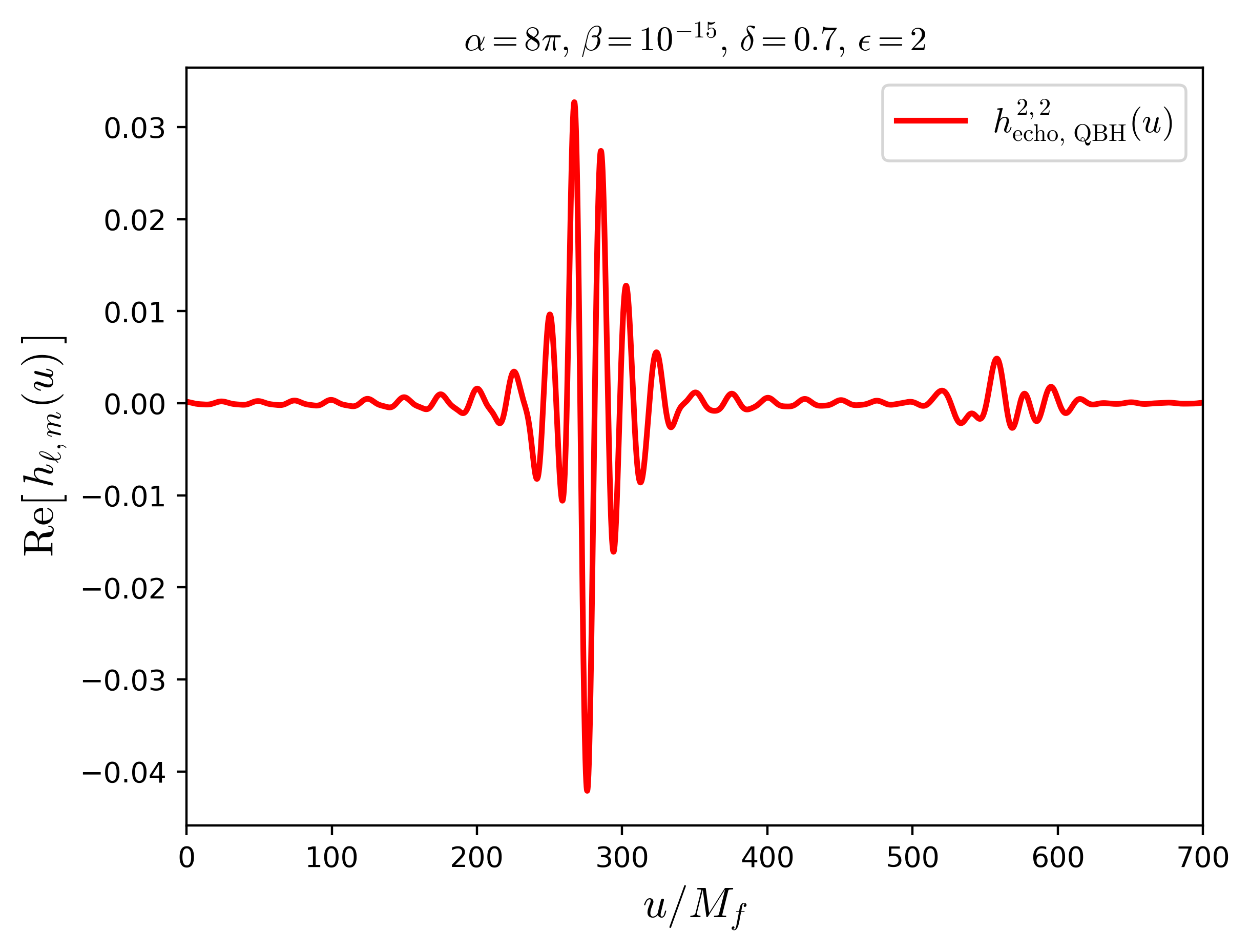}

    \label{fig:sub2}
  \end{subfigure}

  \vspace{0.5cm}

  \begin{subfigure}[b]{0.45\textwidth}
    \centering
    \includegraphics[width=\linewidth]{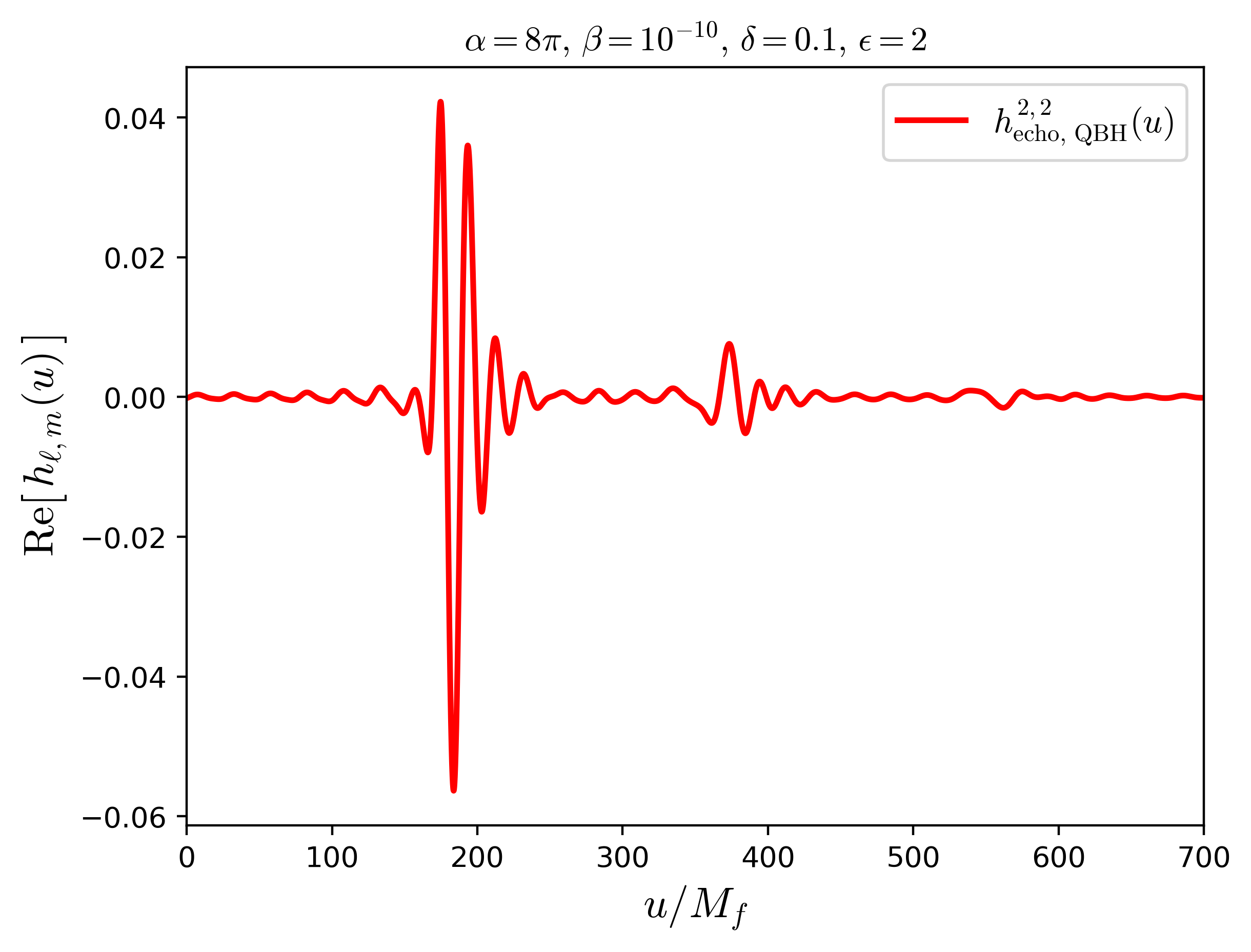}

    \label{fig:sub3}
  \end{subfigure}
  \hfill
  \begin{subfigure}[b]{0.45\textwidth}
    \centering
    \includegraphics[width=\linewidth]{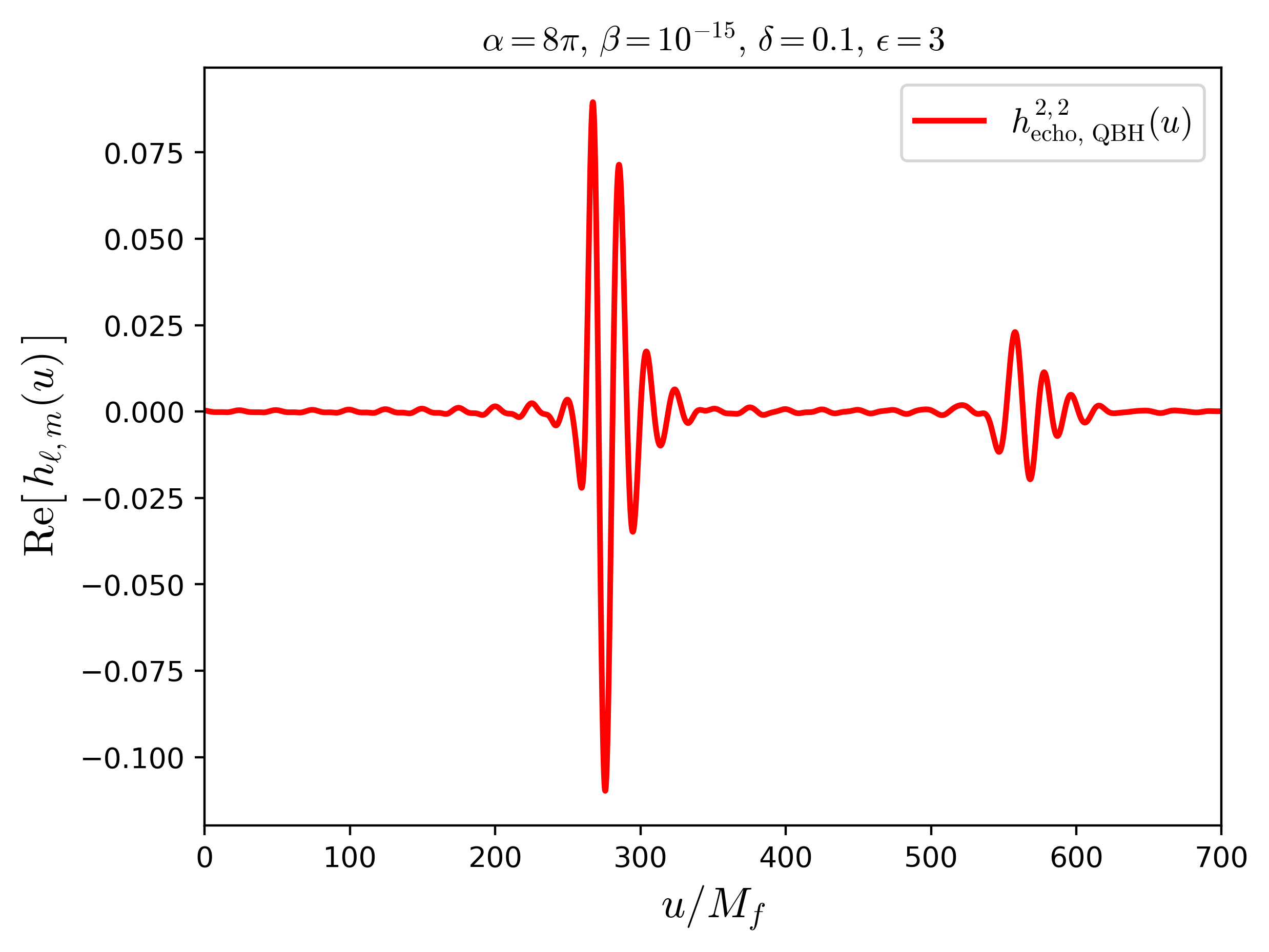}
    \label{fig:sub4}
  \end{subfigure}

  \caption{Pure echo strain $h_{2,2}^\T{echo}$ for the QBH computed using different model parameter and \textit{SXS:BBH:0207}.}
  \label{fig:echo_QBH}
\end{figure}


\end{document}